\documentclass[preprints,article,accept,moreauthors,pdftex]{Definitions/mdpi}
\usepackage{mathrsfs} 
\newcommand{\norm}[1]{\lVert #1 \rVert}
\firstpage{1} 
\pubvolume{1}
\issuenum{1}
\articlenumber{0}
\pubyear{2022}
\copyrightyear{2022}
\datereceived{10 November 2022} 
\dateaccepted{21 December 2022} 
\datepublished{} 
\hreflink{https://doi.org/} % If needed use \linebreak
\usepackage{bm}

\Title{The ngEHT Analysis Challenges}

\TitleCitation{The ngEHT Analysis Challenges}

\Author{Freek Roelofs$^{1,2,}$*, Lindy Blackburn$^{1,2}$, Greg Lindahl$^{1}$, Sheperd S. Doeleman$^{1,2}$, Michael D. Johnson$^{1,2}$, Philipp Arras$^{3,4}$, Koushik Chatterjee$^{1,2}$, Razieh Emami$^{1}$, Christian Fromm$^{5,6,7}$, Antonio Fuentes$^{8}$, Jakob Knollm\"uller$^{9,4}$, Nikita Kosogorov$^{10,11}$, Hendrik M\"uller$^{7}$, Nimesh Patel$^{1}$, Alexander Raymond$^{1,2}$, Paul Tiede$^{1,2}$, Thalia Traianou$^{8}$, Justin Vega$^{1,12}$}

\AuthorNames{Firstname Lastname, Firstname Lastname and Firstname Lastname}

\AuthorCitation{Roelofs, F.; Blackburn, L.; Lindahl, G.; Doeleman, S.S.; Johnson, M.D.; Arras, P.; Chatterjee, K.; Emami, R.; Fromm, C.; Fuentes, A.; Knollm\"uller, J.; Kosogorov, N.; M\"uller, H.; Patel, N.; Raymond, A.; Tiede, P.; Traianou, T.; Vega, J. }
\address{%
$^{1}$ \quad Center for Astrophysics $|$ Harvard \& Smithsonian,  60 Garden Street, Cambridge, MA 02138, USA\\
$^{2}$ \quad Black Hole Initiative, Harvard University, 20 Garden Street, Cambridge, MA 02138, USA\\
$^{3}$ \quad Technical University Munich (TUM), Boltzmannstr. 3, 85748 Garching, Germany\\
$^{4}$ \quad Max-Planck Institute for Astrophysics, Karl-Schwarzschild-Str. 1, 85748 Garching, Germany\\
$^{5}$ \quad Institut f\"ur Theoretische Physik und Astrophysik, Universit\"at W\"urzburg, Emil-Fischer-Str. 31, D-97074 W\"urzburg, Germany\\
$^{6}$ \quad Institut f\"ur Theoretische Physik, Goethe-Universit\"at Frankfurt, Max-von-Laue-Stra\ss e 1, D-60438 Frankfurt am Main, Germany\\
$^{7}$ \quad Max-Planck-Institut f\"ur Radioastronomie, Auf dem H\"ugel 69, D-53121 Bonn, Germany\\
$^{8}$ \quad Instituto de Astrof\'isica de Andaluc\'ia-CSIC, Glorieta de la Astronom\'ia s/n, E-18008 Granada, Spain\\
$^{9}$ \quad Excellence Cluster ORIGINS, Boltzmannstr. 2, 85748 Garching, Germany\\
$^{10}$ \quad Moscow Institute of Physics and Technology, Institutsky per. 9, Dolgoprudny 141700, Russia\\
$^{11}$ \quad Lebedev Physical Institute of the Russian Academy of Sciences, Leninsky prospekt 53, 119991 Moscow, Russia\\
$^{12}$ \quad Northeastern University, 360 Huntington Ave, Boston, MA 02115, USA\\

}

\corres{Correspondence: freek.roelofs@cfa.harvard.edu}

\abstract{The next-generation Event Horizon Telescope (ngEHT) will be a significant enhancement of the Event Horizon Telescope (EHT) array, with $\sim 10$ new antennas and instrumental upgrades of existing antennas. The increased $uv$-coverage, sensitivity, and frequency coverage allow a wide range of new science opportunities to be explored. The ngEHT Analysis Challenges have been launched to inform development of the ngEHT array design, science objectives, and analysis pathways. For each challenge, synthetic EHT and ngEHT datasets are generated from theoretical source models and released to the challenge participants, who analyze the datasets using image reconstruction and other methods. The submitted analysis results are evaluated with quantitative metrics. In this work, we report on the first two ngEHT Analysis Challenges. These have focused on static and dynamical models of M87* and Sgr A*, and shown that high-quality movies of the extended jet structure of M87* and near-horizon hourly timescale variability of Sgr A* can be reconstructed by the reference ngEHT array in realistic observing conditions, using current analysis algorithms. We identify areas where there is still room for improvement of these algorithms and analysis strategies. Other science cases and arrays will be explored in future challenges.}

\keyword{very long baseline interferometry; black holes; active galactic nuclei; radio astronomy; imaging; instrument design; telescopes; algorithms; data analysis} 

\begin{document}

%%%%%%%%%%%%%%%%%%%%%%%%%%%%%%%%%%%%%%%%%%
%\setcounter{section}{-1} %% Remove this when starting to work on the template.

\section{Introduction} \label{sec:intro}

\subsection{The ngEHT}

The Next-Generation Event Horizon Telescope \citep[ngEHT;][]{Doeleman2019, Doeleman2022} will build on the success of the Event Horizon Telescope (EHT), the mm VLBI array which has imaged the black hole shadows of M87* and Sgr A* \citep{M87PaperI, M87PaperII, M87PaperIII, M87PaperIV, M87PaperV, M87PaperVI, M87PaperVII, M87PaperVIII, PaperI, PaperII, PaperIII, PaperIV, PaperV, PaperVI}. The array will be transformatively enhanced with the current design envisioning $\sim10$ additional stations, a quadrupled bandwidth, and frequency coverage including 86 \citep{Issaoun2022}, 230, and 345 GHz. Multiple operating modes will make it suitable for a wide array of science cases. The primary science goals will involve making movies of M87* and Sgr A* resolving the plasma dynamics on event horizon scales, providing black hole photon ring measurements sufficiently accurate to put constraints on black hole spin, and increasing the sample of imaged black hole shadows \citep{Pesce2021, Pesce2022}.

\subsection{Challenge motivation}
End-to-end science simulations, which cover the full source physics, observation, calibration, and analysis processes, are of great value for the design and optimization of new instrumentation in astrophysics. These simulations realistically predict what the capabilities of the new instrument will be and which science questions it will be able to answer, and can help guide the instrument design and analysis algorithm development. The ngEHT Analysis Challenges aim to provide such end-to-end simulations, bringing together expertise in all relevant areas to be applied to a well-defined set of problems. The challenge concept was inspired by the EHT Imaging Challenges \citep{Bouman2017}. In these challenges, EHT imaging experts imaged synthetic EHT datasets of different source models, which led to rapid development of imaging algorithms and strategies tailored to the specifics of EHT datasets. While the EHT Imaging Challenges were aimed at maximizing the image quality that can be obtained from a known instrument, the ngEHT Analysis Challenges aim to help guide the development of a new instrument. Additionally, the ngEHT concept allows to expand the imaging in two new dimensions, which are frequency (the ngEHT will operate at 2-3 distinct frequency bands simultaneously) and time (movie making). While not the focus of the challenges reported in this work, we aim to extend the ngEHT Analysis Challenges to model fitting and parameter estimation as well.

\subsection{Challenge procedure}
For each challenge, we generate synthetic datasets from a set of source models. The source models \citep[see also][]{Chatterjee2022} are representative for a specific ngEHT science case, may be static or time-variable, and may be generated for different (potential) ngEHT frequencies (86, 230, and 345 GHz in this work). The synthetic datasets are generated for different arrays (here, the 2022 EHT array and an ngEHT reference array), and contain different levels of data complexities (e.g., systematic weather or instrument noise). For each challenge, the synthetic datasets and other information and instructions are released to the challenge participants through the ngEHT Analysis Challenge website\footnote{\url{https://challenge.ngeht.org/}}. Participants then upload their analysis results through the same website before a pre-set deadline. Image and movie reconstructions are then uniformly plotted for visual comparison, and evaluated using quantitative metrics (see Sec. \ref{sec:metrics}). Participation is open to anyone, with access to the downloads provided upon request to the organizers. 

\subsection{Outline}
In this work, we report on the first two ngEHT Analysis Challenges. Section \ref{sec:methods} details the reconstruction algorithms used by the challenge participants, and Section \ref{sec:metrics} describes the submission evaluation metrics. The Challenge 1 and 2 source models, synthetic data generation, and results are presented in Sections \ref{sec:challenge1} and \ref{sec:challenge2}, respectively, and the conclusions and outlook are discussed in Section \ref{sec:conclusions}.

\section{Reconstruction methods}

\label{sec:methods}
In radio interferometry, image reconstruction is an underconstrained problem, as the finite number of telescopes and baselines cause only a limited number of Fourier components of the image (visibilities) to be measured. Hence, an infinite number of images could fit the data, and additional assumptions need to be made in order to arrive at a unique image solution. Different image reconstruction algorithms tackle this problem in different ways. The algorithms can be divided into inverse modeling, regularized maximum likelihood (RML), and Bayesian methods. The section below describe the algorithms used for the image reconstructions in this work, separated in methods reconstructing static images and methods reconstructing movies. 

An alternative method to reconstruct the sky brightness distribution is fitting (geometrical) models to the interferometric data. Since such reconstructions have not been submitted for the challenges described in this paper, we do not discuss them here. We aim to explore these methods in future challenges, aimed at measuring specific black hole and accretion parameters.

\subsection{Static imaging}
\subsubsection{CLEAN}
%\textcolor{red}{(Thalia; To be revised)} 
\label{sec:clean}
The {\tt CLEAN} algorithm, is a well-known inverse modeling imaging technique. The basic algorithm was developed by \citet{1974A&AS...15..417H} with other variants developed later. {\tt CLEAN} deconvolves a sampling function (known as dirty beam) from the measured brightness (or dirty map) of a radio source. The imaging procedure via {\tt CLEAN} involves a number of iterations, where in each iteration the algorithm creates a point-source component, the {\tt CLEAN} component, at the position of the brightness peak in the dirty image. Then, it convolves the {\tt CLEAN} component with the dirty beam, subtracting it from the dirty image and transferring it to the clean map \citep{1974A&AS...15..417H}. The cleaning iterations continue until a specific cleaning halting requirement is met. In the case of noisy data, the user can steer the process by limiting the searching area with {\tt CLEAN} windows. Finally, the generated set of {\tt CLEAN} components is convolved with a Gaussian restoring beam. The image quality can be further enhanced via self-calibration \citep[][and references thereafter]{1977Natur.269..764W}, which corrects the amplitude and phase information using the current image estimate. The residual dirty image, representing the image noise level, may be added to the clean map as the final step.

During the last decades, this technique has been widely used for imaging astronomical targets, as well as for a broad range of other applications \citep{2009A&A...500...65C}. Together with {\tt eht-imaging} and {\tt SMILI} (Sec. \ref{sec:ehtim}), it was one of the methods used for reconstructing the first EHT images of M87* \citep{M87PaperI,M87PaperII,M87PaperIII,M87PaperIV,M87PaperV,M87PaperVI} and Sgr A* \citep[][]{PaperI,PaperII,PaperIII,PaperIV,PaperV,PaperVI}.

The strategy followed for the ngEHT Analysis Challenges used a semi-scripted approach, in a similar fashion to the one in \cite{PaperIII}, employing the {\tt CLEAN} algorithm via the software DIFMAP \citep[][and references thereafter]{1995BAAS...27..903S}.

\subsubsection{RML methods: eht-imaging and SMILI}
%\textcolor{red}{(Thalia; To be revised)} 
\label{sec:ehtim}

RML methods calculate each pixel of the source image $\mathbf{I}$ by fitting directly to the data $\mathbf{D}$, with the fidelity of the final image to be adjusted by specific regularization terms \citep[e.g.,][]{Frieden:72,10.1093/mnras/stw2159, doi:10.1146/annurev.aa.24.090186.001015}. The data $\mathbf{D}$ consists of separate dataproducts $d$. These are typically visibility amplitudes, closure phases, or (log) closure amplitudes \citep[see, e.g.,][]{Chael2018}. These regularizers $R$ could entail the entropy, sparsity, smoothness, or other properties of the image. For more details of regularizer definitions, see Appendix\,A of \citet{M87PaperIV}. RML methods find an image which minimizes a specified objective function,   
\begin{equation} \label{eq::objfunc}
J(\mathbf{I}) = \sum_{d} \alpha_{d} \chi^2_{d}\left(\mathbf{I}\right) - \sum_{R} \beta_{R} S_{R}\left(\mathbf{I}\right), 
\end{equation}
consisting of goodness-of-fit ($\chi^2_d$) and regularization ($S_R$) terms, weighted by hyperparameters ($\alpha_d$ and $\beta_R$). 

Both the {\tt eht-imaging} \citep{Chael2016, Chael2018} and {\tt SMILI} \citep{Akiyama2017a, Akiyama2017b} frameworks are suitable for directly using the closure phases and (closure) amplitudes, making them ideal for high-frequency interferometric imaging \citep{2019ascl.soft04004C}. Like for {\tt CLEAN}, multiple rounds of self-calibration are often performed.

Various submitters used different regularizers and weights for producing reconstructions of Challenge 1 and 2 data. For the M87 datasets, an informed prior was often used consisting of a small Gaussian with most of the flux, corresponding  to the core, and a large disk with little flux to capture the extended emission from the jet.  For the Sgr A$^\star$ images, a disk or Gaussian prior was often used, deblurring the data and in some cases applying a constant noise floor to mitigate the intergalactic scattering before imaging. {\tt eht-imaging} was also used to produce multi-frequency images, regularizing the spectral index map \citep{Chael2022}.

\subsection{Dynamical imaging}

\subsubsection{eht-imaging}

The dynamical imaging module of {\tt eht-imaging} (abbreviated to {\tt ehtim-di} in this work) generalizes static imaging using a regularized maximum likelihood approach to reconstruct movies of time-variable sources \cite{Johnson2018}. Specifically, the reconstruction consists of a series of $N_{\rm t}$ images (movie), $\mathbf{M}=\lbrace \mathbf{I}_1, \mathbf{I}_2, ..., \mathbf{I}_{N_t} \rbrace$. These images are determined by minimizing an objective function, 
\begin{align}
\label{eq::Objective_Function}
%J \equiv \chi^2(\{ \mathbf{I}_j \},\mathbf{d}) - \alpha_{\rm S} \left[ \frac{1}{N_{\rm t}}  \sum_{j=1}^{N_{\rm t}} S(\mathbf{I}_j) \right] + \alpha_x \mathcal{R}_x(\{ \mathbf{I}_j \}).
J \equiv  \sum_{d}\alpha_d\chi^2_d(\{ \mathbf{I}_j \}) - \sum_{R}\beta_{\rm R} \left[ \frac{1}{N_{\rm t}}  \sum_{j=1}^{N_{\rm t}} S_R(\mathbf{I}_j) \right] + \sum_{x}\gamma_x \mathcal{R}_x(\{ \mathbf{I}_j \}).
\end{align}
The objective function consists of three components:
\begin{itemize}
    \item Like for static RML imaging, a data term which defines the log-likelihood of the reconstruction with respect to whatever data products are fit. 
    \item A spatial regularization term, where for each regularizer we compute a weighted sum over individual image regularization terms, $S_R(\mathbf{I}_j)$. 
    \item A dynamical regularization term with temporal regularizers $\mathcal{R}_x(\{ \mathbf{I}_j \})$ with associated hyperparameters $\gamma_x$. This term computes a penalty function that can be used to favor reconstructions that evolve smoothly in time ($\mathcal{R}_{\Delta t}$), that have small variations relative to the mean ($\mathcal{R}_{\Delta I}$), or that evolve according to fluid motion with a steady flow ($\mathcal{R}_{\rm flow}$).
\end{itemize}

For the dynamical imaging reconstructions in the analysis challenges, we first fit a simple geometrical model to the full dataset \citep[a thick ``$m$-ring''; see][]{Johnson2020,SgrA_PaperIV}. We then used this model as both a prior (for relative entropy of individual images) and initialization of reconstructed movies, with a typical frame separation of 1~minute for Sgr~A$^*$. We fit amplitudes and closure phases, with iterative self-calibration of the visibility amplitudes. The imaging was performed using gradient descent with the limited-memory Broyden–Fletcher–Goldfarb–Shanno (BFGS) algorithm \cite{BFGS}, as implemented in {\tt Scipy} \cite{Jones2001}.

\subsubsection{StarWarps}
The {\tt StarWarps} algorithm \cite{2017arXiv171101357B} reconstructs time-variable sources by simultaneously reconstructing both the image and its time evolution. {\tt StarWarps} reconstructs $N_{\rm t}$ images $\mathbf{M}=\lbrace \mathbf{I}_1, \mathbf{I}_2, ..., \mathbf{I}_{N_t} \rbrace$, using the observational data snapshots $\mathbf{D} = \lbrace \mathbf{D}_1, \mathbf{D}_2, ..., \mathbf{D}_{N_t} \rbrace $ at the corresponding timestamps. It employs a dynamical imaging model $\varphi$ at each timestamp $j$:
\begin{align}
\label{starwarps1}
\varphi_{\mathbf{D}_j |\mathbf{I}_j} &= \mathcal{N}_{\mathbf{D}_j}(f_j(\mathbf{I}_j), \mathbf{R}_j), \\
\label{static2}
\varphi_{\mathbf{I}_j} &= \mathcal{N}_{\mathbf{I}_1}(\boldsymbol{\mu}_j, \bm{\Lambda}_j), \\
\label{dynamic}
\varphi_{\mathbf{I}_j |\mathbf{I}_{j-1}} &= \mathcal{N}_{\mathbf{I}_j}(\mathbf{A}\mathbf{I}_{j-1},\mathbf{Q}),
\end{align}
where $\mathcal{N}_{\mathbf{D}_j}(f_j(\mathbf{I}_j), R_j)$ refers to the multivariate normal distribution of $\mathbf{D}_j$ with mean $f_j(\mathbf{I}_j)$ and the covariance $R_j$.
$\bm{\Lambda}_i = {\rm diag}[\boldsymbol{\mu}_i]^T \bm{\Lambda}' {\rm diag}[\boldsymbol{\mu}_i]$, where $\bm{\Lambda}'$ is defined in terms of the priors, see Eq. 13 of \cite{2017arXiv171101357B} for more details.
$\boldsymbol{\mu}_i$ is the mean of a multivariate Gaussian distribution and $\bm{\Lambda}$ describes the covariance, setting the spatial regularization. $f_j(\mathbf{I}_j)$ describes the functional relationship between the source image $\mathbf{I}_j$ and the contemporaneous observed data $\mathbf{D}_j$. The global time evolution of the source between timestamps $j-1$ and $j$ is described by the evolution matrix $\mathbf{A}$, so that $\mathbf{I}_j \approx \mathbf{A} \mathbf{I}_{j-1}$, and any additional perturbations of the source are constrained by the covariance matrix $\mathbf{Q}$. The process hence reduces to static imaging for $(\mathbf{A} = \mathbf{1}, \mathbf{Q} = \mathbf{0})$. The joint probability distribution is then given by:
\begin{equation}
p(\lbrace\mathbf{I}_j\rbrace, \mathbf{D} ;\mathbf{A}) \propto \prod_{j=1}^{N_t} \varphi_{\mathbf{D}_j | \mathbf{I}_j} \prod_{j=1}^{N_t} \varphi_{\mathbf{I}_j} \prod _{j=2}^{N_t} \varphi_{\mathbf{I}_j | \mathbf{I}_{j-1}}. 
\end{equation}
In StarWarps, we jointly solve for the image reconstructions as well as $\mathbf{A}$. First, we learn $\mathbf{A}$ using the Expectation-Maximization (EM) algorithm, and then reconstruct the images with that $\mathbf{A}$.

For the analysis challenge image reconstruction with {\tt StarWarps}, we used the visibility amplitudes, log closure amplitudes, and the bispectrum (triple amplitudes and closure phases) as our data products, with 2\% added systematic noise. We used the EHT 2017 image of Sgr A* \citep{PaperI} blurred with a 25 $\mu$as Gaussian kernel as a prior \citep[see also][]{Emami2022}. 

\subsubsection{Resolve} 
The algorithm {\tt resolve}\footnote{\url{https://gitlab.mpcdf.mpg.de/ift/resolve}} approaches the imaging task for the (ng)EHT from a probabilistic, Bayesian perspective.
It is based on Bayes' theorem:
\begin{align}
	\mathcal P(\mathbf{M}| \mathbf{D}) = \frac{\mathcal P (\mathbf{D} | \mathbf{M}) \,\mathcal P (\mathbf{M})}{\mathcal P (\mathbf{D})},
\end{align}
where $\mathbf{D}$ refers to the measured data and $\mathbf{M}$ denotes the time-varying sky brightness distribution.
The quantity $\mathcal P(\mathbf{M}| \mathbf{D})$ is called \emph{posterior probability density} and contains all information on $\mathbf{M}$ after taking the information from the data $\mathbf{D}$ into account.
In the case of the dynamic Sgr A* model we consider $\mathbf{M}$ to be a discretized quantity with spatial dimensions $200 \times 200$ and a temporal axis of length $720$, in total $2.88 \cdot 10^7$ degrees of freedom.
Therefore, the posterior can be considered to be a function $\mathbb R^{28\,800\,000} \to \mathbb R^{>0}$.

The \emph{prior probability density $\mathcal P(\mathbf{M})$} represents our knowledge on the source before the data is considered.
Since the sky brightness distribution represents a flux density, we can safely assume that its values are non-negative.
Additionally, we know a priori that the emission is correlated in both spatial and temporal direction.
Thus, we assume generic homogeneous and isotropic spatial and temporal correlation structures whose specific form is learned from the data alongside with $\mathbf{M}$.
For more details on the prior refer to \citet{iftclean,iftm87}.

The \emph{likelihood $\mathcal P(\mathbf{D}|\mathbf{M})$} encodes our knowledge on the meaurement process.
In general, the calibration pipeline of the (ng)EHT provides visibilities whose phases suffer from temporally uncorrelated station-based effects.
The amplitudes of the visibilities are approximately correct and only subject to small time-correlated station-based effects.
Therefore, we use closure phases and self-calibrated (non-closure) amplitudes in the likelihood.

After combining prior and likelihood, our best guess for the time-variable behaviour of the source is given by the expectation value of the posterior: $\int \mathbf{M} \mathcal P(\mathbf{M} | \mathbf{D}) \,d\mathbf{M}$.
Since evaluating such high-dimensional integrals directly is virtually impossible, we use \emph{Metric Gaussian Variational Inference} \citep{iftmgvi} that provides approximate solutions to Bayes' theorem efficiently.
As a result we obtain a collection of approximate posterior samples that can be averaged to obtain an approximate posterior mean.
Additionally, the variability of the approximate posterior samples represents the uncertainty of the computed solution.
This uncertainty could be propagated to downstream analyses of the time-variable reconstruction of the source.

For the ngEHT analysis challenges, we adopted the implementation of {\tt resolve} of \citet{iftm87} where {\tt resolve} has been applied to the 2017 EHT observation of M87*.
For evaluating the interferometry measurement equation (i.e.\ the non-equidistant Fourier transform), we use the implementation presented in \citet{iftresponse}.
Variations of {\tt resolve} have been verified and tested against standard methods in various contexts \citep{iftclean,iftm87}.
To enable comparisons to results of algorithms that do not quantify uncertainties, we will depict only the posterior mean of the sky brightness distribution $s$ in the following.

\subsubsection{DoG-HiT}
{\tt DoG-HiT} is a multi-scale RML imaging algorithm \citep{Mueller2022a}. {\tt DoG-HiT} models the image by a set of wavelets (a dictionary $\Gamma$) constructed by the difference of Gaussian method: $\mathbf{I} = \Gamma \mathscr{I}$ \citep{Mueller2022a, Mueller2022b}. With {\tt DoG-HiT} we aim to recover the array of wavelet coefficients $\mathscr{I}$ that represents the true sky brightness distribution best. The wavelets define filters in the Fourier domain that are ring-like. Hence, every wavelet compresses the spatial information from a specific band of baselines in the uv-coverage. For {\tt DoG-HiT} the scales are fitted to the uv-coverage, thus giving rise to wavelets most sensitive to gaps in the uv-coverage and wavelets most sensitive to Fourier coefficients sampled by baselines. In the spirit of compressed sensing {\tt DoG-HiT} utilizes a sparsity promoting penalization by a $l_0$ penalty term on the wavelet coefficients. In detail we solve the following optimization problem consisting of data fidelity terms for the closure quantities ($\chi^2_\mathrm{cp}$ and $\chi^2_\mathrm{camp}$), the $l_0$ penalty term and a total flux constraint by an updated forward-backward splitting approach \citep{Mueller2022a}:
\begin{align}
    \hat{\mathscr{I}} \in \mathrm{argmin}_\mathscr{I} &\left[  \chi^2_\mathrm{cp}(\Gamma \mathscr{I}) + \chi^2_\mathrm{camp}(\Gamma \mathscr{I}) + \alpha \cdot \norm{\mathscr{I}}_\mathrm{l_0} + R_\mathrm{flux} (\mathscr{I}, f) \right],
\end{align}
where $R_\mathrm{flux}$ is a total flux indicator function with flux $f$, and $\Gamma$ denotes the wavelet dictionary. {\tt DoG-HiT} is a data-driven, automatic imaging pipeline that depends on only one hyper-parameter (the relative weighting of the penalization term $\alpha$). It has been demonstrated to produce high-quality, super-resolved reconstructions for static sources in relatively short time with minimal manual interaction and without the need of extensive parameter surveys \citep[e.g. compare the Challenge 1 reconstructions with DoG-HiT in Sec. 6 of][]{Mueller2022a}. 

The dynamic reconstructions are based rather straightforwardly upon the success of this static imaging. We utilize the automatic static imaging pipeline to construct a mean image from the full length of the observation without taking the dynamics of the source into account. {\tt DoG-HiT} computes a set of statistically significant wavelet coefficients from the mean image as byproduct (the multiresolution support). Then the mean image (with a relatively bad fit to the data due to not respecting the source dynamics) is subtracted from the self-calibrated visibilities and the observation is cut into frames of six minutes. The residuals are minimized frame by frame with {\tt StarWarps} with implicit dynamic variability imposed by {\tt StarWarps}. A small {\tt StarWarps} internal regularization parameter is used, but, in constrast to {\tt StarWarps}, the reconstruction is done in a multiscalar constrained minimization framework (multiresolution support constraint), i.e. only the wavelet coefficients classified as significant during the static image reconstruction are allowed to vary. This introduces correlation between frames and consistency to the mean static image.

DoG-HiT is still under development and is currently extended to dynamic, polarimetric reconstructions \citep{Mueller2022c} with promising first results on synthetic data (see upcoming Challenge 3 reconstructions). A finer set of directional-dependent wavelet functions \citep{Mueller2022b} allows for dynamic reconstructions in a constrained minimization reconstruction on frames independently, thus replacing {\tt StarWarps} during the current {\tt DoG-HiT} dynamic imaging pipeline and relying on a completely unsupervised, automatic wavelet approach only. On one hand such an unsupervised, automatic imaging procedure is desired as it reduces the human bias in the reconstruction, on the other hand driving by an astronomer could be crucial to address data issues, in particular for challenging data sets such as will be produced by the ngEHT. 

\section{Submission evaluation metrics}
\label{sec:metrics}
Submitted reconstructions were evaluated with several quantitative quality metrics. These metrics, which all probe different aspects of what makes a high-quality reconstruction, are summarized below.

\subsection{Data fit quality}
The goodness-of-fit of the submitted reconstructions to the provided synthetic data was quantified by computing the reconstruction visibilities using the synthetic data $uv$-coverage and then calculating the $\chi^2$-metric on closure quantities, $\chi^2_{\mathrm{cphase}}$ and $\chi^2_{\mathrm{lcamp}}$. These quantities are the closure phases, which are the sum of visibility phases measured simultaneously on a closed triangle of baselines, and the (log) closure amplitudes, respectively, which are ratios of visibility amplitudes on a baseline quadrangle \citep[see, e.g.,][]{TMS2017}. Closure quantities are robust against station-based calibration errors.

\subsection{Ground truth image similarity}
Apart from the goodness-of-fit to the synthetic data, another important quality metric is the similarity of the reconstruction to the ground truth source model. We quantify this similarity using the normalized cross-correlation. The normalized cross-correlation between two images $X$ and $Y$ is

\begin{equation}
\label{eq:nxcorr}
\rho_{\mathrm{NX}} = \frac{1}{N} \sum_{i=1}^{N} \frac{\left( X_i - \langle X \rangle \right) \left( Y_i - \langle Y \rangle \right)}{\sigma_X \sigma_Y}.
\end{equation}
Here, $N$ is the number of pixels in the images, $X_i$ and $Y_i$ are the pixel values of images $X$ and $Y$, respectively, $\langle\ldots\rangle$ denotes an average, and $\sigma_X$ and $\sigma_Y$ are the standard deviations of the pixel values of images $X$ and $Y$, respectively. The value of $\rho_{\mathrm{NX}}$ will be equal to 1 for identical images (maximal correlation), 0 for completely uncorrelated images, and -1 for perfectly anticorrelated images. Using the implementation in \texttt{eht-imaging}, the two images are regridded to contain the same number of pixels with an equal pixel size, and aligned to maximize $\rho_{\mathrm{NX}}$. For M87 reconstructions, we are often most interested in the arrays' ability to reconstruct of the large-scale and low-surface brightness jet emission. Therefore, we also compute $\rho_{\mathrm{NX}}$ on the log pixel values ($\rho_{\mathrm{NX, log}}$). In order to suppress the influence of low-surface brightness image noise, which may appear at a certain flux level depending on the particularities of the reconstruction algorithm, we limit the dynamic range of the ground truth and reconstructed images to $10^4$ in this case.

\subsection{Effective resolution}
$\rho_{\mathrm{NX}}$ also provides a way to compute the effective angular resolution obtained by the reconstructed image. Following \citet{M87PaperIV}, we blur the ground truth model images with a circular Gaussian with varying FWHM and calculate $\rho_{\mathrm{NX, FWHM}}$ with respect to the ground truth model images using Equation \ref{eq:nxcorr}. For a submitted reconstruction with a particular $\rho_{\mathrm{NX, rec}}$ with respect to the ground truth, the effective resolution $\theta_{\mathrm{eff}}$ is then the (interpolated) FWHM for which $\rho_{\mathrm{NX, rec}}$ is equal to $\rho_{\mathrm{NX, FWHM}}$.

\subsection{Dynamic range}
Dynamic range is usually defined as the ratio between the brightest and dimmest pixel value in an image, and has been frequently used in radio astronomy to assess the ability of an array to reconstruct low-surface brightness features. For images reconstructed with {\tt CLEAN} algorithms, the dynamic range can be naturally calculated as the ratio between the brightest {\tt CLEAN} component and the noise floor (Sec. \ref{sec:clean}). However, for images reconstructed with other algorithms (e.g., RML-based approaches), formally defining a dynamic range metric that works universally and reflects our intuitive sense of dynamic range is non-trivial. This difficulty has two main causes. First, not all imaging methods naturally produce a noise floor like {\tt CLEAN}, and have many (near)-zero pixel values. Second, many imaging algorithms produce spurious structures due to, e.g., sparse $uv$-coverage, so that the lowest reconstructed pixel brightness cannot be used to robustly define a dynamic range metric. 

To evaluate the challenge reconstructions, we use a dynamic range proxy following \citet{Bustamante2022}. This metric considers the ratios between the brightest pixel of the ground truth image $\mathbf{I}_{\mathrm{ground truth}}$ and the absolute pixel residuals of the reconstructed image $\mathbf{I}_{\mathrm{reconstructed}}$ with respect to the ground truth,
\begin{equation}
   \bm{\mathcal{D}} = \frac{\mathrm{max}(\mathbf{I}_{\mathrm{ground truth}} \ast \bm{\mathcal{G}}^{2\mathrm{D}}_{\theta_{\mathrm{eff}}})}{\left|\mathbf{I}_{\mathrm{reconstructed}}-\mathbf{I}_{\mathrm{ground truth}} \ast \bm{\mathcal{G}}^{2\mathrm{D}}_{\theta_{\mathrm{eff}}}\right|}.
\end{equation}

Here, $\ast$ denotes convolution, $\left|\ldots\right|$ indicates that we take the absolute values, and $\bm{\mathcal{G}}^{2\mathrm{D}}_{\theta_{\mathrm{eff}}}$ represents a two-dimensional circular Gaussian with a FWHM equal to the effective resolution $\theta_{\mathrm{eff}}$ of the reconstructed image. From $\bm{\mathcal{D}}$, which has the form of an image, we can calculate a dynamic range proxy by selecting the $q$th quantile of the pixel values:

\begin{equation}
    \mathcal{D}_q = \mathrm{quantile}(\bm{\mathcal{D}}, q).
\end{equation}

By using the residuals, this metric penalizes spurious structures in the reconstructed image, and does not rely on a noise floor being calculated as part of the imaging process. A disadvantage is that it requires the ground truth image, and hence cannot be applied to real data. Setting $q$ too low will make the metric dominated by outliers in the residual image, while setting it too high will not penalize high residuals strongly enough. We set $q=0.1$ for our dynamic range proxy $\mathcal{D}_{0.1}$. Because of its sensitivity to $q$ and the chosen blurring kernel, we emphasize that this metric should not be regarded as giving \emph{the} dynamic range of the image, but rather as a dynamic range proxy which can be used to compare different reconstructions of the same source model.

\section{Challenge 1}
\label{sec:challenge1}
\subsection{Rationale and charge}
The primary objectives of the first challenge were to set up a framework for the generation of synthetic ngEHT data based on theoretical source models, to conduct the organized submission and cross-comparison of reconstruction results from multiple people, and to get a first idea of the benefits and challenges of ngEHT datasets as compared to the current EHT. The model and data properties were therefore kept relatively simple. Participants were asked to submit image reconstructions for each provided synthetic dataset. The challenge was not blind, i.e. the participants had access to the input source models and synthetic data generation script. The challenge was launched on 18 June 2021, and the submission deadline was 16 July 2021. It was advertised to the ngEHT simulations group. All information is available on the challenge website: \url{https://challenge.ngeht.org/challenge1/}.

\subsection{Source models}
\begin{figure*}
%\begin{adjustwidth}{-\extralength}{0cm}
\centering
\includegraphics[width=0.45\textwidth]{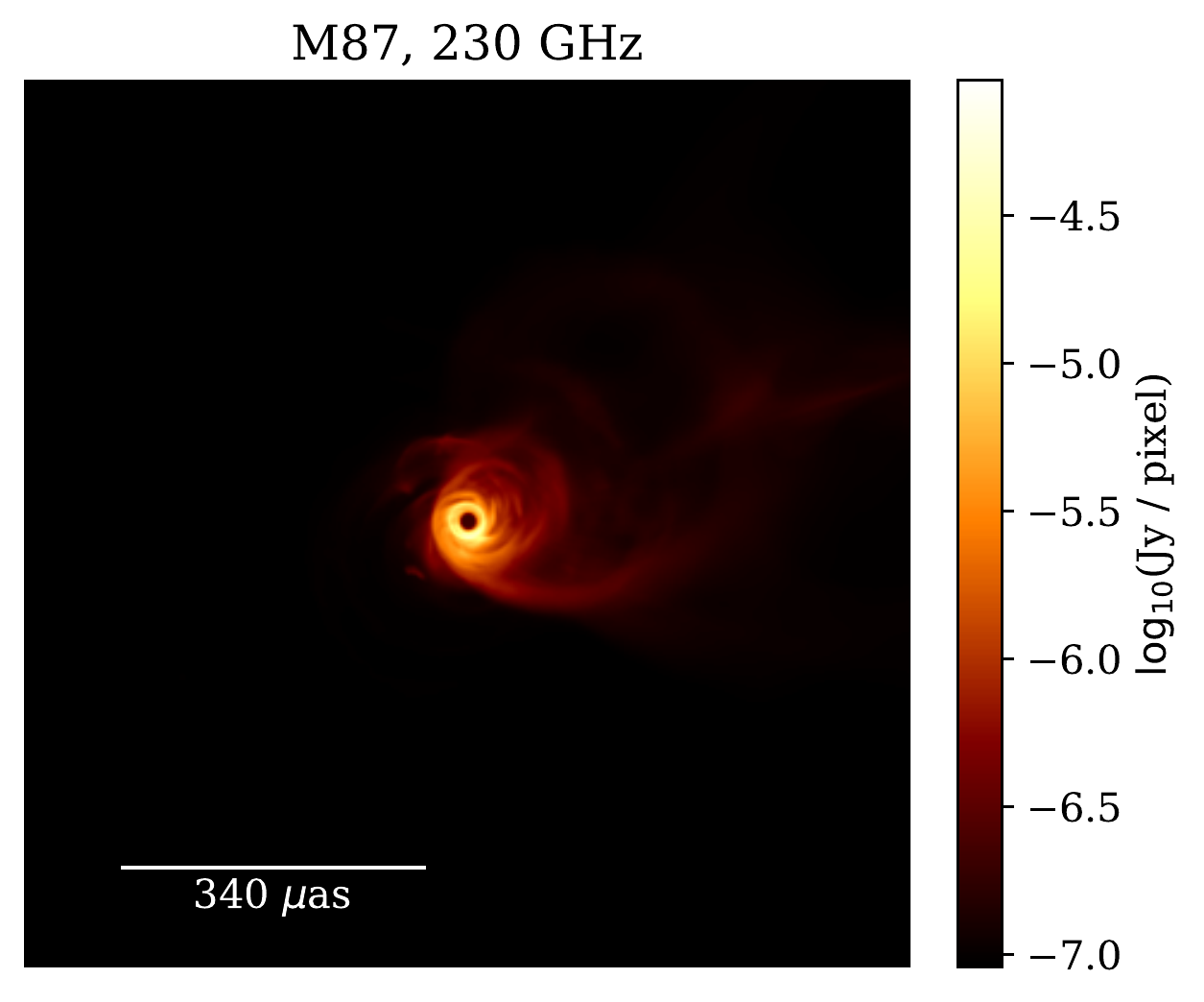}
\includegraphics[width=0.45\textwidth]{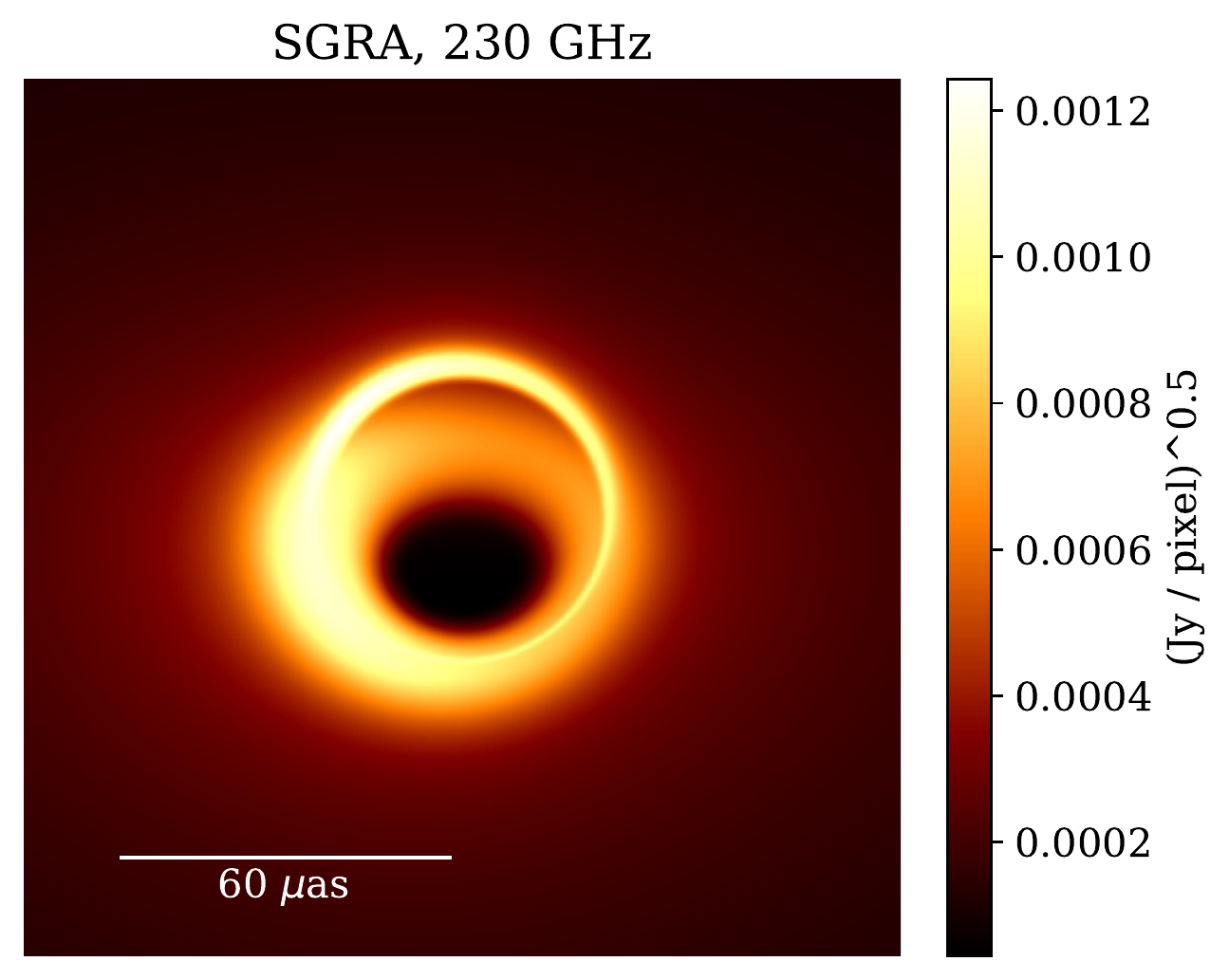} \\
\includegraphics[width=0.45\textwidth]{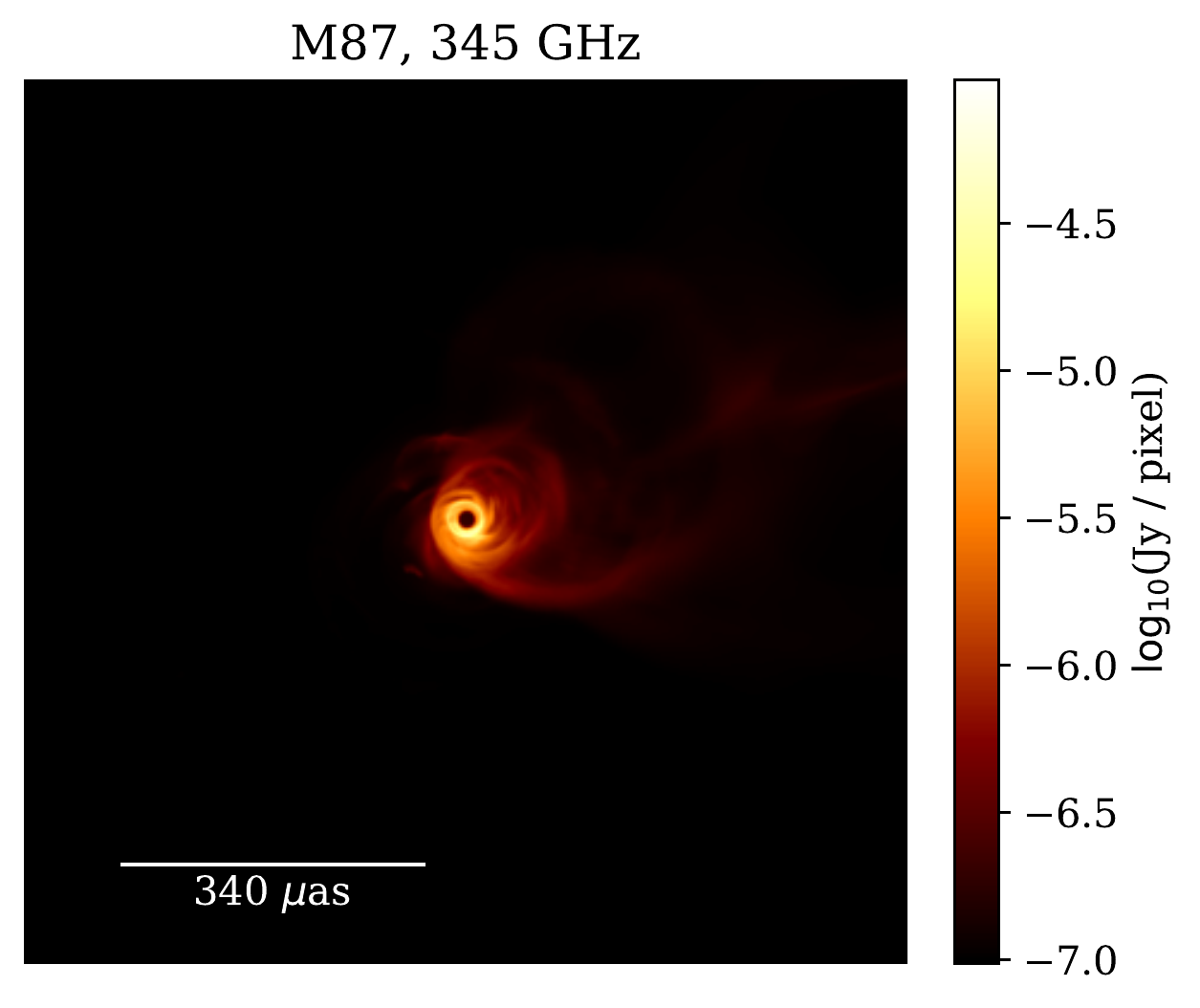}
\includegraphics[width=0.45\textwidth]{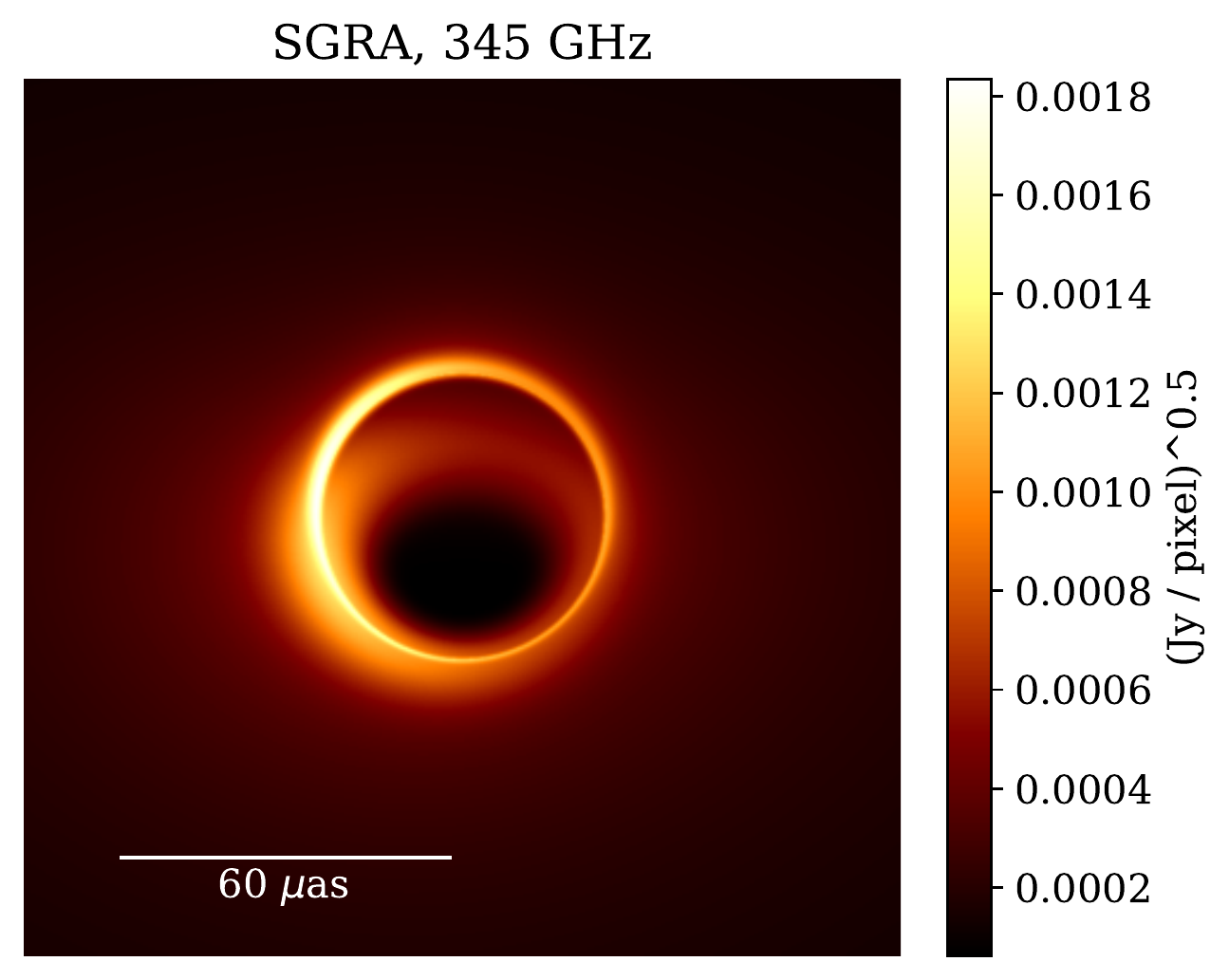}
%\end{adjustwidth}
  \caption{Source models used for Challenge 1.}
     \label{fig:ch1_models}
\end{figure*}

For Challenge 1, we used two static, unpolarized models of M87 and Sgr A*, respectively. Both models are displayed in Figure \ref{fig:ch1_models} and described below. More detailed descriptions and comparisons of the source models used for Challenge 1 and 2 can be found in \citet{Chatterjee2022}.

\subsubsection{M87}
The Challenge 1 M87 model is a magnetically arrested disk (MAD) general relativistic magnetohydrodynamics (GRMHD) frame from a rapid spinning black hole $a_*=0.94$ with electron thermodynamics from reconnection heating \citep[see][for details]{Mizuno2021}. The GRMHD simulation was performed with the \texttt{BHAC} code \citep{Porth2017} using three levels of adaptive mesh refinement (AMR) in logarithm Kerr-Schild coordinates. The numerical grid covers 384x192x192 cells in radial, azimuthal and theta direction and is extending up to 2500 gravitational radii ($GM/c^2$, where $G$ is Newton's gravitational constant, $M$ is the black hole mass, and $c$ is the speed of light) in radial direction. The mass accretion rate and MAD parameter \citep[see][]{Tchekhovskoy2012} were monitored and after obtaining a steady state we performed the general relativistic radiative transfer (GRRT) calculations with the radiative transfer code BHOSS \citep{Younsi2012, Younsi2020}.

During the radiative transport we included non-thermal particles via the kappa electron distribution \citep[see][]{Pandya2016} in the jet sheath while excluding the highly magnetized spine by using a cut in the magnetization at a value of 1 (typically referred to as a sigma cut). The power-law slope of the kappa distribution was set by a particle in cell (PIC) motivated sub-grid model depending on the local magnetization and plasma-beta following \citet{Ball2018}. In addition, we included a fraction of the magnetic energy density to accelerate the non-thermal particles \citep[see][]{Davelaar2019, Fromm2022}. In the jet wind and disk region we used a thermal electron distribution, where the electron temperature is directly obtained from the GRMHD simulation. In order to guarantee capturing small scale structure on the horizon scale and at the same time the large scale jet structure, we used a field of view (FOV) of 1 mas using a resolution of 4096x4096 pixels. Since the GRMHD simulations are scale free, we normalized our GRRT simulations by setting the mass ($6.5\times10^9 M_{\odot}$) and distance (16.9 Mpc) of the black hole in M87 and iterated the mass accretion rate until a compact flux density of 0.8 Jy at 230 GHz is obtained. 

\subsubsection{Sgr A*}
The Challenge 1 Sgr A* model is a semi-analytic stationary radiatively inefficient accretion flow (RIAF) model \citep[e.g.][]{Broderick2006}. This model can be used to test the capabilities of next generation arrays in precision modeling of black hole parameters. High resolution is needed to capture the unique signature from subring structure. This model does not capture any variability due to turbulence in the system. The basic model has is $a_*=0$ (Schwarzschild) at an inclination of i=130 deg and includes non-thermal particles. The model includes disk height \citep[following][]{Pu2018}, sub-Keplerian flow properties ($\kappa$=0.5, $\alpha$=0.5), following the notation of \citet{Tiede2020} -- e.g. Eq. 10 and 11, and fitted to the observed data of \citep{Zhao2003, Bower2015, Bower2019, Liu2016}. Images were ray-traced at 230 and 345 GHz with 4096x4096 pixels and a FOV of 128$GM/c^2$, using a distance of 8.178 kpc and mass of $4.14M_{\odot}$ \citep{Gravity2019}. Finally, the 230 GHz and 345 GHz images were scattered with the same realization of the \citet{Johnson2018} interstellar scattering model before generating the synthetic data.

\subsection{Synthetic data}
\label{sec:challenge1_synthdata}
\subsubsection{Station locations}
\begin{figure*}
\centering
\includegraphics[width=\textwidth]{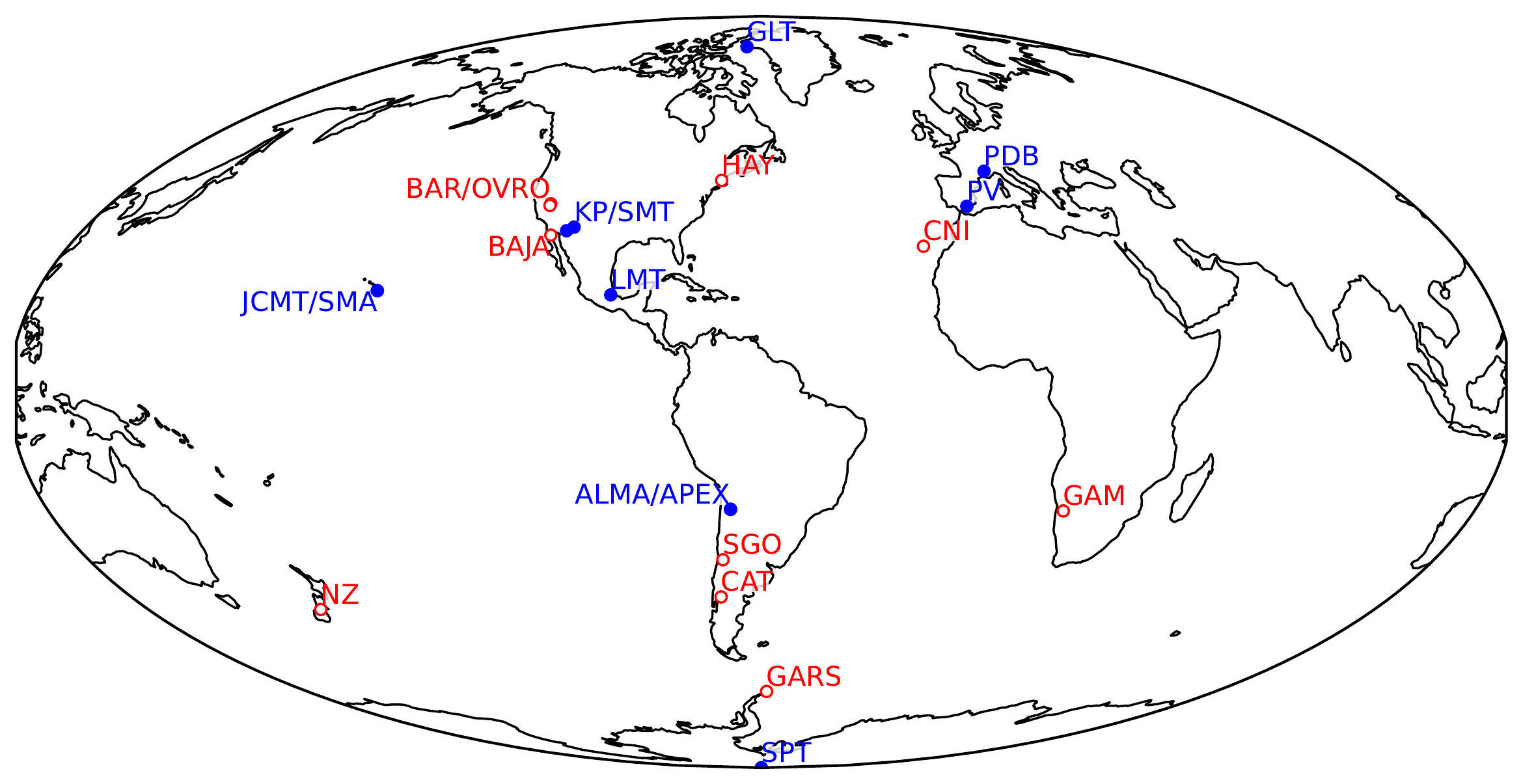}
  \caption{Station locations for Challenge 1 and 2. Stations in blue form the EHT2022 array, and stations in red are added to EHT2022 to form a reference array ngEHT1. The new station locations are near the National Astronomical Observatory in Baja California, Mexico (BAJA); Barcroft Field Station in California, USA (BAR); Cerro Catedral in R'io Negro in Argentina (CAT), La Palma, part of the Canary Islands, Spain (CNI); the Gamsberg in Namibia (GAM), the German Antarctic Receiving Station O'Higgins in Antarctica (GARS); Haystack Observatory in Westford, MA, USA; Canterbury, New Zealand (NZ); Owens Valley Radio Observatory in California, USA (OVRO); and Santiago, Chile (SGO). See \citet{Raymond2021} for details, e.g., weather statistics for each site.}
     \label{fig:ch1_arrays}
\end{figure*}

Two arrays were used to generate the Challenge 1 synthetic data. EHT2022 consists of the 11 stations that participated in the 2022 EHT observations. In ngEHT reference array 1 (ngEHT1), 10 stations are added to this array. The station locations were chosen based on a $uv$-coverage analysis (A. Raymond, priv. comm.), investigating which combination of sites from \citet{Raymond2021} provided optimal $uv$-coverage by performing a Monte Carlo simulation involving telescope dropouts due to weather conditions. The LMT, SPT, and KP were not included in the 345 GHz observations with EHT2022. The station locations are shown in Figure \ref{fig:ch1_arrays}.

\subsubsection{Data properties}
A 24-hour observing track was simulated for each array, source, and frequency, resulting in eight separate datasets. Each track consists of 10-minute scans interleaved with 10-minute gaps and is identical for each dataset. A single frequency channel with a time resolution of 10 seconds was provided. Thermal noise expected from the receiver and atmospheric opacity were added to the complex visibilities calculated using \texttt{eht-imaging} \citep{Chael2016, Chael2018}. The following assumptions were made for all sites:
\begin{itemize}
    \item Receiver temperature: 60 K for 230 GHz; 100 K for 345 GHz
    \item Aperture efficiency: 0.68 for 230 GHz; 0.42 for 345 GHz
    \item Bandwidth: 8 GHz
    \item Quantization efficiency: 0.88
    \item Dish diameter: 6 m for new sites, actual diameter for existing sites
    \item Opacity: median values in April as extracted from the MERRA-2 data by \citet{Raymond2021}, at 30 degrees elevation. The opacities were set constant throughout and across the different datasets, but are frequency-dependent.
\end{itemize}

\begin{figure*}
\centering
\includegraphics[width=0.45\textwidth]{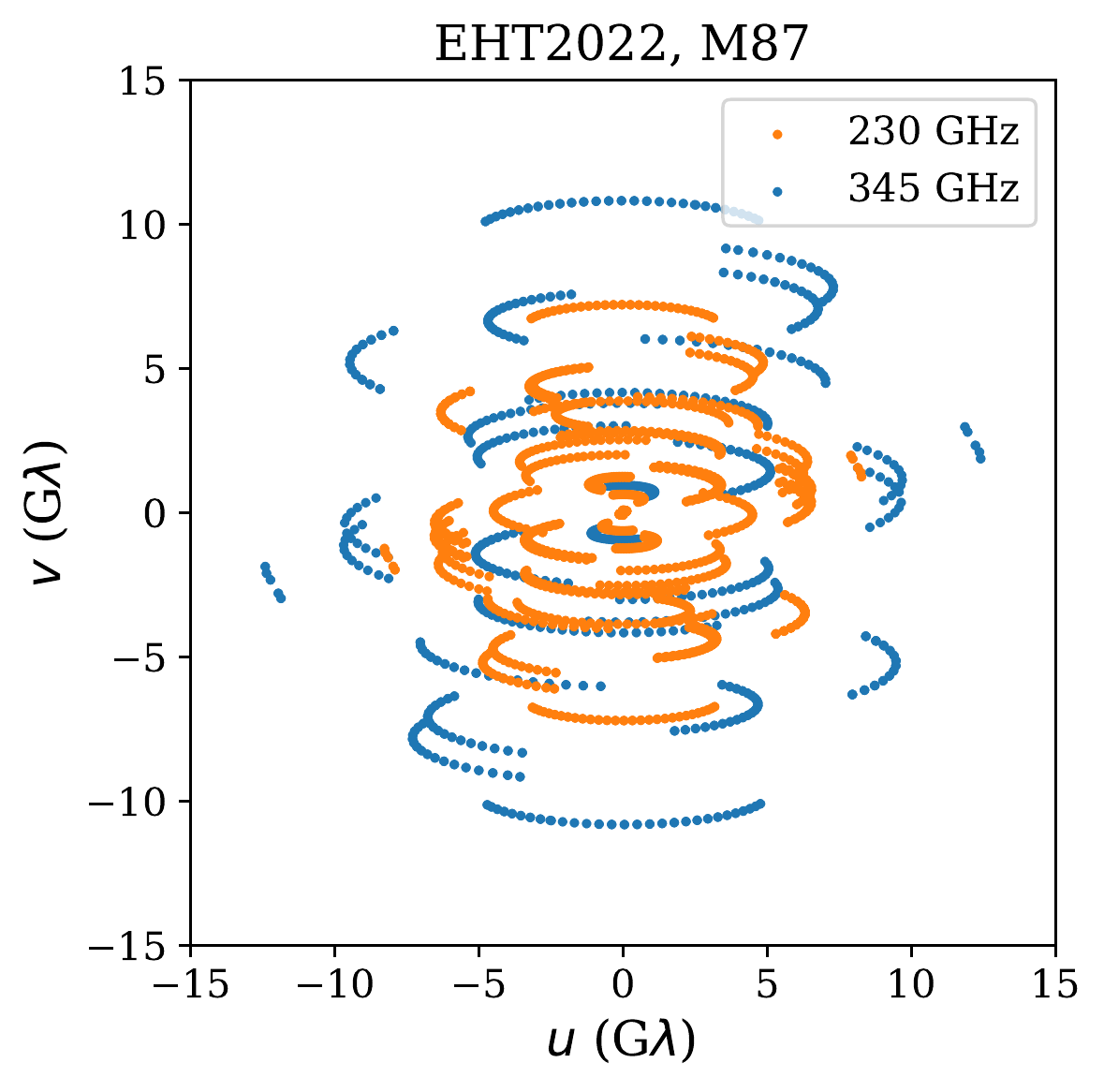}
\includegraphics[width=0.45\textwidth]{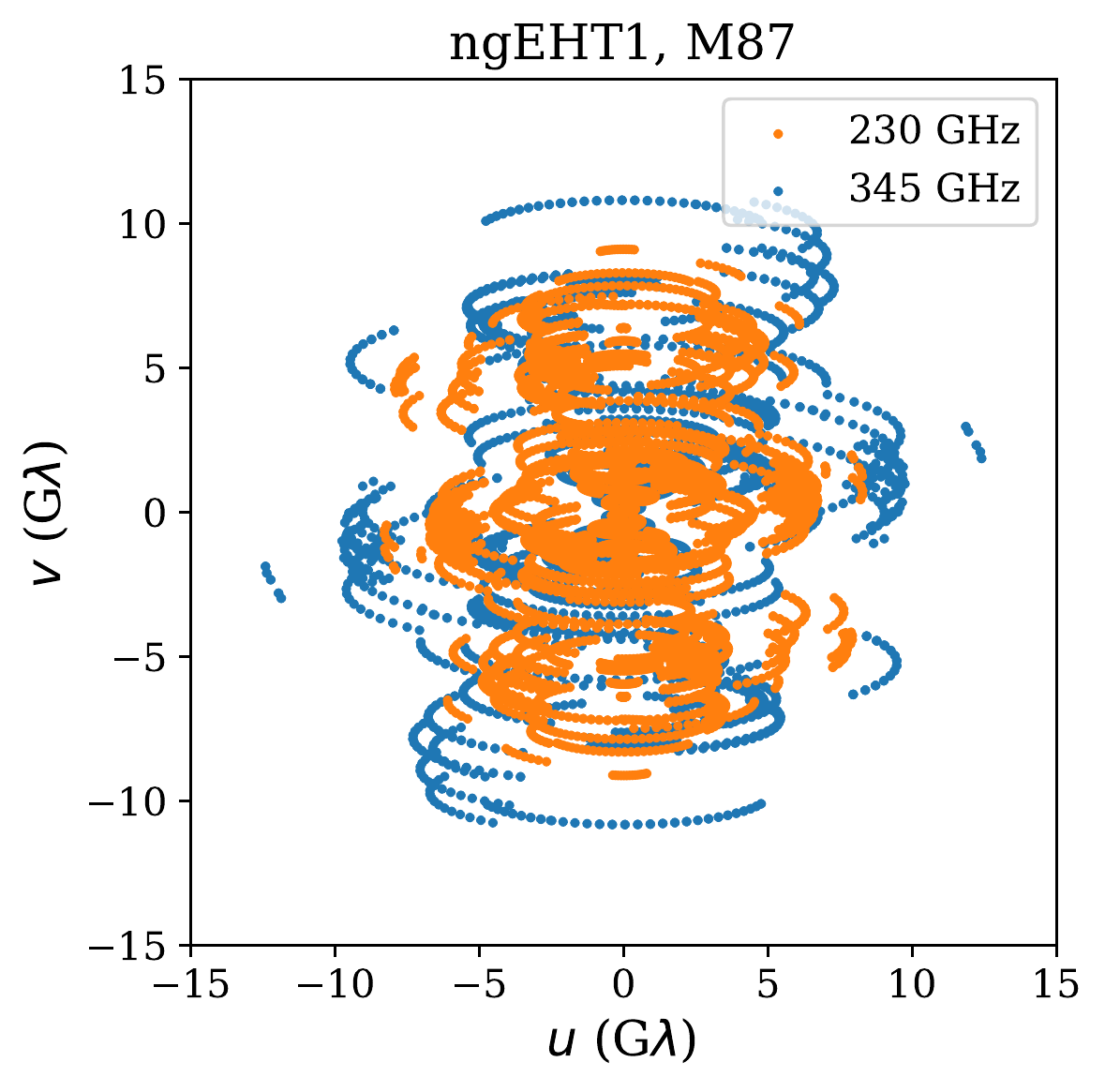} \\
\includegraphics[width=0.45\textwidth]{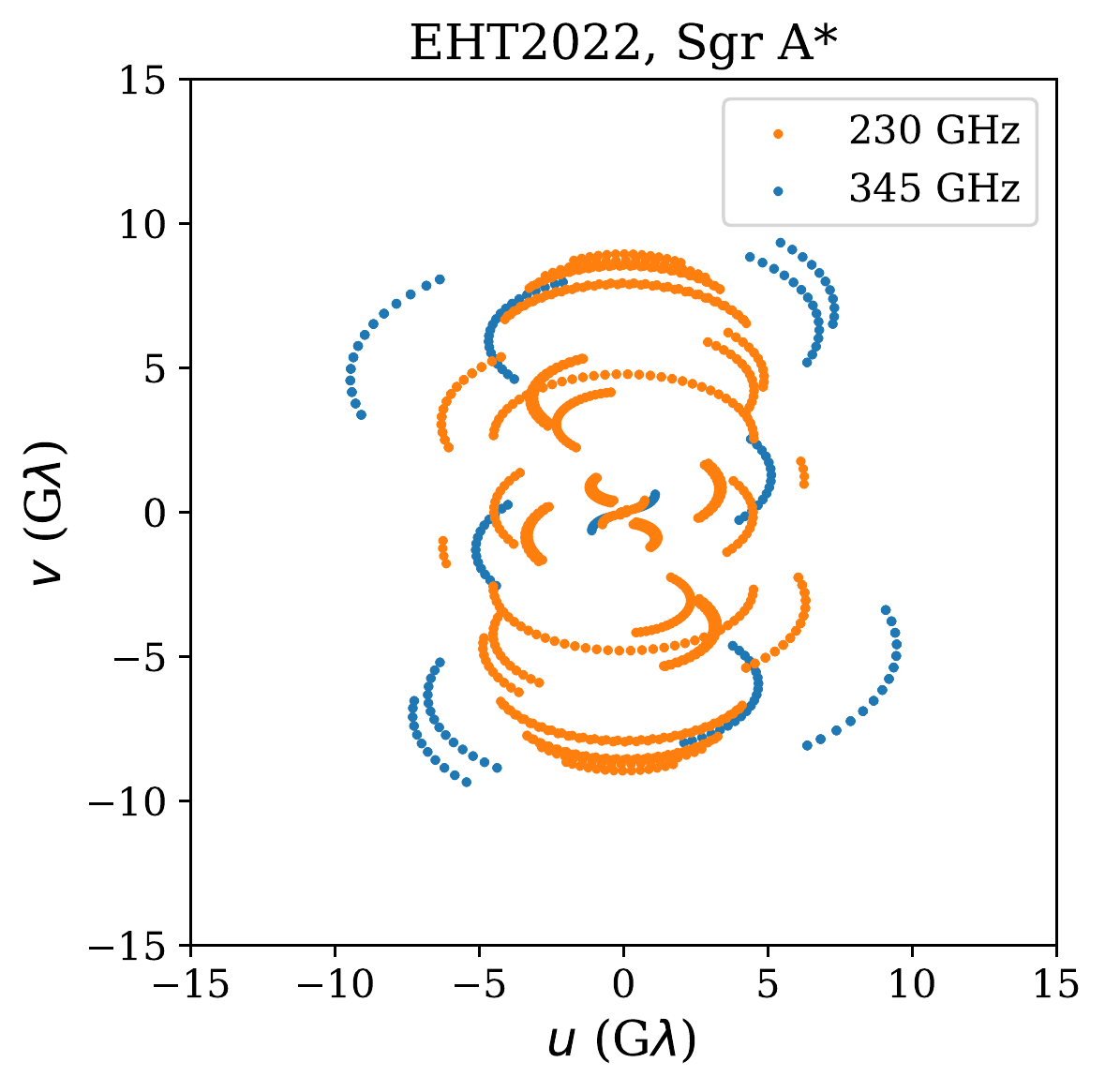}
\includegraphics[width=0.45\textwidth]{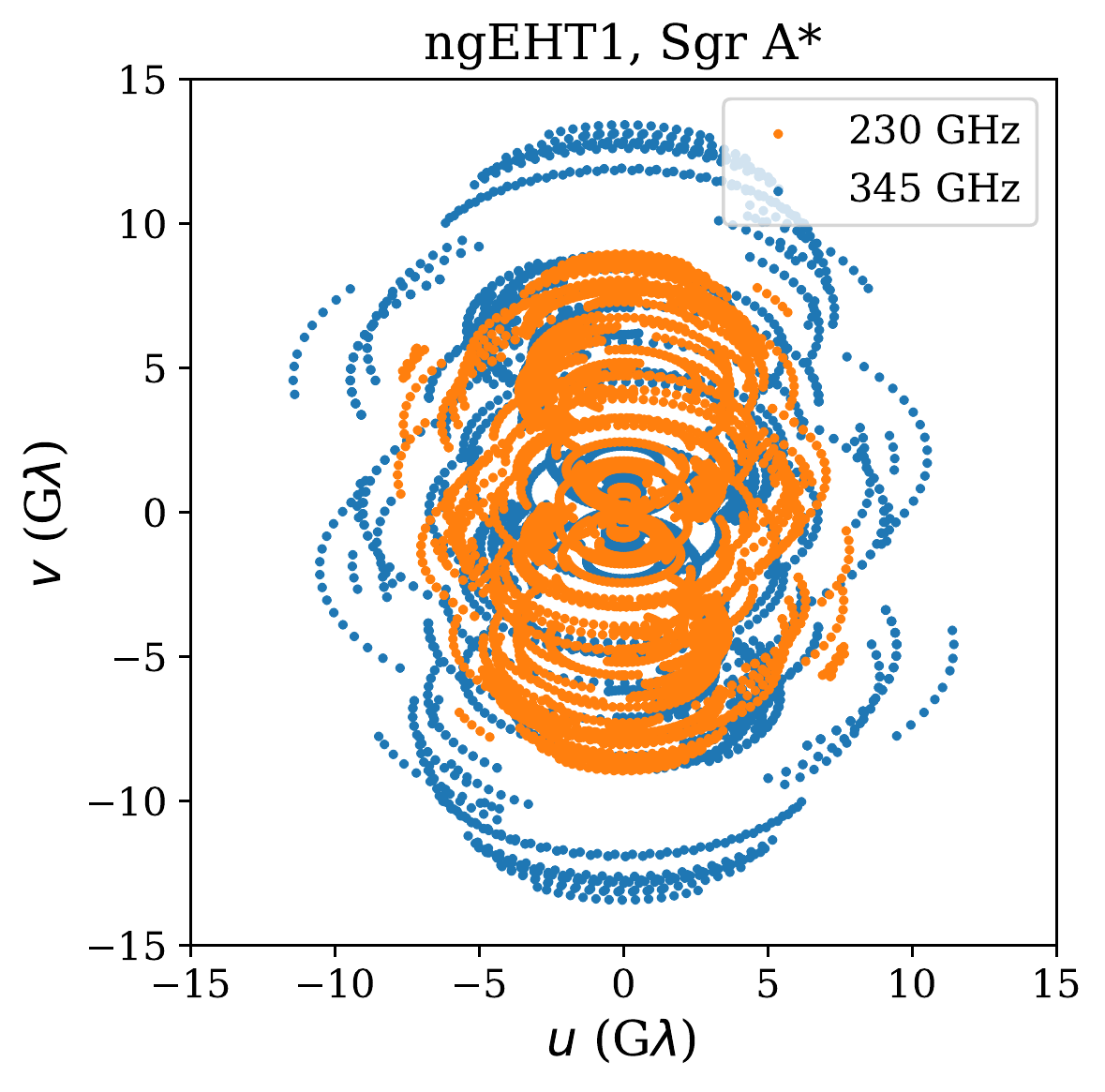} 
  \caption{$uv$-coverage for the Challenge 1 datasets.}
     \label{fig:ch1_uvcov}
\end{figure*}

Visibility phases were scrambled, but stabilized across scans. No further systematic errors were added. After data generation, data points with a signal-to-noise ratio less than 1 were flagged. Figure \ref{fig:ch1_uvcov} shows the resulting $uv$-coverage for all datasets.

\subsection{Results}
\begin{figure*}
\begin{adjustwidth}{-\extralength}{0cm}
\setlength{\lineskip}{0pt}
\centering
\includegraphics[width=25mm]{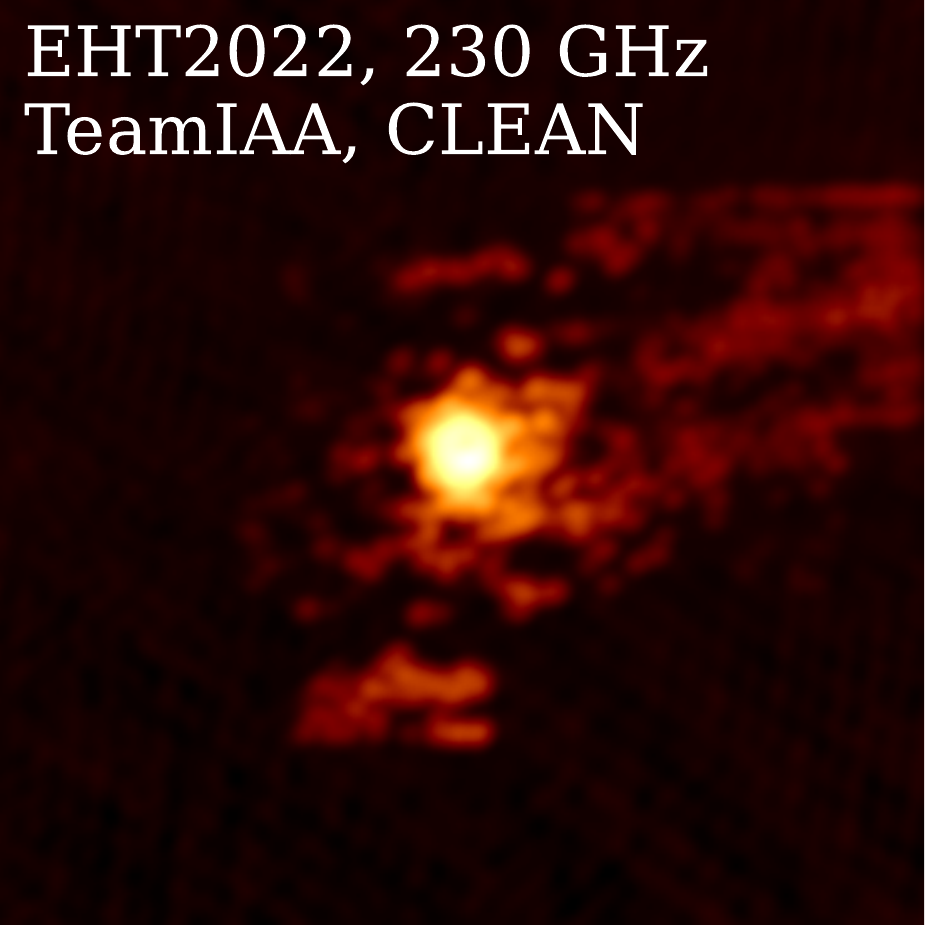}% 
\includegraphics[width=25mm]{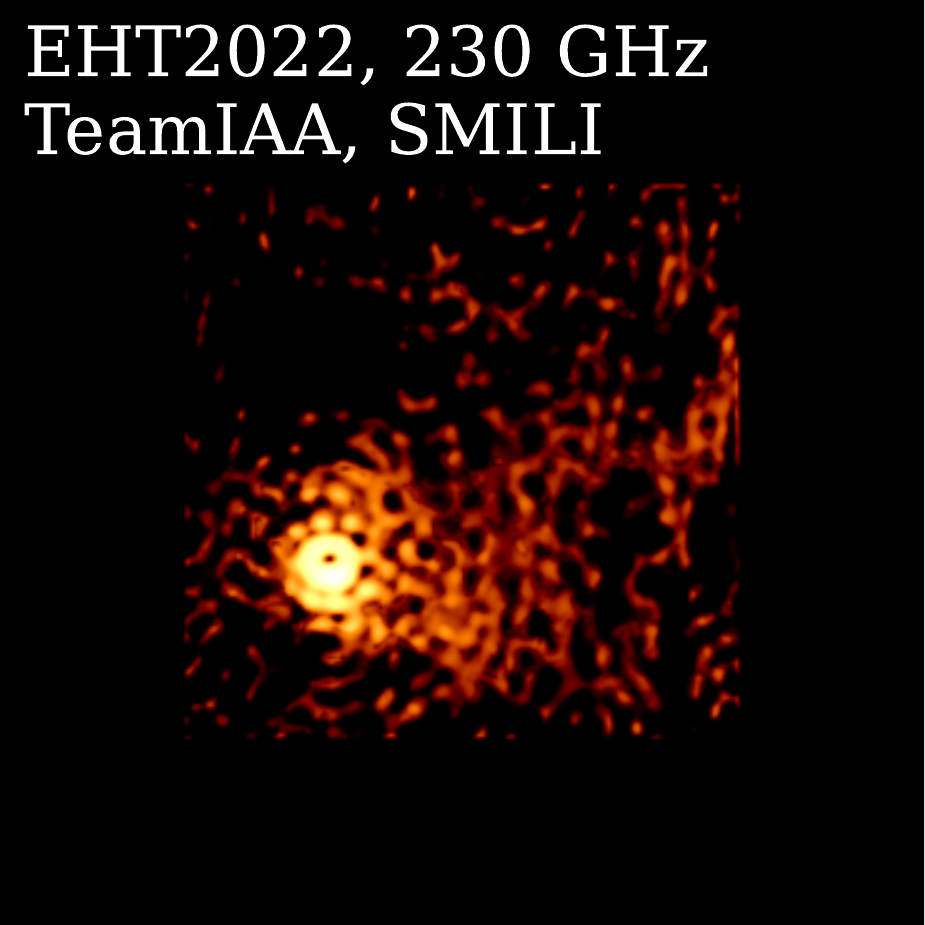}%
\includegraphics[width=25mm]{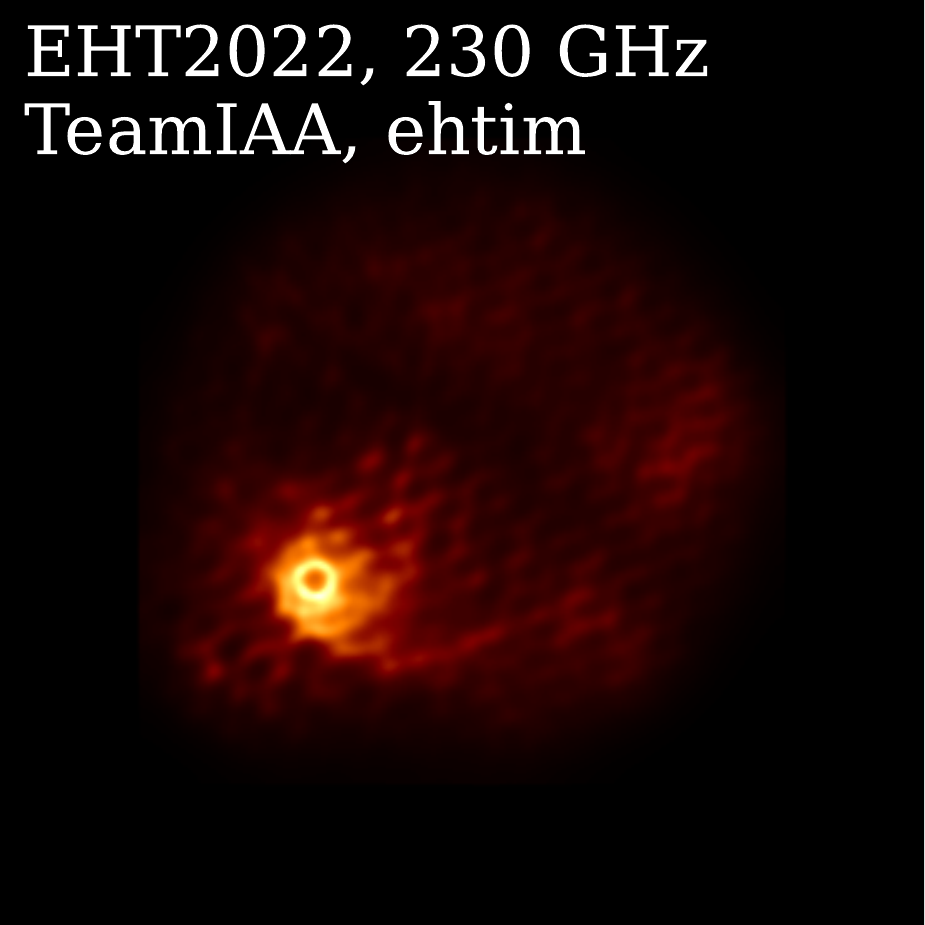}%
\includegraphics[width=25mm]{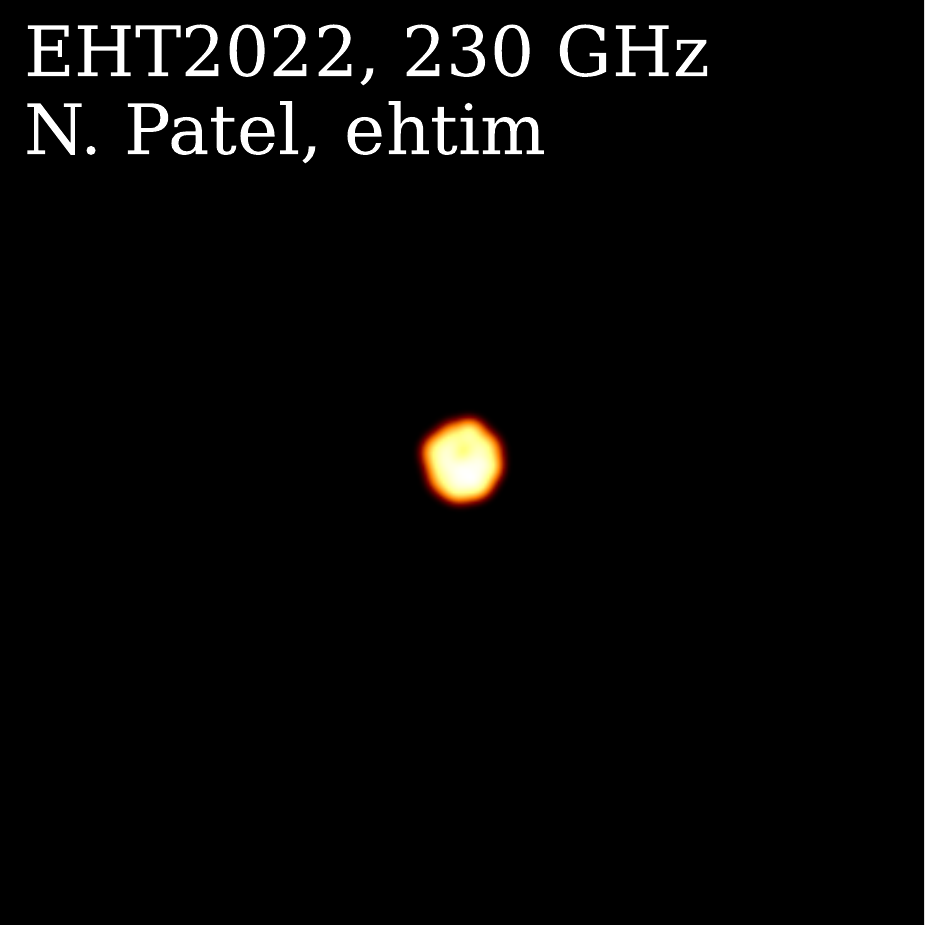}%
\includegraphics[width=25mm]{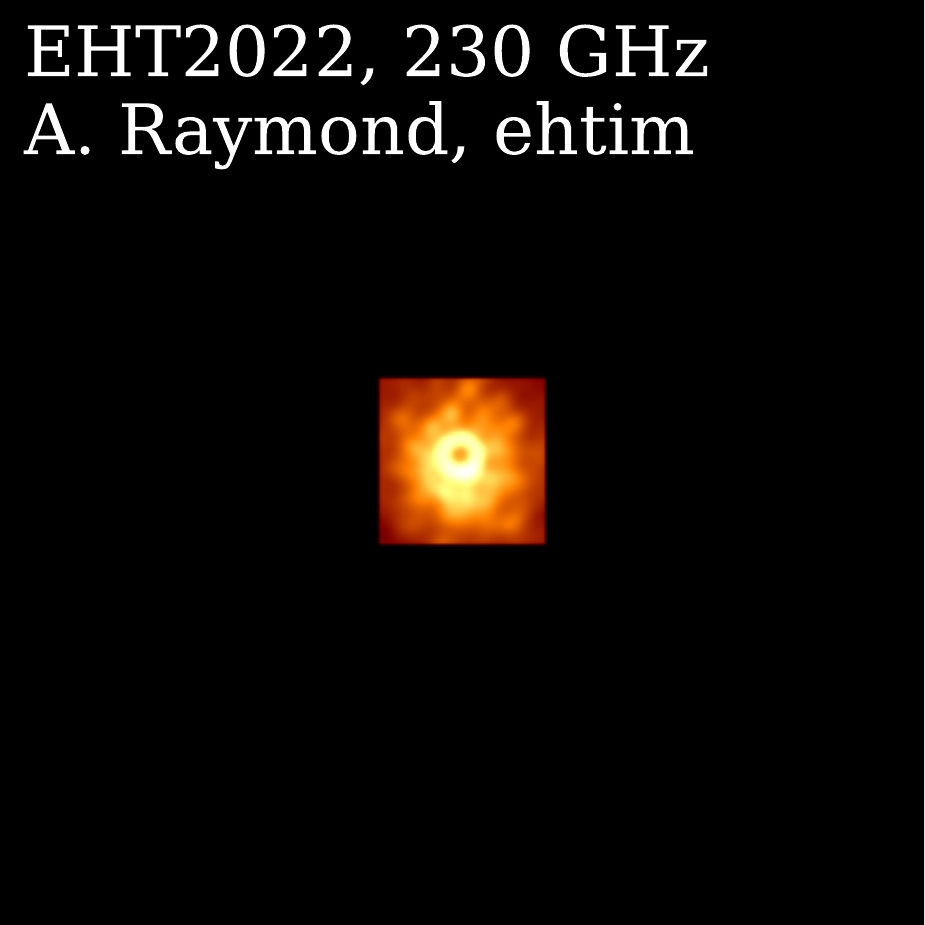}%
\includegraphics[width=25mm]{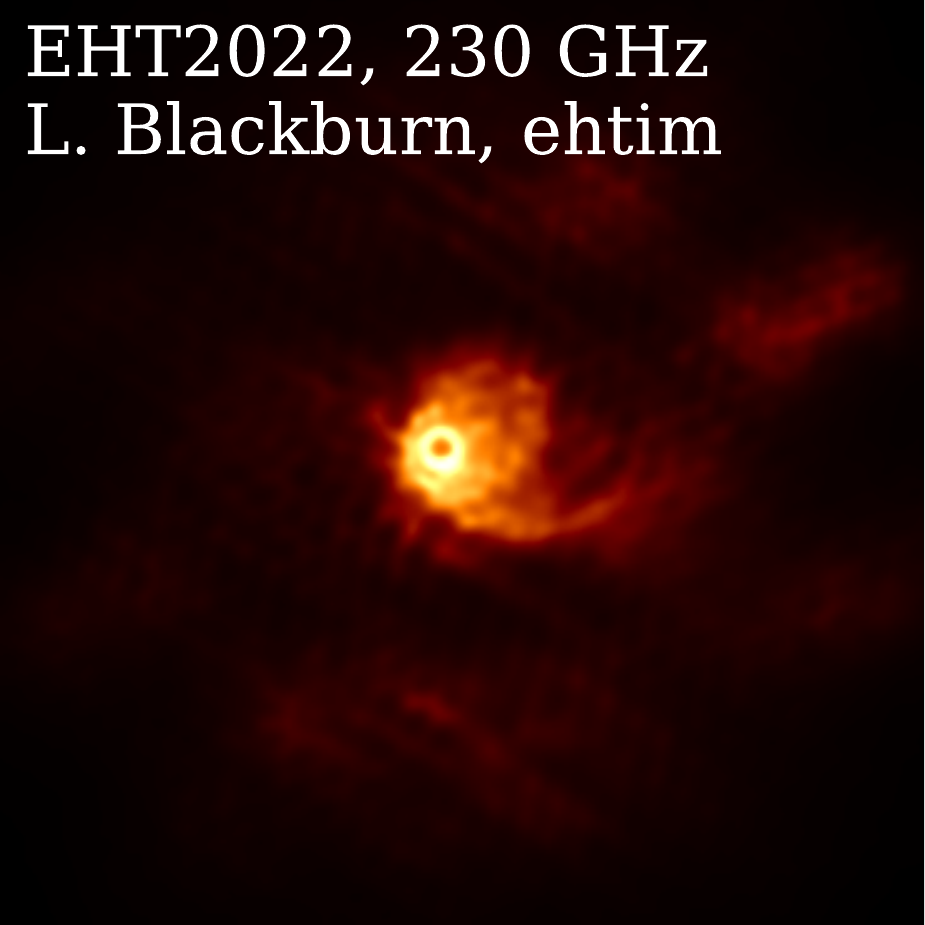}%
\includegraphics[width=25mm]{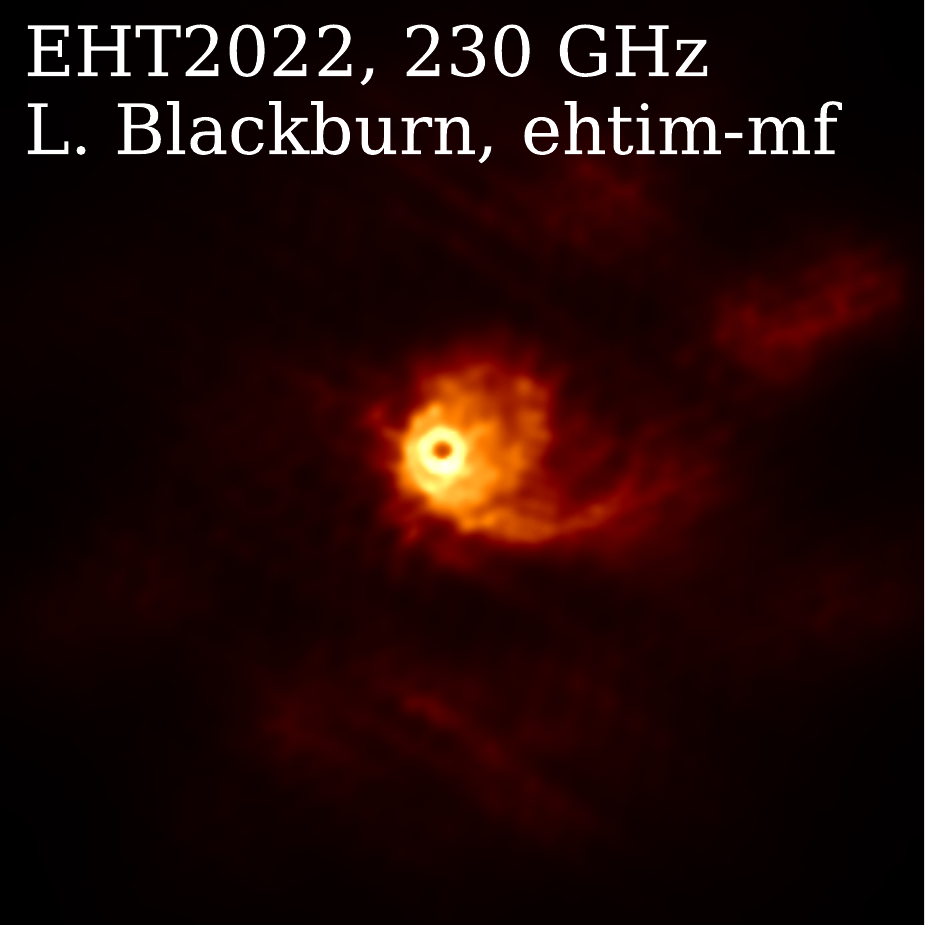} \\
\includegraphics[width=25mm]{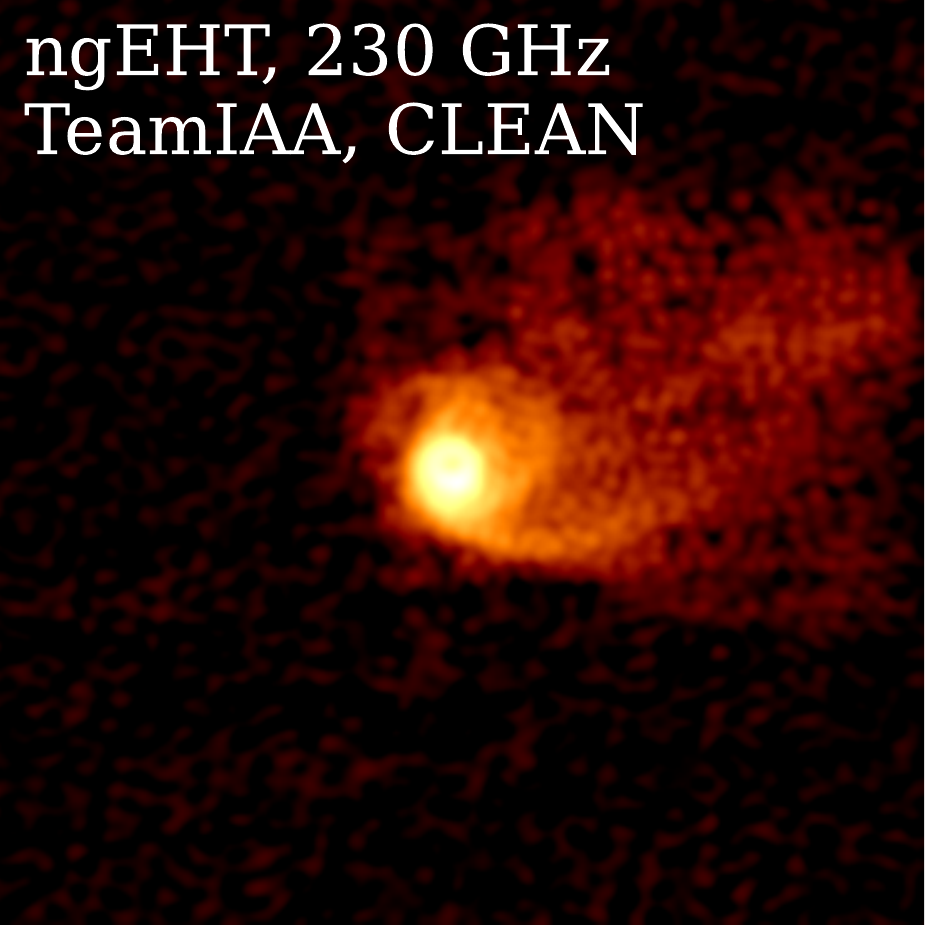}% 
\includegraphics[width=25mm]{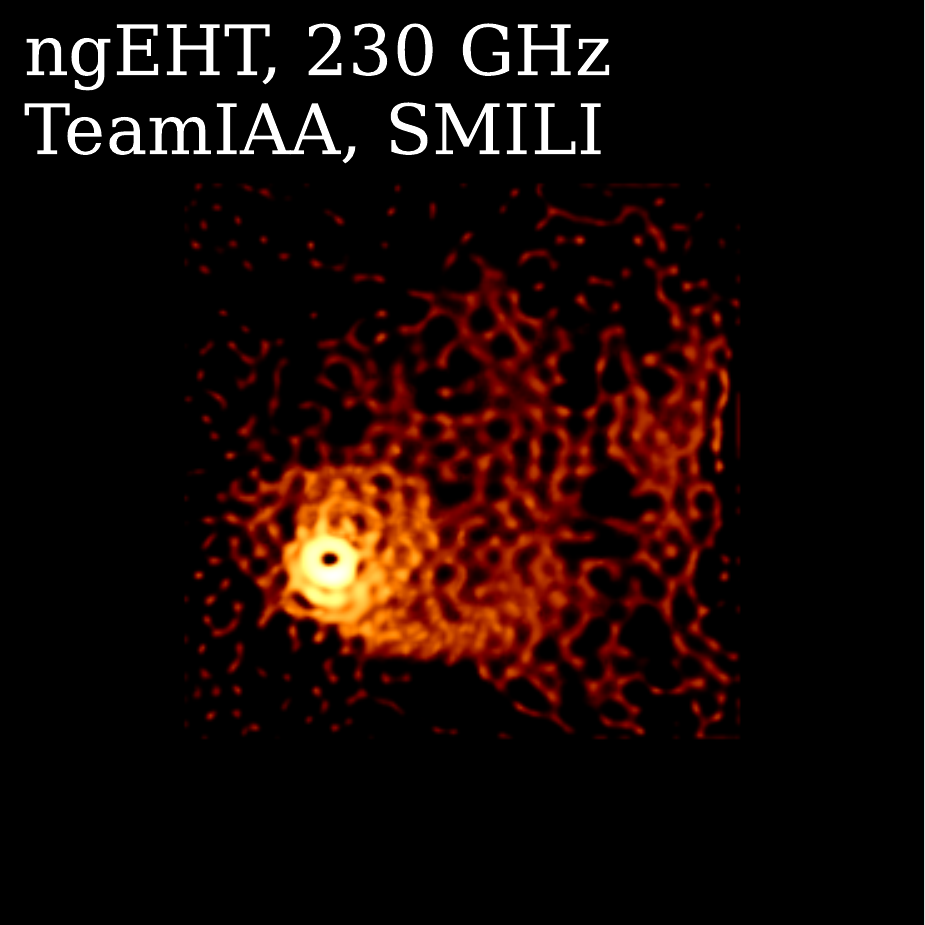}%
\includegraphics[width=25mm]{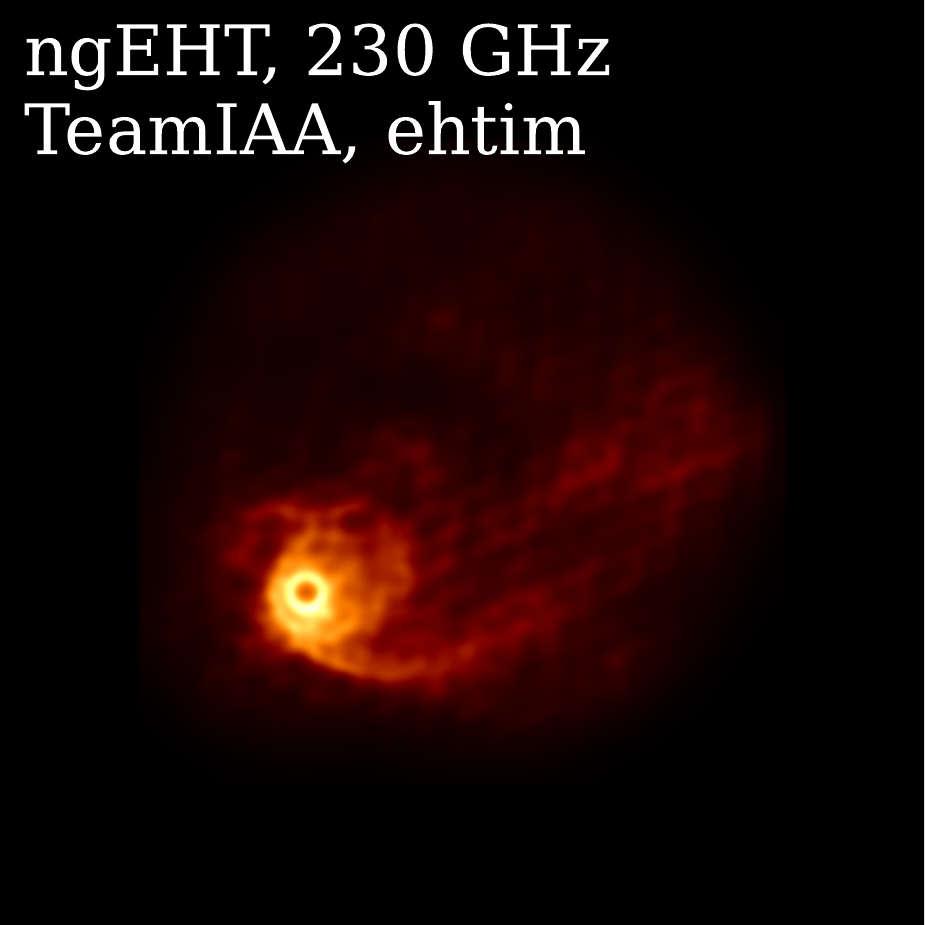}%
\includegraphics[width=25mm]{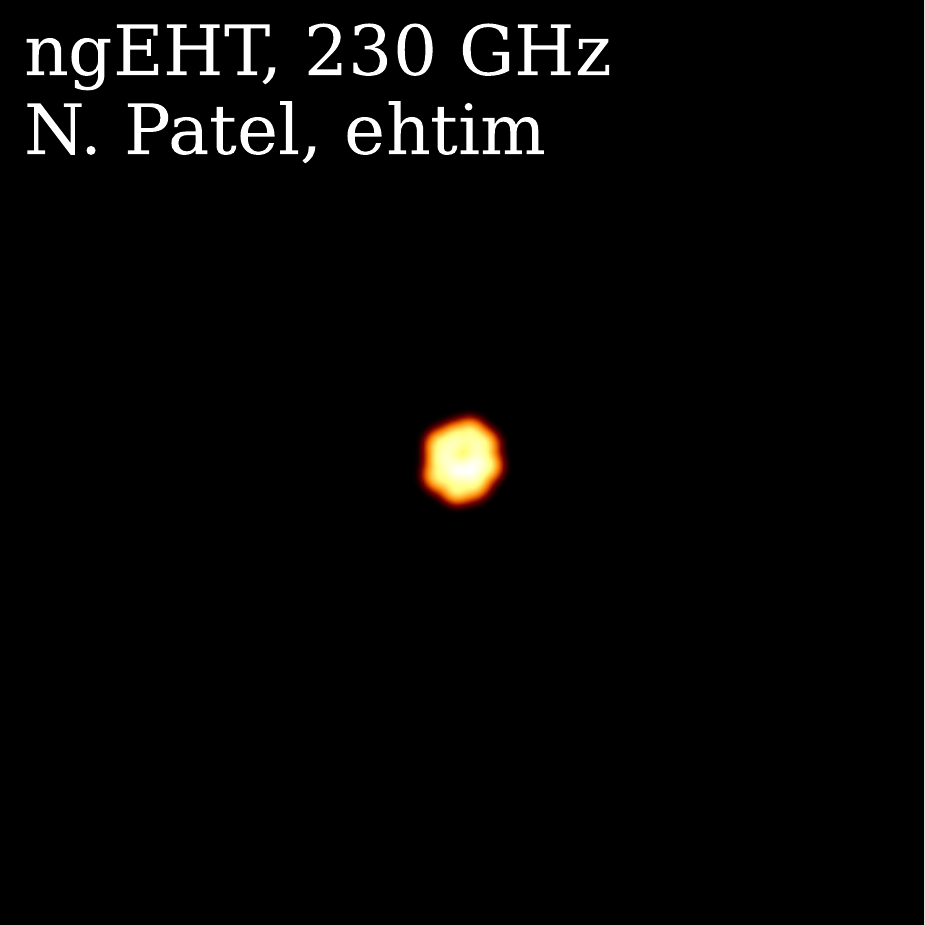}%
\includegraphics[width=25mm]{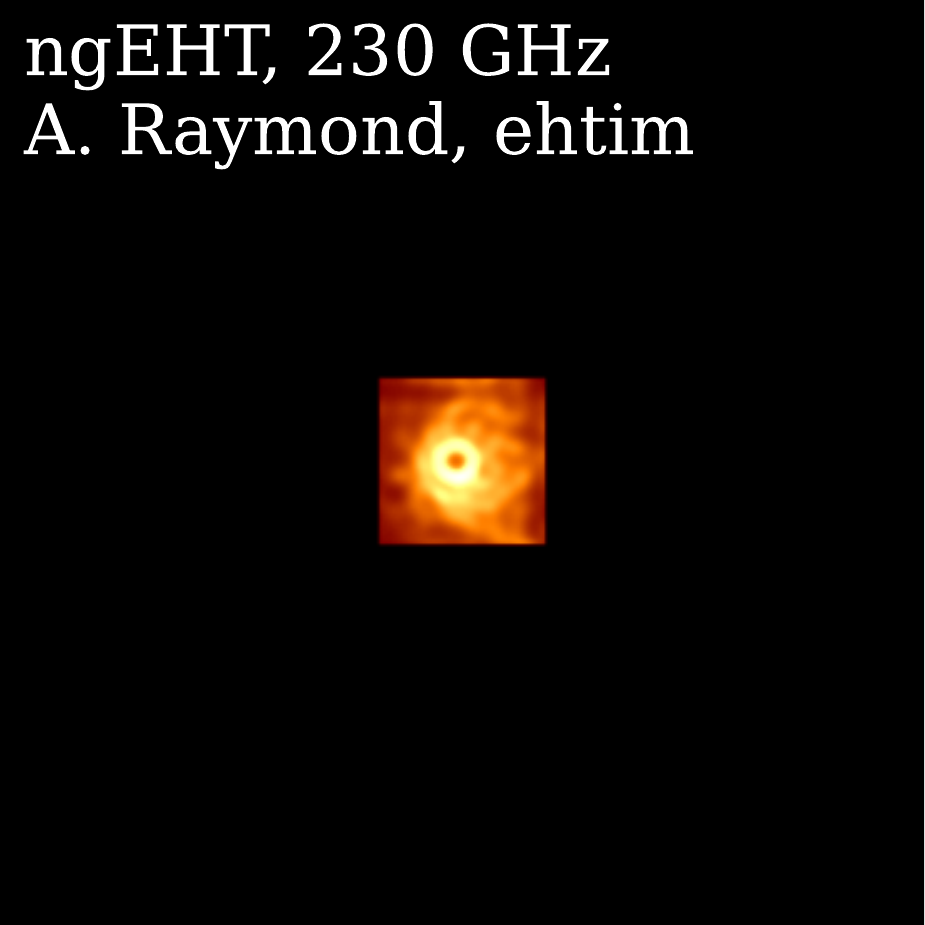}%
\includegraphics[width=25mm]{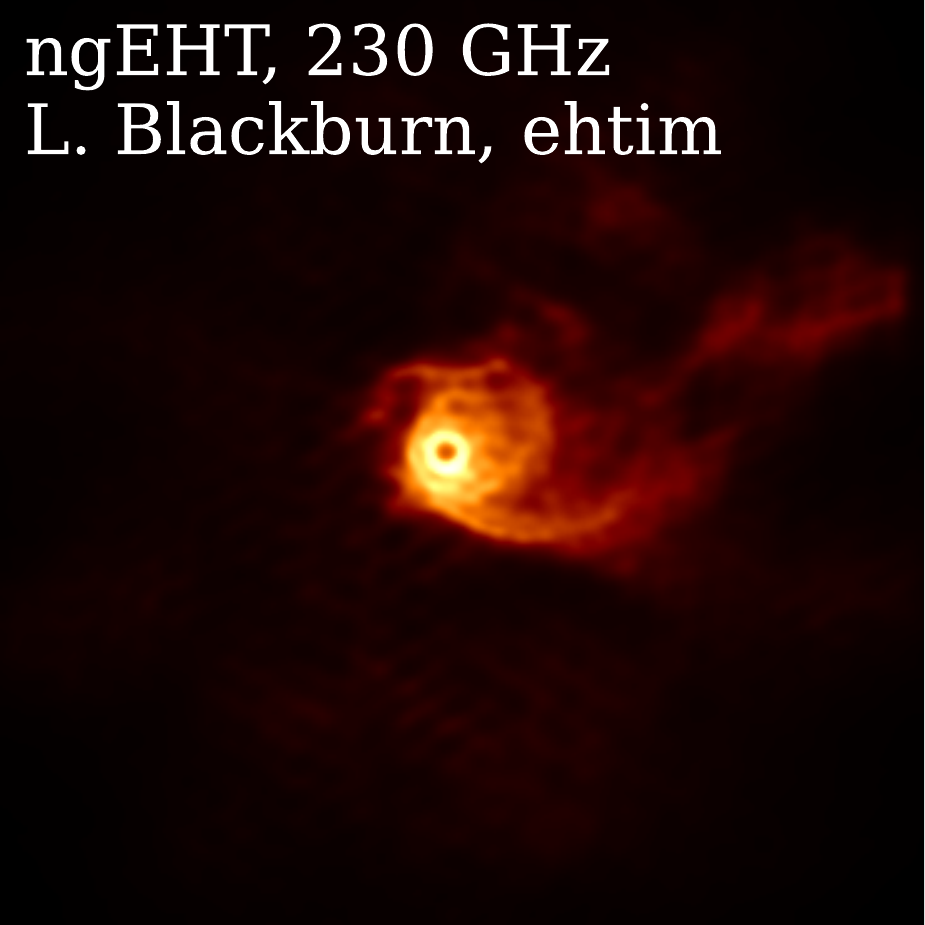}% 
\includegraphics[width=25mm]{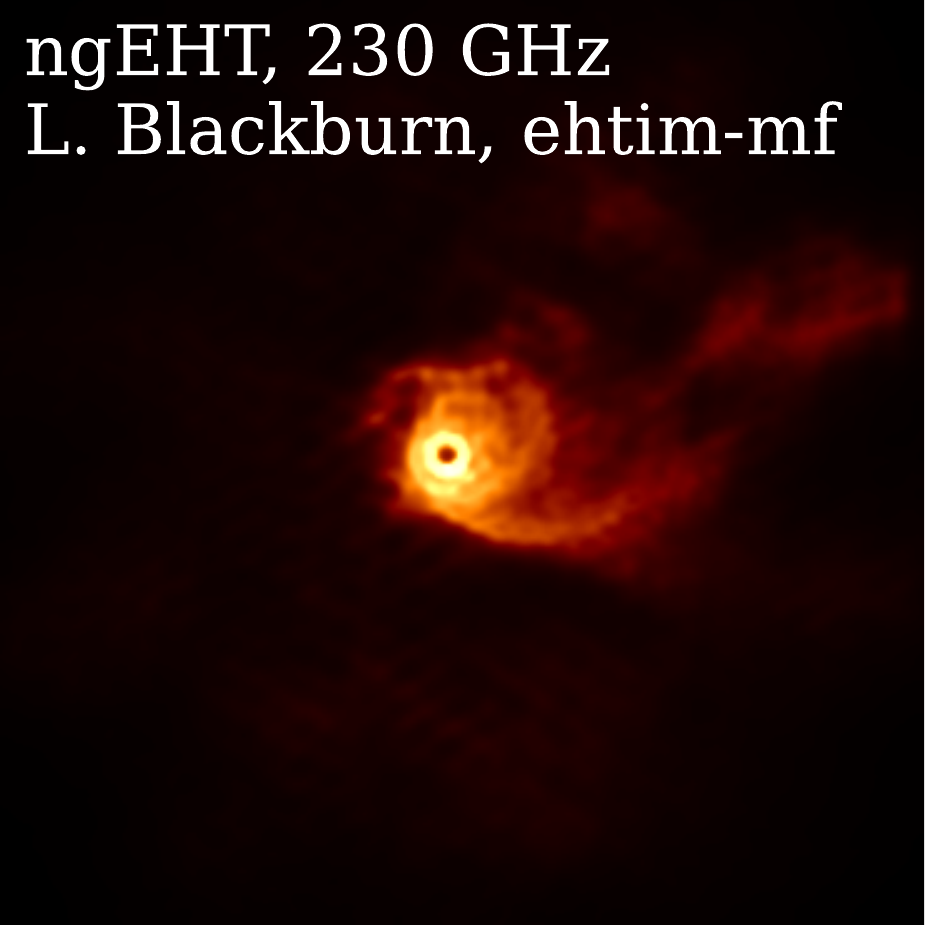} \\
\includegraphics[width=25mm]{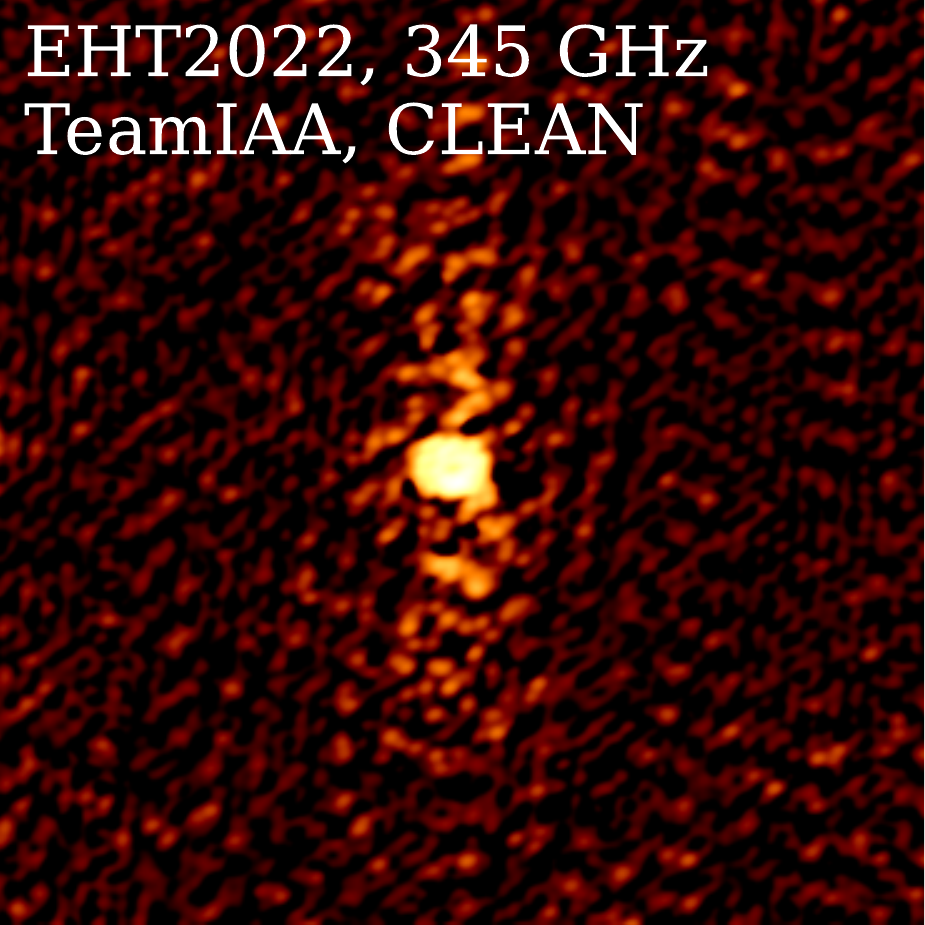}% 
\includegraphics[width=25mm]{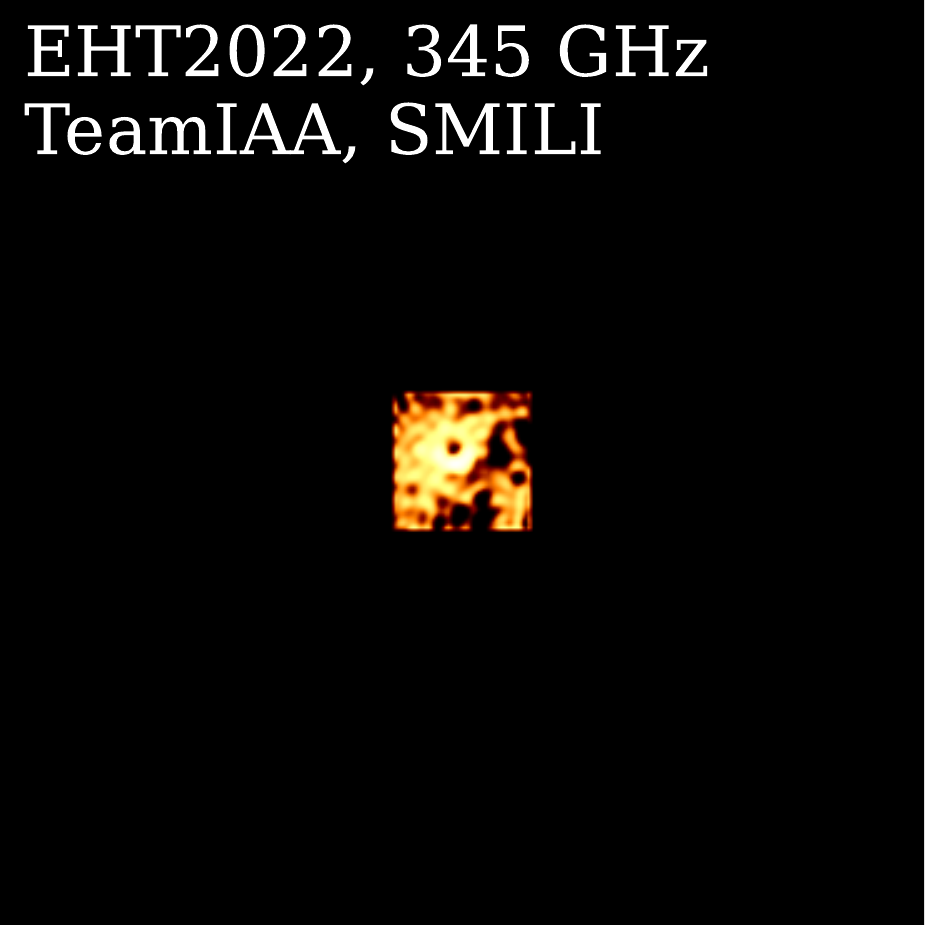}%
\includegraphics[width=25mm]{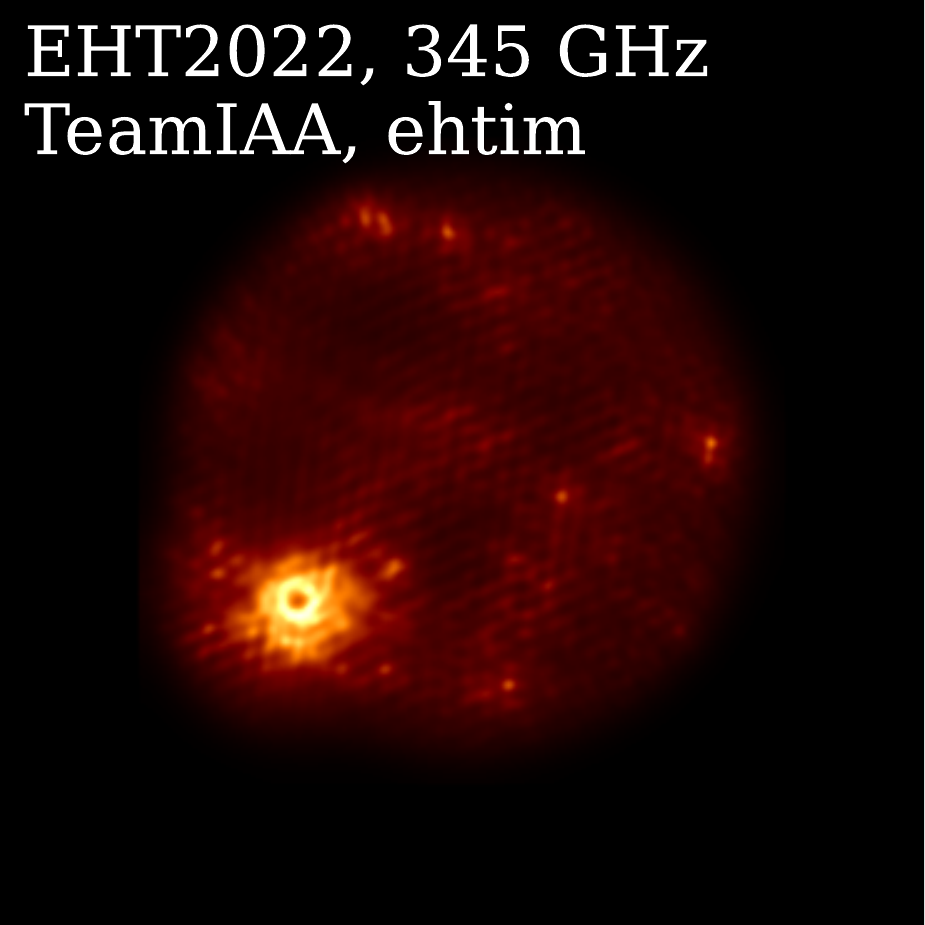}%
\includegraphics[width=25mm]{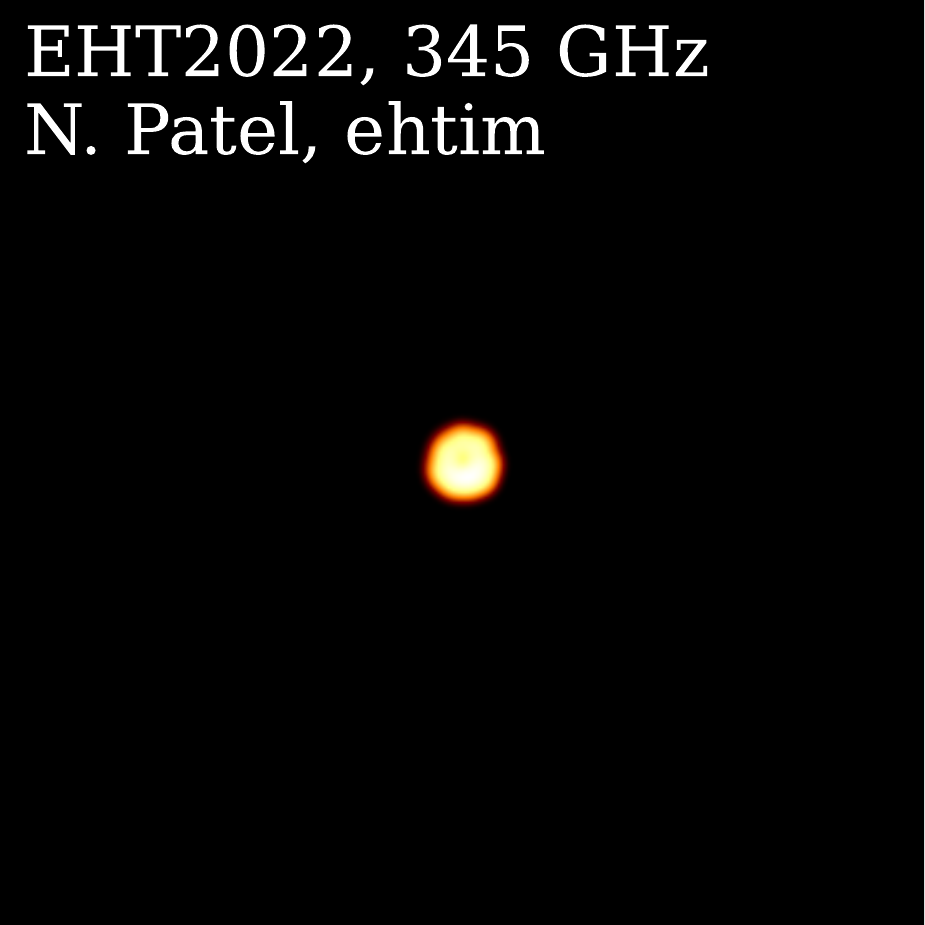}%
\includegraphics[width=25mm]{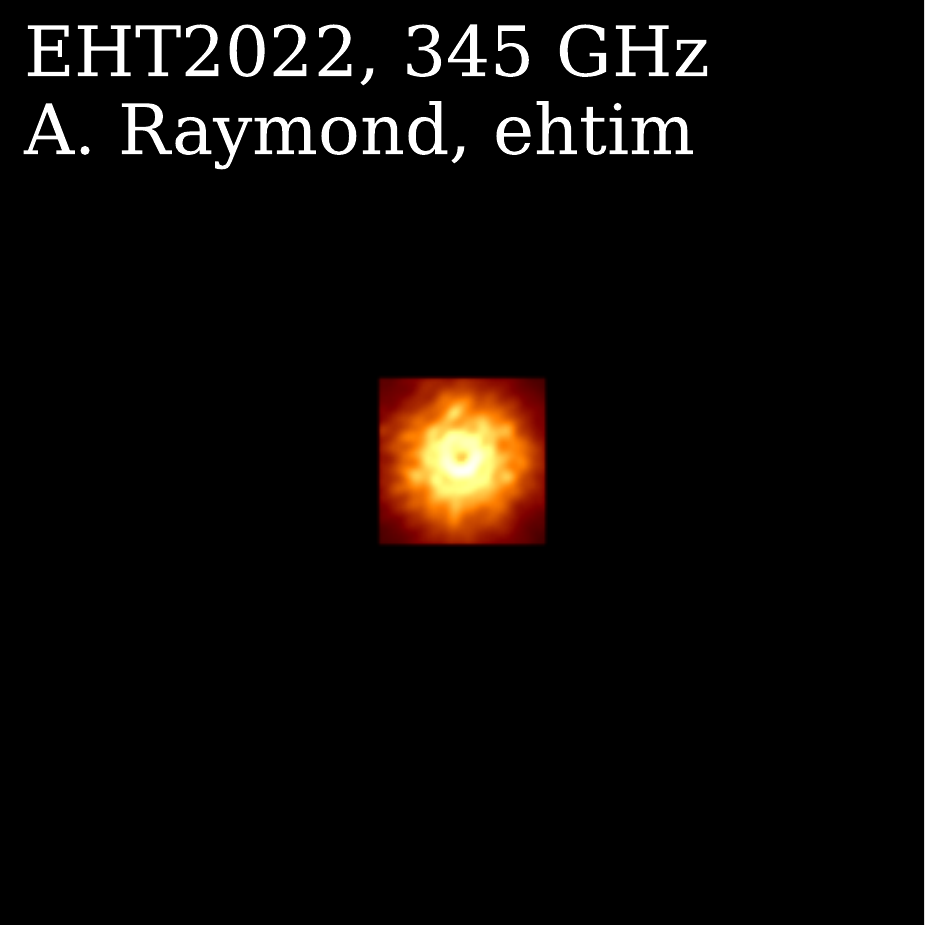}%
\includegraphics[width=25mm]{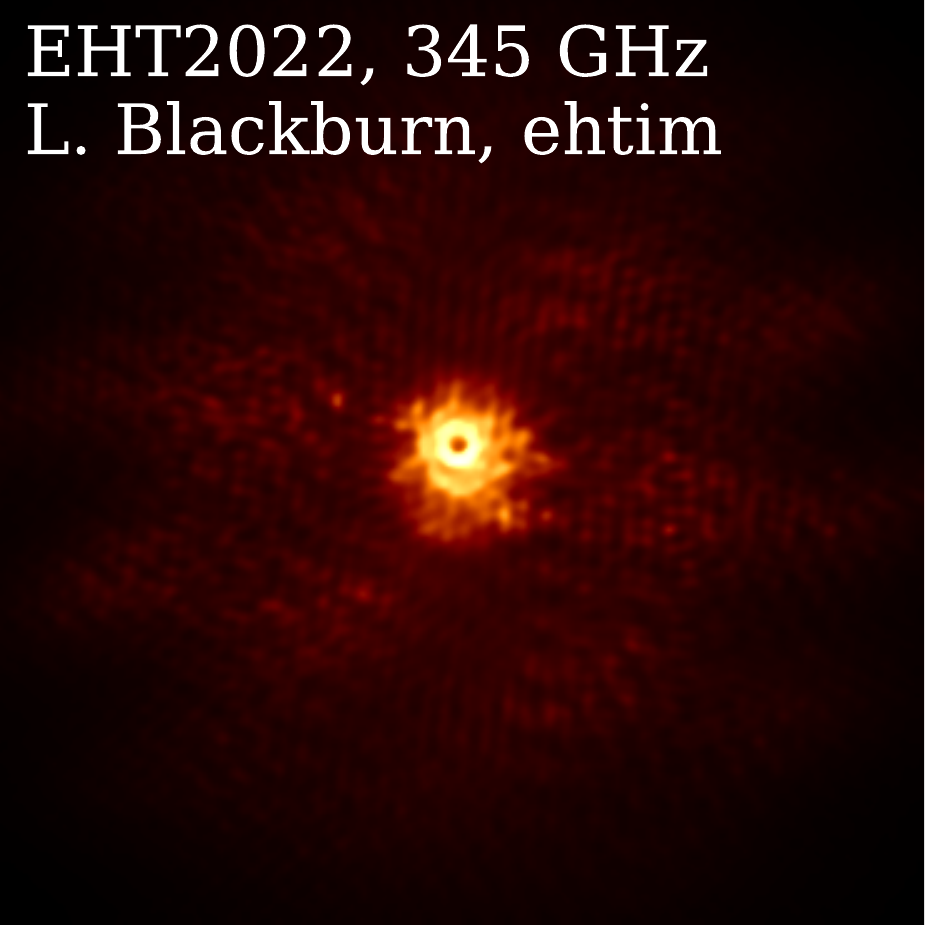}%
\includegraphics[width=25mm]{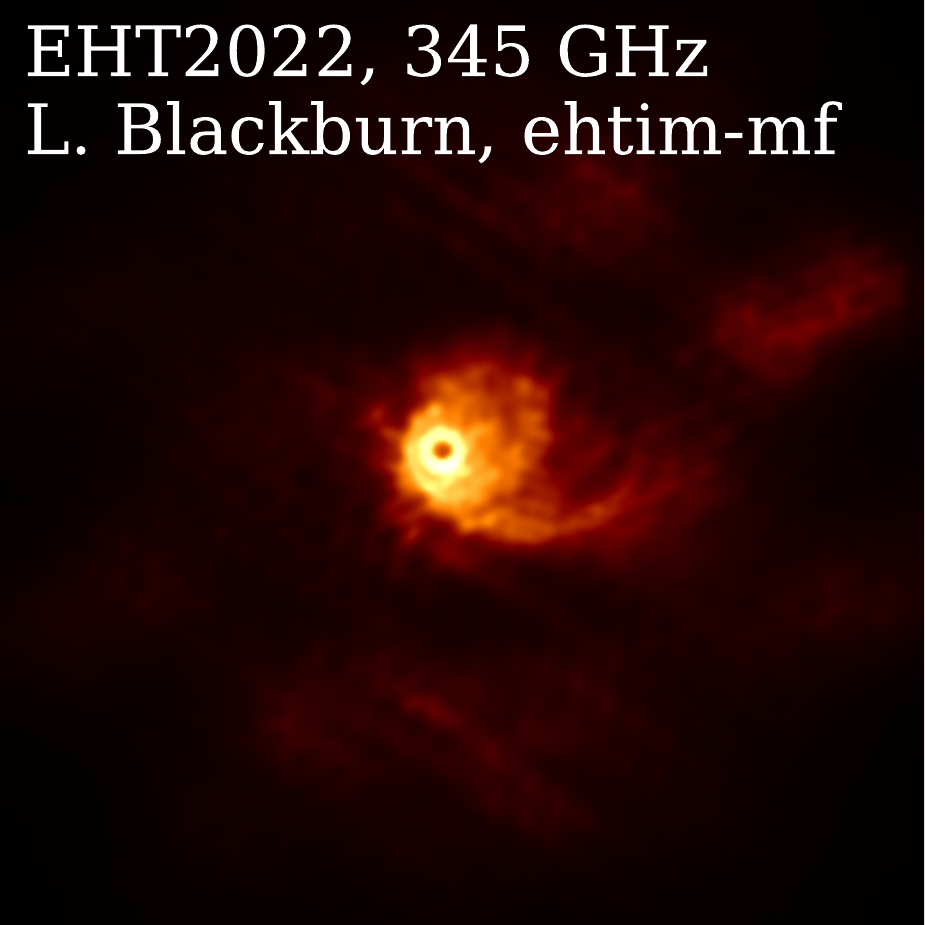} \\
\includegraphics[width=25mm]{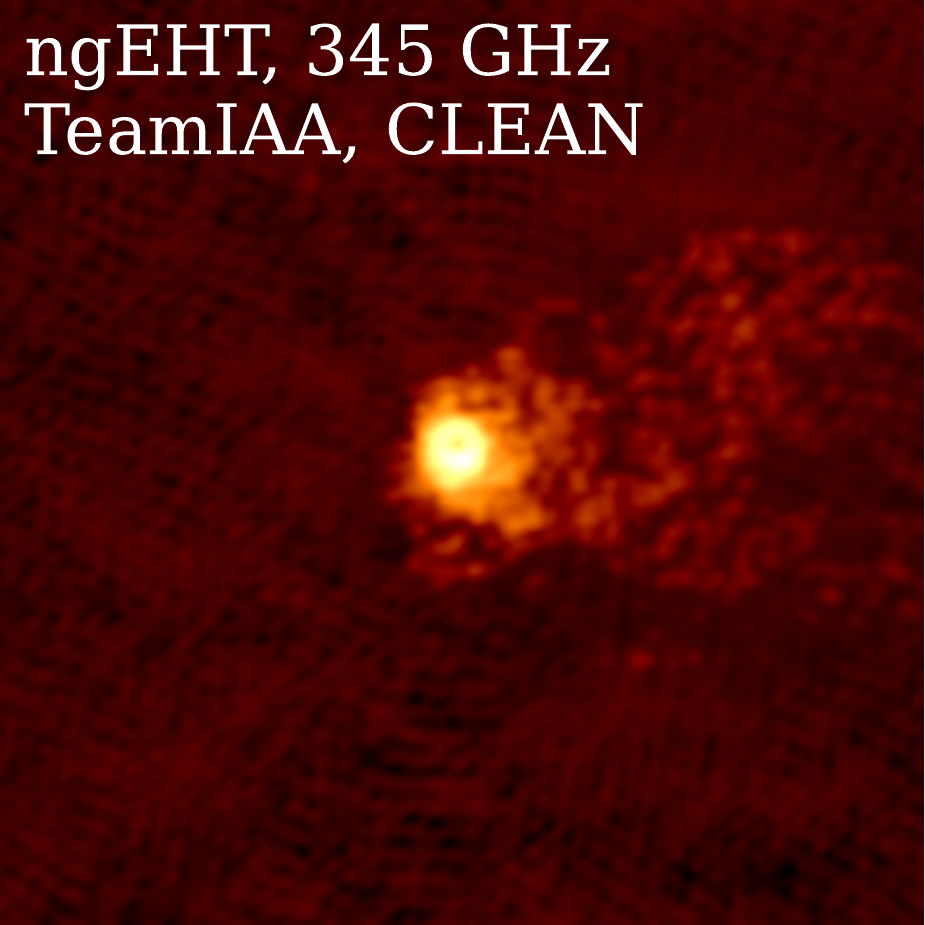}% 
\includegraphics[width=25mm]{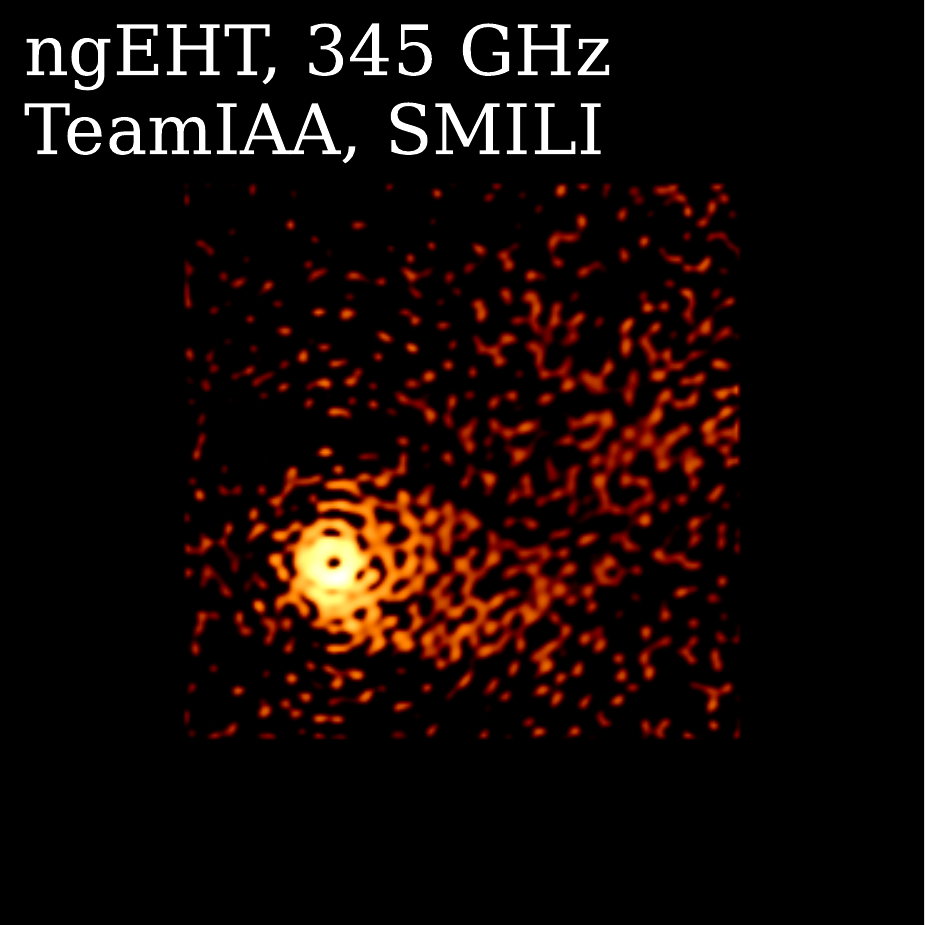}%
\includegraphics[width=25mm]{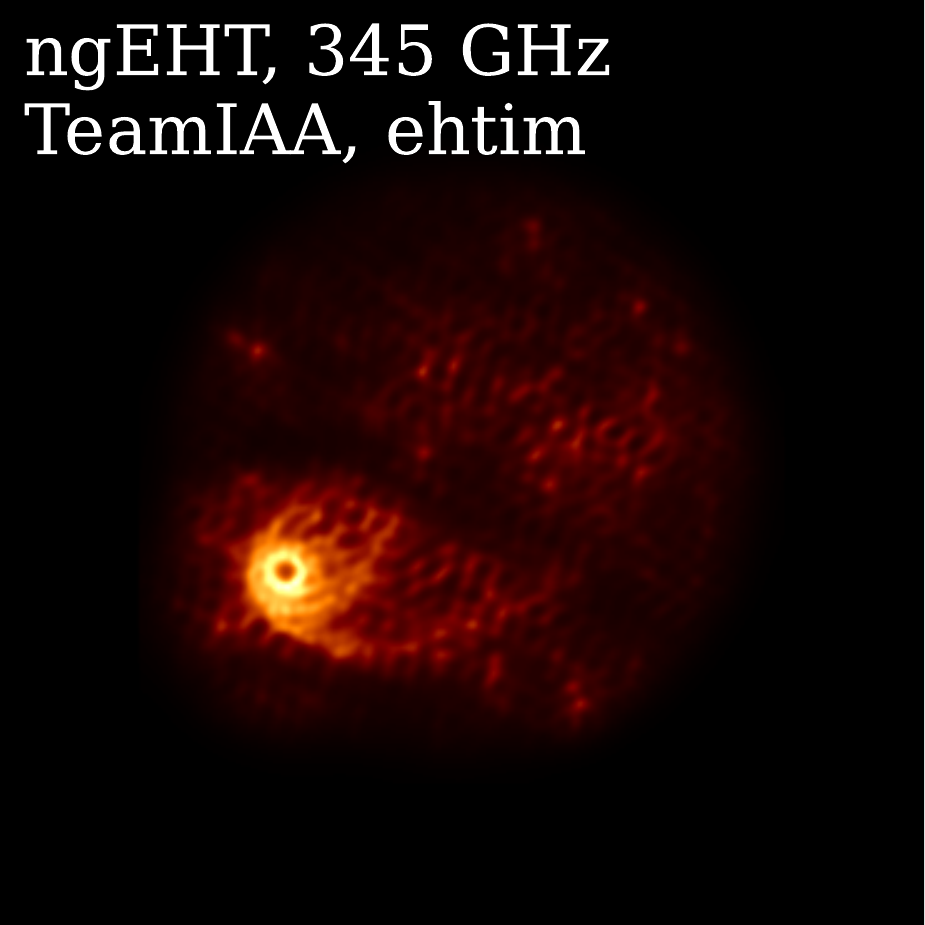}%
\includegraphics[width=25mm]{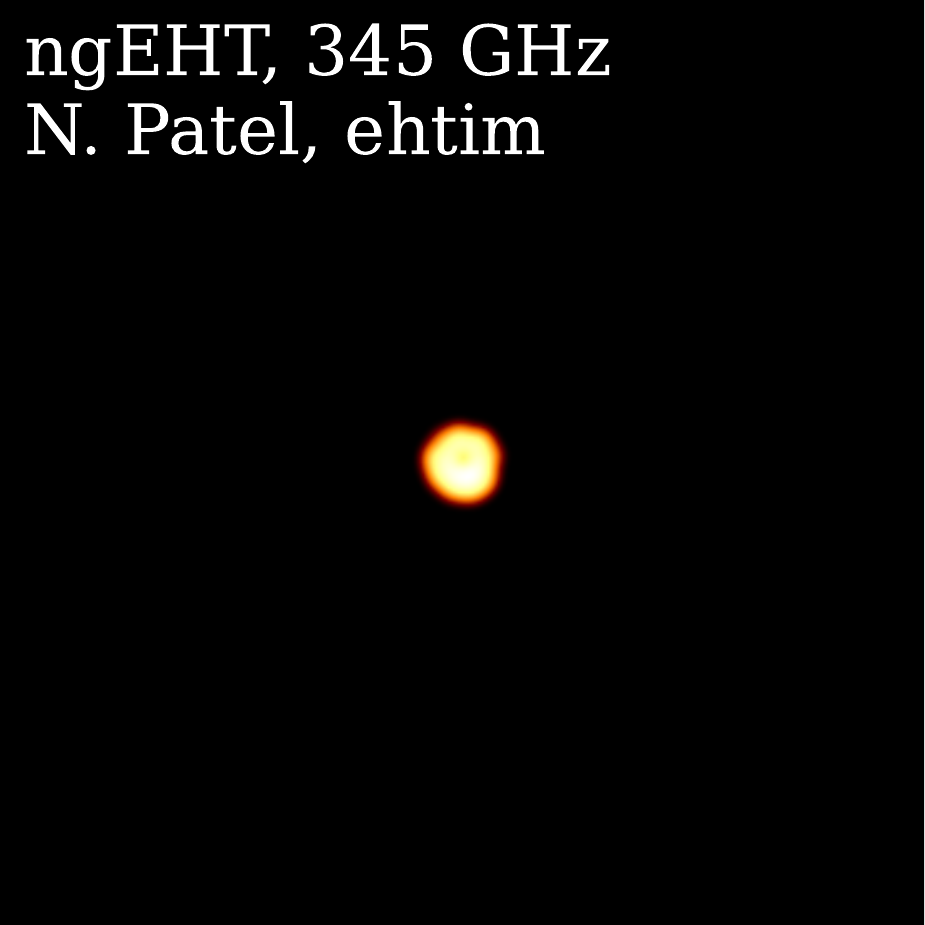}%
\includegraphics[width=25mm]{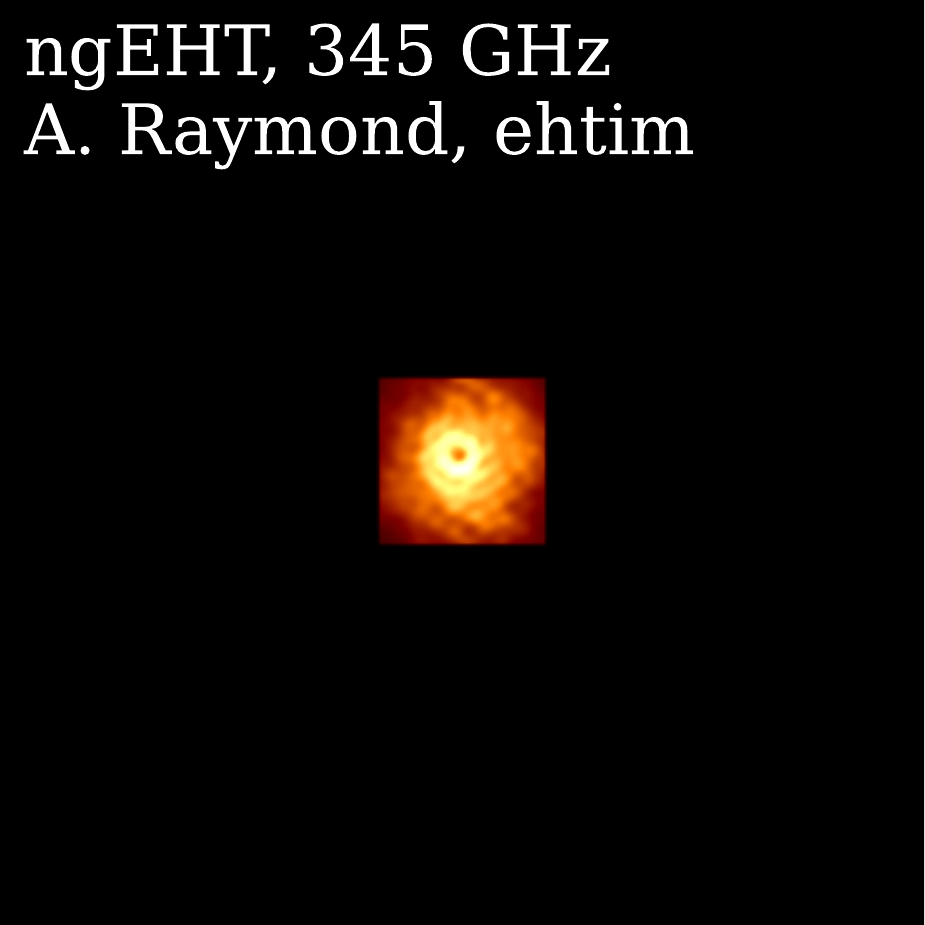}%
\includegraphics[width=25mm]{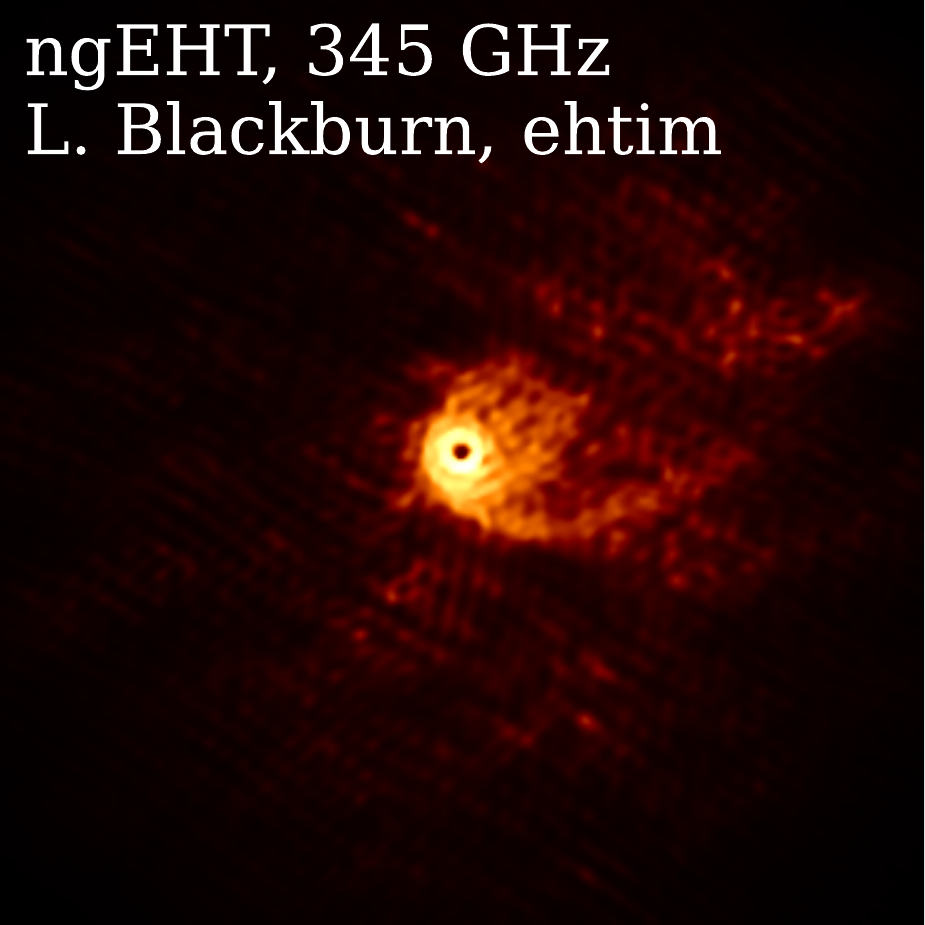}%
\includegraphics[width=25mm]{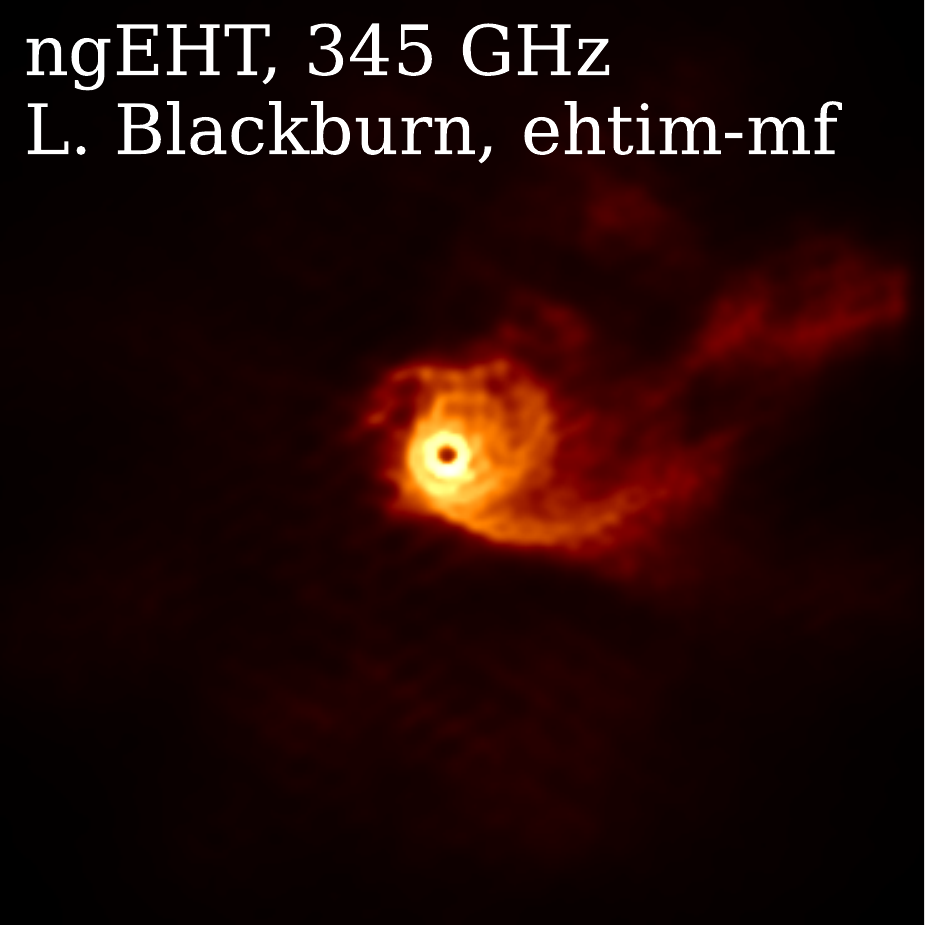} \\
\vspace{0.5cm}
\includegraphics[width=25mm]{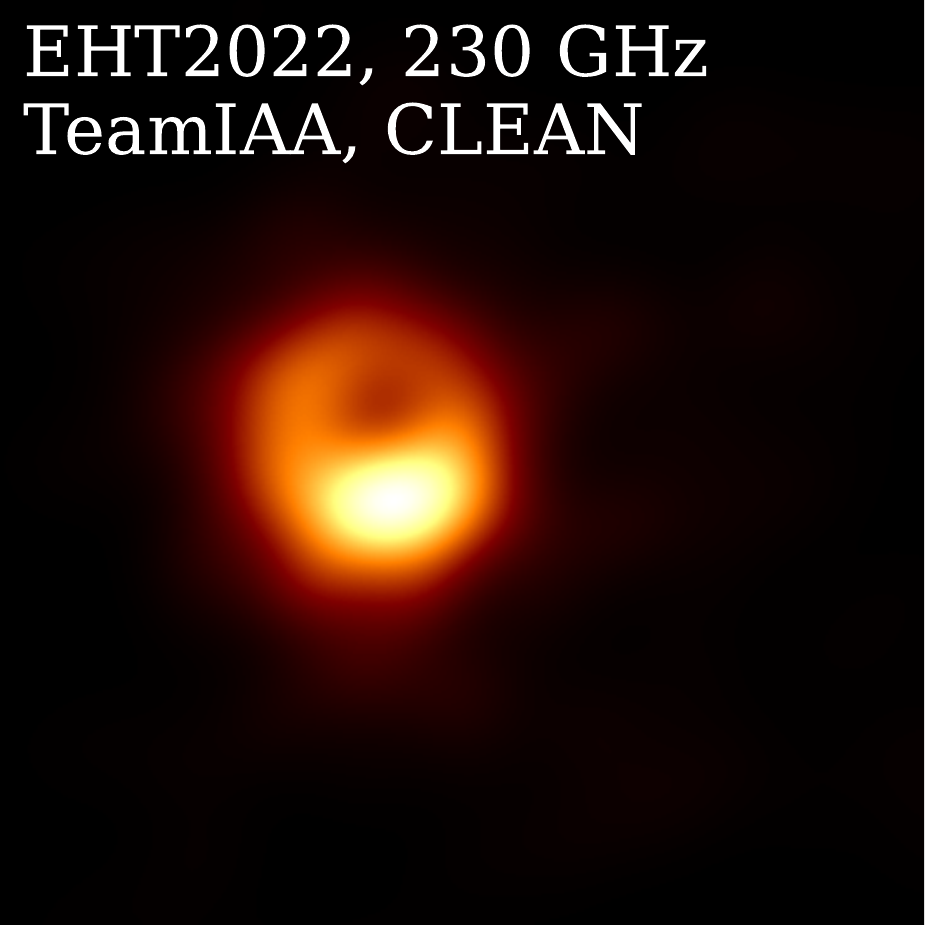}% 
\includegraphics[width=25mm]{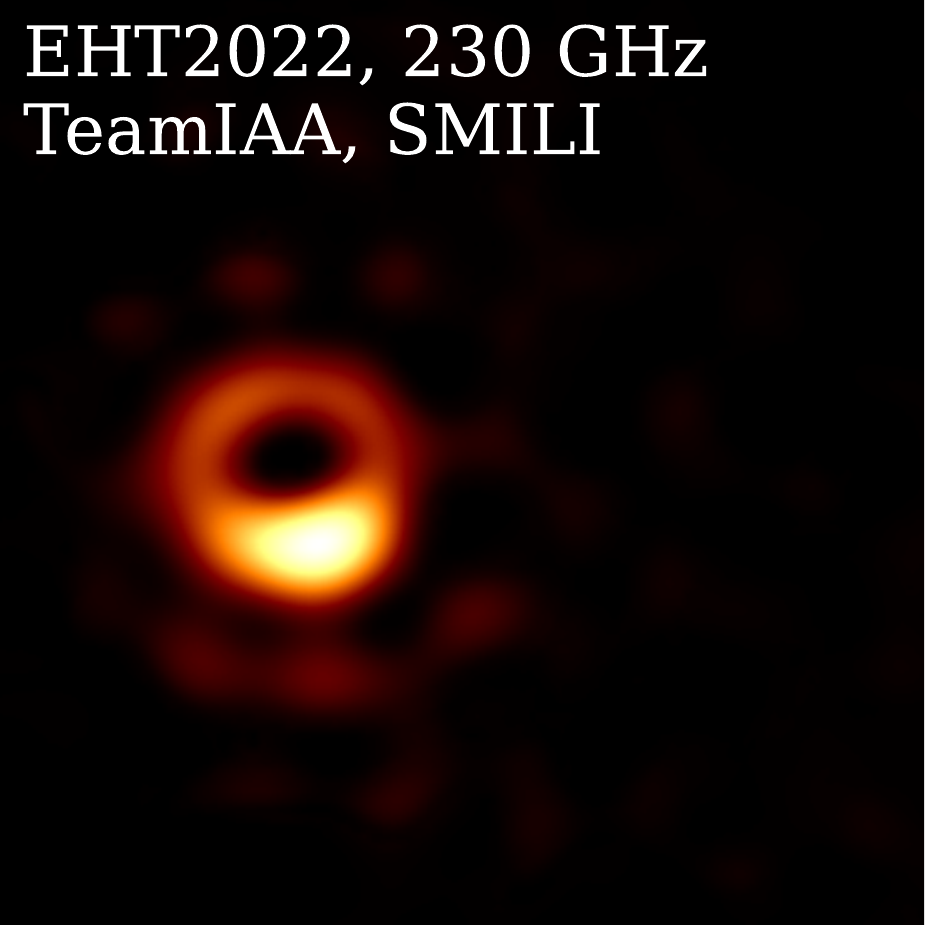}%
\includegraphics[width=25mm]{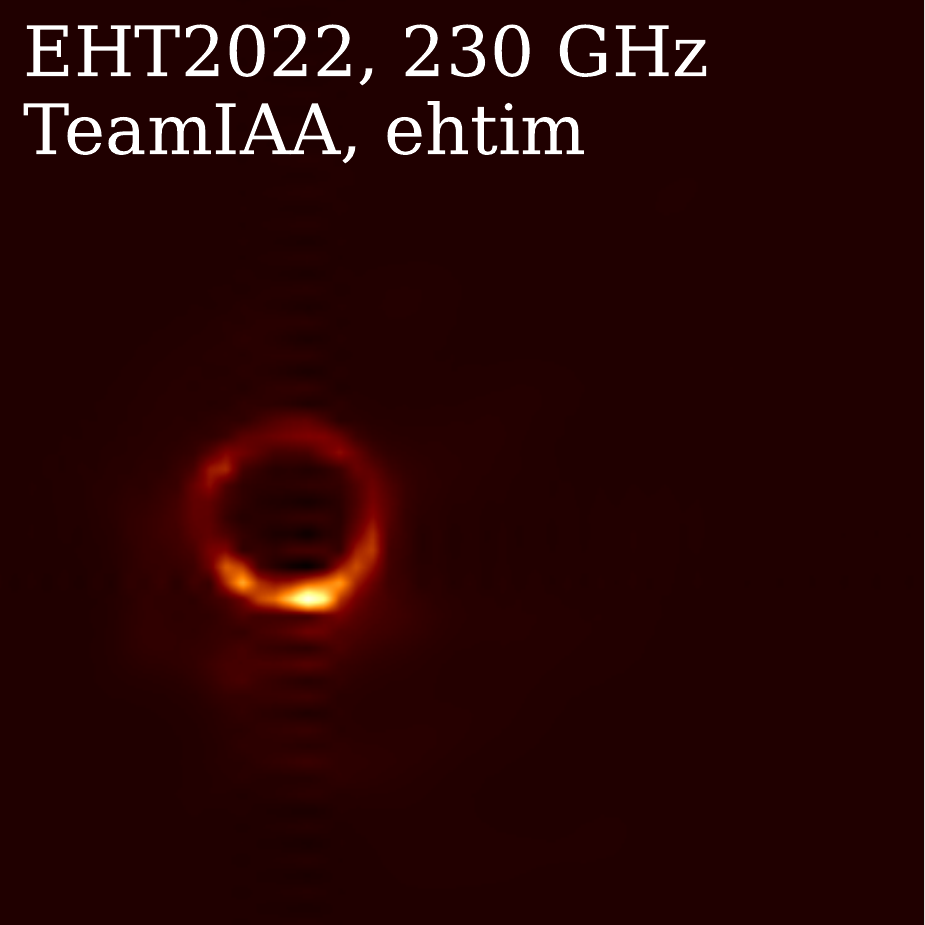}%
\includegraphics[width=25mm]{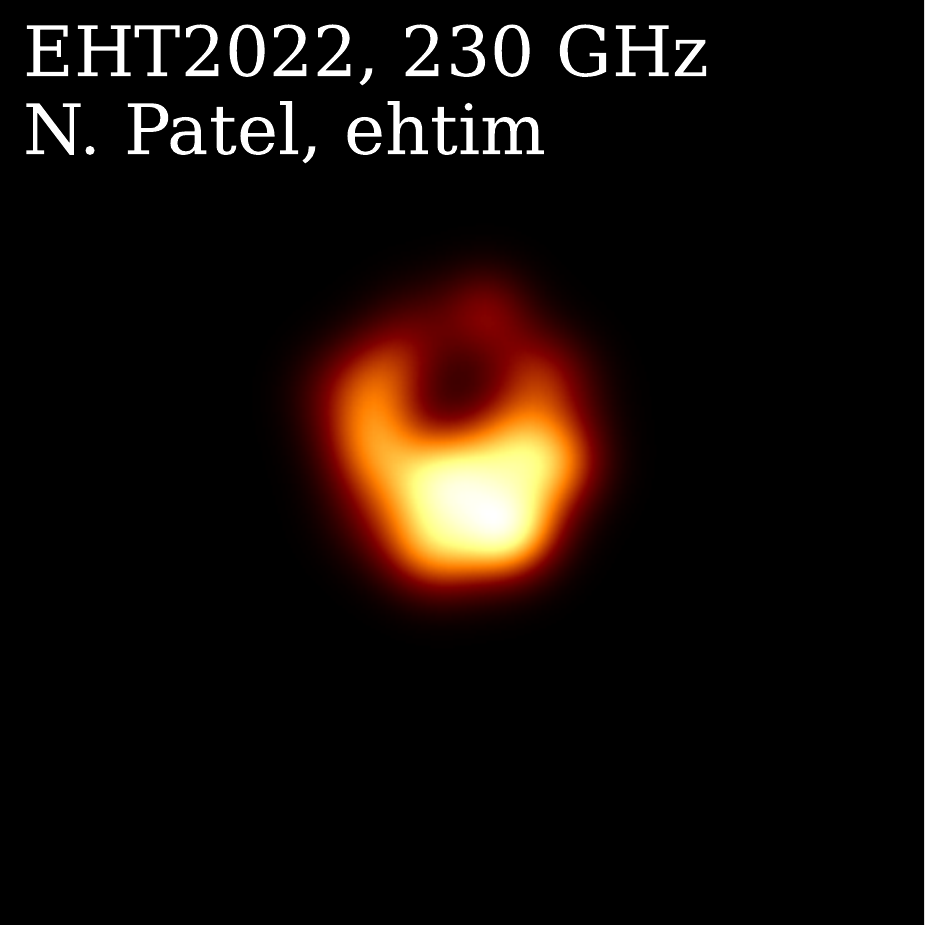}%
\includegraphics[width=25mm]{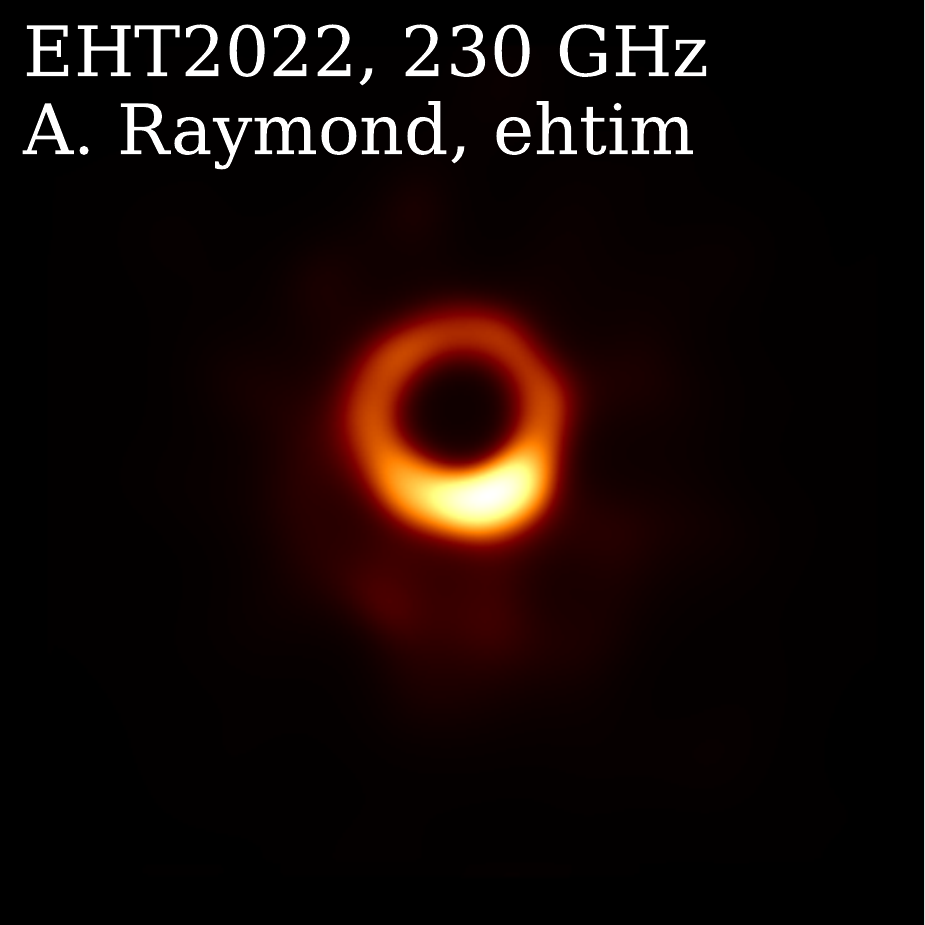}%
\includegraphics[width=25mm]{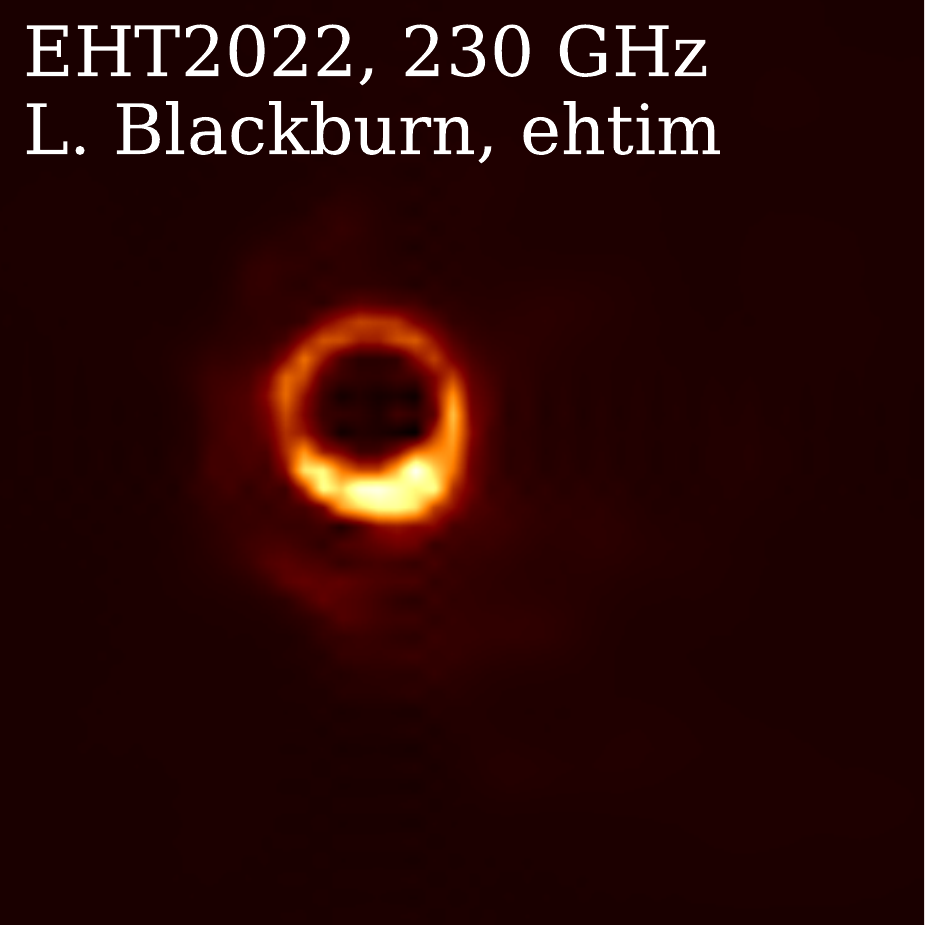}%
\includegraphics[width=25mm]{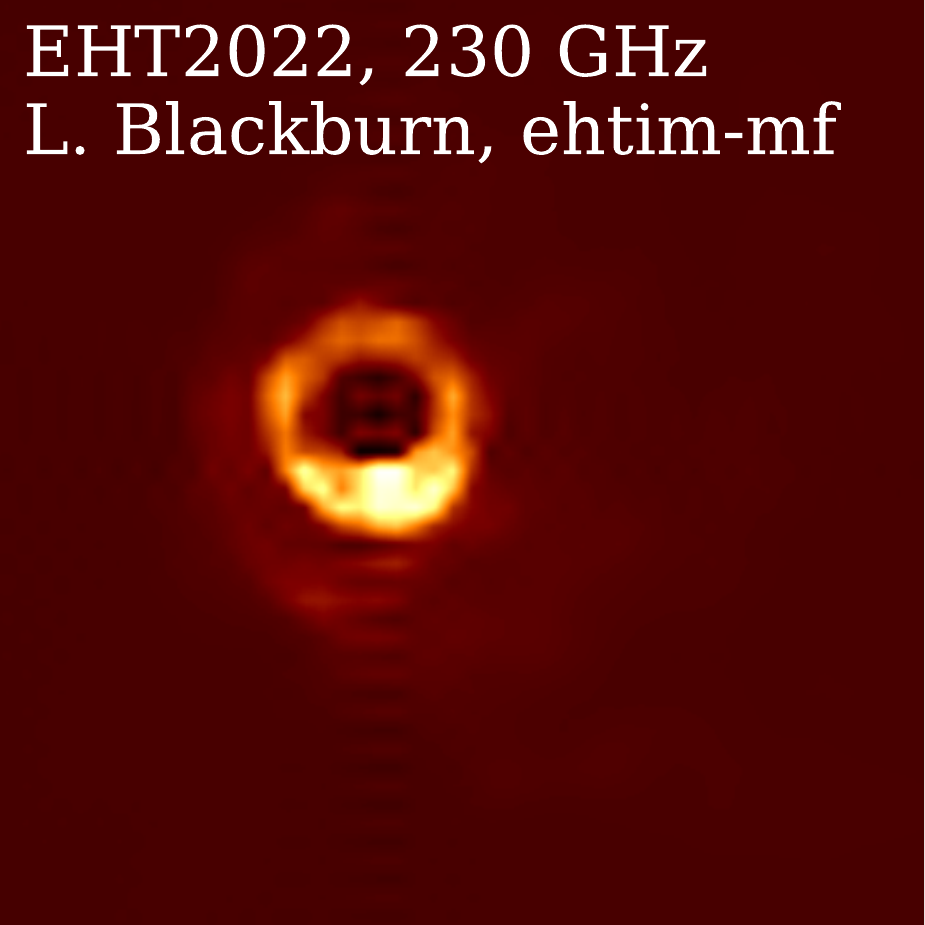} \\
\includegraphics[width=25mm]{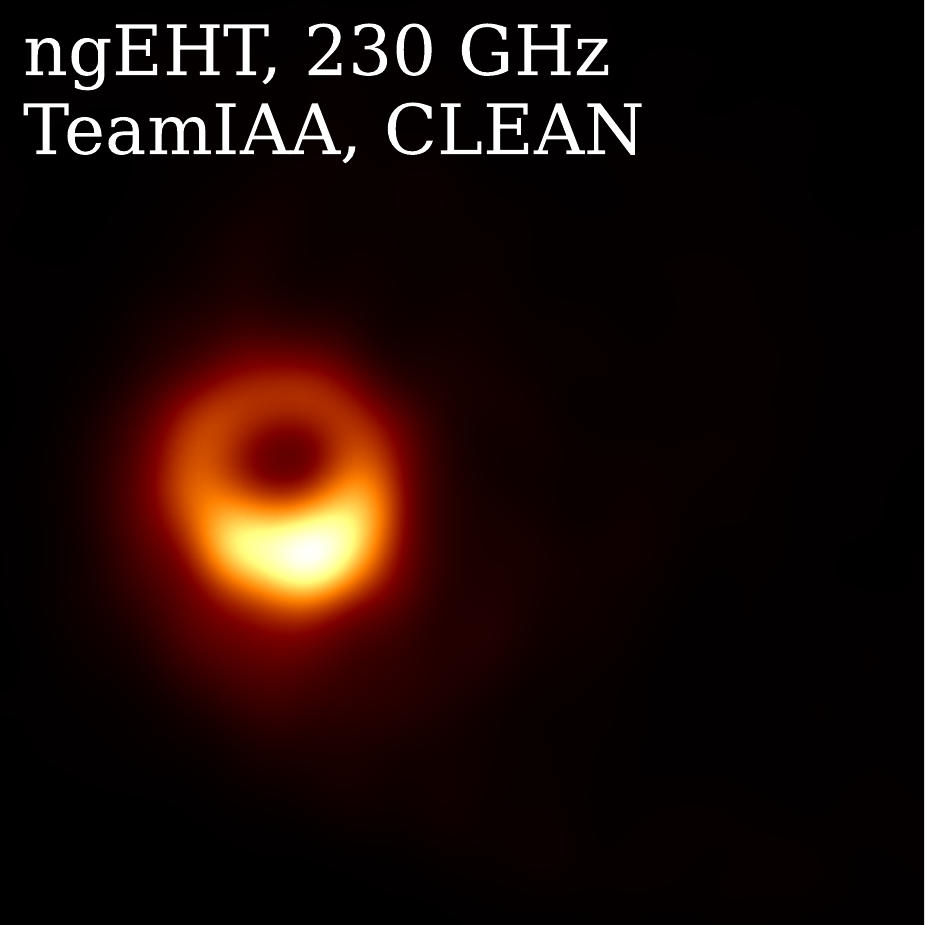}% 
\includegraphics[width=25mm]{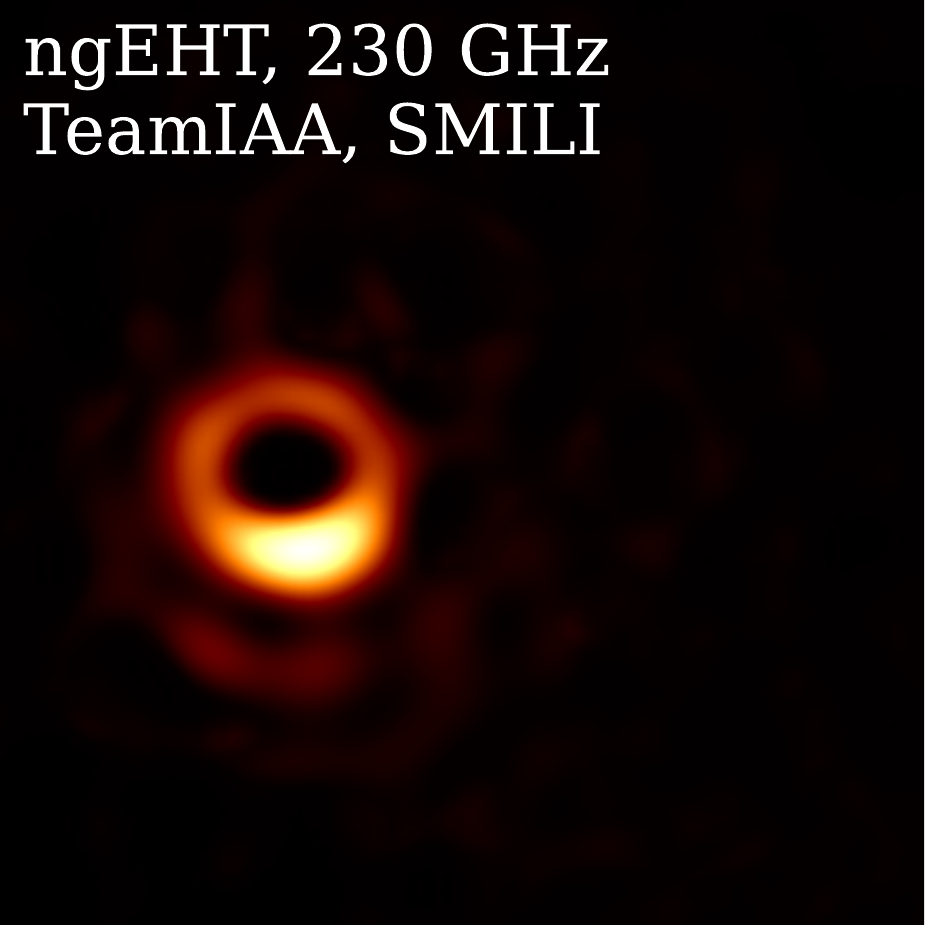}%
\includegraphics[width=25mm]{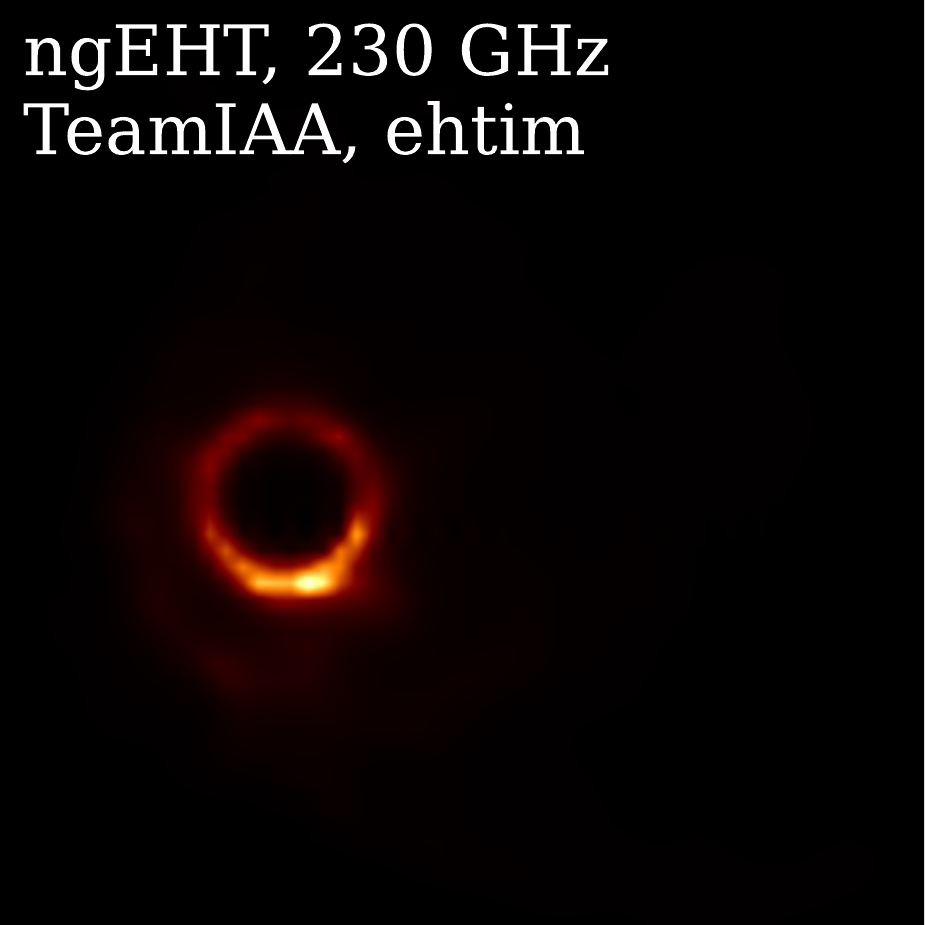}%
\includegraphics[width=25mm]{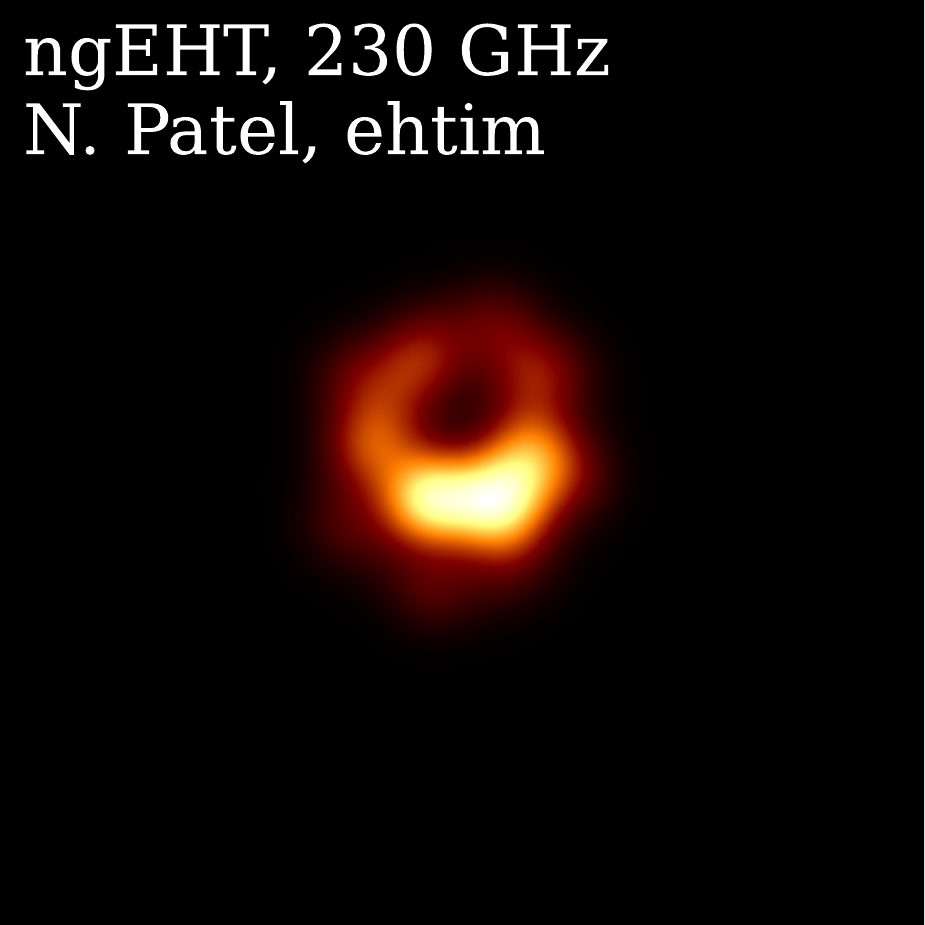}%
\includegraphics[width=25mm]{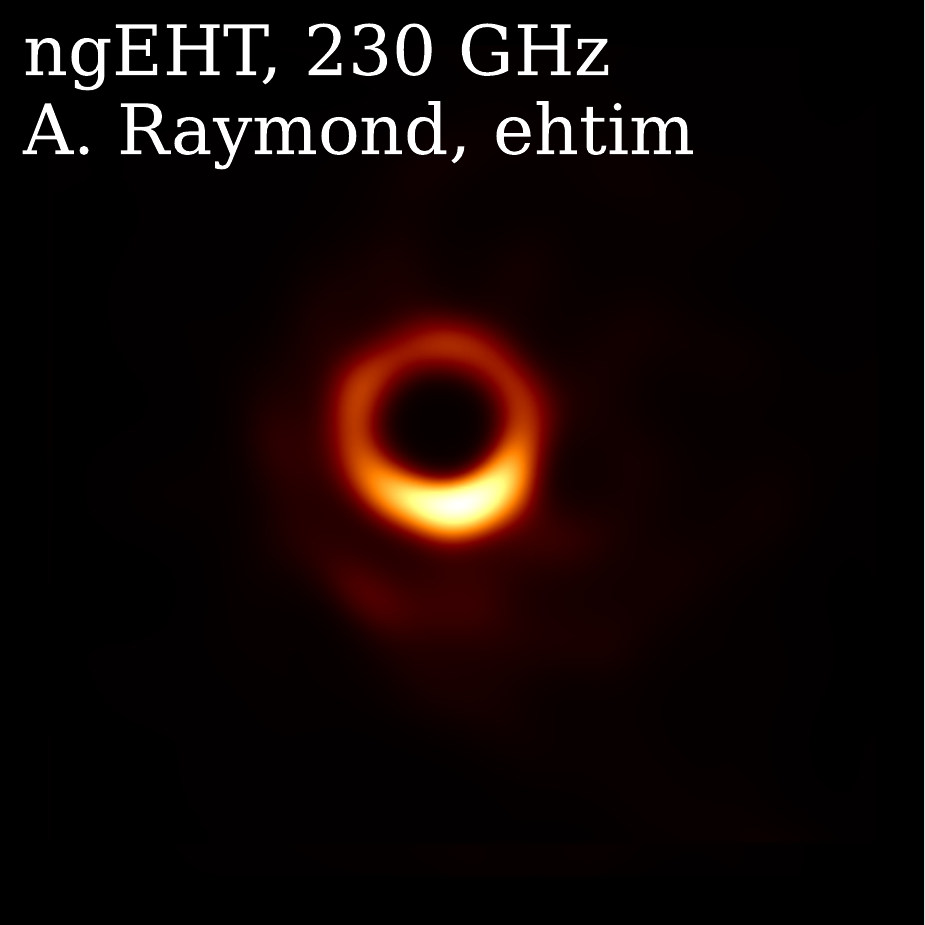}%
\includegraphics[width=25mm]{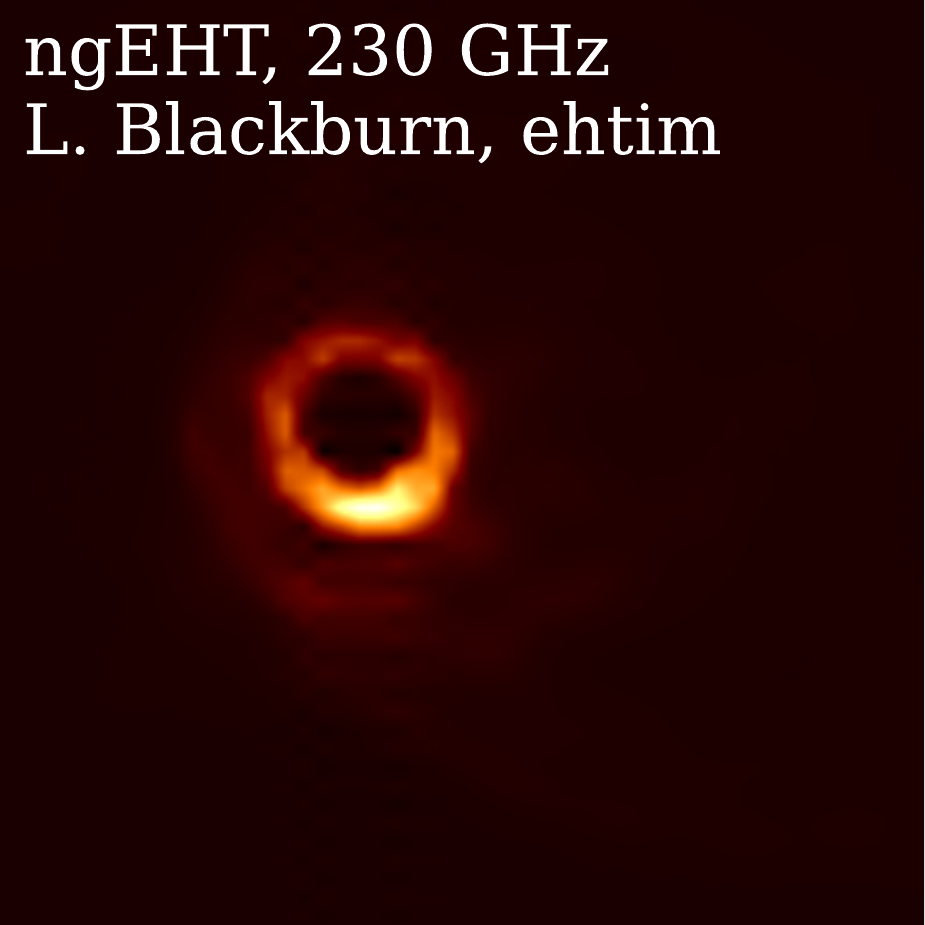}% 
\includegraphics[width=25mm]{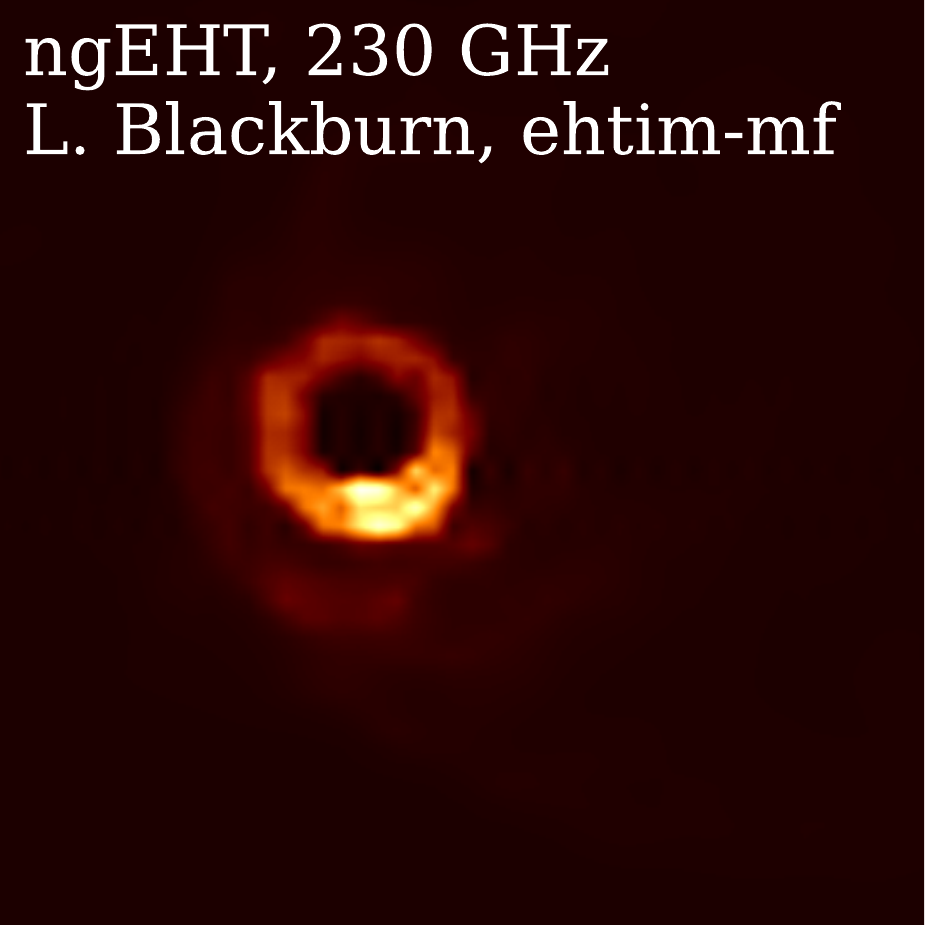} \\
\includegraphics[width=25mm]{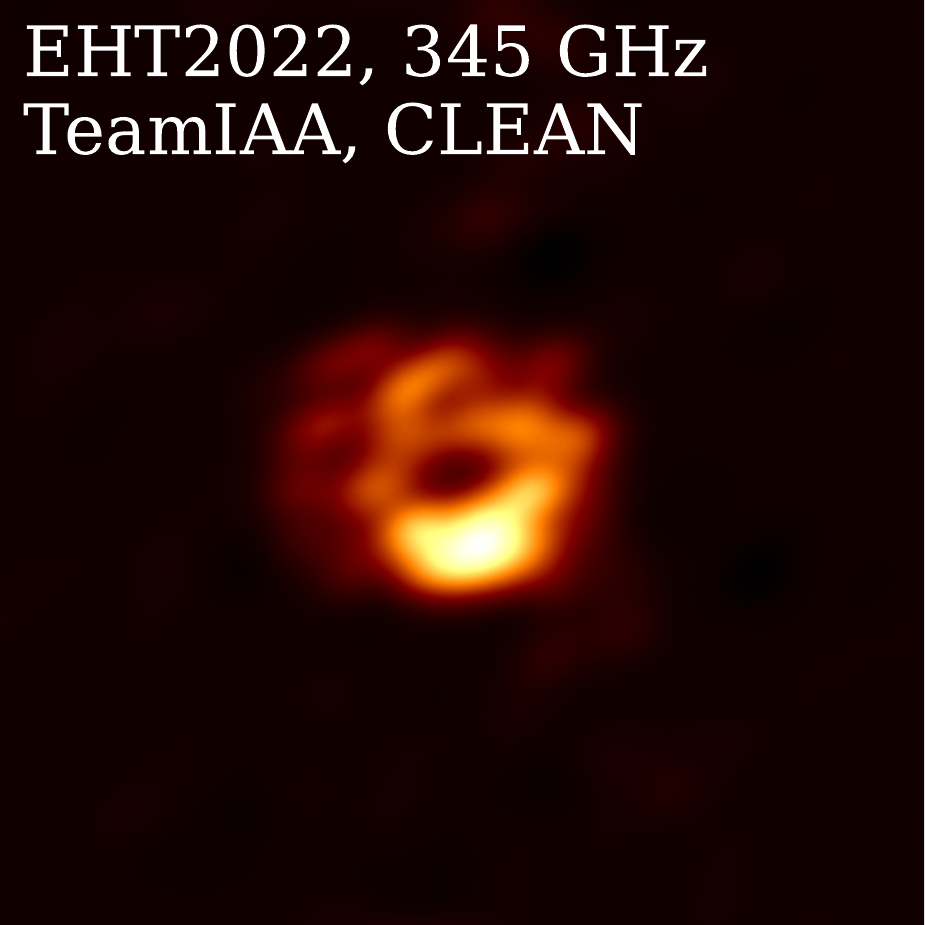}% 
\includegraphics[width=25mm]{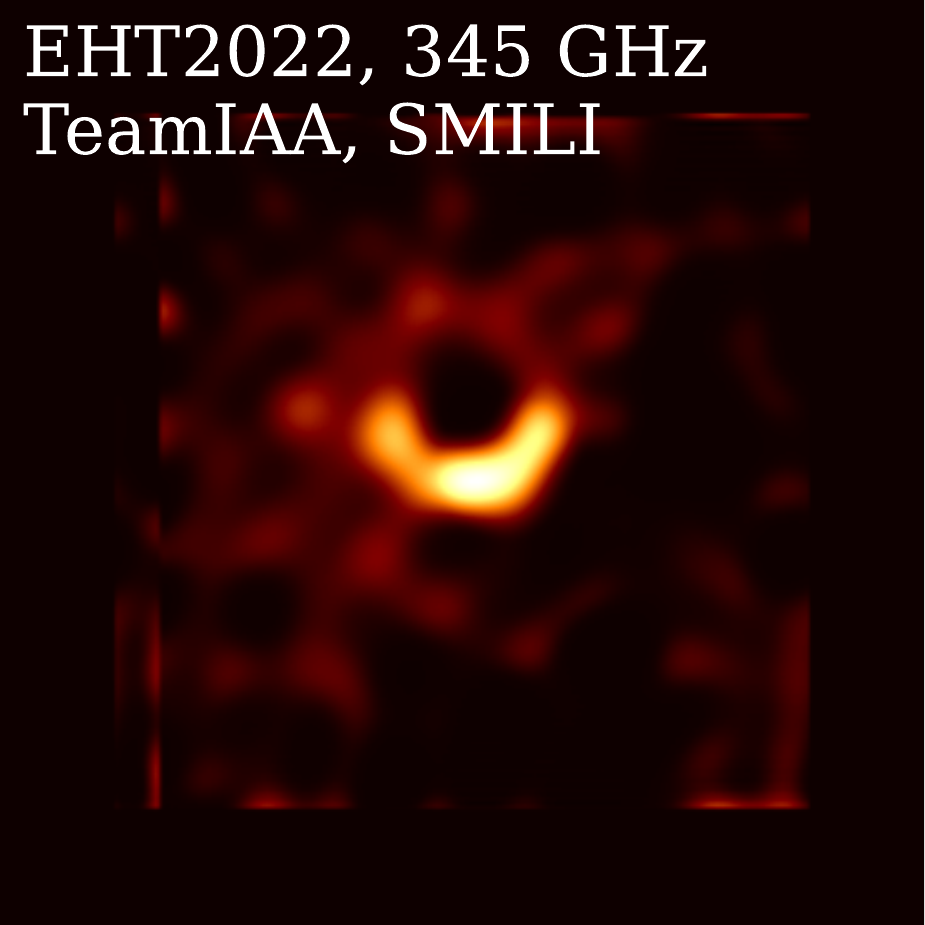}%
\includegraphics[width=25mm]{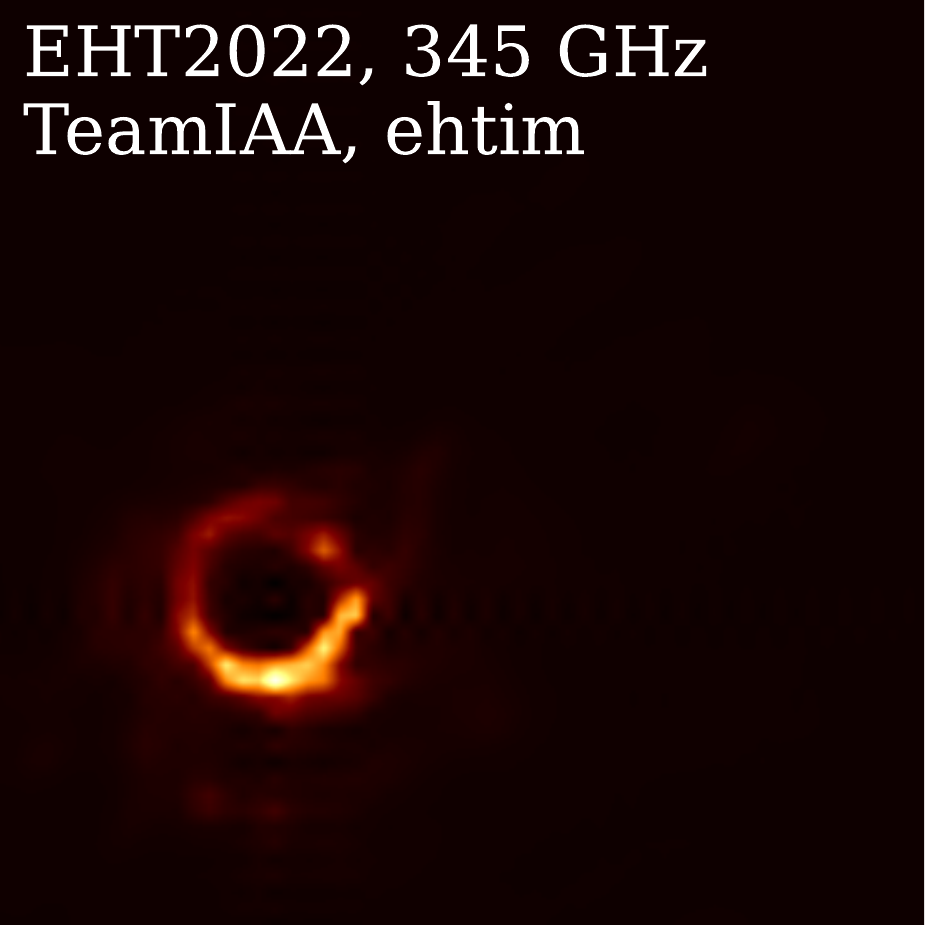}%
\includegraphics[width=25mm]{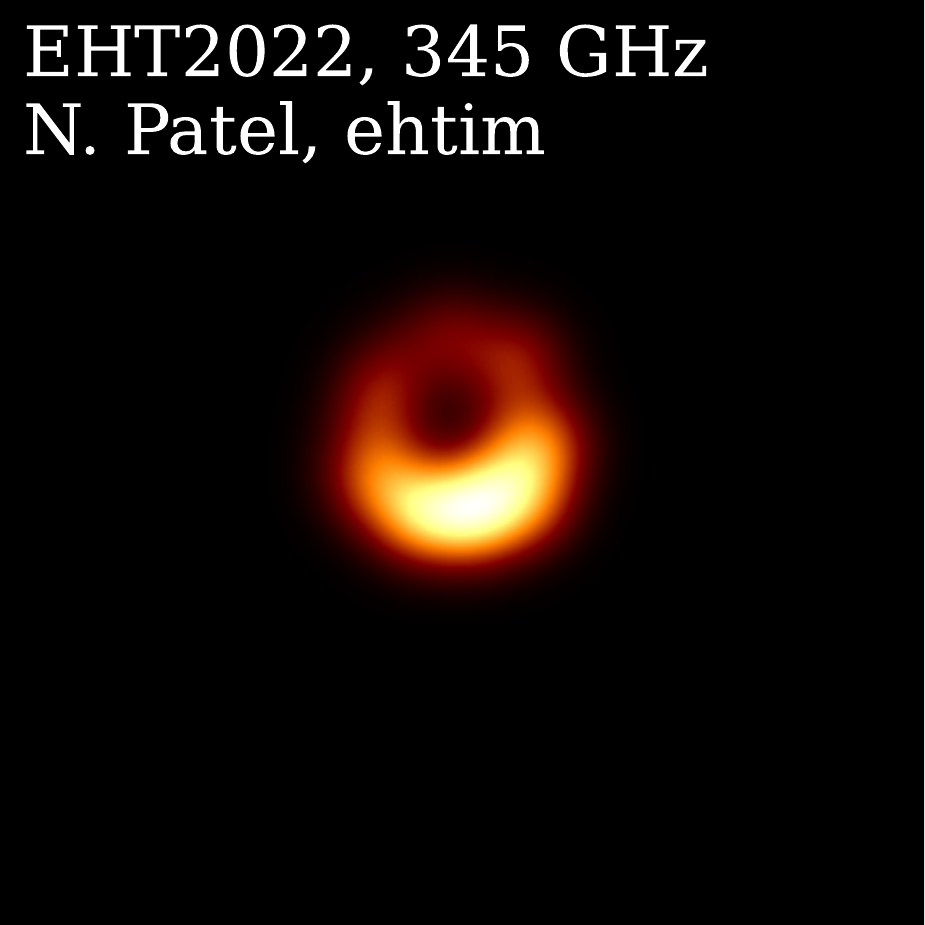}%
\includegraphics[width=25mm]{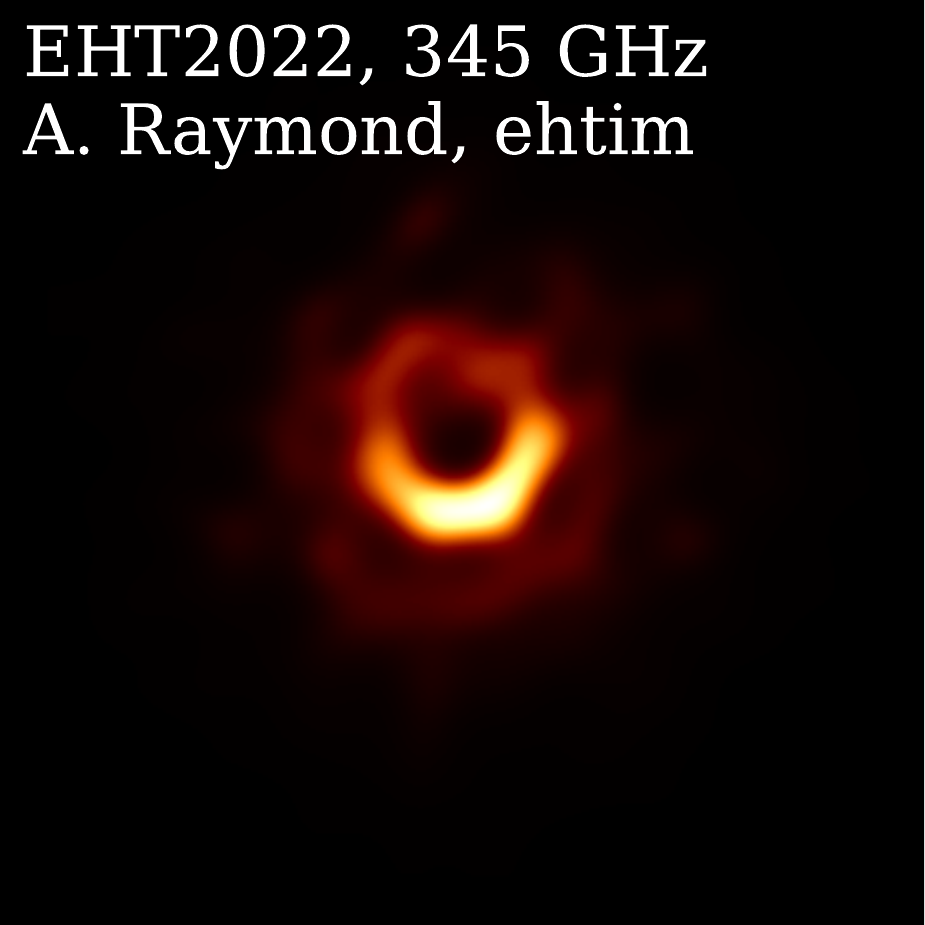}%
\includegraphics[width=25mm]{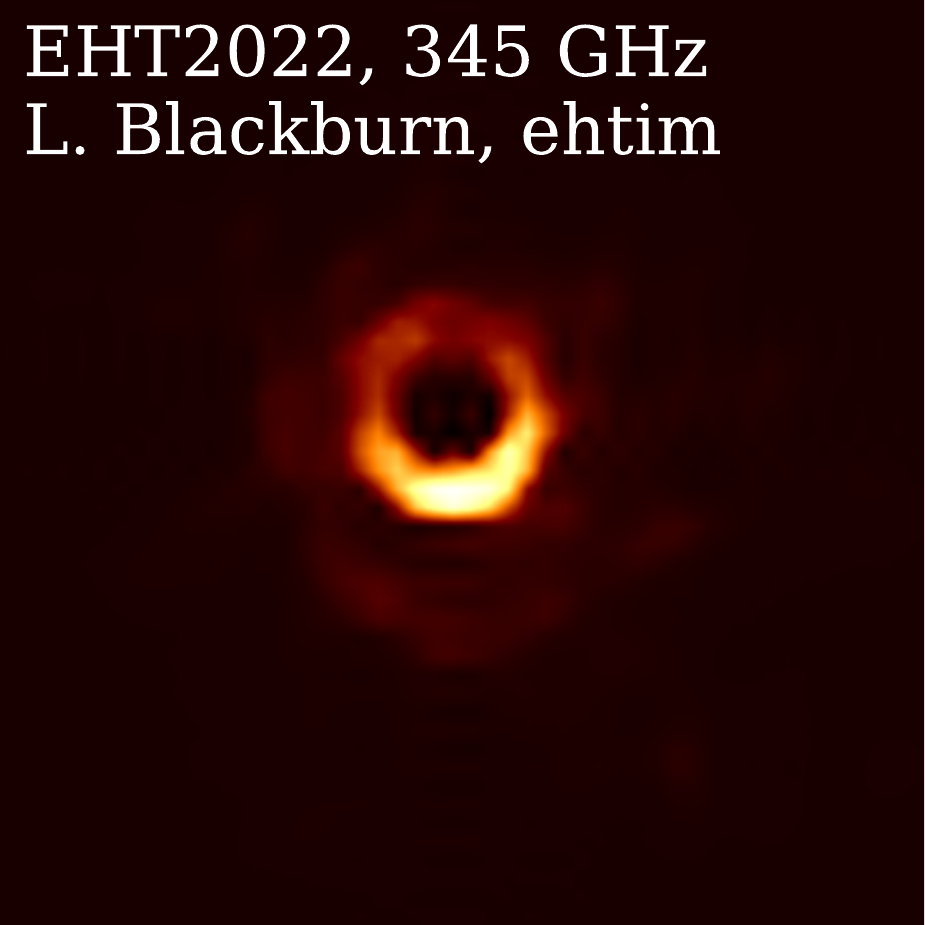}%
\includegraphics[width=25mm]{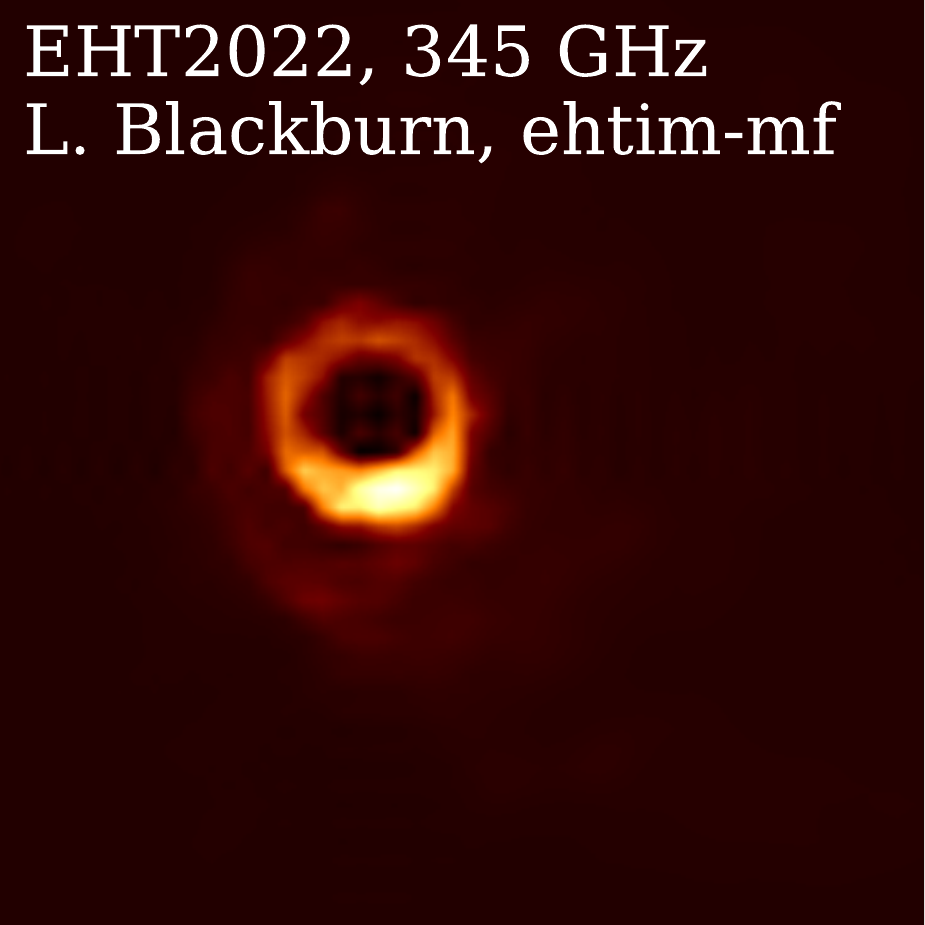} \\
\includegraphics[width=25mm]{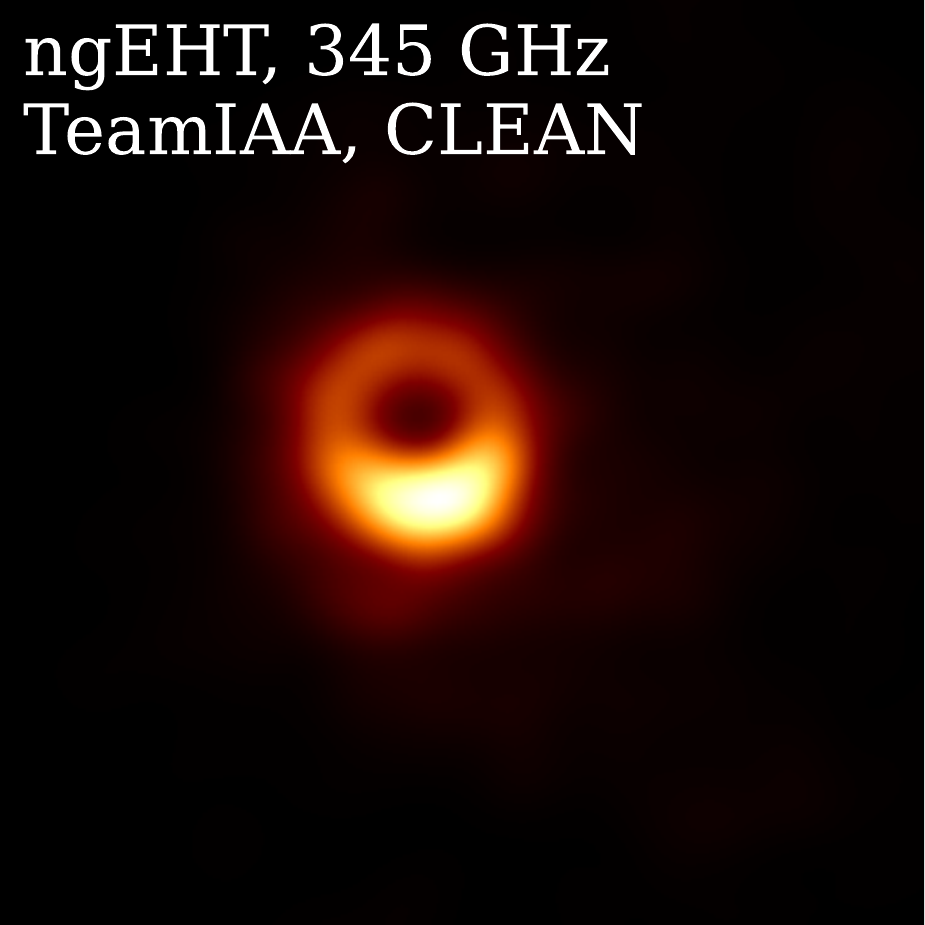}% 
\includegraphics[width=25mm]{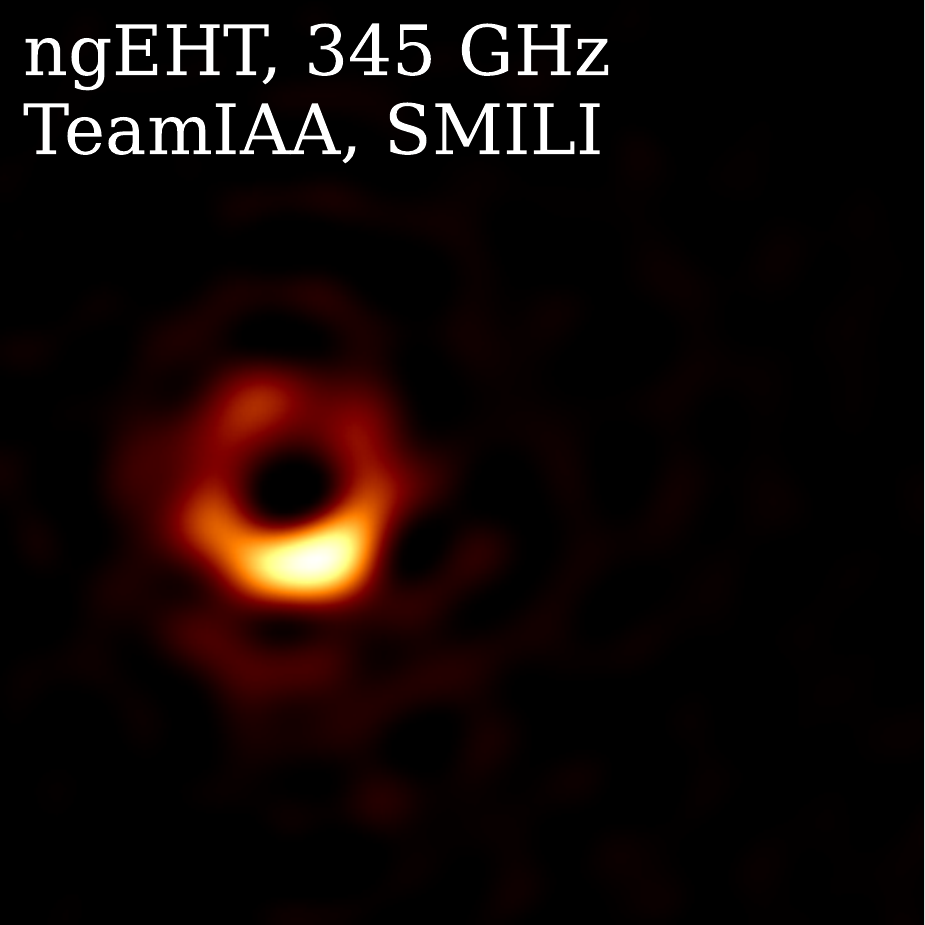}%
\includegraphics[width=25mm]{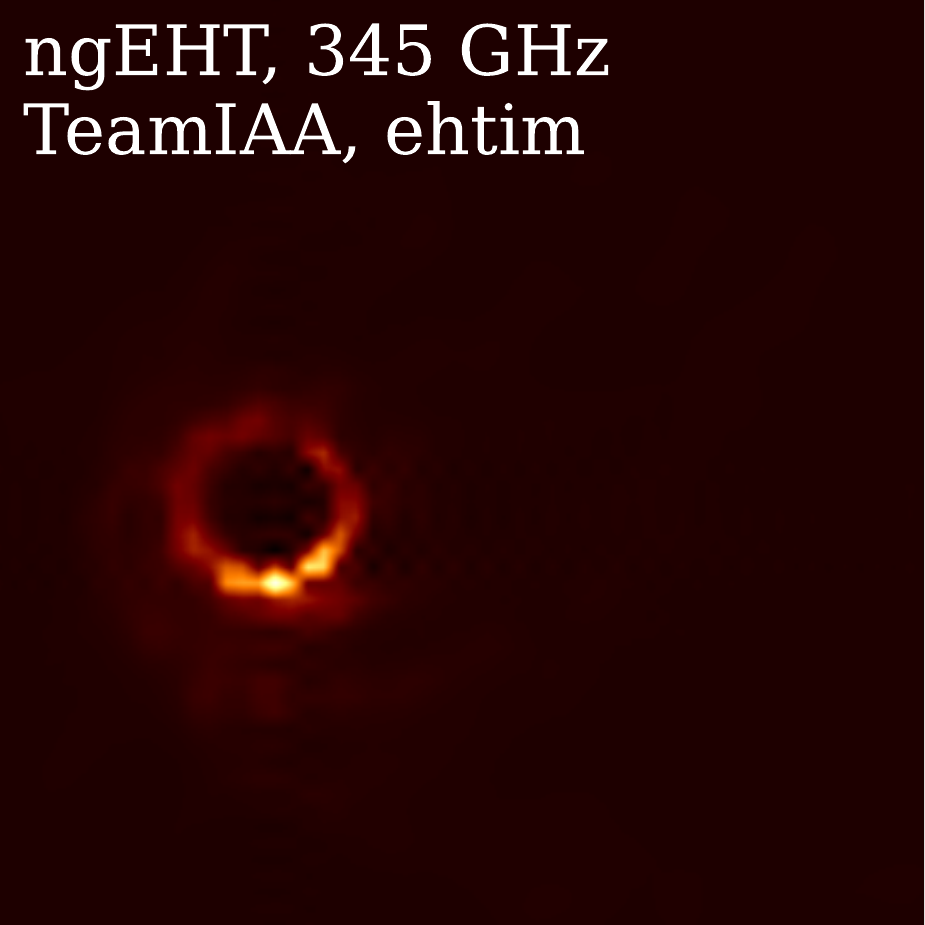}%
\includegraphics[width=25mm]{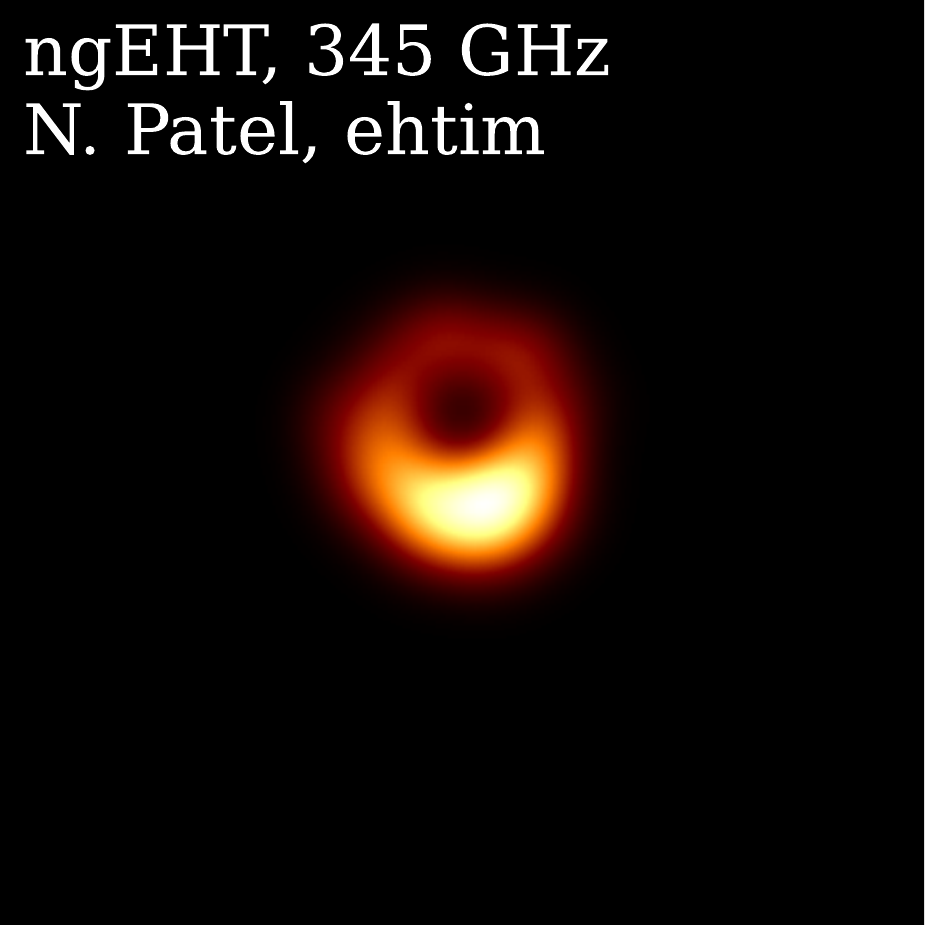}%
\includegraphics[width=25mm]{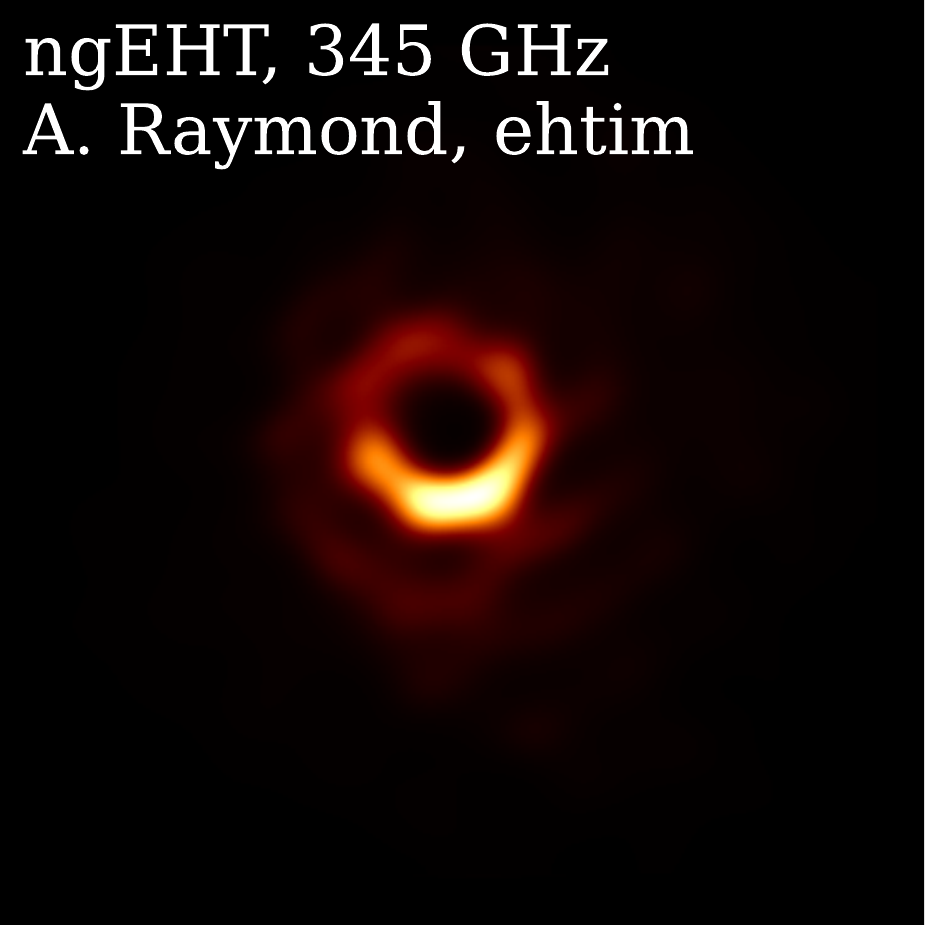}%
\includegraphics[width=25mm]{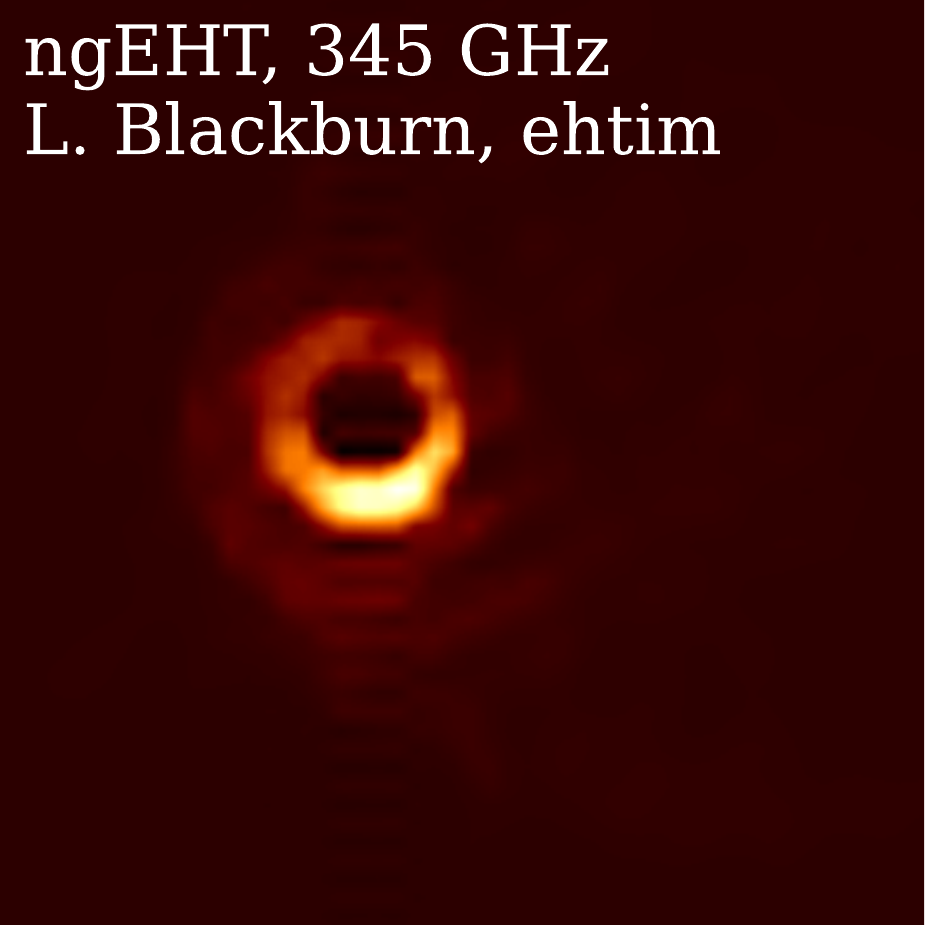}%
\includegraphics[width=25mm]{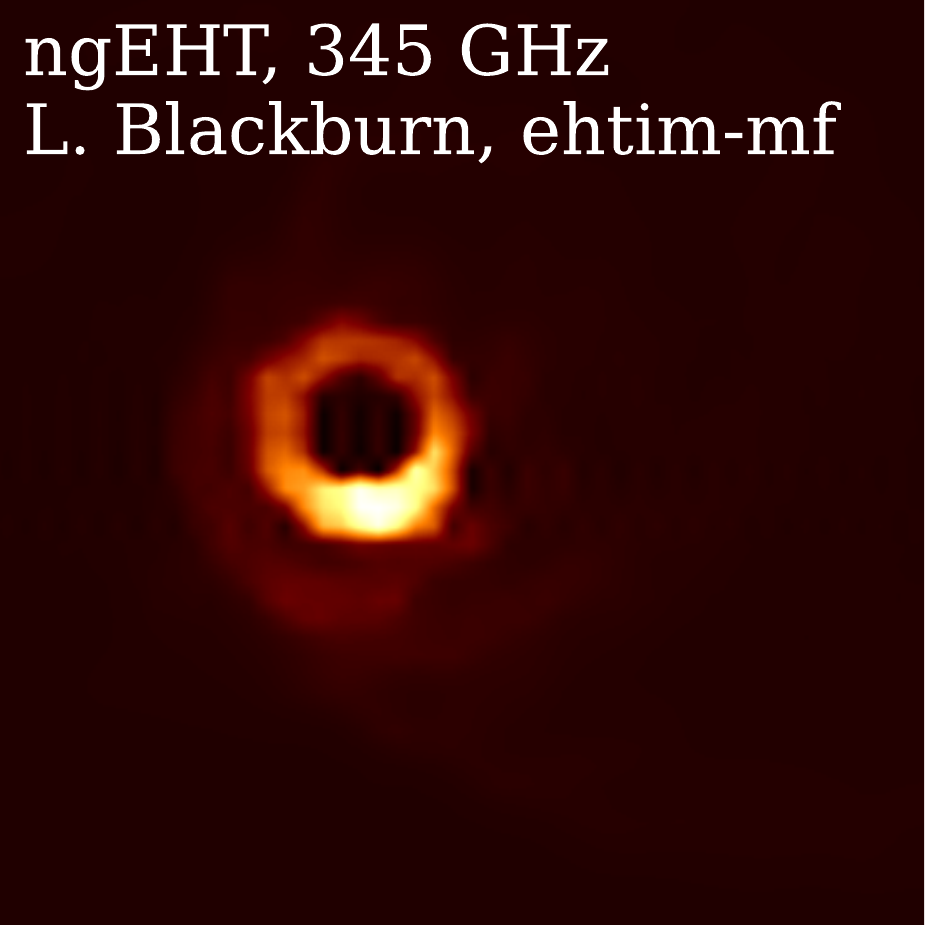}
\end{adjustwidth}
  \caption{M87 reconstructions submitted for Challenge 1. Images are shown on a log scale with a 1 mas field of view in the top set of panels. The same images are shown on a linear scale with a 200 $\mu$as field of view in the bottom set of panels.}
     \label{fig:ch1_reconstructions_m87}
\end{figure*}

\begin{figure*}
\setlength{\lineskip}{0pt}
\includegraphics[width=25mm]{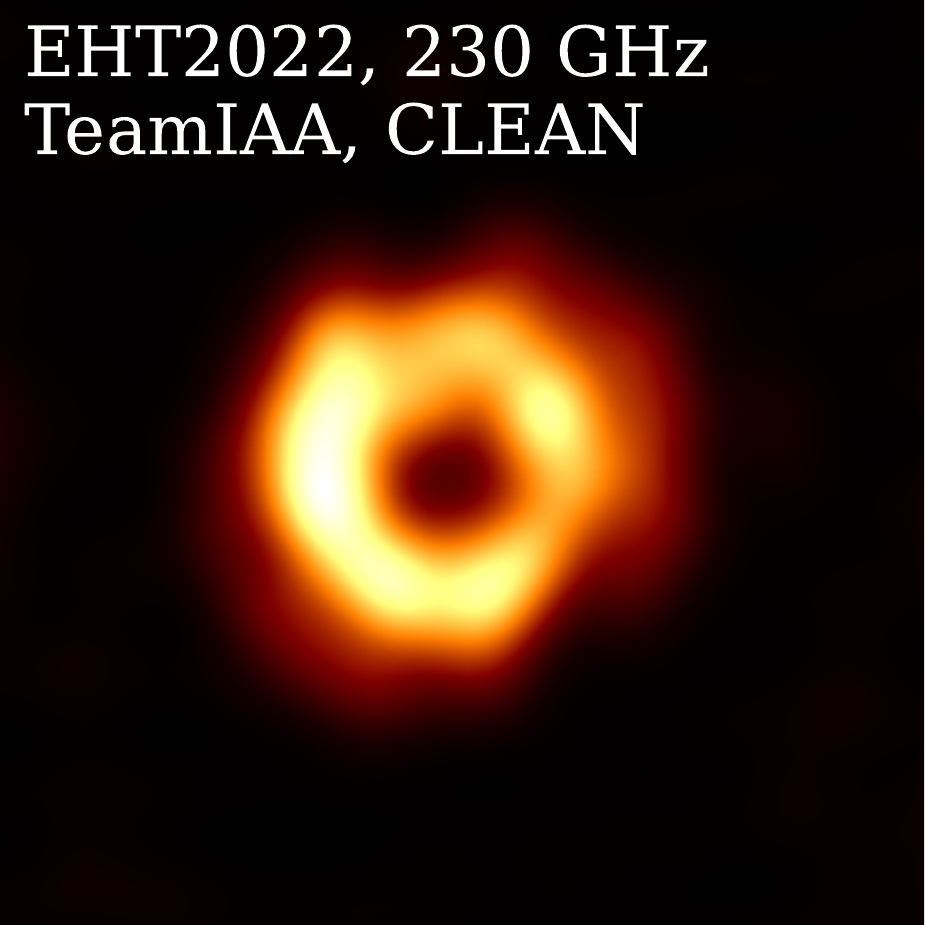}%
\includegraphics[width=25mm]{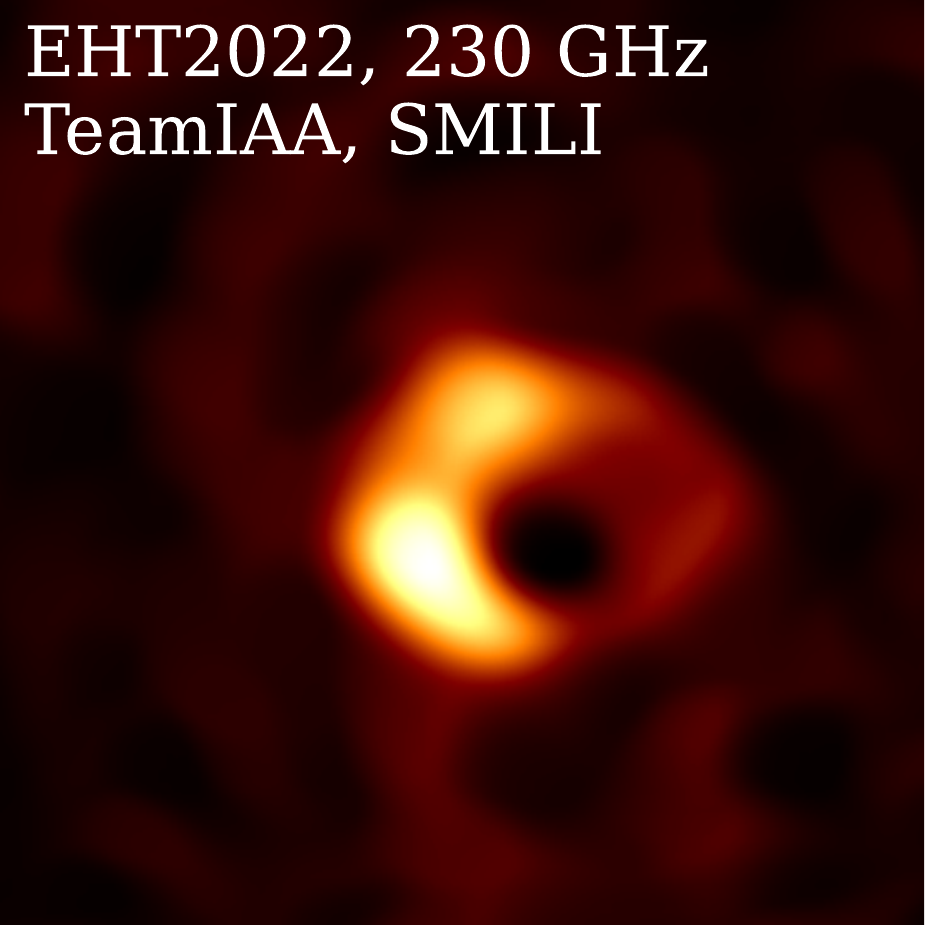}%
\includegraphics[width=25mm]{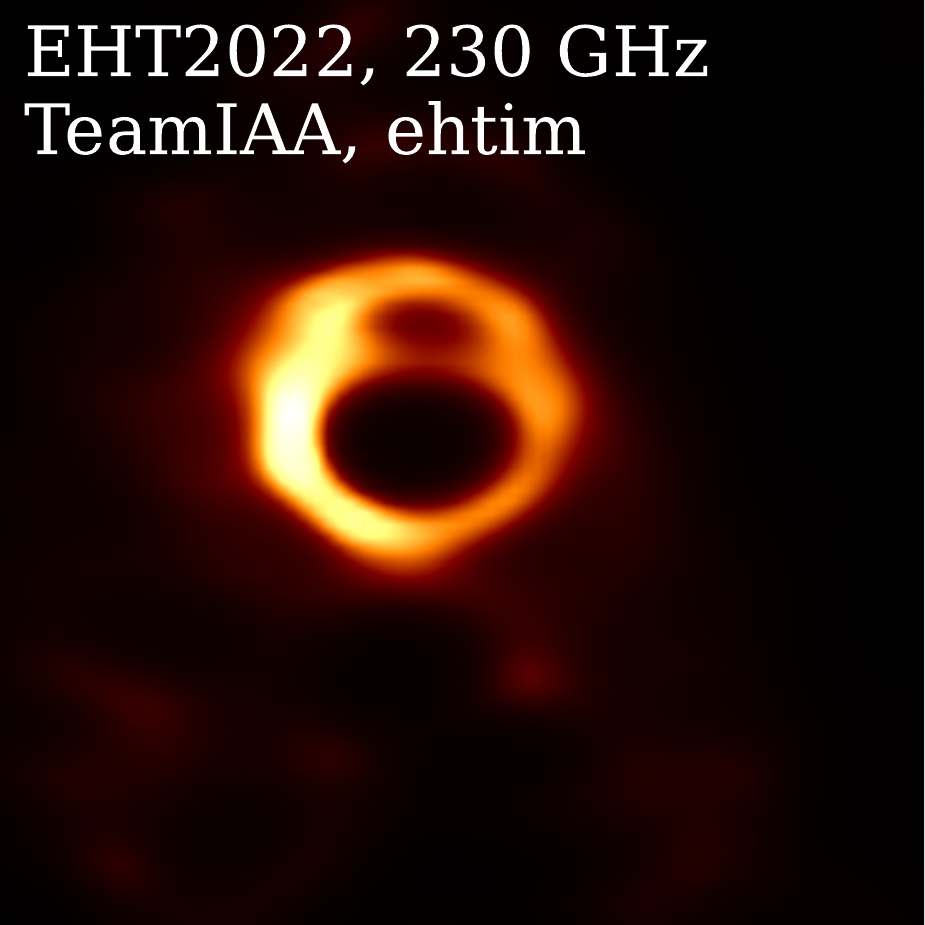}%
\includegraphics[width=25mm]{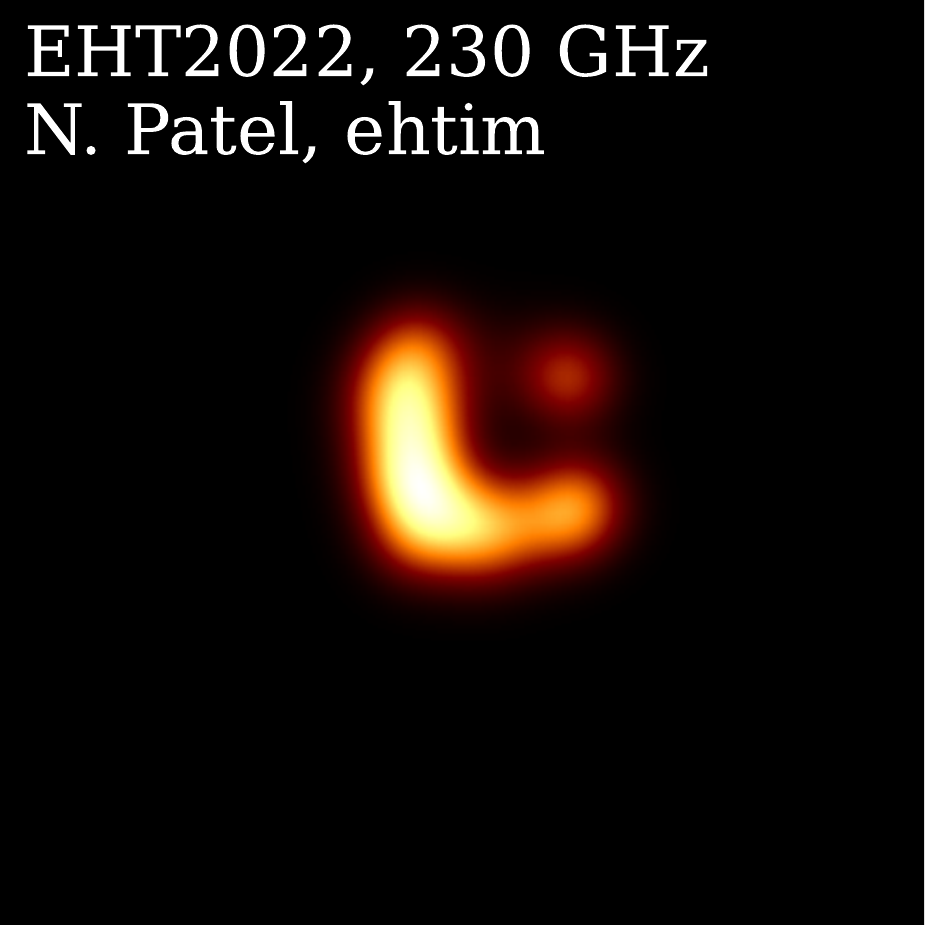}%
\includegraphics[width=25mm]{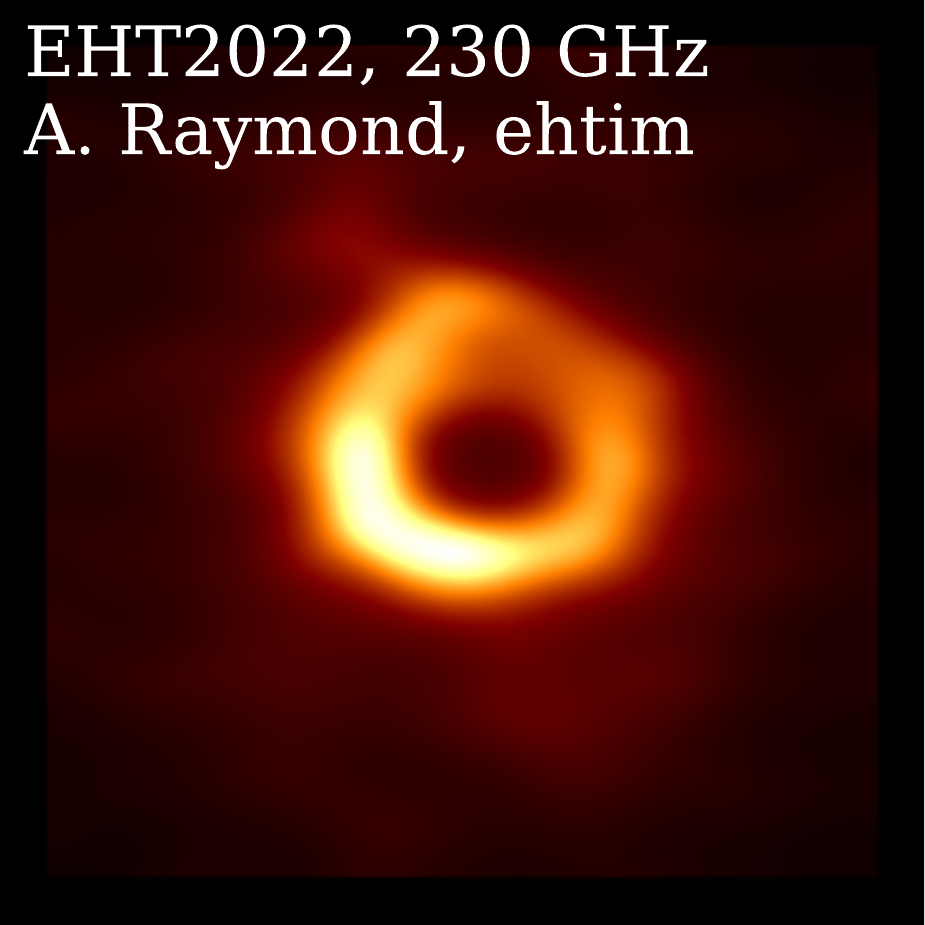} \\
\includegraphics[width=25mm]{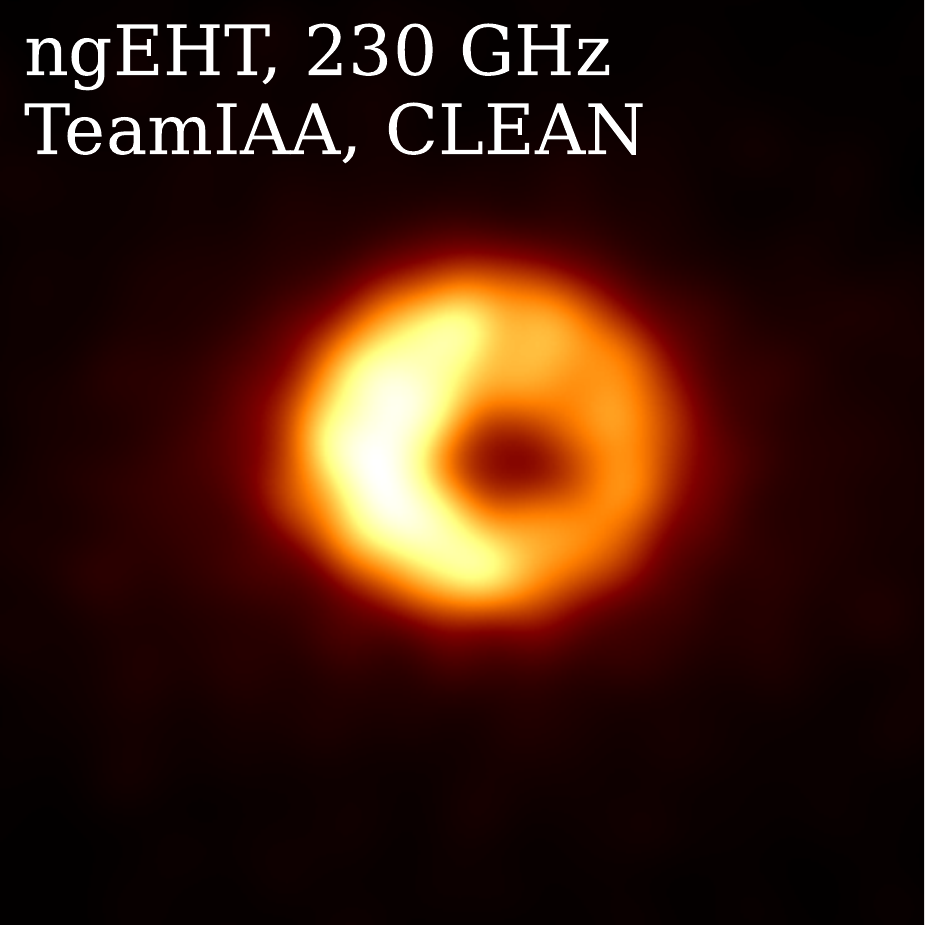}% 
\includegraphics[width=25mm]{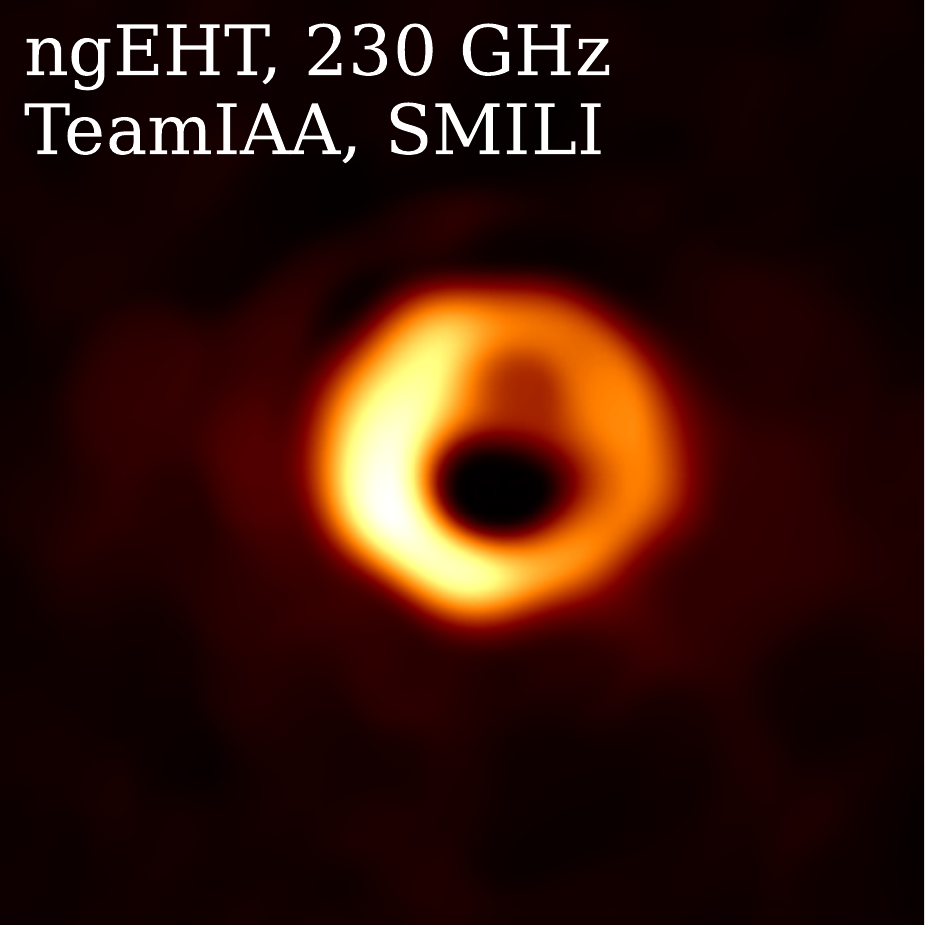}%
\includegraphics[width=25mm]{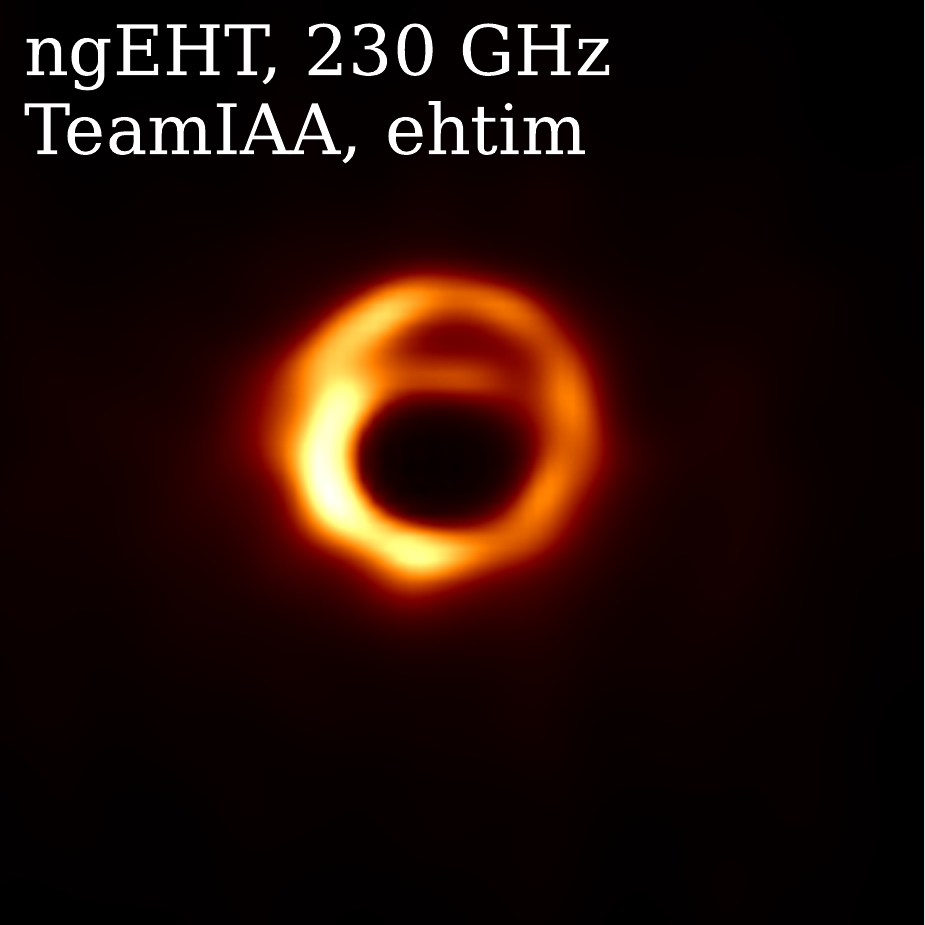}%
\includegraphics[width=25mm]{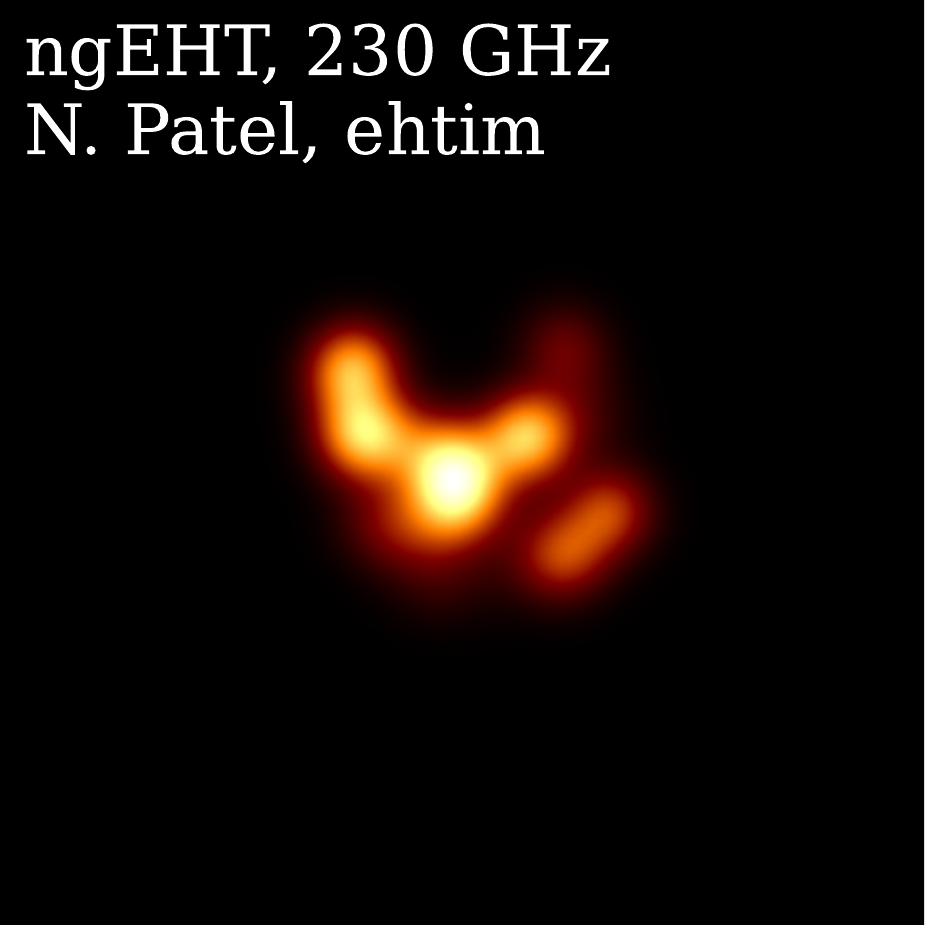}%
\includegraphics[width=25mm]{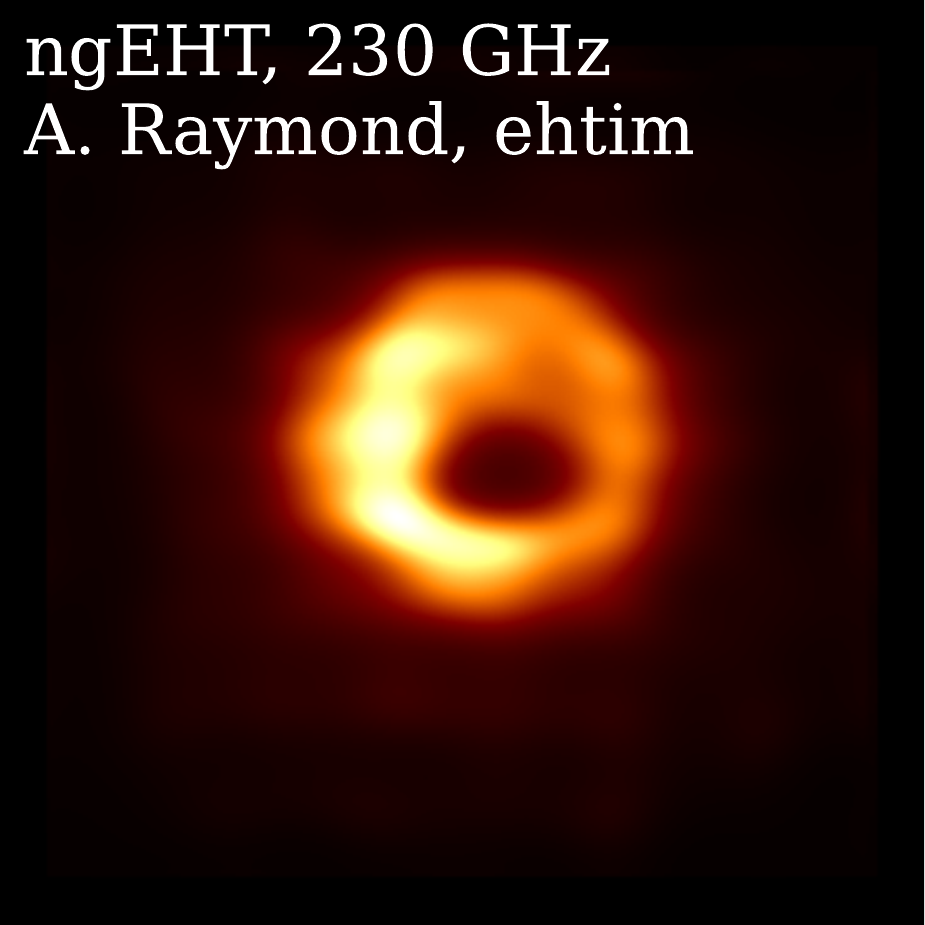} \\
\includegraphics[width=25mm]{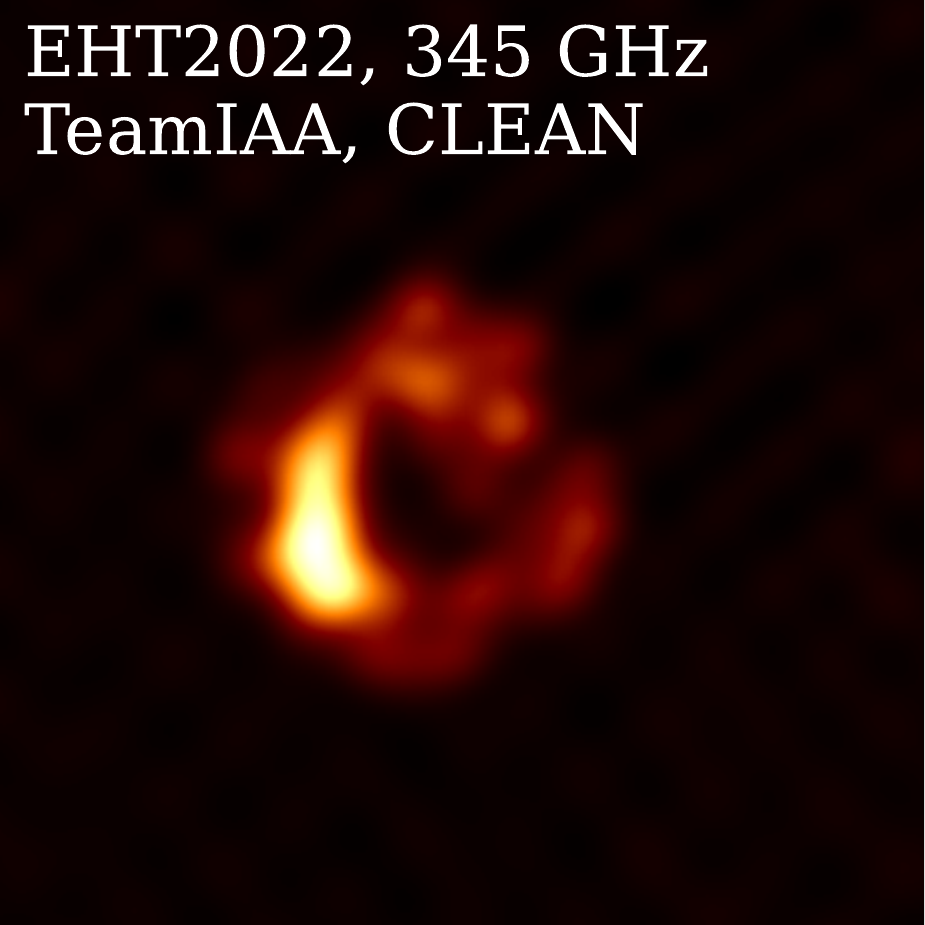}% 
\includegraphics[width=25mm]{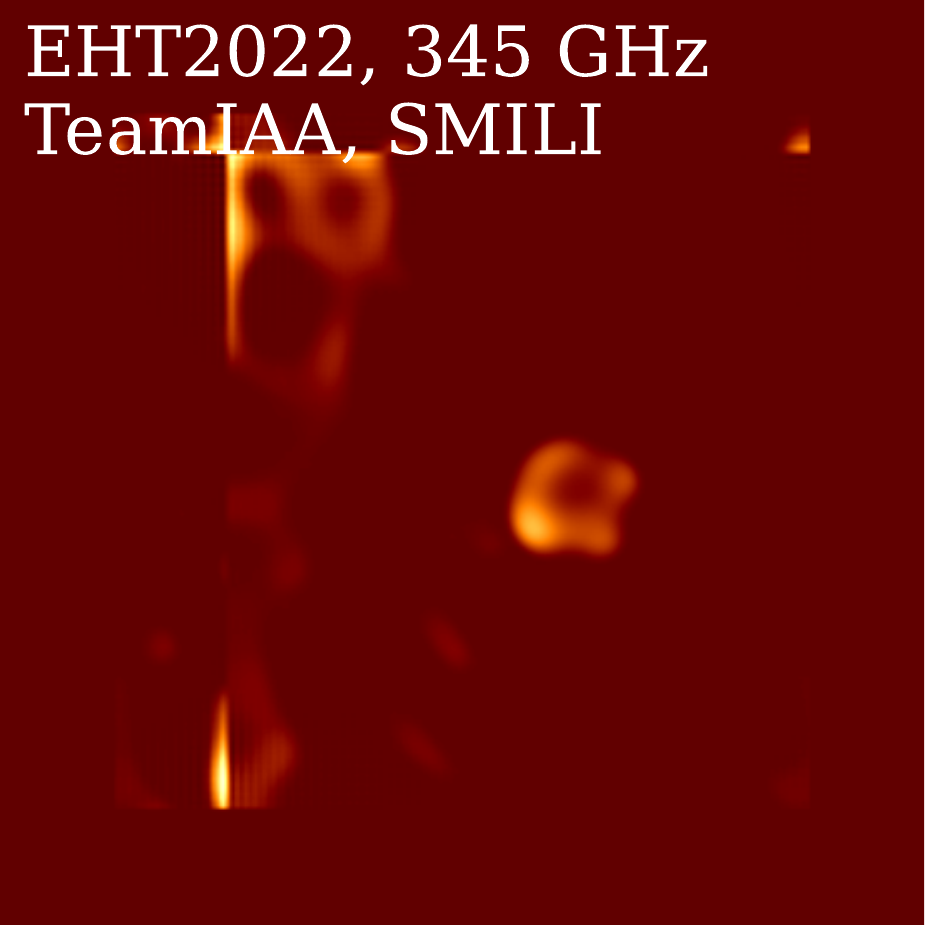}%
\includegraphics[width=25mm]{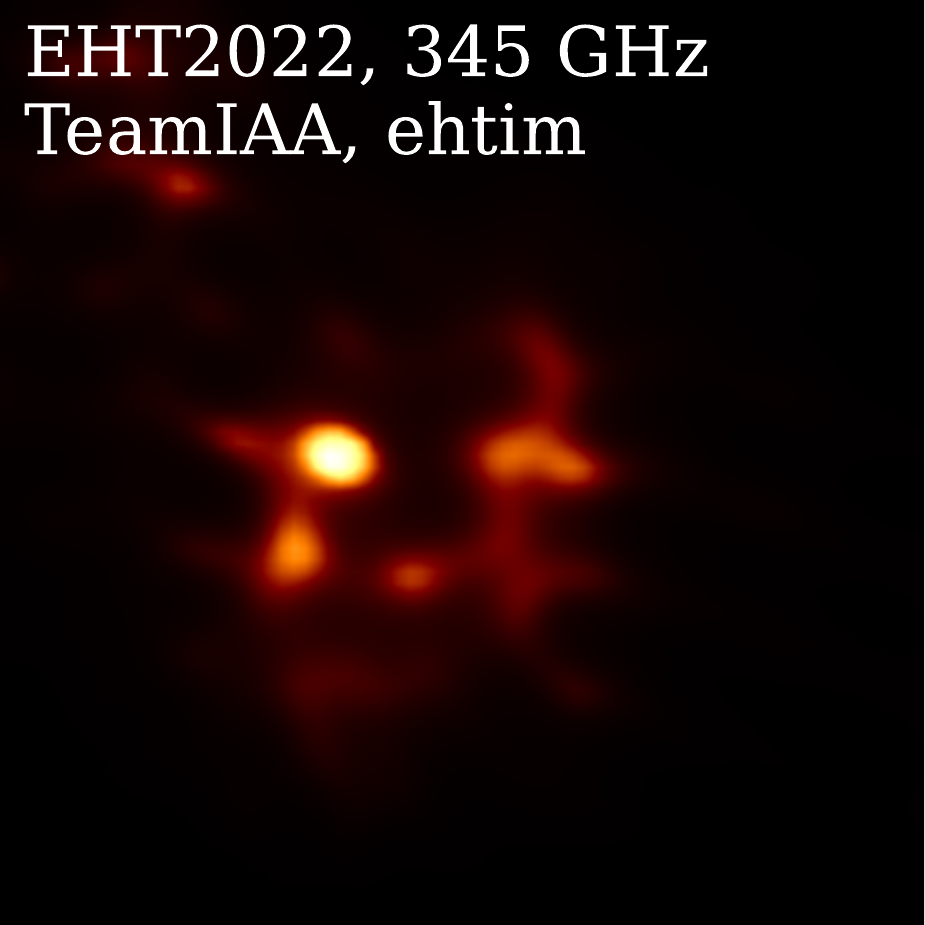}%
\includegraphics[width=25mm]{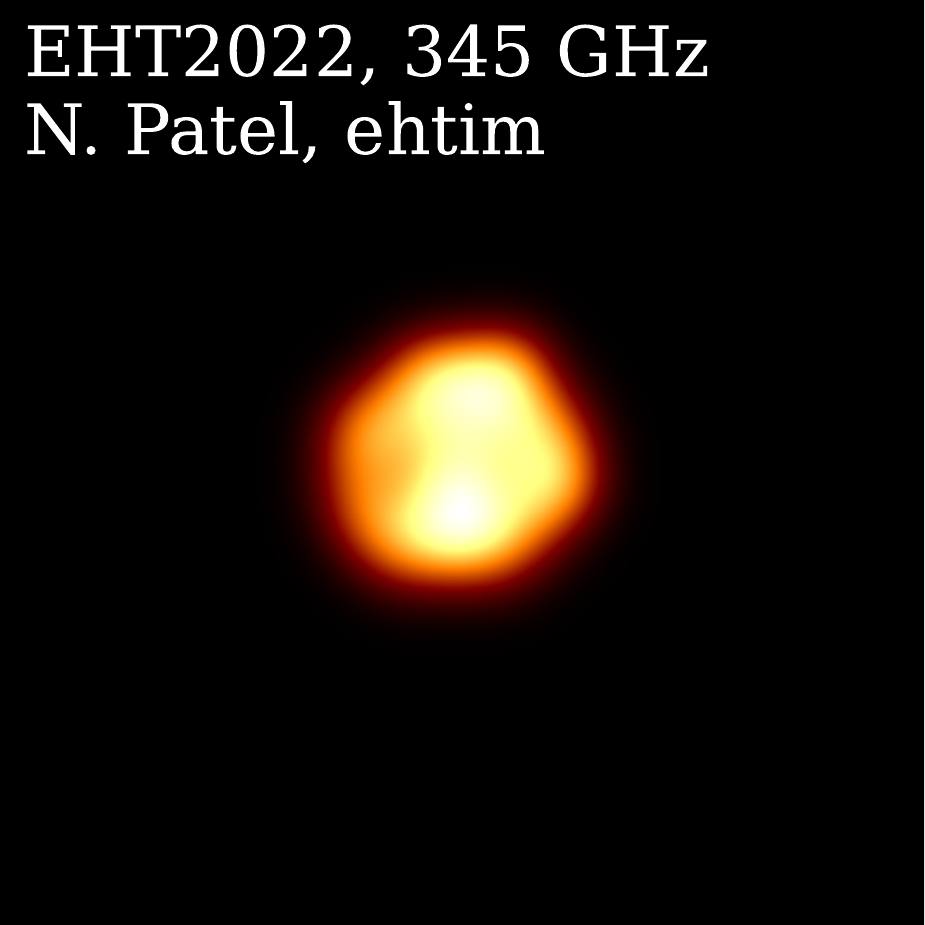}%
\includegraphics[width=25mm]{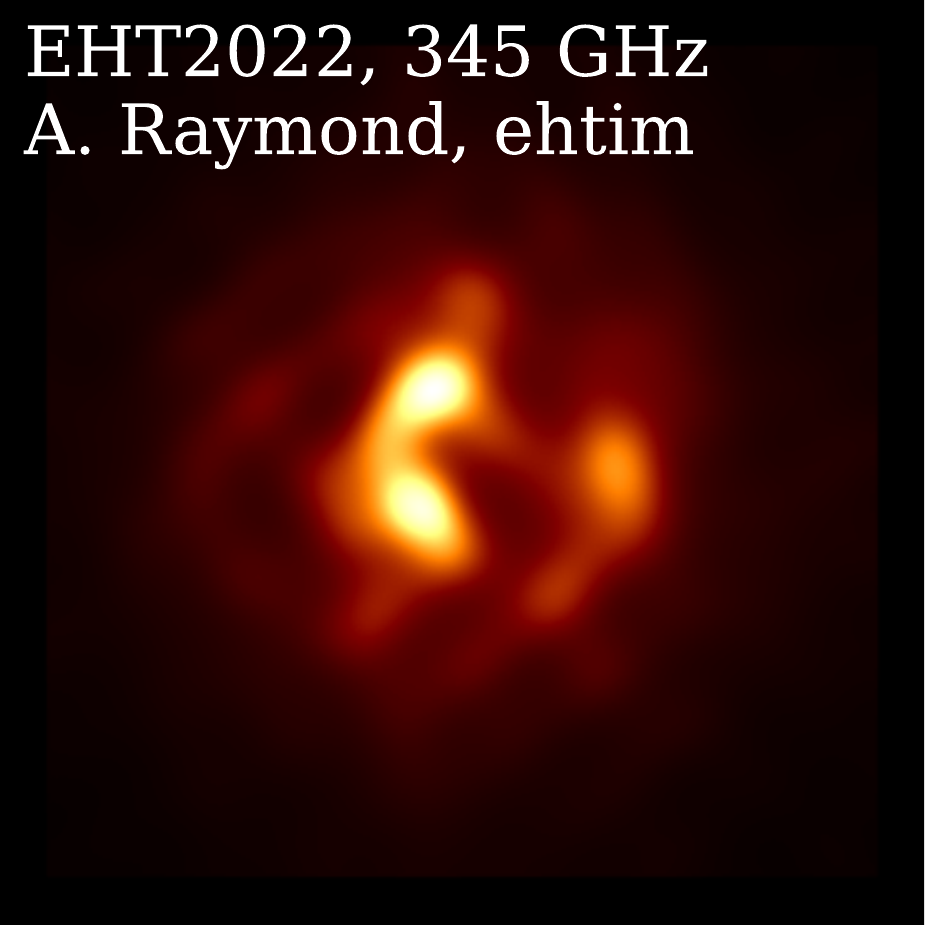} \\
\includegraphics[width=25mm]{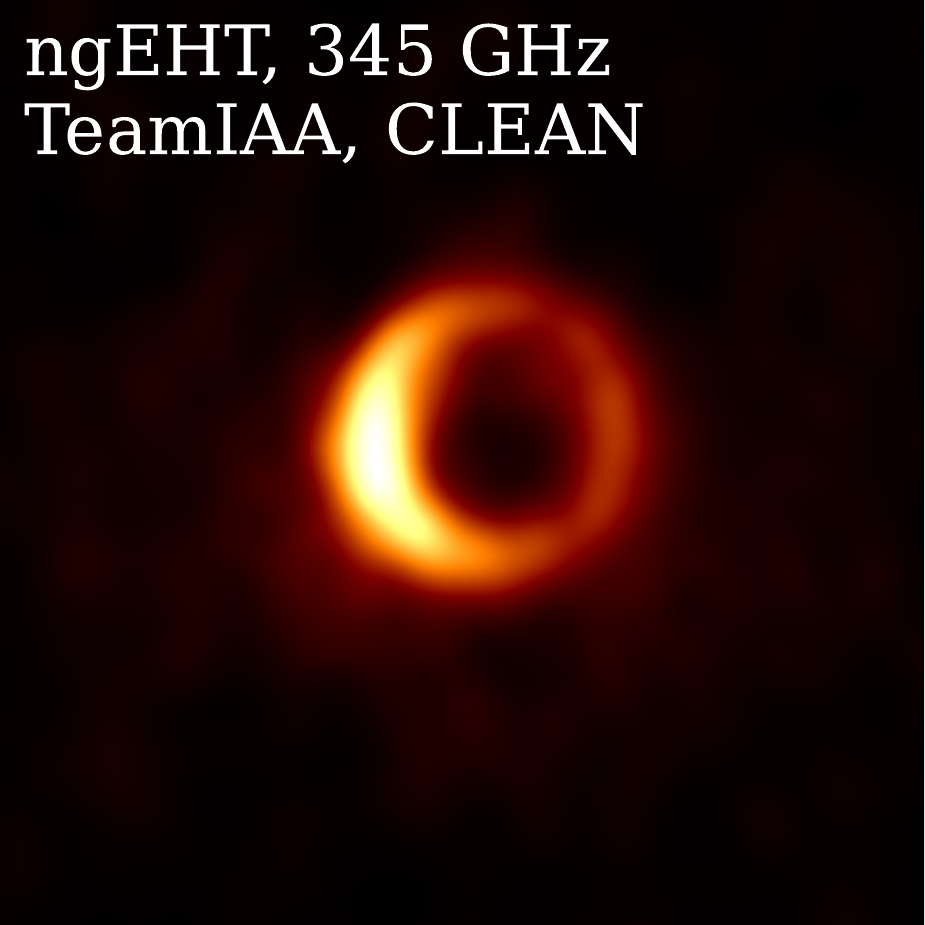}% 
\includegraphics[width=25mm]{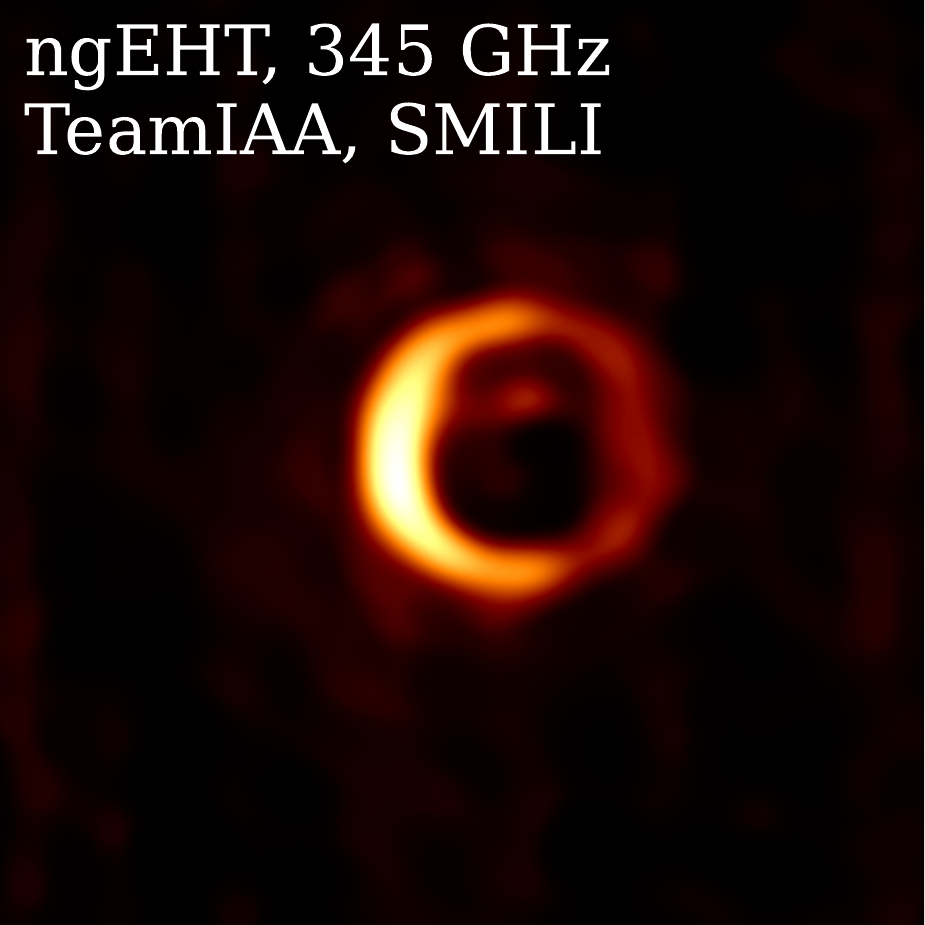}%
\includegraphics[width=25mm]{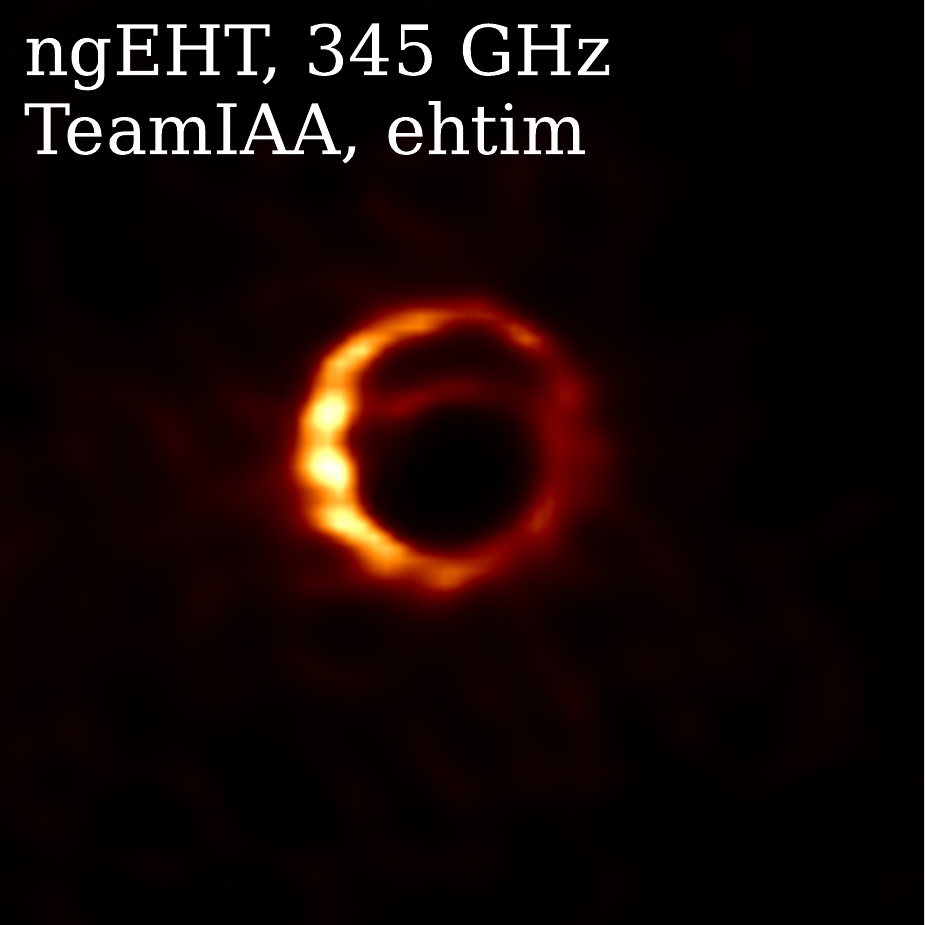}%
\includegraphics[width=25mm]{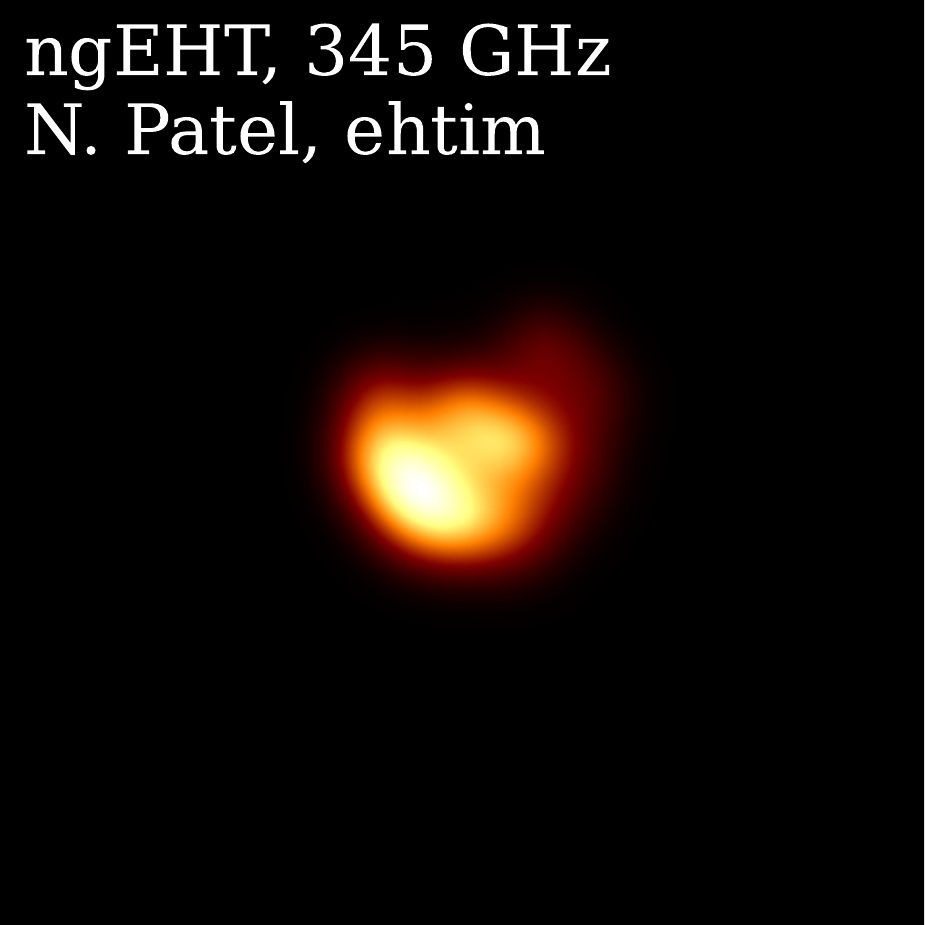}%
\includegraphics[width=25mm]{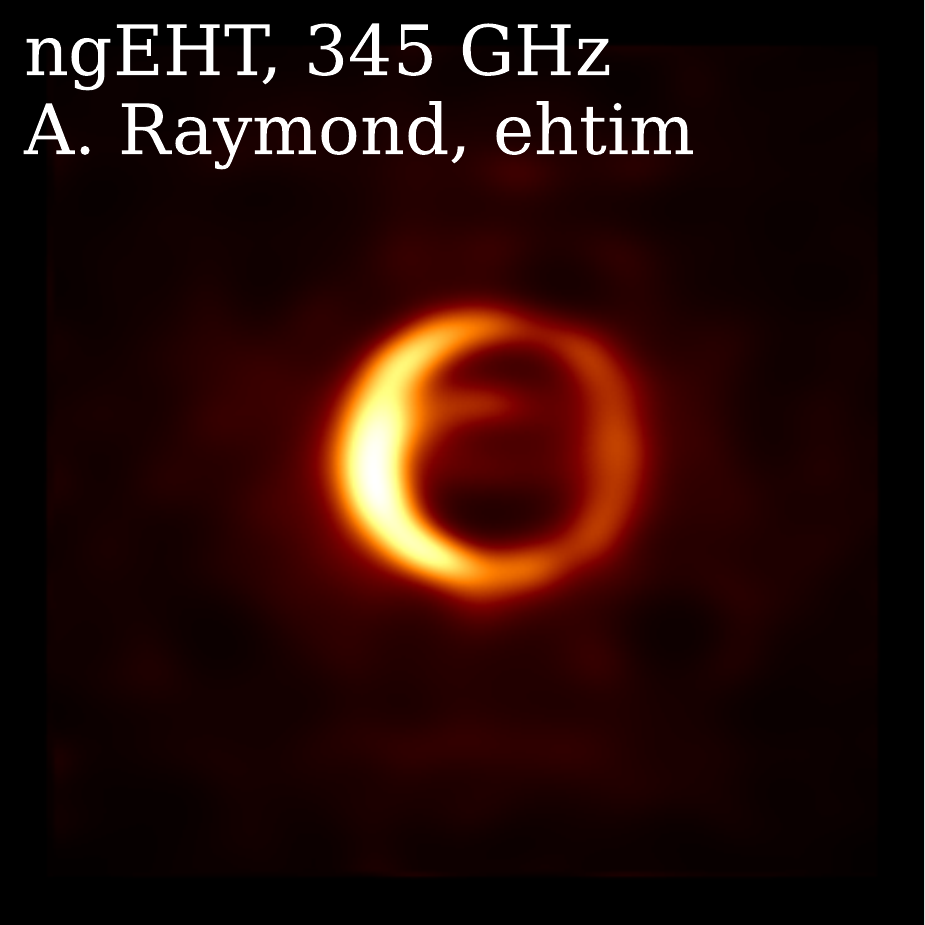}%
  \caption{Sgr A* reconstructions submitted for Challenge 1. Images are shown on a linear scale with a 200 $\mu$as field of view.}
     \label{fig:ch1_reconstructions_sgra}
\end{figure*}

Challenge 1 image reconstructions were submitted by three individual submitters and one team, using {\tt CLEAN}, \texttt{SMILI}, and \texttt{eht-imaging}. For M87, one submitter performed a multi-frequency image reconstruction with \texttt{eht-imaging}, using information from the 230 GHz data to reconstruct the image at 345 GHz, and vice versa. All submitted reconstructions are displayed in Figures \ref{fig:ch1_reconstructions_m87} and \ref{fig:ch1_reconstructions_sgra} for M87 and Sgr A*, respectively. Reconstruction quality metrics (Sec. \ref{sec:metrics}) are shown in Table \ref{tab:metrics_challenge1}. While it should be kept in mind that the synthetic data was idealized in certain aspects (no systematic amplitude errors, and a static Sgr A* source model), these results show some interesting trends.

All M87 reconstructions recover the black hole shadow, whereas the jet features are only recovered by some. The low surface brightness structure in the M87 jet is already visible in some EHT2022 reconstructions. The jet reconstructions improve significantly with ngEHT1 coverage, at attested by both visual inspection of the images and the quality metrics. The \texttt{eht-imaging} submissions perform best, although the $\rho_{\mathrm{NX}}$ ranking of individual submissions changes depending on whether the linear or log-scale images are used for the comparison. $\rho_{\mathrm{NX,log}}$ and $\mathcal{D}_{0.1}$ are more sensitive to the reconstruction of the extended jet structure, and generally show a clearer improvement of ngEHT1 versus EHT2022 reconstructions. The {\tt CLEAN} and \texttt{SMILI} reconstructions show poorer jet structure recovery than \texttt{eht-imaging}, although they may potentially be improved by adapting the specific scripts used for these reconstructions. The reconstruction quality is generally better for 230 GHz than for 345 GHz due to the higher flux and better $uv$-coverage at 230 GHz. Reconstructions with relatively high $\chi^2$ values often have relatively low $\rho_{\mathrm{NX}}$ values. The multi-frequency analysis ({\tt ehtim-mf})is an exception with relatively good reconstruction quality (especially as shown by $\rho_{\mathrm{NX,log}}$ and and $\mathcal{D}_{0.1}$) for relatively high $\chi^2$ values. The multi-frequency analysis is especially useful for reconstructing the jet features at 345 GHz, as these are reconstructed significantly more poorly with other methods.

For the Sgr A* model, the black hole shadow is recovered by all arrays except the EHT2022 array at 345 GHz: the $uv$-coverage is too sparse for a high-fidelity image reconstruction in this case (Kitt Peak, the LMT, and the SPT cannot observe at 345 GHz yet). ngEHT reconstructions at 345 GHz are significantly sharper than EHT2022 and ngEHT reconstructions at 230 GHz. ngEHT reconstructions at 230 GHz are generally less noisy than EHT2022 reconstructions at the same frequency, but for Sgr A* the real value of ngEHT coverage will be in dynamical reconstructions (Sec. \ref{sec:challenge2}). The $\chi^2_{\mathrm{lcamp}}$ are generally high for Sgr A* reconstructions, which is likely due to the comparison to the provided synthetic data, which includes interstellar scattering, while submitters may have deblurred the visibility amplitudes in the reconstruction process.

\begin{table*}[h]
\caption{Reconstruction quality metrics for Challenge 1.}
\begin{adjustwidth}{-\extralength}{0cm}
\begin{tabular}{lllll|llllll}
Source&Array&$\nu$ (GHz)&Submitter&Method&$\chi^2_{\mathrm{cp}}$&$\chi^2_{\mathrm{lcamp}}$&$\rho_{\mathrm{NX}}$&$\rho_{\mathrm{NX,log}}$&$\theta_\mathrm{eff}$&$\mathcal{D}_{0.1}$ \\
\hline
M87&EHT2022&230&L. Blackburn&{\tt ehtim}&1.1&1.01&0.93&0.87&5.4&856 \\
M87&EHT2022&230&L. Blackburn&{\tt ehtim-mf}&5.17&4.36&0.88&0.9&9.8&797 \\
M87&EHT2022&230&N. Patel&{\tt ehtim}&3.66&1159.56&0.77&0.52&21.2&418 \\
M87&EHT2022&230&TeamIAA&{\tt SMILI}&0.99&1.06&0.83&0.79&14.6&409 \\
M87&EHT2022&230&TeamIAA&{\tt CLEAN}&2.94&879.77&0.8&0.8&17.7&529 \\
M87&EHT2022&230&TeamIAA&{\tt ehtim}&1.79&1.03&0.89&0.91&8.9&564 \\
M87&EHT2022&230&A. Raymond&{\tt ehtim}&2.28&1.77&0.9&0.72&8.0&291 \\
M87&ngEHT&230&L. Blackburn&{\tt ehtim-mf}&2.62&1.43&0.89&0.96&8.9&1681 \\
M87&ngEHT&230&L. Blackburn&{\tt ehtim}&1.07&1.01&0.93&0.95&5.4&1604 \\
M87&ngEHT&230&N. Patel&{\tt ehtim}&3.5&89.74&0.83&0.52&14.6&640 \\
M87&ngEHT&230&TeamIAA&{\tt SMILI}&1.01&1.03&0.87&0.85&10.8&708 \\
M87&ngEHT&230&TeamIAA&{\tt CLEAN}&1.32&138.45&0.84&0.91&13.6&1828 \\
M87&ngEHT&230&TeamIAA&{\tt ehtim}&1.08&1.01&0.91&0.97&7.1&1727 \\
M87&ngEHT&230&A. Raymond&{\tt ehtim}&1.65&2.14&0.92&0.73&6.2&532 \\
M87&EHT2022&345&L. Blackburn&{\tt ehtim-mf}&2.36&1.06&0.91&0.87&5.7&1403 \\
M87&EHT2022&345&L. Blackburn&{\tt ehtim}&1.19&0.62&0.91&0.72&5.7&984 \\
M87&EHT2022&345&N. Patel&{\tt ehtim}&1.2&7.29&0.79&0.53&16.7&734 \\
M87&EHT2022&345&TeamIAA&{\tt SMILI}&1.19&0.62&0.79&0.66&16.7&645 \\
M87&EHT2022&345&TeamIAA&{\tt ehtim}&1.22&0.62&0.88&0.81&8.2&700 \\
M87&EHT2022&345&TeamIAA&{\tt CLEAN}&3.34&2.77&0.82&0.38&13.7&320 \\
M87&EHT2022&345&A. Raymond&{\tt ehtim}&1.19&0.62&0.88&0.74&8.2&546 \\
M87&ngEHT&345&L. Blackburn&{\tt ehtim}&1.15&0.97&0.92&0.89&4.9&1570 \\
M87&ngEHT&345&L. Blackburn&{\tt ehtim-mf}&1.25&1.13&0.91&0.94&5.7&2244 \\
M87&ngEHT&345&N. Patel&{\tt ehtim}&1.2&9.99&0.79&0.54&16.7&853 \\
M87&ngEHT&345&TeamIAA&{\tt CLEAN}&1.31&4.39&0.84&0.75&11.8&651 \\
M87&ngEHT&345&TeamIAA&{\tt SMILI}&1.16&1.0&0.85&0.71&10.9&766 \\
M87&ngEHT&345&TeamIAA&{\tt CLEAN}&1.31&4.39&0.84&0.75&11.8&651 \\
M87&ngEHT&345&TeamIAA&{\tt ehtim}&1.16&0.98&0.9&0.92&6.5&1638 \\
M87&ngEHT&345&A. Raymond&{\tt ehtim}&1.17&1.0&0.91&0.75&5.7&782 \\
Sgr A*&EHT2022&230&N. Patel&{\tt ehtim}&6.08&347.88&0.8&-&45.5&- \\
Sgr A*&EHT2022&230&TeamIAA&{\tt ehtim}&1.11&33.13&0.95&-&14.3&- \\
Sgr A*&EHT2022&230&TeamIAA&{\tt CLEAN}&140.97&130.2&0.9&-&23.4&- \\
Sgr A*&EHT2022&230&TeamIAA&{\tt SMILI}&1.47&23.19&0.85&-&32.6&- \\
Sgr A*&EHT2022&230&A. Raymond&{\tt ehtim}&3.02&8.27&0.89&-&25.2&- \\
Sgr A*&ngEHT&230&N. Patel&{\tt ehtim}&20.23&122.65&0.65&-&100.0&- \\
Sgr A*&ngEHT&230&TeamIAA&{\tt SMILI}&1.4&8.81&0.95&-&14.3&- \\
Sgr A*&ngEHT&230&TeamIAA&{\tt CLEAN}&2.3&23.3&0.9&-&23.4&- \\
Sgr A*&ngEHT&230&TeamIAA&{\tt ehtim}&1.06&10.61&0.97&-&10.1&- \\
Sgr A*&ngEHT&230&A. Raymond&{\tt ehtim}&1.14&1.87&0.93&-&18.1&- \\
Sgr A*&EHT2022&345&N. Patel&{\tt ehtim}&1.03&20.32&0.64&-&61.9&- \\
Sgr A*&EHT2022&345&TeamIAA&{\tt CLEAN}&71.44&66.33&0.79&-&24.5&- \\
Sgr A*&EHT2022&345&TeamIAA&{\tt ehtim}&1.03&1.95&0.65&-&57.8&- \\
Sgr A*&EHT2022&345&TeamIAA&{\tt SMILI}&1.63&1.7&0.34&-&100.0&- \\
Sgr A*&EHT2022&345&A. Raymond&{\tt ehtim}&1.03&0.85&0.78&-&26.0&- \\
Sgr A*&ngEHT&345&N. Patel&{\tt ehtim}&2.18&15.58&0.64&-&61.9&- \\
Sgr A*&ngEHT&345&TeamIAA&{\tt ehtim}&1.14&1.19&0.93&-&7.5&- \\
Sgr A*&ngEHT&345&TeamIAA&{\tt CLEAN}&2.24&4.48&0.87&-&14.0&- \\
Sgr A*&ngEHT&345&TeamIAA&{\tt SMILI}&1.17&1.23&0.89&-&11.7&- \\
Sgr A*&ngEHT&345&A. Raymond&{\tt ehtim}&1.14&1.15&0.9&-&10.6&- \\
\hline
\end{tabular}
\end{adjustwidth}
\label{tab:metrics_challenge1}
\end{table*}

\section{Challenge 2}
\label{sec:challenge2}
\subsection{Rationale and charge}
\label{sec:challenge2_rationale}
With the challenge infrastructure and initial participant imaging efforts set up in the first challenge, the second challenge was more realistic and science oriented, and different from the first challenge in two aspects. 

First, the ground truth source models were dynamic instead of static. For Sgr A*, with variability on timescales of $\sim$ minutes, the charge to the participants was to reconstruct a movie of the source evolving across a single day of observations. We used two source models to test reconstruction capabilities for different variability properties: a GRMHD model with turbulent variability, and a shearing hotspot in a RIAF disk, exhibiting more coherent variable structure. Hints of such coherent variability of Sgr A* at 230 GHz consistent with an orbiting hotspot have been observed by \citet{Wielgus2022}. For M87*, with variability on timescales of $\sim$ days, we used a bright jet GRMHD model like in the first challenge, but evolved it over a period of five months, simulating a full-day observation every week. The charge was to reconstruct a movie of the large-scale and low-surface brightness jet emission, connecting it to the dynamics near the black hole shadow.

The second aspect in which this challenge differed from the previous one is that the synthetic observations included significantly more realistic effects. Contrary to the idealized data generated for Challenge 1, which only includes thermal noise, the Challenge 2 data sets have been generated under the assumption of realistic observing conditions, and include data systematics originating from weather, instrumental, and calibration effects (see Section \ref{sec:challenge2_synthdata}). The results of this challenge thus reflect what would actually be seen by an array built with the described specifics, using current reconstruction algorithms.

Challenge 2 was launched on 25 October 2021. It was advertized more broadly than the first challenge, to the full ngEHT community. The first submission comparisons were done in January 2022, and due to the complexity of the datasets and on-going development of reconstruction algorithms, reconstructions were submitted until August 2022.

\subsection{Source models}
This section describes the dynamical source models used for Challenge 2. See \citet{Chatterjee2022} for more detailed model descriptions and comparisons. 
\subsubsection{M87}
\begin{figure*}
%\centering
\setlength{\lineskip}{0pt}
\includegraphics[width=30mm]{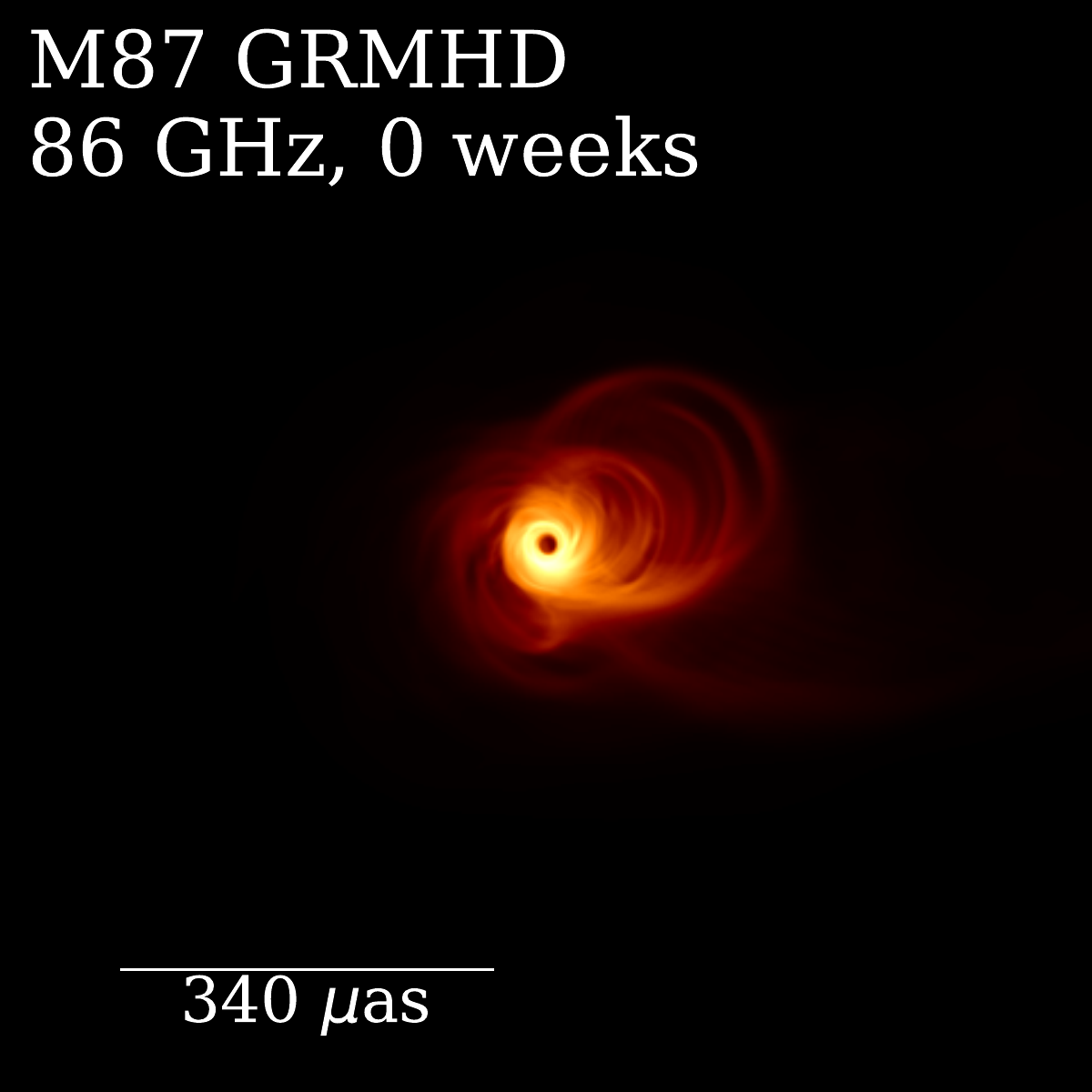}%
\includegraphics[width=30mm]{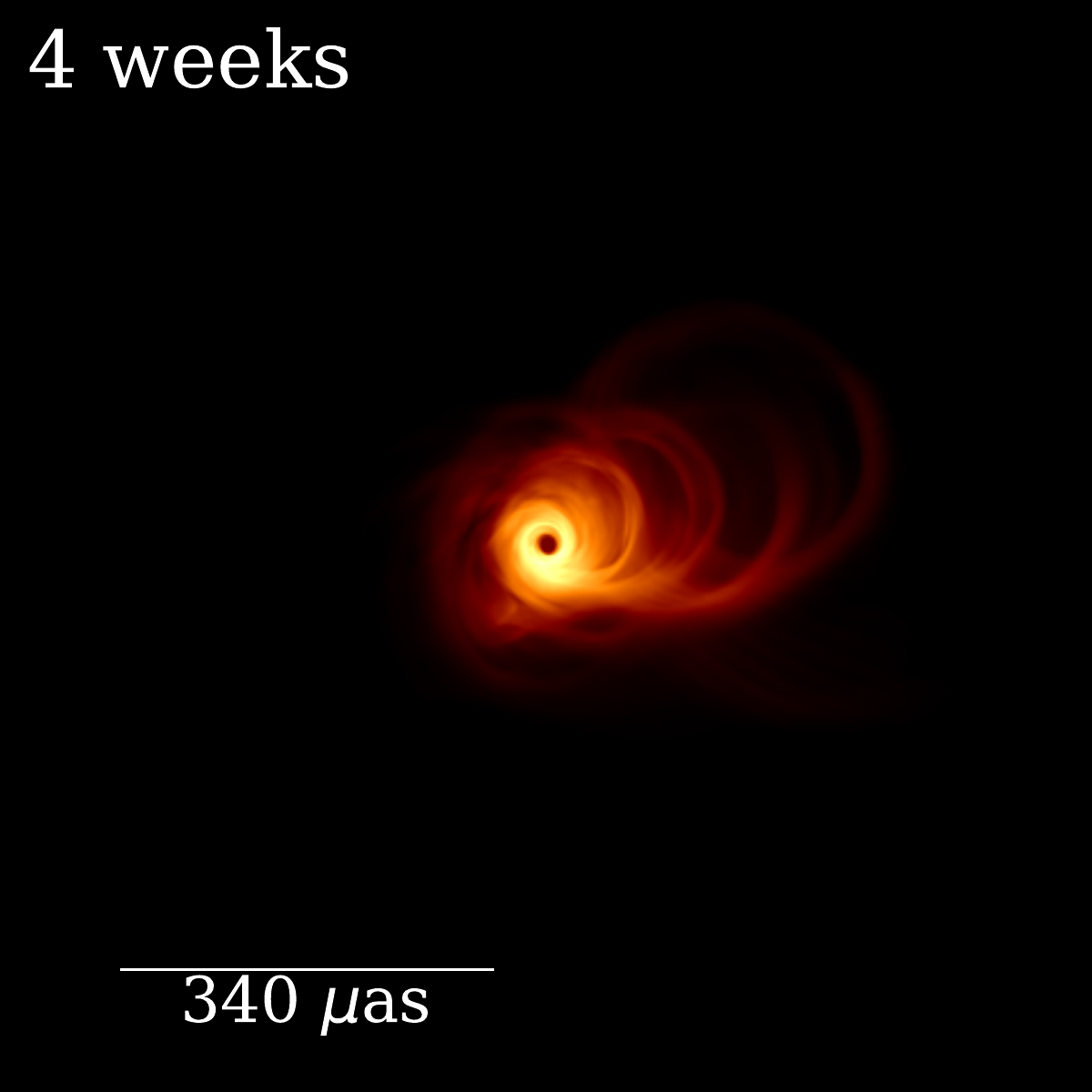}%
\includegraphics[width=30mm]{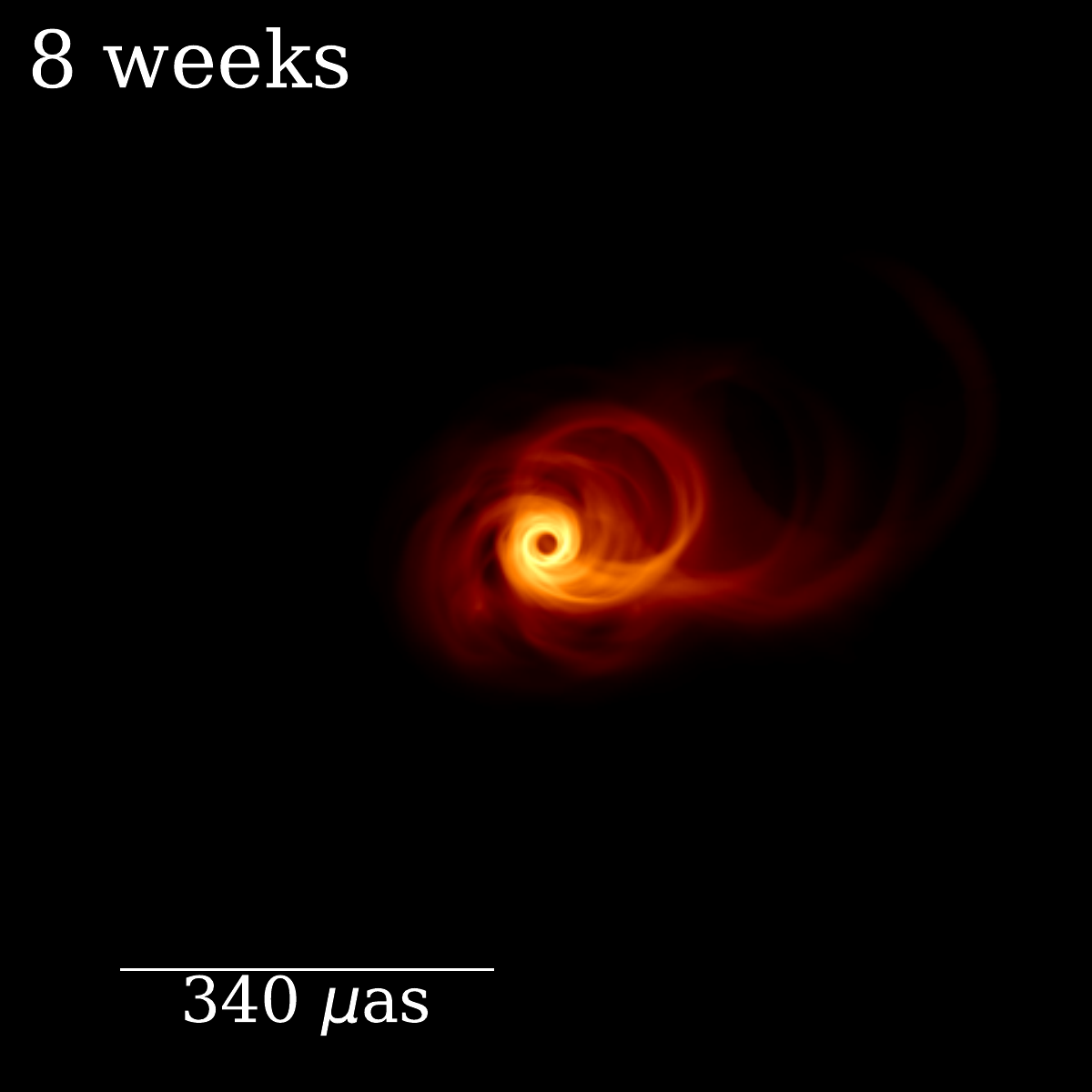} \\
\includegraphics[width=30mm]{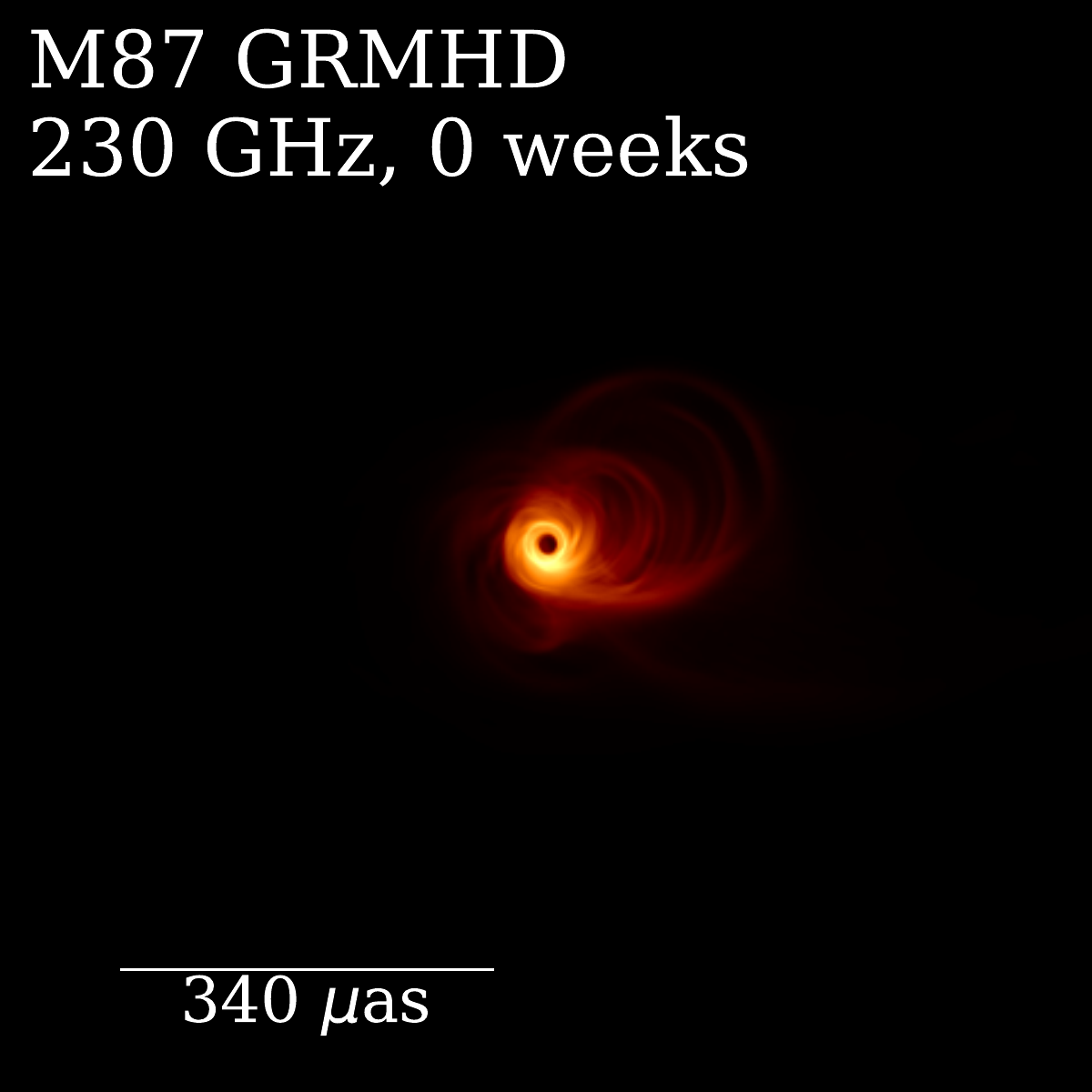}%
\includegraphics[width=30mm]{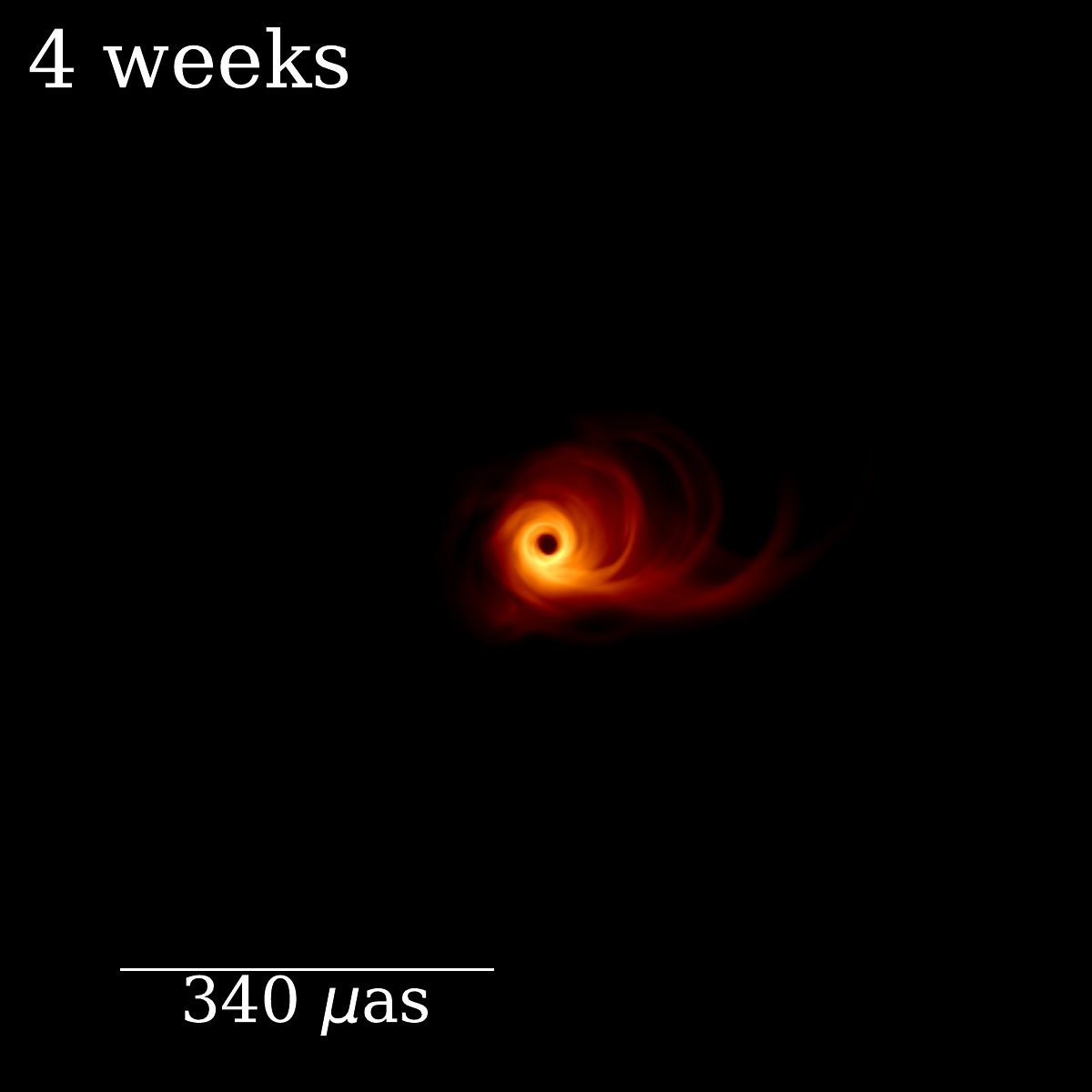}%
\includegraphics[width=30mm]{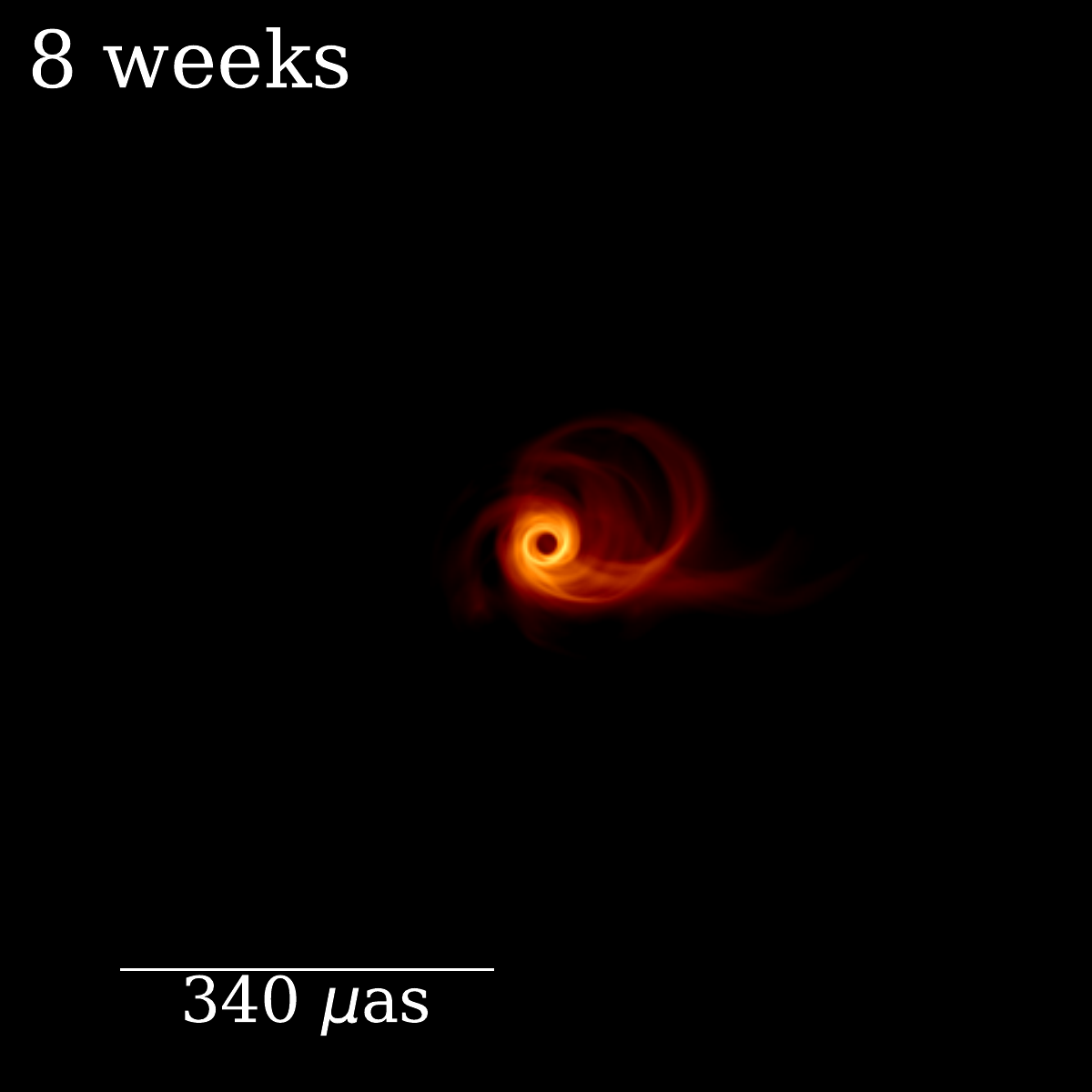} \\
\includegraphics[width=30mm]{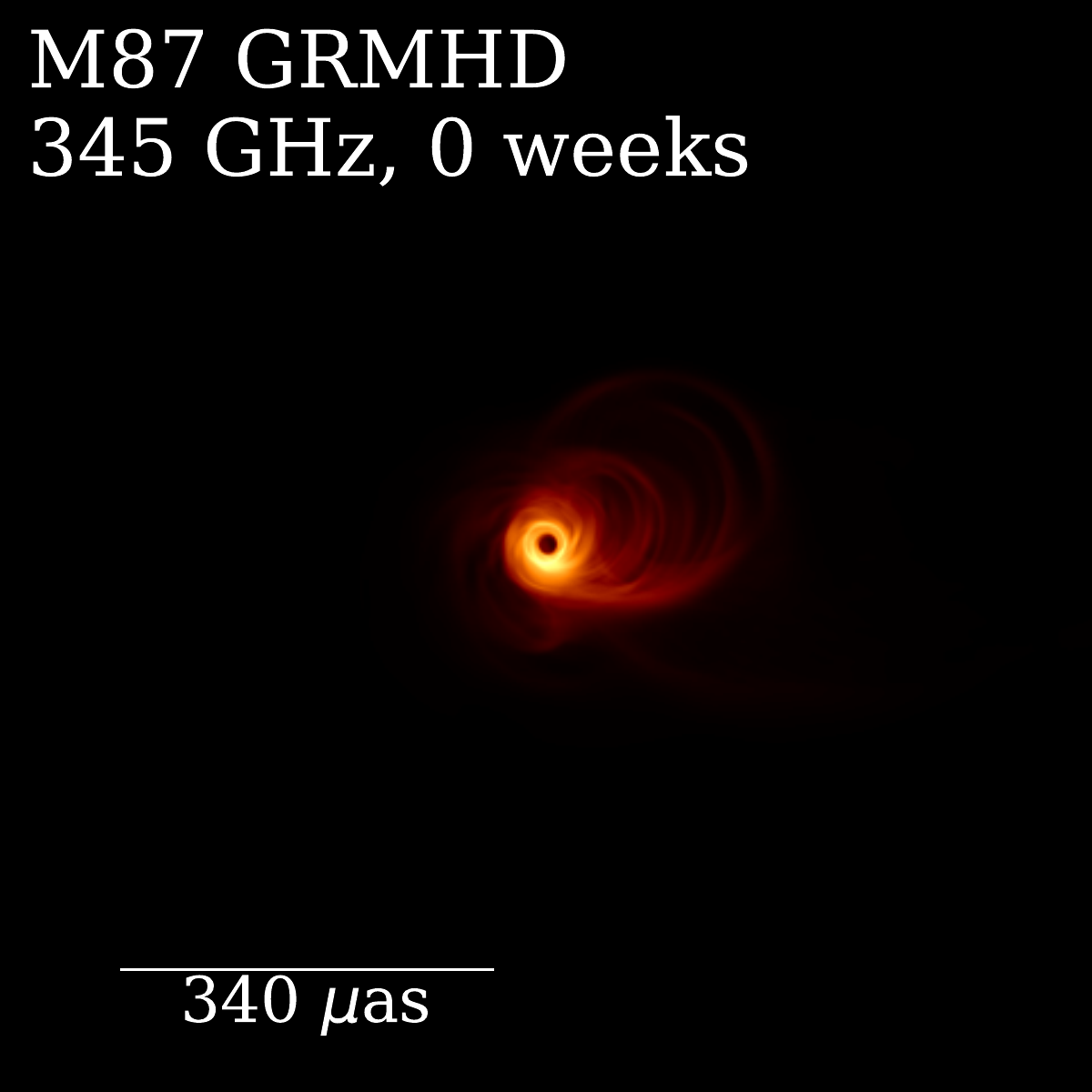}%
\includegraphics[width=30mm]{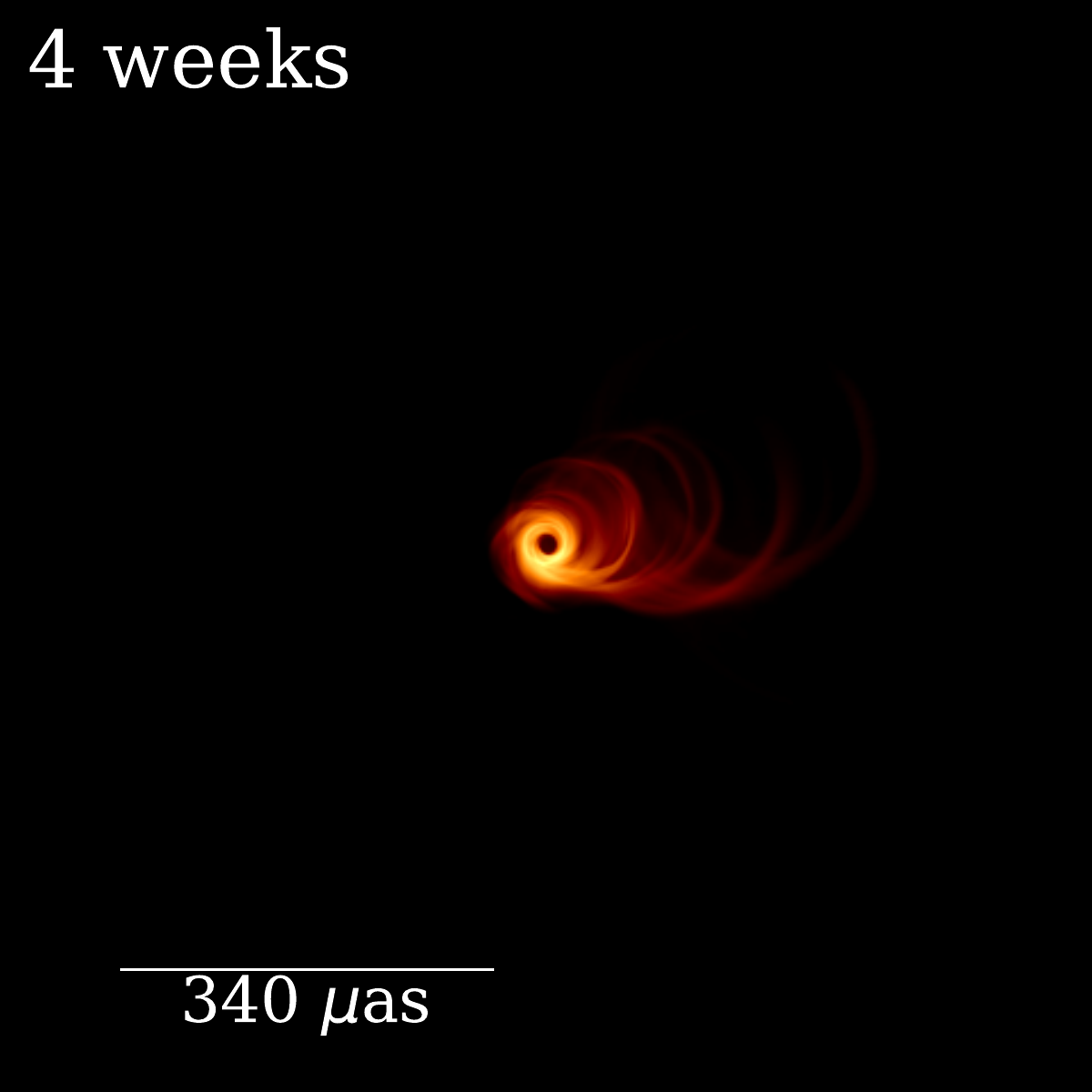}%
\includegraphics[width=30mm]{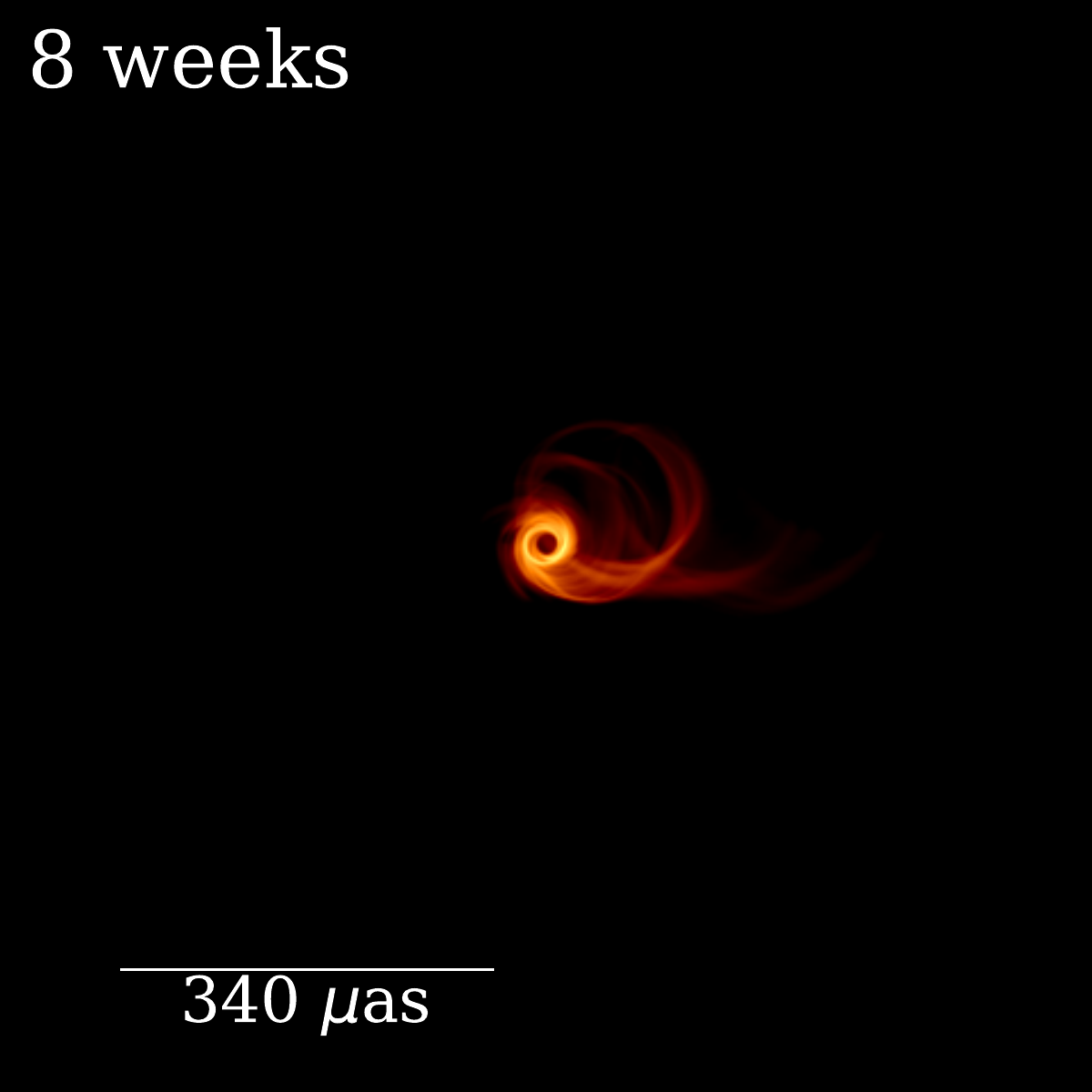}

  \caption{Ground truth M87 source models used for Challenge 2. For each frequency, three movie frames are shown. Images are shown on a log scale, which is normalized to the brightest pixel value across each set of three movie frames, with a dynamic range of $10^{3.5}$.}
     \label{fig:ch2_models_m87}
\end{figure*}

The Challenge 2 M87 model is a GRMHD movie with 20 frames that are spaced 20$GM/c^3$ ($\sim$1 week) apart. The pixel resolution is 2048x2048, with a field of view of 1 mas. The images were ray-traced from a H-AMR \citep{Liska2019} simulation (MAD, spin 0.94) using \texttt{ipole} \citep{Moscibrodzka2018}. $R_{\mathrm{high}}$ was set to 160 and accelerated electron heating was included, setting $\kappa=3.5$ \citep[e.g.][]{Davelaar2019}. We only use the Stokes I information from the model. The model is shown in Figure \ref{fig:ch2_models_m87}.

\subsubsection{Sgr A*}
\begin{figure*}
\begin{adjustwidth}{-\extralength}{0cm}
\setlength{\lineskip}{0pt}
\centering
\includegraphics[width=30mm]{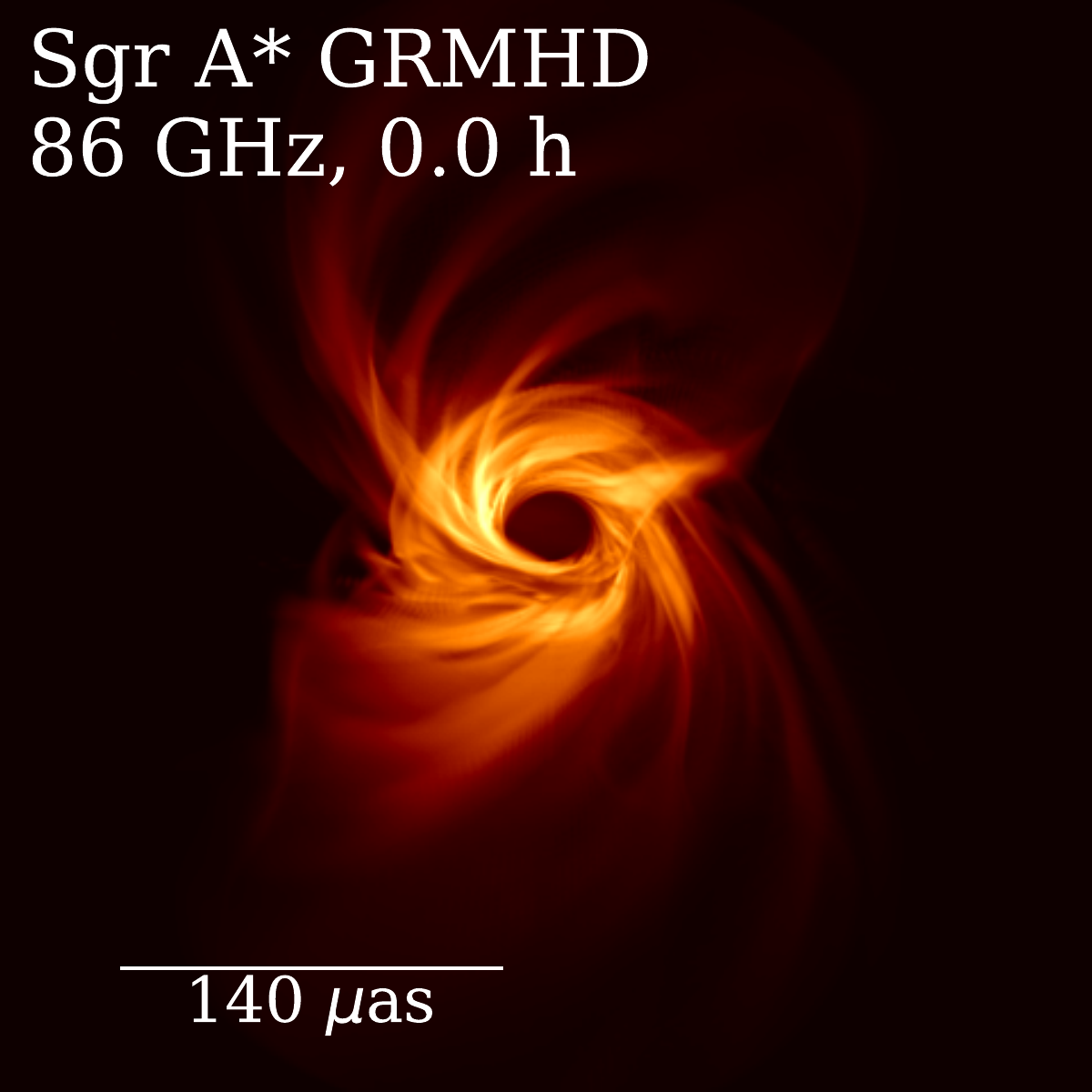}%
\includegraphics[width=30mm]{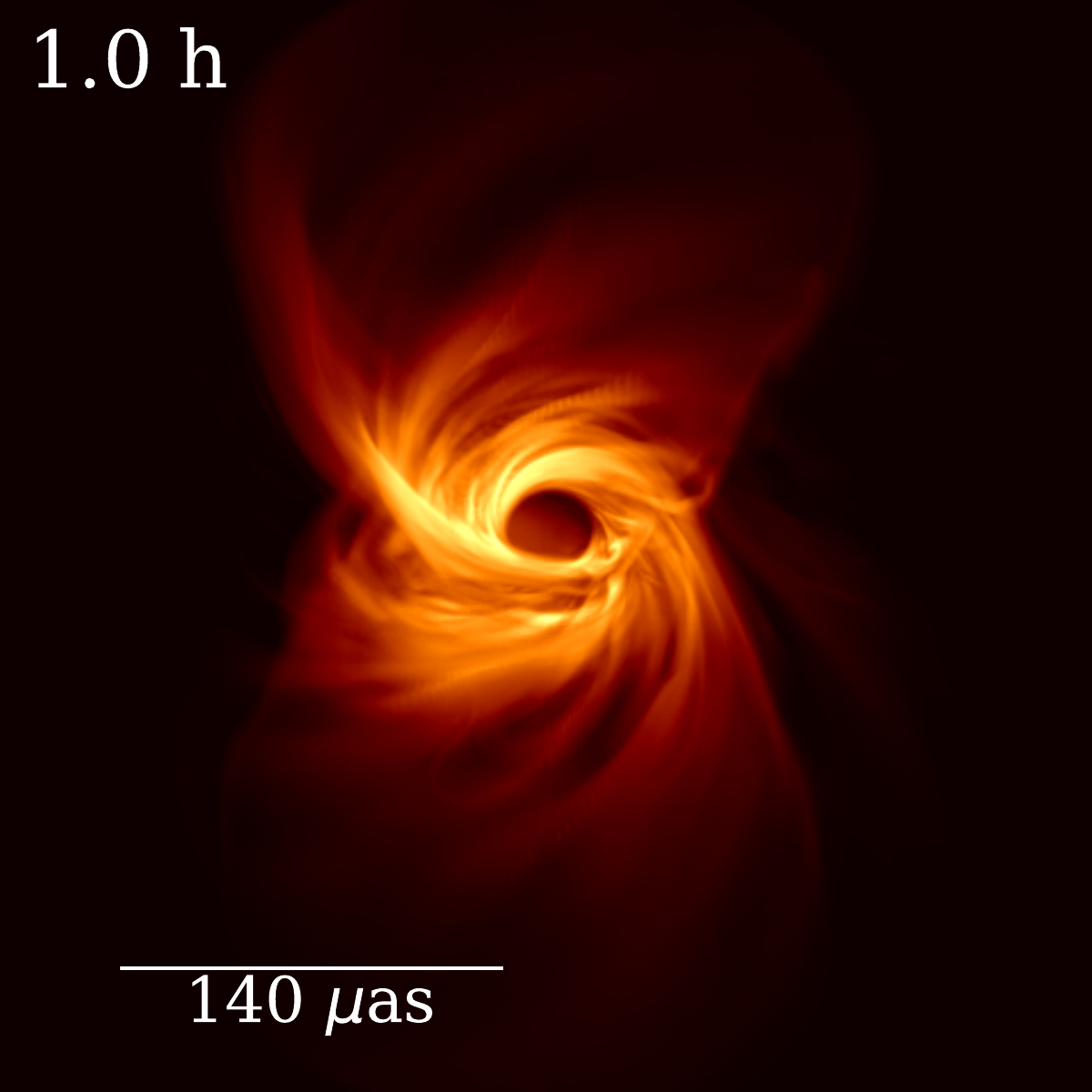}%
\includegraphics[width=30mm]{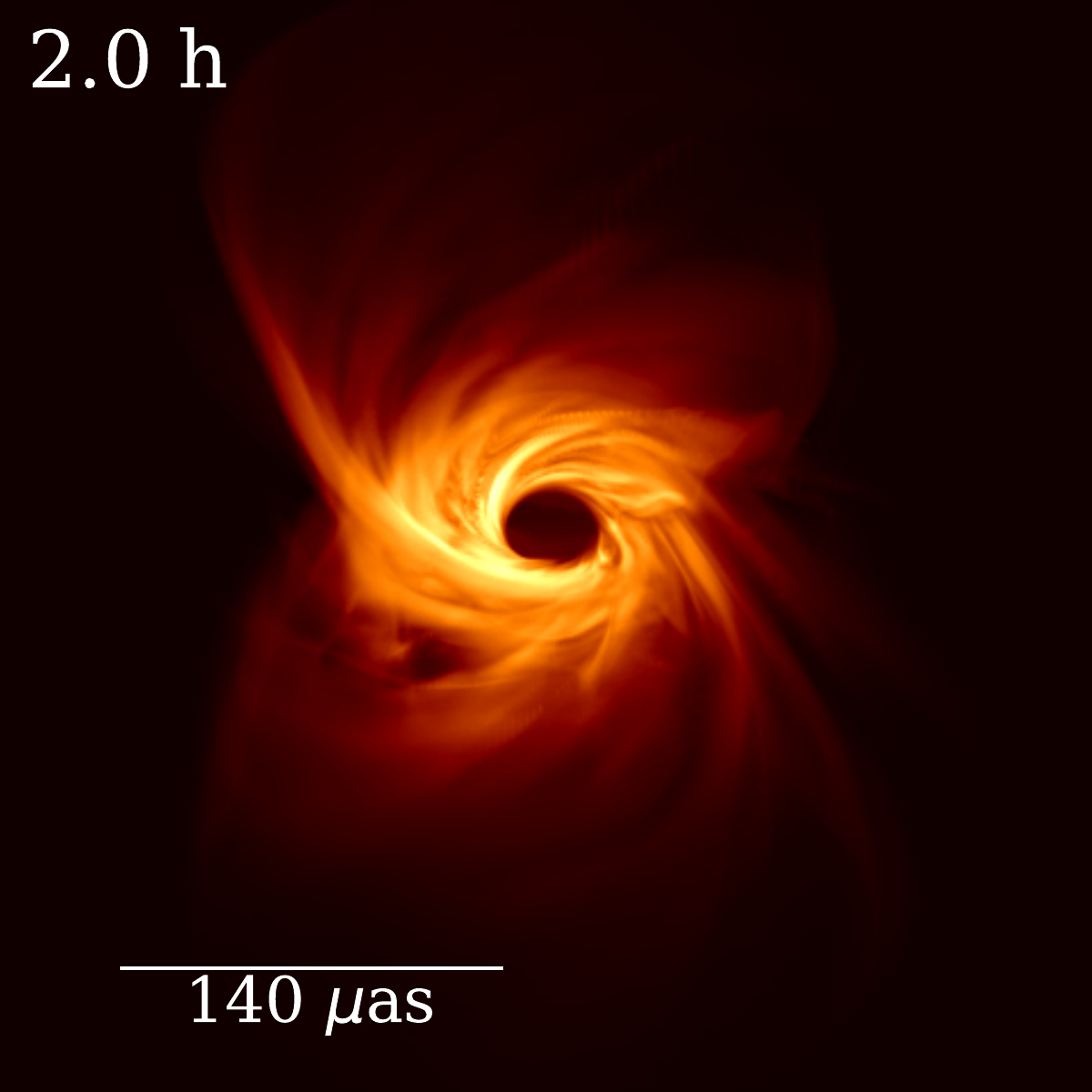}%
\includegraphics[width=30mm]{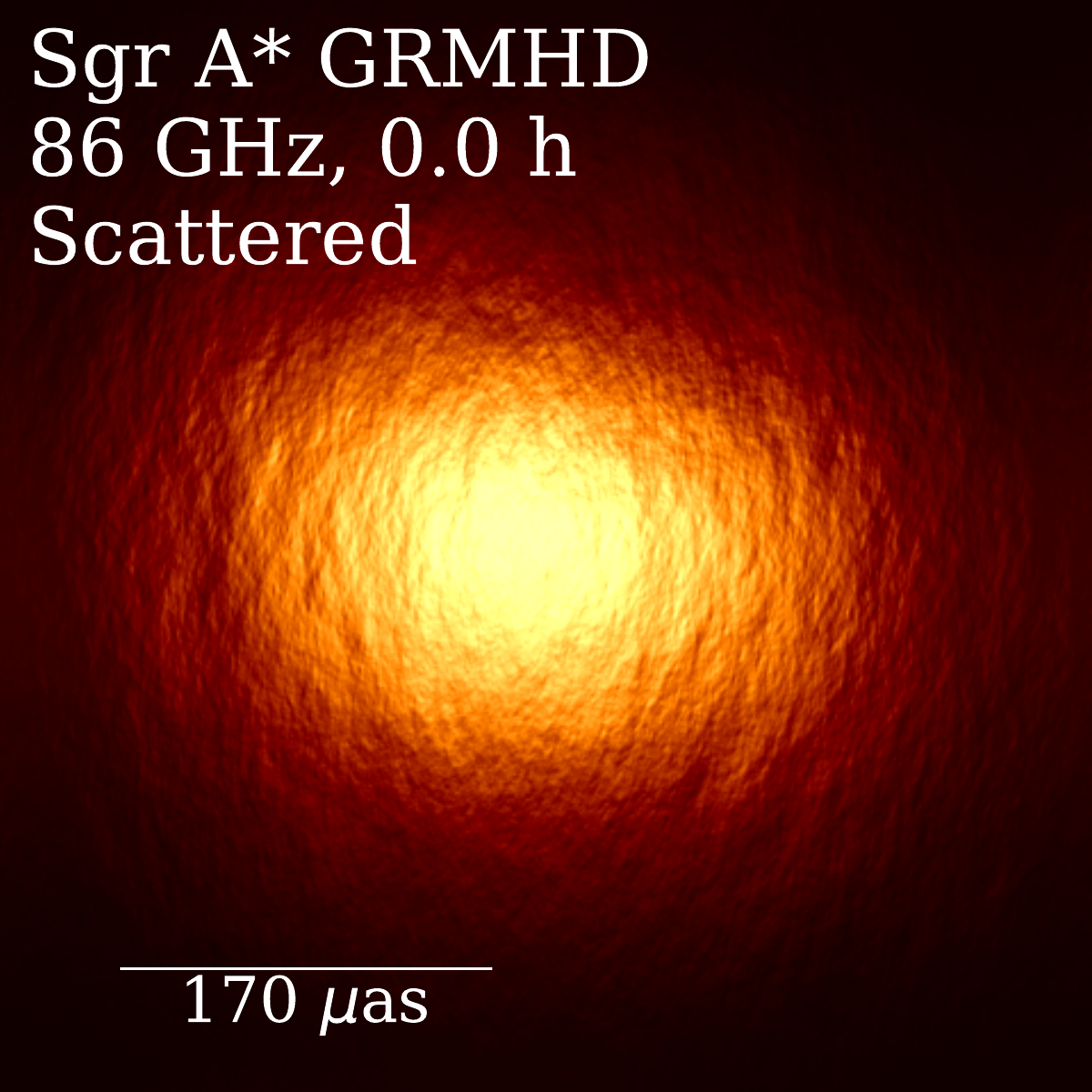}%
\includegraphics[width=30mm]{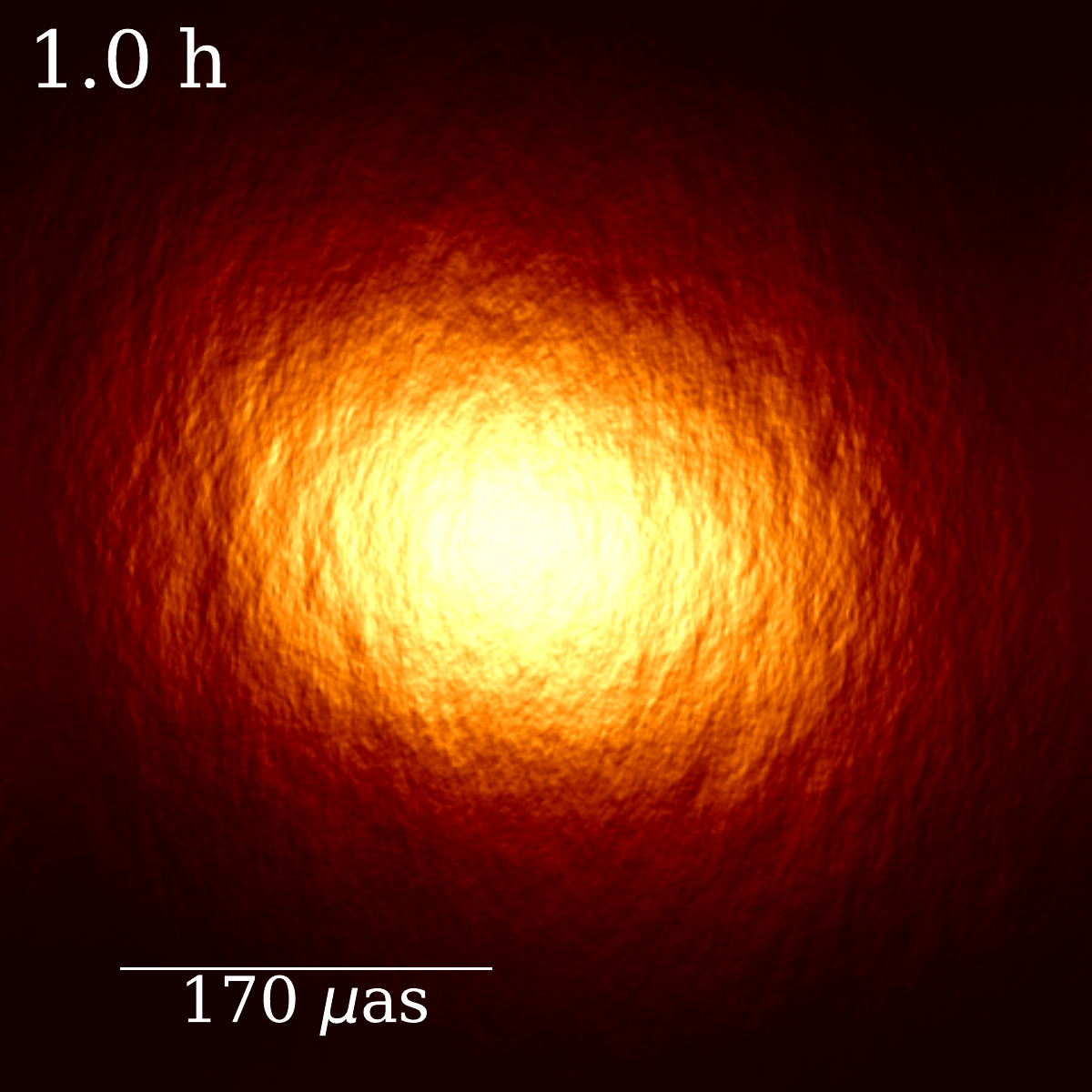}%
\includegraphics[width=30mm]{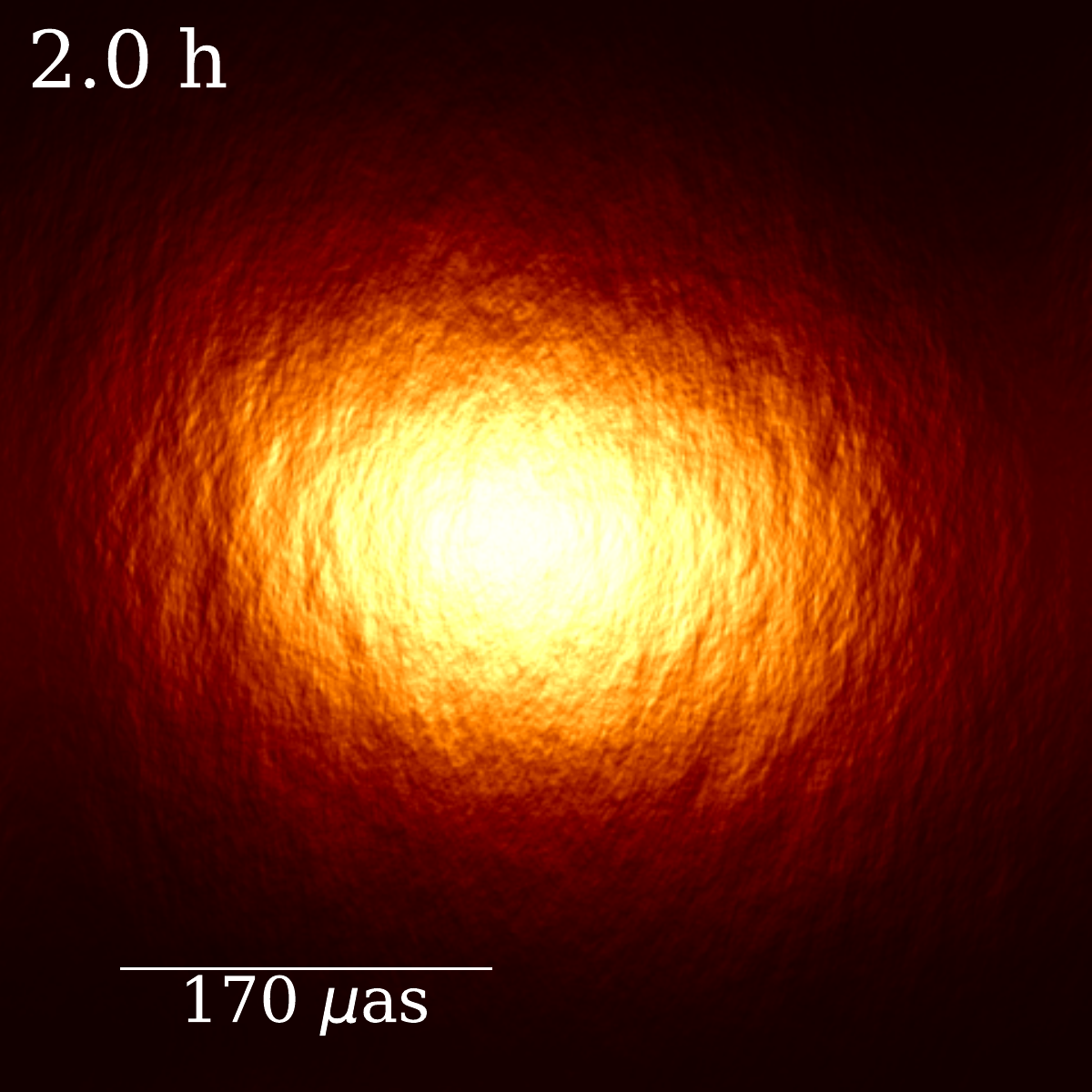} \\
\includegraphics[width=30mm]{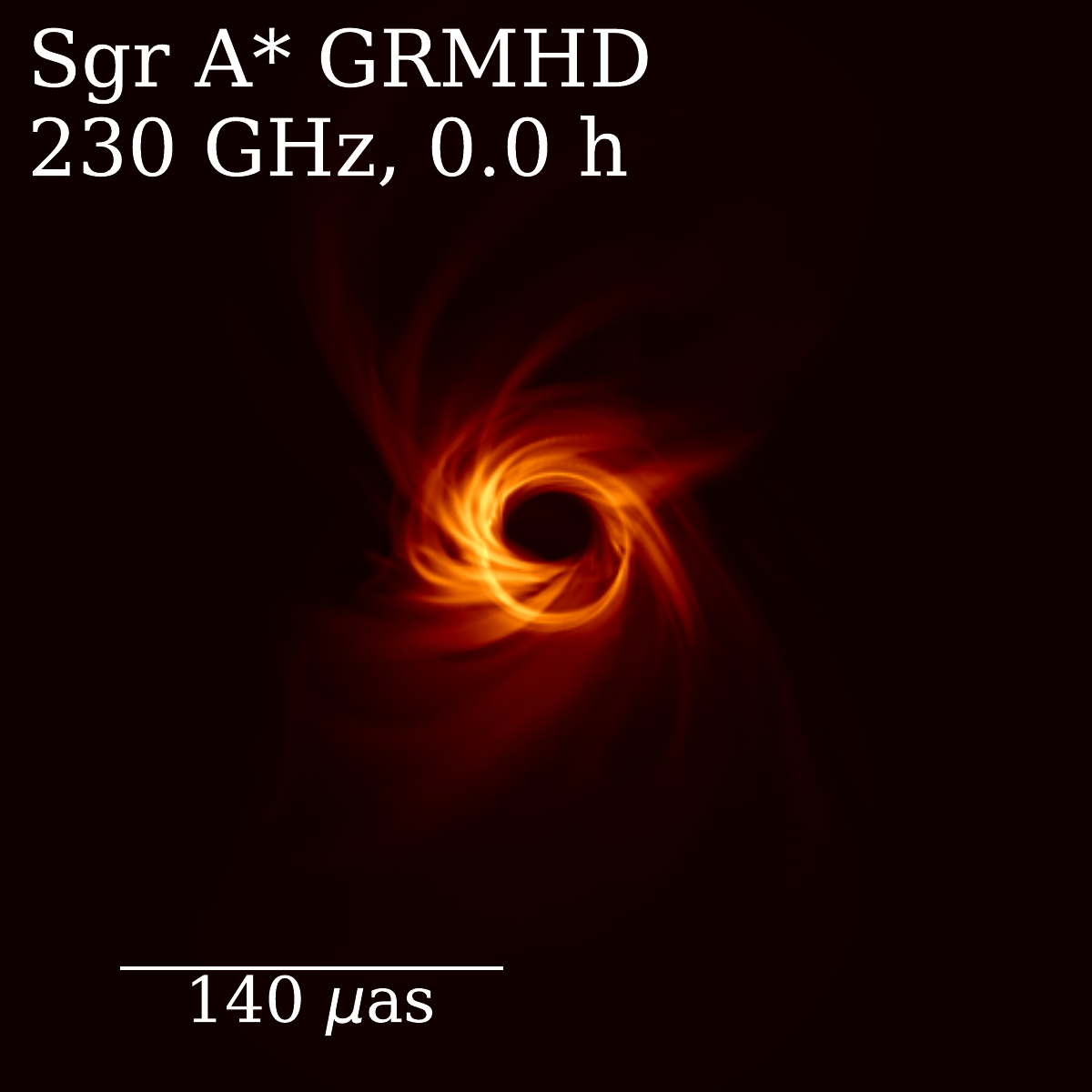}%
\includegraphics[width=30mm]{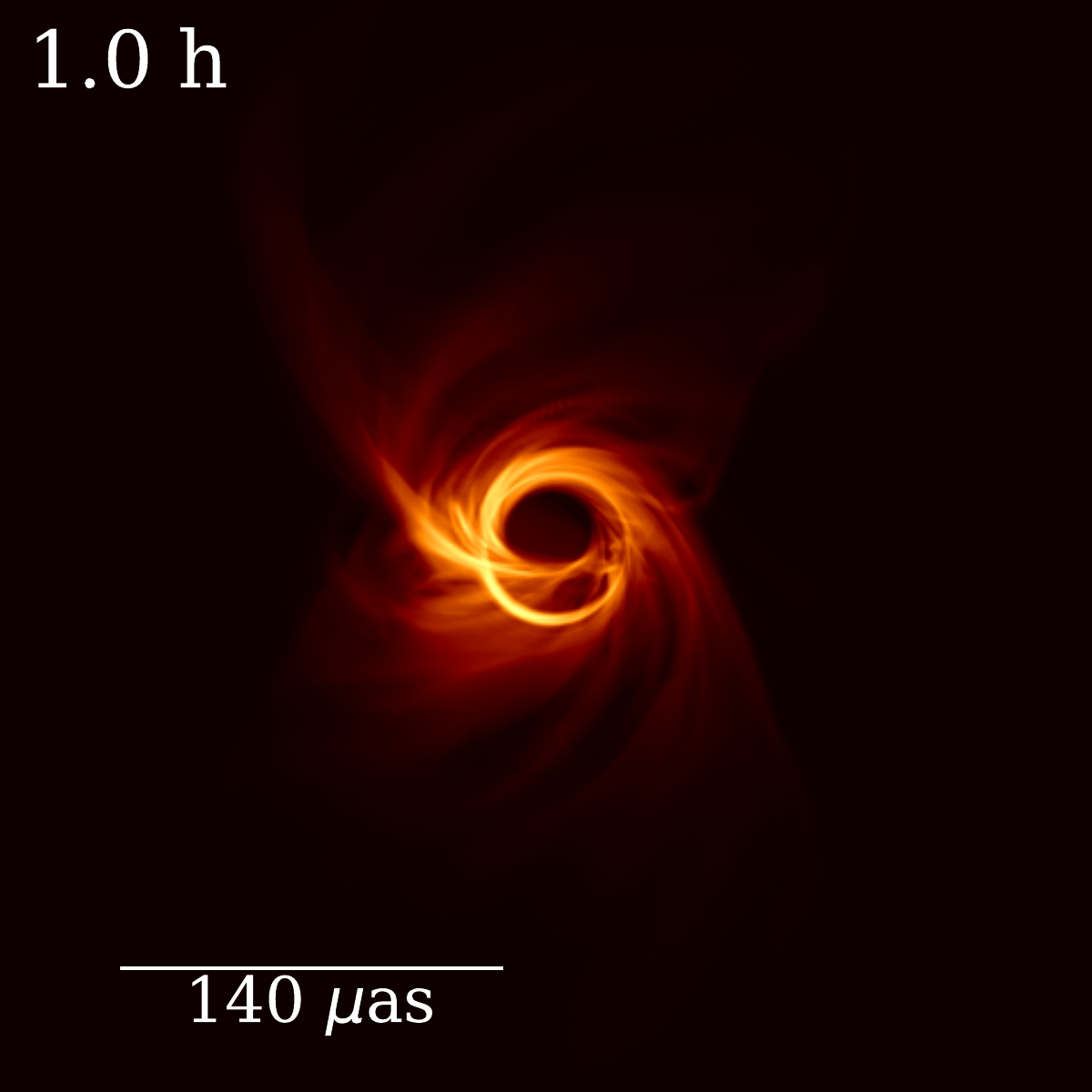}%
\includegraphics[width=30mm]{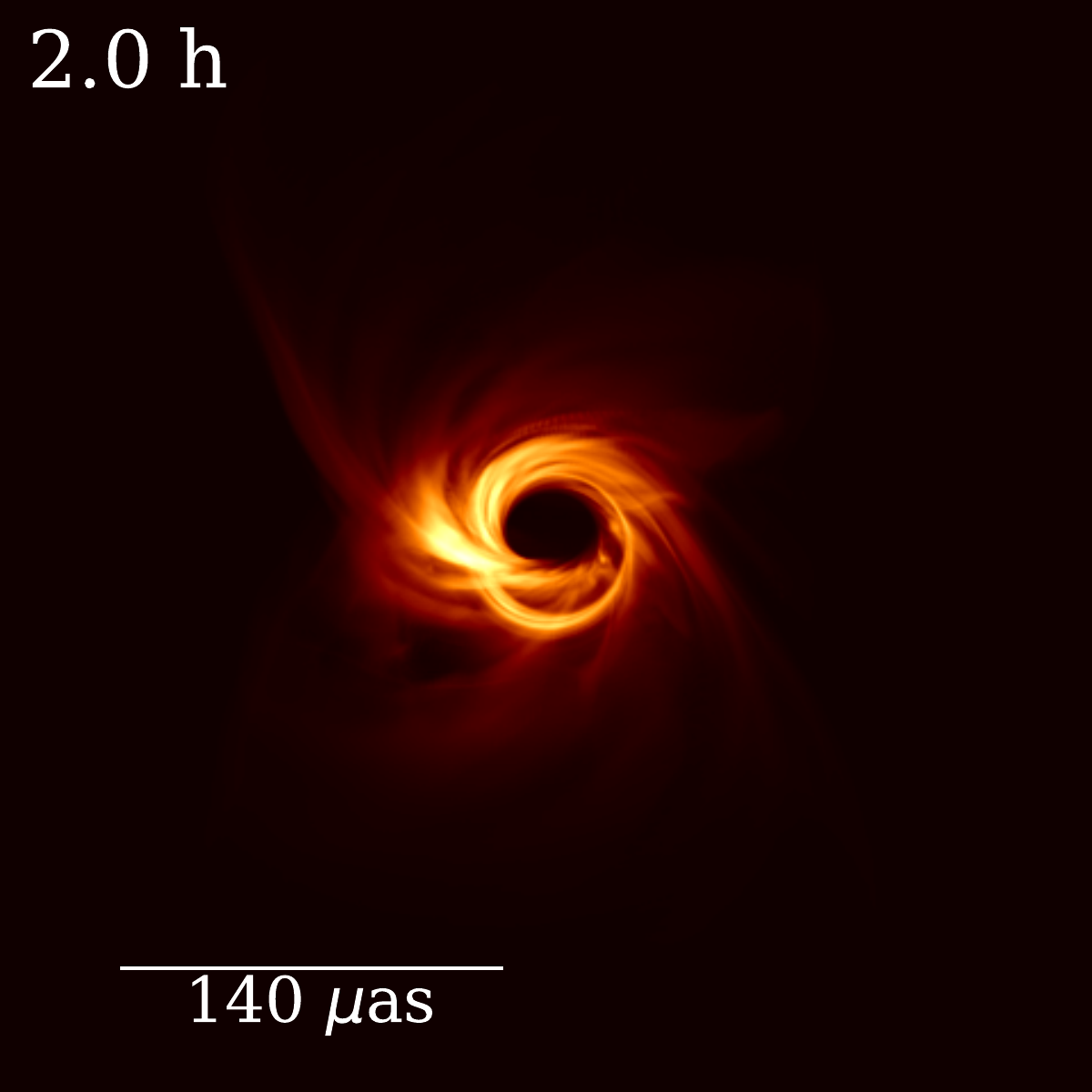}%
\includegraphics[width=30mm]{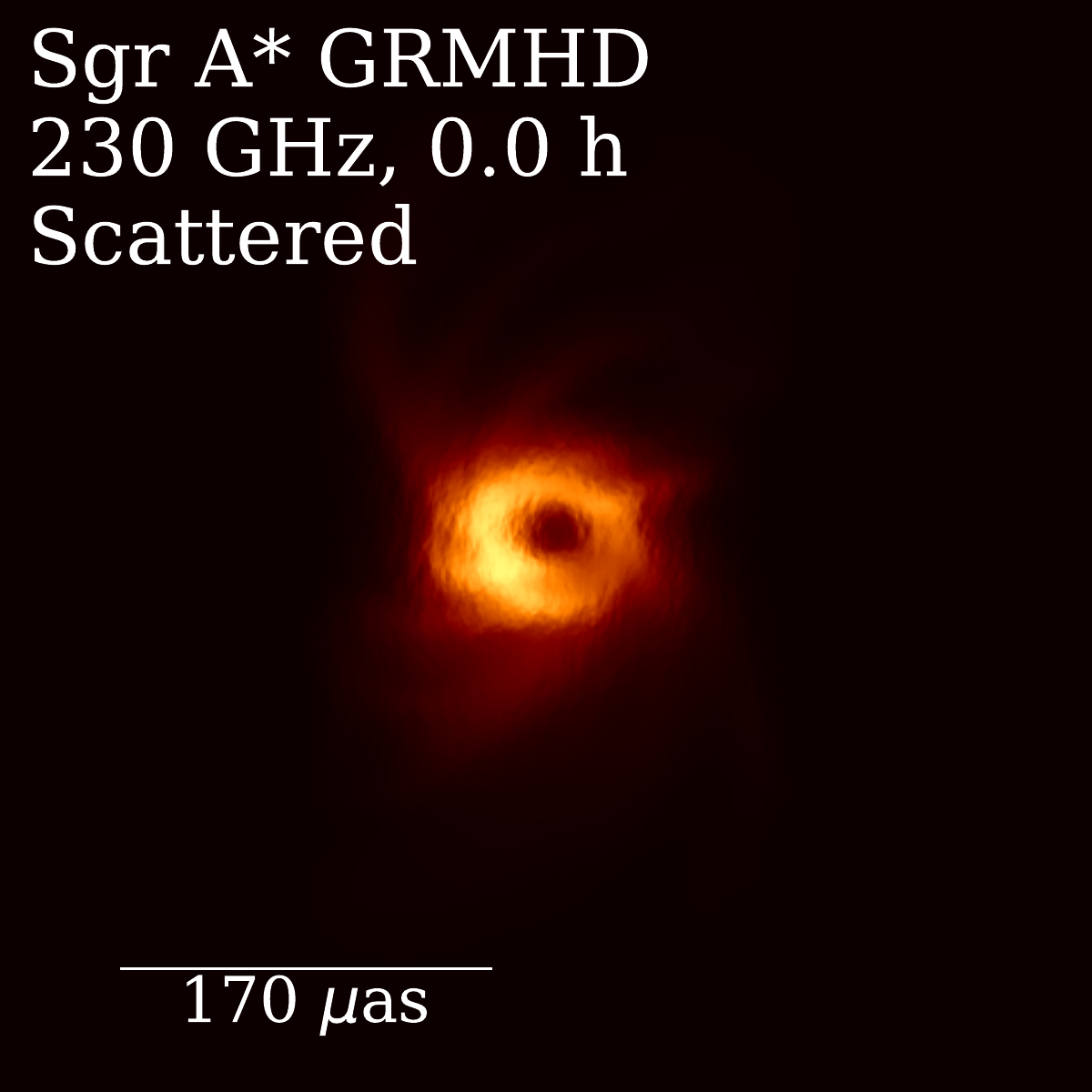}% 
\includegraphics[width=30mm]{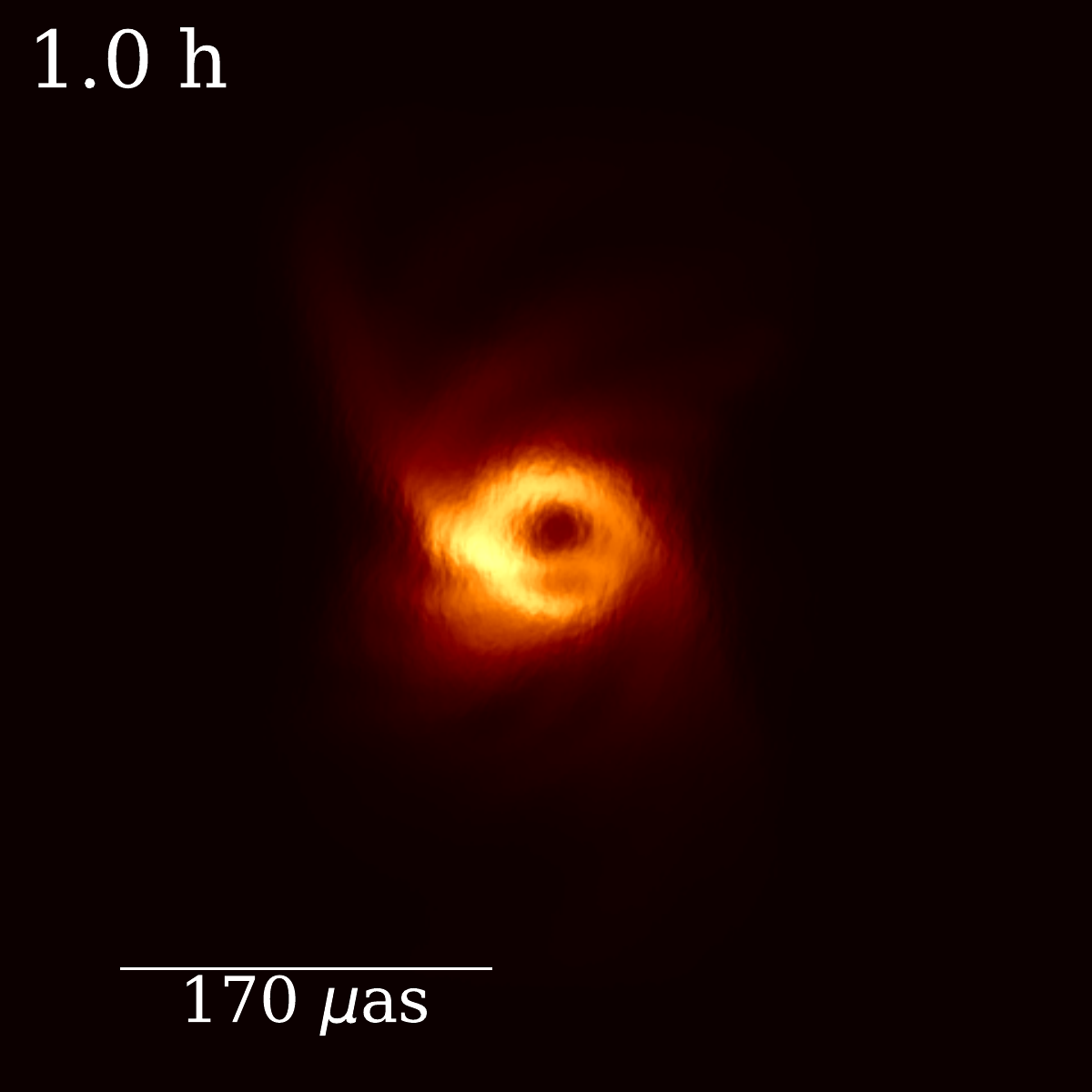}%
\includegraphics[width=30mm]{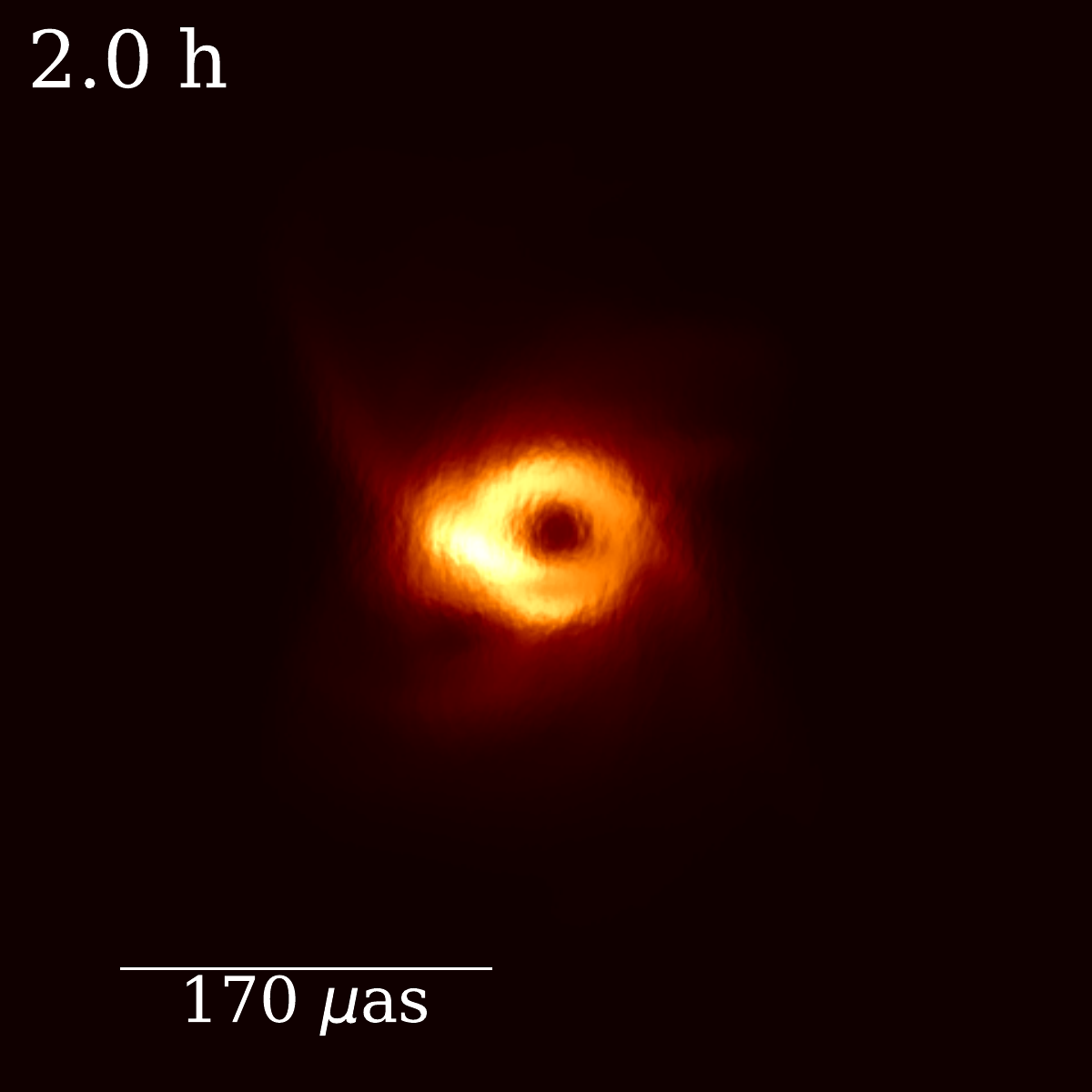} \\
\includegraphics[width=30mm]{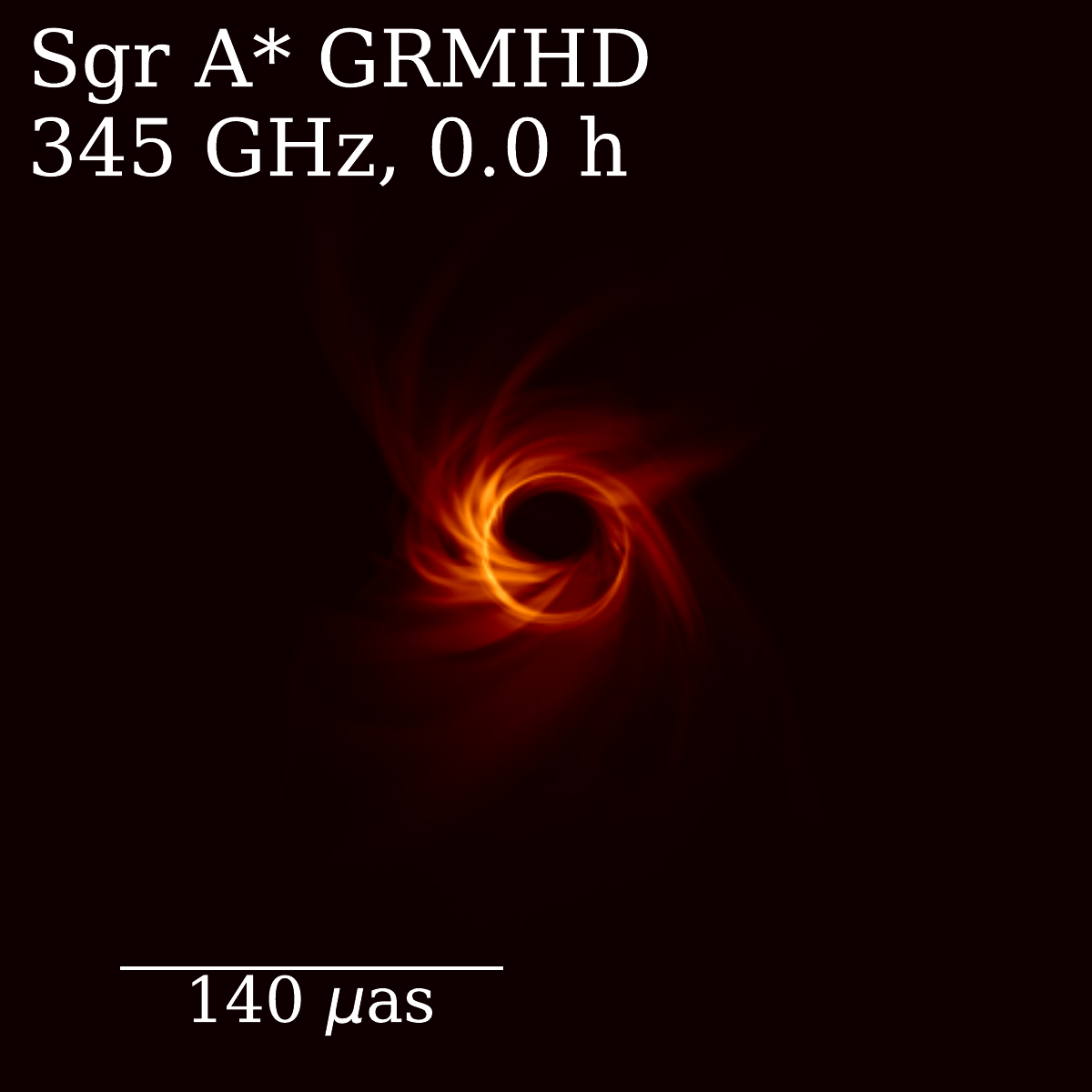}%
\includegraphics[width=30mm]{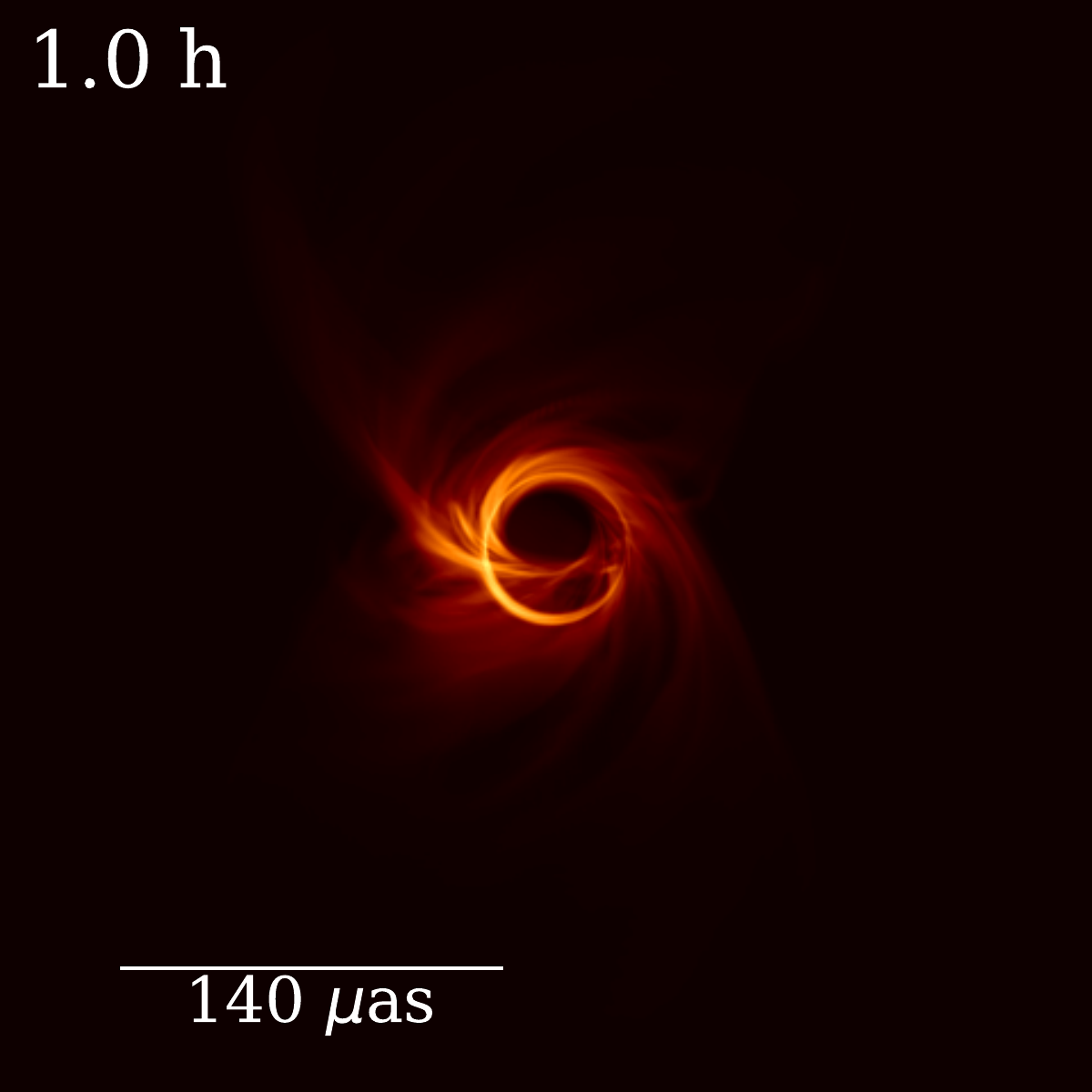}%
\includegraphics[width=30mm]{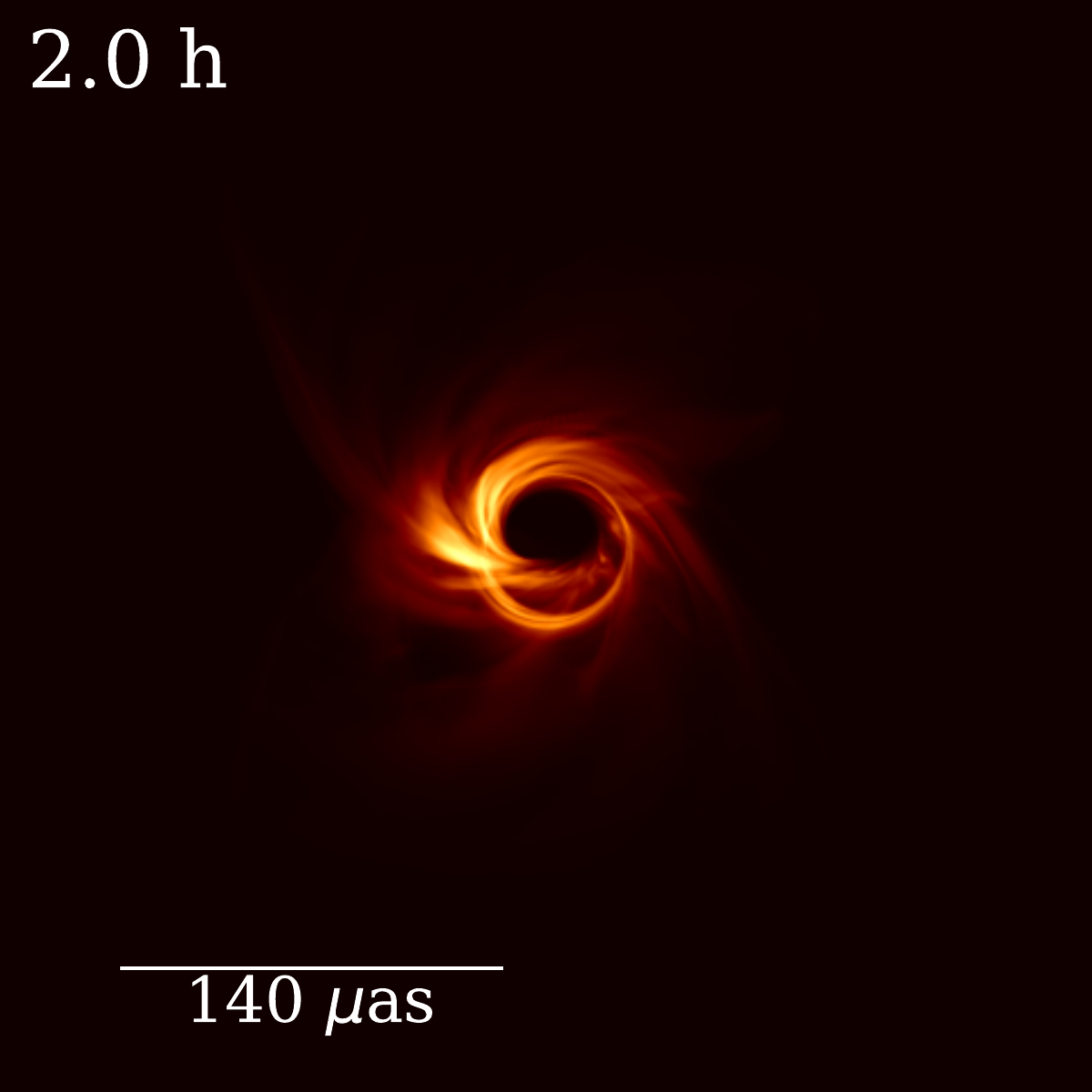}%
\includegraphics[width=30mm]{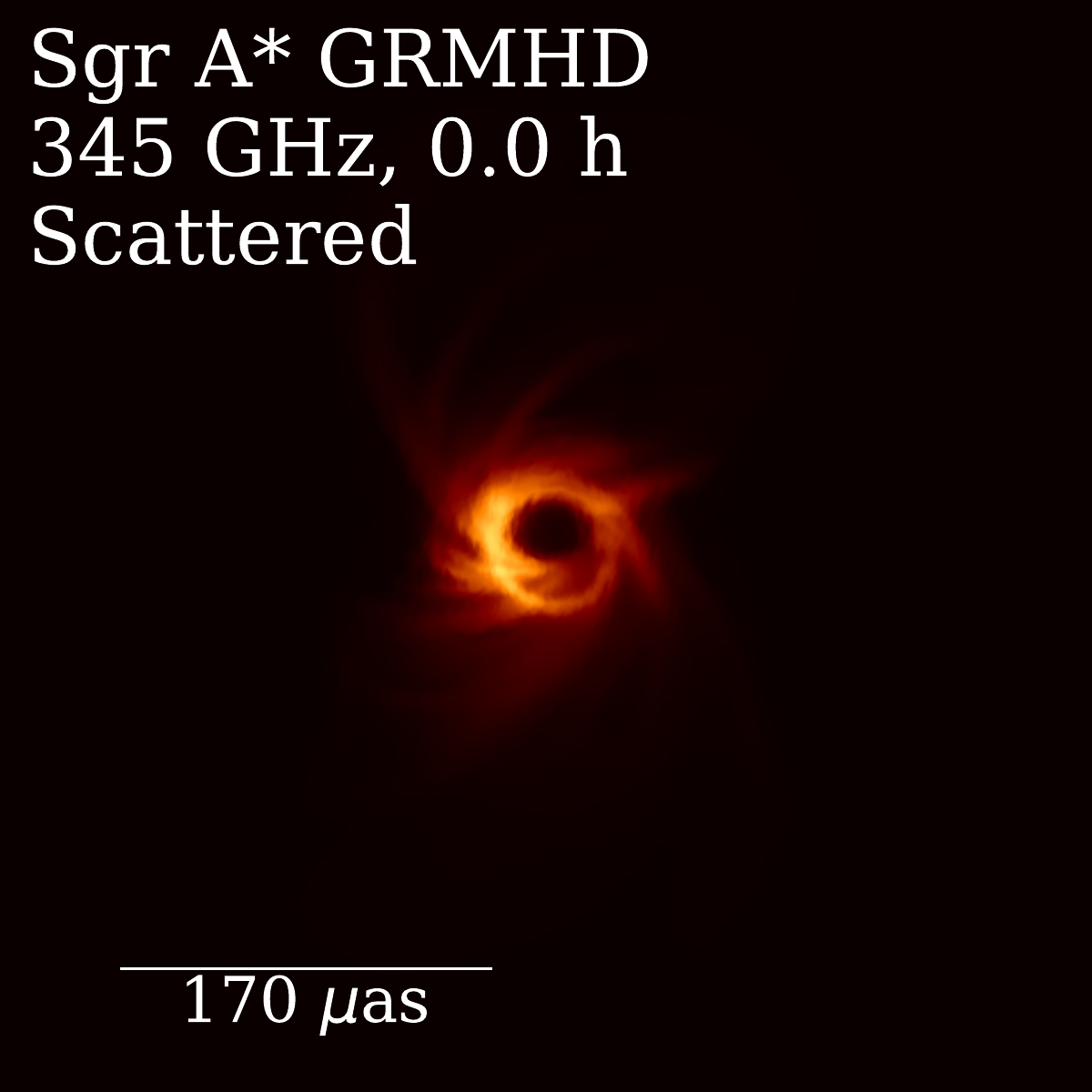}% 
\includegraphics[width=30mm]{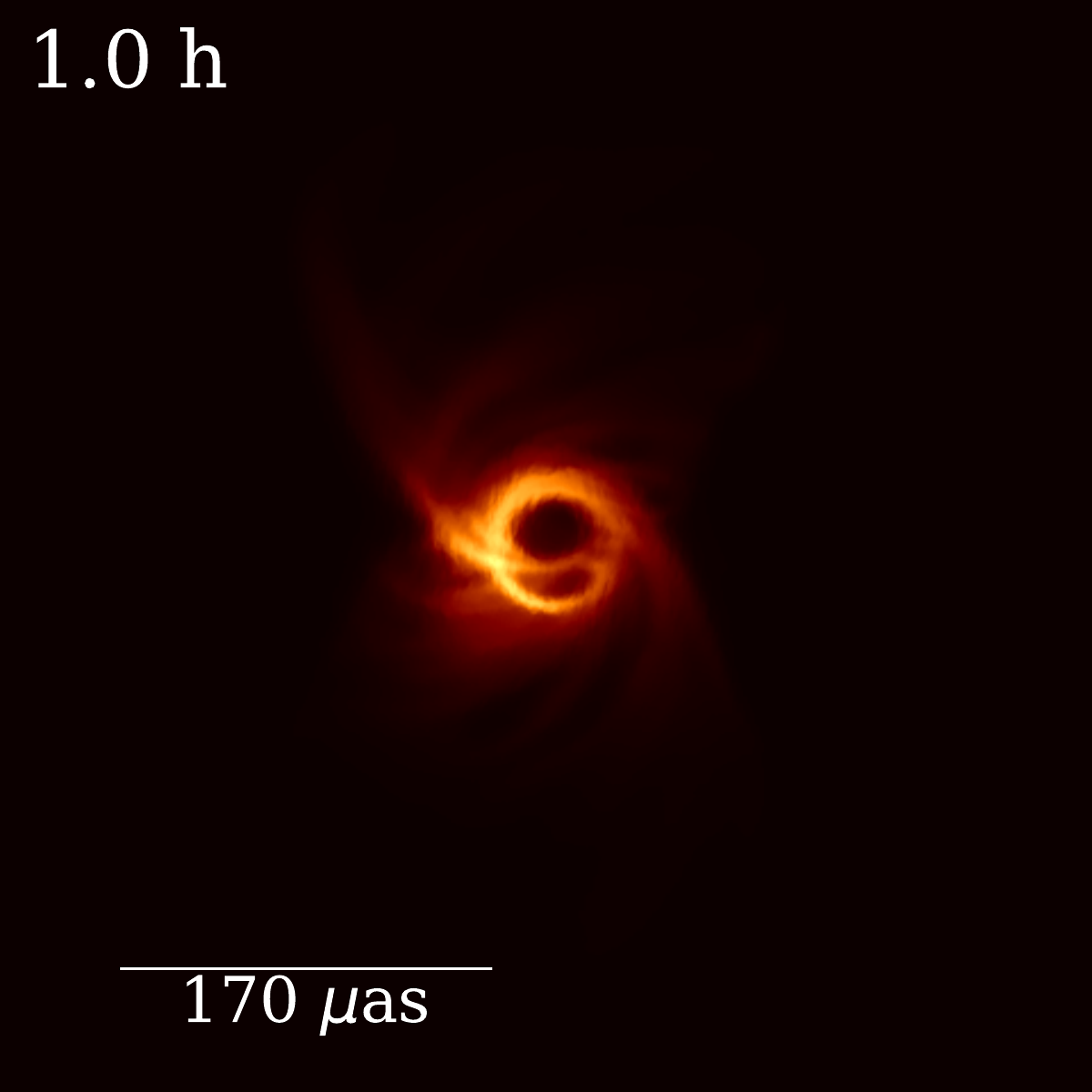}%
\includegraphics[width=30mm]{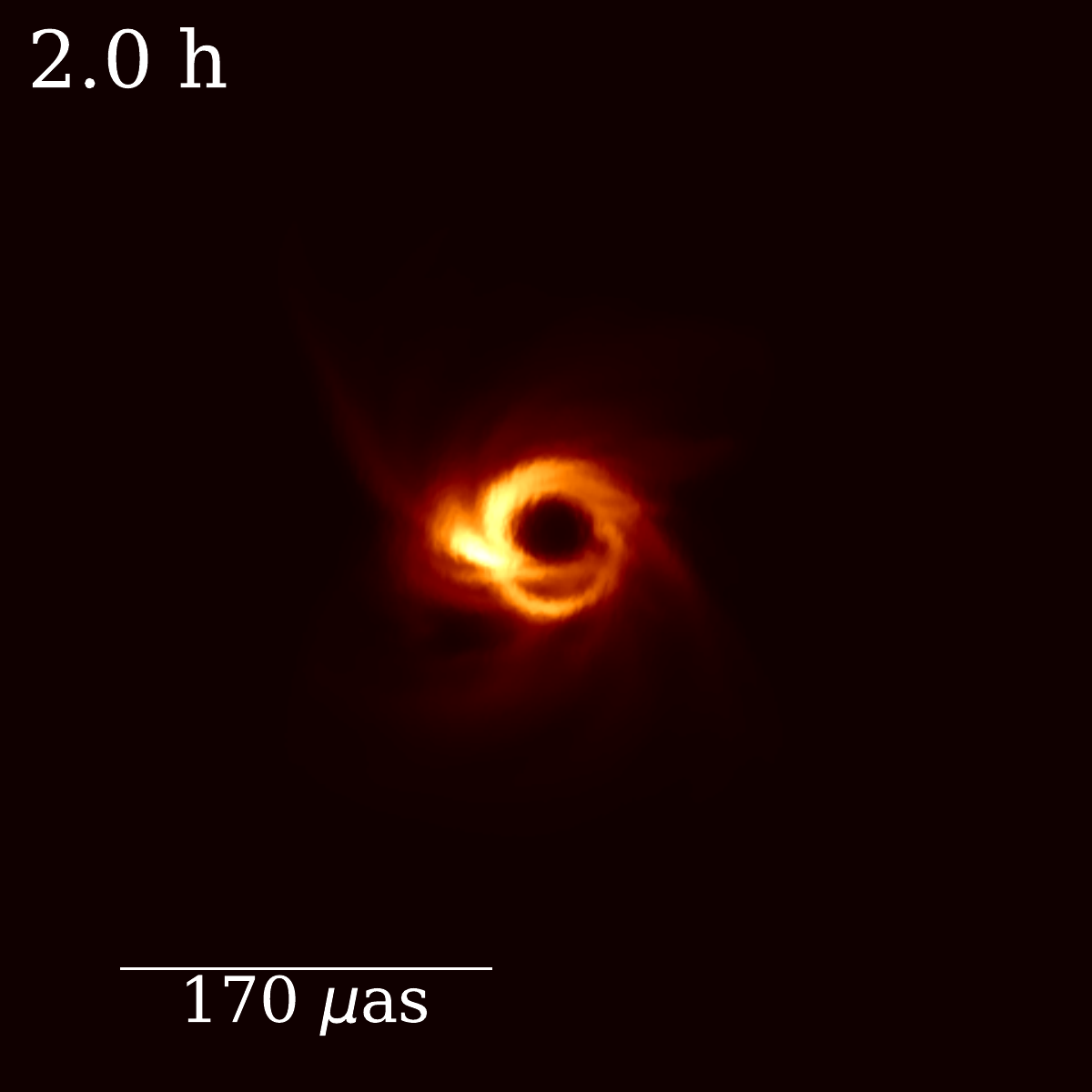} \\
\vspace{2mm}
\includegraphics[width=30mm]{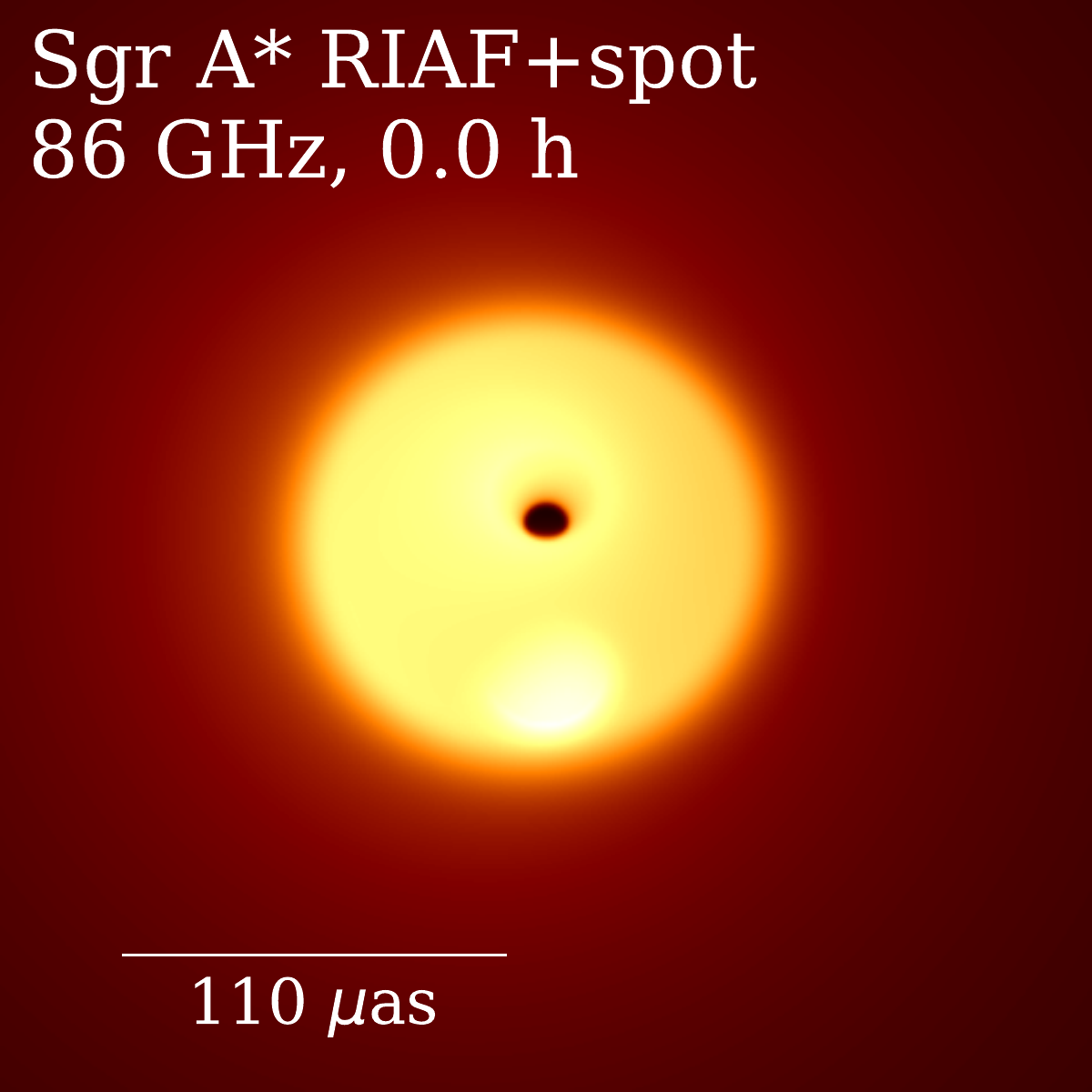}%
\includegraphics[width=30mm]{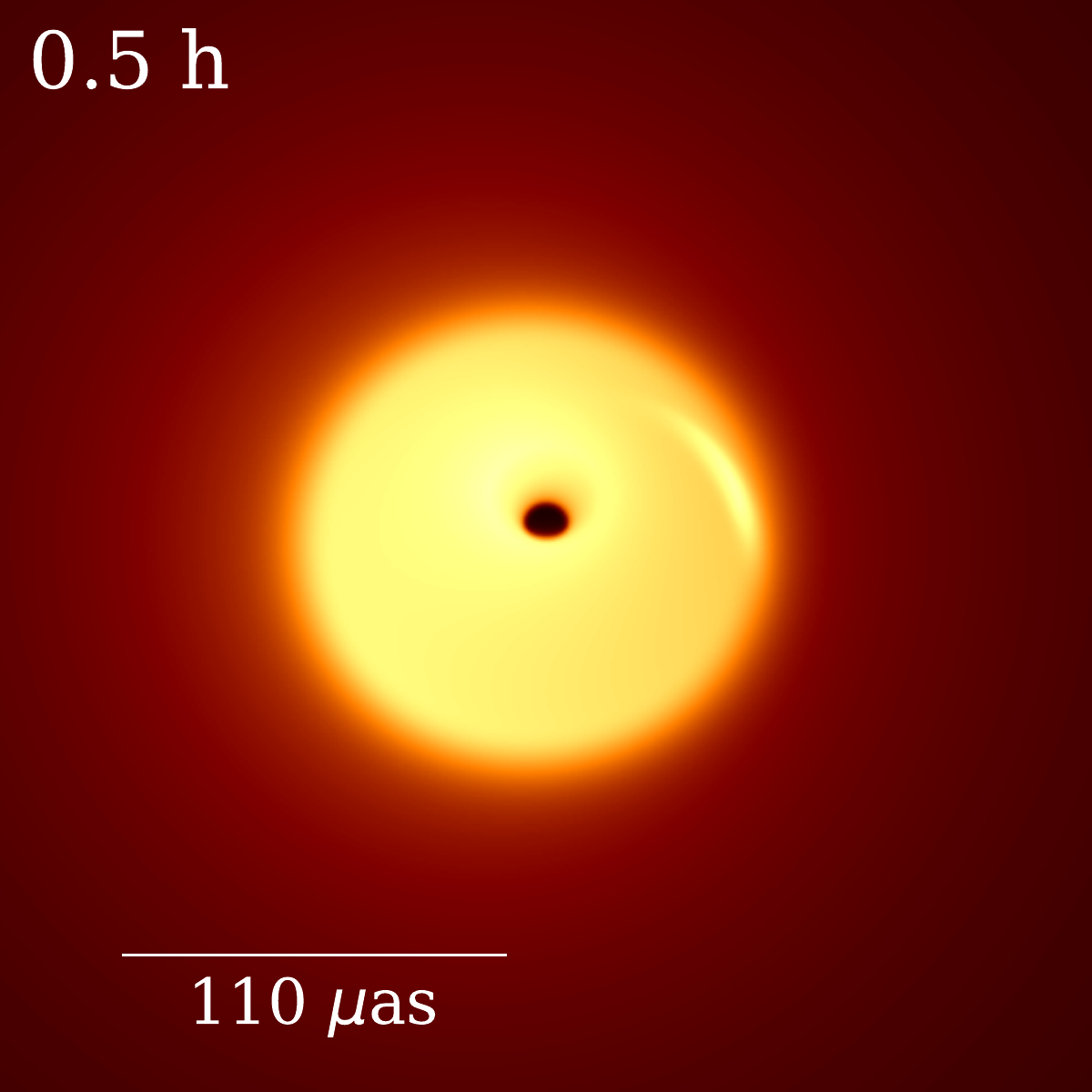}%
\includegraphics[width=30mm]{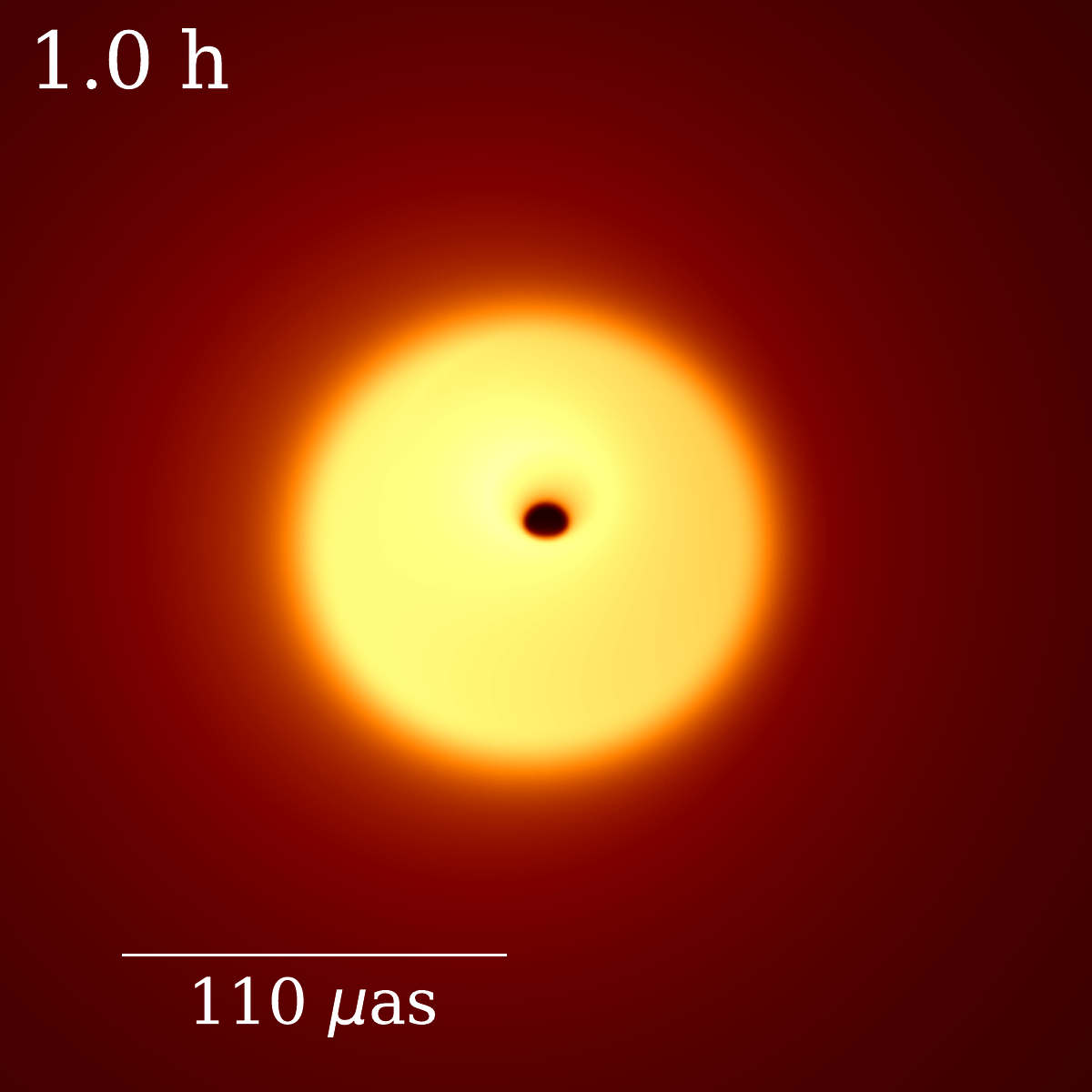}%
\includegraphics[width=30mm]{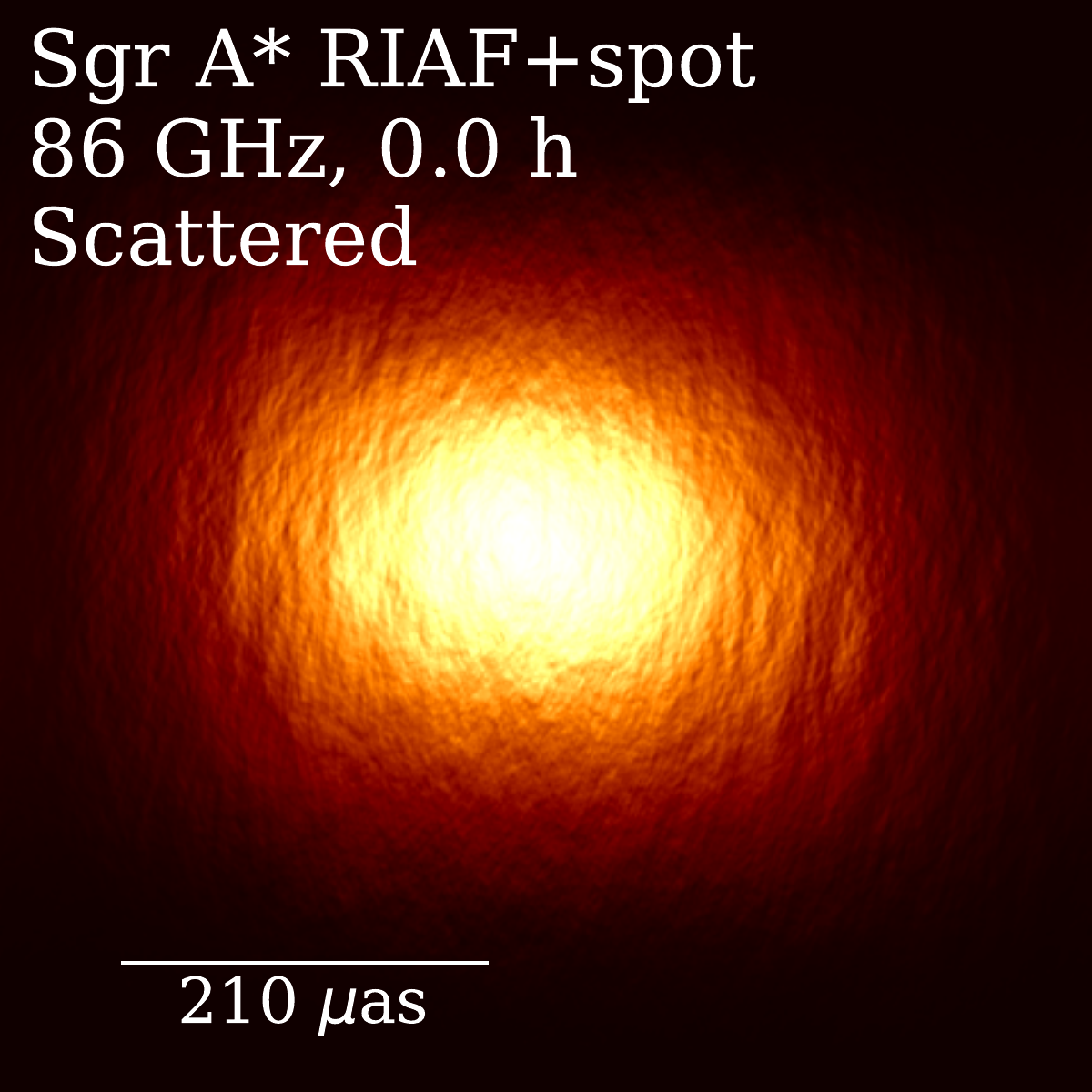}% 
\includegraphics[width=30mm]{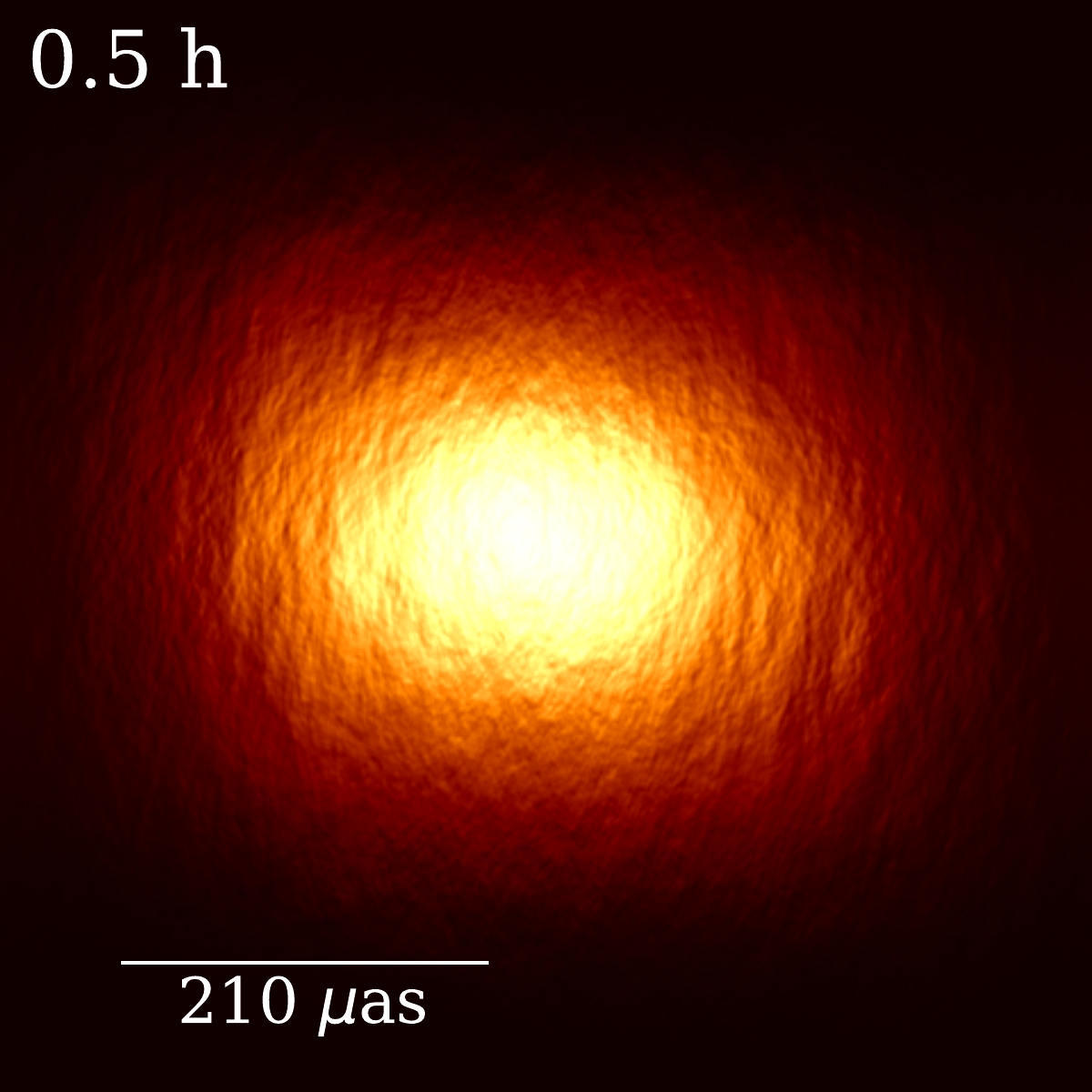}%
\includegraphics[width=30mm]{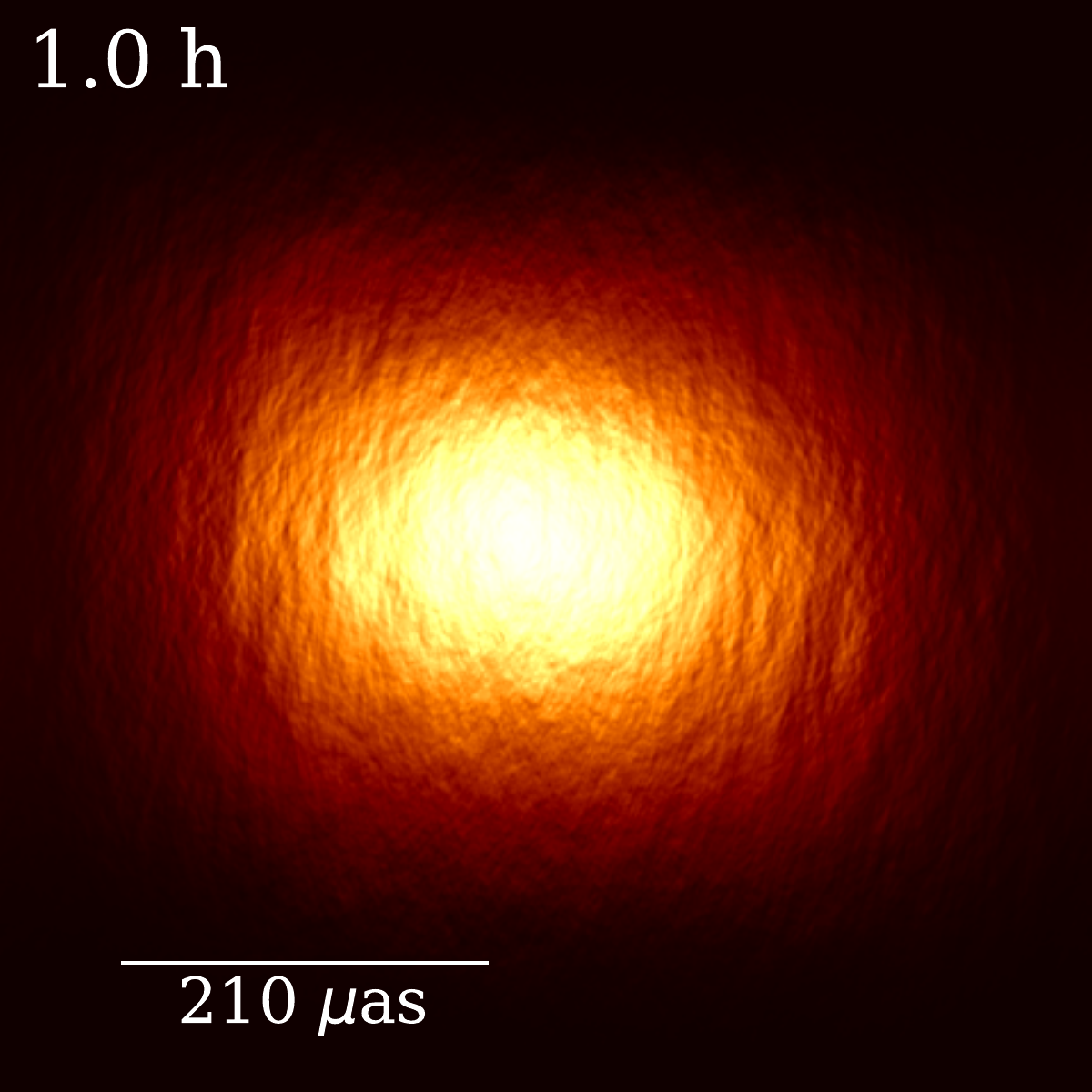} \\
\includegraphics[width=30mm]{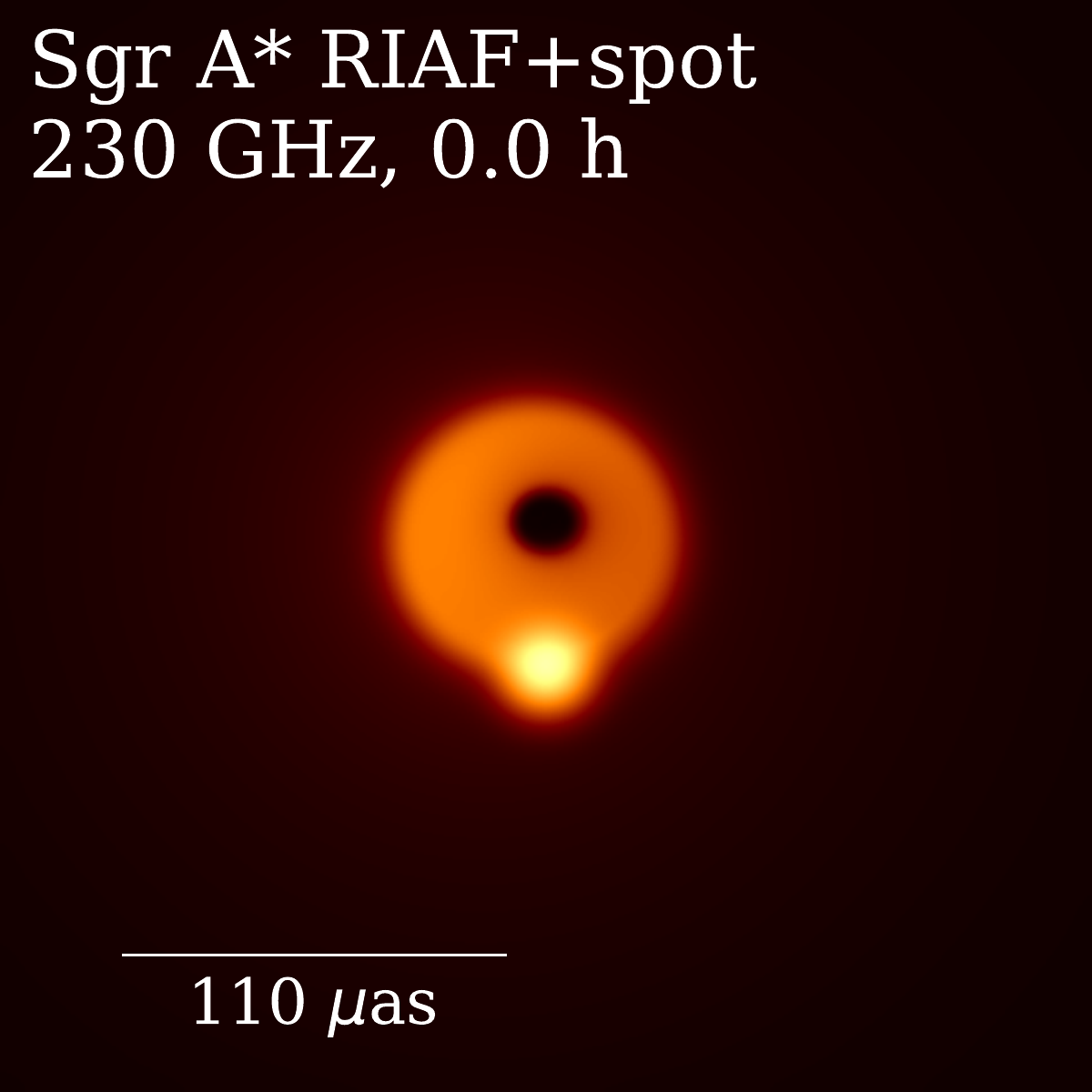}%
\includegraphics[width=30mm]{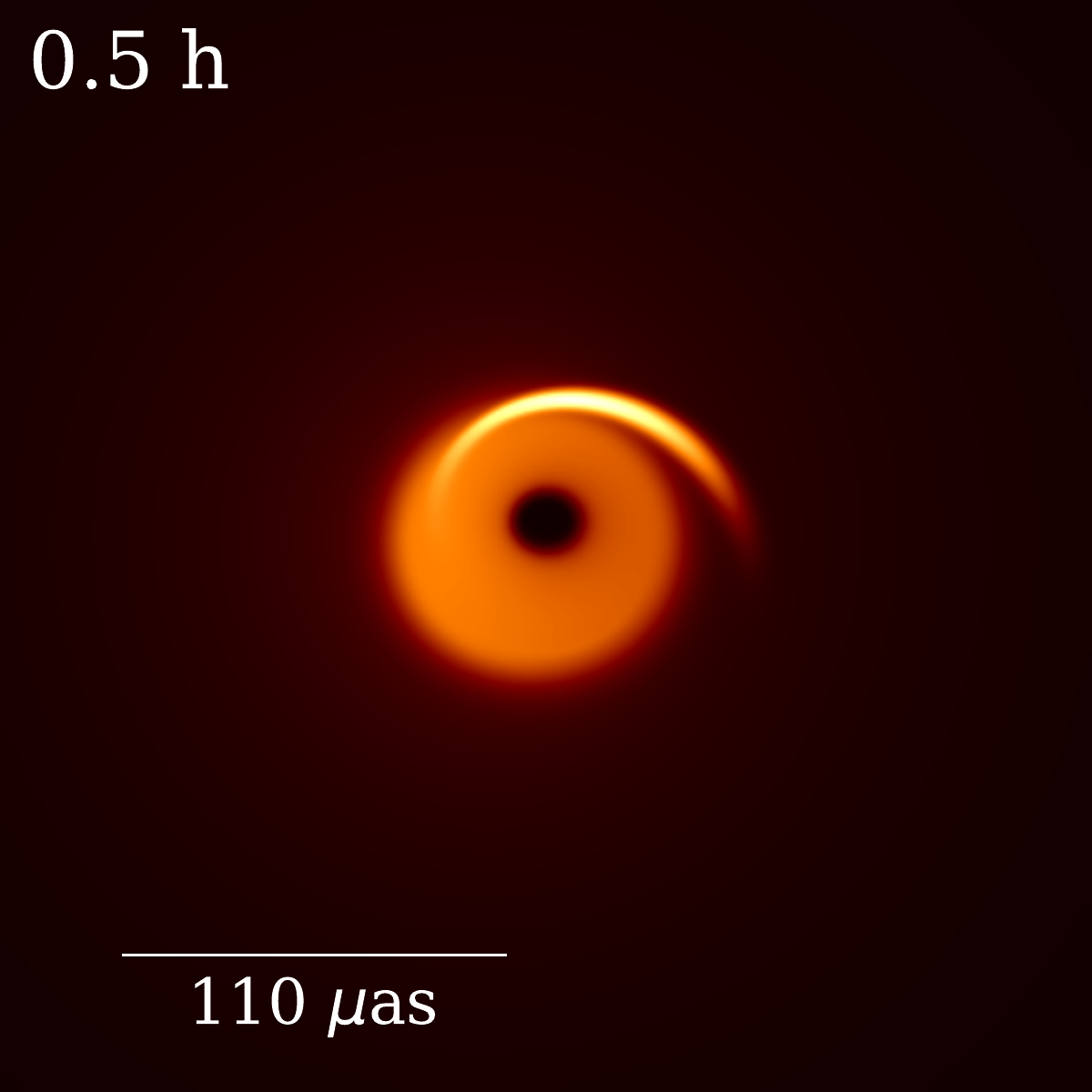}%
\includegraphics[width=30mm]{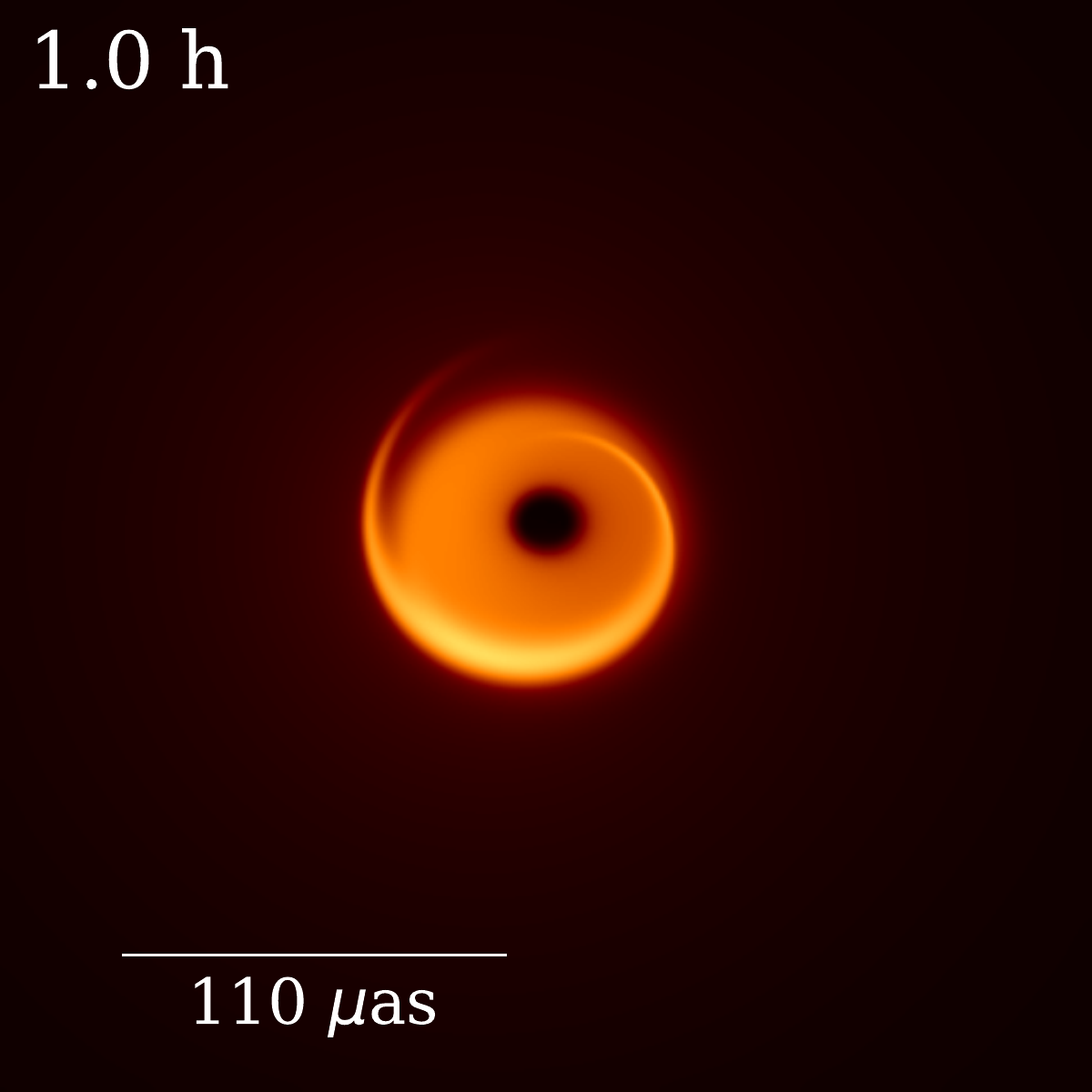}%
\includegraphics[width=30mm]{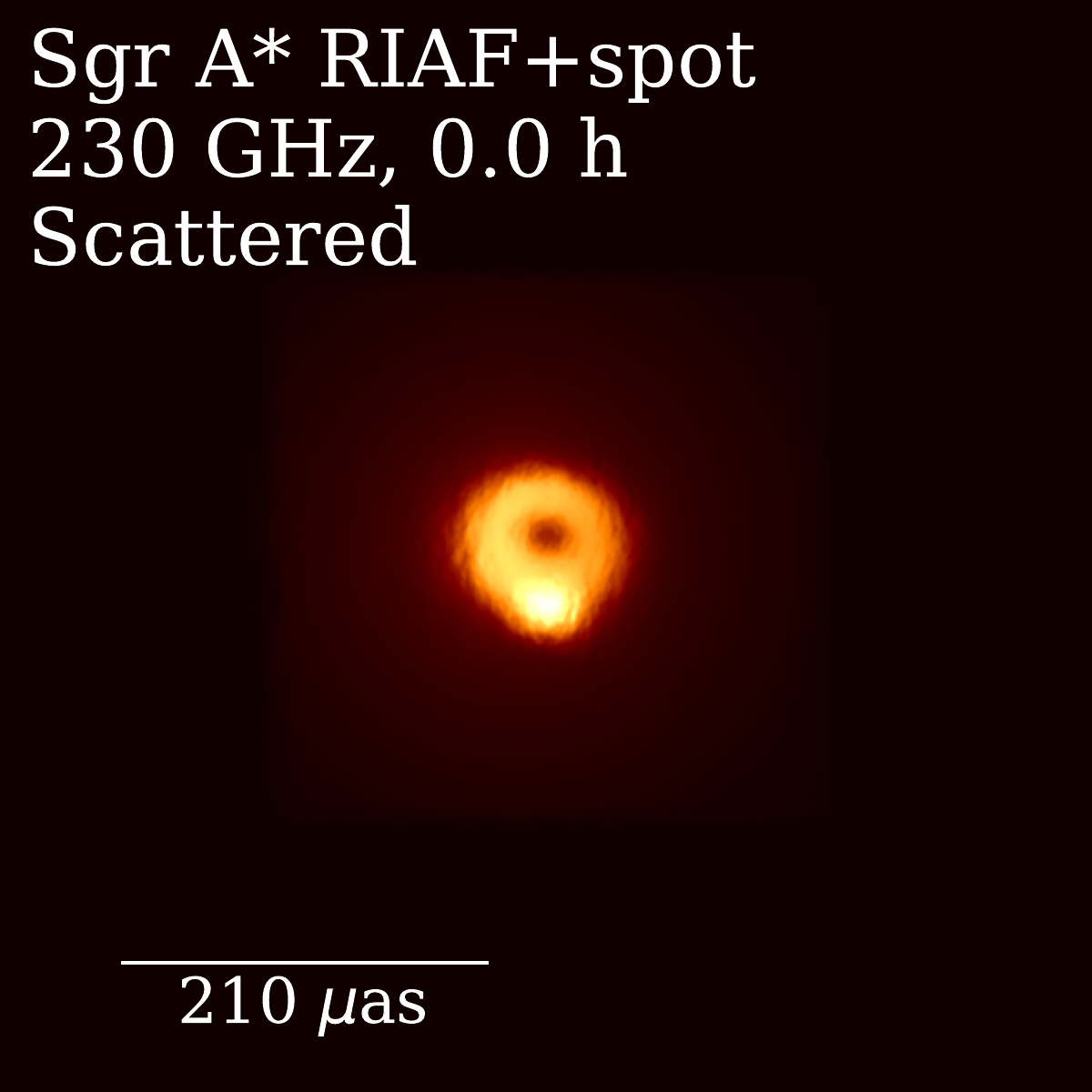}% 
\includegraphics[width=30mm]{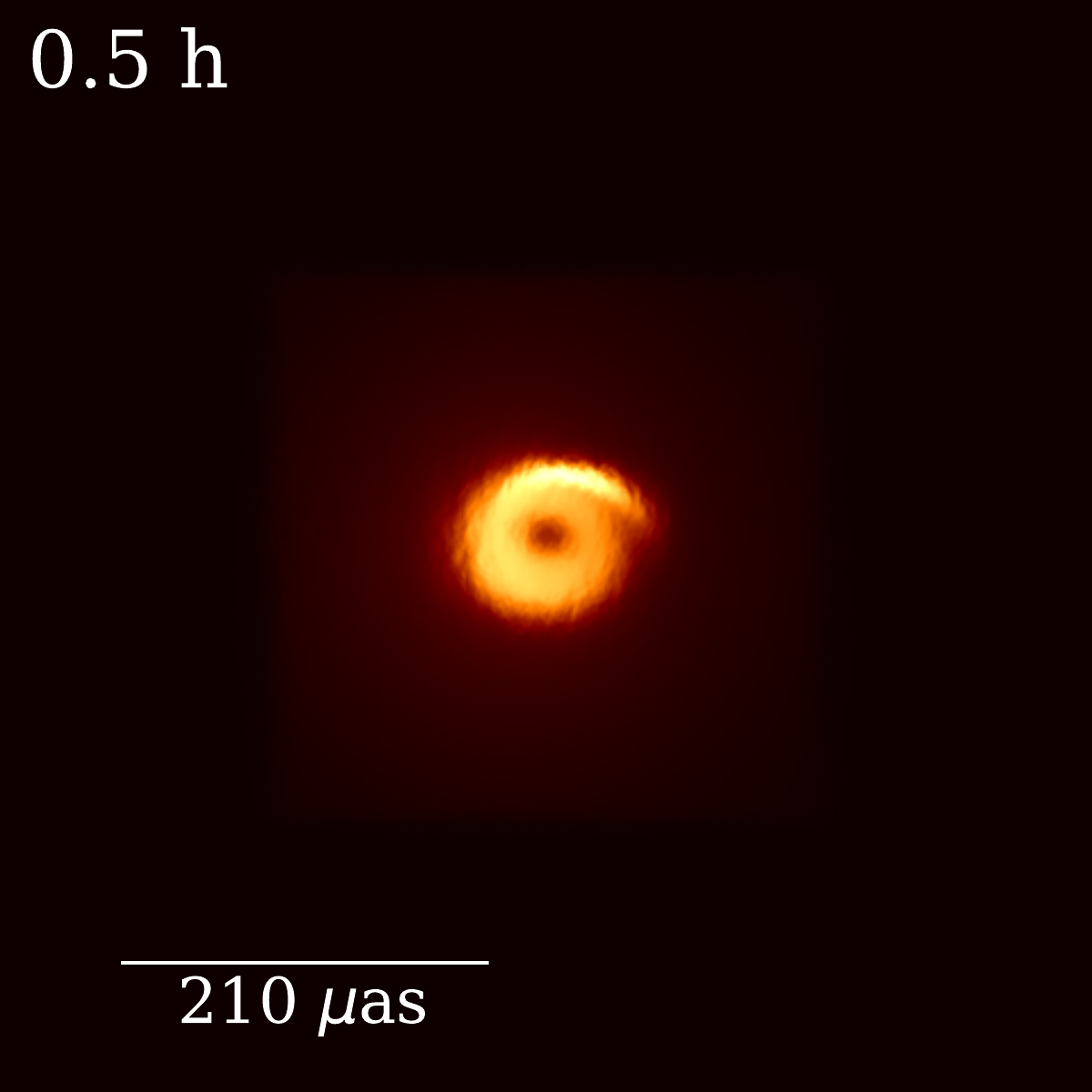}%
\includegraphics[width=30mm]{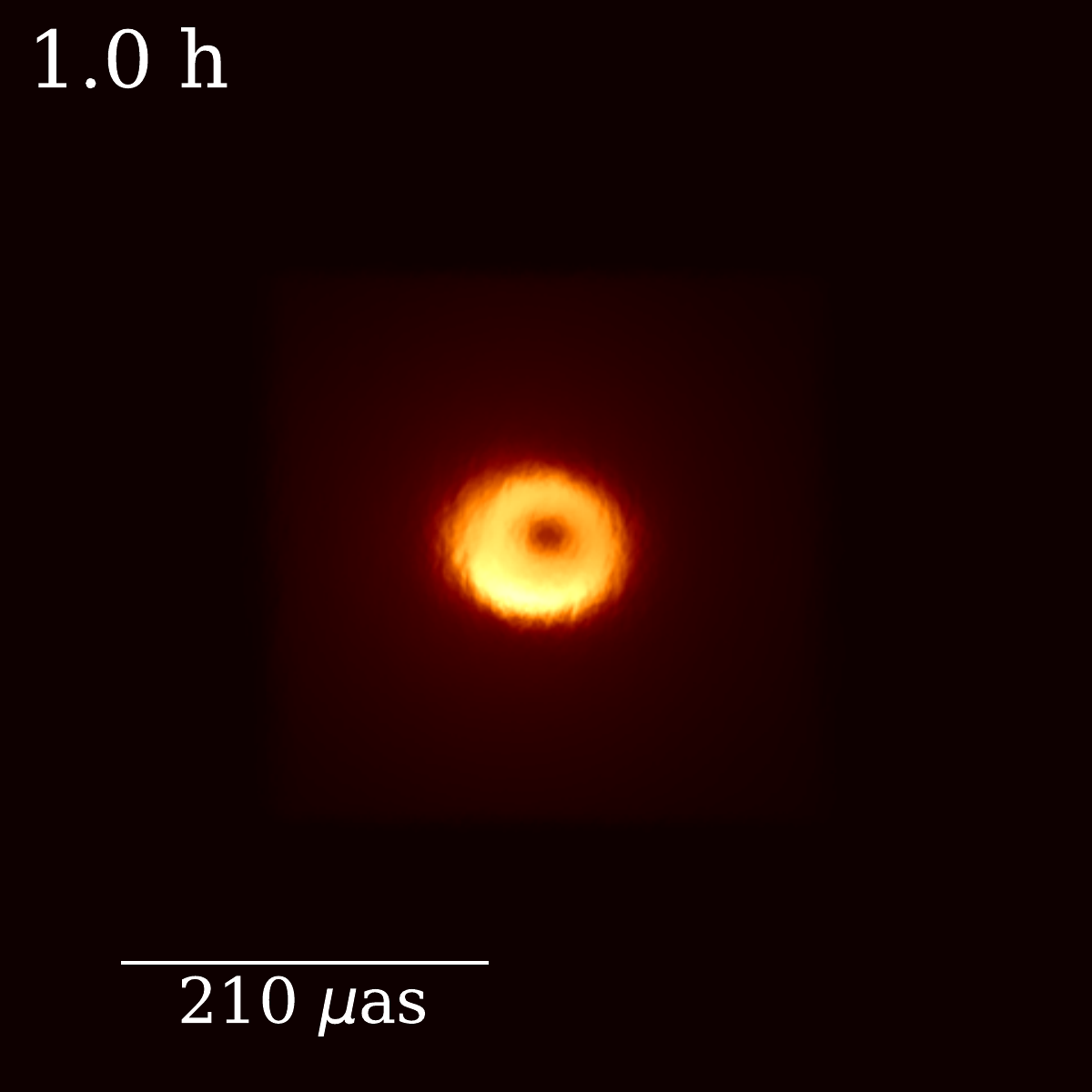} \\
\includegraphics[width=30mm]{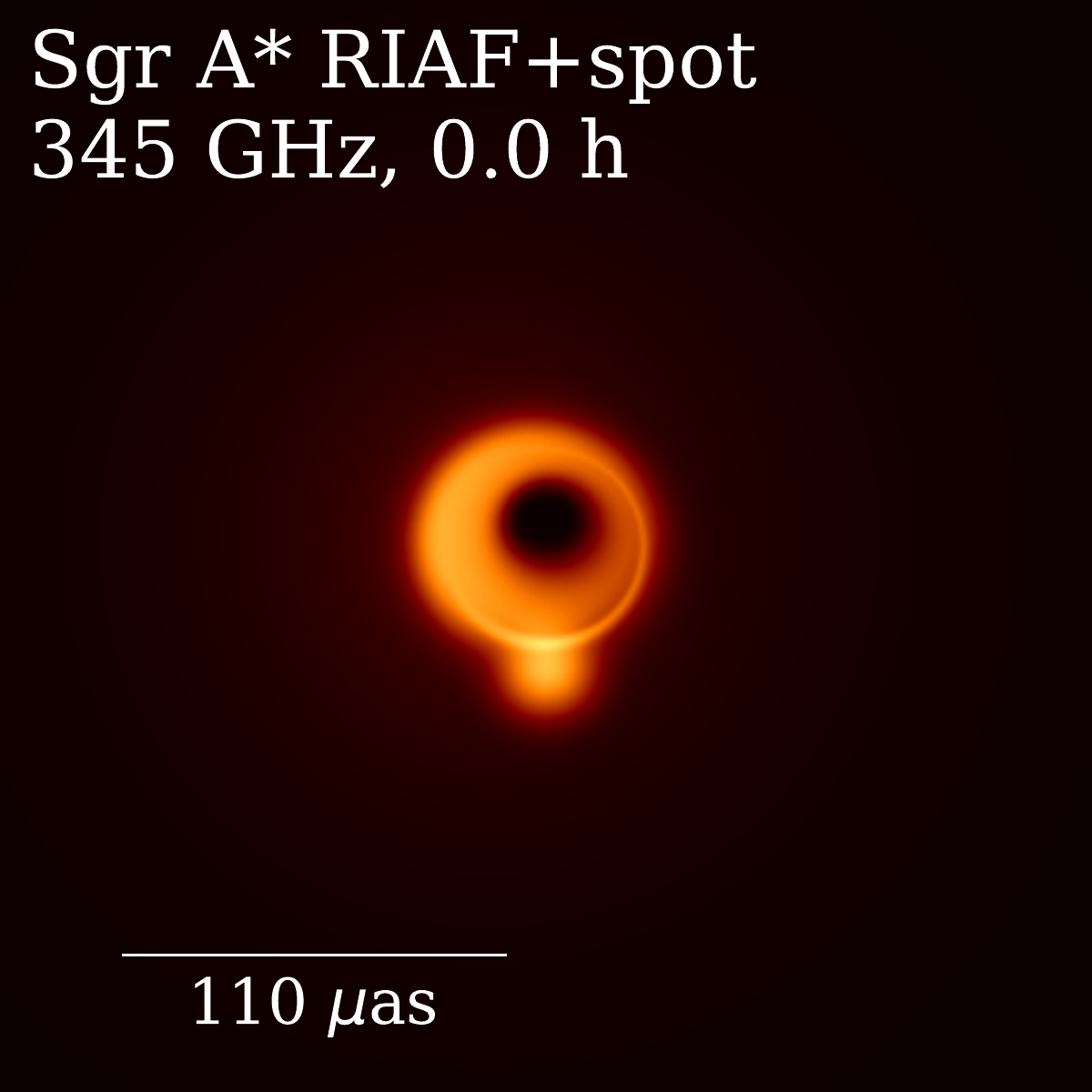}%
\includegraphics[width=30mm]{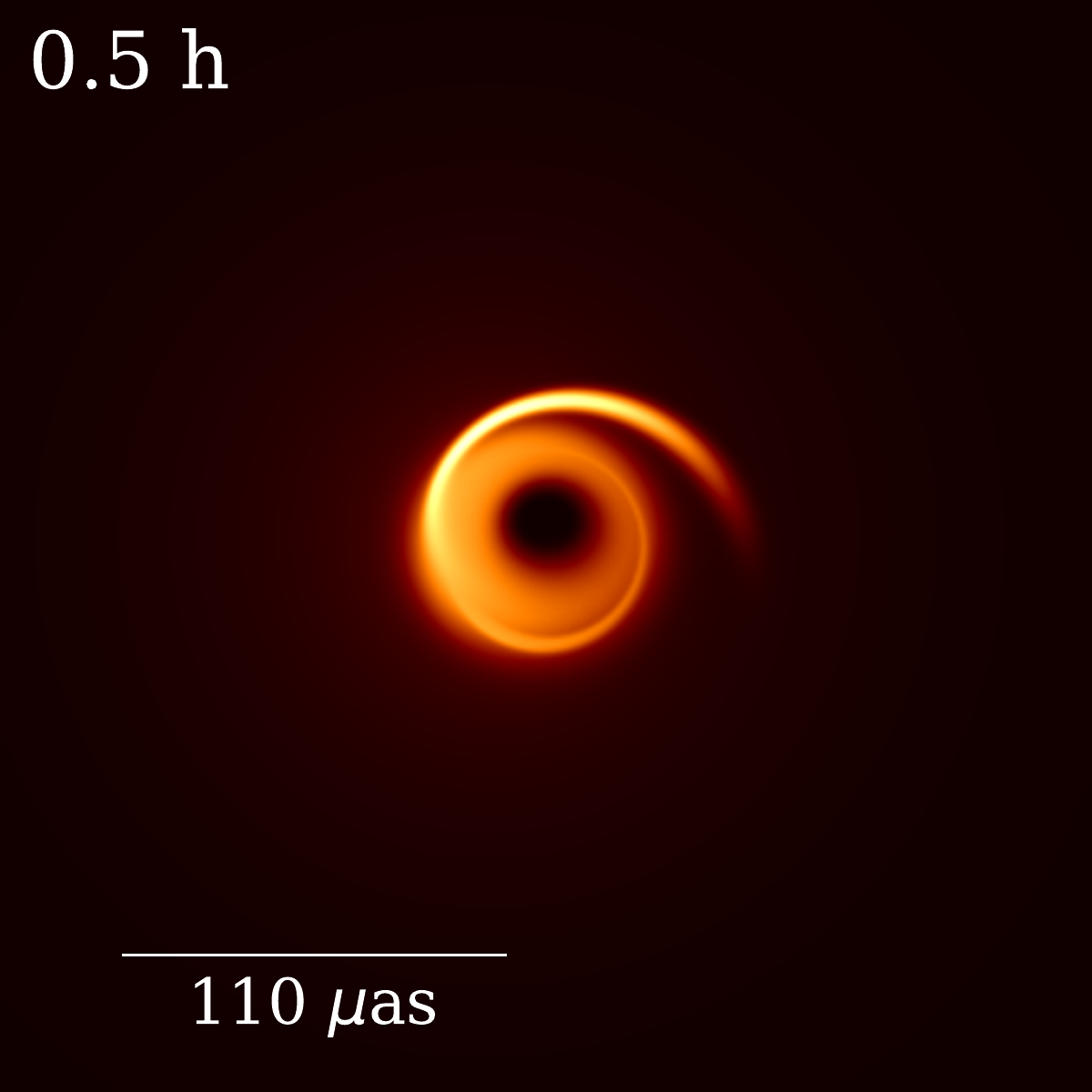}%
\includegraphics[width=30mm]{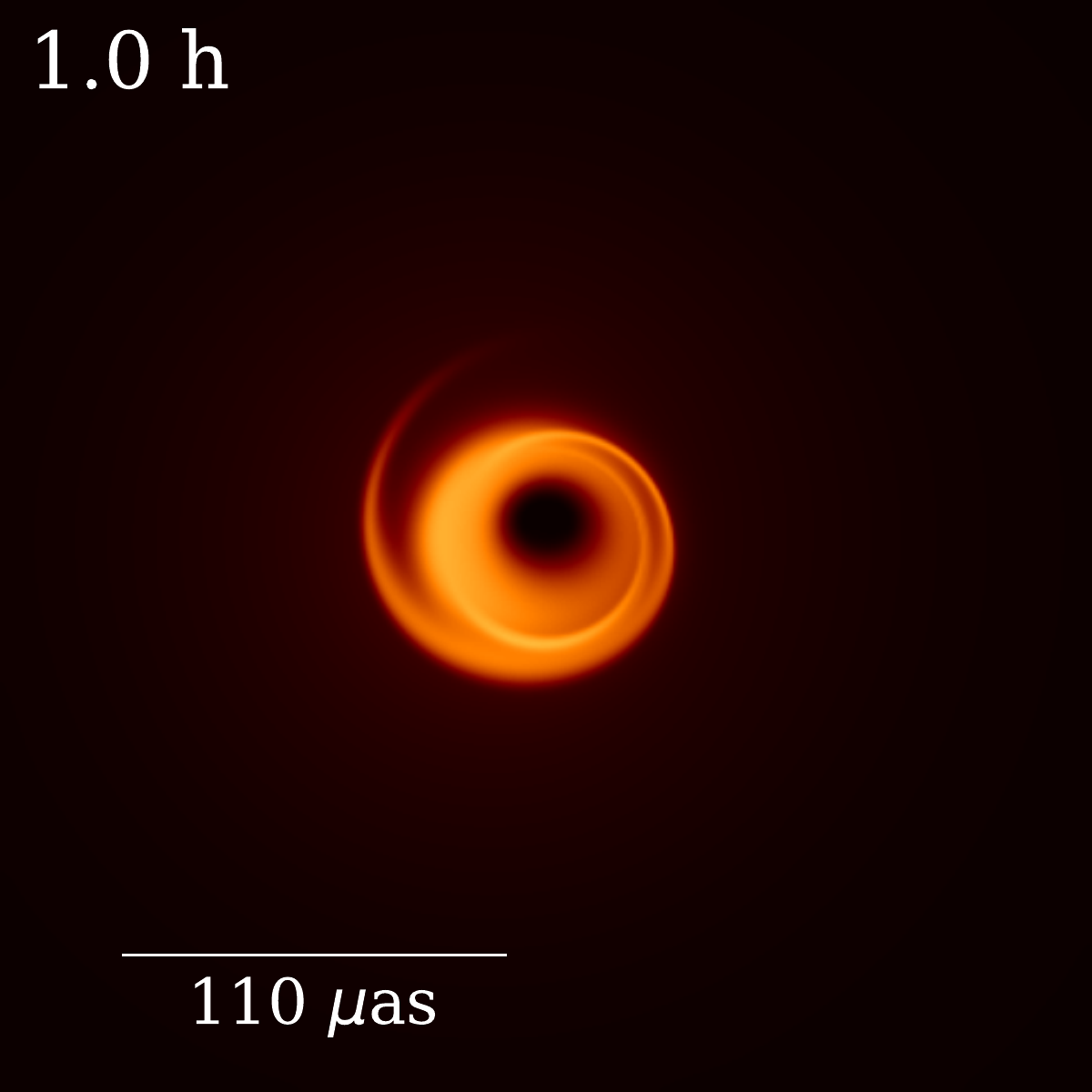}%
\includegraphics[width=30mm]{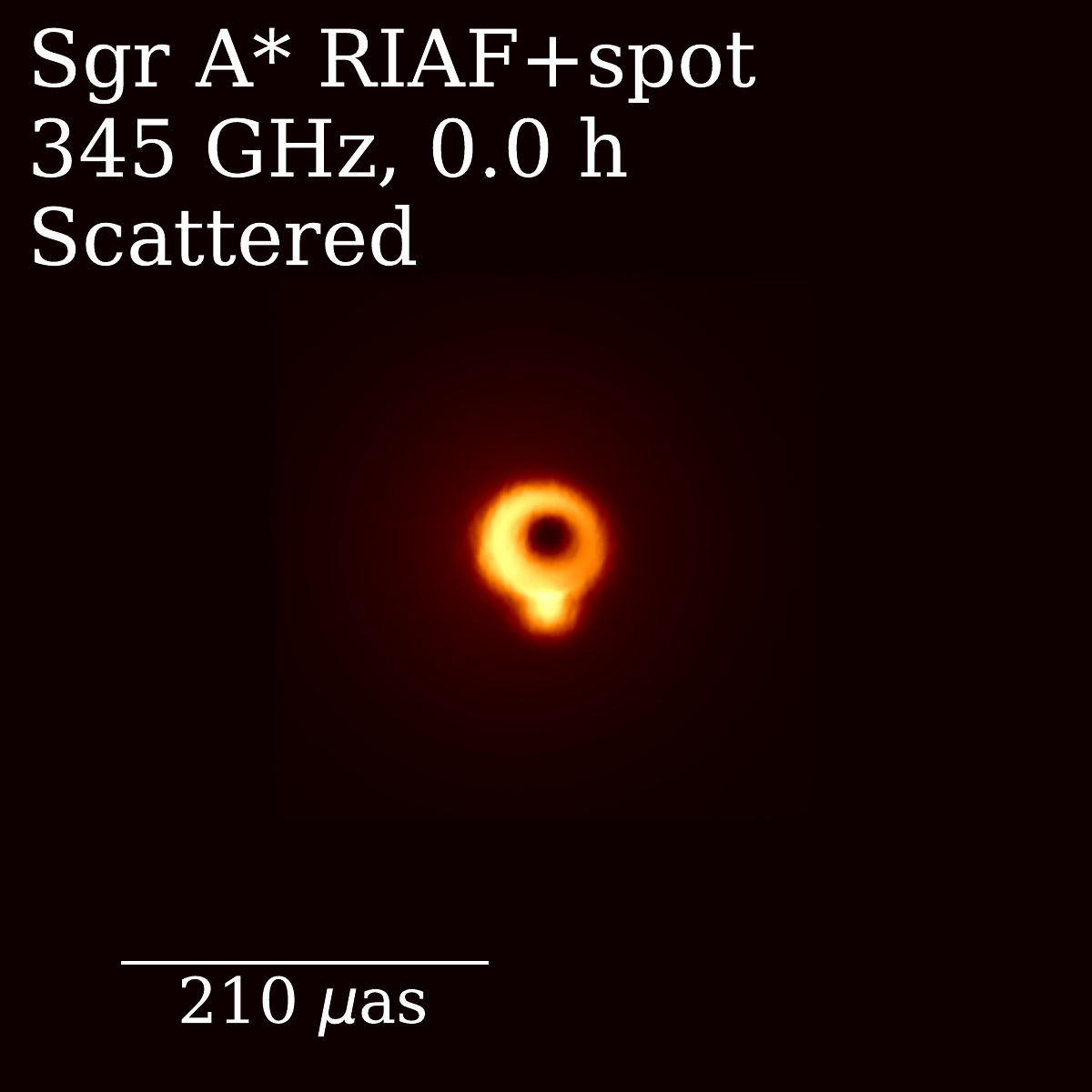}% 
\includegraphics[width=30mm]{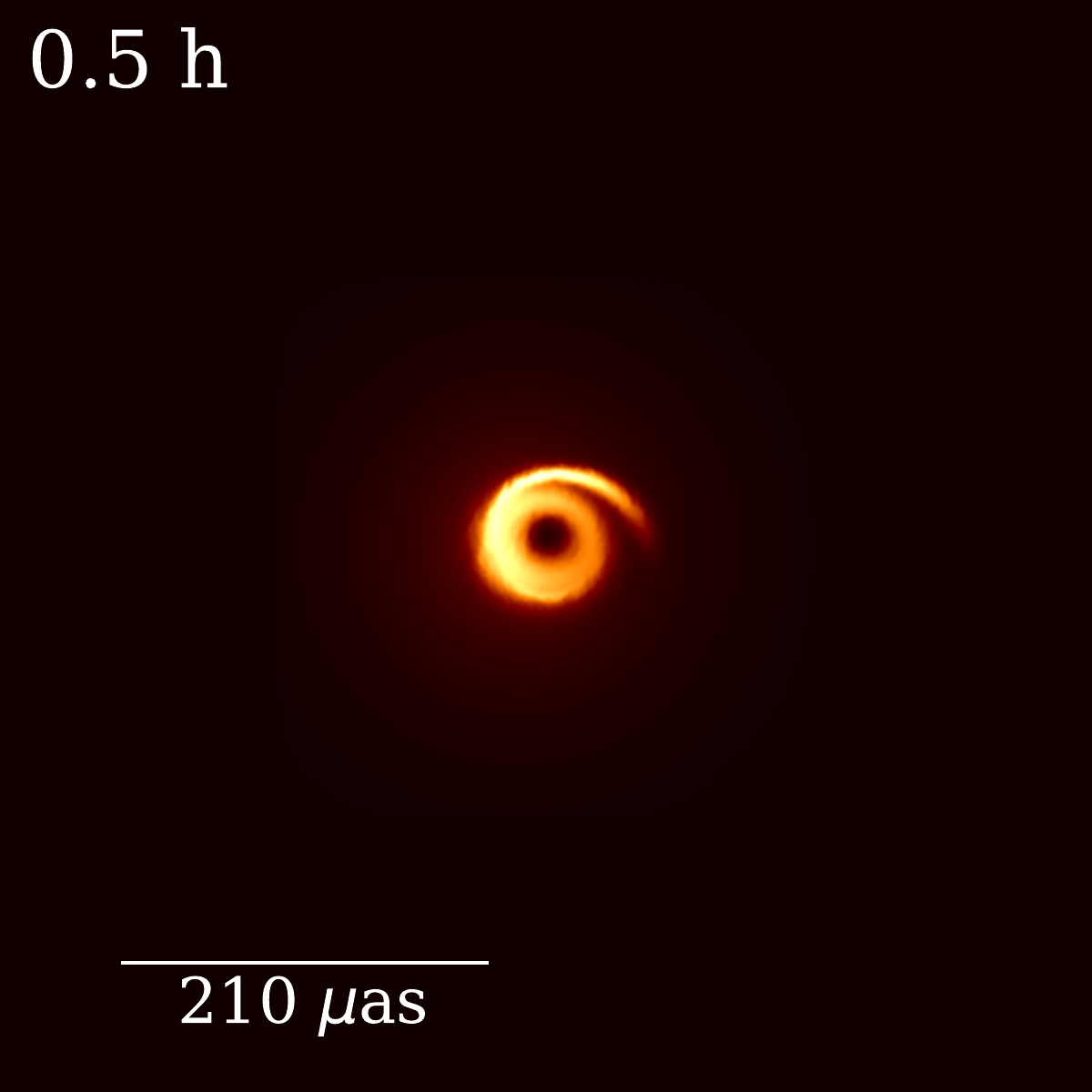}%
\includegraphics[width=30mm]{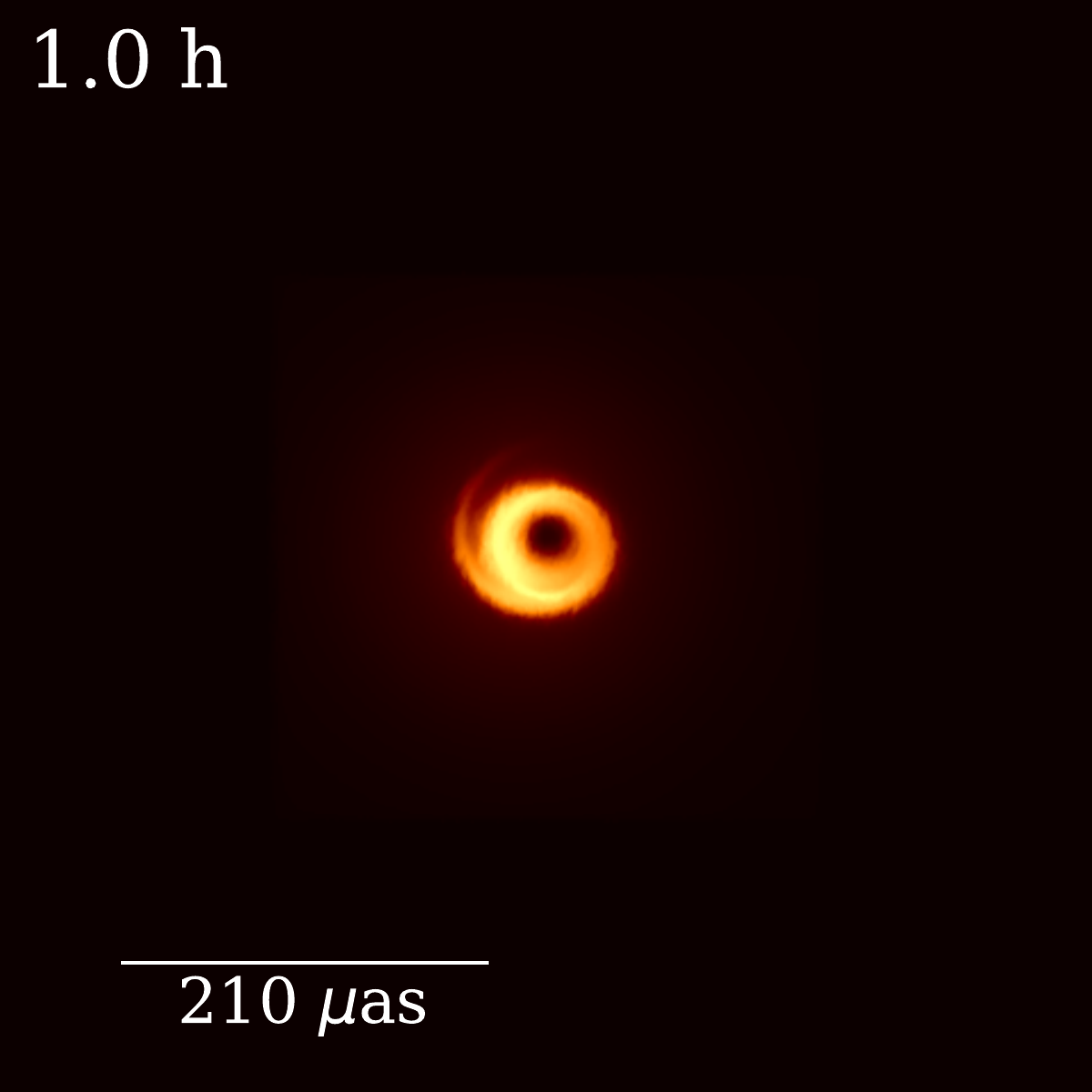} \\

\end{adjustwidth}
  \caption{Ground truth Sgr A* source models used for Challenge 2. For each model and frequency (rows), three movie frames (columns) are shown, with interstellar scattering applied to the first frame in the rightmost column. Images are shown on a square root scale, which is normalized to the brightest pixel value across each set of three movie frames. The scattered movies were used as inputs for the Challenge 2 synthetic data generation.}
     \label{fig:ch2_models_sgra}
\end{figure*}

For Challenge 2, two dynamic source models were used for Sgr\,A*: a GRMHD model exhibiting turbulent variability, and a RIAF + shearing hotspot model with more ordered variability properties. Sample frames of the source models used for Challenge 2 are shown in Figure \ref{fig:ch2_models_sgra}.

The GRMHD model is a MAD model with spin 0.5. The images were ray-traced in Stokes I with BHOSS \citep{Younsi2021} assuming thermal electrons. The 500 frames are spaced 10$GM/c^3$ (221 s) apart. The pixel resolution is 2048x2048, with a field of view of 400 $\mu$as.

The second Sgr A* model is a RIAF \citep{Broderick2016} plus shearing hotspot \citep{Tiede2020} semi-analytical model. The hotspot parameters are inspired by \citep{Gravity2018} and the black hole spin was set to 0.1. The pixel resolution is 313x313, with a field of view of 315 $\mu$as. The frames are spaced 30 seconds apart and form a 4-hour movie of a hotspot shearing and falling in, which is repeated a few times over the course of the 24-hour observation. 

\subsection{Synthetic data}
\label{sec:challenge2_synthdata}
The synthetic data for Challenge 2 includes significantly more systematic effects than Challenge 1 (see also Section \ref{sec:challenge2_rationale}). In the {\tt SYMBA} pipeline \citep{Roelofs2020}, atmospheric absorption, emission, delays and turbulence are simulated and antenna pointing offsets are added to the simulated datasets with \texttt{MeqSilhouette} \citep{Blecher2017, Natarajan2022}. These are then calibrated by performing a fringe fit and a priori amplitude calibration with \texttt{rPICARD} \citep{Janssen2019}, and network calibration with \texttt{eht-imaging}. 

The station locations, dish sizes, aperture efficiencies, and receiver temperatures are identical to those used in Challenge 1 (Figure \ref{fig:ch1_arrays}, Section \ref{sec:challenge1_synthdata}).
For each site, the input precipitable water vapor (PWV), ground temperature, and ground pressure were calculated from the Modern-Era Retrospective Analysis for Research and Applications, version 2 (MERRA-2) from the NASA Goddard Earth Sciences Data and Information Services Center \citep[GES DISC,][]{Gelaro2017}, processed with the \texttt{am} atmospheric model software \citep{Paine2019} \citep[see][for details]{Roelofs2020}. All weather quantities were based on median conditions on 1 April (2000-2020) as registered in the MERRA-2 climatological data. The atmospheric coherence time at 230 GHz was assumed to be 20 s for a PWV of 1 mm, 3 s for a PWV of 15 mm, interpolated linearly between these values for the different sites, and scaled linearly with frequency. Pointing offsets were assumed to be stable across each 10-minute scan, and drawn randomly from a Gaussian distribution with an RMS of 2 arcseconds.

\subsection{Results}
Seven submitters or teams provided dynamical reconstructions of the Challenge 2 datasets. Reconstruction quality metrics for all submissions are shown in Table \ref{tab:metrics_challenge2_M87}.
\subsubsection{M87 GRMHD}
\begin{figure*}
\begin{adjustwidth}{-\extralength}{0cm}
\setlength{\lineskip}{0pt}
\centering
\includegraphics[width=30mm]{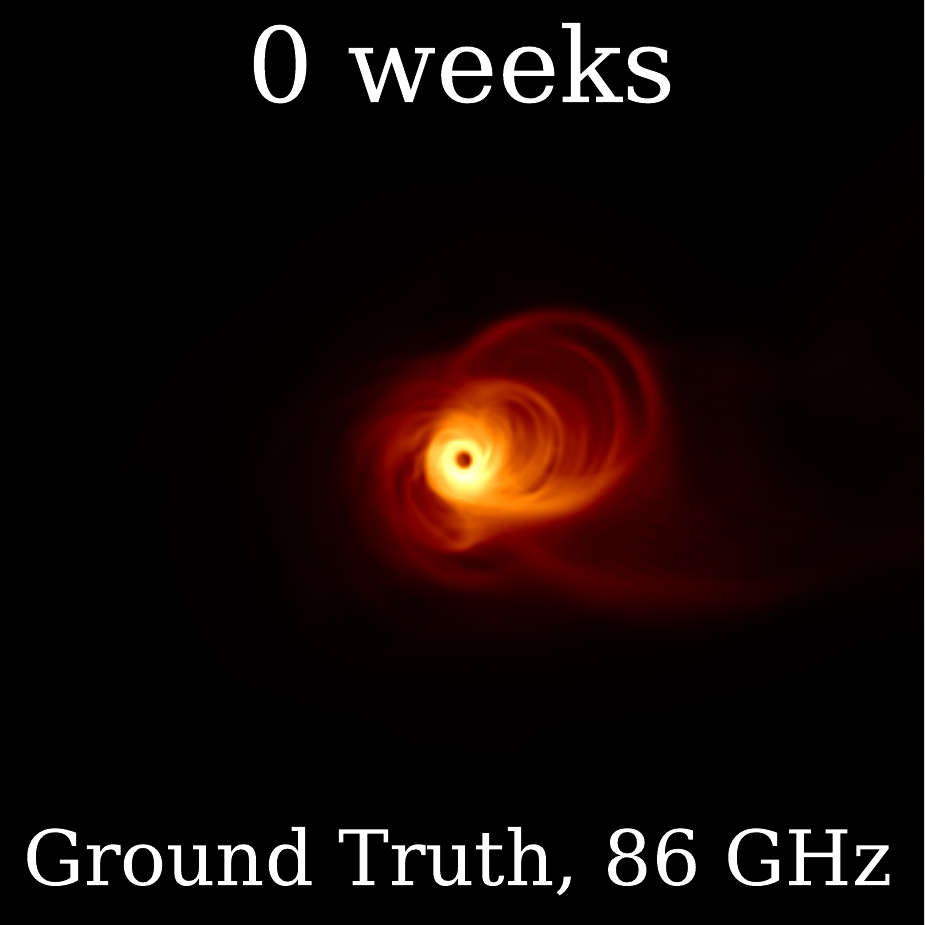}%
\includegraphics[width=30mm]{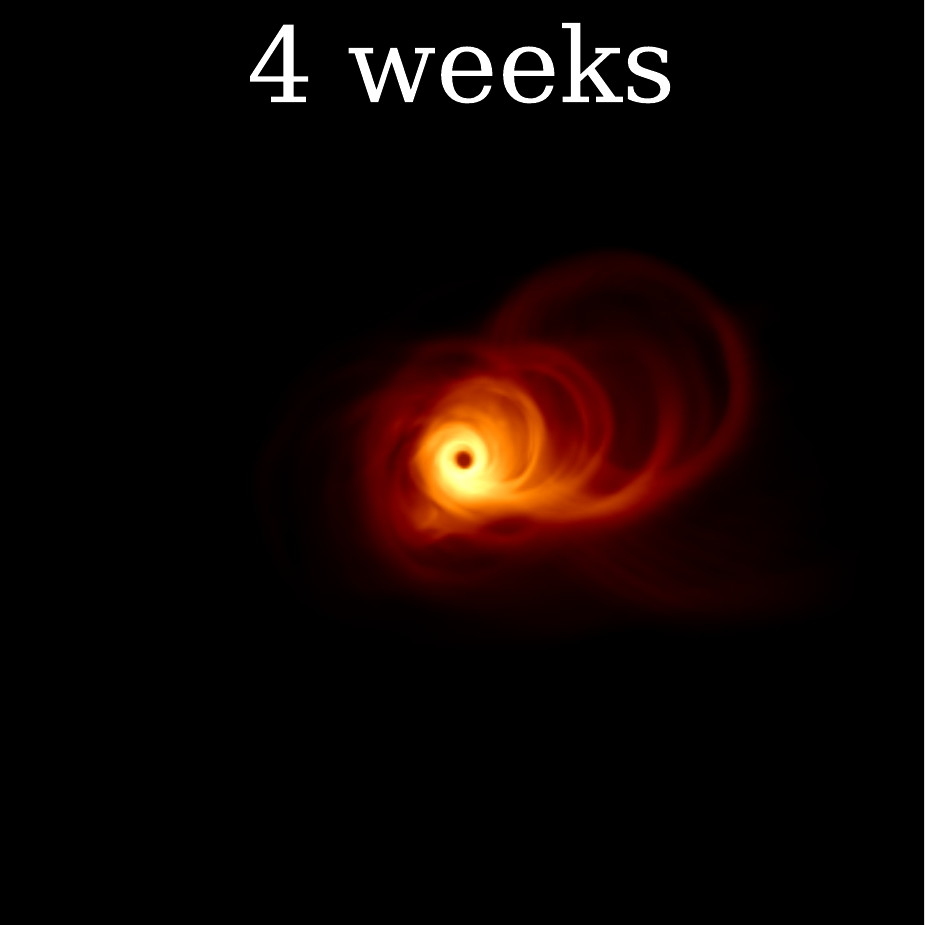}%
\includegraphics[width=30mm]{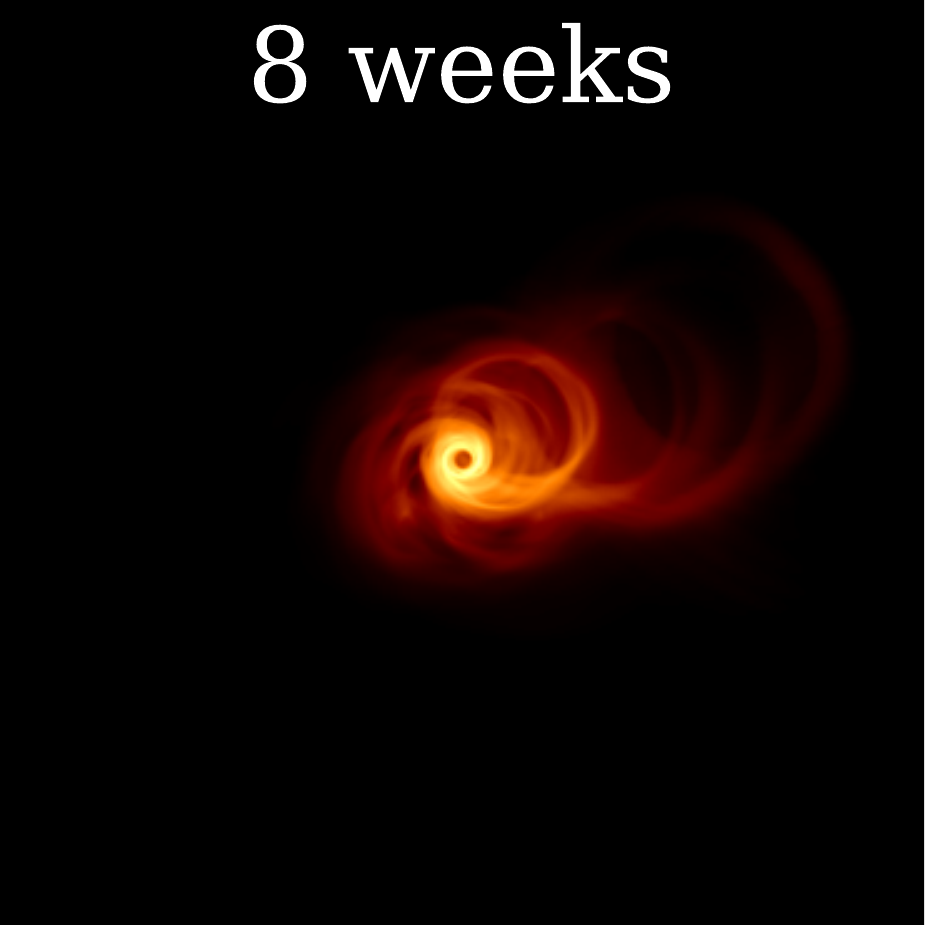}%
\includegraphics[width=30mm]{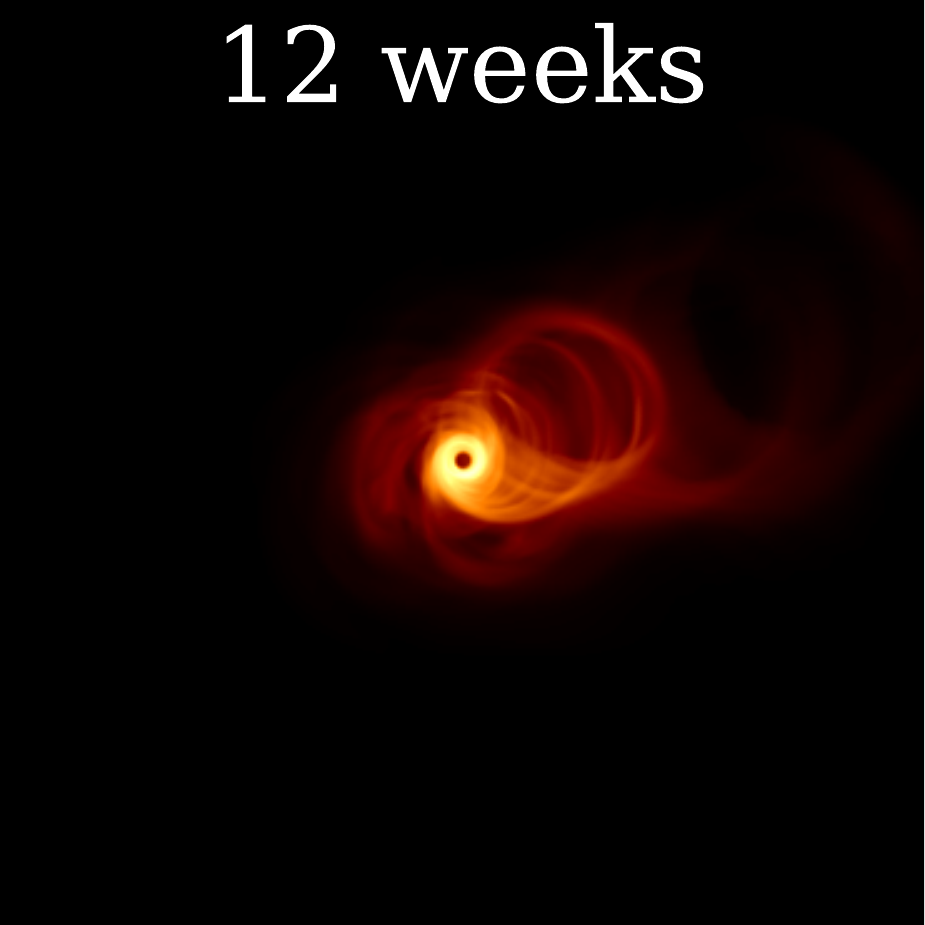}%
\includegraphics[width=30mm]{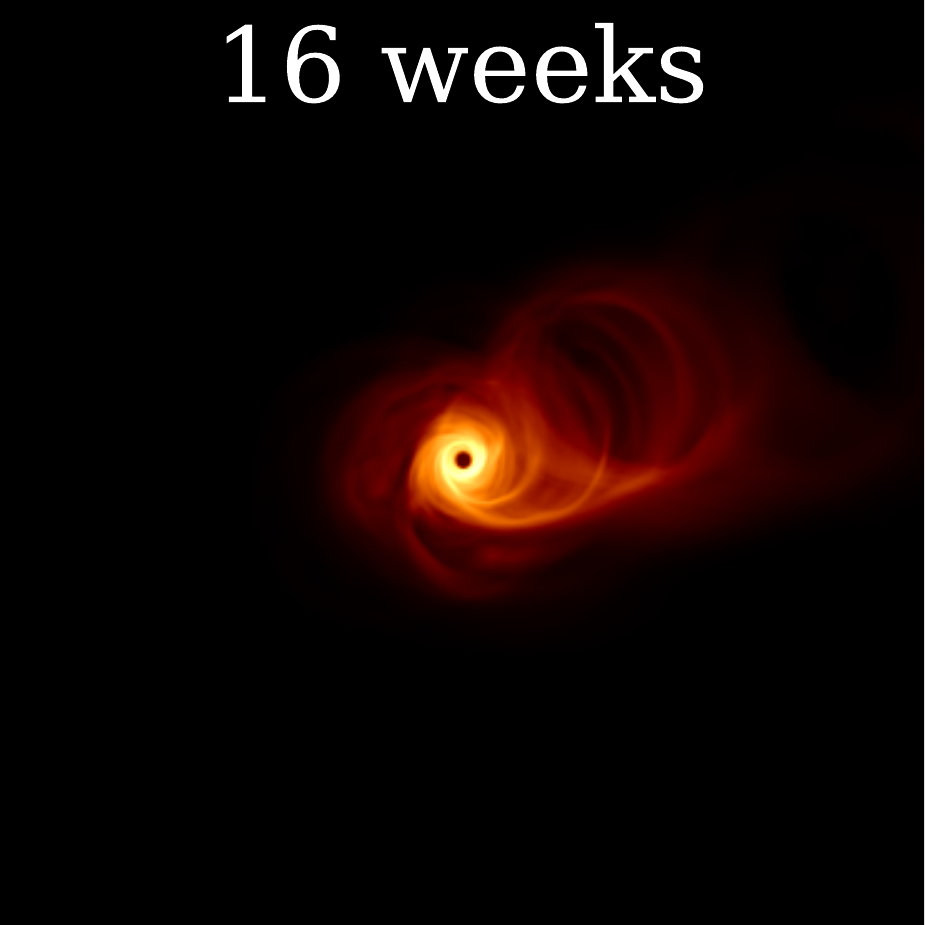}%
\includegraphics[width=30mm]{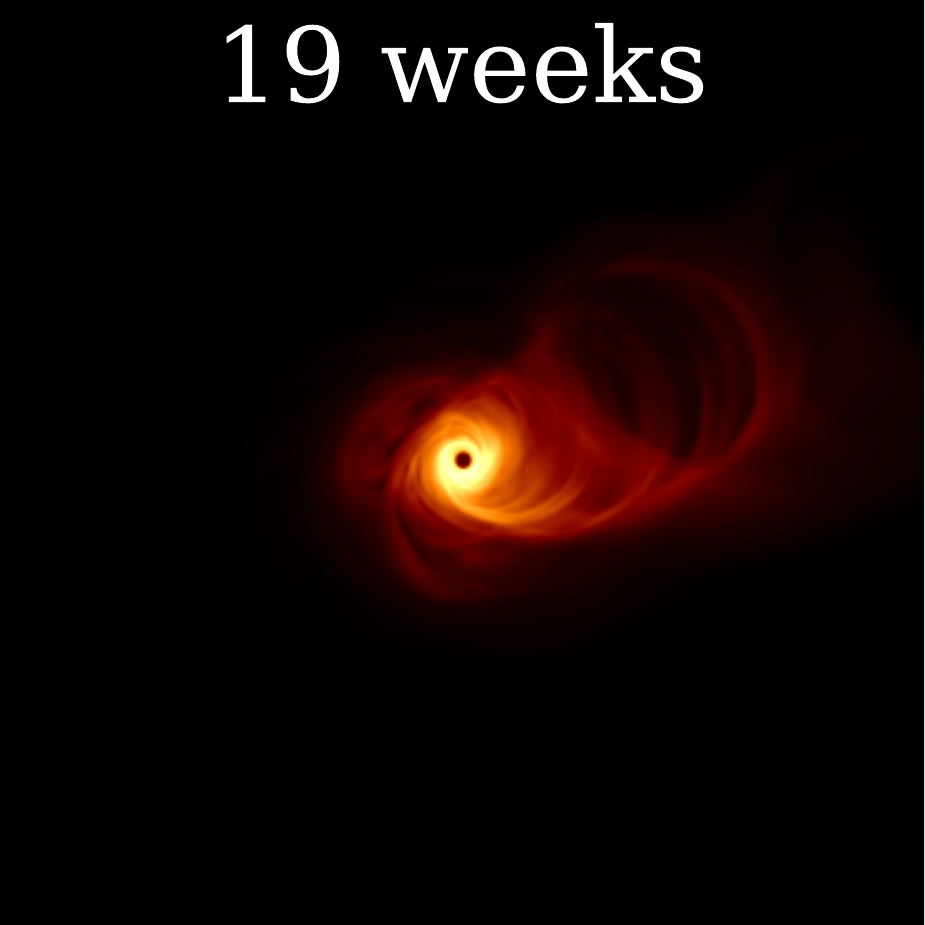} \\
\vspace{2mm}
\includegraphics[width=30mm]{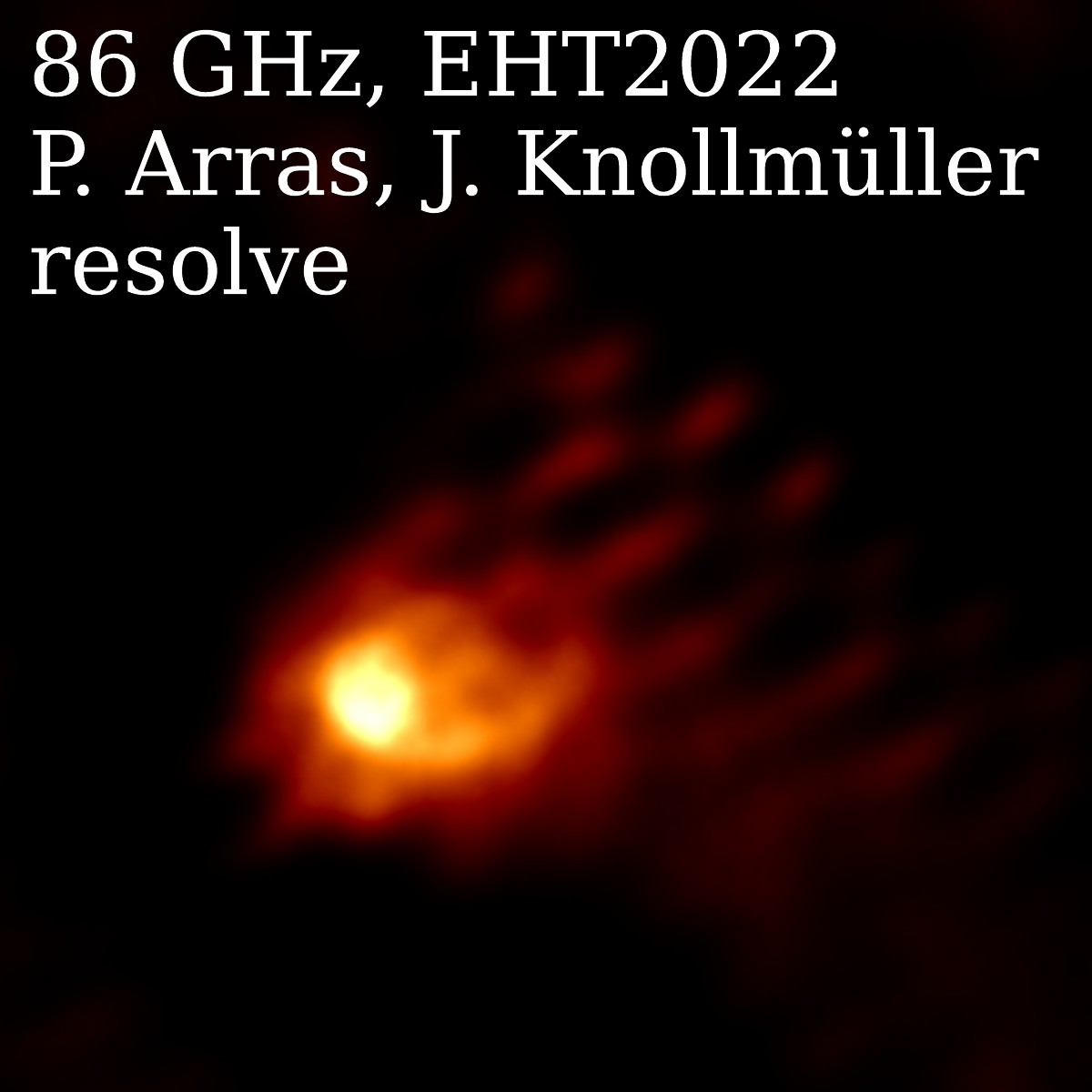}%
\includegraphics[width=30mm]{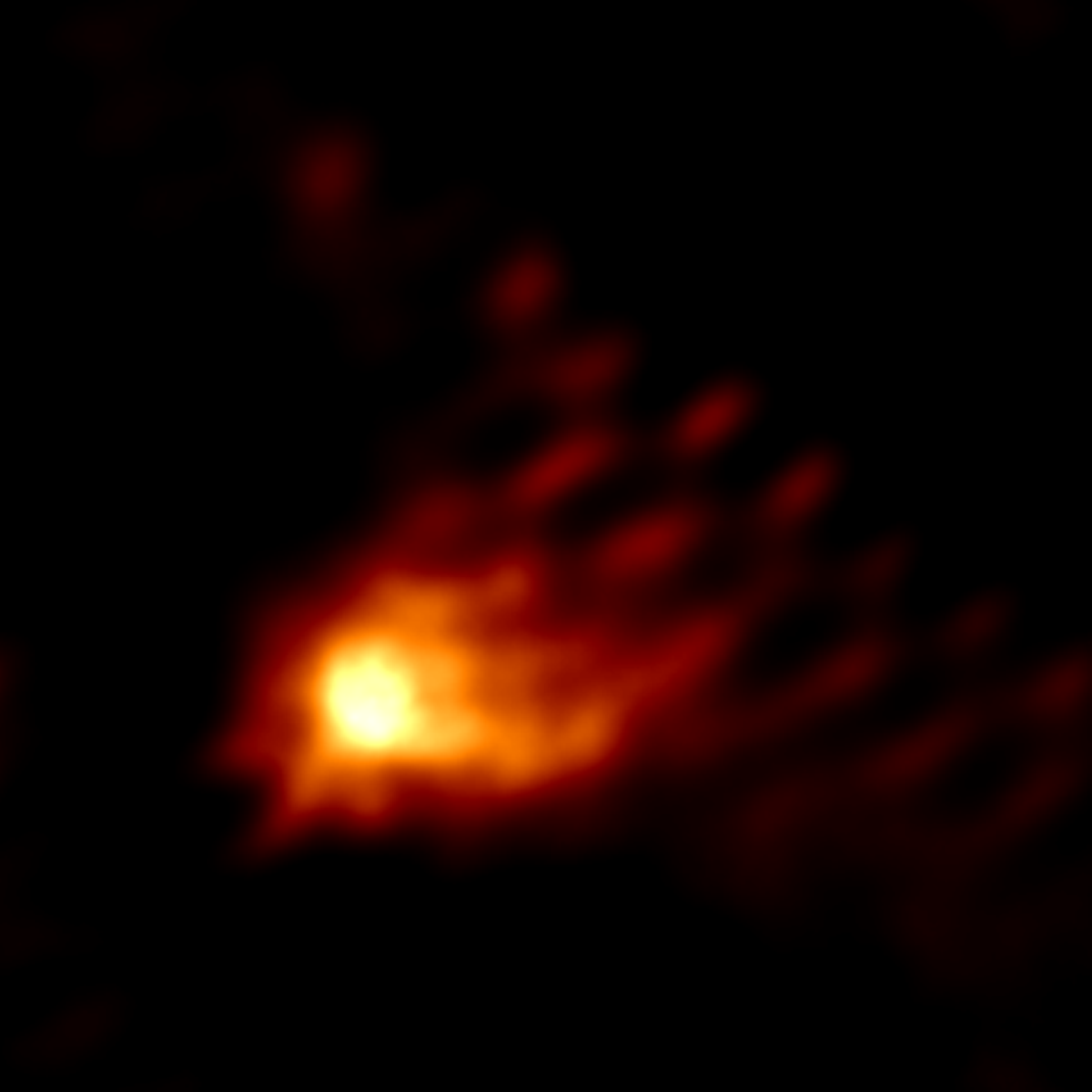}%
\includegraphics[width=30mm]{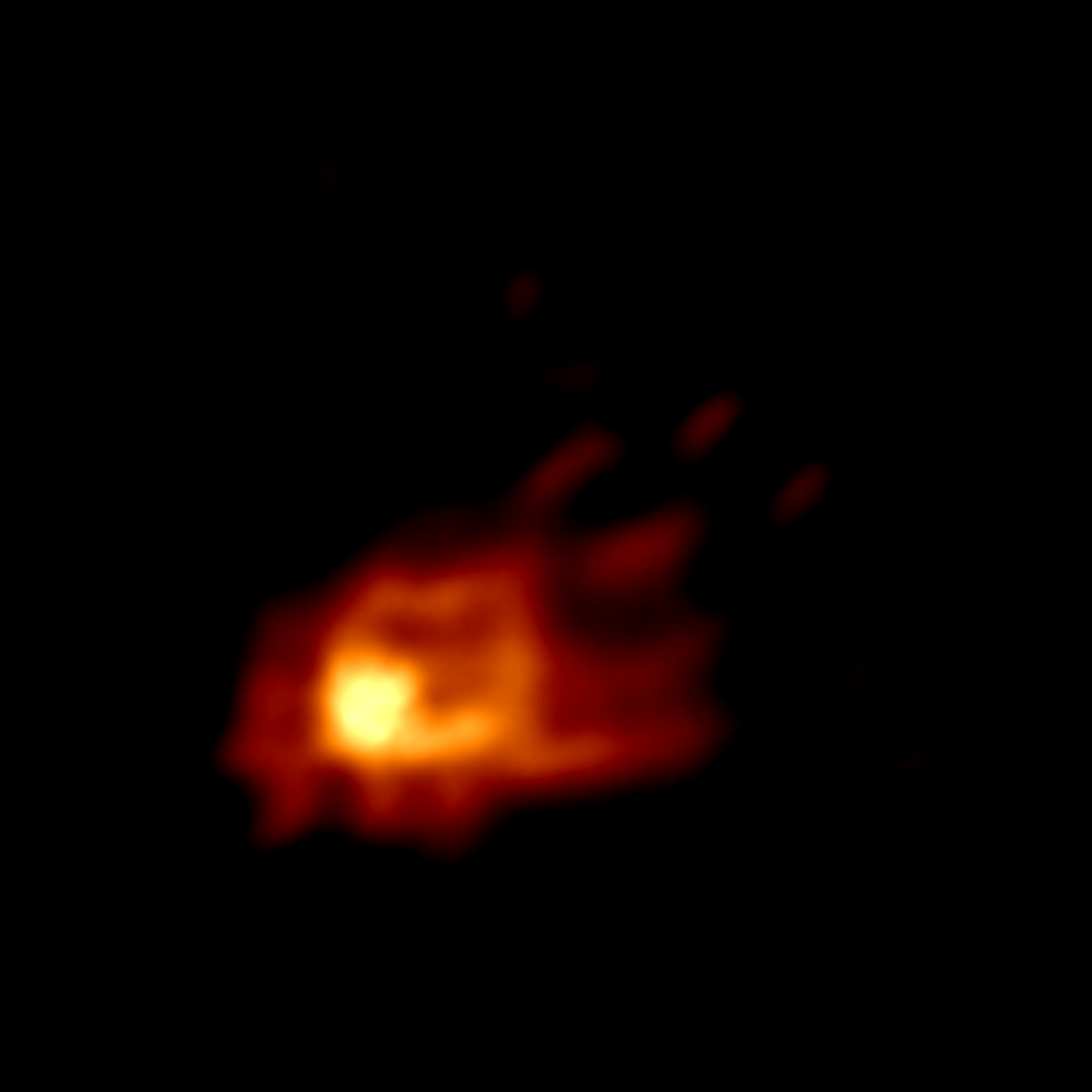}%
\includegraphics[width=30mm]{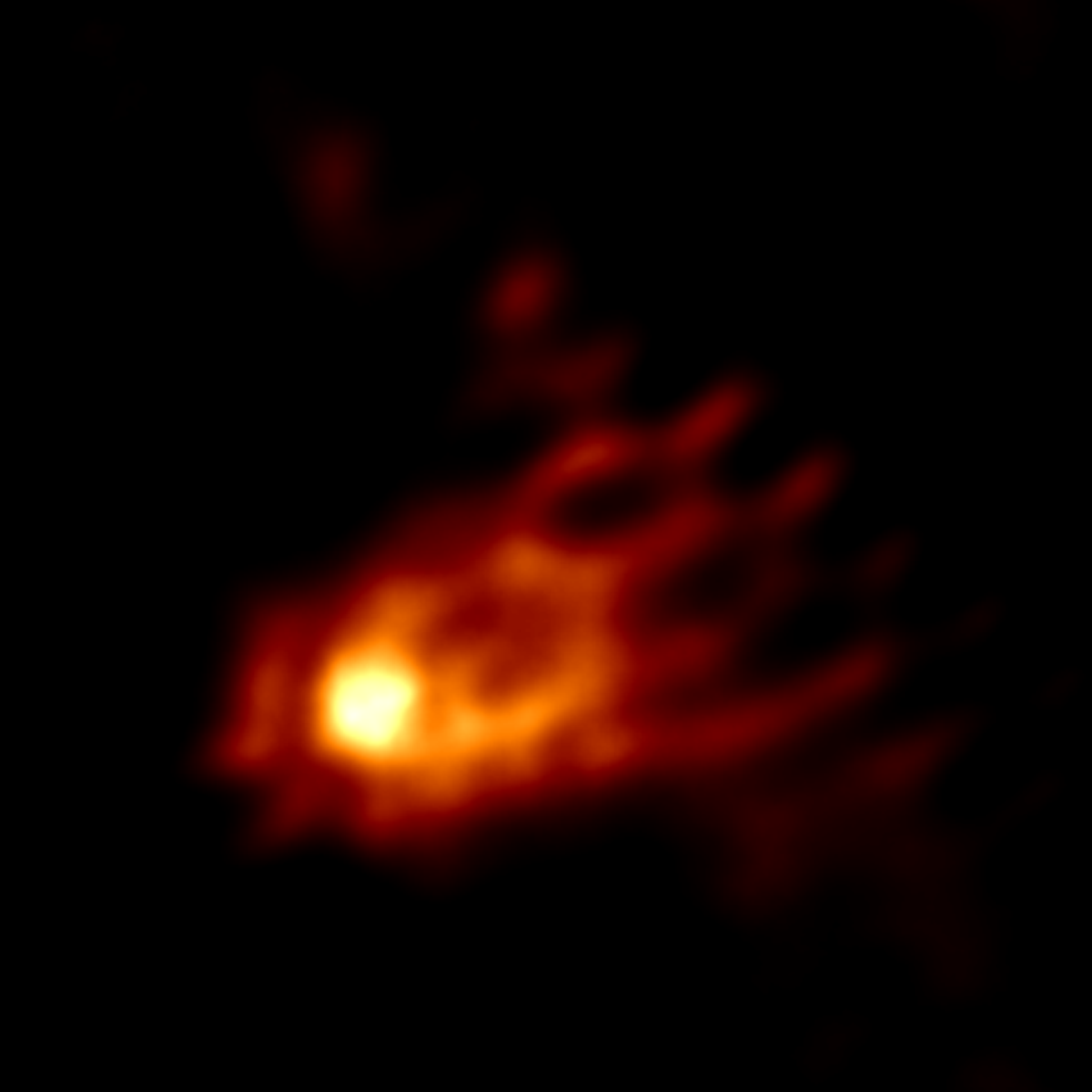}%
\includegraphics[width=30mm]{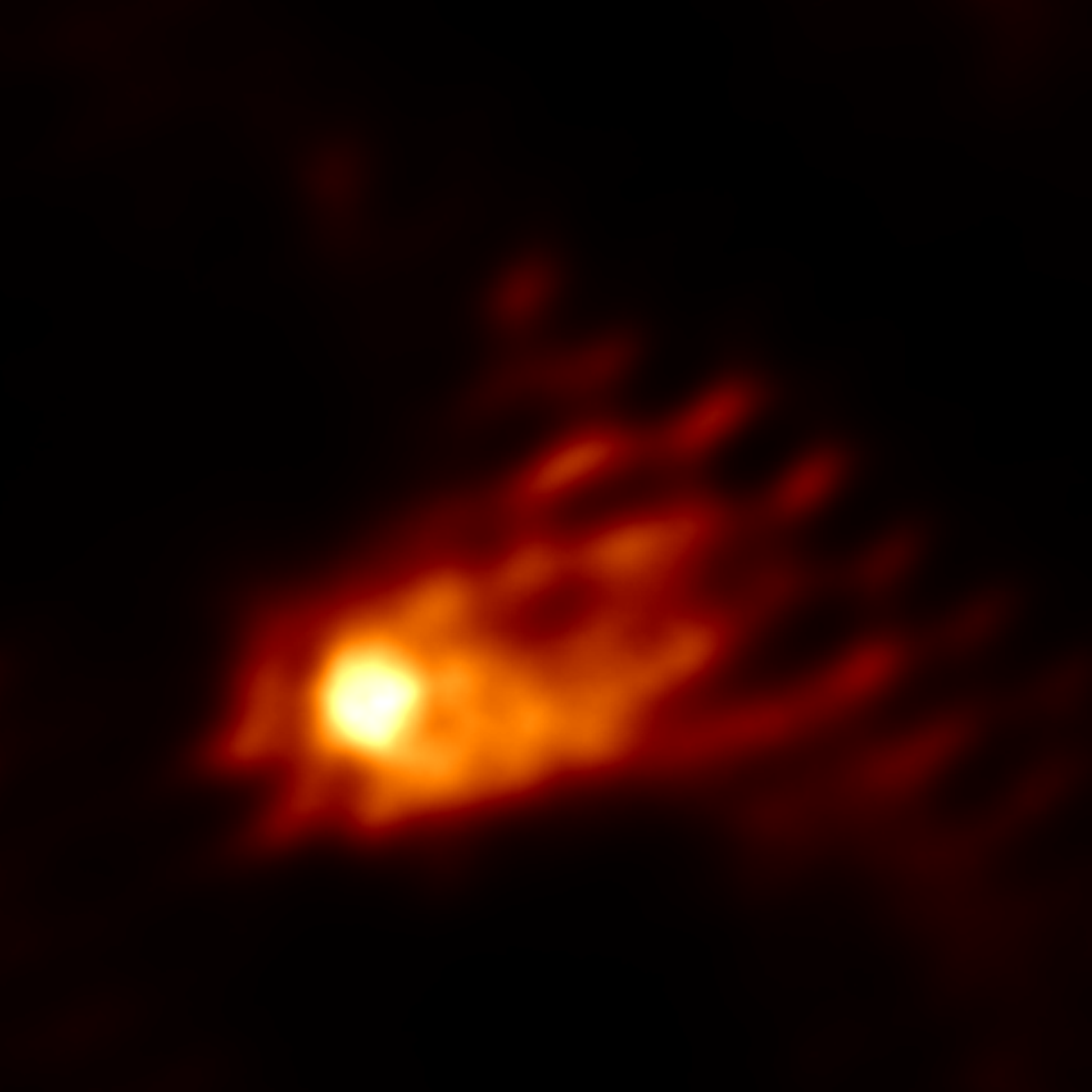}%
\includegraphics[width=30mm]{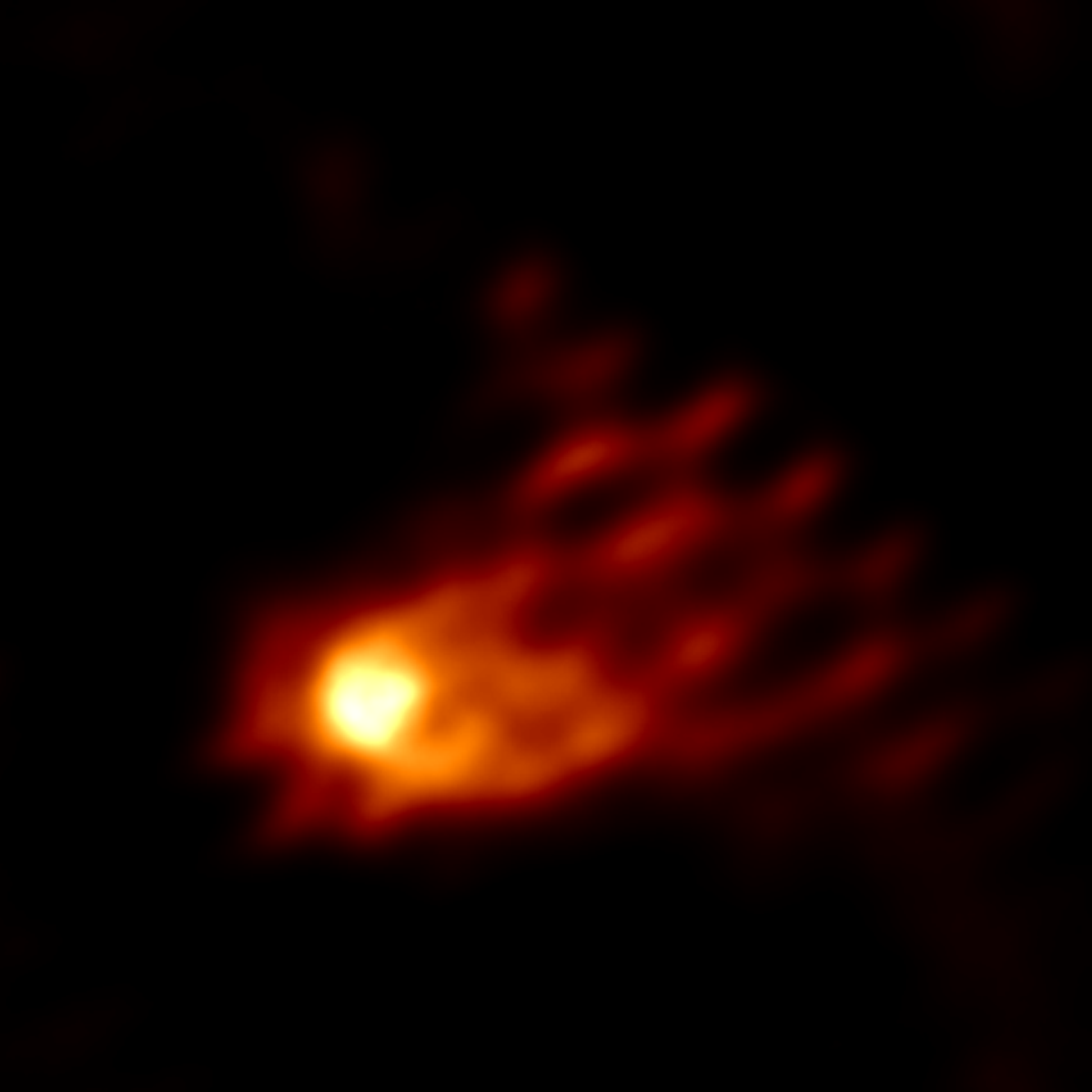} \\
\includegraphics[width=30mm]{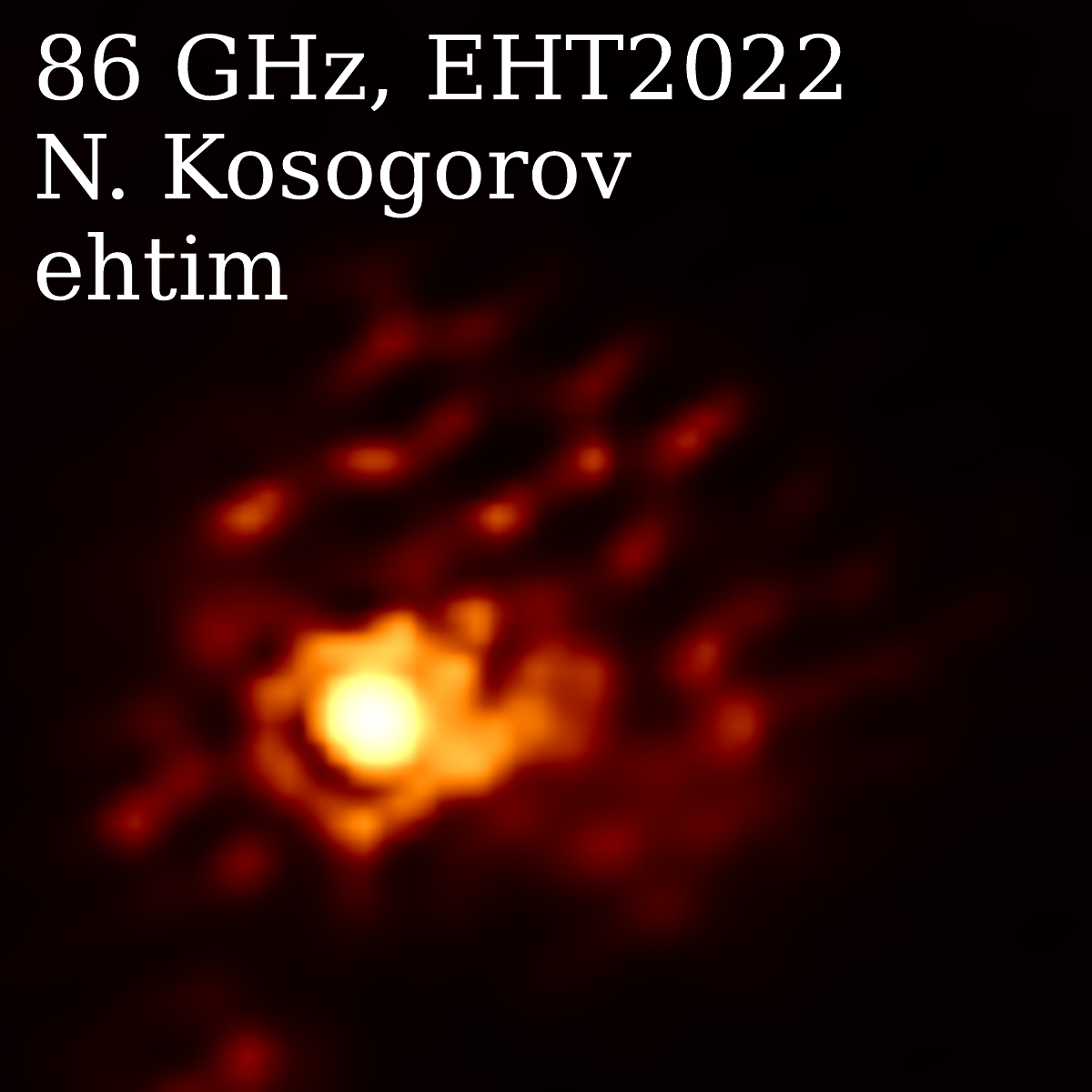}%
\includegraphics[width=30mm]{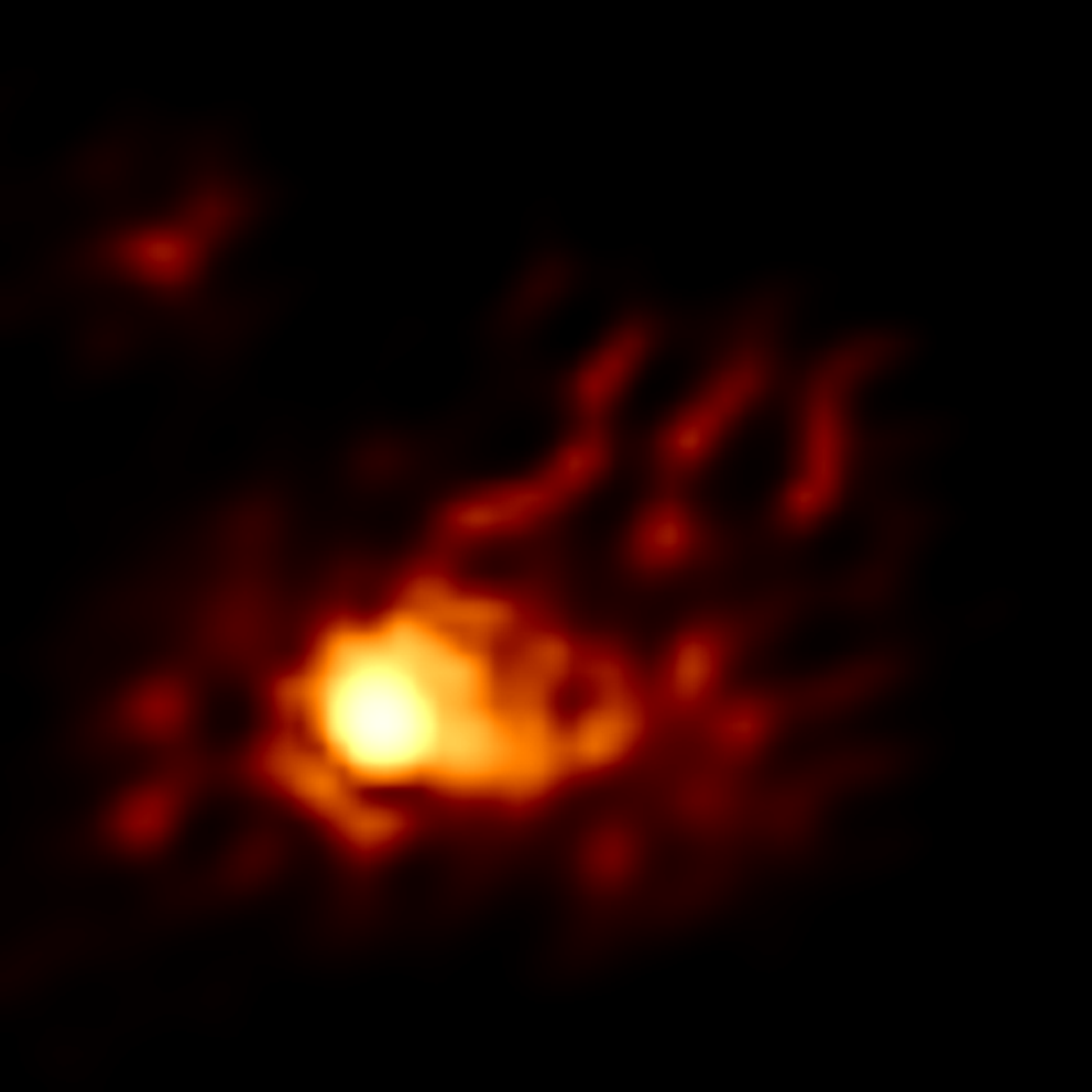}%
\includegraphics[width=30mm]{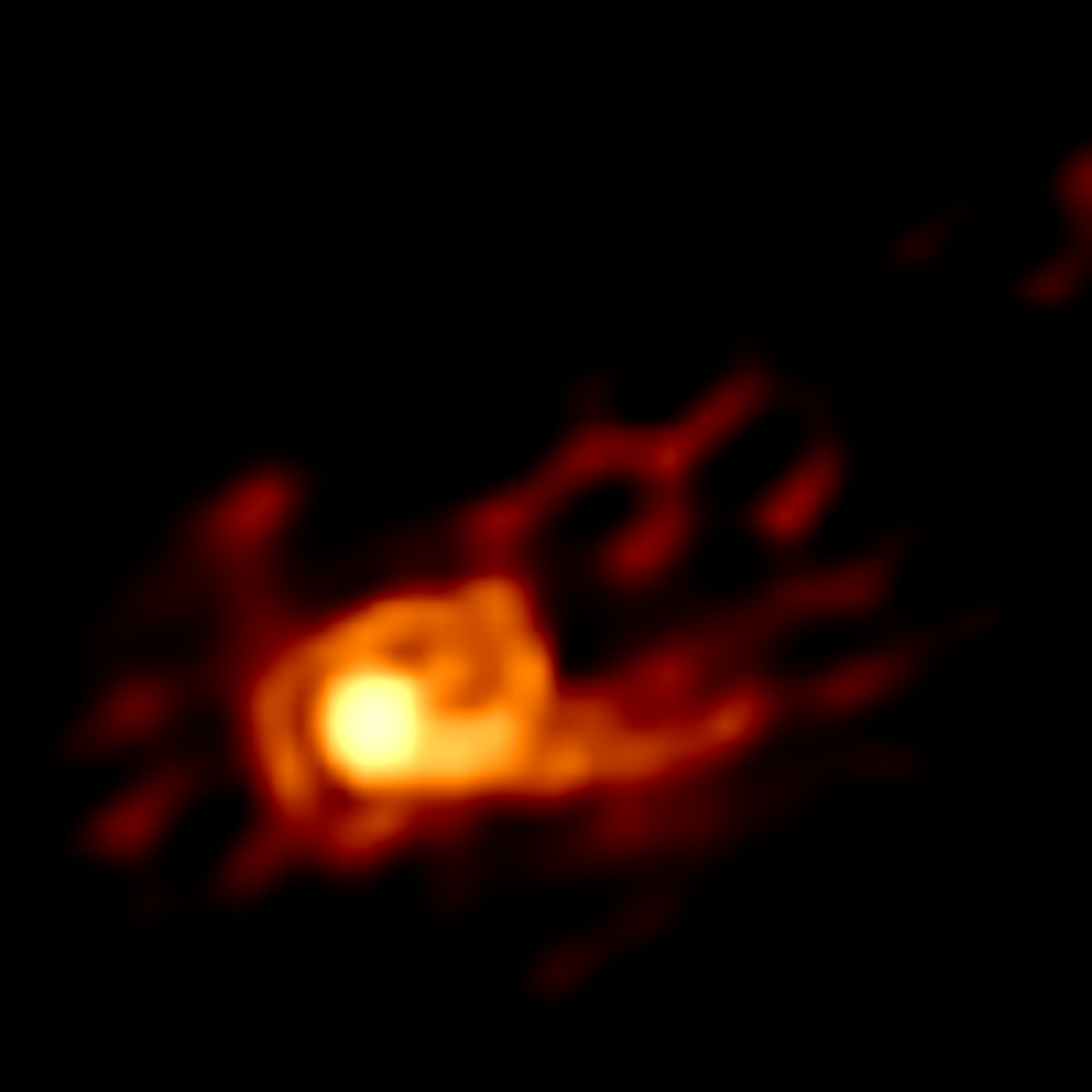}%
\includegraphics[width=30mm]{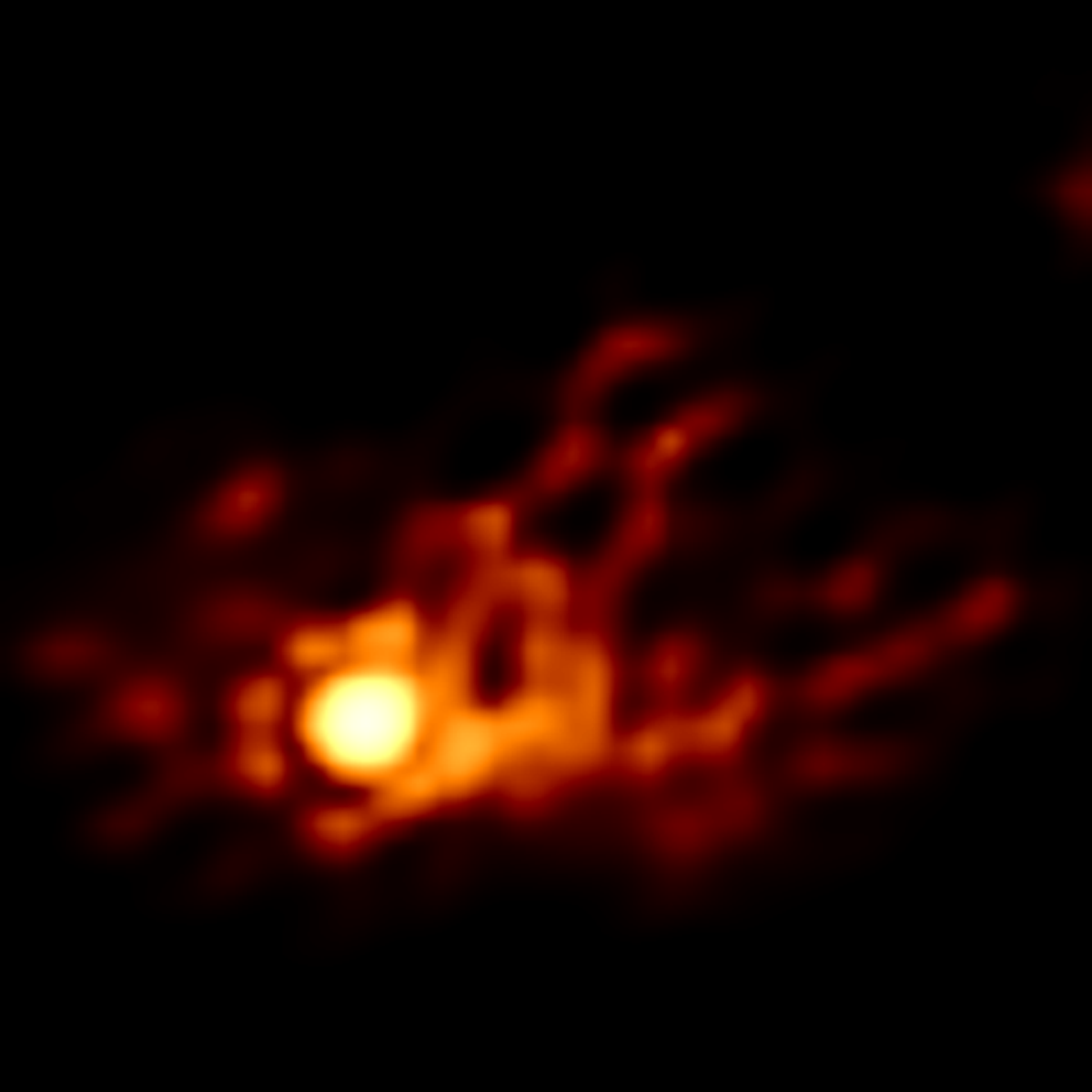}%
\includegraphics[width=30mm]{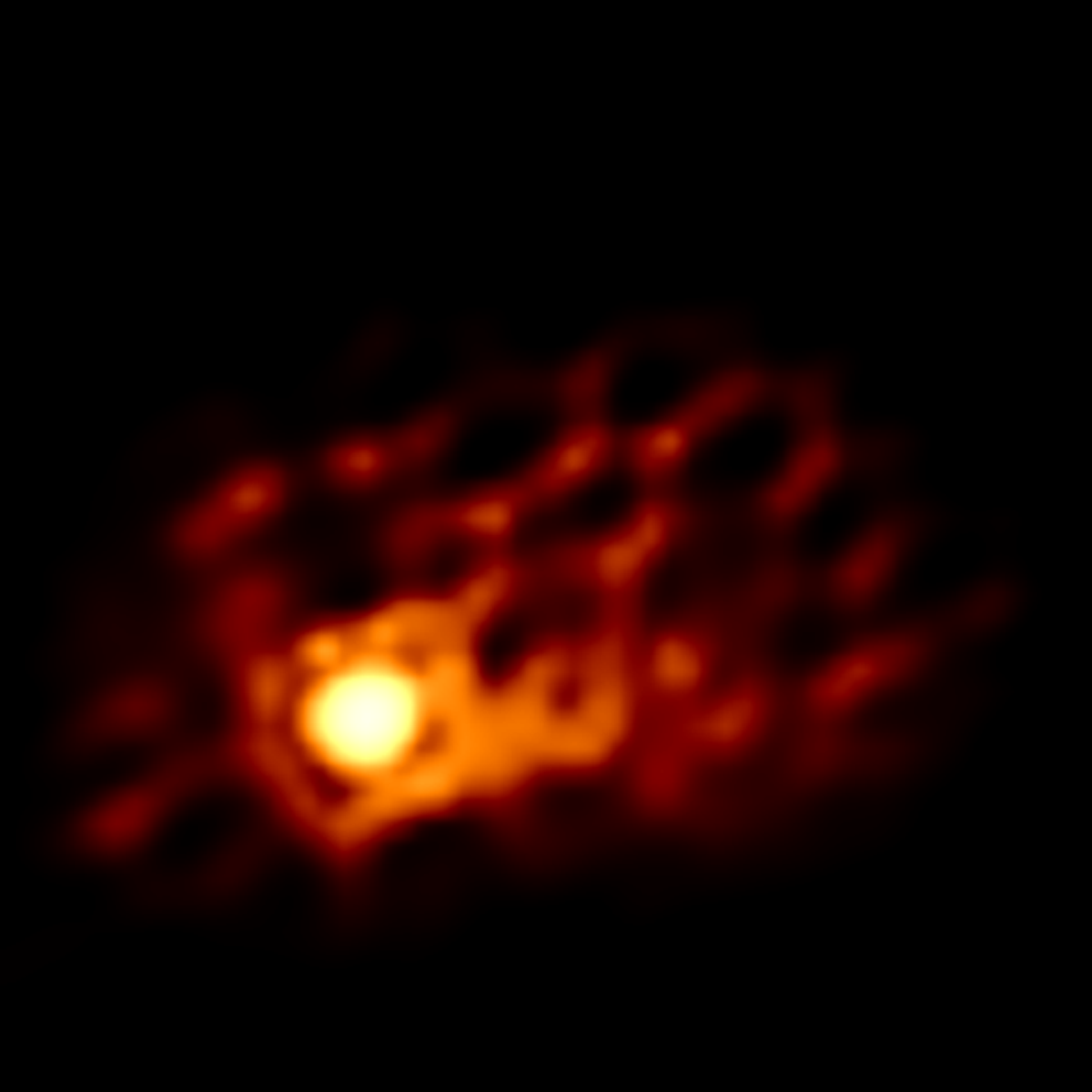}%
\includegraphics[width=30mm]{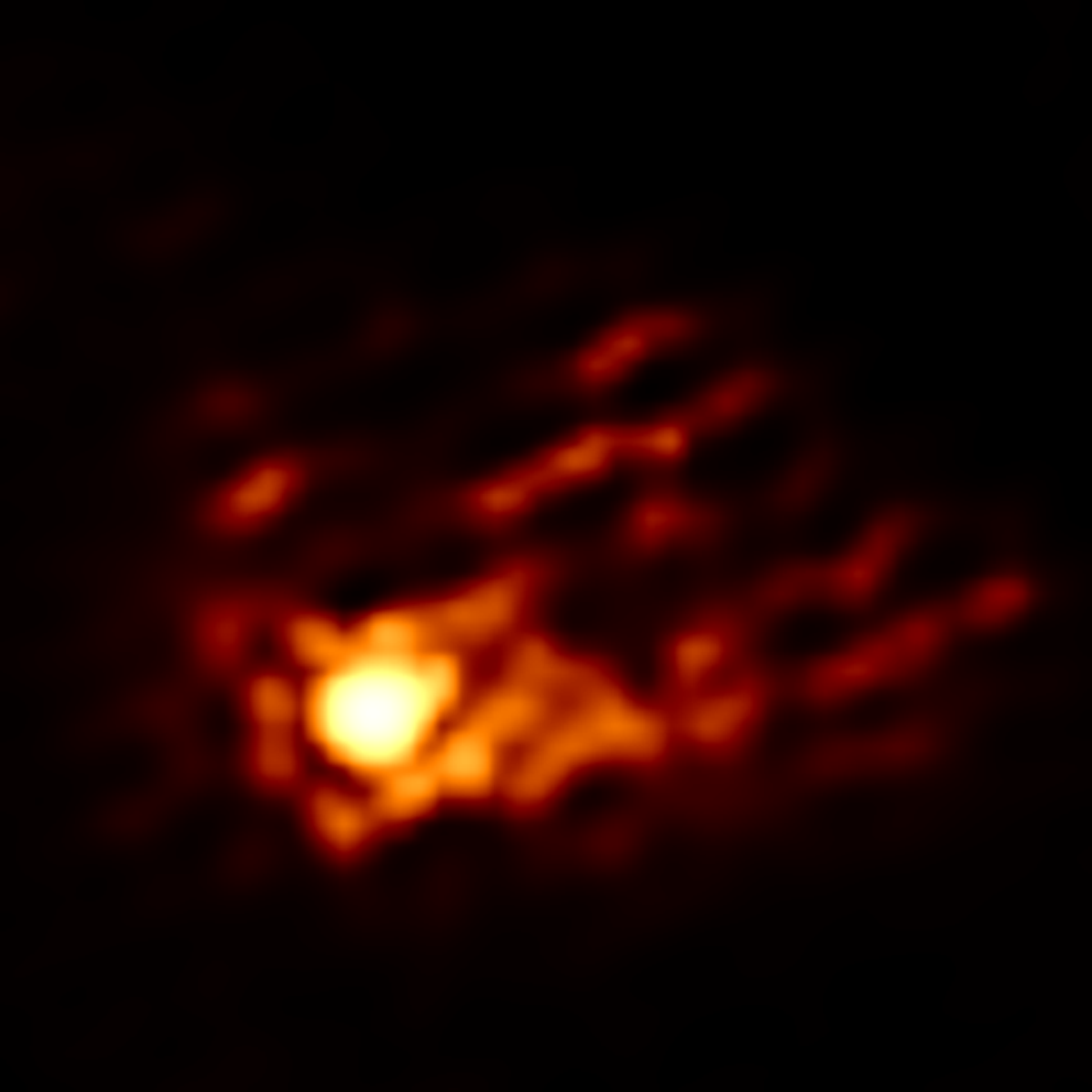} \\
\vspace{2mm}
\includegraphics[width=30mm]{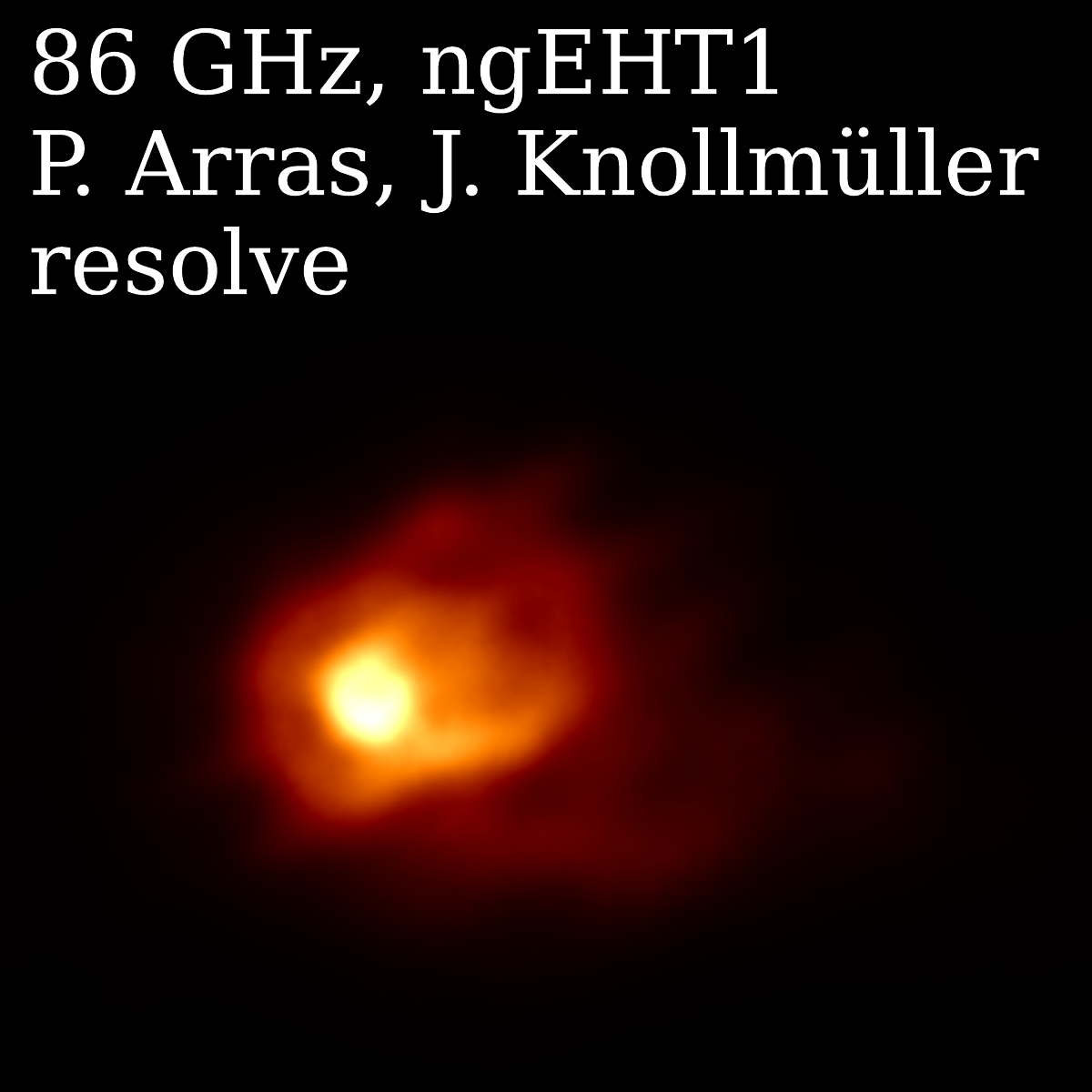}%
\includegraphics[width=30mm]{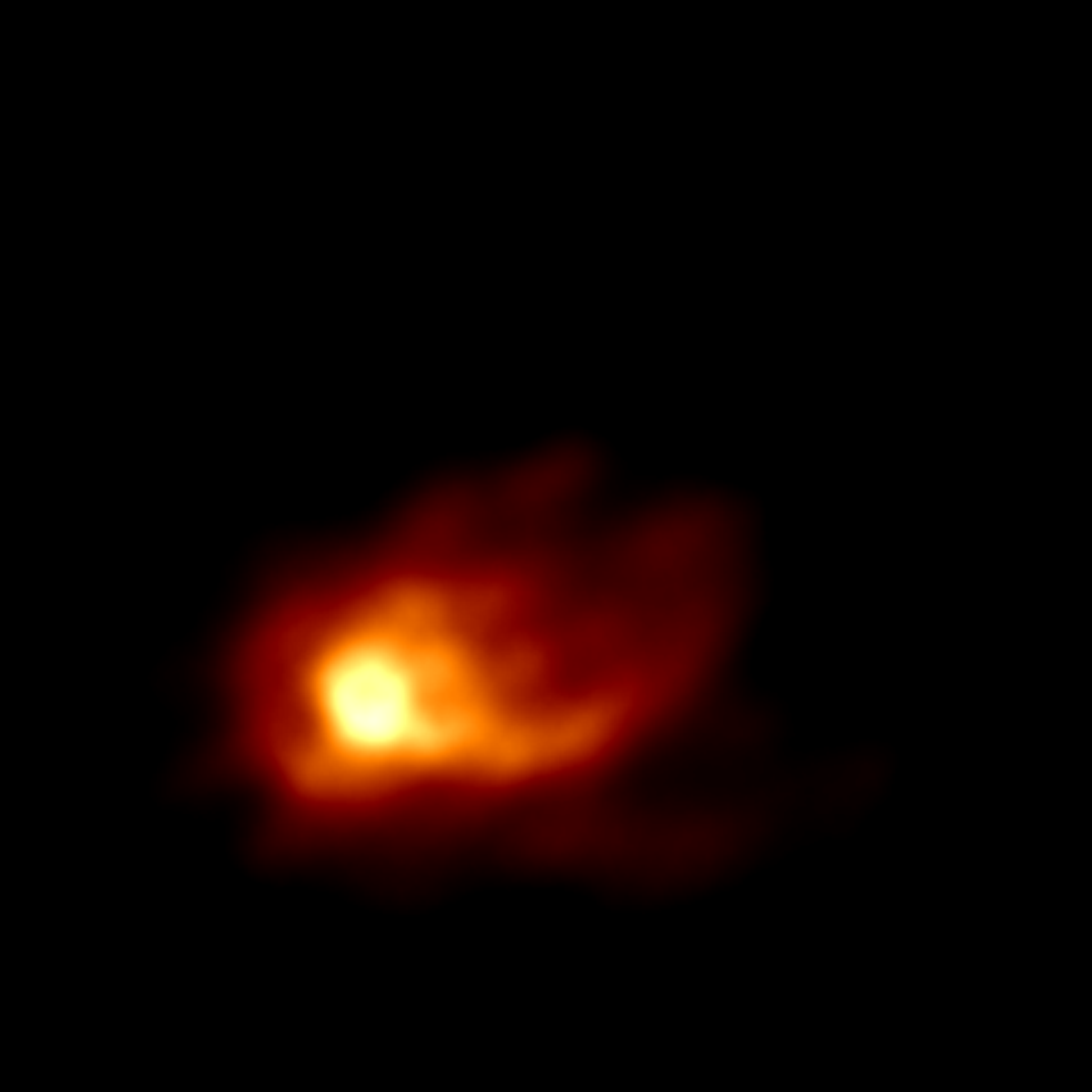}%
\includegraphics[width=30mm]{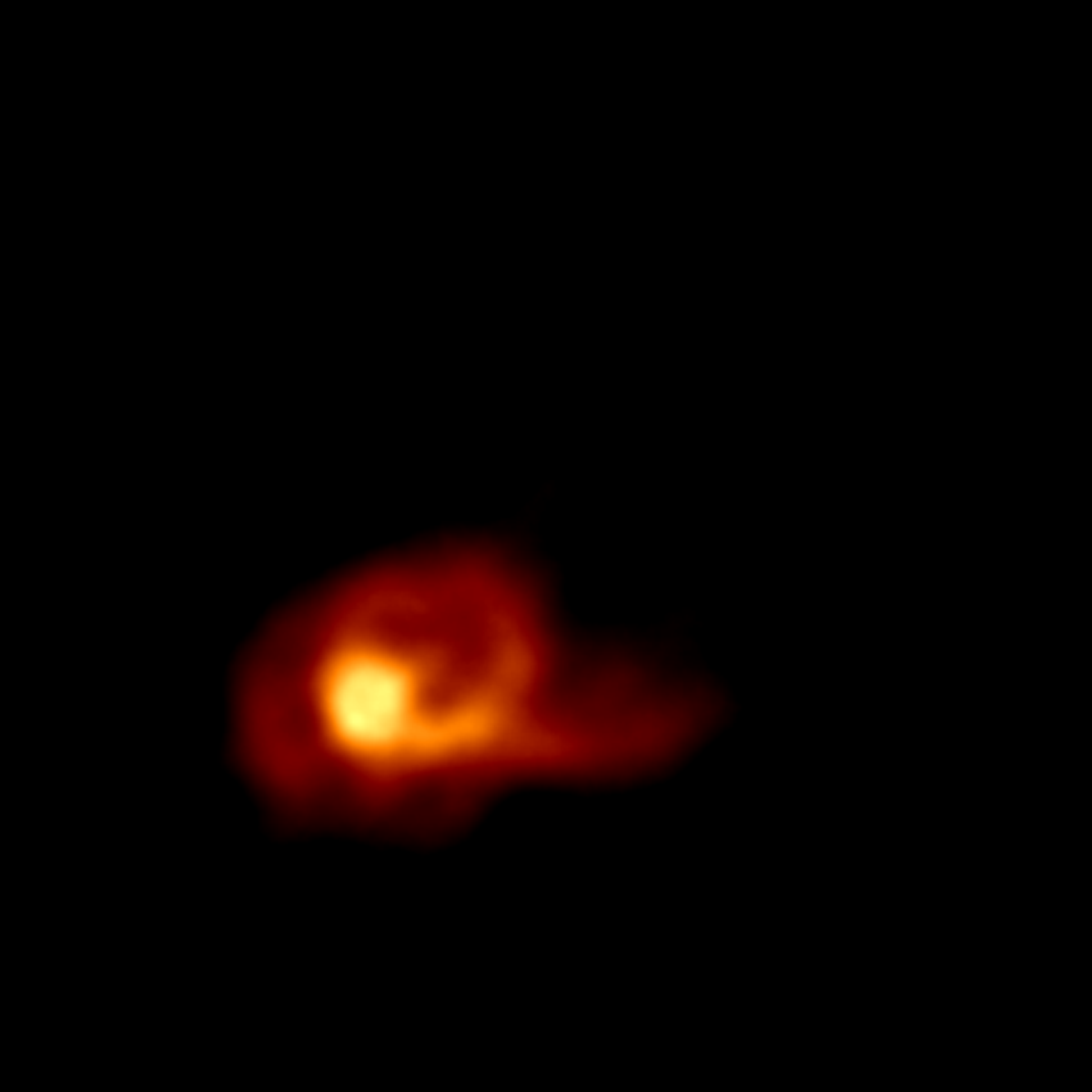}%
\includegraphics[width=30mm]{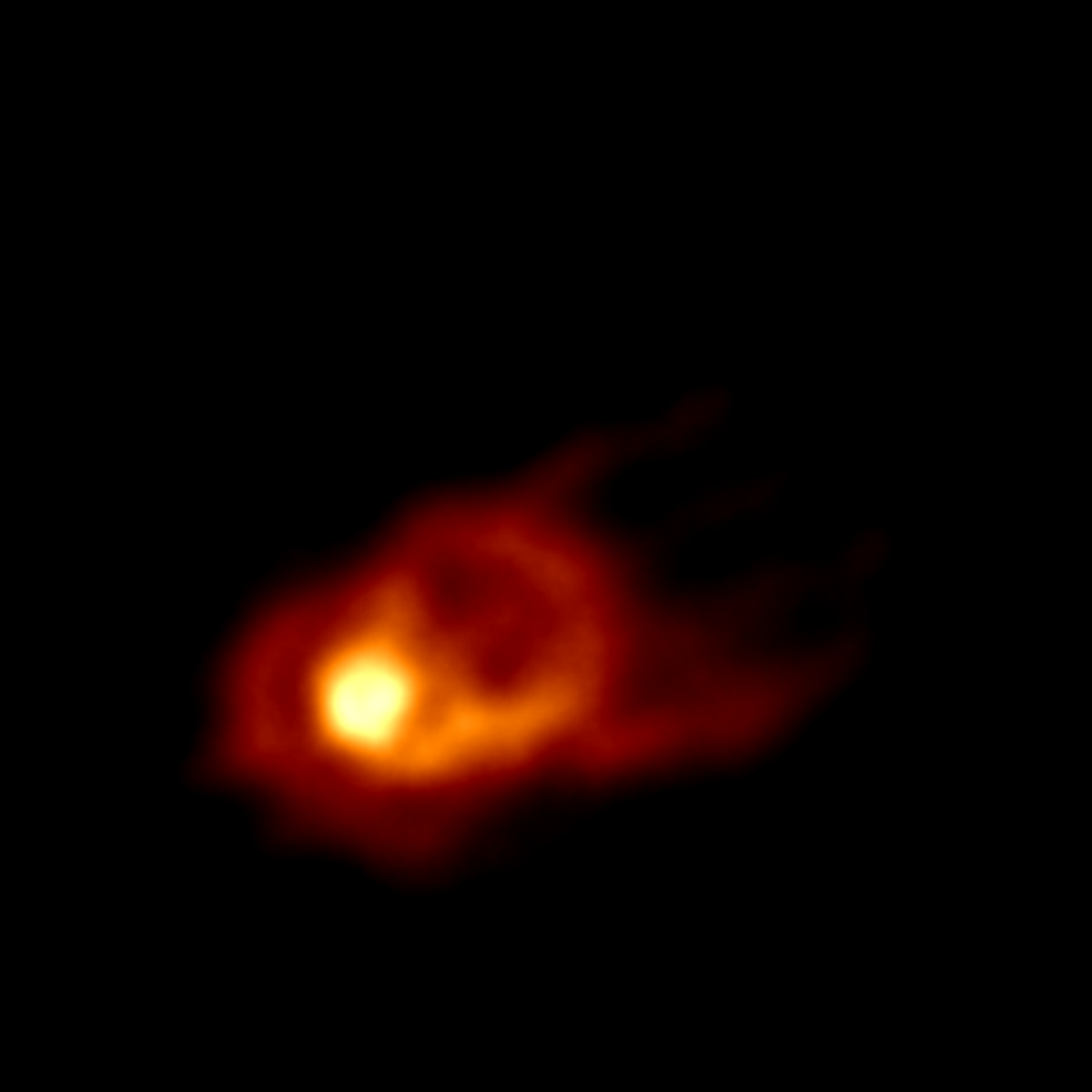}%
\includegraphics[width=30mm]{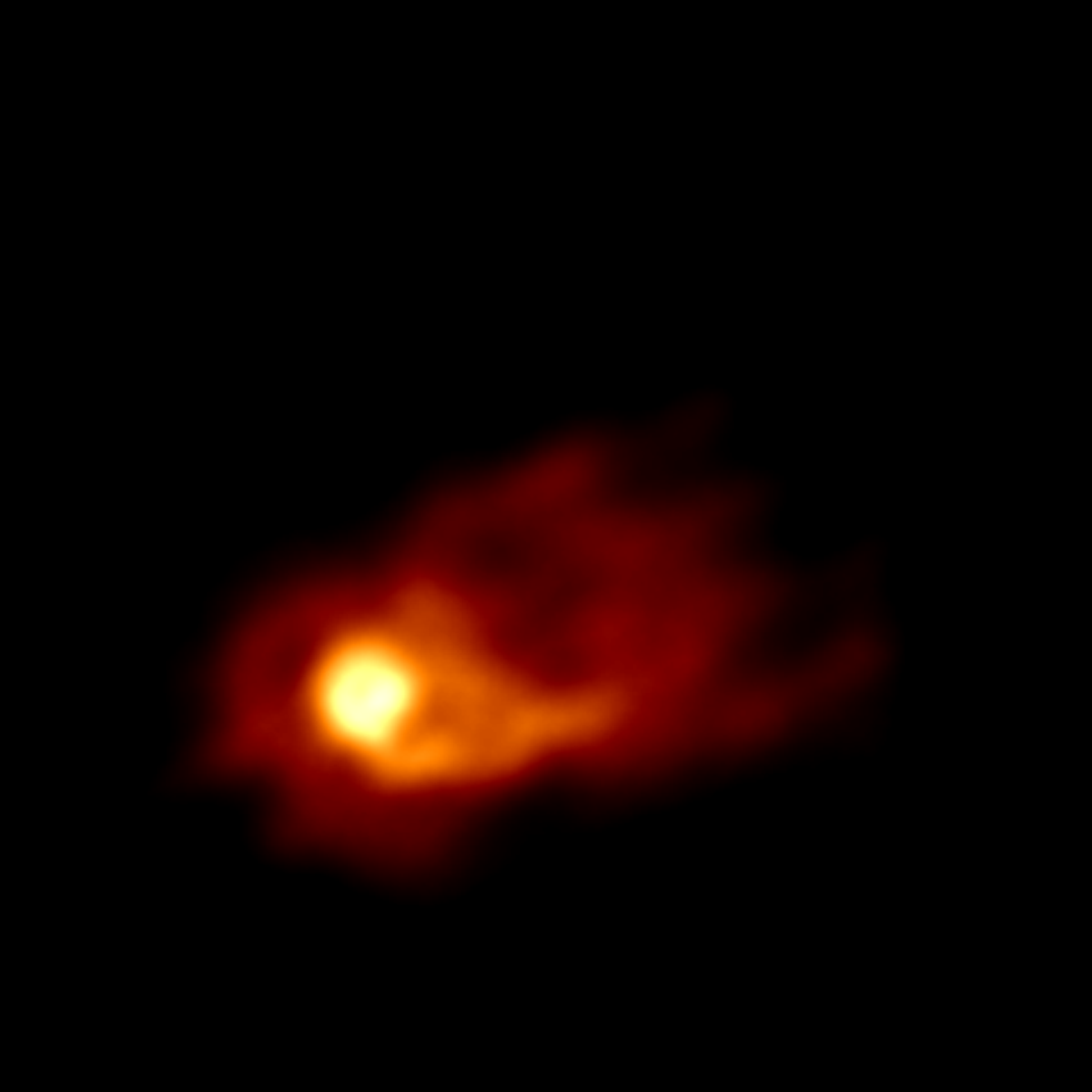}%
\includegraphics[width=30mm]{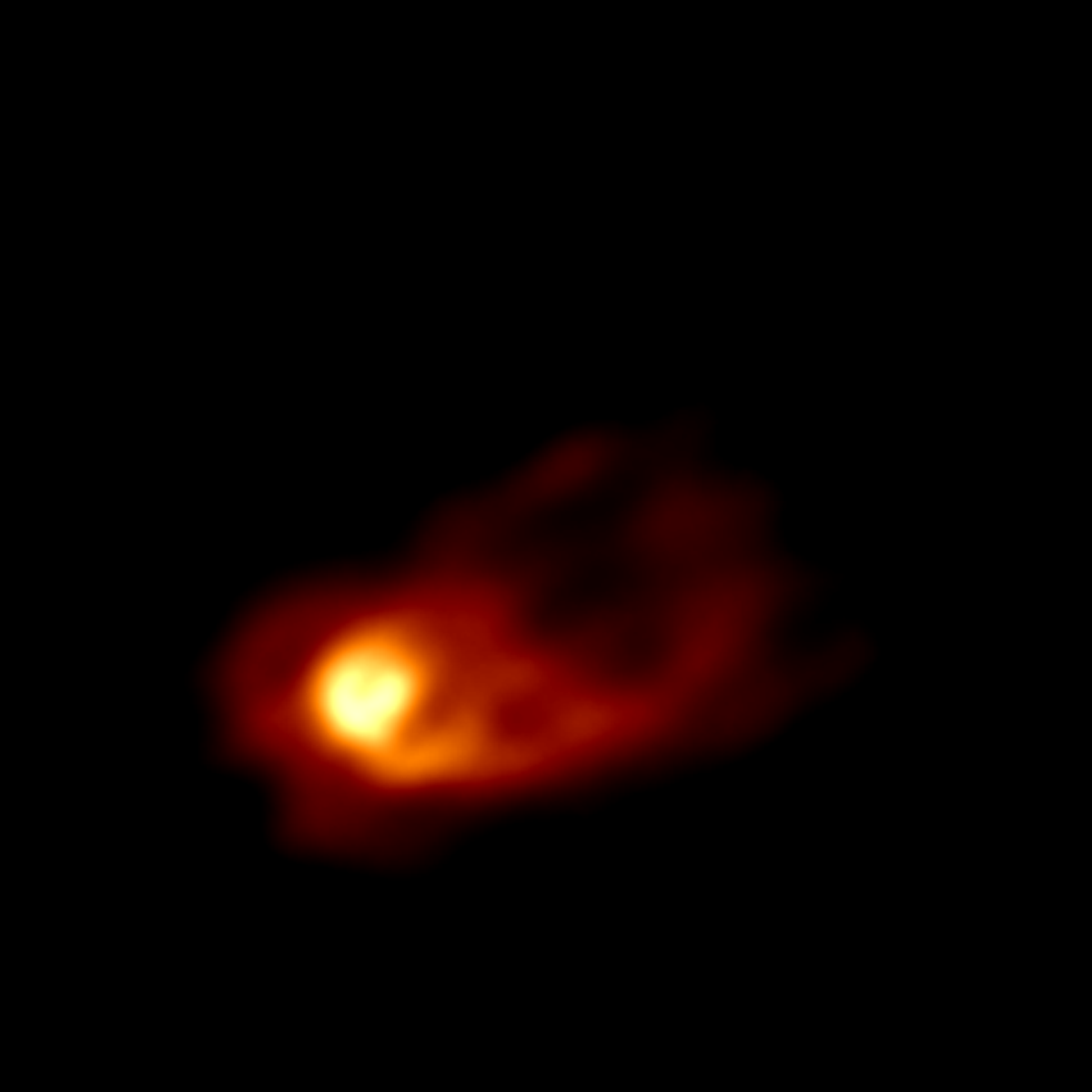} \\
\includegraphics[width=30mm]{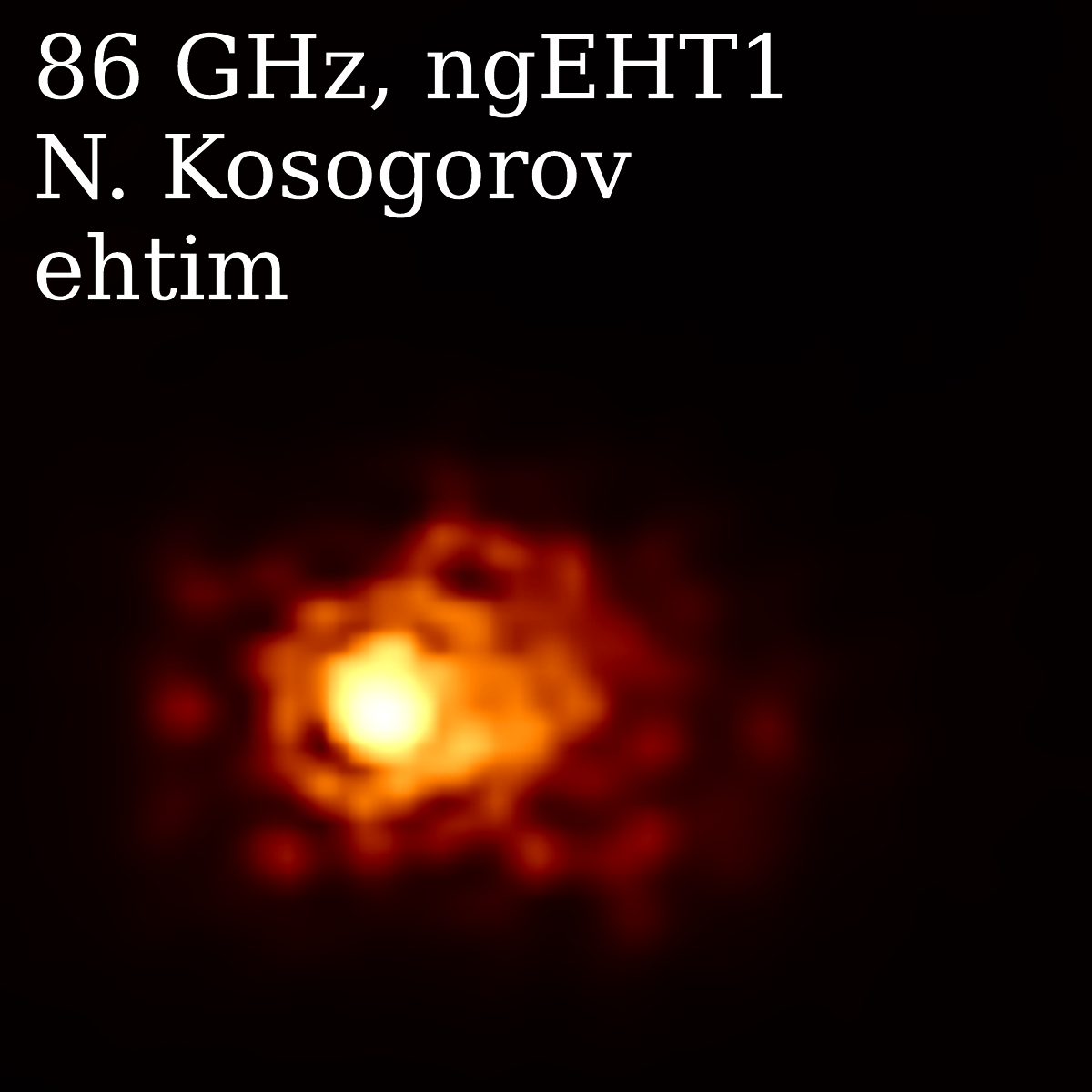}%
\includegraphics[width=30mm]{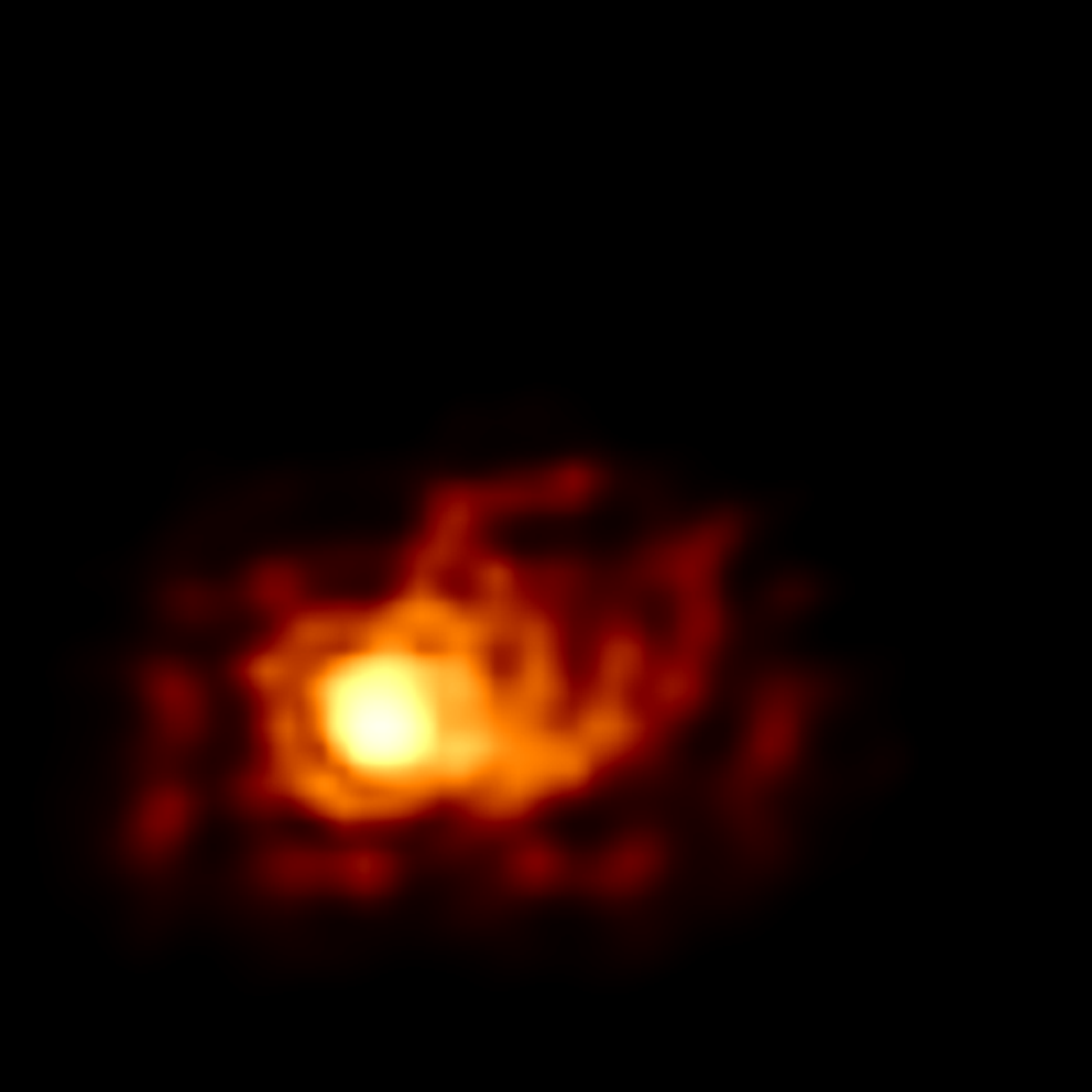}%
\includegraphics[width=30mm]{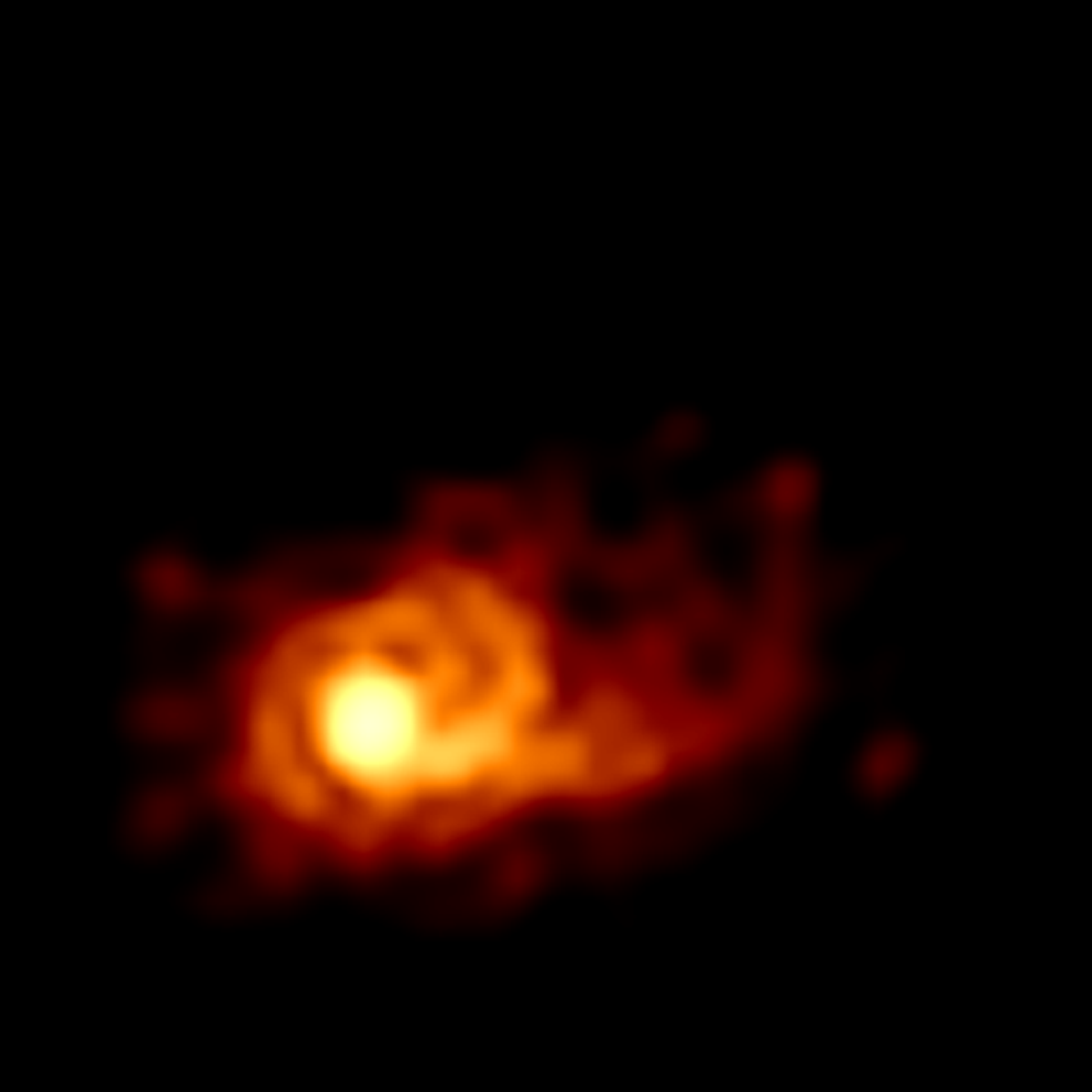}%
\includegraphics[width=30mm]{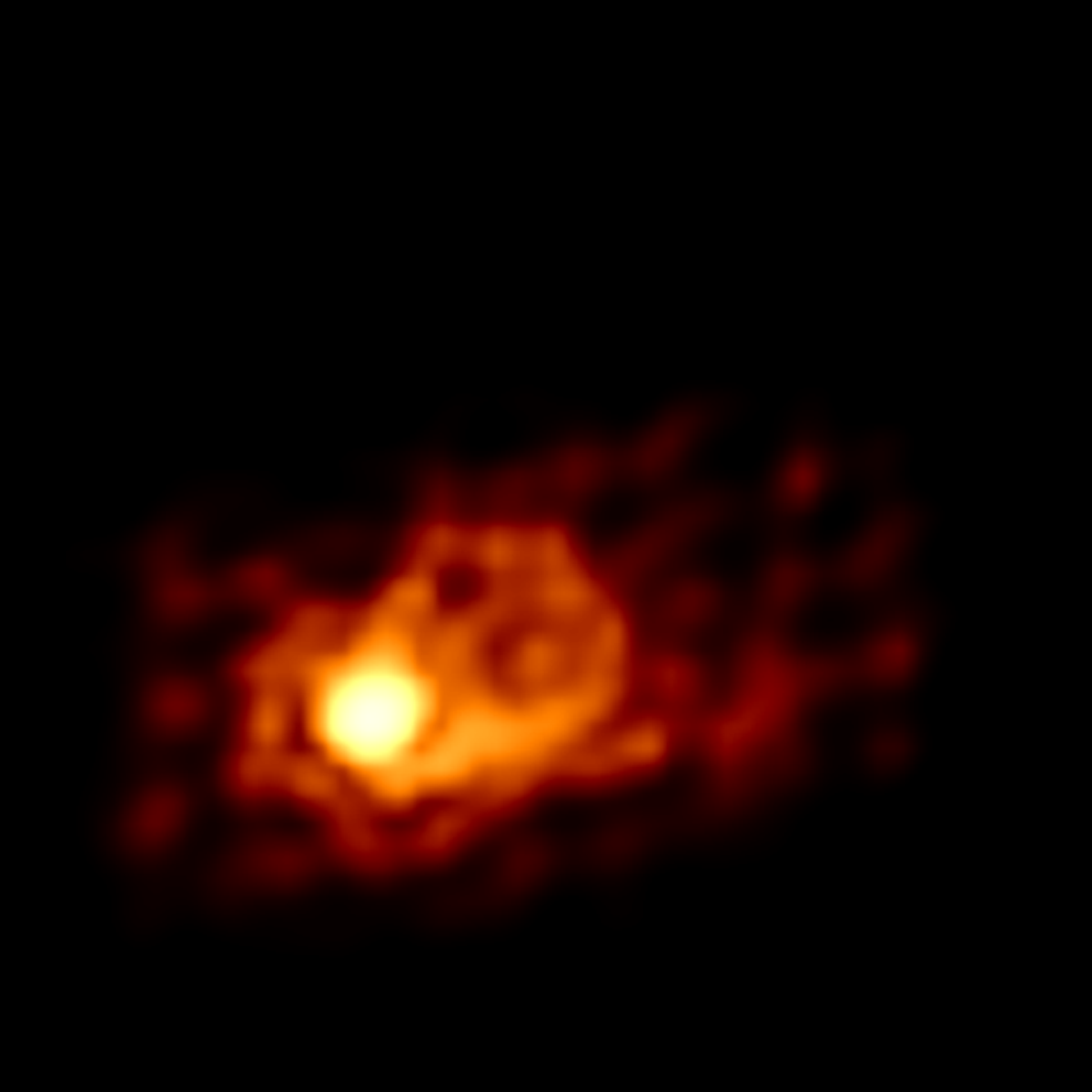}%
\includegraphics[width=30mm]{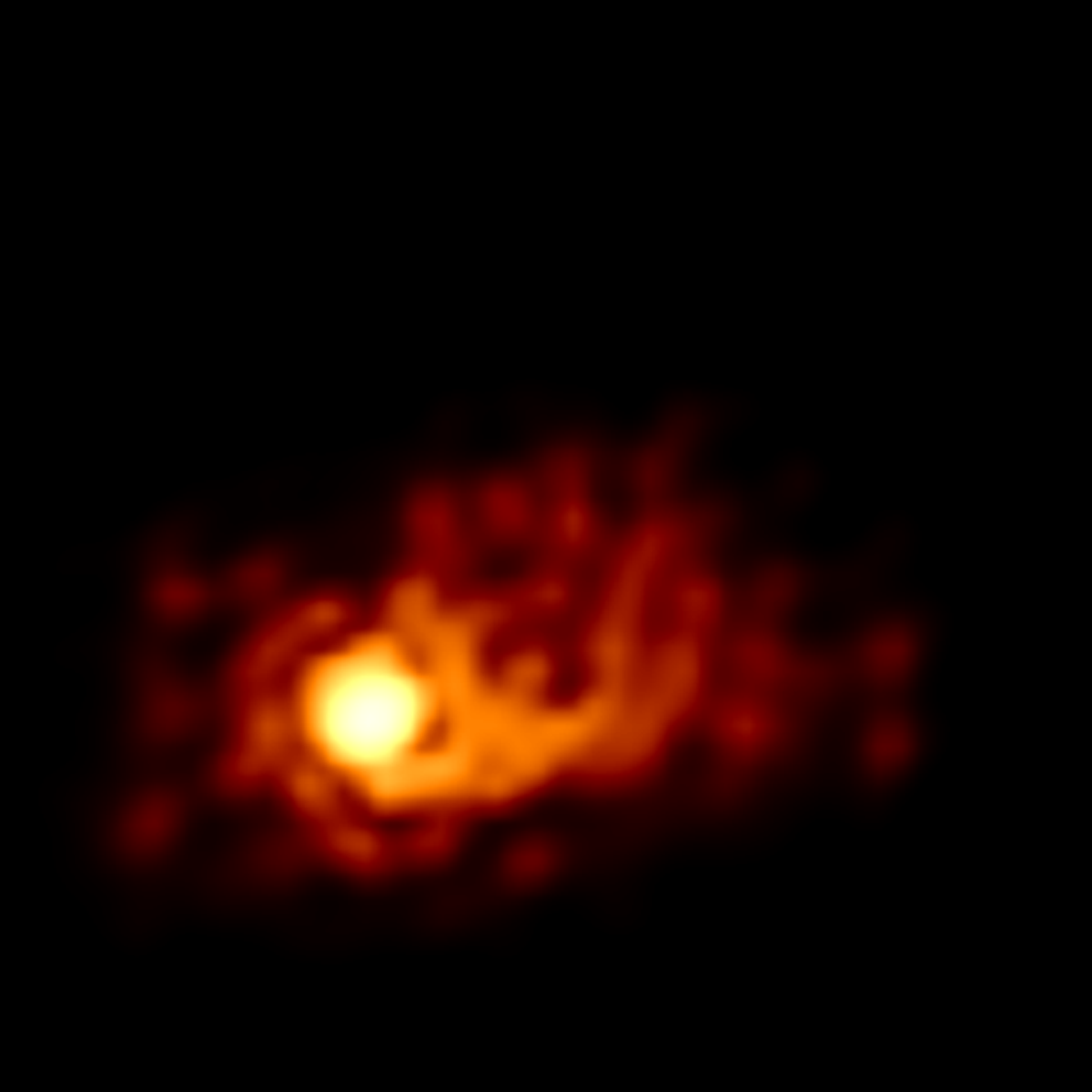}%
\includegraphics[width=30mm]{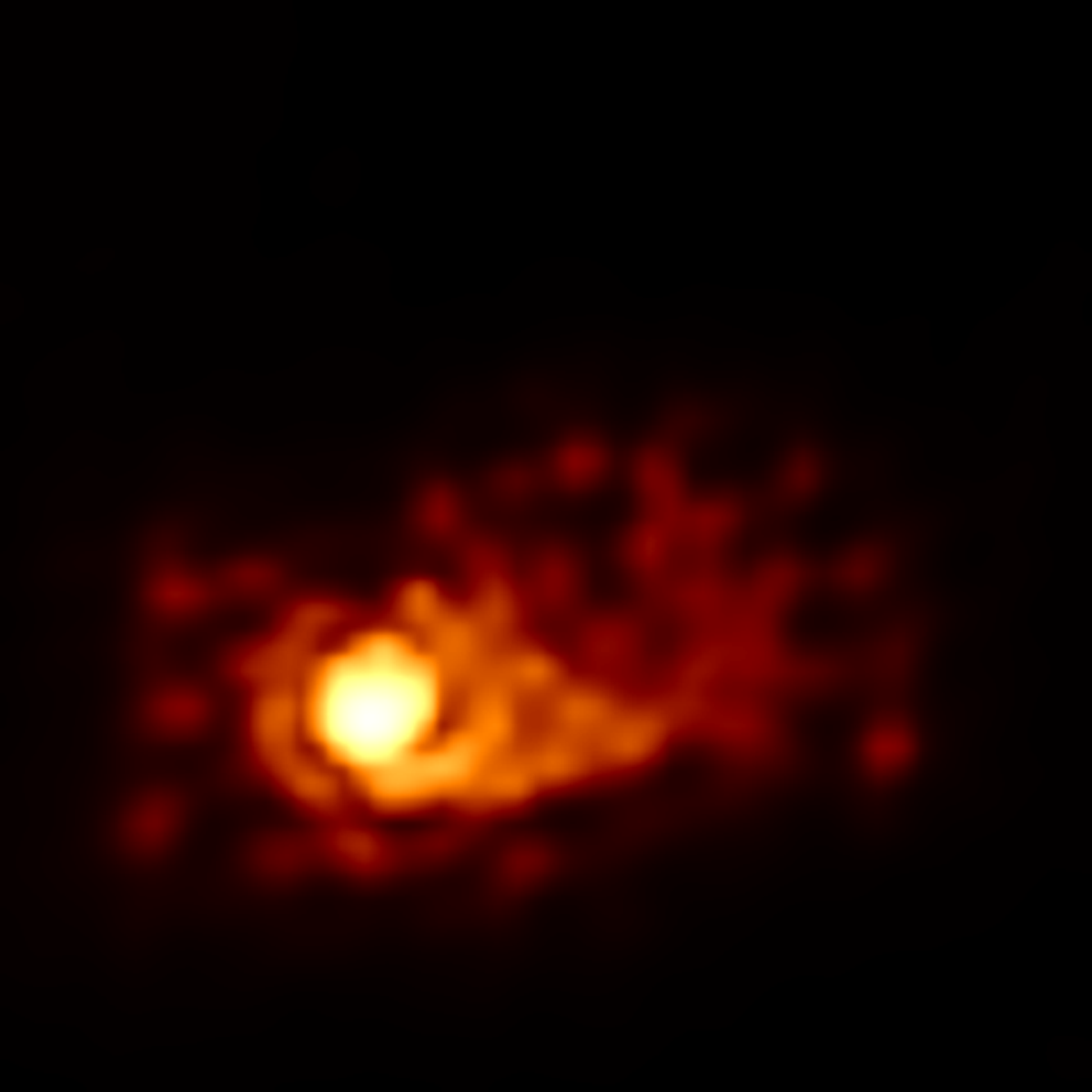} \\
\includegraphics[width=30mm]{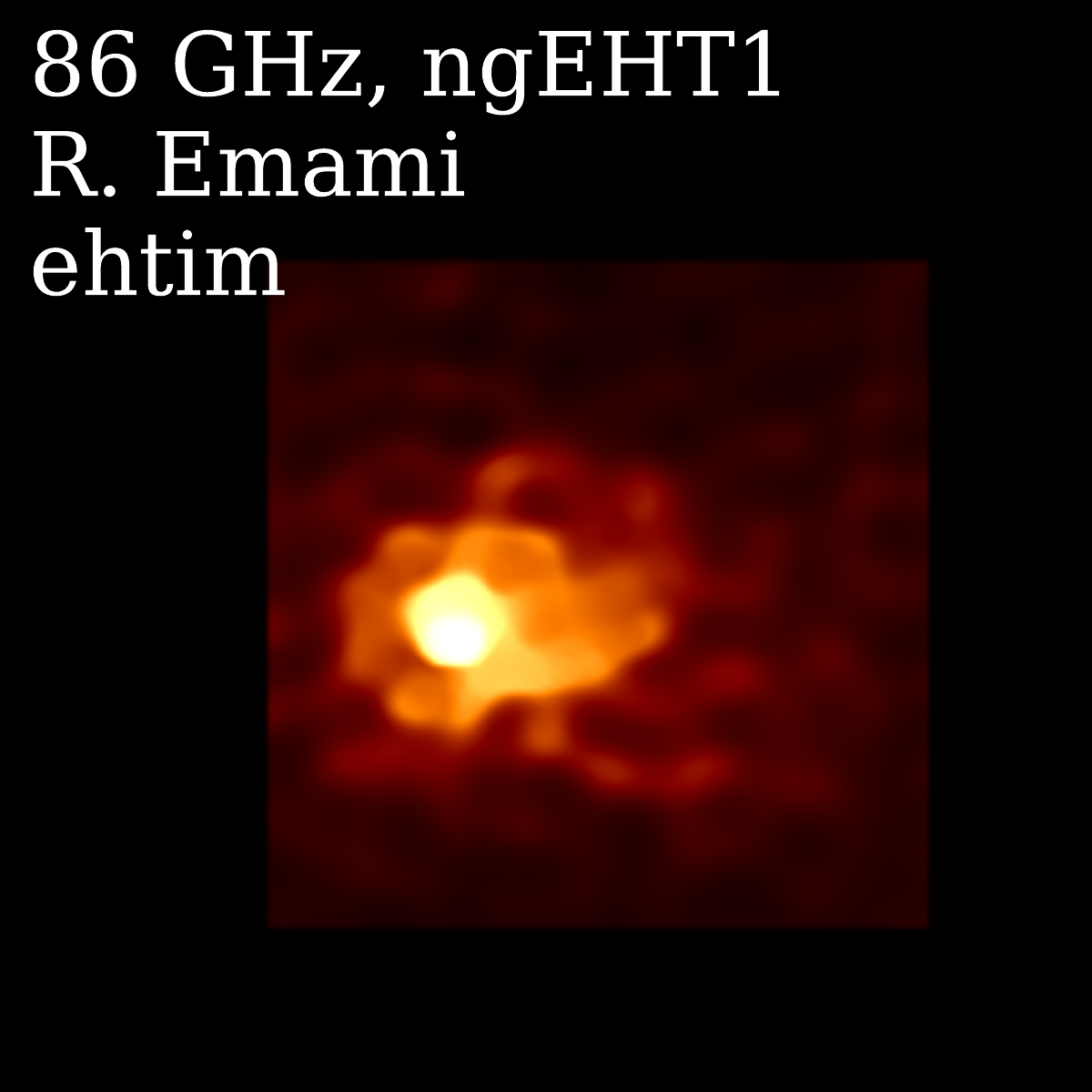}%
\includegraphics[width=30mm]{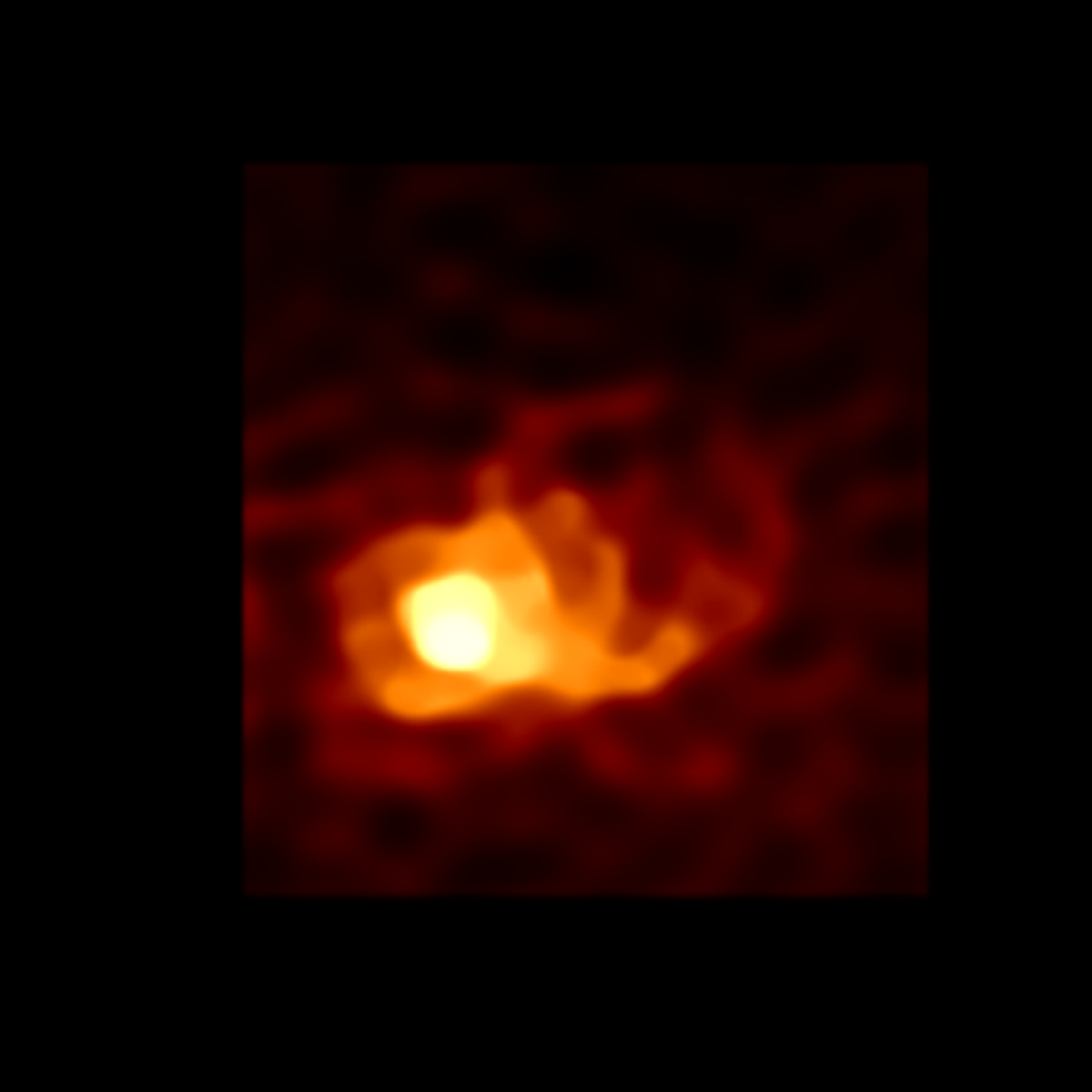}%
\includegraphics[width=30mm]{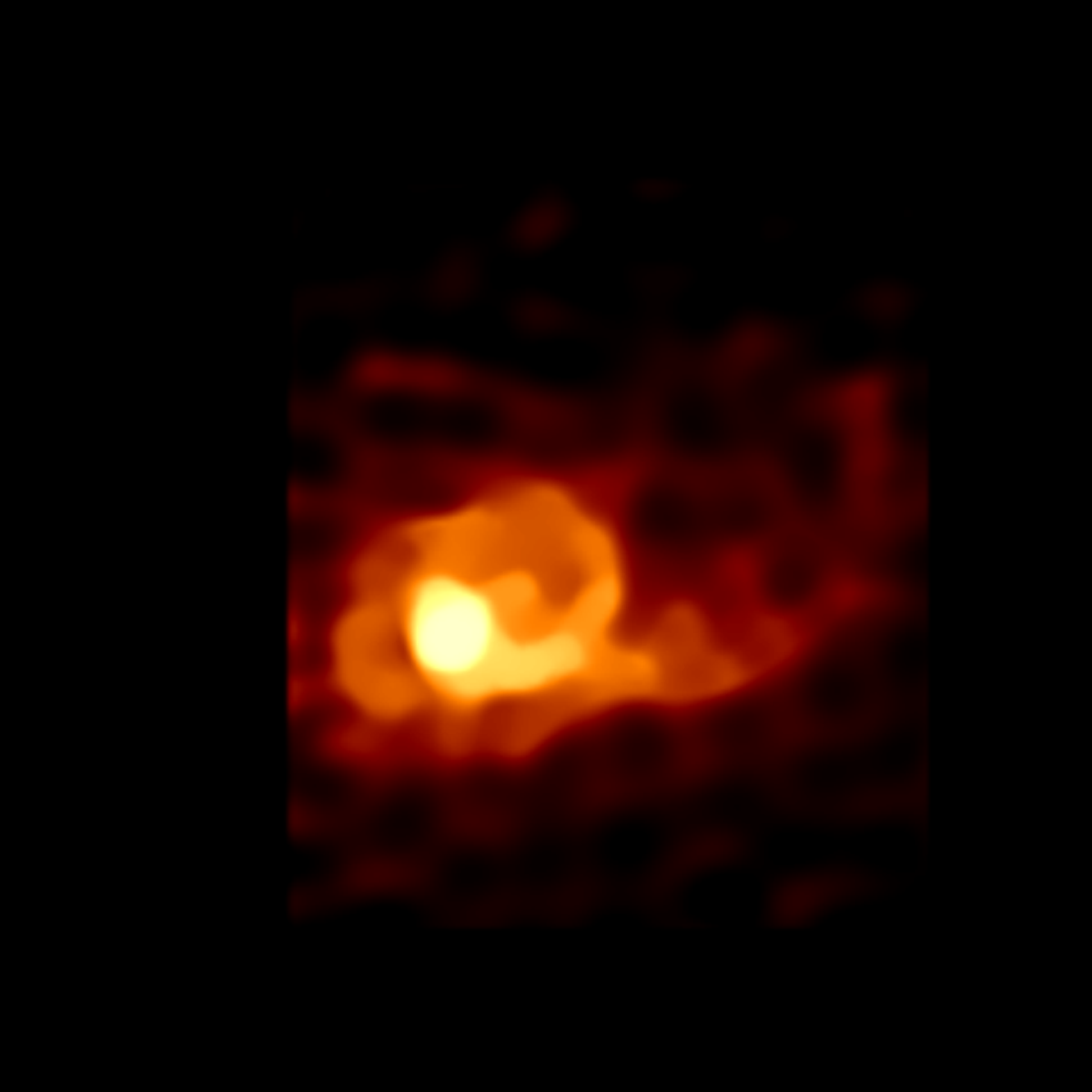}%
\includegraphics[width=30mm]{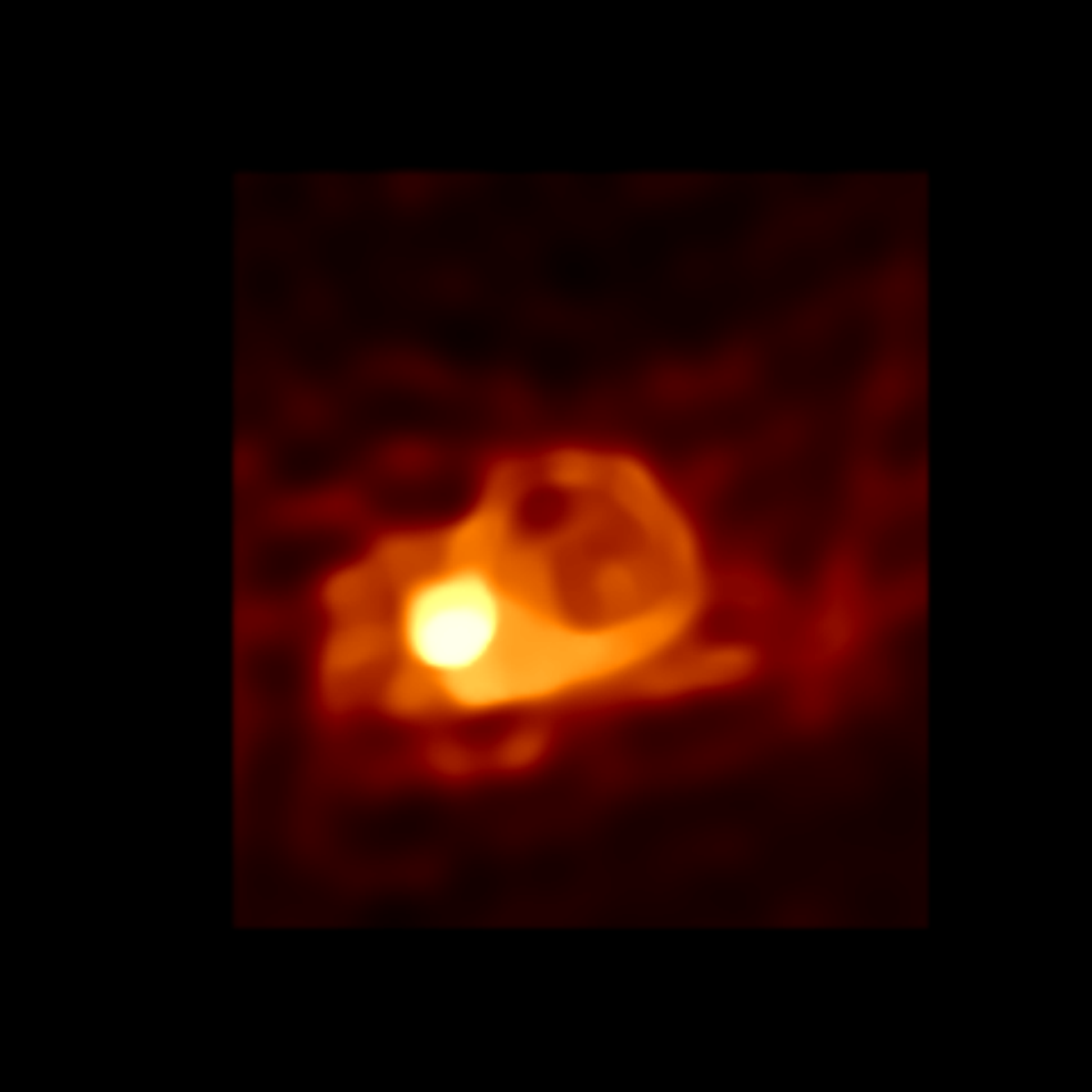}%
\includegraphics[width=30mm]{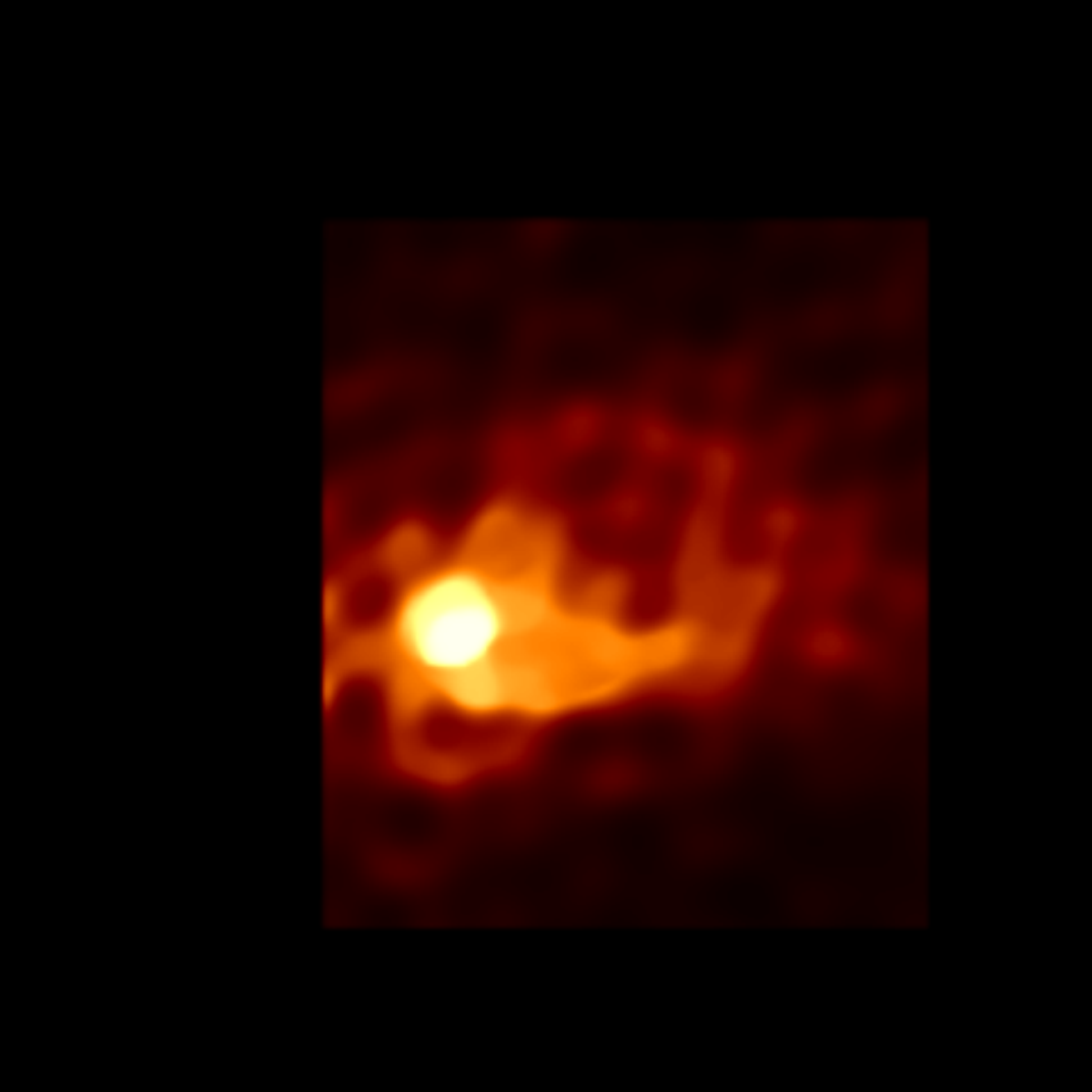}%
\includegraphics[width=30mm]{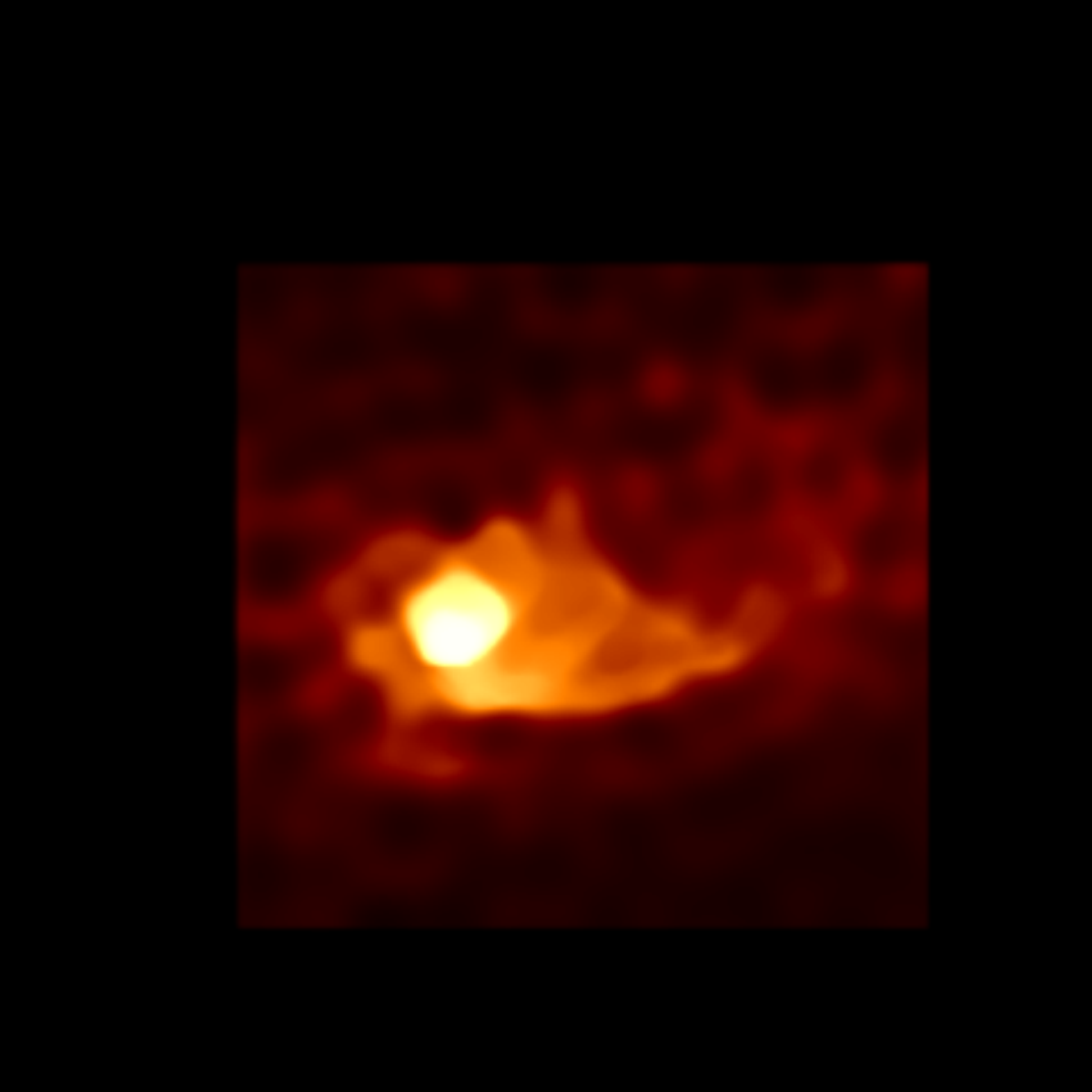} \\
\vspace{5mm}
\end{adjustwidth}
  \caption{Selection of Challenge 2 M87 86 GHz submissions. Images are shown on a log scale, which is normalized to the brightest pixel value across each set of three movie frames, with a dynamic range of $10^{3.5}$ and field of view of 1 mas.}
     \label{fig:ch2_m87_86}
\end{figure*}

\begin{figure*}
\begin{adjustwidth}{-\extralength}{0cm}
\setlength{\lineskip}{0pt}
\centering
\includegraphics[width=30mm]{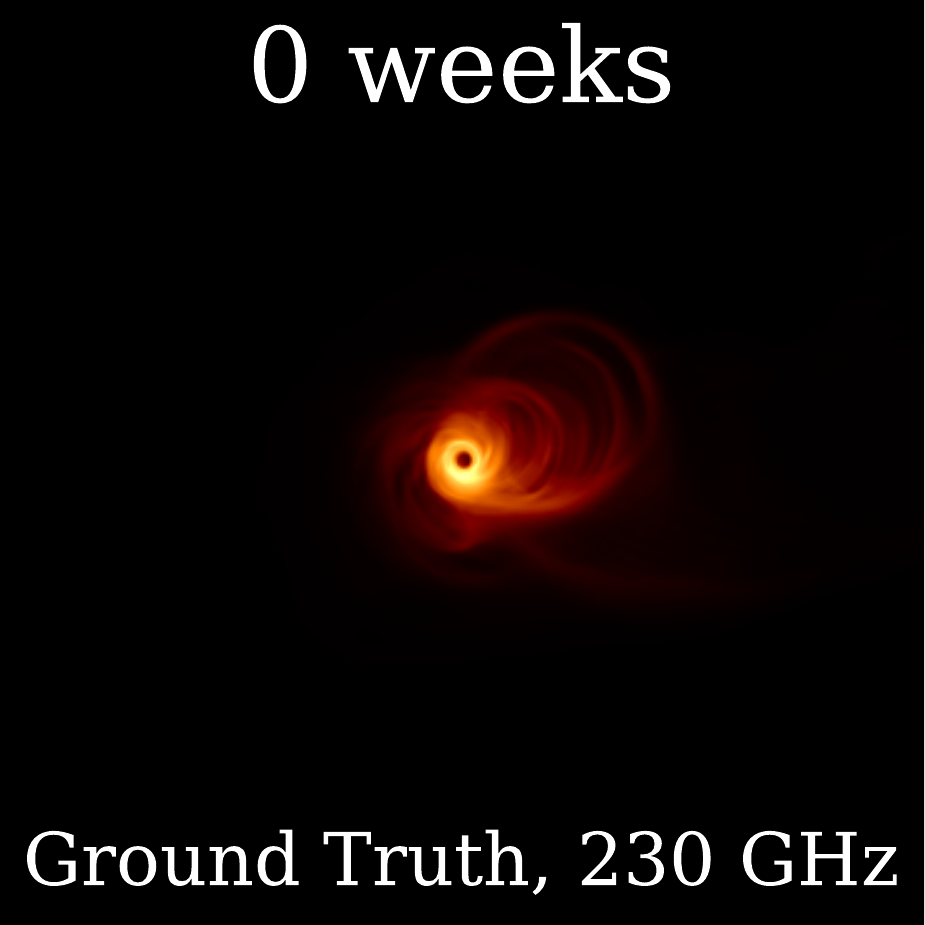}%
\includegraphics[width=30mm]{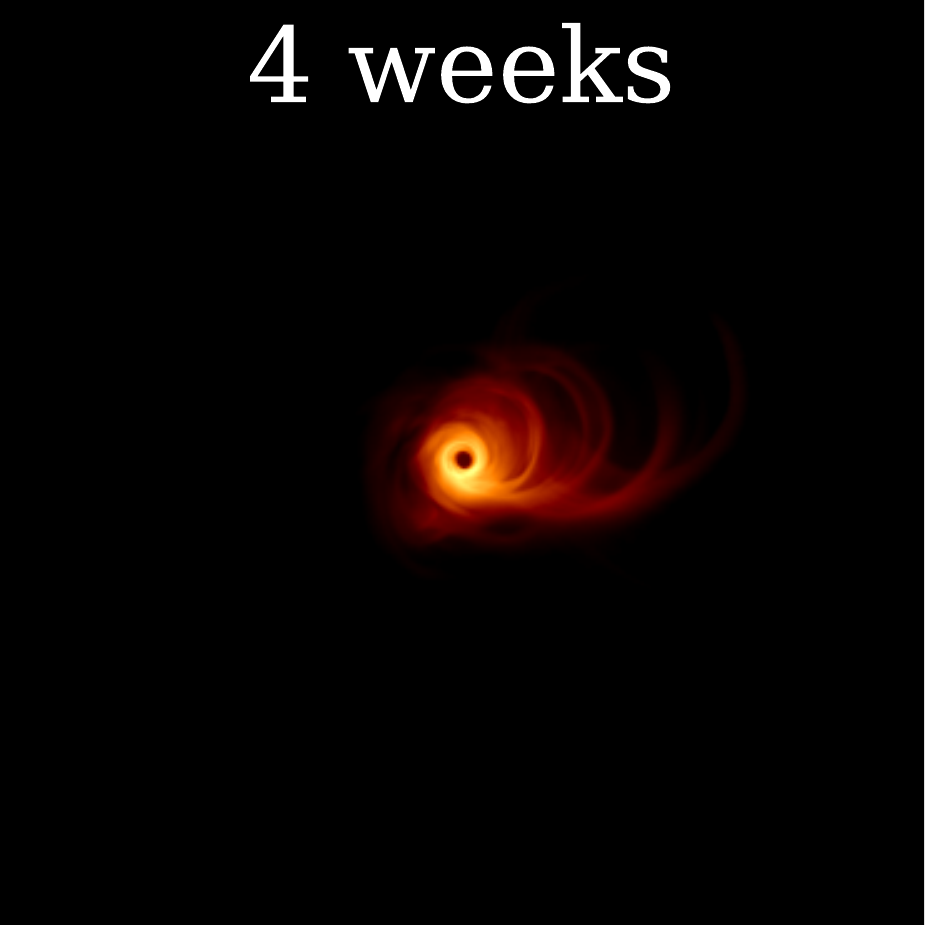}%
\includegraphics[width=30mm]{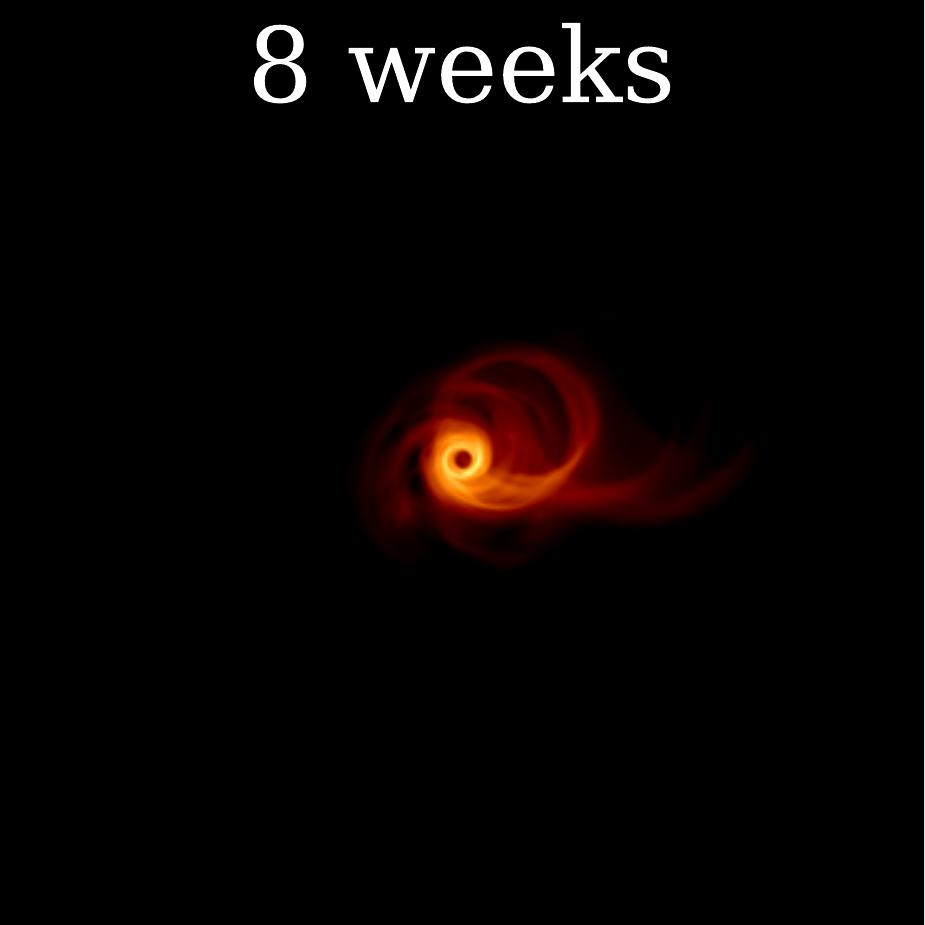}%
\includegraphics[width=30mm]{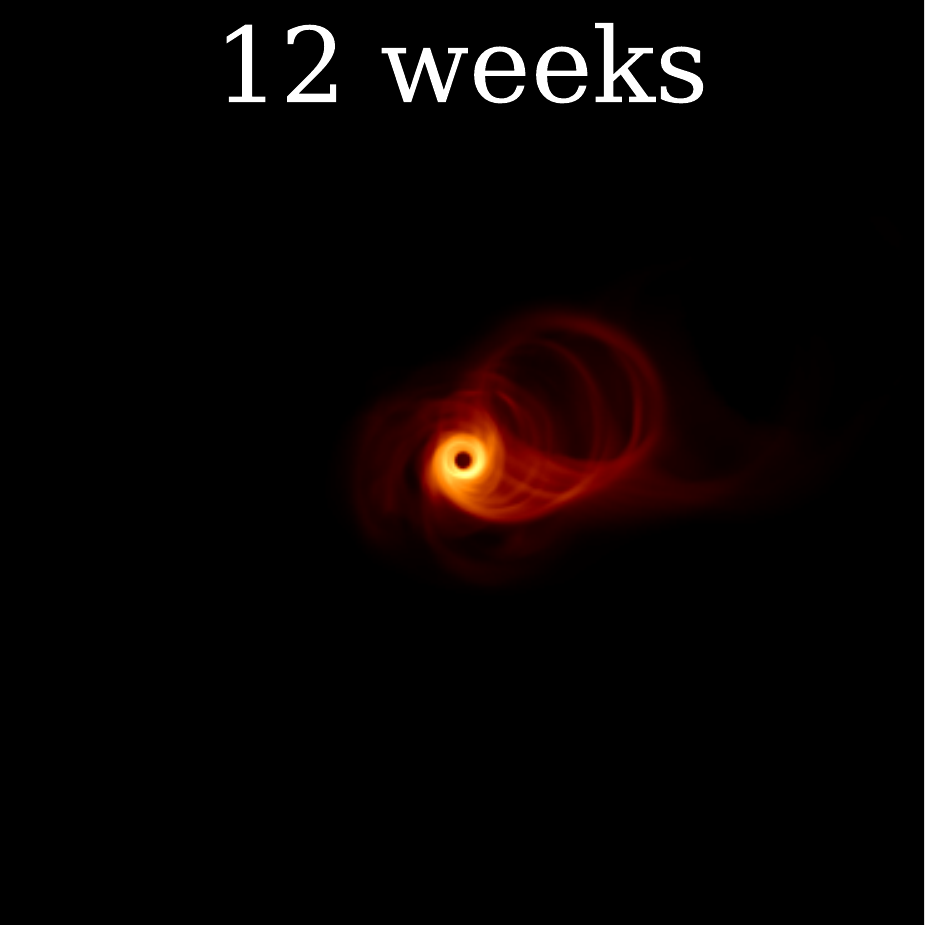}%
\includegraphics[width=30mm]{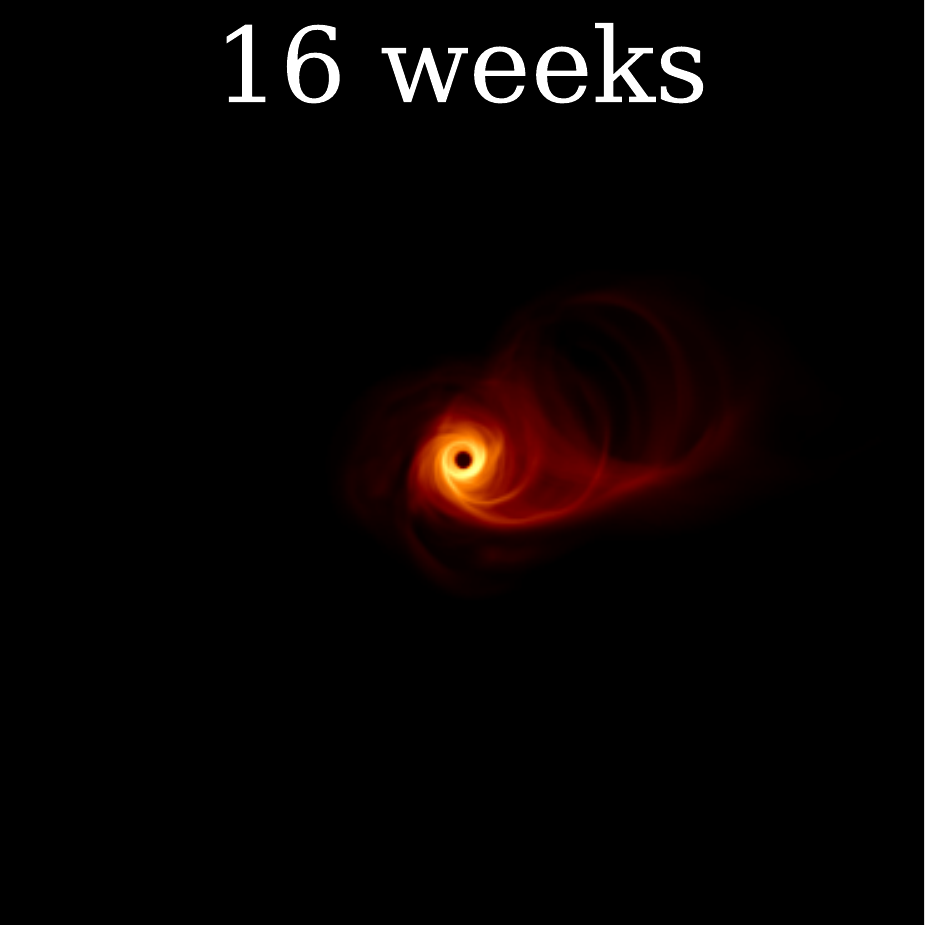}%
\includegraphics[width=30mm]{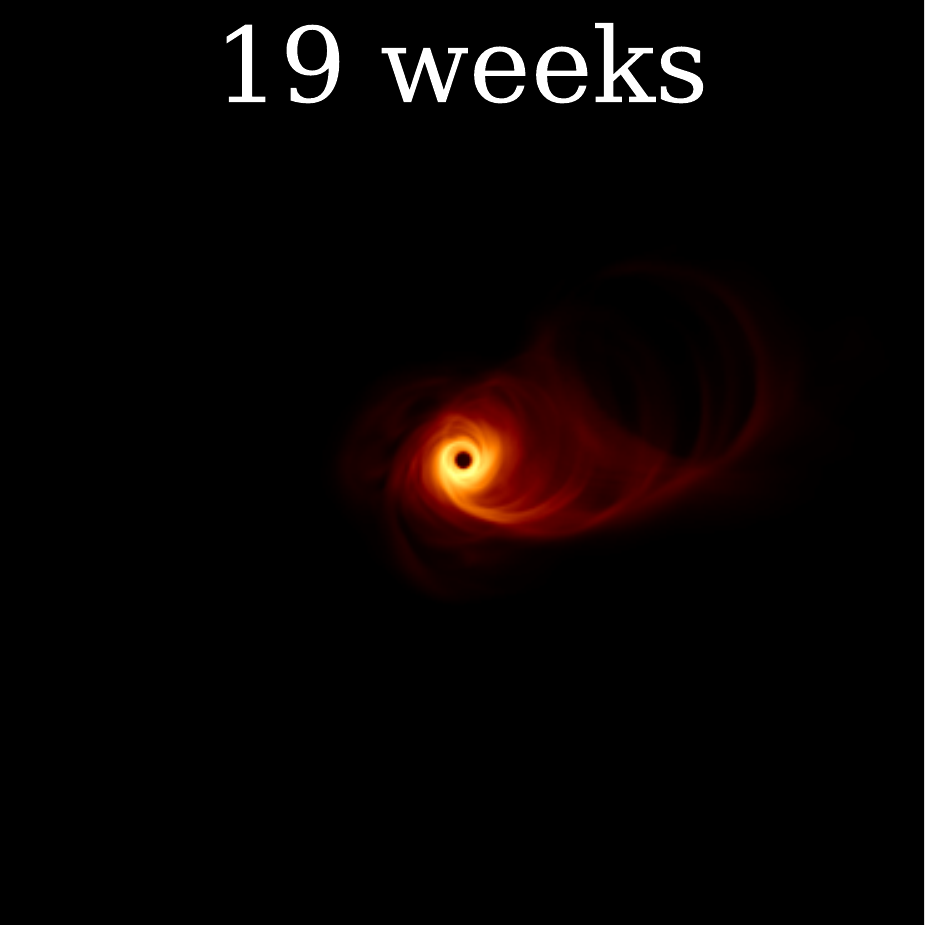} \\
\vspace{2mm}
\includegraphics[width=30mm]{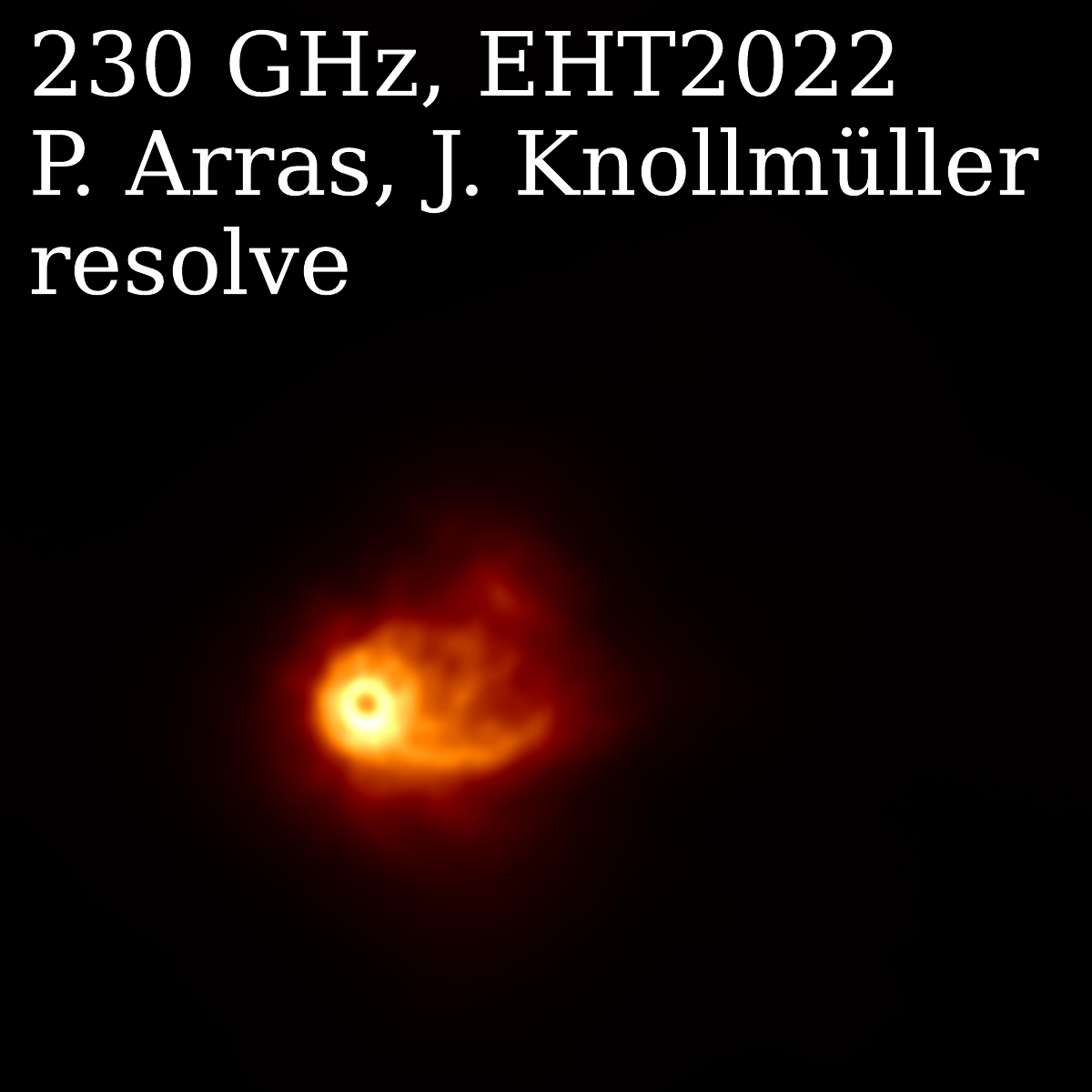}%
\includegraphics[width=30mm]{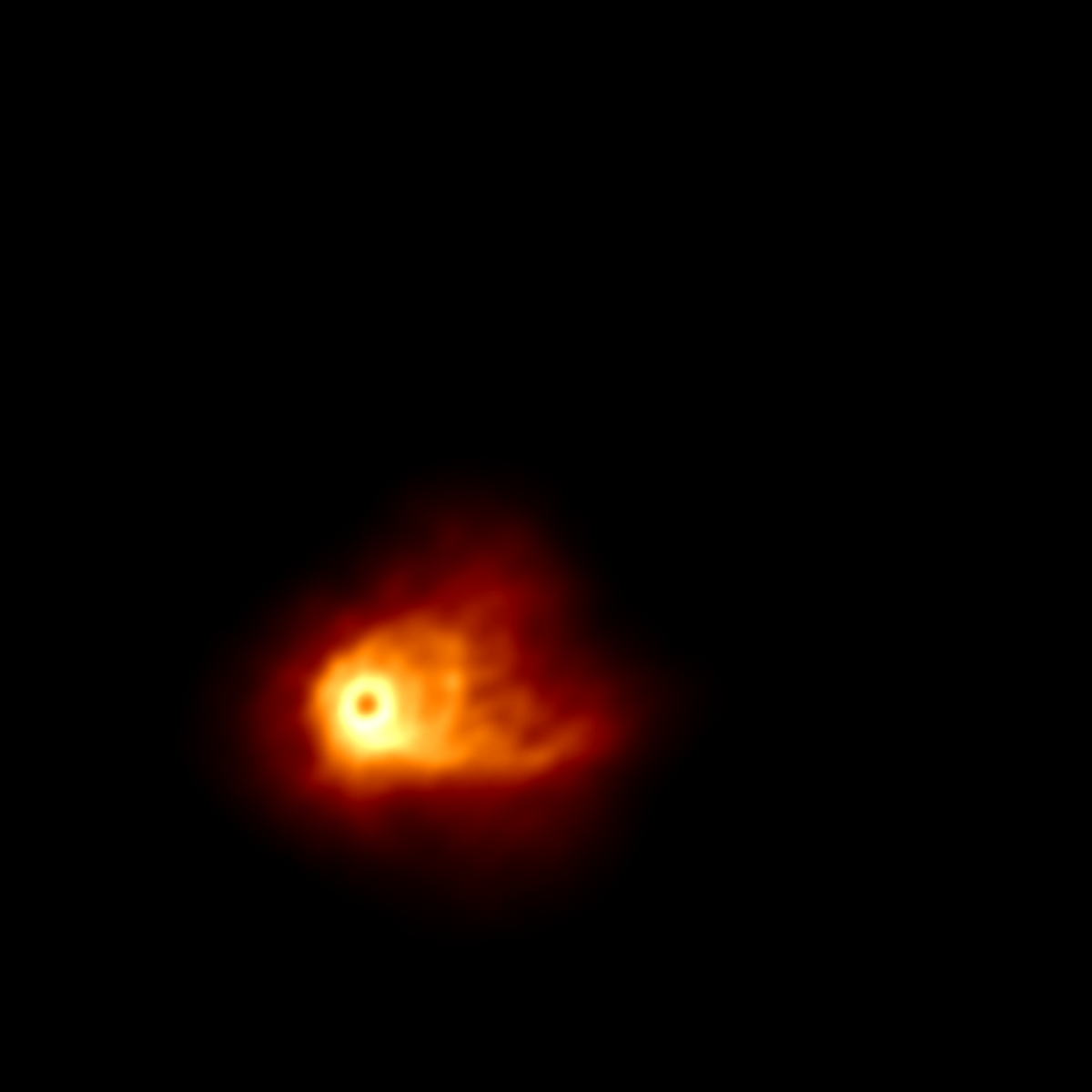}%
\includegraphics[width=30mm]{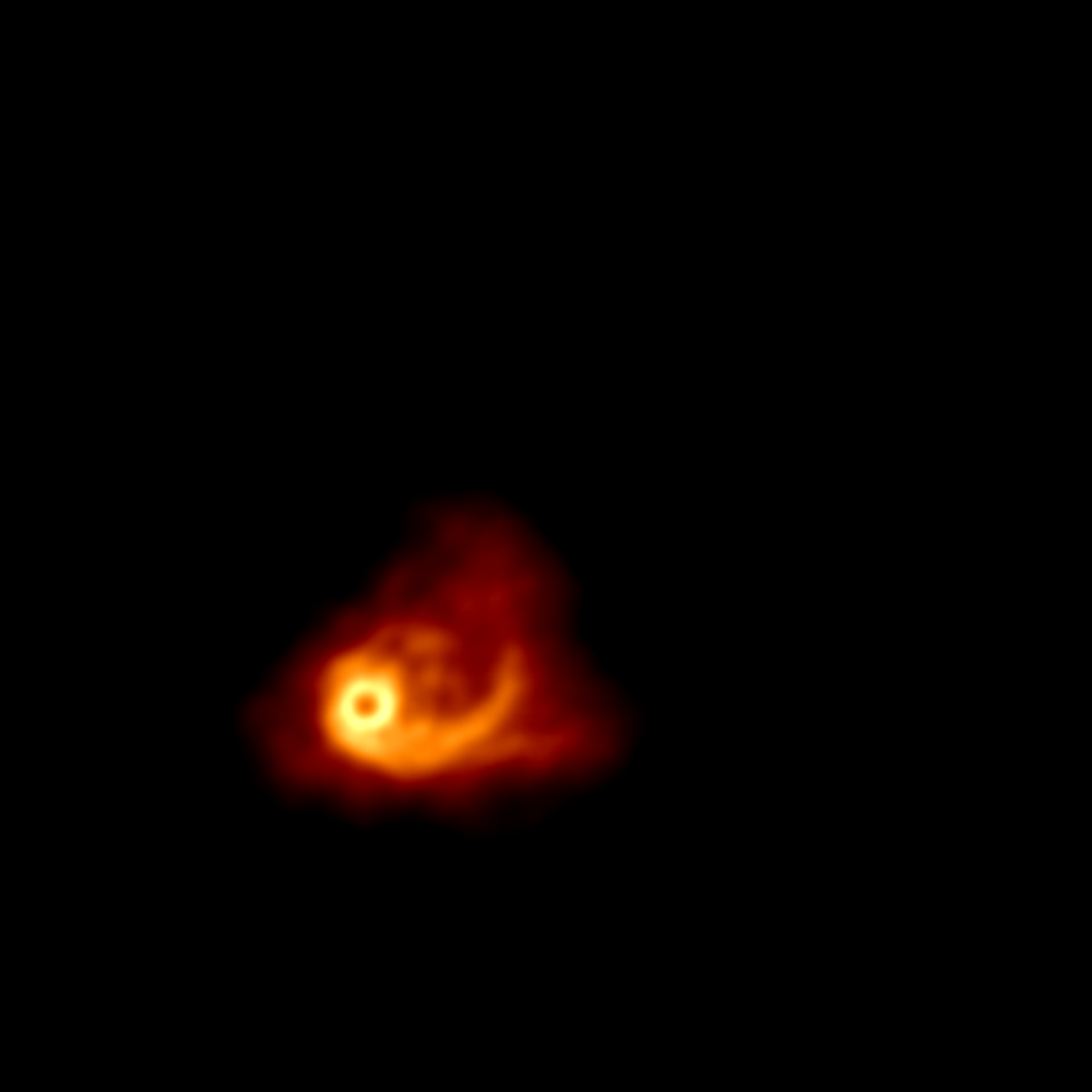}%
\includegraphics[width=30mm]{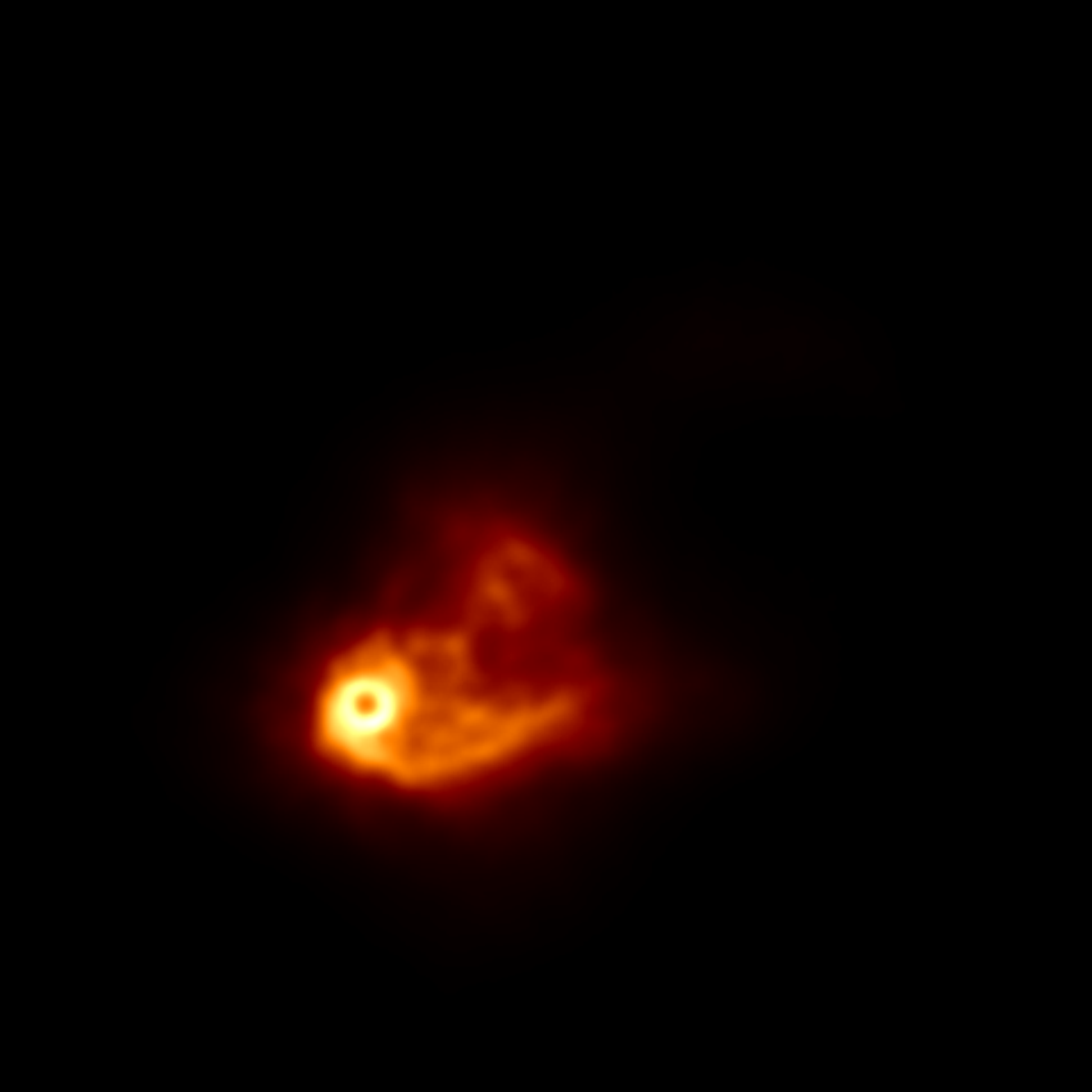}%
\includegraphics[width=30mm]{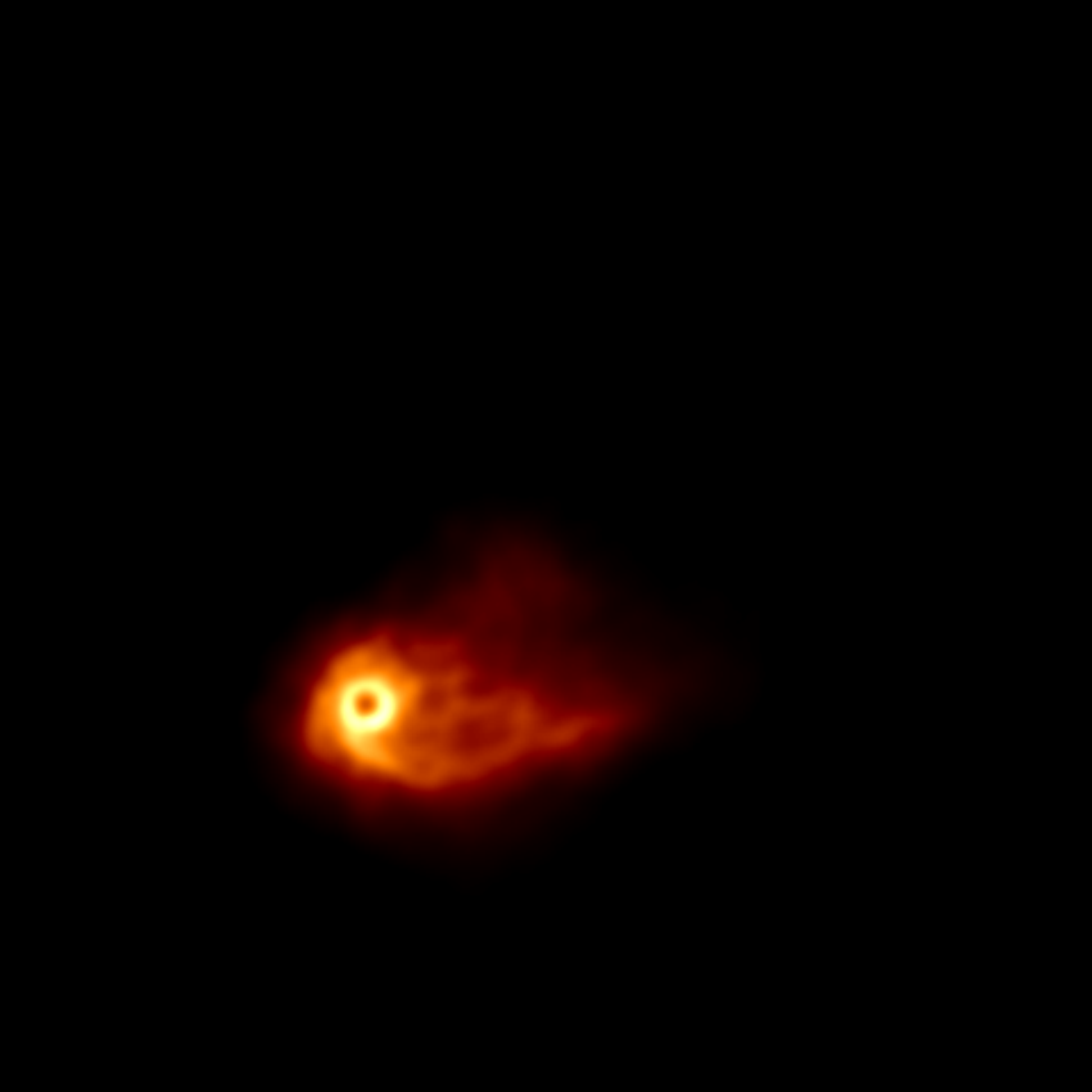}%
\includegraphics[width=30mm]{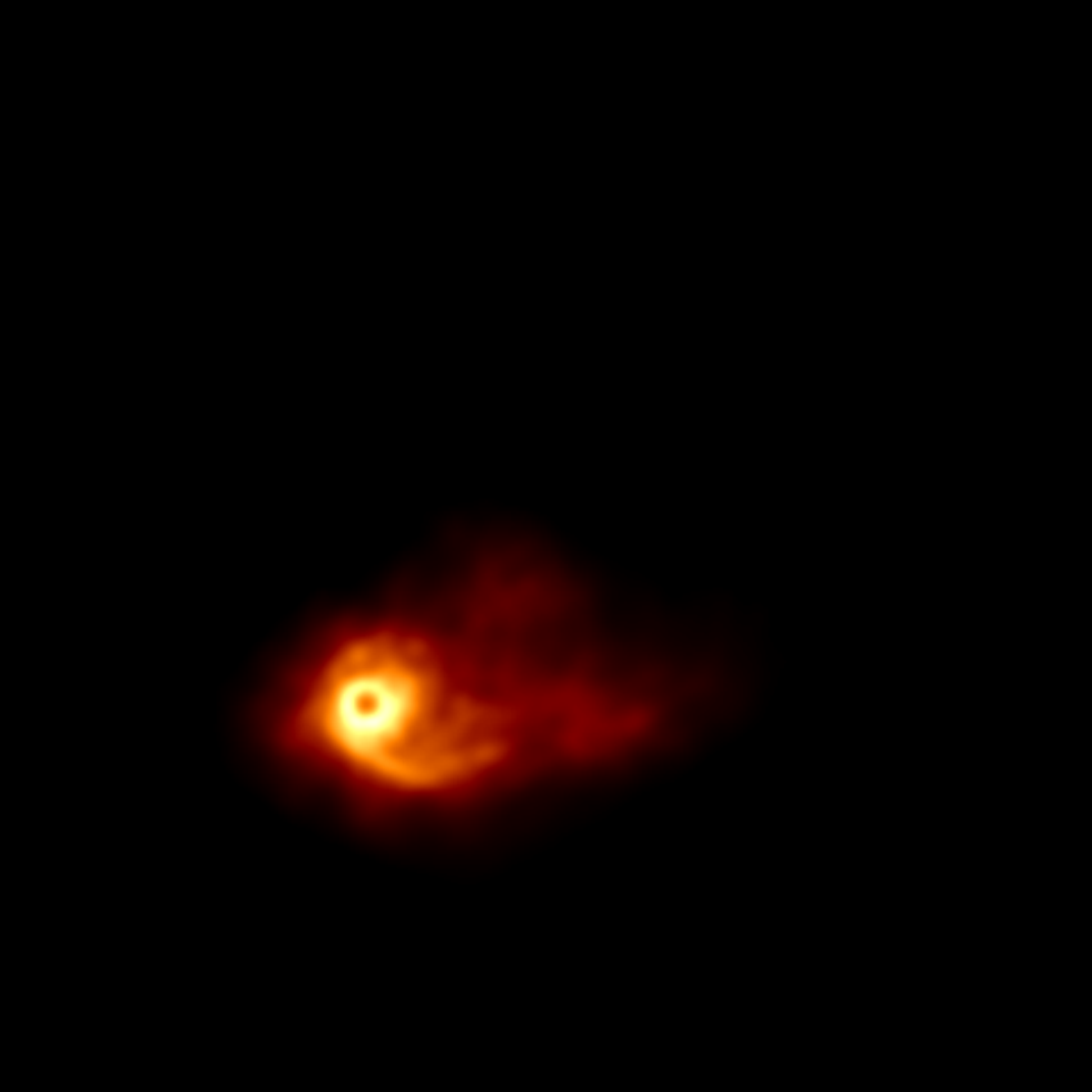} \\
\includegraphics[width=30mm]{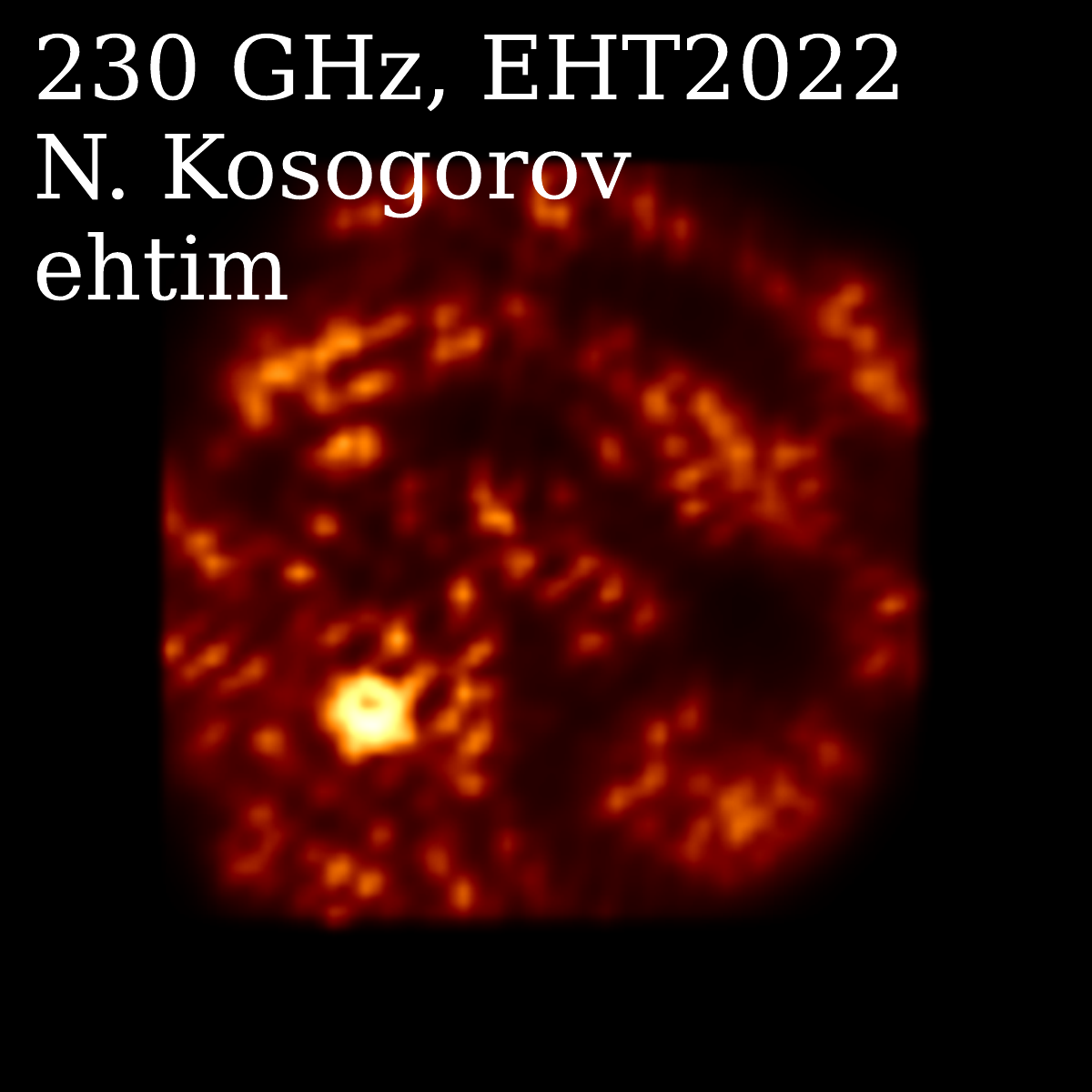}%
\includegraphics[width=30mm]{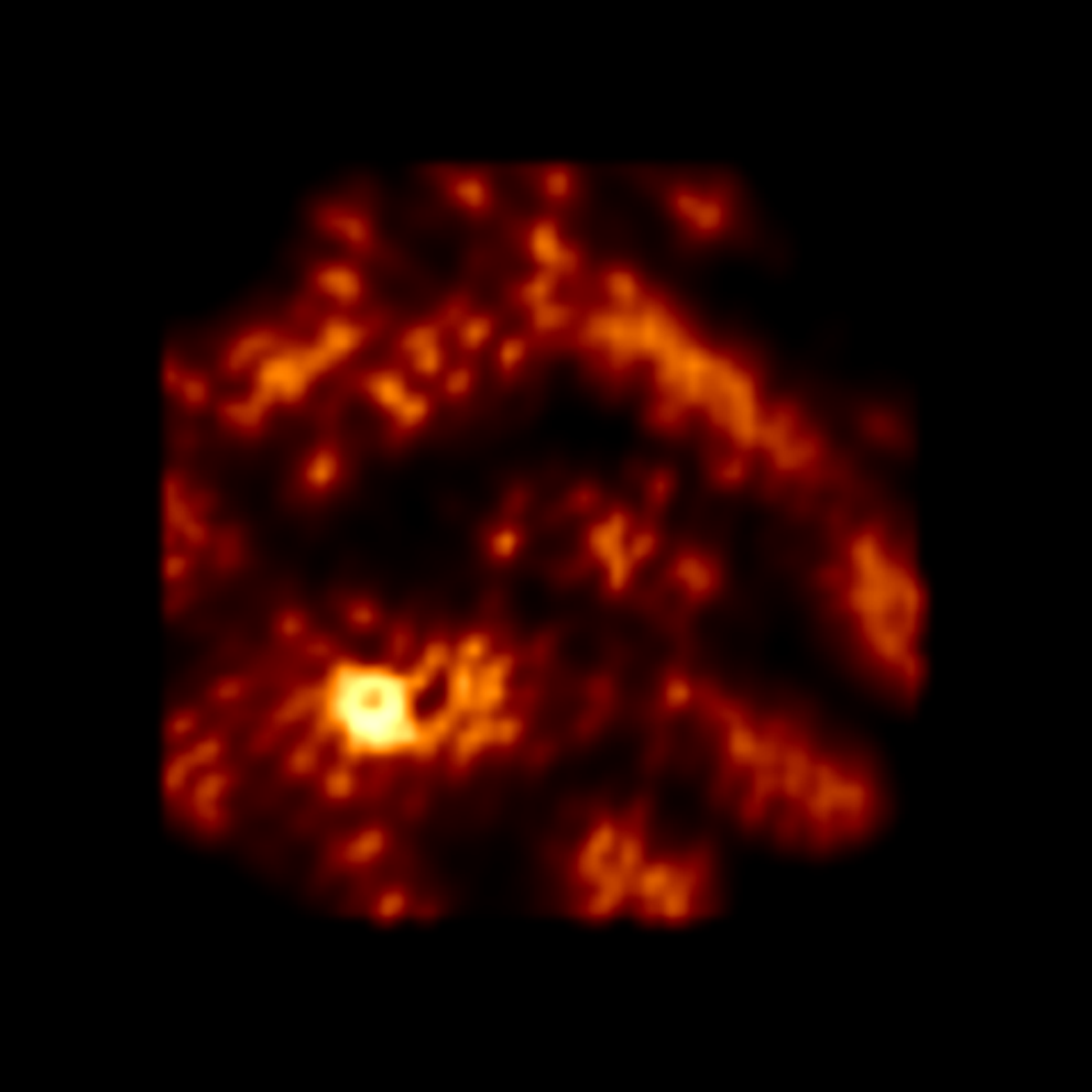}%
\includegraphics[width=30mm]{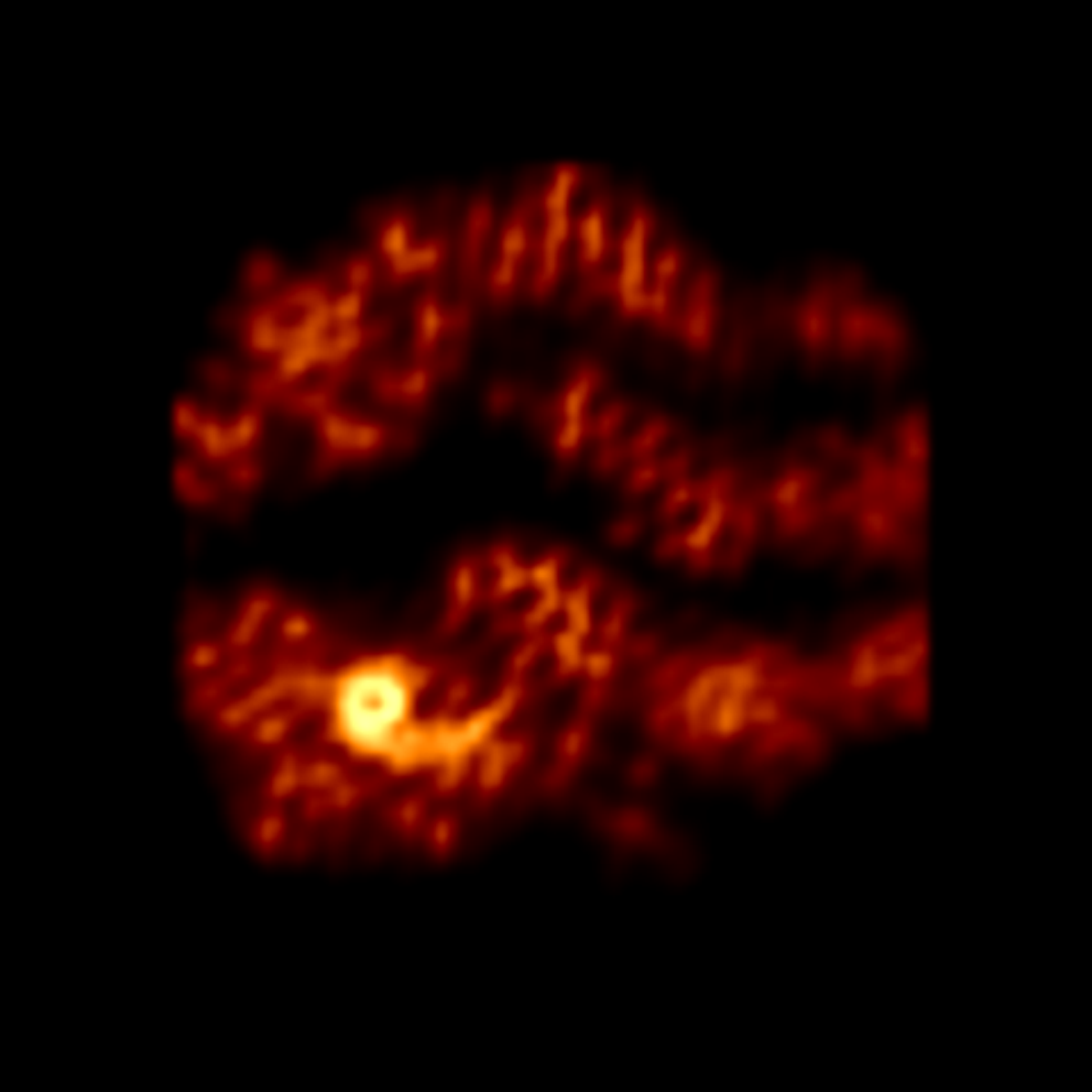}%
\includegraphics[width=30mm]{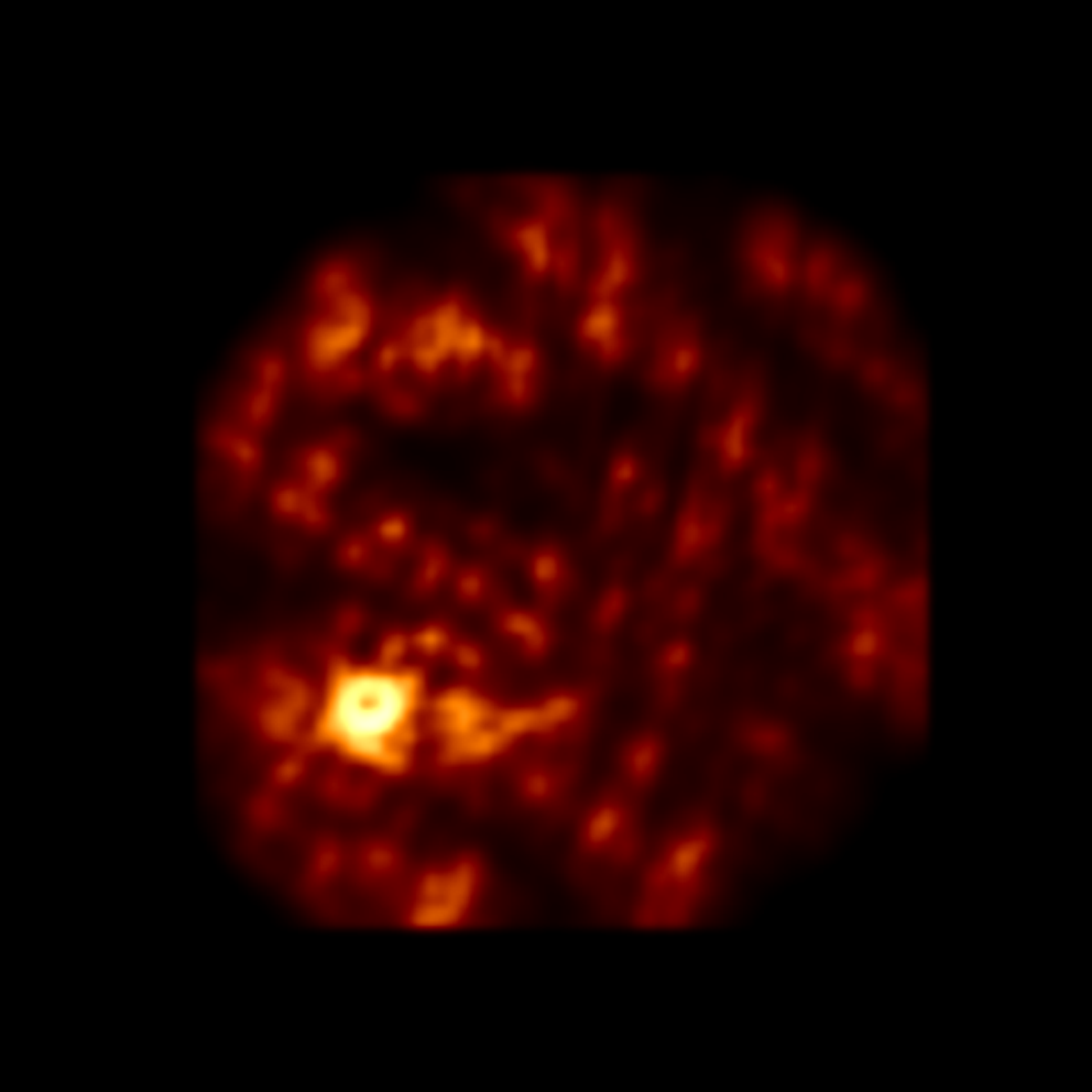}%
\includegraphics[width=30mm]{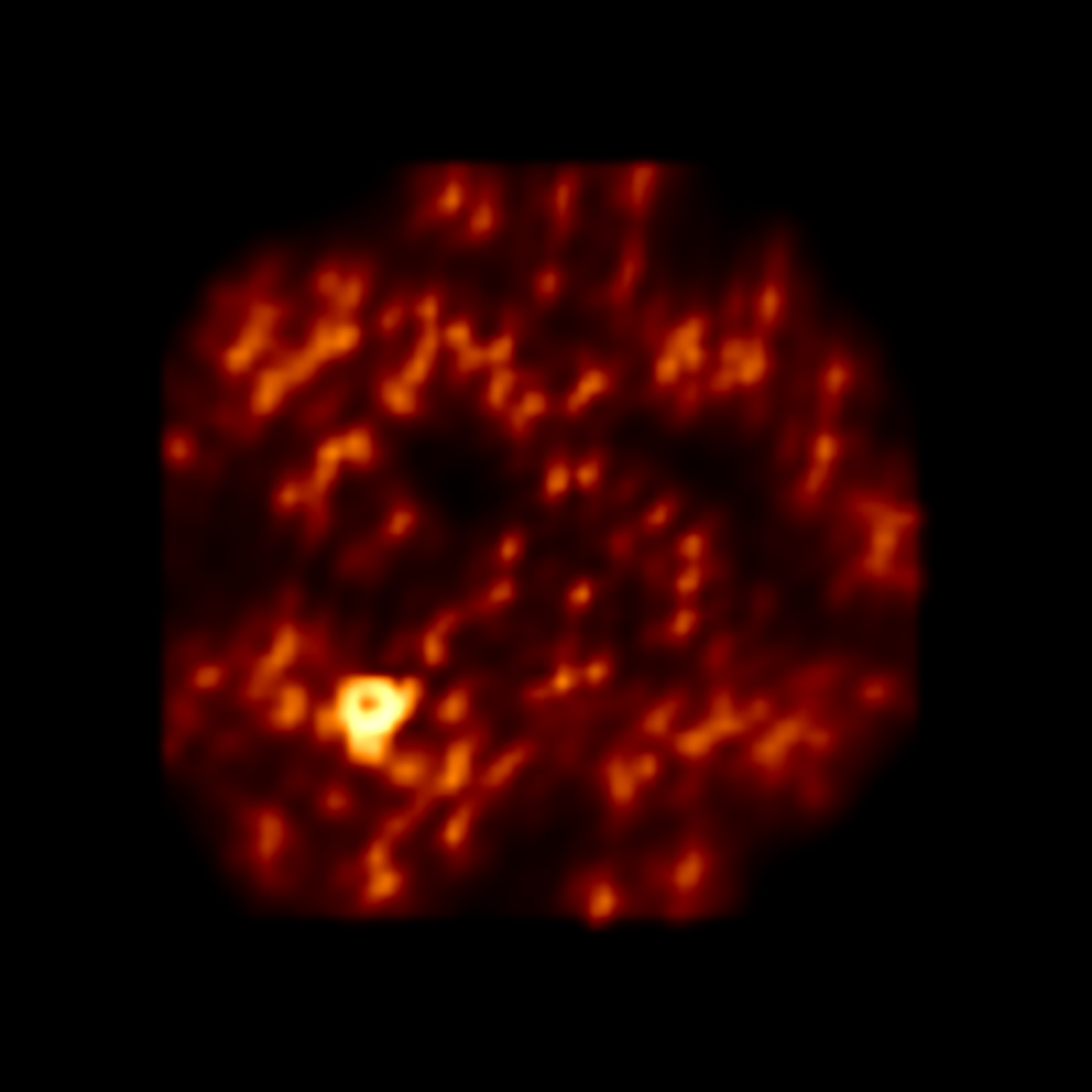}%
\includegraphics[width=30mm]{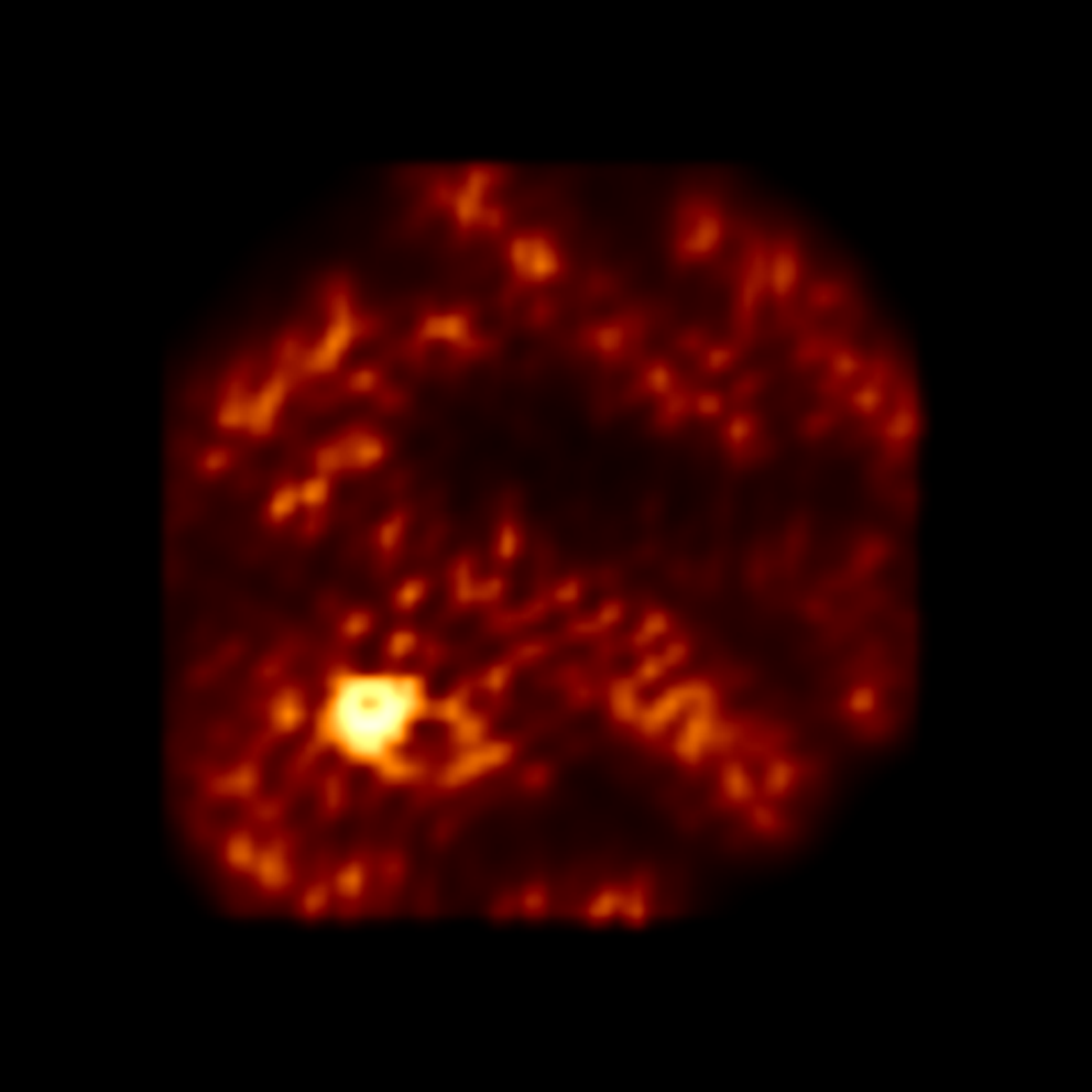} \\
\vspace{2mm}
\includegraphics[width=30mm]{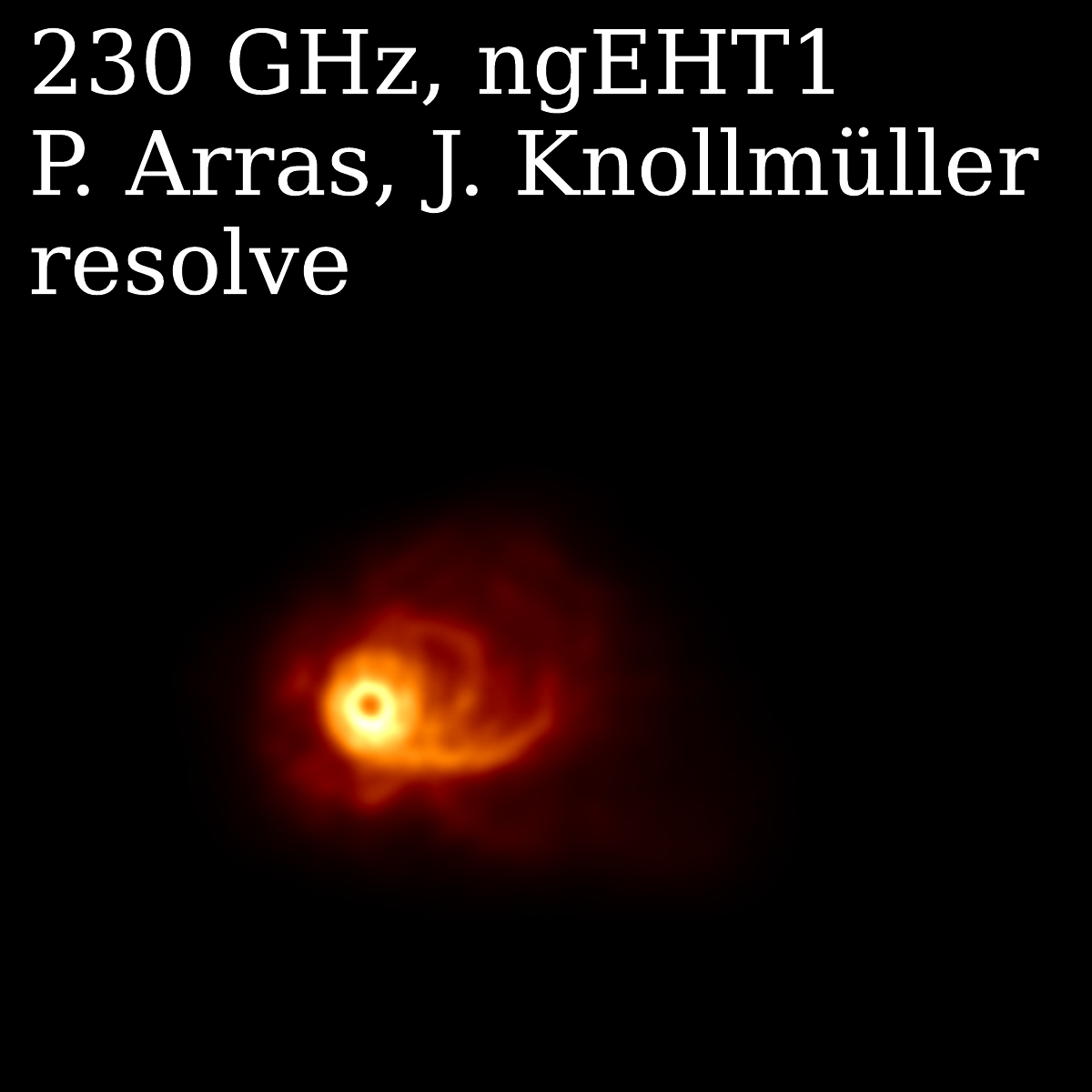}%
\includegraphics[width=30mm]{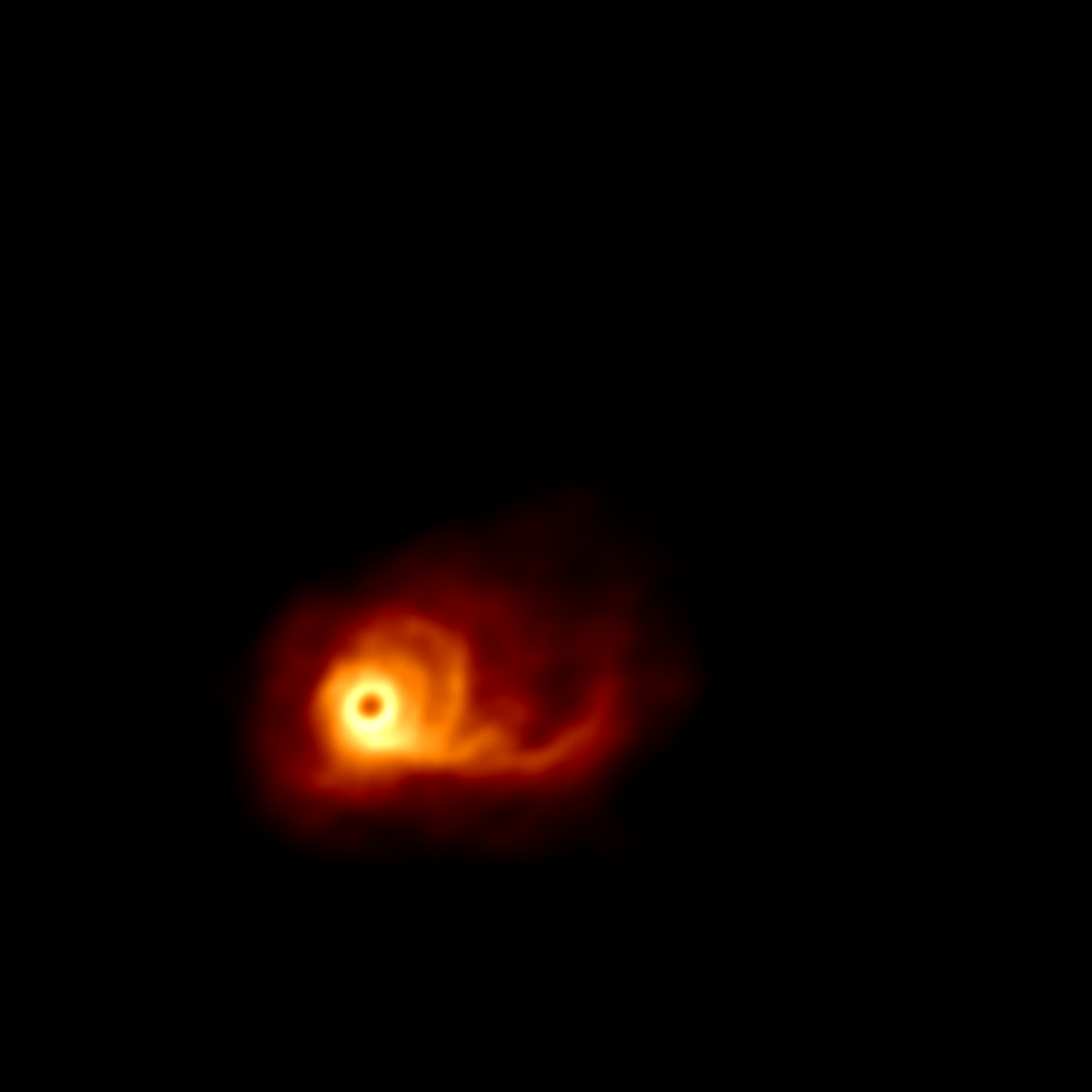}%
\includegraphics[width=30mm]{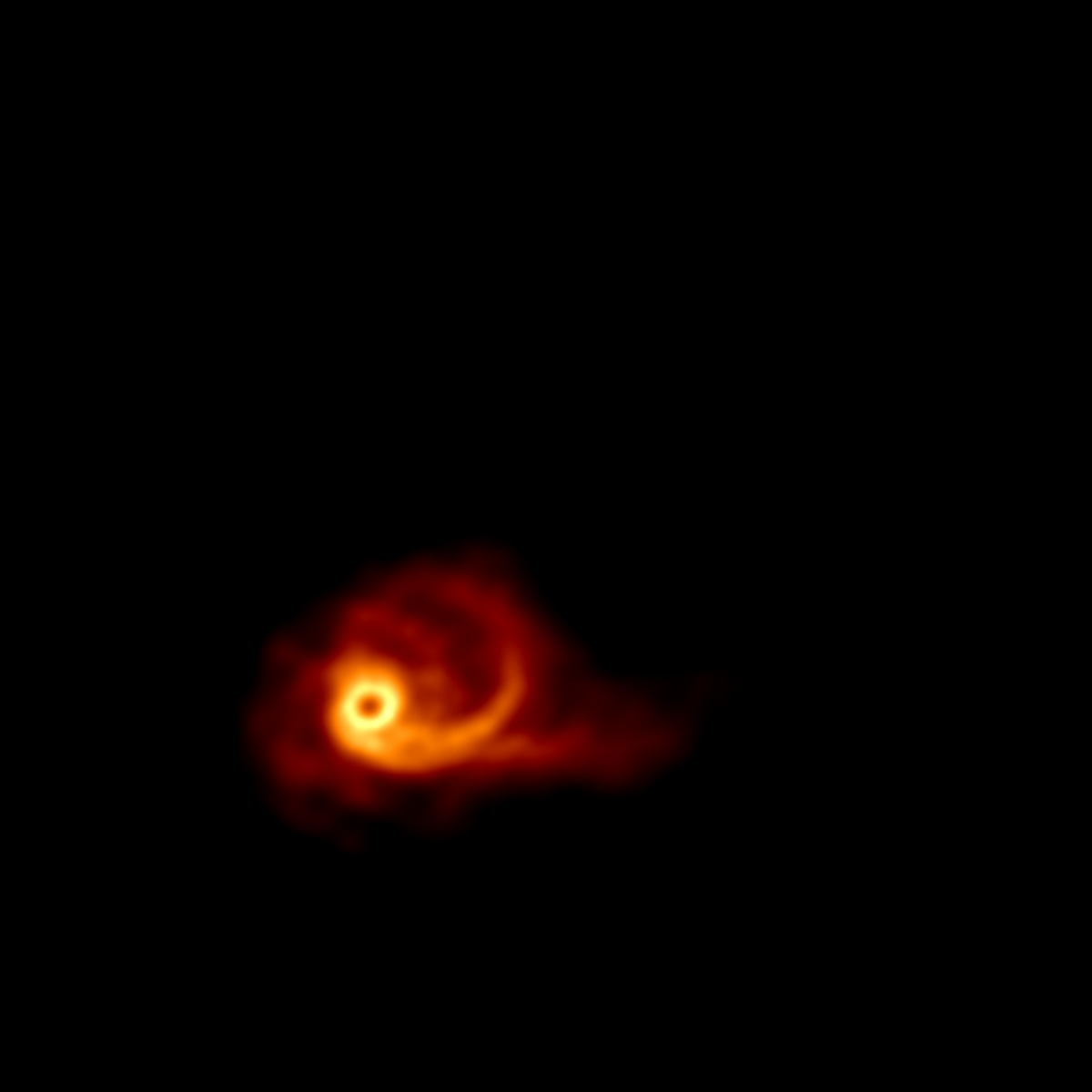}%
\includegraphics[width=30mm]{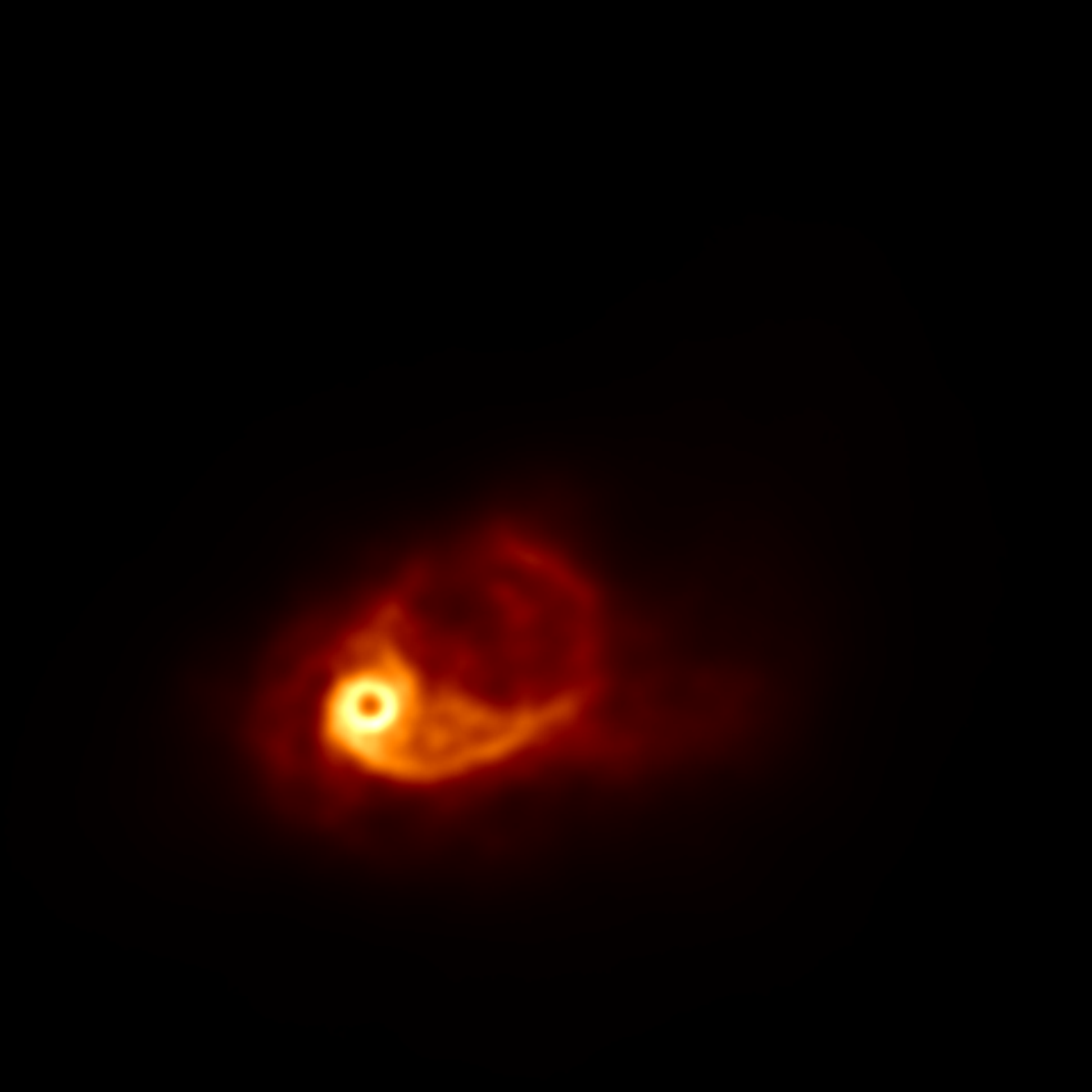}%
\includegraphics[width=30mm]{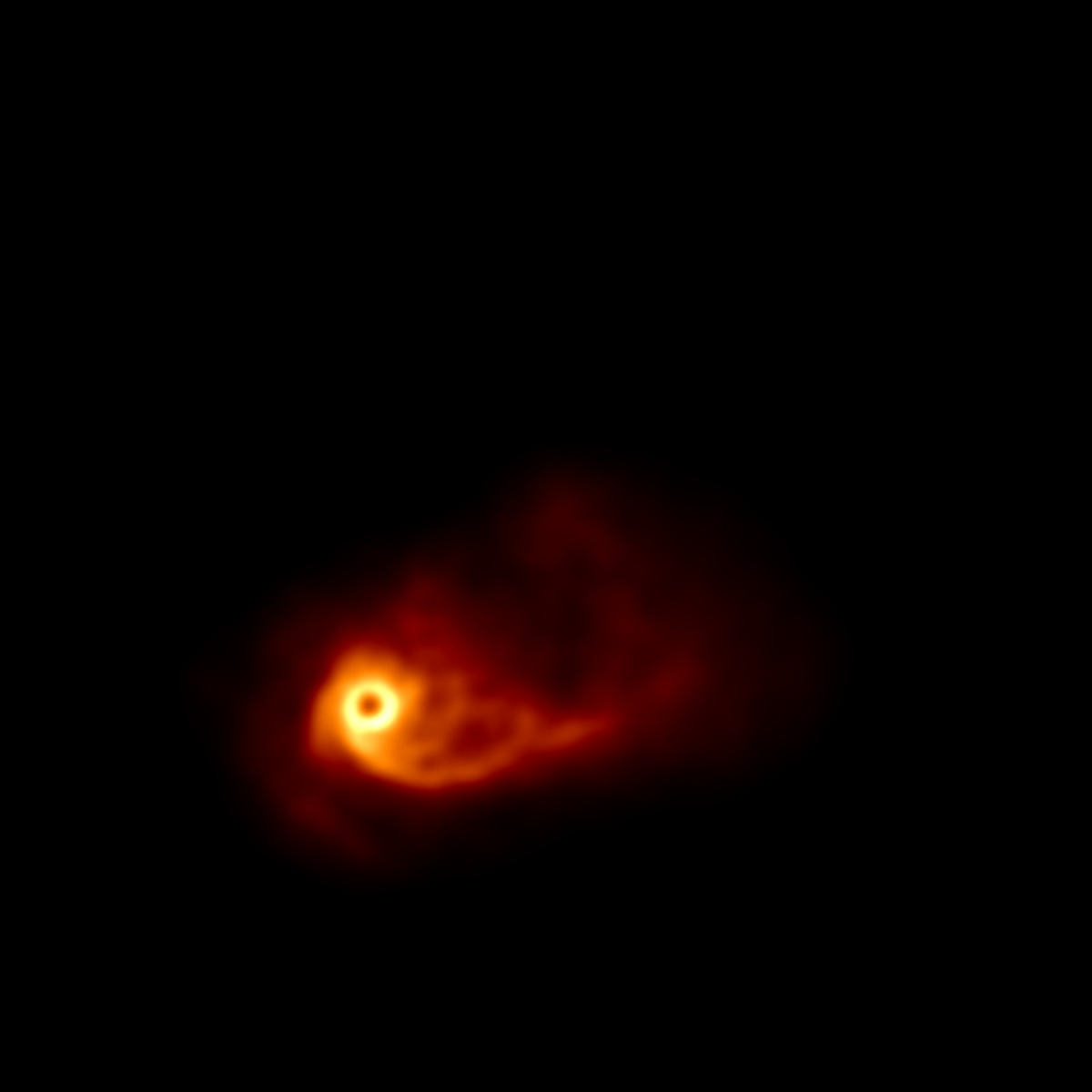}%
\includegraphics[width=30mm]{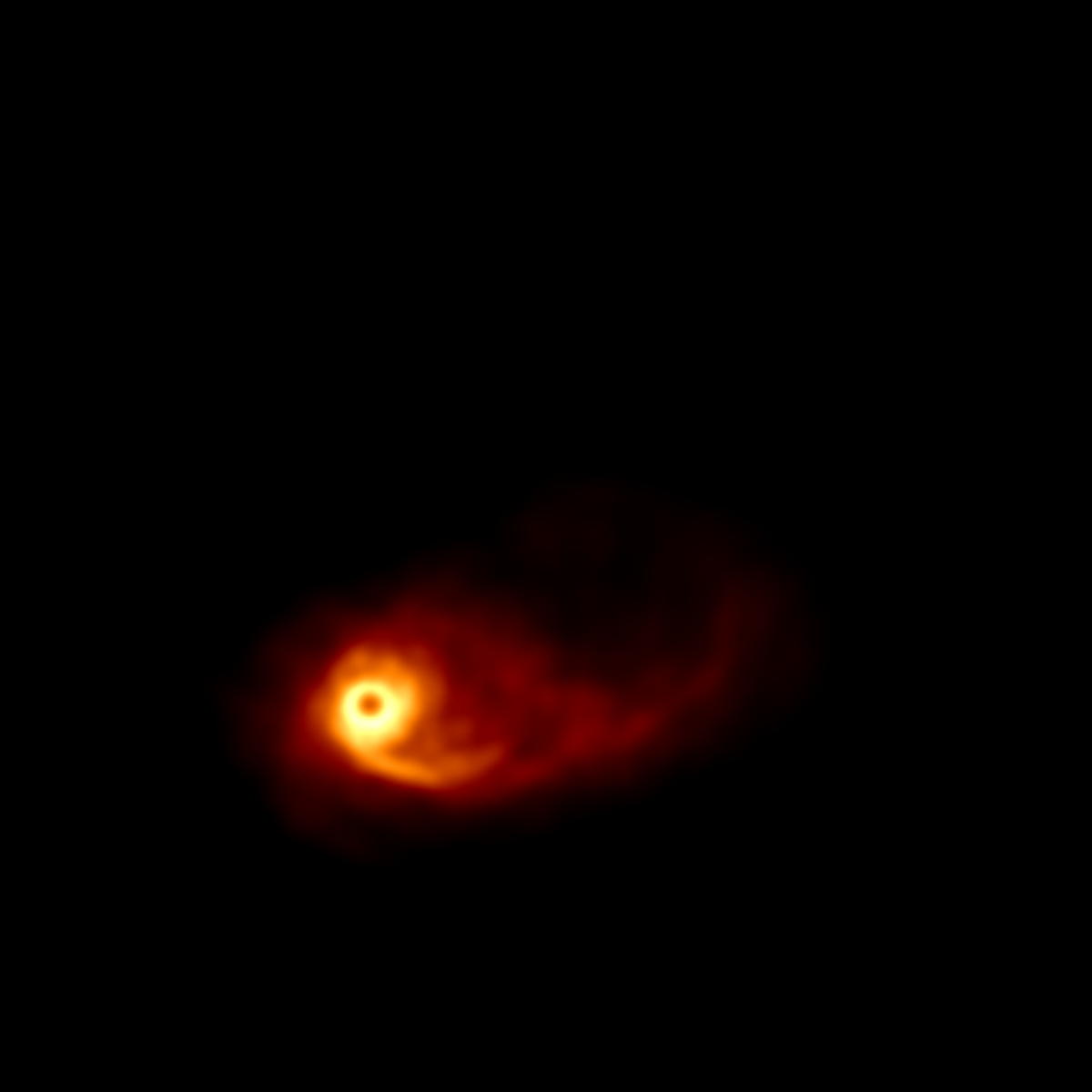} \\
\includegraphics[width=30mm]{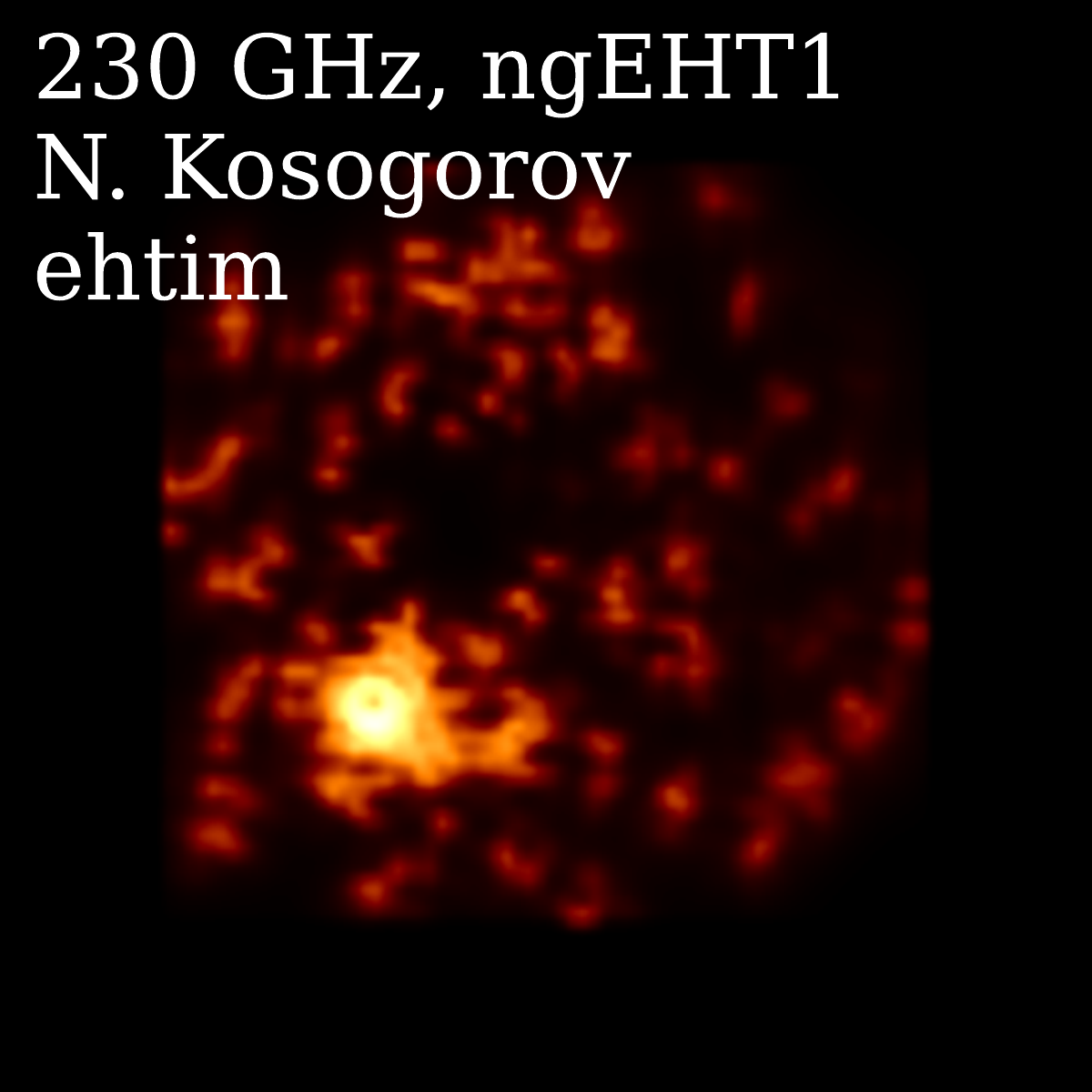}%
\includegraphics[width=30mm]{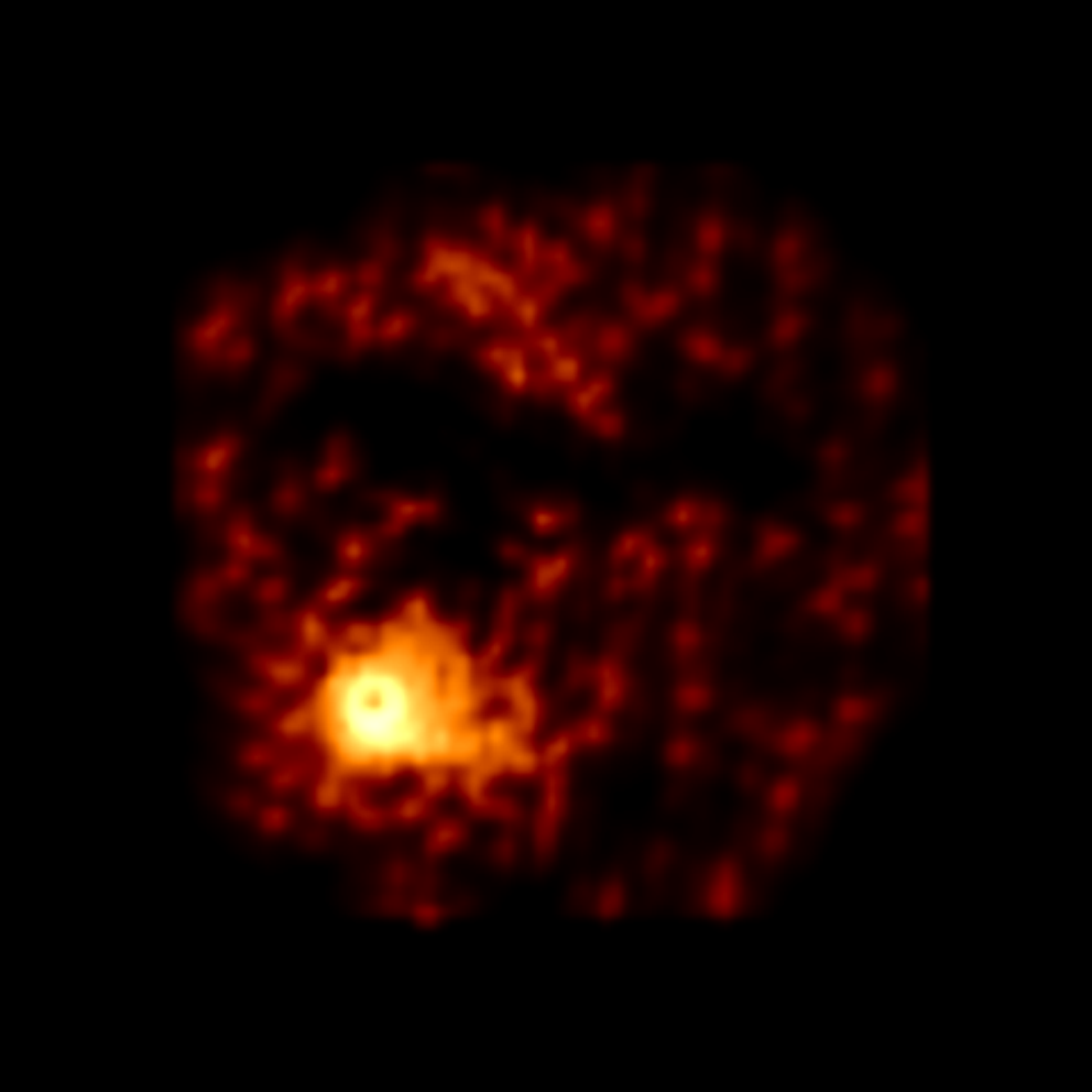}%
\includegraphics[width=30mm]{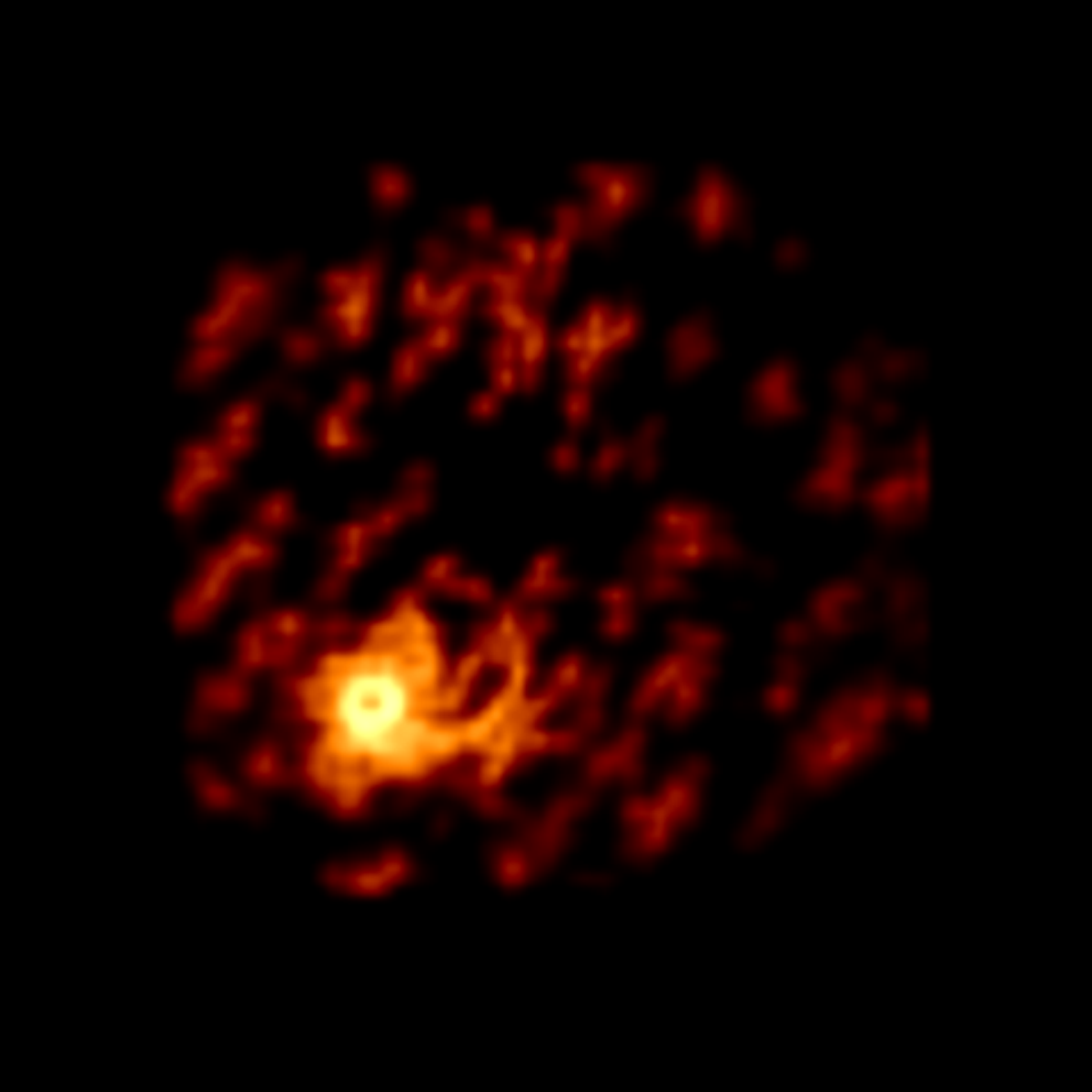}%
\includegraphics[width=30mm]{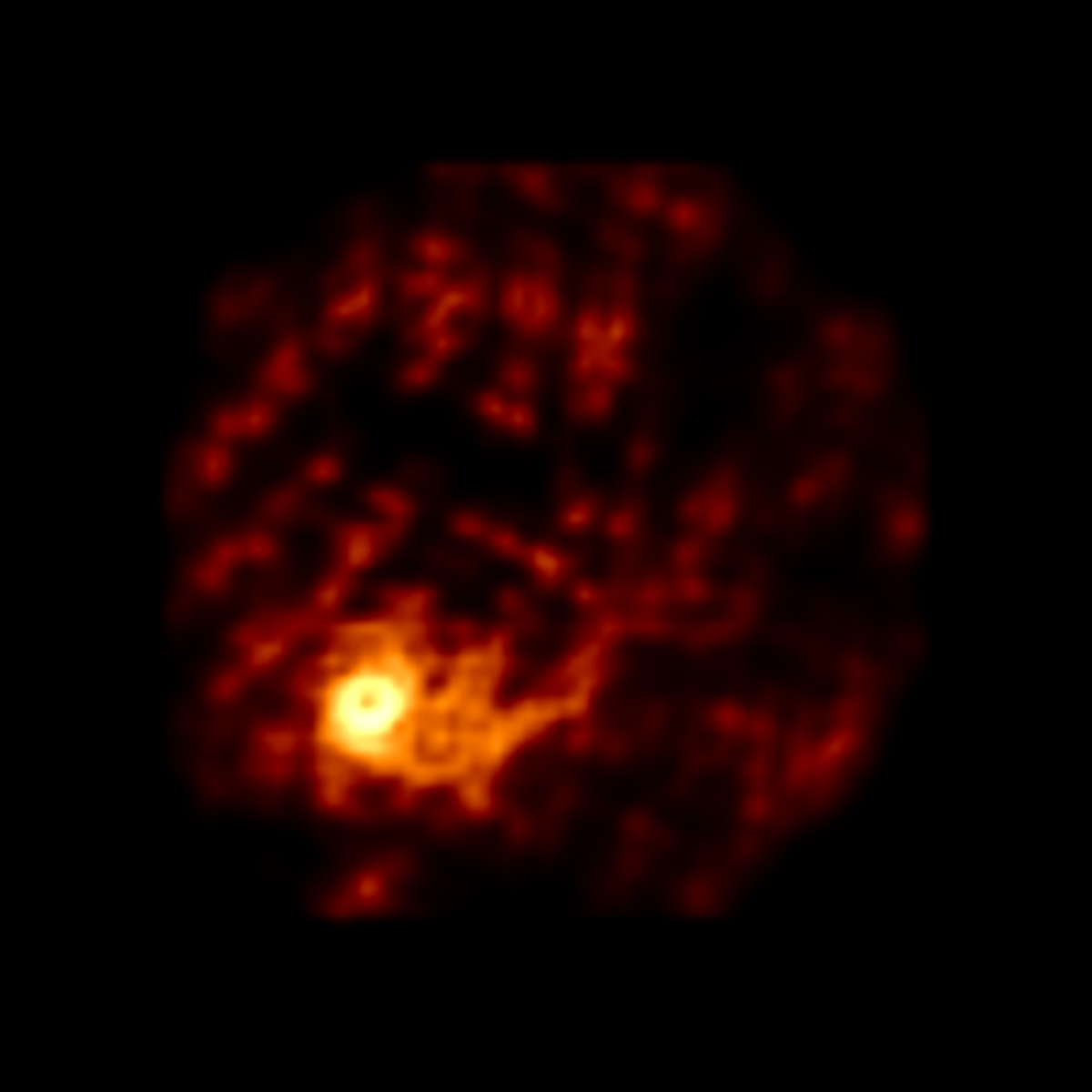}%
\includegraphics[width=30mm]{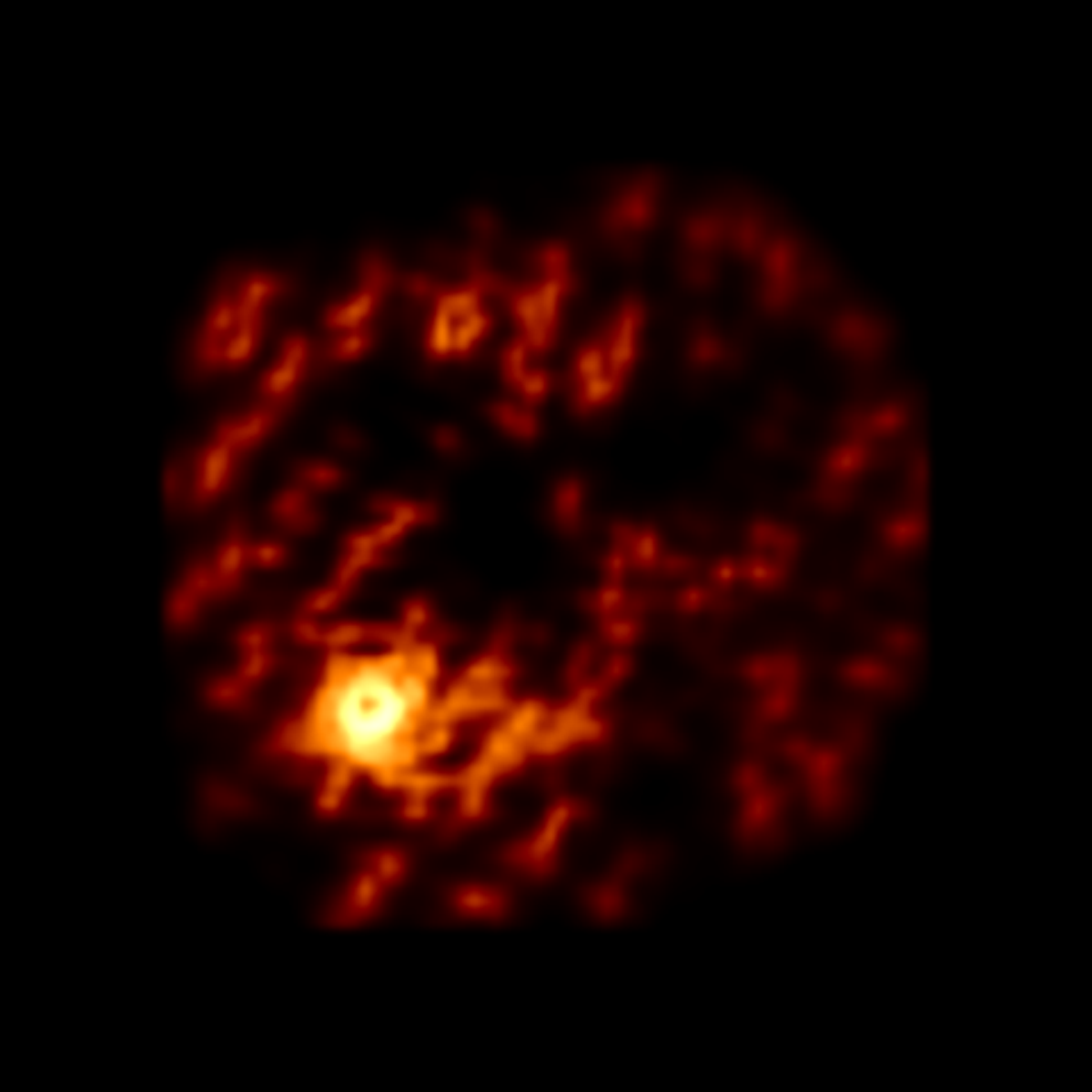}%
\includegraphics[width=30mm]{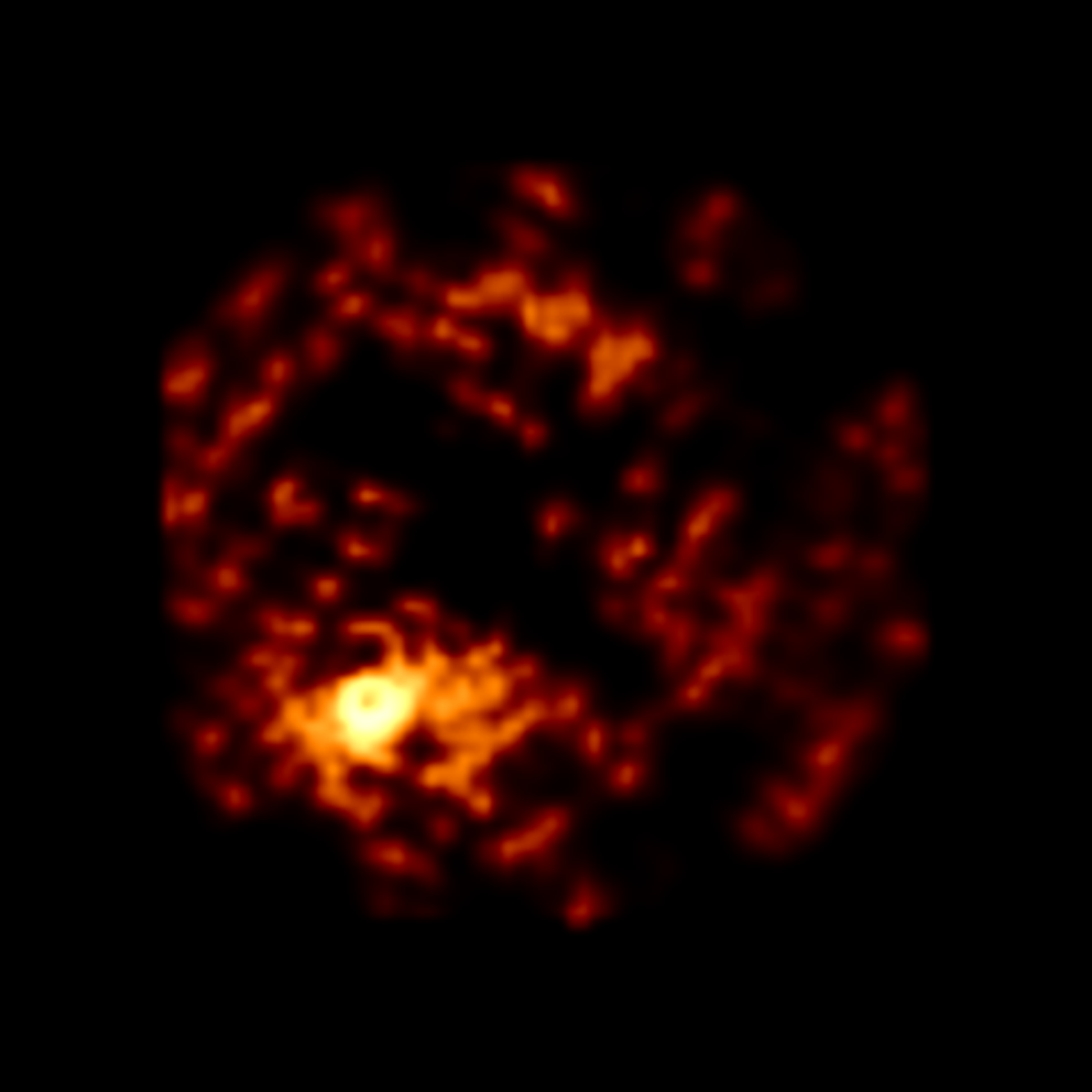} \\
\includegraphics[width=30mm]{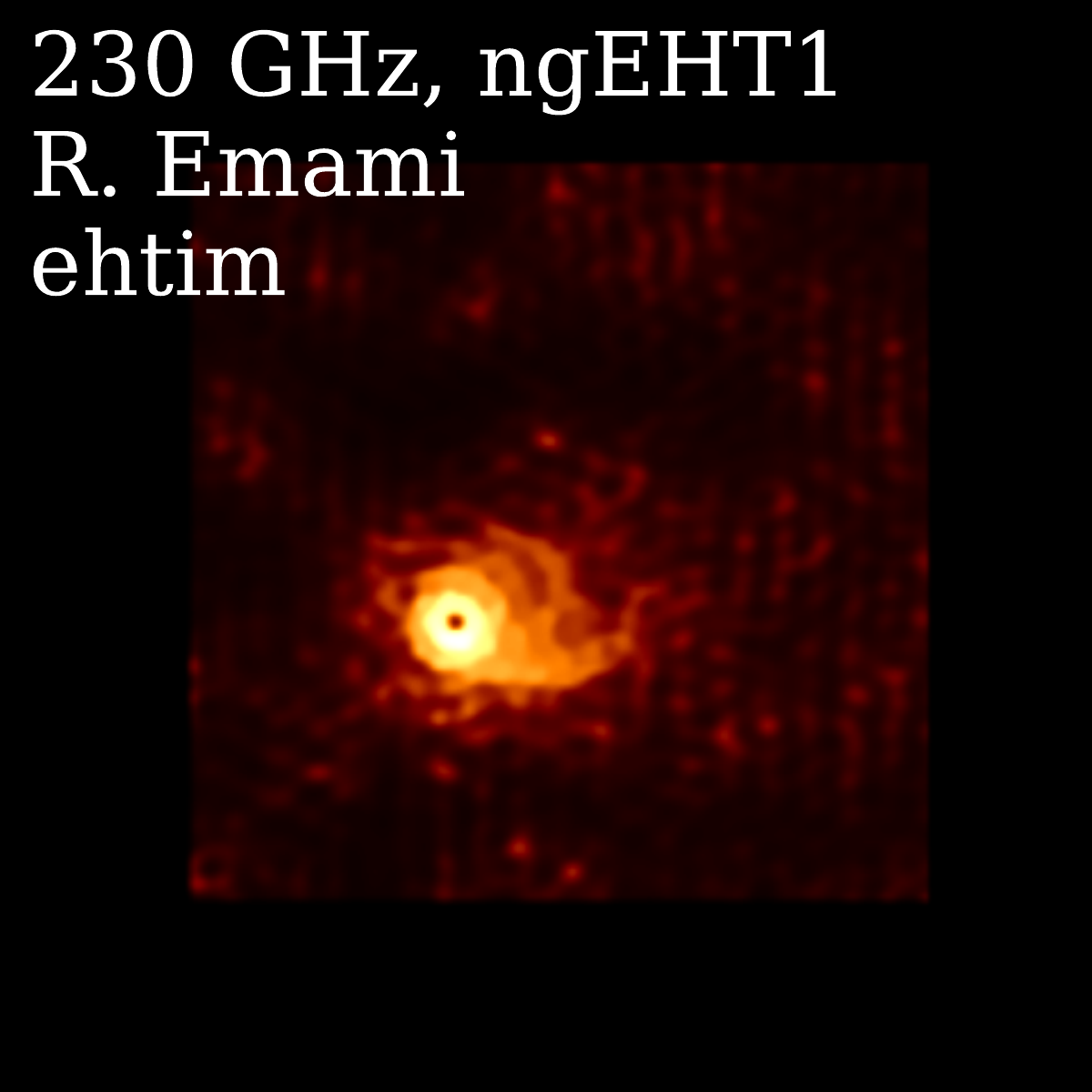}%
\includegraphics[width=30mm]{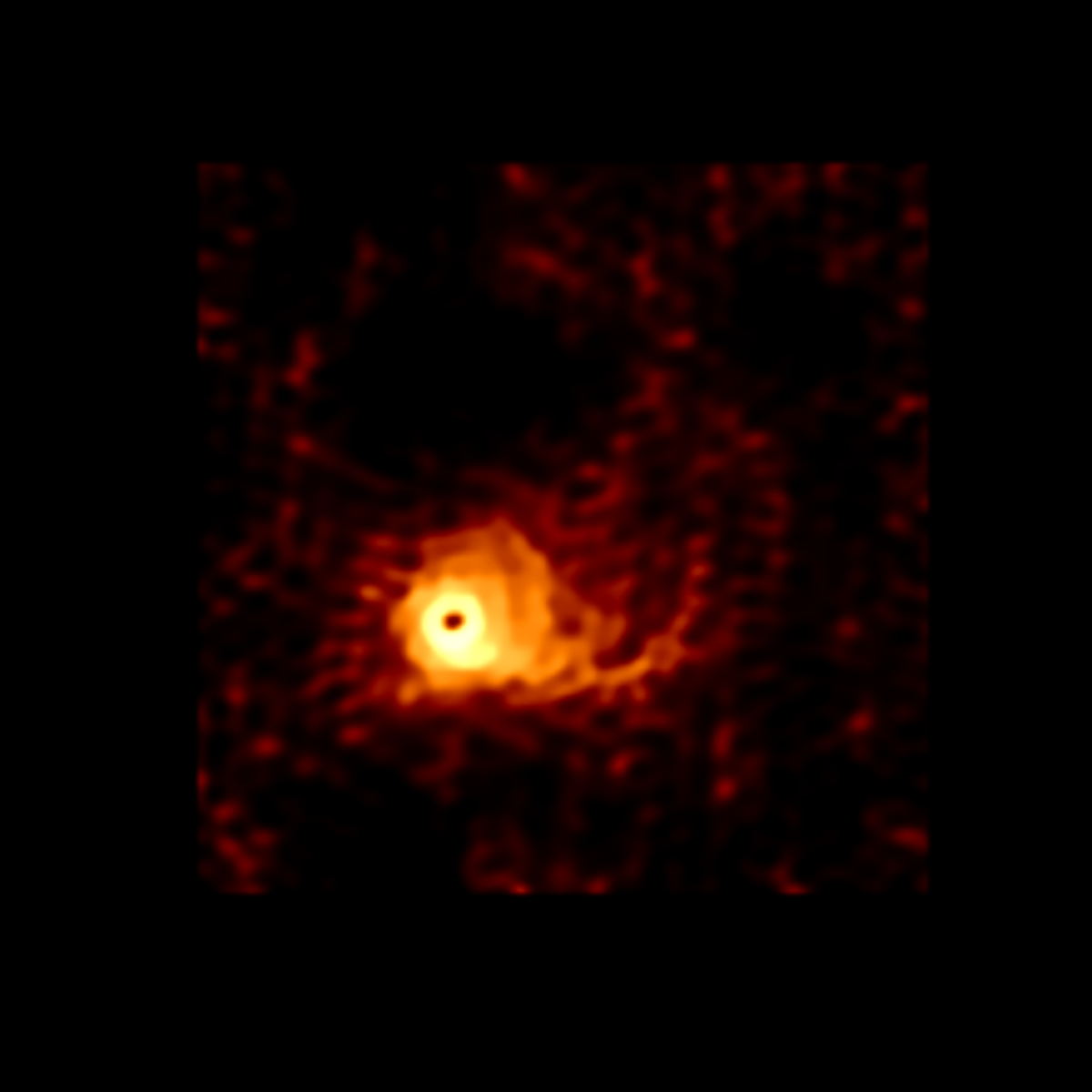}%
\includegraphics[width=30mm]{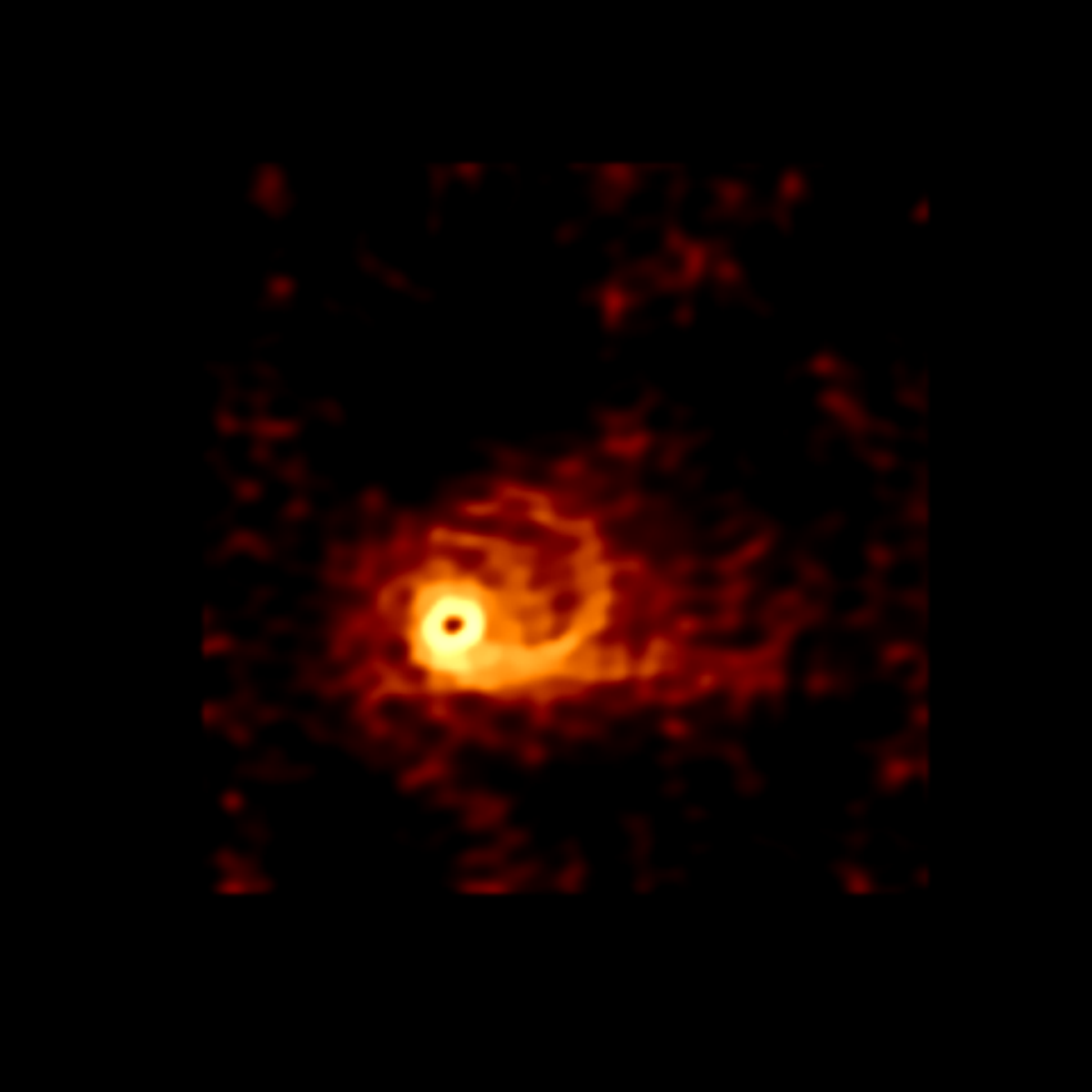}%
\includegraphics[width=30mm]{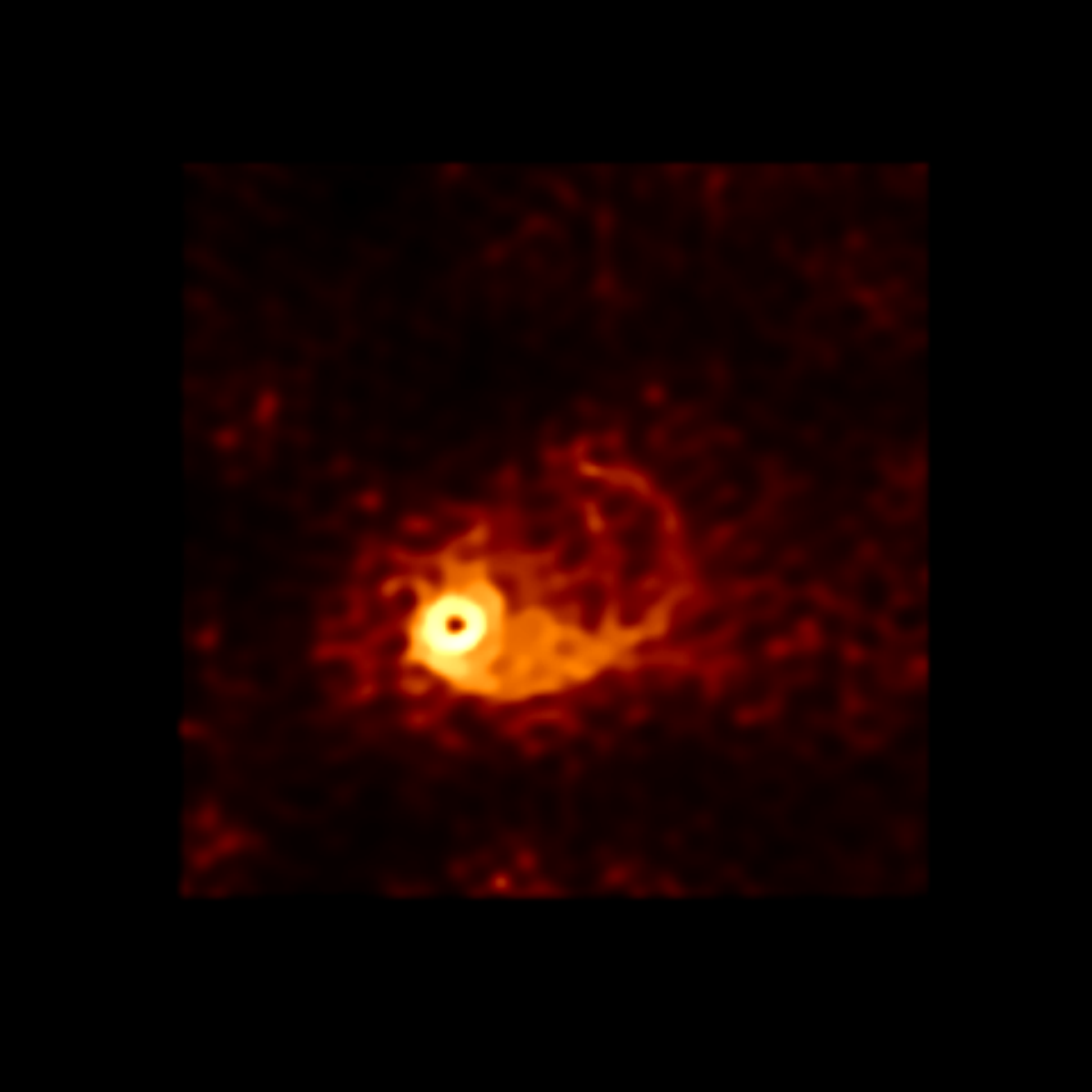}%
\includegraphics[width=30mm]{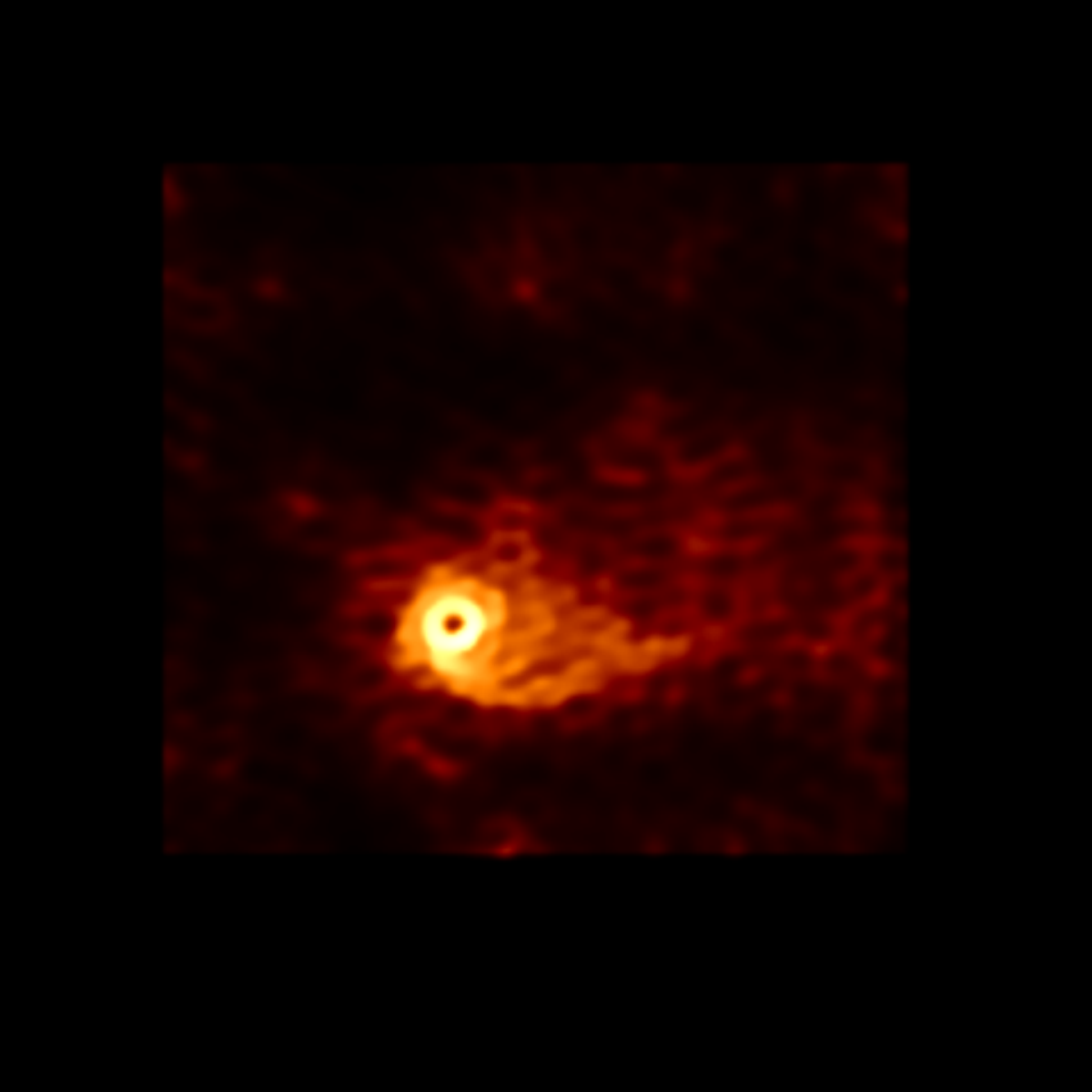}%
\includegraphics[width=30mm]{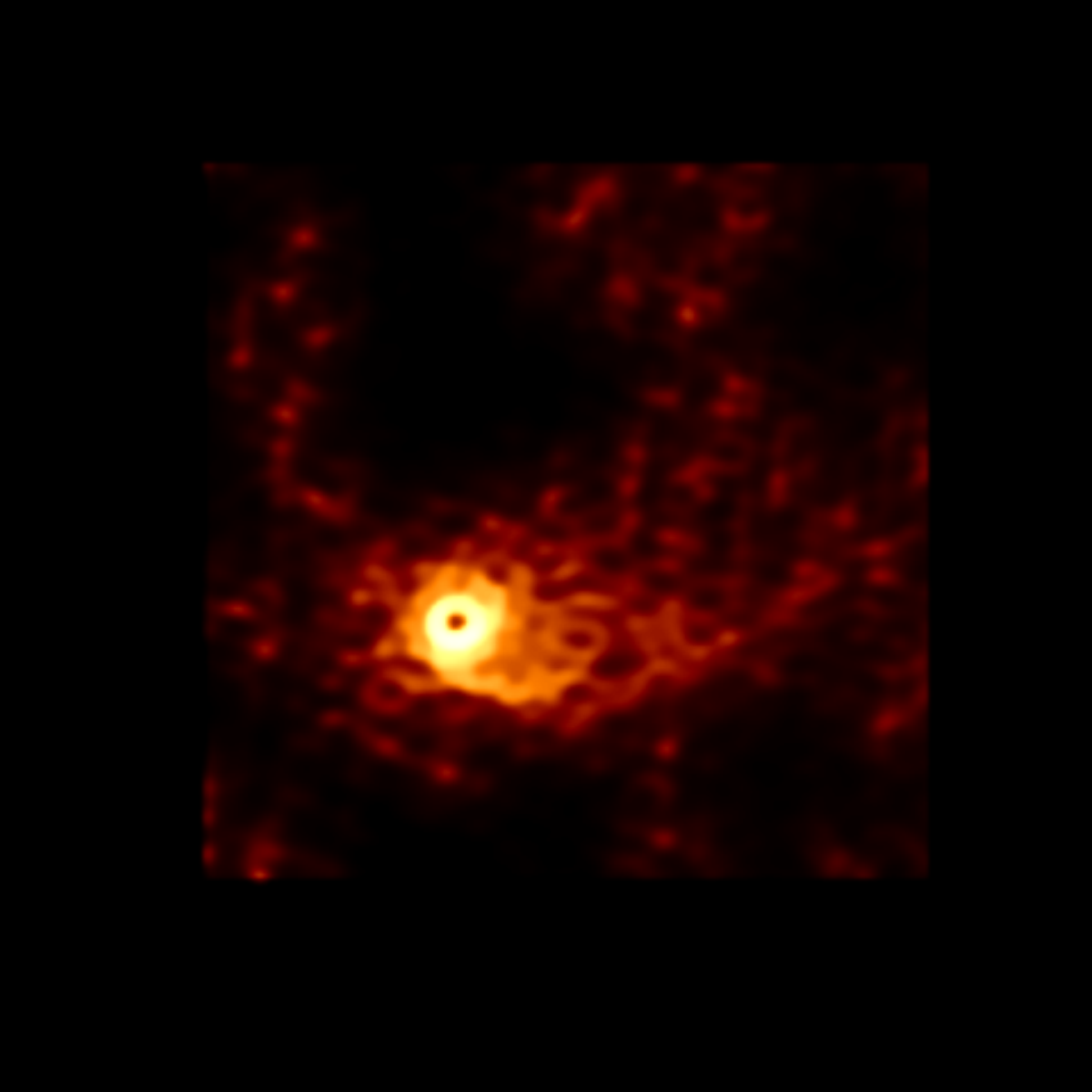} \\
\includegraphics[width=30mm]{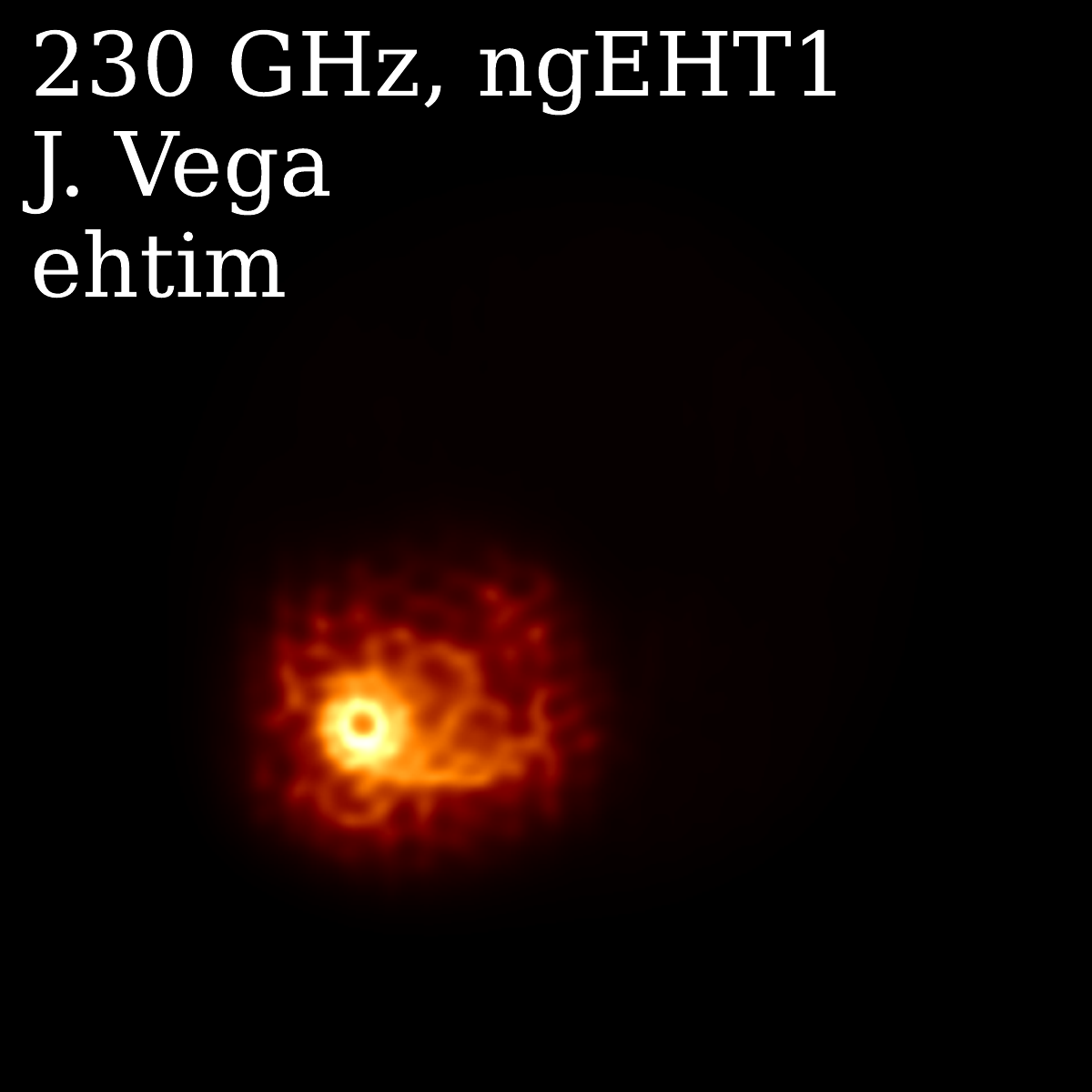}%
\includegraphics[width=30mm]{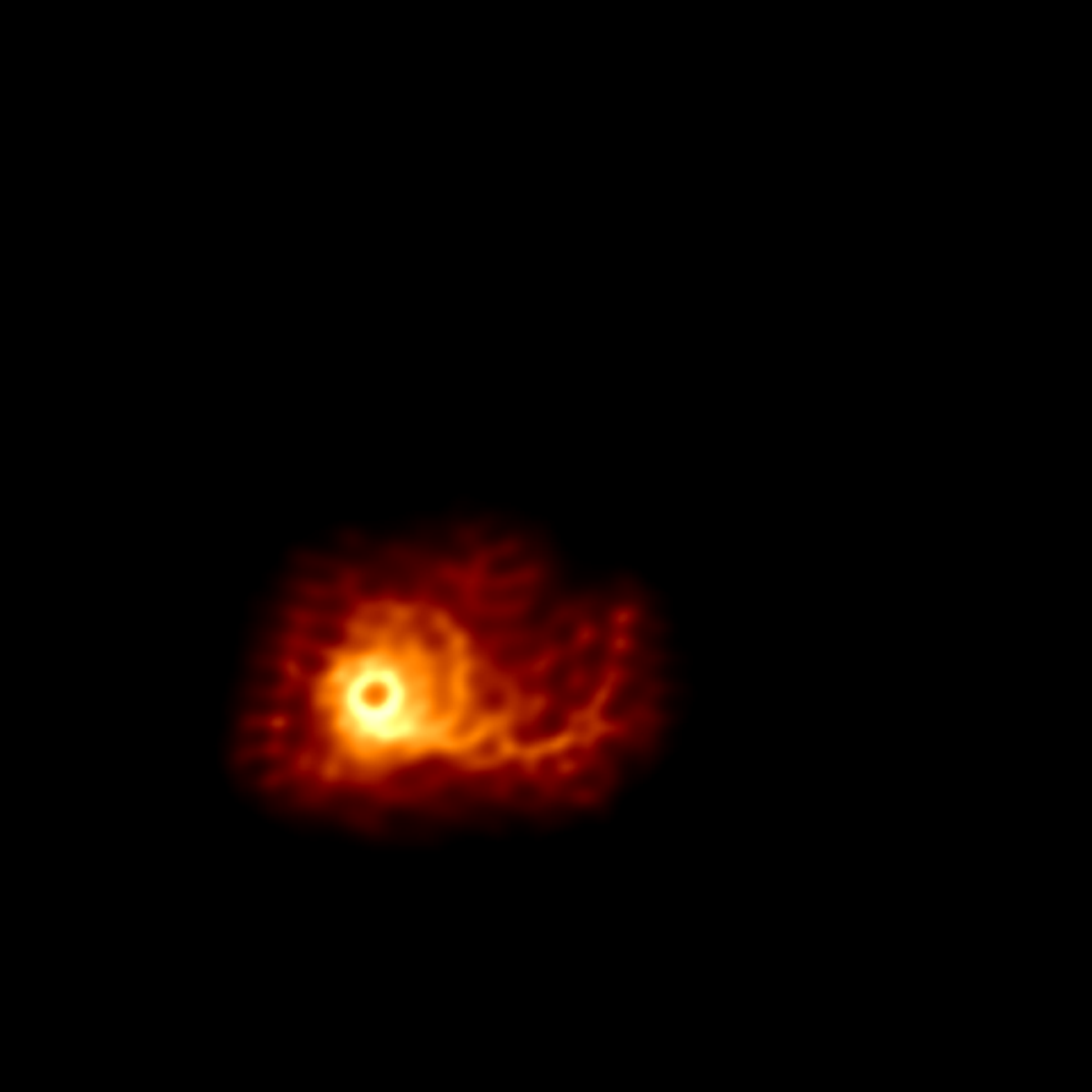}%
\includegraphics[width=30mm]{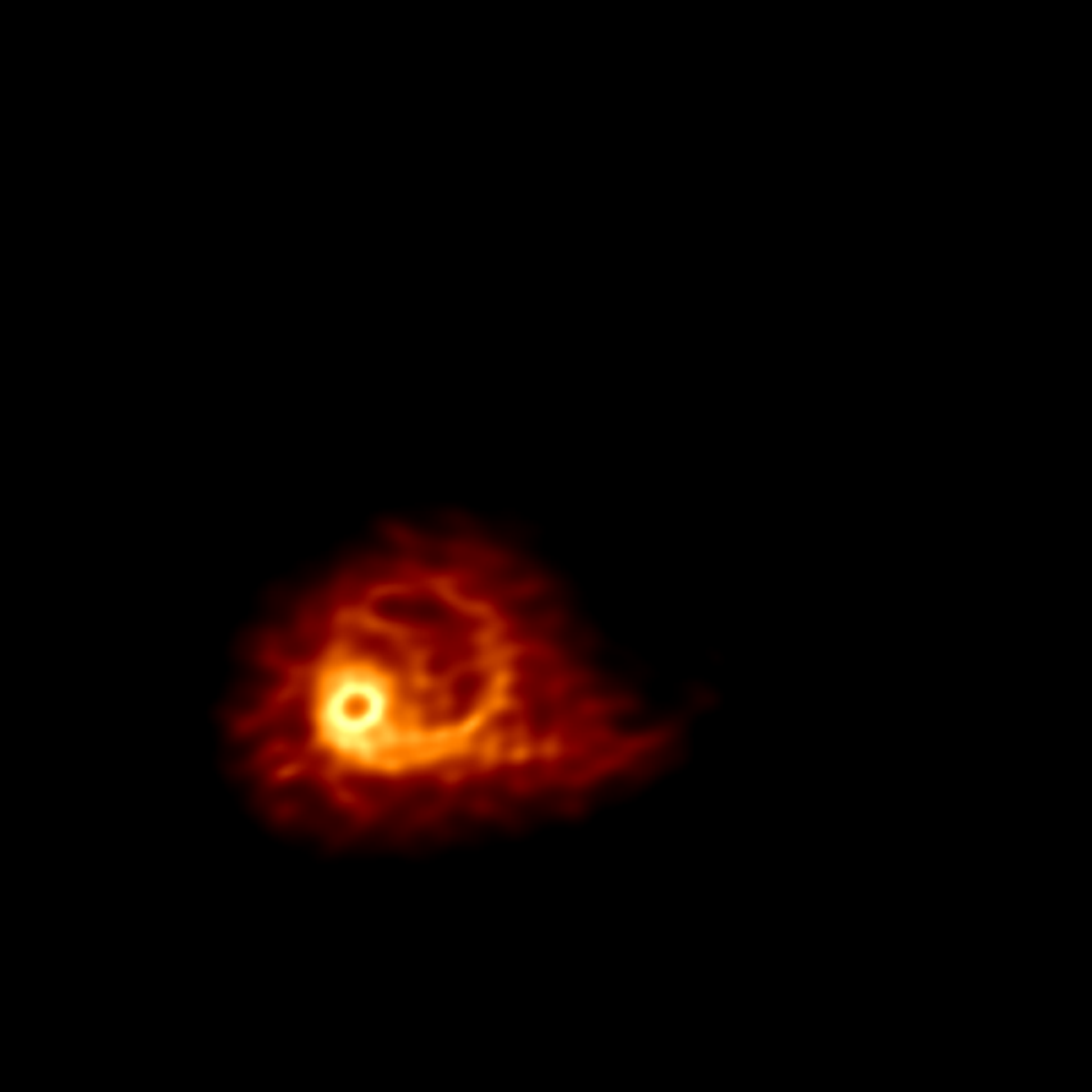}%
\includegraphics[width=30mm]{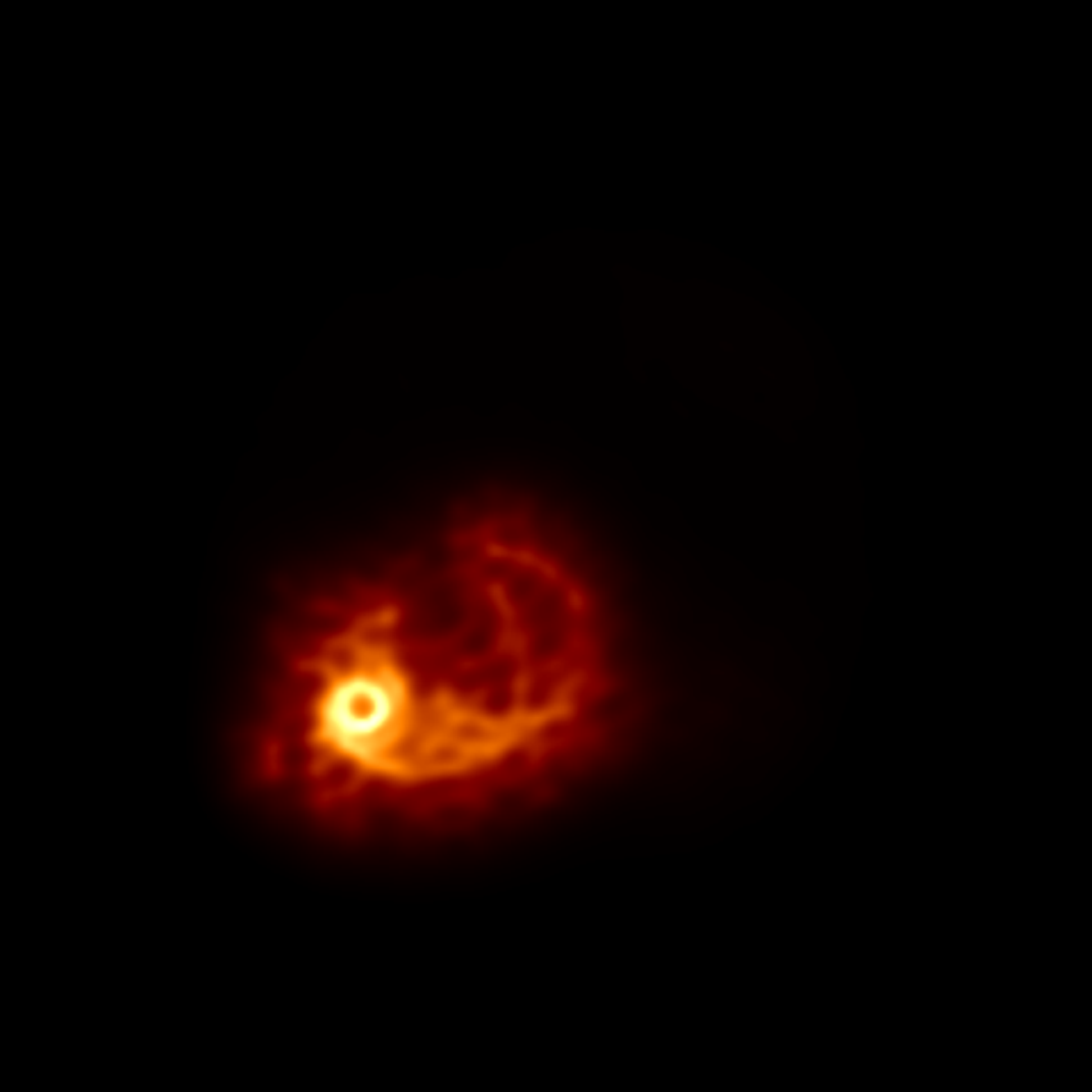}%
\includegraphics[width=30mm]{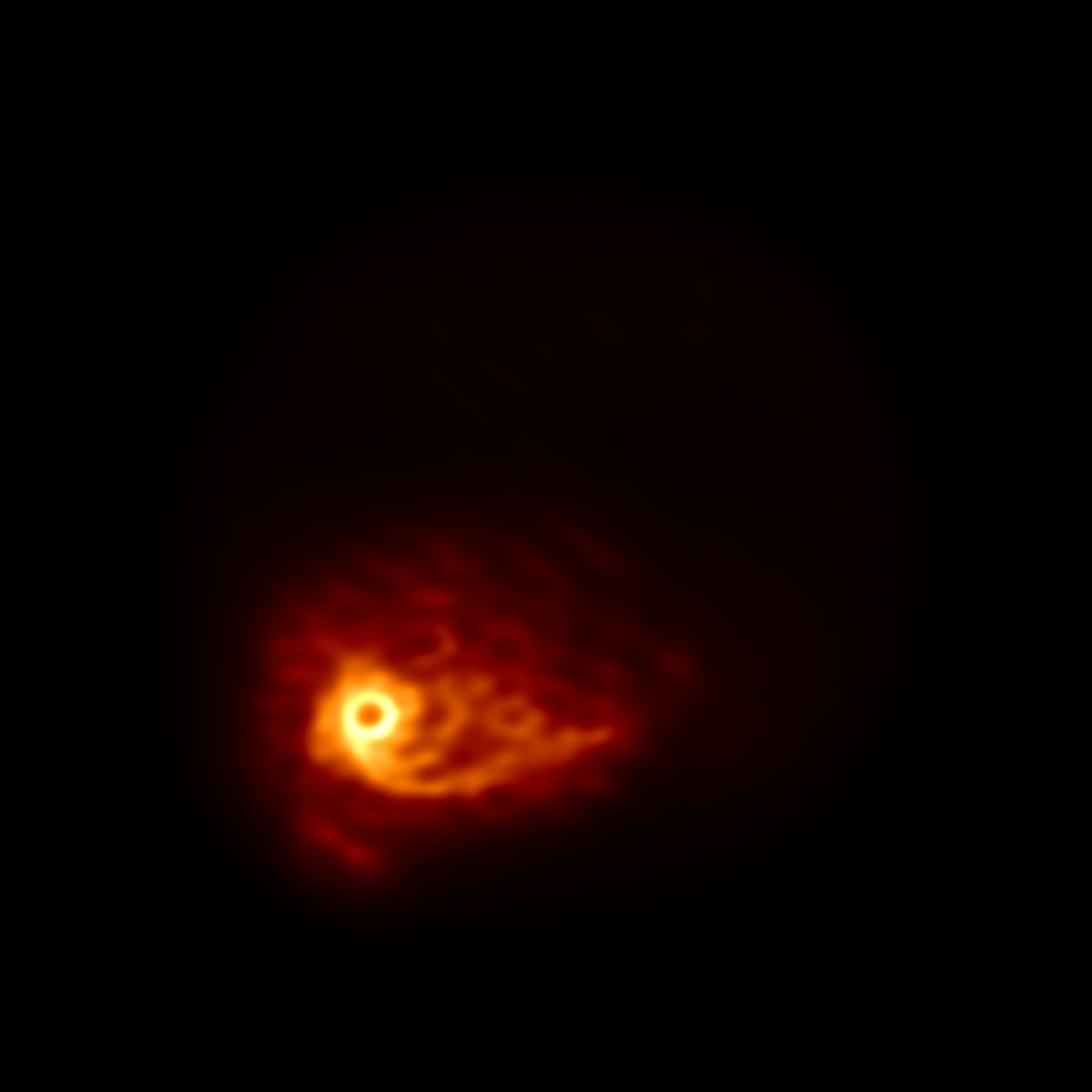}%
\includegraphics[width=30mm]{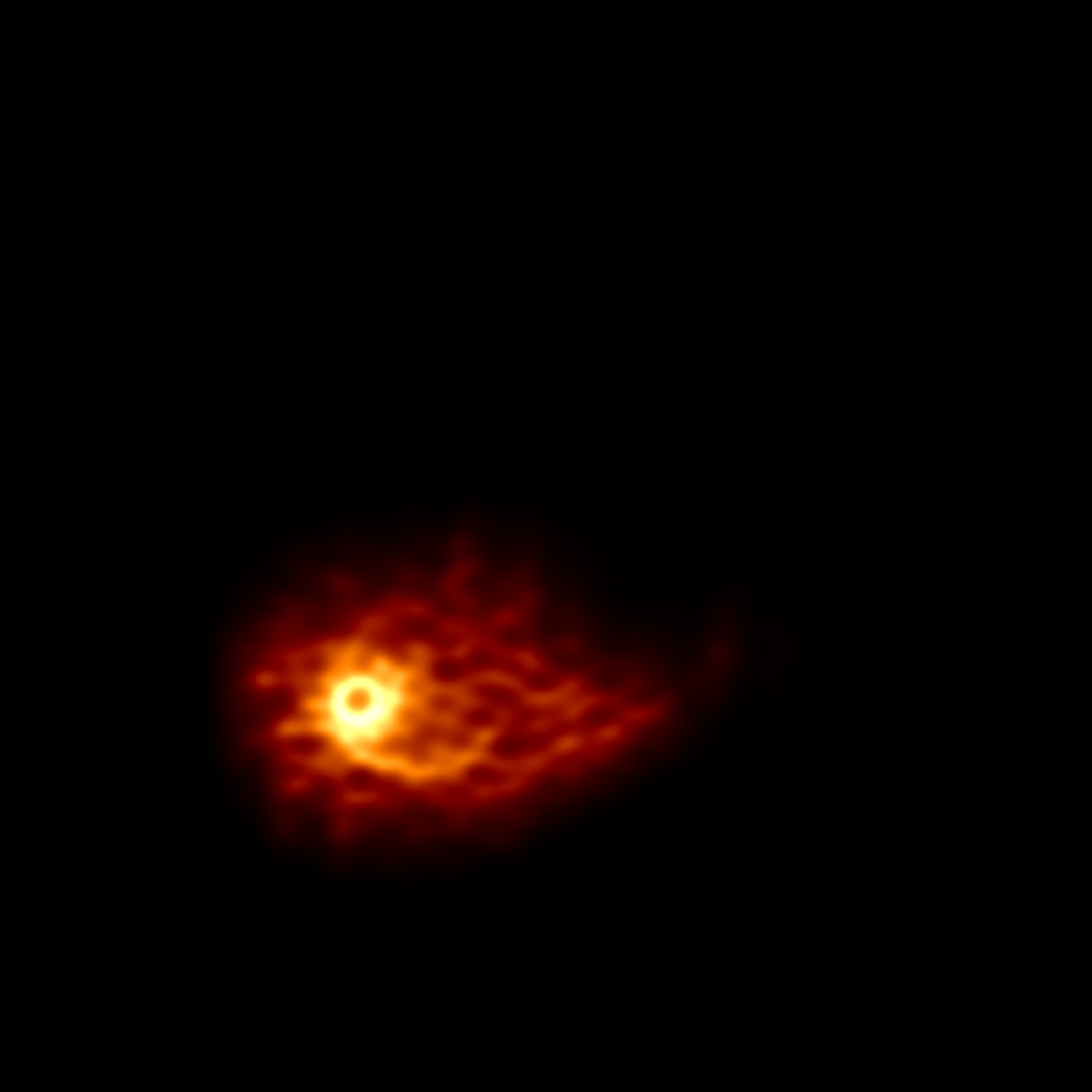} \\
\end{adjustwidth}
  \caption{Selection of Challenge 2 M87 230 GHz submissions. Images are shown on a log scale, which is normalized to the brightest pixel value across each submitted set of movie frames, with a dynamic range of $10^{3.5}$, on a field of view of 1 mas.}
     \label{fig:ch2_m87_230}
\end{figure*}

\begin{figure*}
\begin{adjustwidth}{-\extralength}{0cm}
\setlength{\lineskip}{0pt}
\centering
\includegraphics[height=55mm]{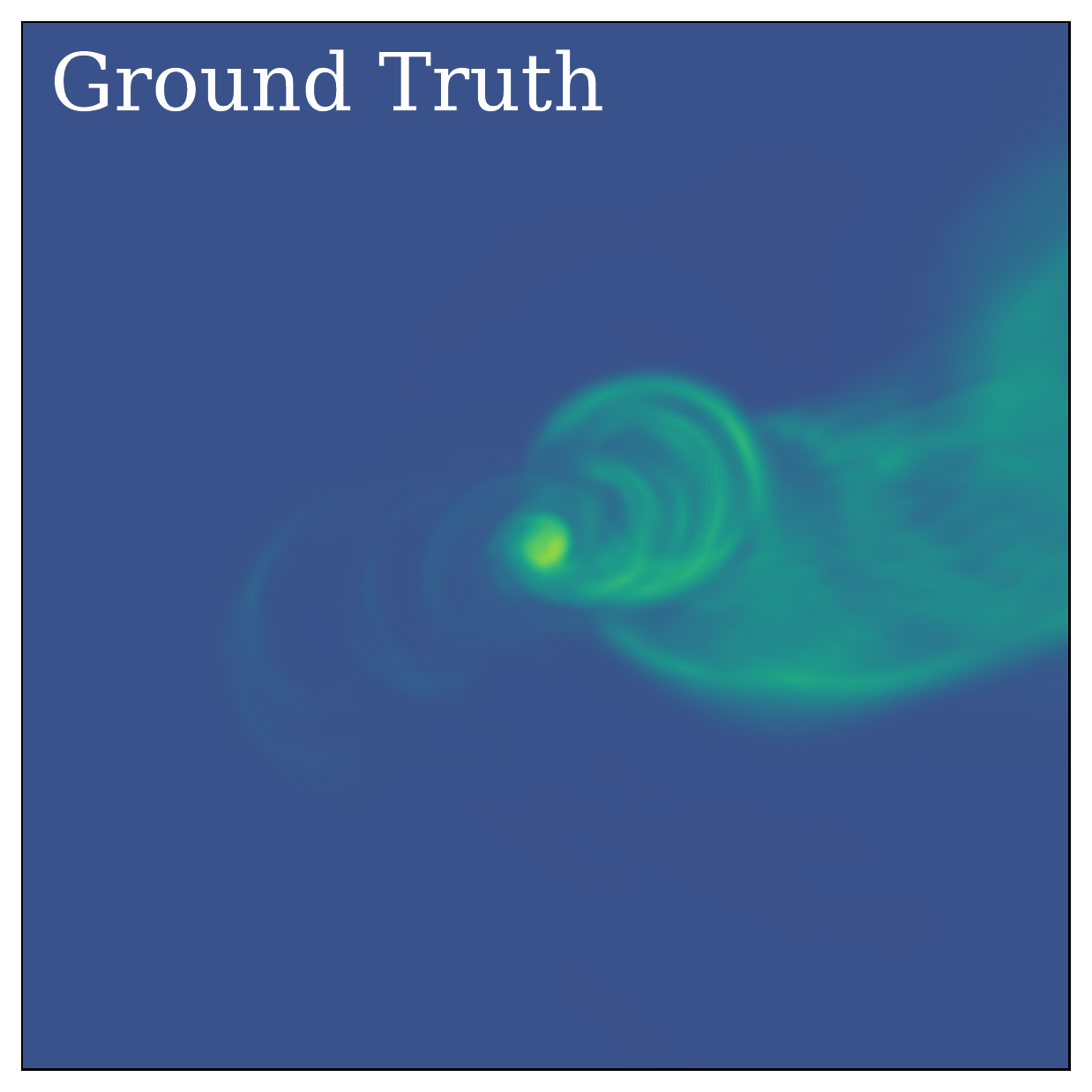}%
\includegraphics[height=55mm]{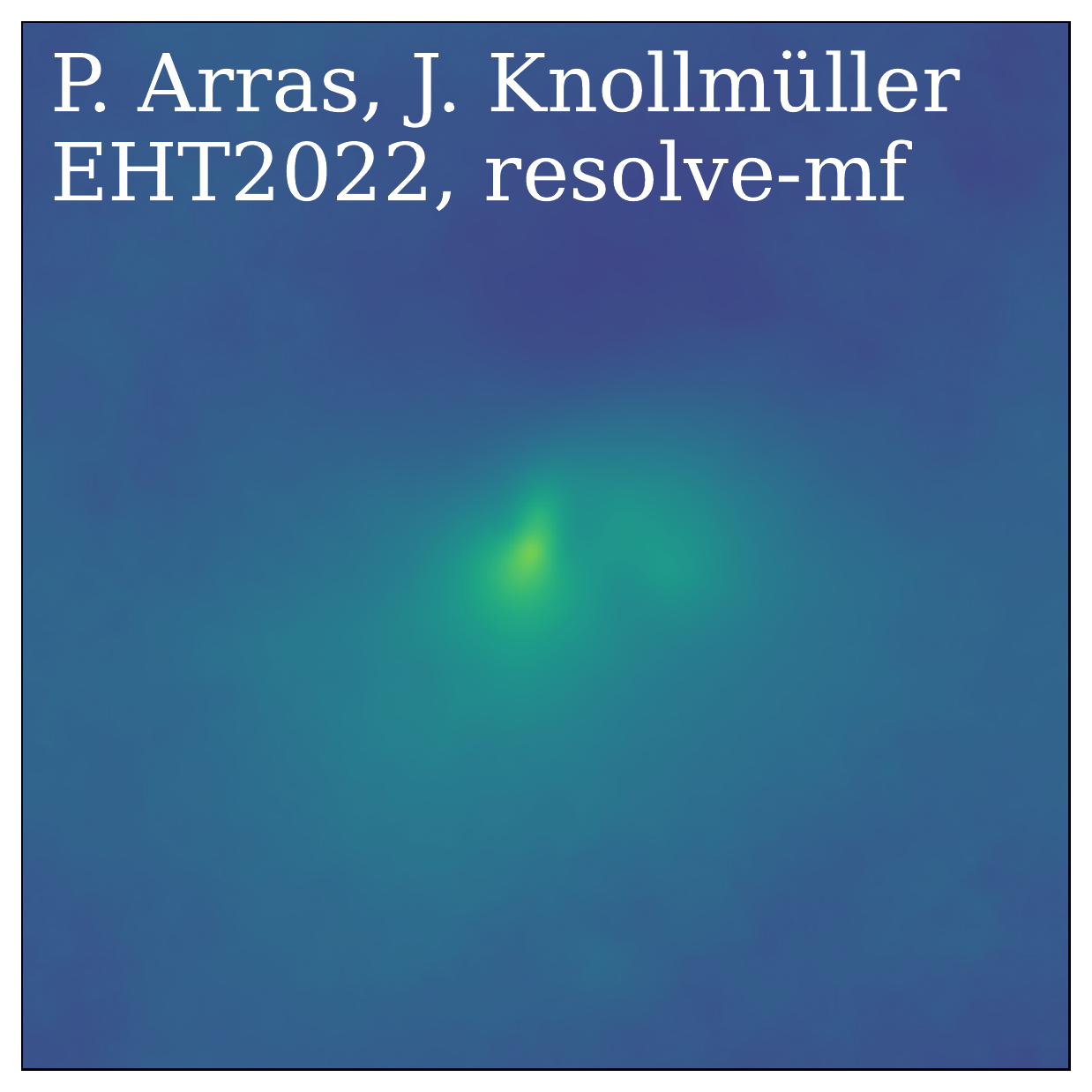}%
\includegraphics[height=55mm]{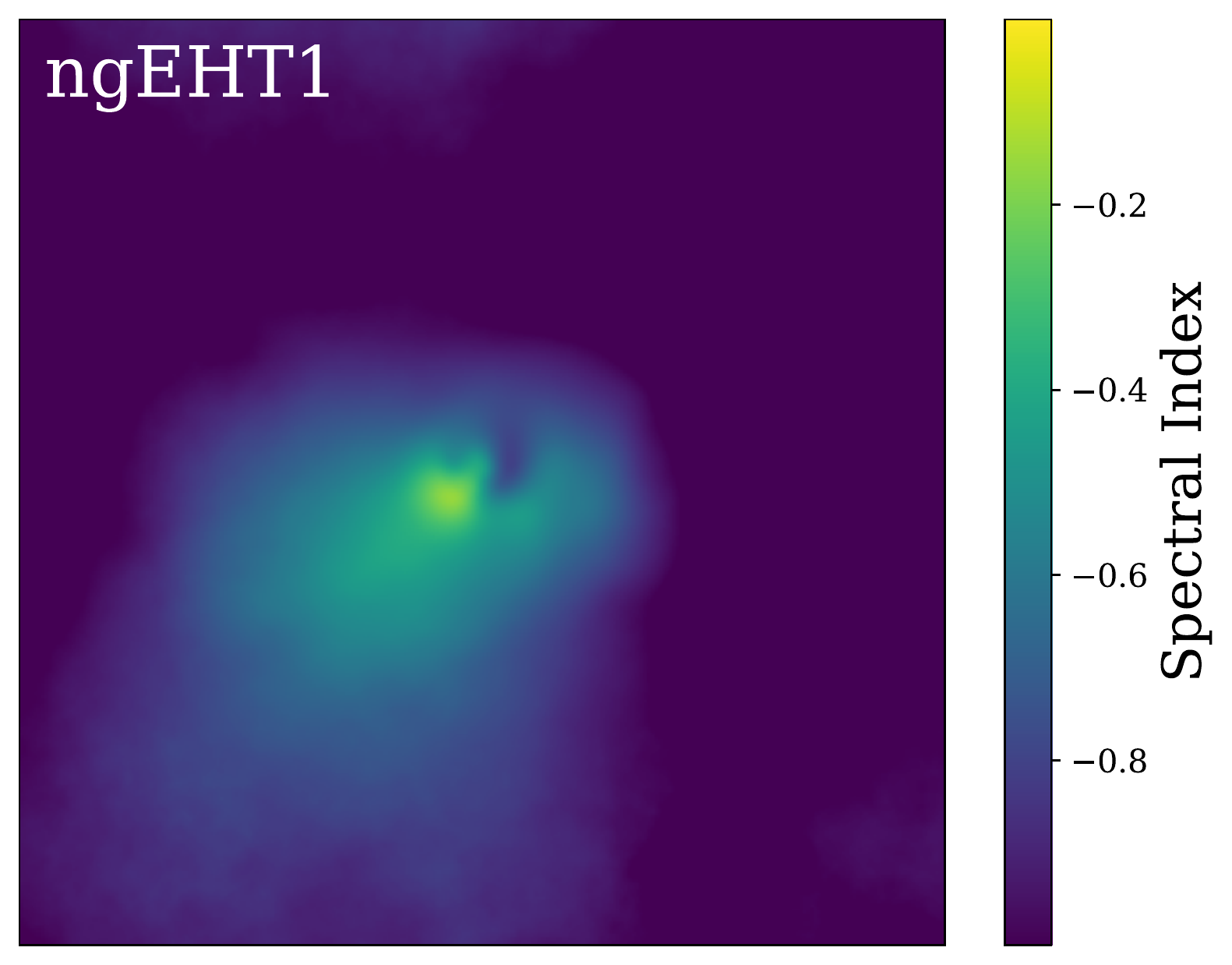} \\

\end{adjustwidth}
  \caption{Spectral index maps of the Challenge 2 M87 ground truth model (first frame) and {\tt resolve} reconstructions of the spectral index map with the EHT2022 and ngEHT1 arrays. The ground truth spectral index map was blurred with a Gaussian with a FWHM of 9.4 $\mu$as.}
     \label{fig:ch2_m87_resolve_si}
\end{figure*}

\begin{figure*}
\begin{adjustwidth}{-\extralength}{0cm}
\setlength{\lineskip}{0pt}
\centering

\includegraphics[width=50mm]{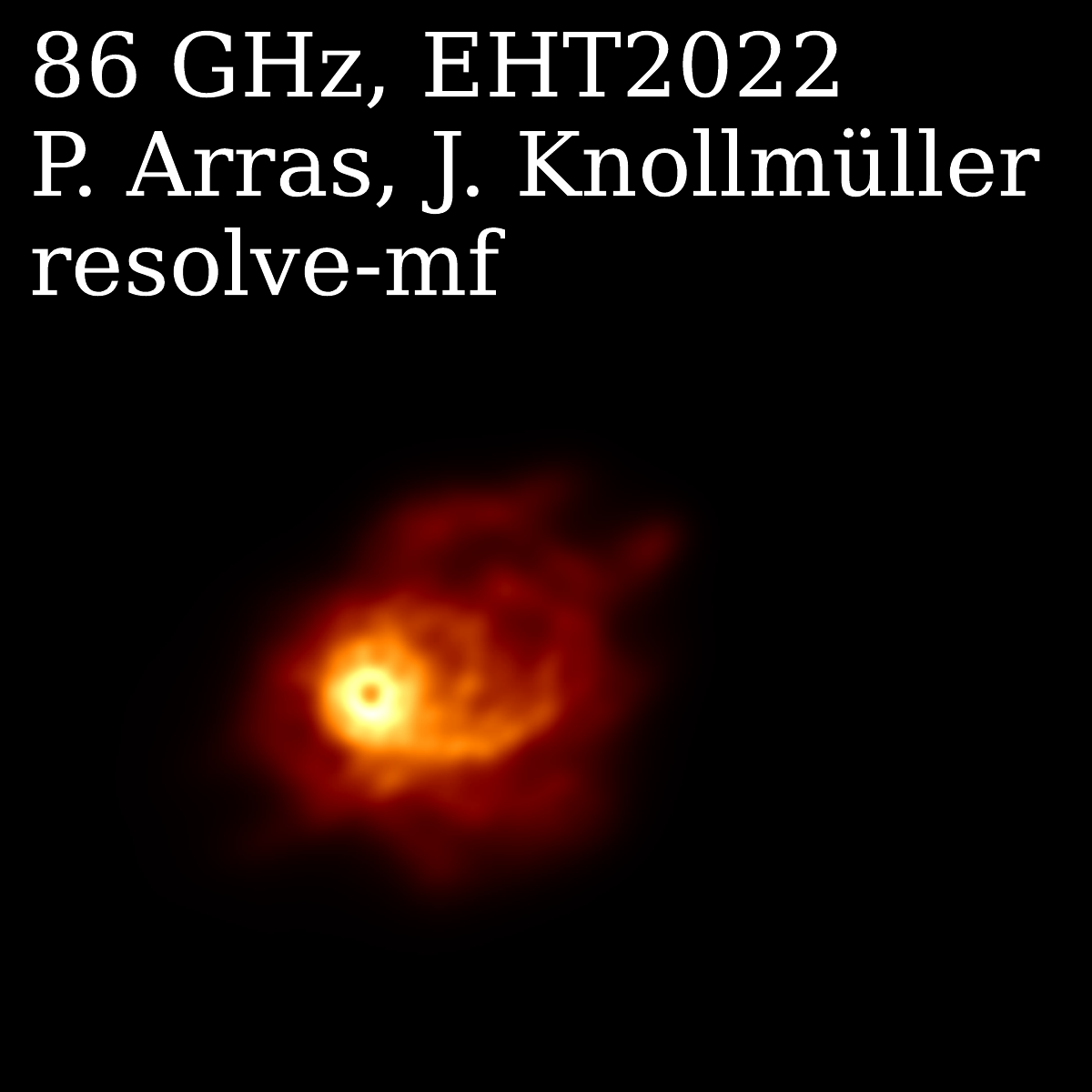}%
\includegraphics[width=50mm]{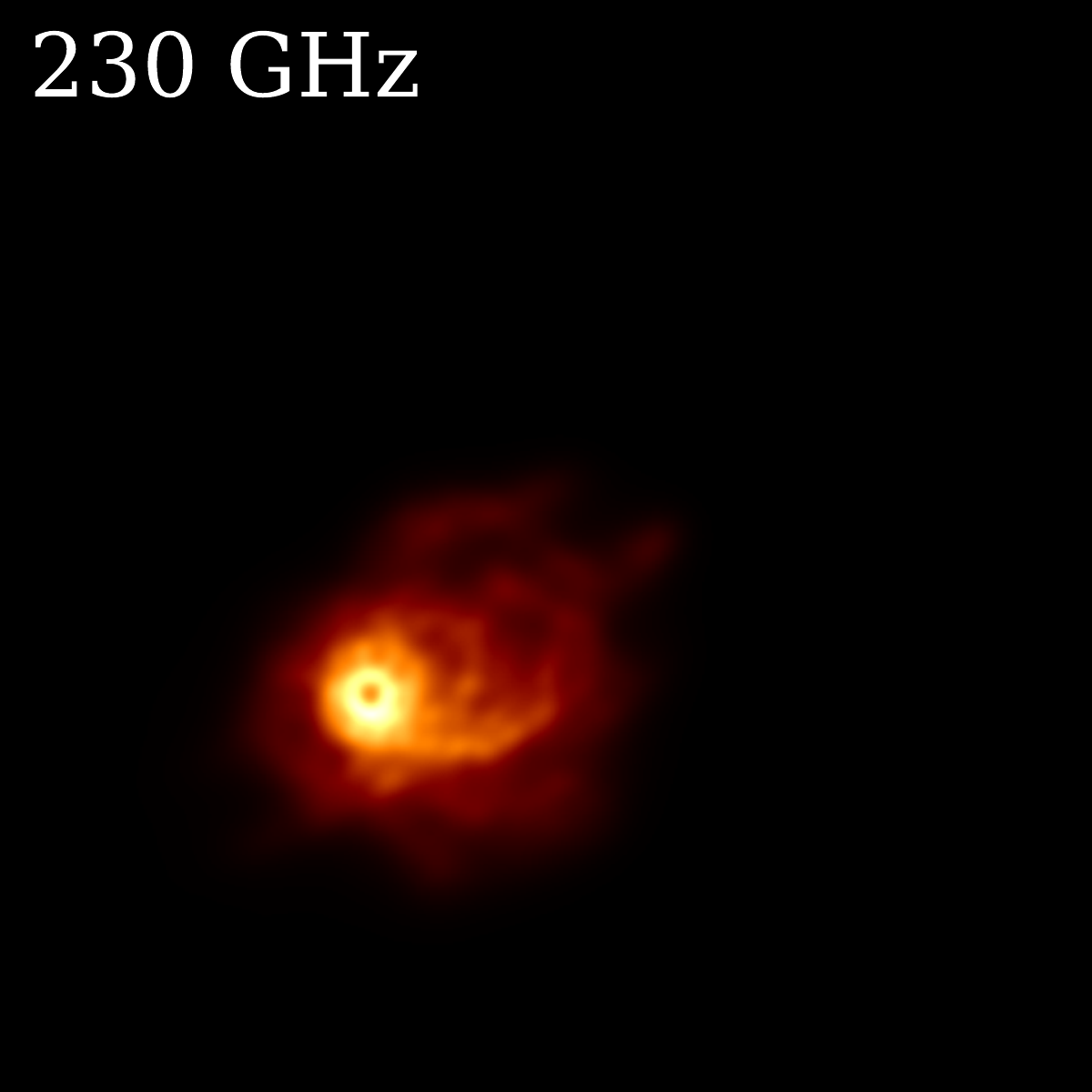}%
\includegraphics[width=50mm]{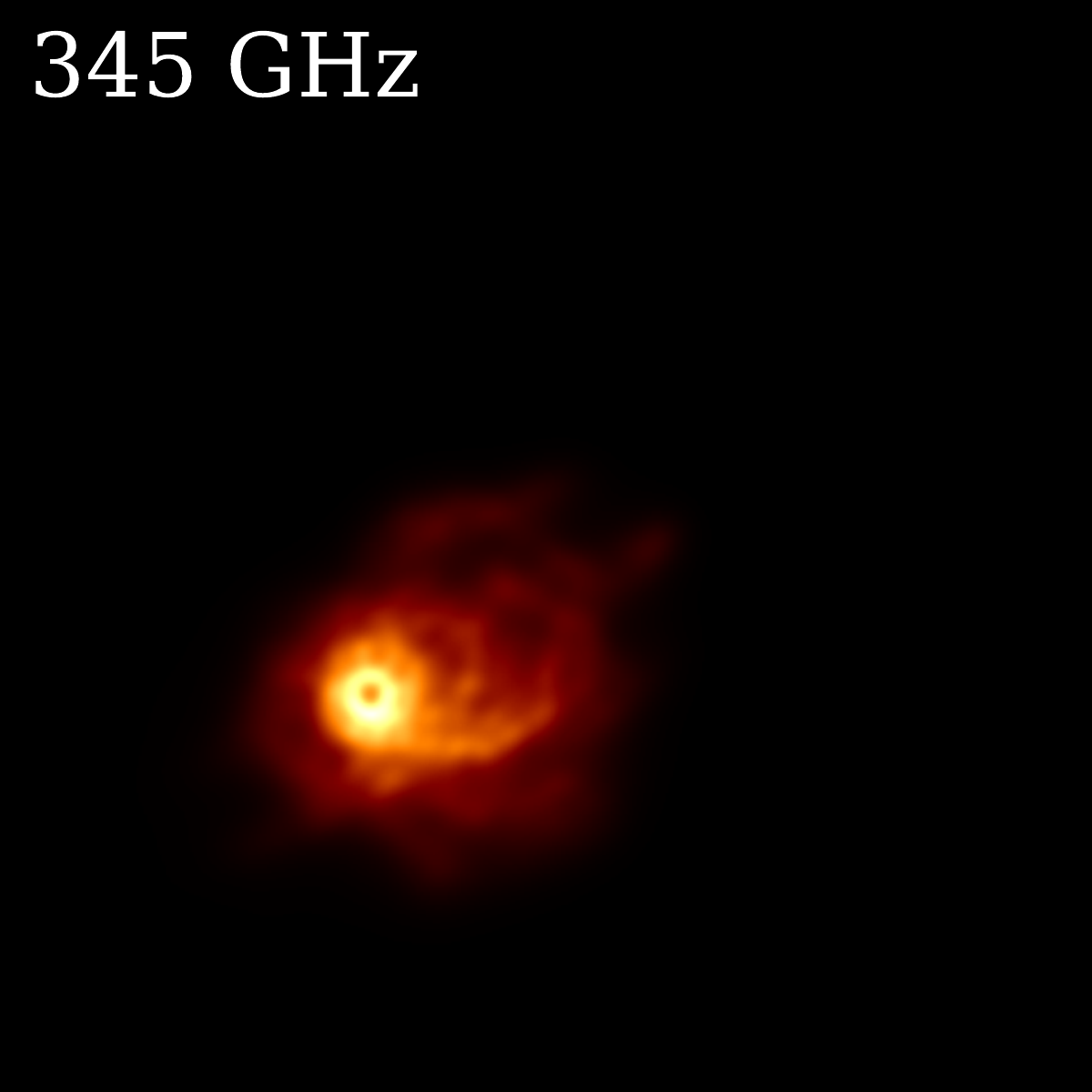} \\
\includegraphics[width=50mm]{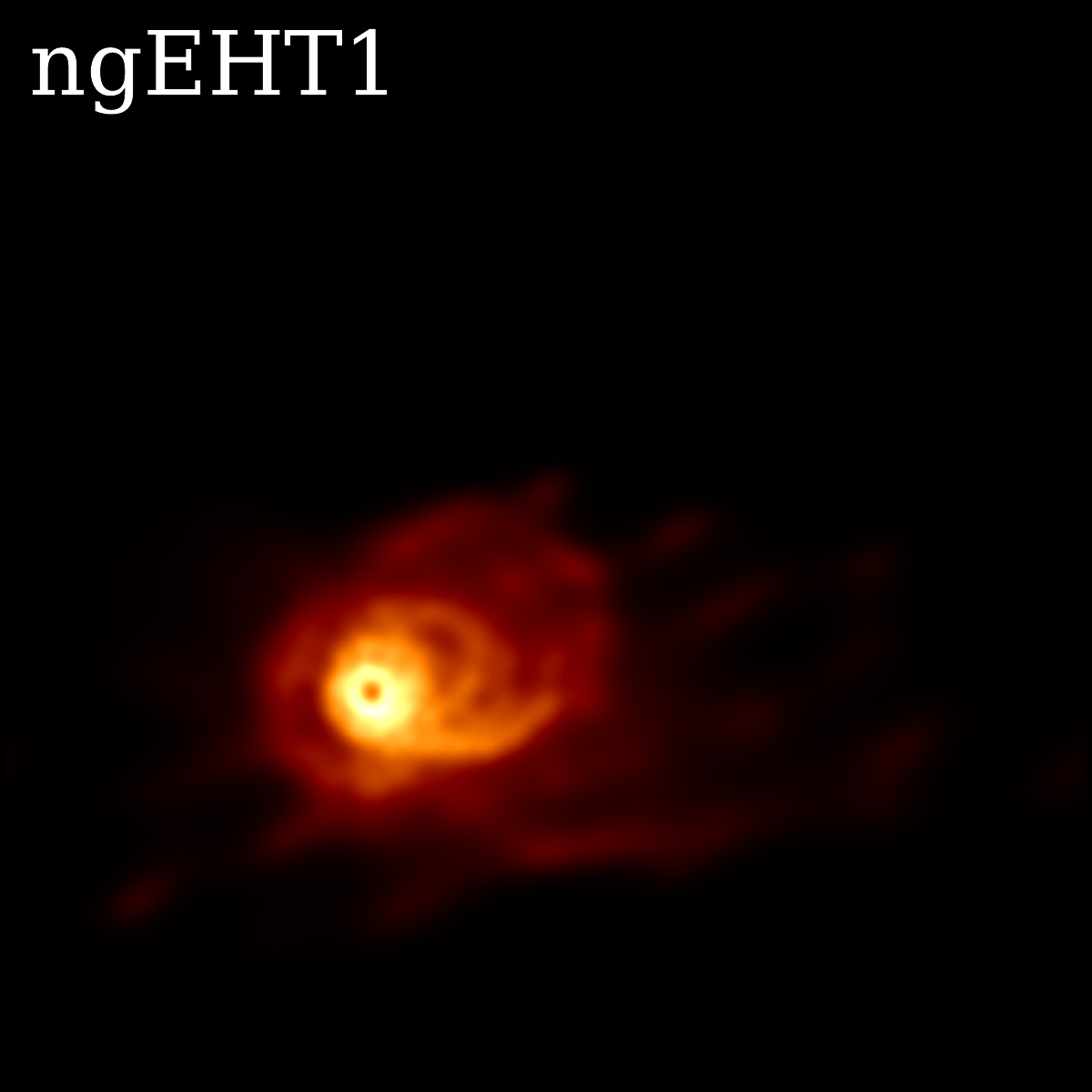}%
\includegraphics[width=50mm]{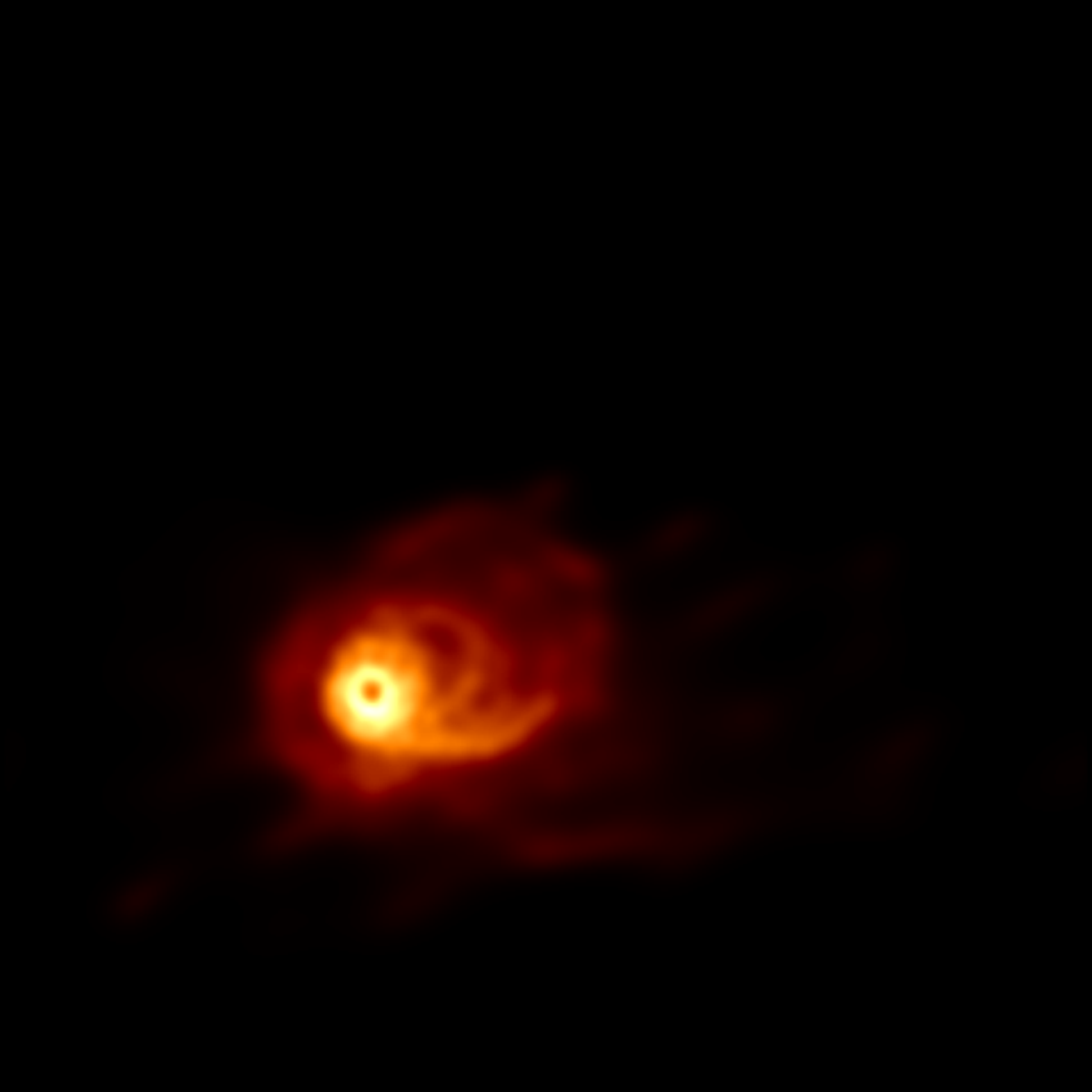}%
\includegraphics[width=50mm]{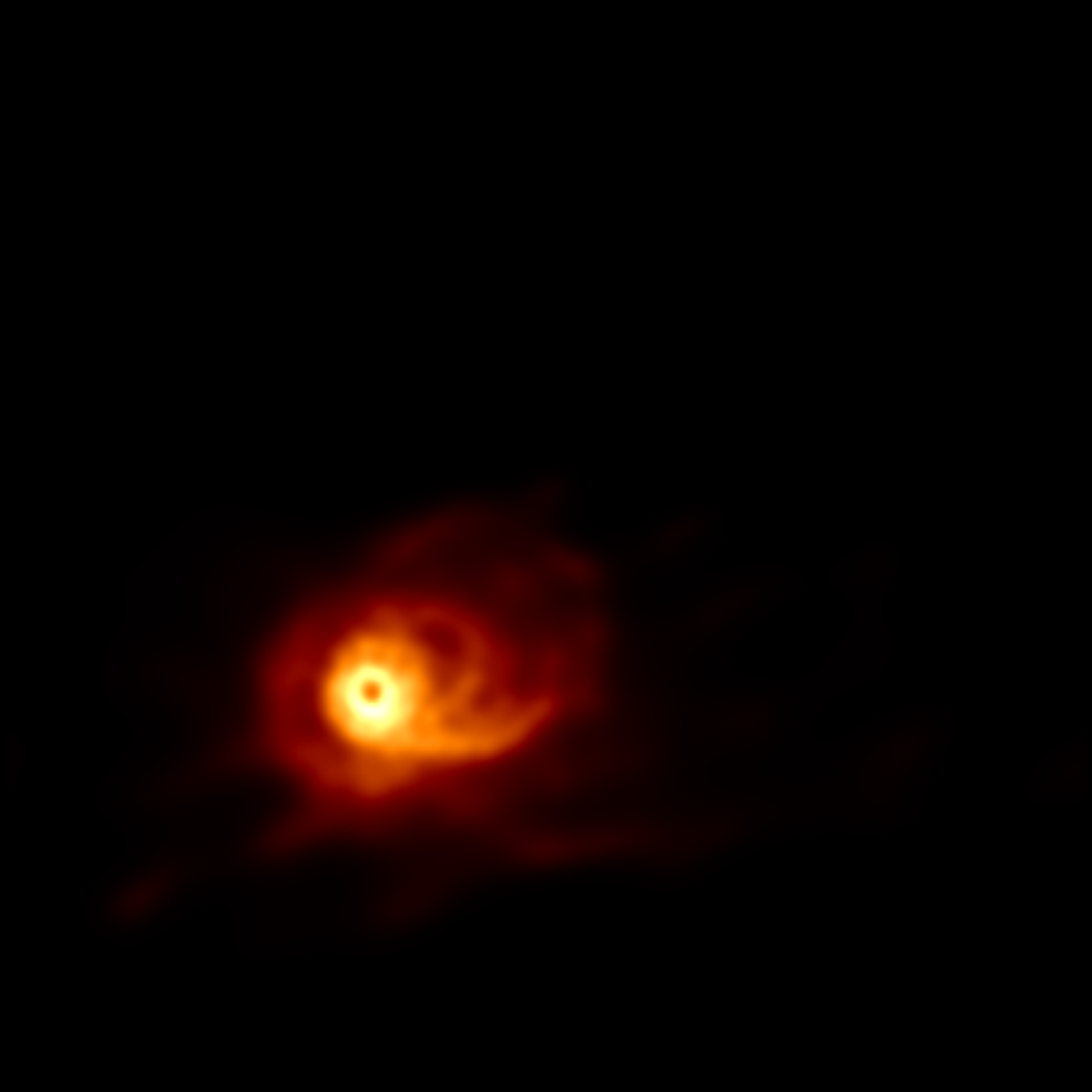}% \\

\end{adjustwidth}
  \caption{Multi-frequency {\tt resolve} reconstructions of the Challenge 2 M87 model at 86, 230, and 345 GHz (first frame) with the EHT2022 and ngEHT1 arrays.}
     \label{fig:ch2_m87_resolve_images}
\end{figure*}

 Six frames of the 86 and 230 GHz M87 reconstructions are shown in Figures \ref{fig:ch2_m87_86} and \ref{fig:ch2_m87_230}, respectively. At 86 GHz, EHT2022 coverage allows to reconstruct the central component and overall shape of the extended jet emission, but all reconstructions contain spurious artefacts. With ngEHT1 coverage, these artefacts become far less severe or disappear completely, and the jet dynamics can be imaged as the jet features move outwards over the course of several weeks. 230 GHz reconstructions provide significantly more detail, both in the extended jet and in the visibility of the black hole shadow. The reconstructions again show strong improvement of ngEHT1 compared to EHT2022, although the EHT2022 reconstructions from the {\tt resolve} algorithm already show some jet dynamics with EHT2022 coverage. Figures \ref{fig:ch2_m87_resolve_si} and \ref{fig:ch2_m87_resolve_images} show spectral index and individual frequency image reconstructions, respectively, from {\tt resolve} when solving for all frequencies simultaneously and imposing a prior on the spectral index map ({\tt resolve-mf}), for the first movie frame only. This method leads to remarkably high-quality images for all frequencies and arrays, even showing the black hole's central brightness depression at 86 GHz. These results demonstrate that utilizing information from simultaneous multi-frequency observations can significantly boost the reconstruction quality \citep[see also][]{Issaoun2022}.

The reconstruction quality metrics show that images with low $\rho_{\mathrm{NX}}$ or $\rho_{\mathrm{NX, log}}$ often have relative high $\chi^2$ and low $\theta_\mathrm{eff}$ and $\mathcal{D}_{0.1}$, with the {\tt resolve} and especially the {\tt resolve-mf} reconstructions performing best overall, with the caveat that the {\tt resolve-mf} reconstructions were only done for the first movie frame. For single-frequency reconstructions, $\theta_\mathrm{eff}$ reaches 21.2 $\mu$as at 86 GHz and 6.6 $\mu$as at 230 GHz (median values across the 20 reconstructed frames); the superresolution with respect to the nominal array resolution (60 and 23 $\mu$as for 86 and 230 GHz, respectively), is significant (up to a factor 3.5) for most reconstructions. For multi-frequency reconstructions, the supperresolution factor increases even further, up to 8.6 at 86 GHz. $\chi^2$-values generally increase as a function of frequency, which is likely due to increased data complexity with more severe systematics. 345 GHz reconstructions often showed difficulty in reconstructing the extended jet structure, which could be due to the sparser coverage and more severe corruptions and noise.

\subsubsection{Sgr A* RIAF+hotspot}
\begin{figure*}
\begin{adjustwidth}{-\extralength}{0cm}
\setlength{\lineskip}{0pt}
\centering

\includegraphics[width=23mm]{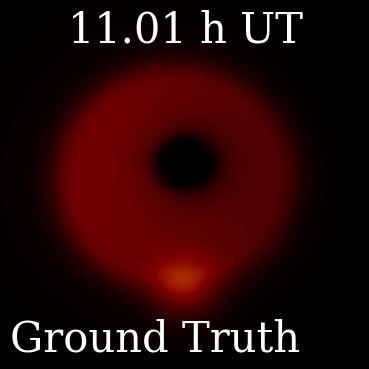}%
\includegraphics[width=23mm]{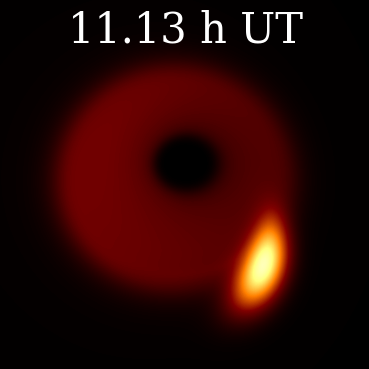}%
\includegraphics[width=23mm]{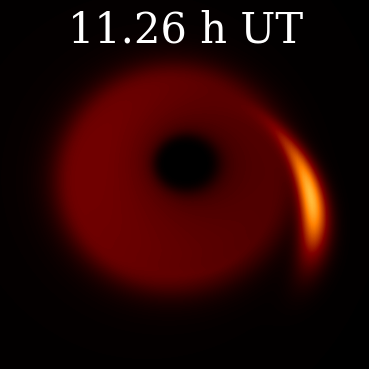}%
\includegraphics[width=23mm]{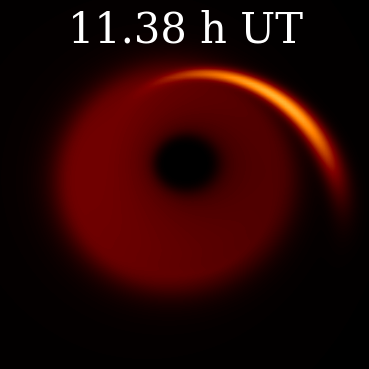}%
\includegraphics[width=23mm]{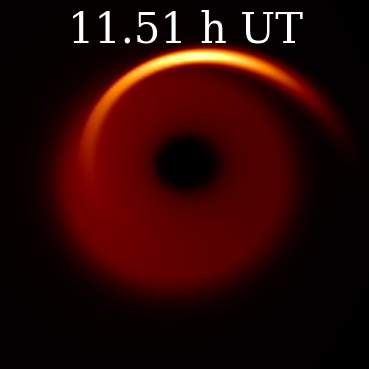}%
\includegraphics[width=23mm]{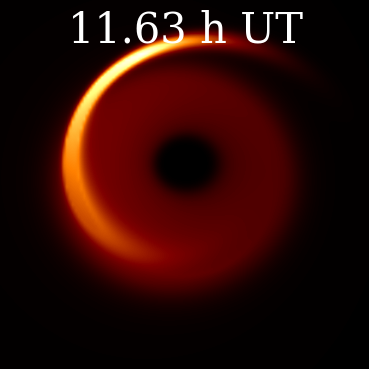}%
\includegraphics[width=23mm]{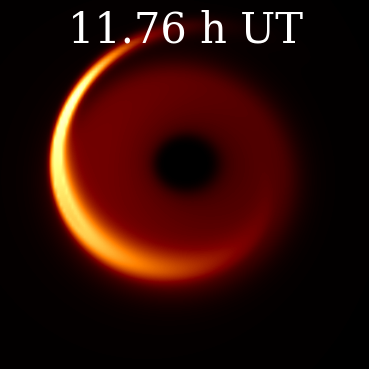}%
\includegraphics[width=23mm]{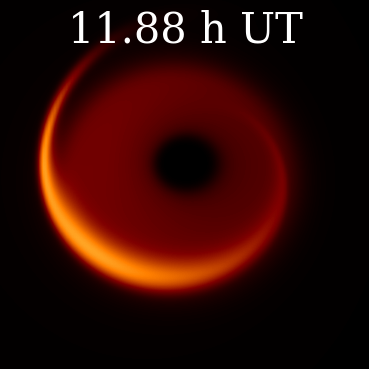}% \\
\vspace{2mm}

\includegraphics[width=23mm]{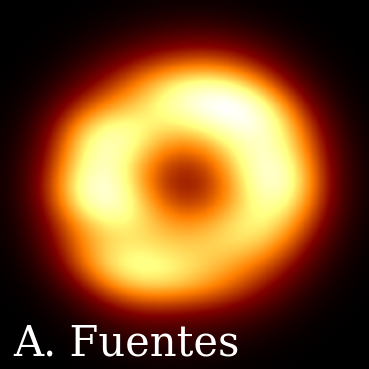}%
\includegraphics[width=23mm]{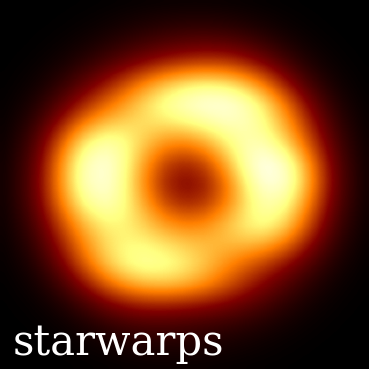}%
\includegraphics[width=23mm]{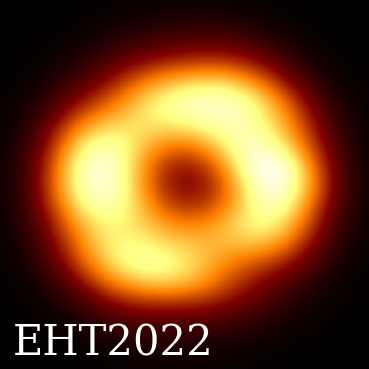}%
\includegraphics[width=23mm]{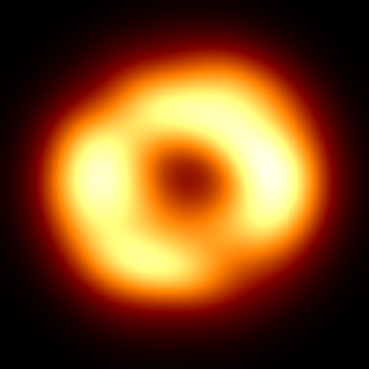}%
\includegraphics[width=23mm]{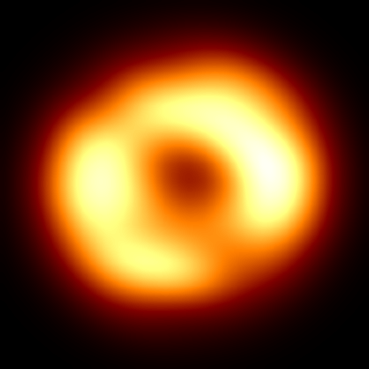}%
\includegraphics[width=23mm]{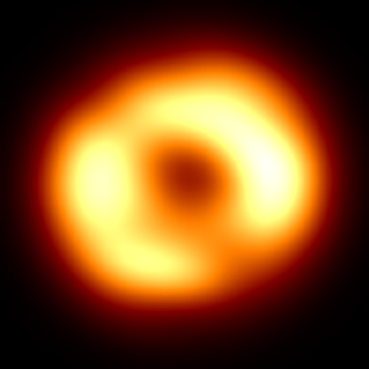}%
\includegraphics[width=23mm]{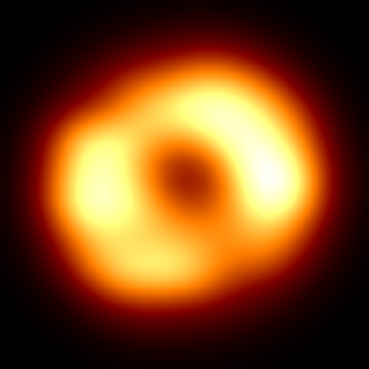}%
\includegraphics[width=23mm]{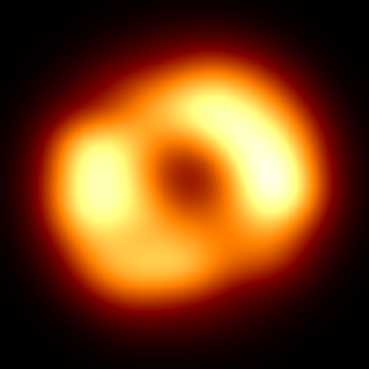}% \\

\includegraphics[width=23mm]{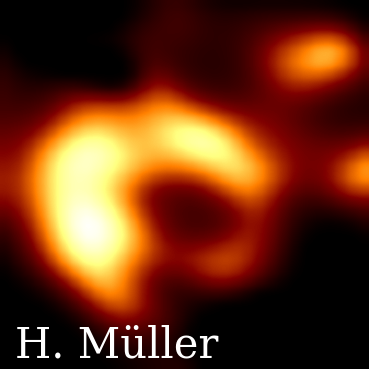}%
\includegraphics[width=23mm]{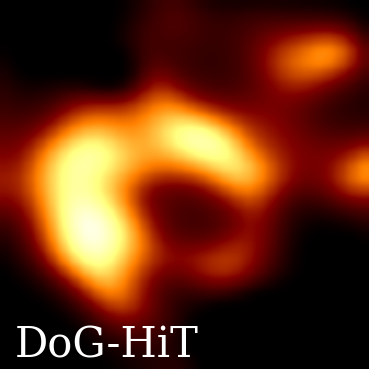}%
\includegraphics[width=23mm]{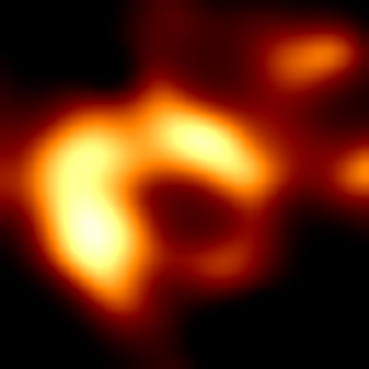}%
\includegraphics[width=23mm]{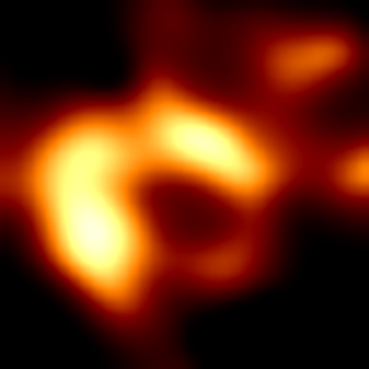}%
\includegraphics[width=23mm]{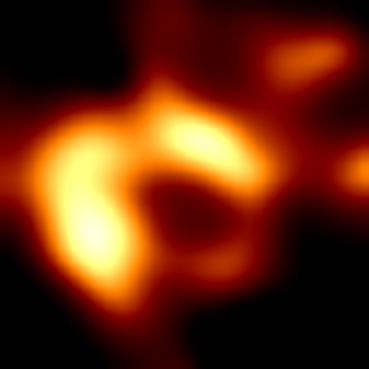}%
\includegraphics[width=23mm]{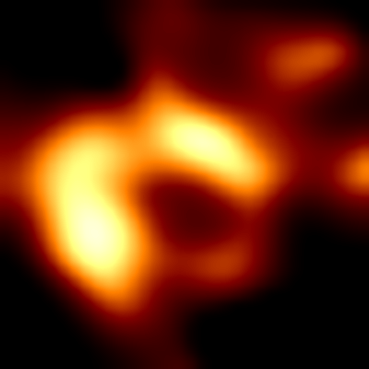}%
\includegraphics[width=23mm]{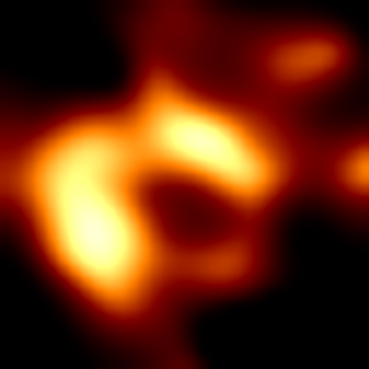}%
\includegraphics[width=23mm]{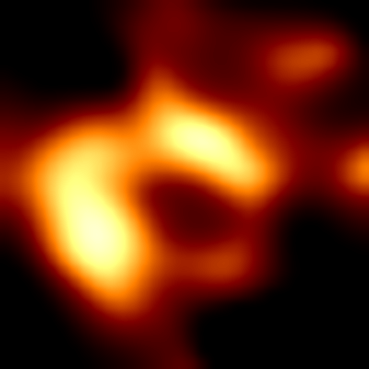}% \\
\vspace{2mm}

\includegraphics[width=23mm]{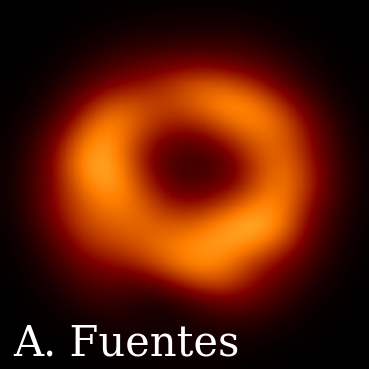}%
\includegraphics[width=23mm]{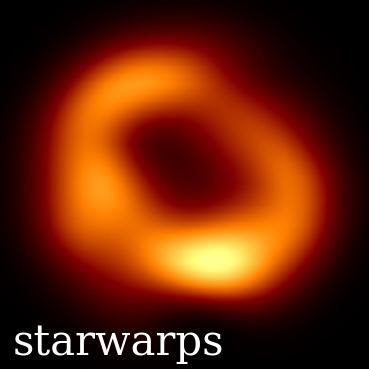}%
\includegraphics[width=23mm]{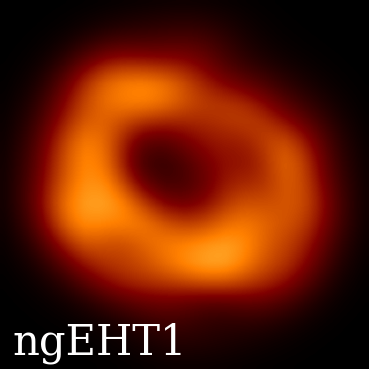}%
\includegraphics[width=23mm]{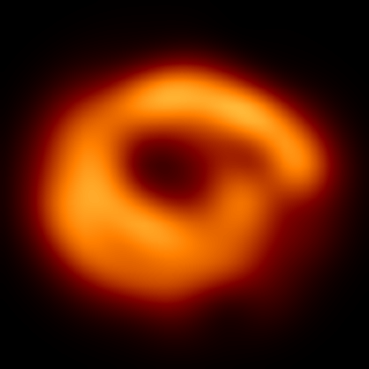}%
\includegraphics[width=23mm]{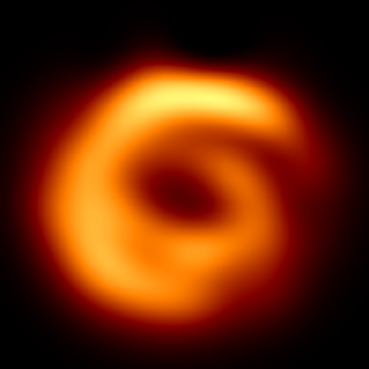}%
\includegraphics[width=23mm]{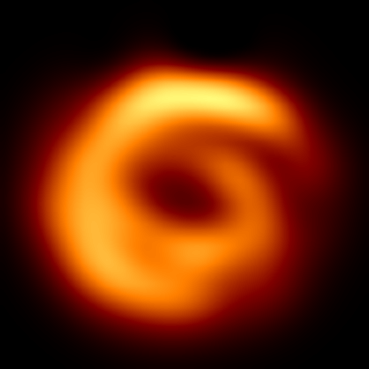}%
\includegraphics[width=23mm]{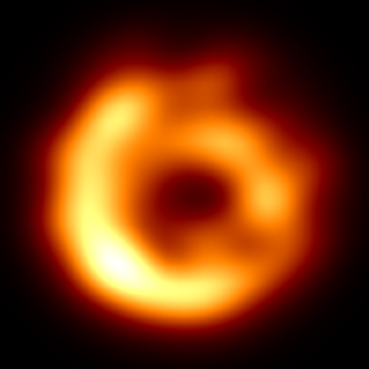}%
\includegraphics[width=23mm]{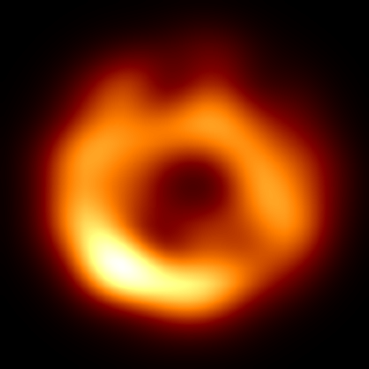}% \\

\includegraphics[width=23mm]{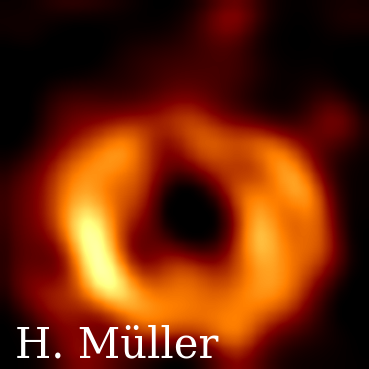}%
\includegraphics[width=23mm]{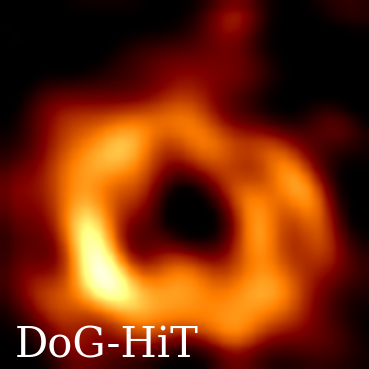}%
\includegraphics[width=23mm]{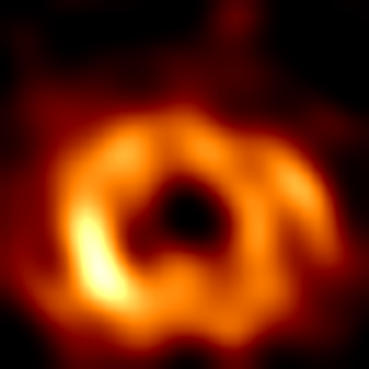}%
\includegraphics[width=23mm]{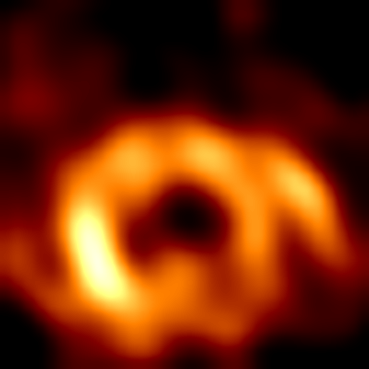}%
\includegraphics[width=23mm]{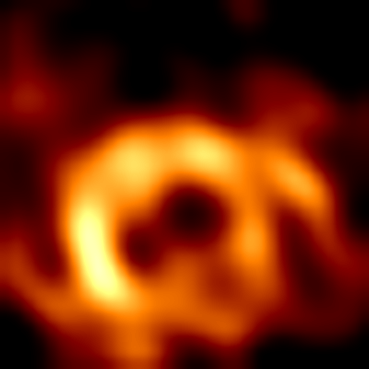}%
\includegraphics[width=23mm]{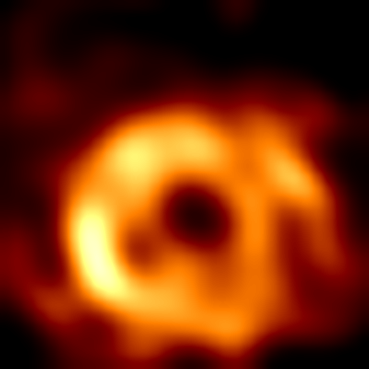}%
\includegraphics[width=23mm]{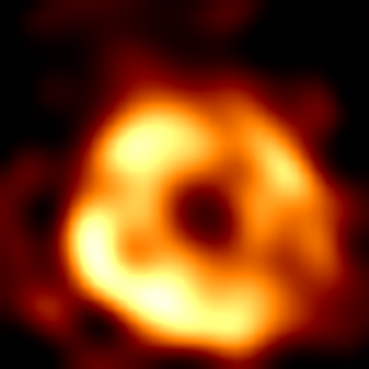}%
\includegraphics[width=23mm]{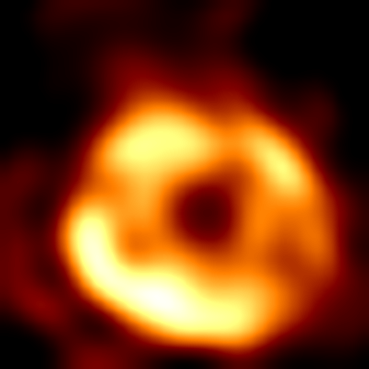}% \\

\includegraphics[width=23mm]{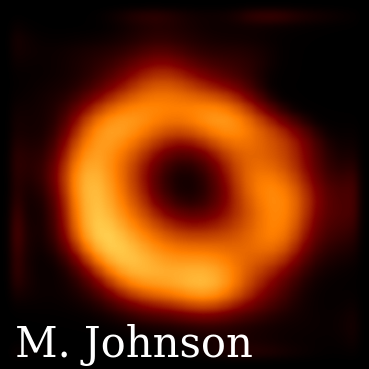}%
\includegraphics[width=23mm]{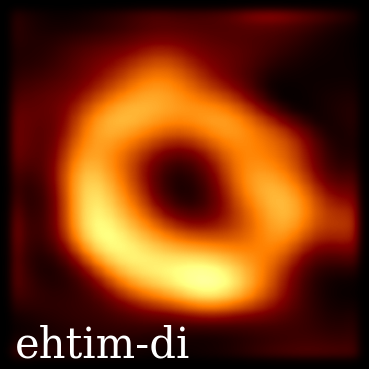}%
\includegraphics[width=23mm]{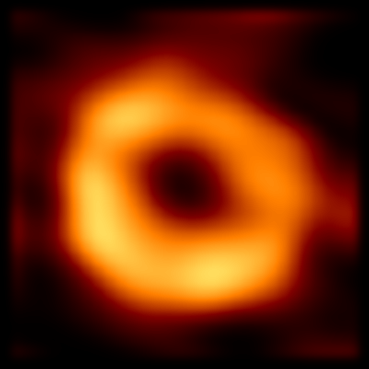}%
\includegraphics[width=23mm]{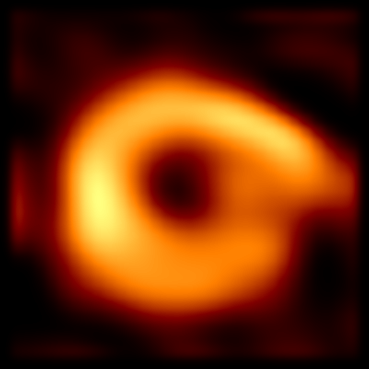}%
\includegraphics[width=23mm]{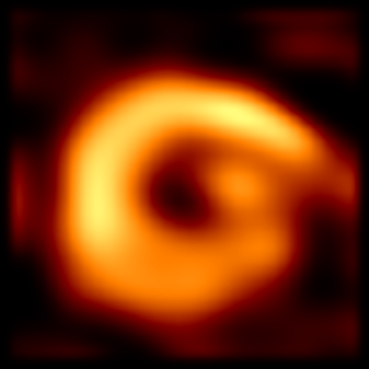}%
\includegraphics[width=23mm]{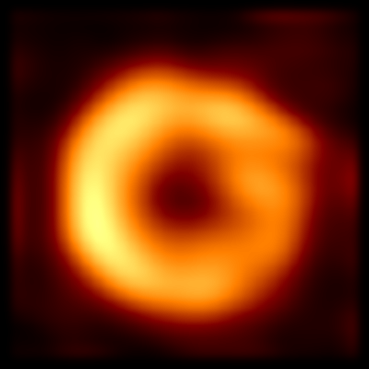}%
\includegraphics[width=23mm]{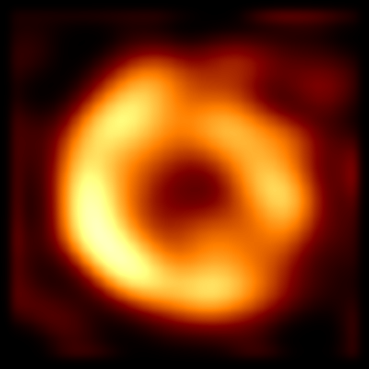}%
\includegraphics[width=23mm]{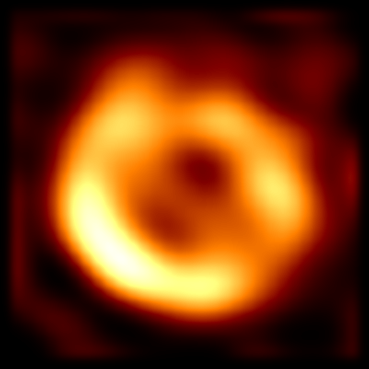}% \\

\includegraphics[width=23mm]{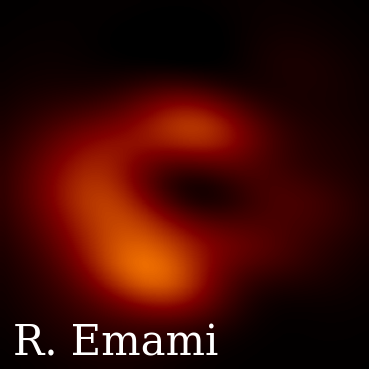}%
\includegraphics[width=23mm]{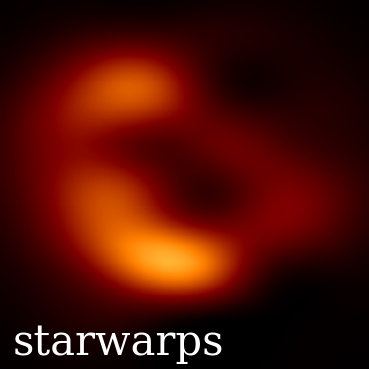}%
\includegraphics[width=23mm]{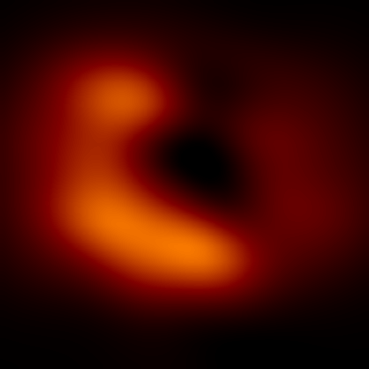}%
\includegraphics[width=23mm]{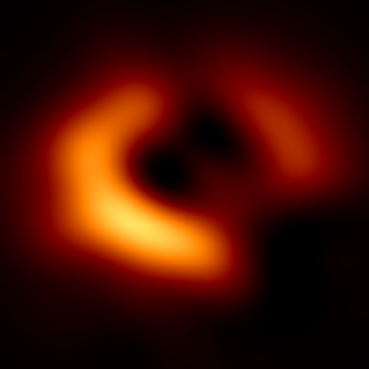}%
\includegraphics[width=23mm]{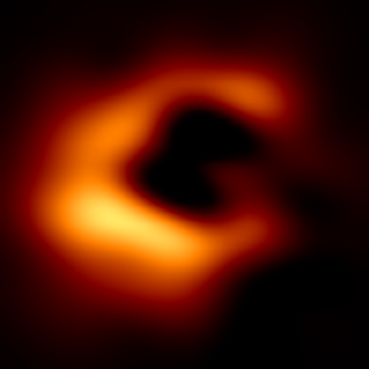}%
\includegraphics[width=23mm]{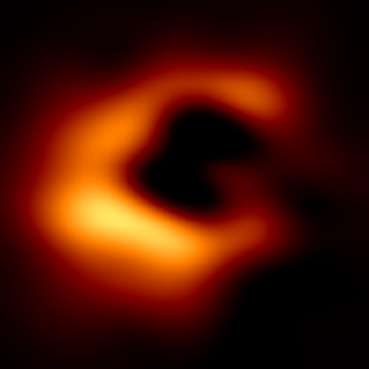}%
\includegraphics[width=23mm]{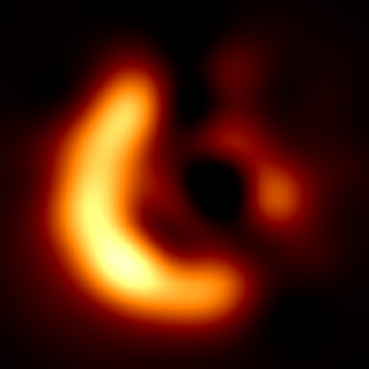}%
\includegraphics[width=23mm]{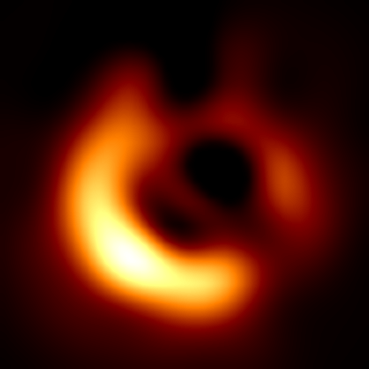}%\\
\end{adjustwidth}
  \caption{Challenge 2 Sgr A* RIAF+hotspot 230 GHz submissions. Images are shown on a linear scale, which is normalized to the brightest pixel value across each submitted set of movie frames, on a field of view of 126 $\mu$as.}
     \label{fig:ch2_sgra_riafspot}
\end{figure*}

Figure \ref{fig:ch2_sgra_riafspot} shows eight frames of all Sgr A* RIAF+hotspot submissions at 230 GHz. These frames span the 11-12 h UT window. This time window was chosen for the visual and metric submission comparisons as it corresponds to the first full rotation of the hotspot after it forms, and it is the hotspot rotation we aimed to reconstruct for this model. The 11-12 UT window also has strong overlap with the ngEHT1 ``best times'' \citep[e.g.][]{Farah2022} window for Sgr A*, with 14 stations observing Sgr A* simultaneously from 11.3 until 13.5 h UT. Also, all submissions contained frames in this window, whereas the total UT ranges reconstructed varied strongly between submissions.

None of the EHT2022 reconstructions show a significantly variable source structure, although a ring-like structure is recovered. The ngEHT1 reconstructions vary in quality, with especially the StarWarps and \texttt{eht-imaging} dynamical imaging algorithms recovering the shearing hotspot. This particular hotspot model was particularly challenging because of its quick shearing. Also, the data sampling with 10-minute scans interleaved with 10-minute gaps was relatively sparse compared to the hotspot period, giving only $\sim3$ scans per full hotspot rotation. These features make the reconstruction quality obtained by some methods remarkable. 

The $\chi^2$ metrics (Table \ref{tab:metrics_challenge2_M87}) are remarkably high for the Sgr A* reconstructions, which has several causes. First, in order to provide a uniform comparison, the metrics were calculated on the 11-12 h UT window only. The $\chi^2$ are lower when considering the full UT ranges submitted ($\sim3-4$ for the highest-quality reconstructions). Considering that the source is particularly variable and hence more difficult to reconstruct during the 11-12 UT window, it is not surprising that the $\chi^2$ are higher here. Second, the $\chi^2$ were calculated with respect to the original synthetic data, which have a 10-second resolution within 10-minute scans. Many submitters added systematic noise and time-averaged the data down to $\sim$minutes before imaging, which included averaging of rapidly variable structures in the visibility domain. For example, closure phases may swing by well over 120 degrees within a single 10-minute scan. Combined with the high signal-to-noise ratios of the ngEHT visibilities, the fit quality to the original input data is then significantly poorer than seen during the imaging process. Finally, the sumbissions are compared to the data which includes interstellar scattering, while many submitters deblurred the data before imaging. This process does not affect the closure phases, but the closure amplitudes are affected. The effective resolution $\theta_{\mathrm{eff}}$ for the best reconstructions is comparable to the nominal array resolution, reflecting the increased difficulty of recovering intraday time-variable structures compared to static reconstructions, which often reached significant superresolution.

\begin{figure*}
%\centering
\setlength{\lineskip}{0pt}
\includegraphics[width=100mm]{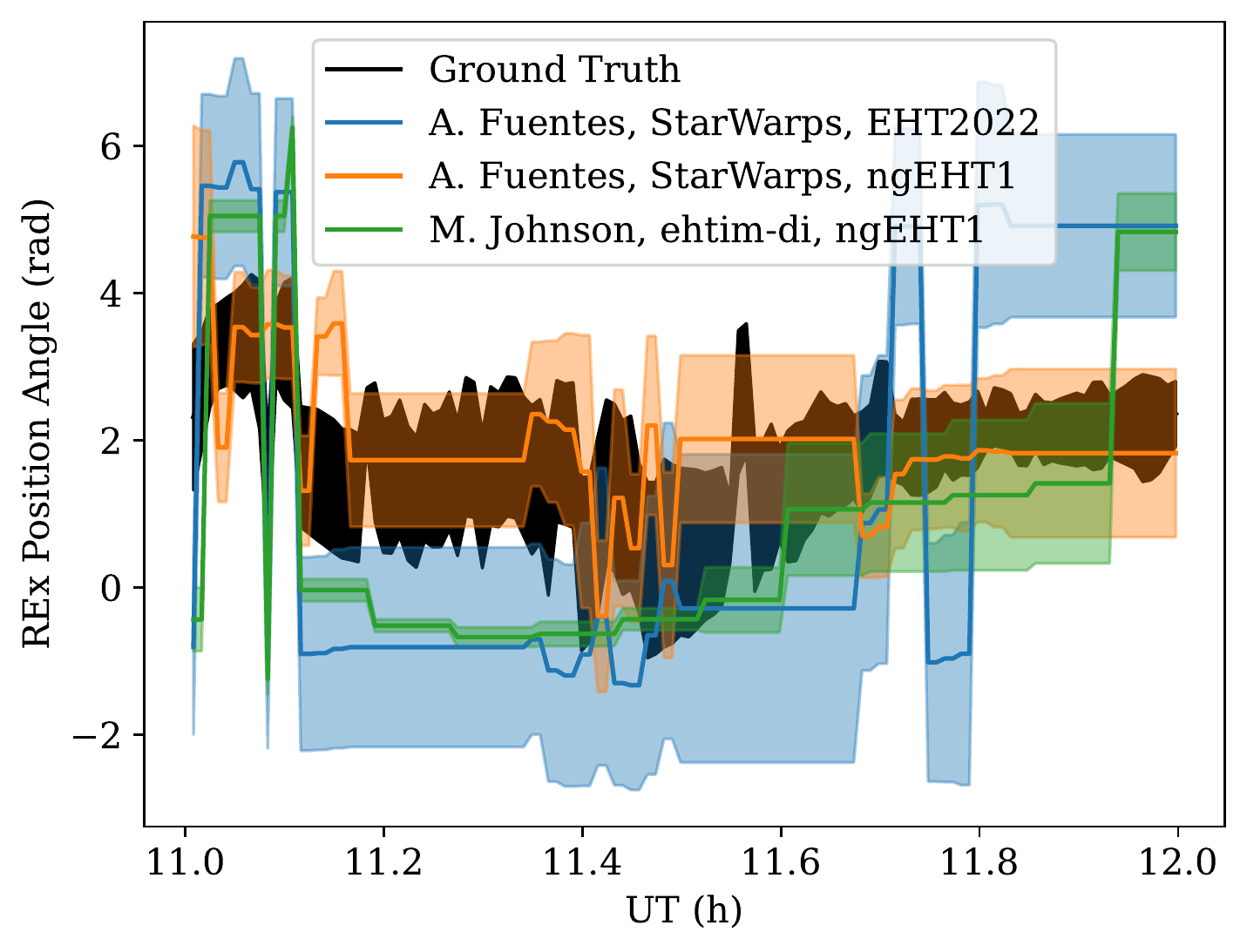}

  \caption{\texttt{REx} position angle fits with 1$\sigma$ uncertainties for three Challenge 2 Sgr A* RIAF+hotspot 230 GHz submissions, compared to the ground truth.}
     \label{fig:posangle_riafspot}
\end{figure*}

Figure \ref{fig:posangle_riafspot} shows the average image position angle as calculated by the Ring Extractor algorithm \texttt{REx} \citep{Chael2019, SgrA_PaperIV}, which characterizes the properties of ring-like images, for the ground truth and a few reconstructions in the 11-12 UT window. Especially the ngEHT1 StarWarps reconstruction shows excellent agreement with the ground truth, and the ngEHT1 {\tt ehtim-di} reconstruction performs well after about 11.4 UT. The EHT2022 StarWarps reconstruction shows a stable position angle (note the $2\pi$ ambiguity) that is generally offset from the ground truth.

\subsubsection{Sgr A* GRMHD}
\begin{figure*}
\begin{adjustwidth}{-\extralength}{0cm}
\setlength{\lineskip}{0pt}
\centering
\includegraphics[width=17mm]{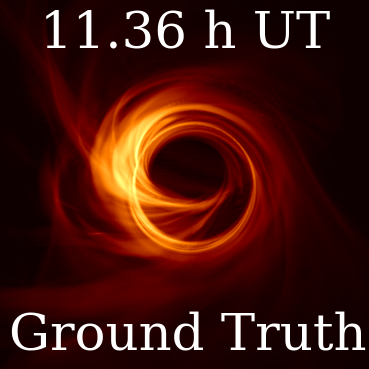}%
\includegraphics[width=17mm]{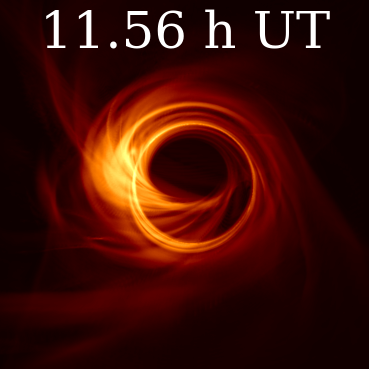}%
\includegraphics[width=17mm]{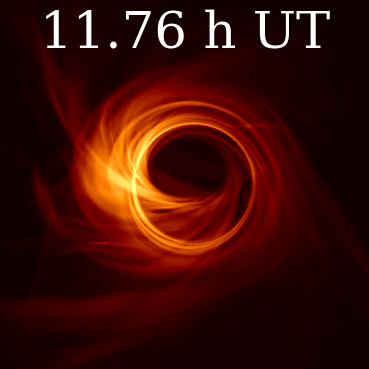}%
\includegraphics[width=17mm]{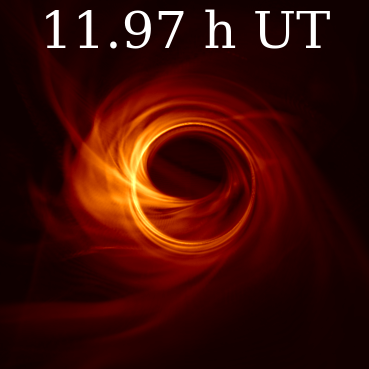}%
\includegraphics[width=17mm]{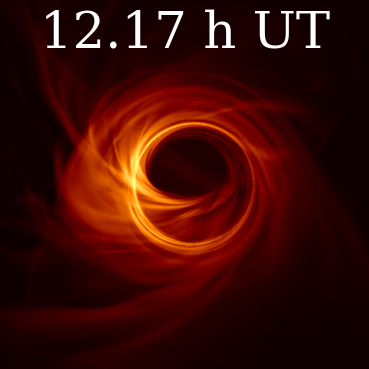}%
\includegraphics[width=17mm]{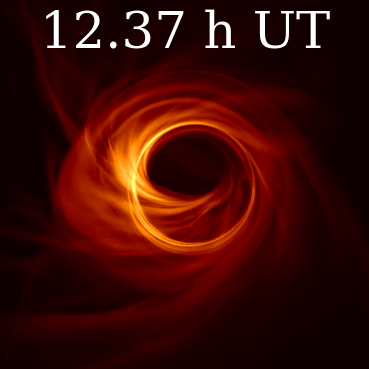}%
\includegraphics[width=17mm]{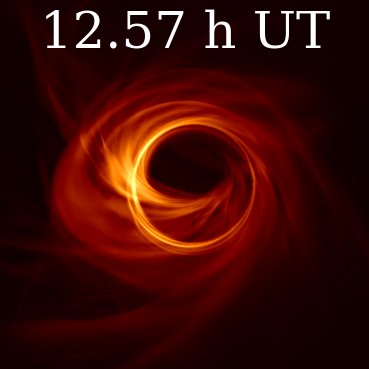}%
\includegraphics[width=17mm]{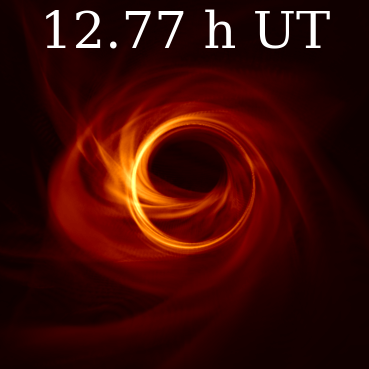}% 
\includegraphics[width=17mm]{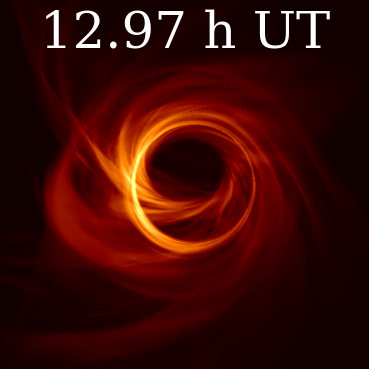}% 
\includegraphics[width=17mm]{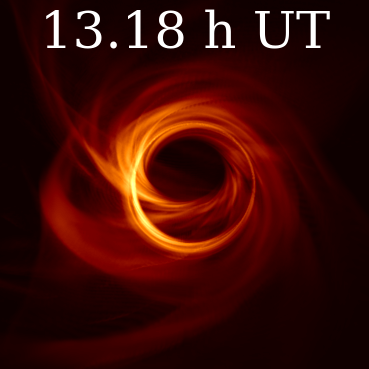}% 
\includegraphics[width=17mm]{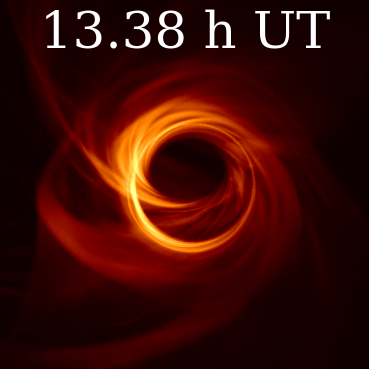}%  \\
\vspace{2mm}

\includegraphics[width=17mm]{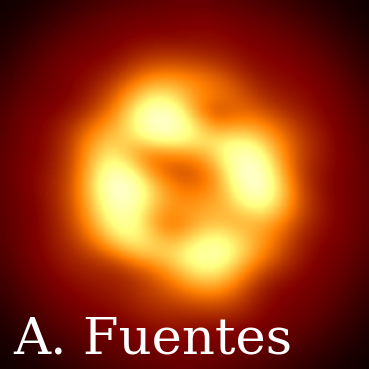}%
\includegraphics[width=17mm]{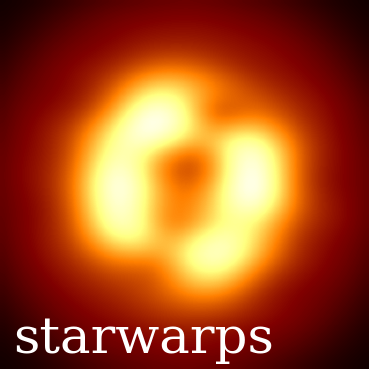}%
\includegraphics[width=17mm]{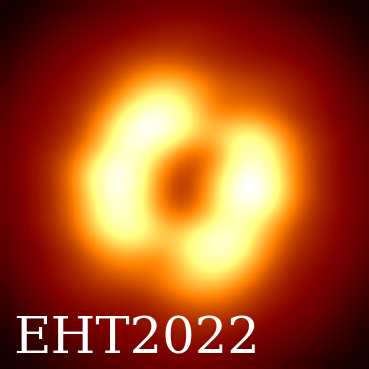}%
\includegraphics[width=17mm]{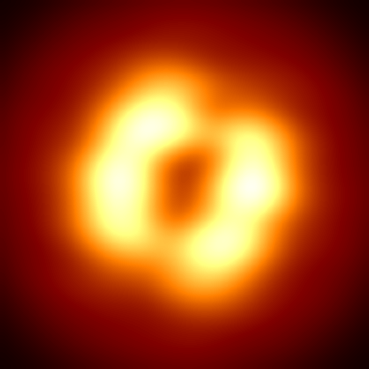}%
\includegraphics[width=17mm]{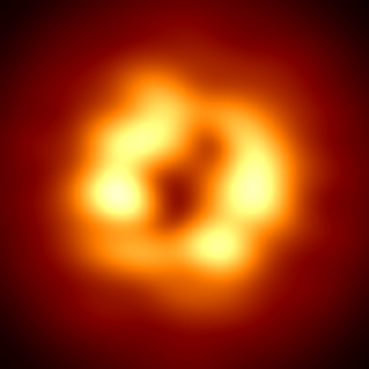}%
\includegraphics[width=17mm]{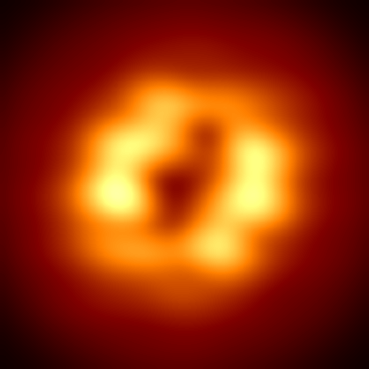}%
\includegraphics[width=17mm]{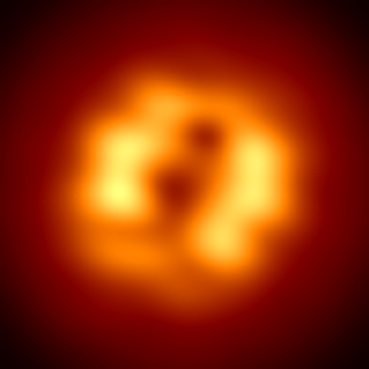}%
\includegraphics[width=17mm]{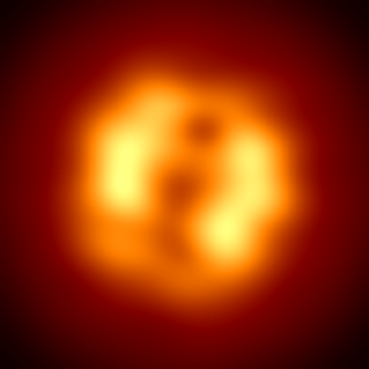}% 
\includegraphics[width=17mm]{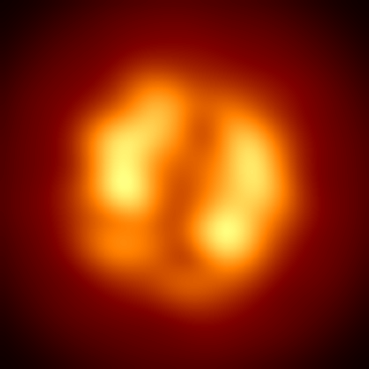}% 
\includegraphics[width=17mm]{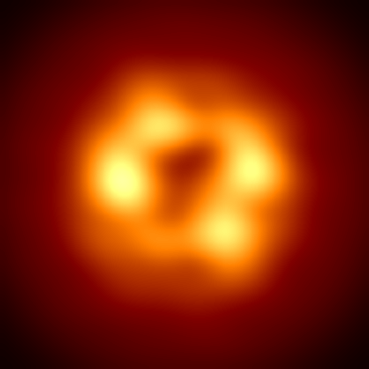}% 
\includegraphics[width=17mm]{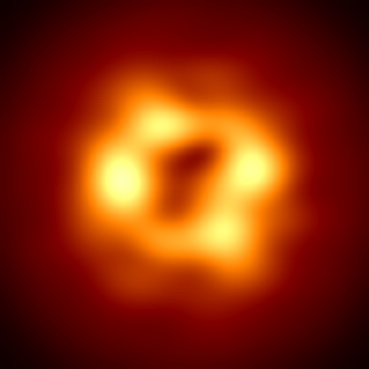}%  \\

\includegraphics[width=17mm]{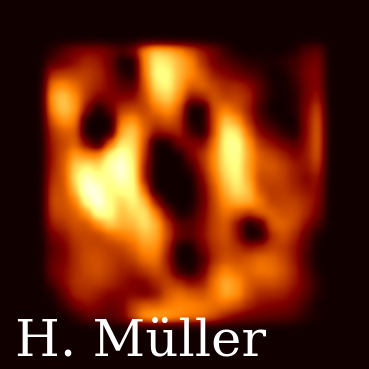}%
\includegraphics[width=17mm]{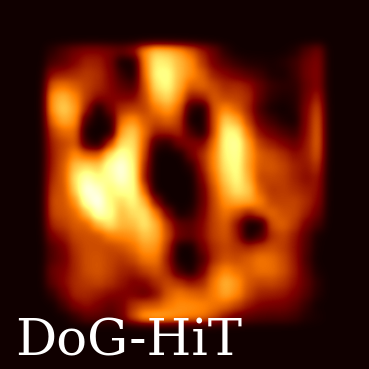}%
\includegraphics[width=17mm]{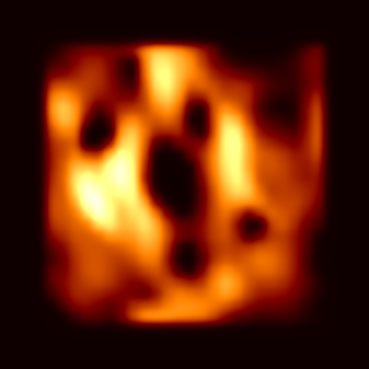}%
\includegraphics[width=17mm]{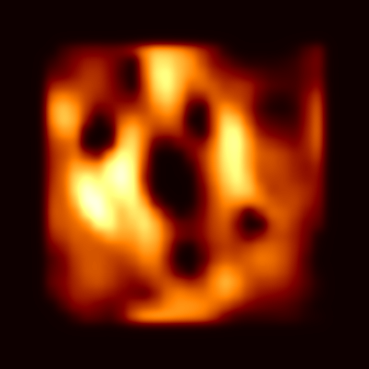}%
\includegraphics[width=17mm]{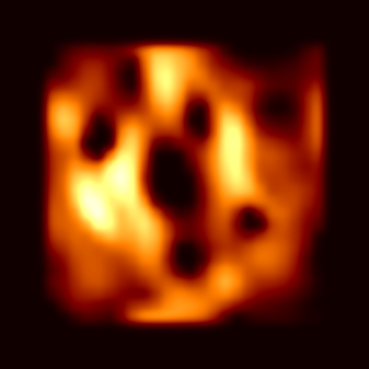}%
\includegraphics[width=17mm]{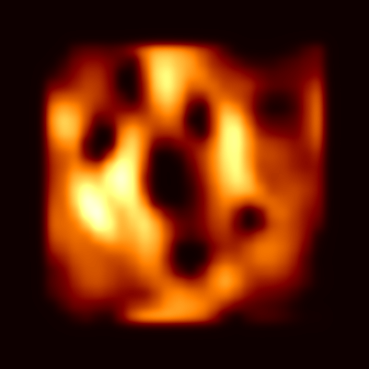}%
\includegraphics[width=17mm]{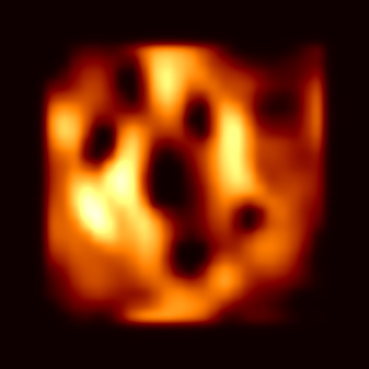}%
\includegraphics[width=17mm]{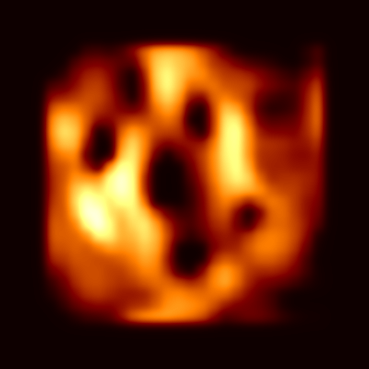}% 
\includegraphics[width=17mm]{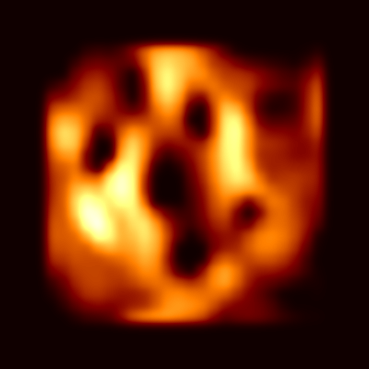}% 
\includegraphics[width=17mm]{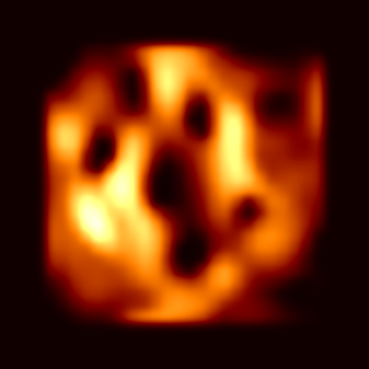}% 
\includegraphics[width=17mm]{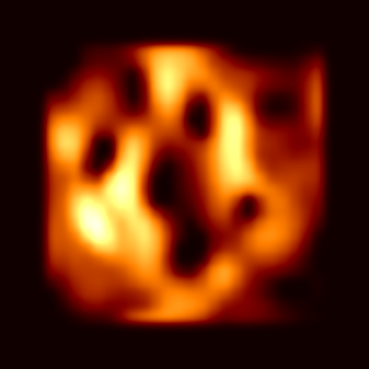}% \\
\vspace{2mm}

\includegraphics[width=17mm]{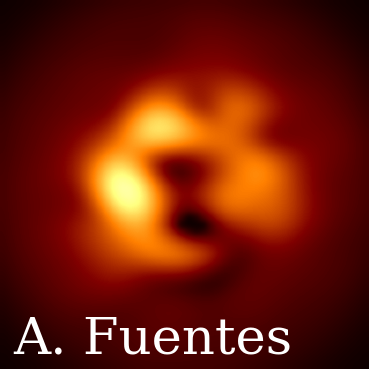}%
\includegraphics[width=17mm]{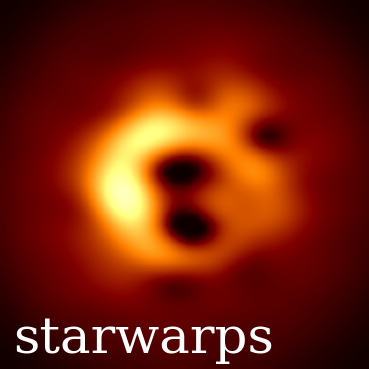}%
\includegraphics[width=17mm]{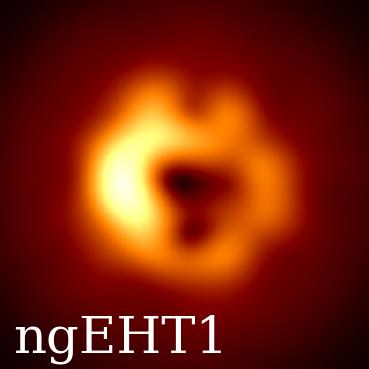}%
\includegraphics[width=17mm]{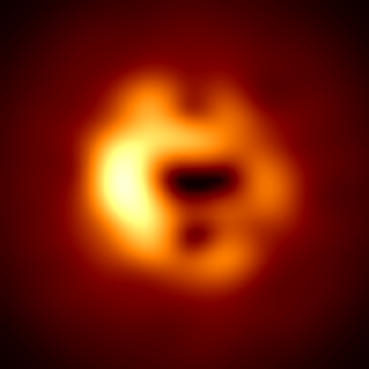}%
\includegraphics[width=17mm]{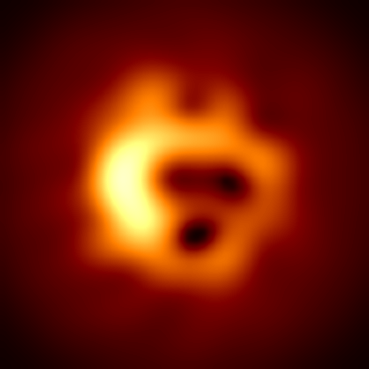}%
\includegraphics[width=17mm]{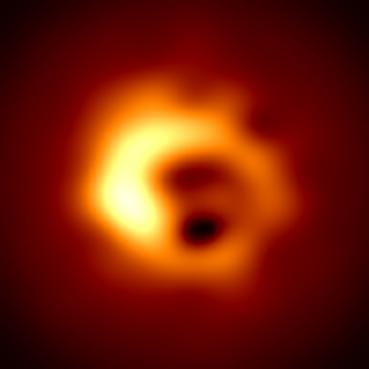}%
\includegraphics[width=17mm]{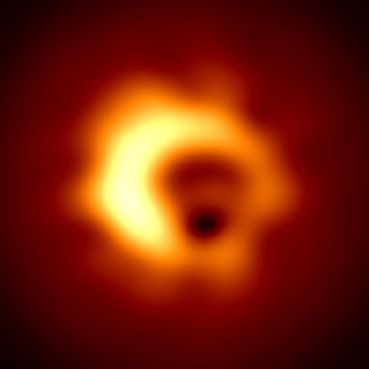}%
\includegraphics[width=17mm]{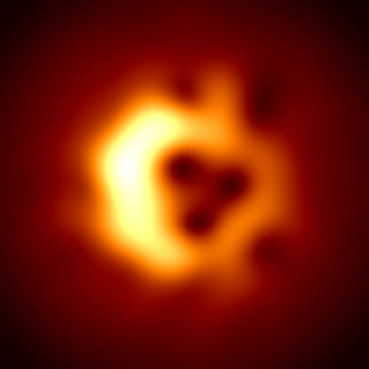}% 
\includegraphics[width=17mm]{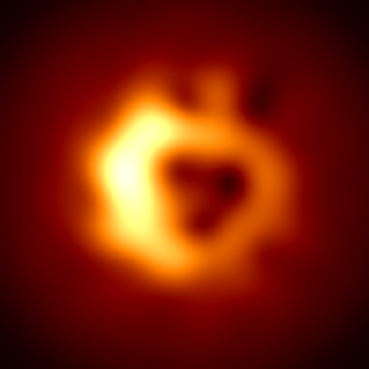}% 
\includegraphics[width=17mm]{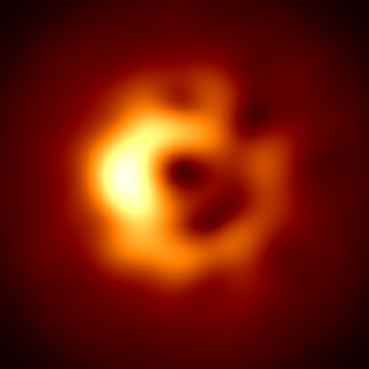}% 
\includegraphics[width=17mm]{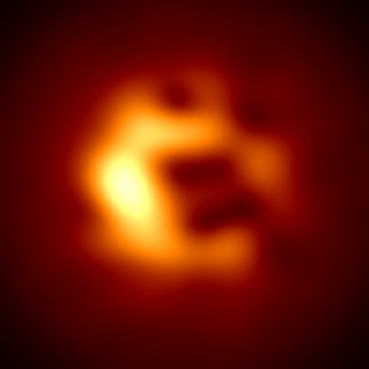}% \\

\includegraphics[width=17mm]{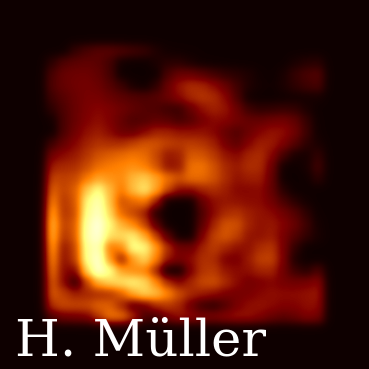}%
\includegraphics[width=17mm]{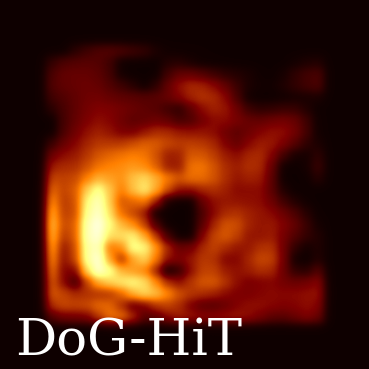}%
\includegraphics[width=17mm]{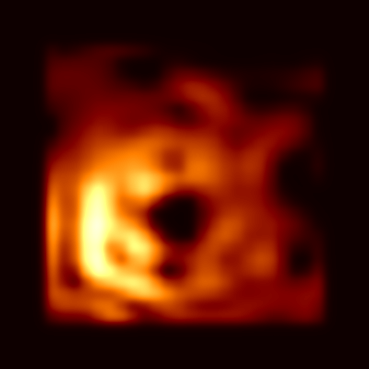}%
\includegraphics[width=17mm]{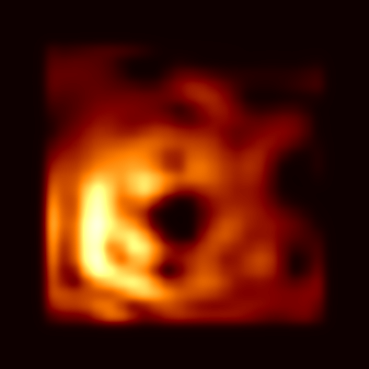}%
\includegraphics[width=17mm]{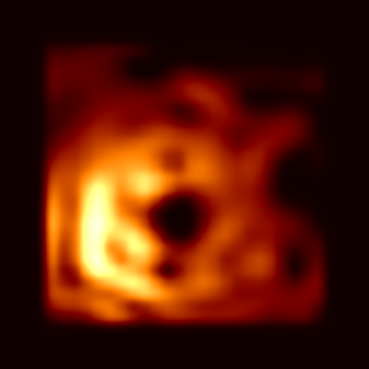}%
\includegraphics[width=17mm]{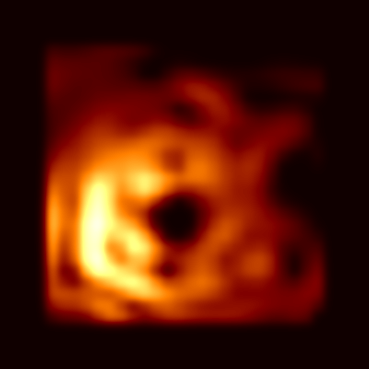}%
\includegraphics[width=17mm]{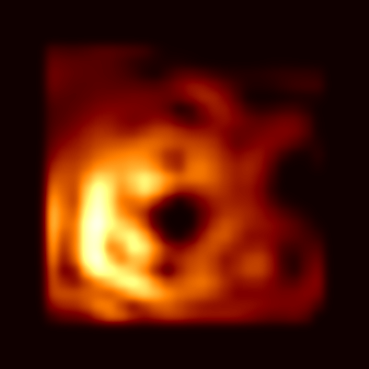}%
\includegraphics[width=17mm]{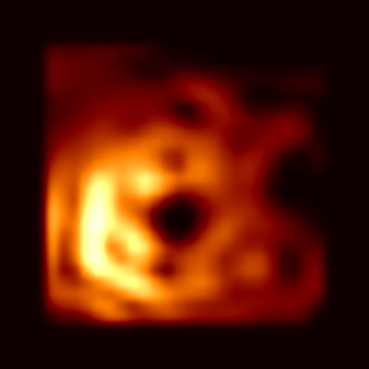}% 
\includegraphics[width=17mm]{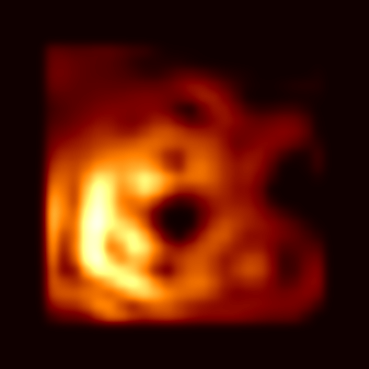}% 
\includegraphics[width=17mm]{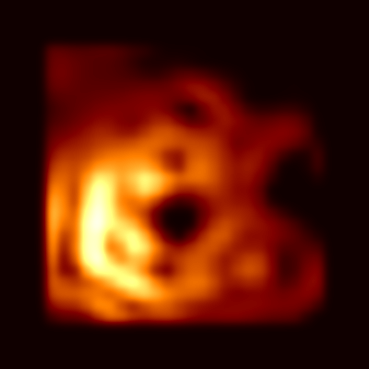}% 
\includegraphics[width=17mm]{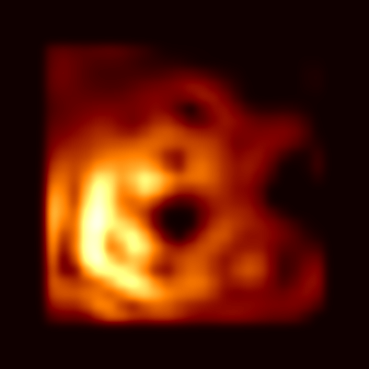}% \\

\includegraphics[width=17mm]{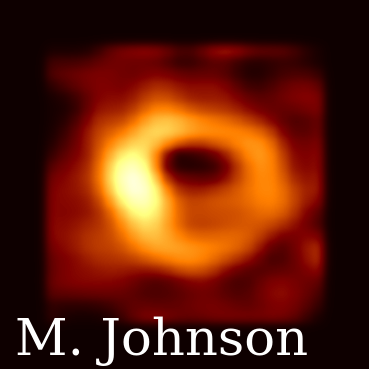}%
\includegraphics[width=17mm]{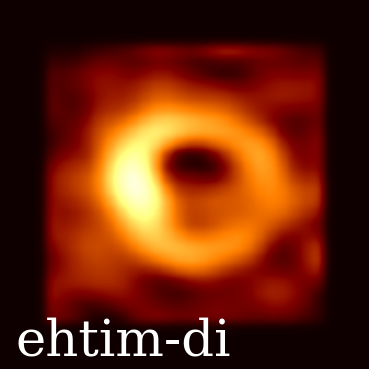}%
\includegraphics[width=17mm]{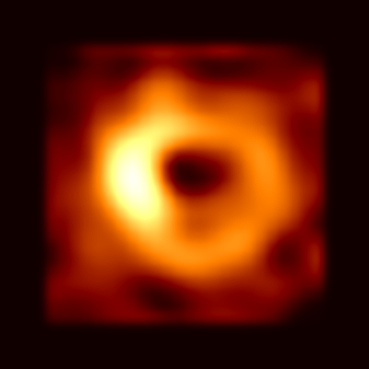}%
\includegraphics[width=17mm]{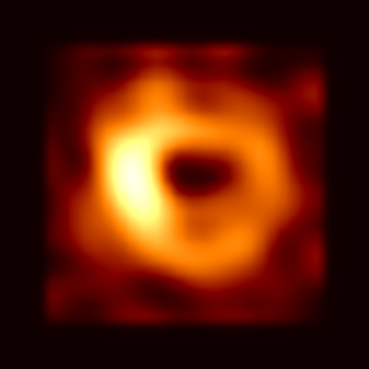}%
\includegraphics[width=17mm]{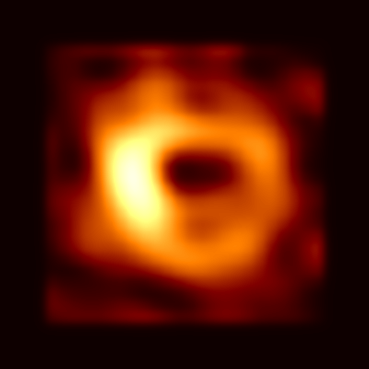}%
\includegraphics[width=17mm]{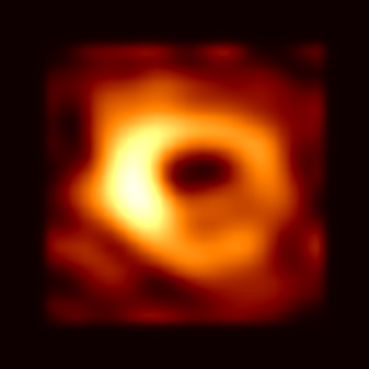}%
\includegraphics[width=17mm]{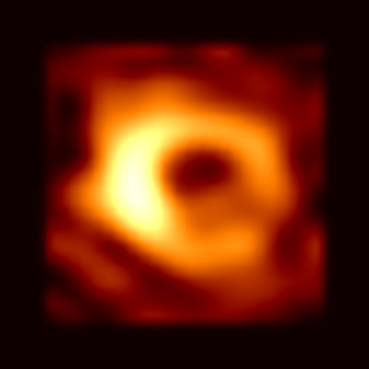}%
\includegraphics[width=17mm]{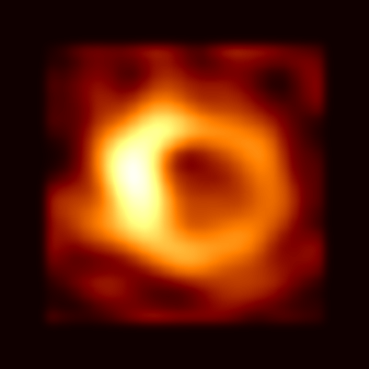}% 
\includegraphics[width=17mm]{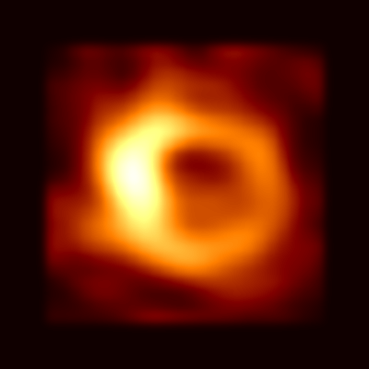}% 
\includegraphics[width=17mm]{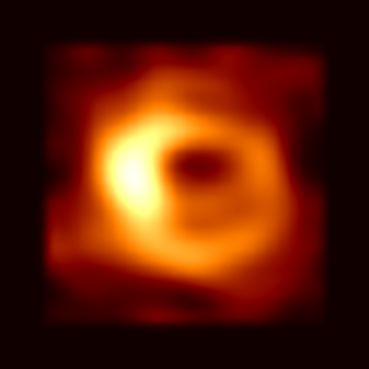}% 
\includegraphics[width=17mm]{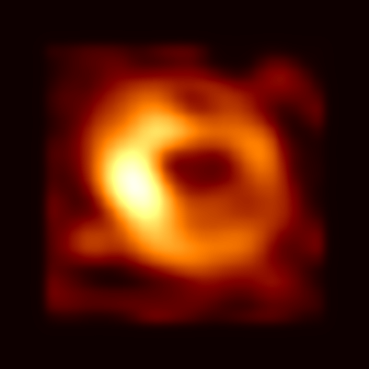}% \\

\includegraphics[width=17mm]{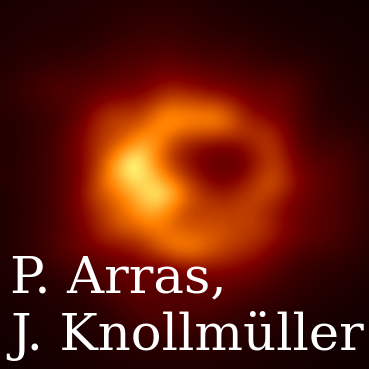}%
\includegraphics[width=17mm]{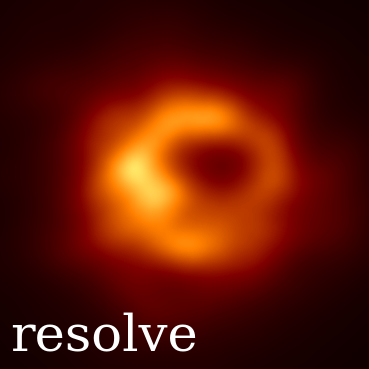}%
\includegraphics[width=17mm]{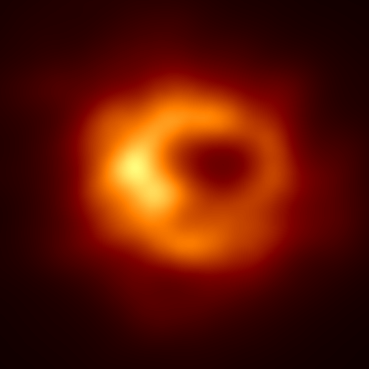}%
\includegraphics[width=17mm]{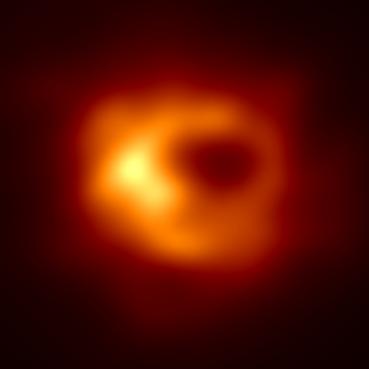}%
\includegraphics[width=17mm]{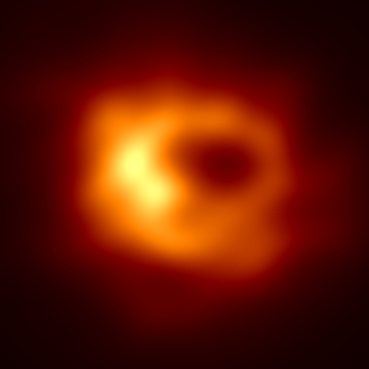}%
\includegraphics[width=17mm]{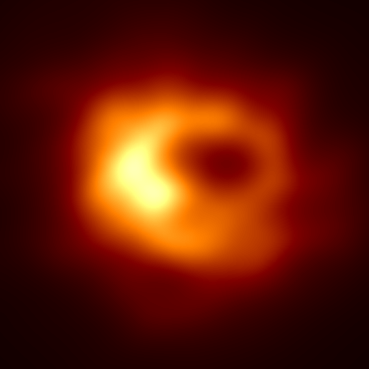}%
\includegraphics[width=17mm]{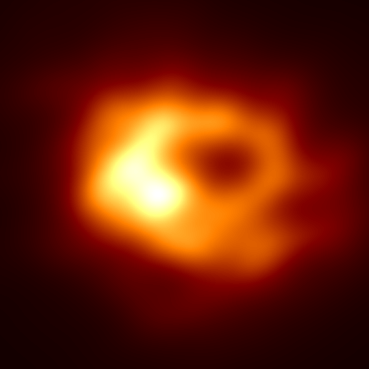}%
\includegraphics[width=17mm]{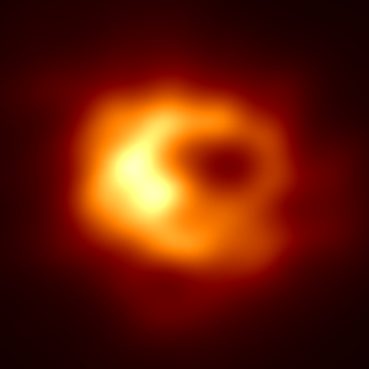}% 
\includegraphics[width=17mm]{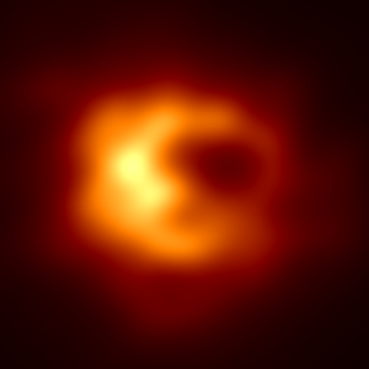}% 
\includegraphics[width=17mm]{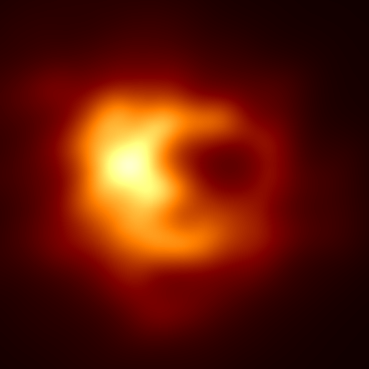}% 
\includegraphics[width=17mm]{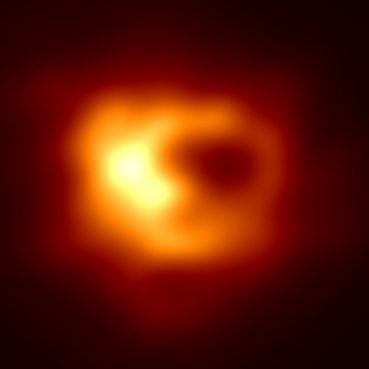}% \\

\includegraphics[width=17mm]{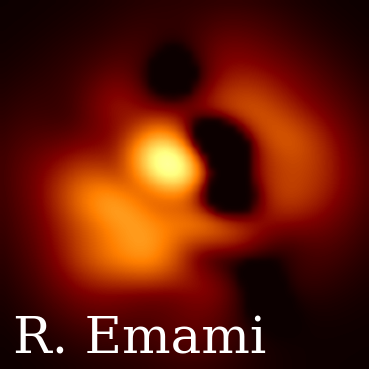}%
\includegraphics[width=17mm]{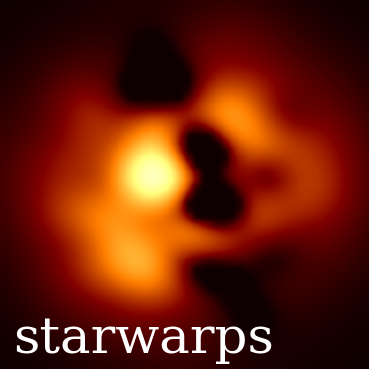}%
\includegraphics[width=17mm]{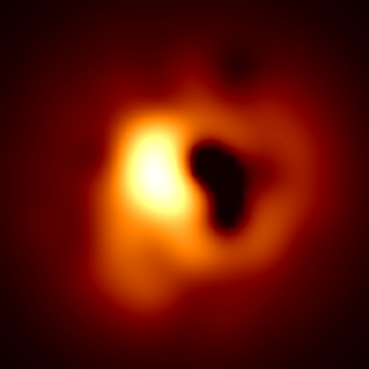}%
\includegraphics[width=17mm]{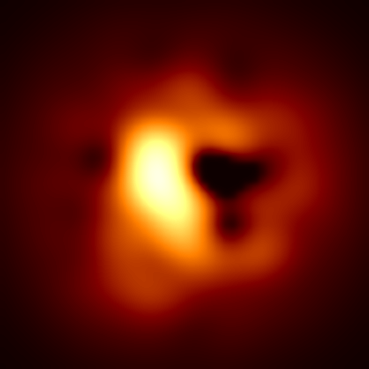}%
\includegraphics[width=17mm]{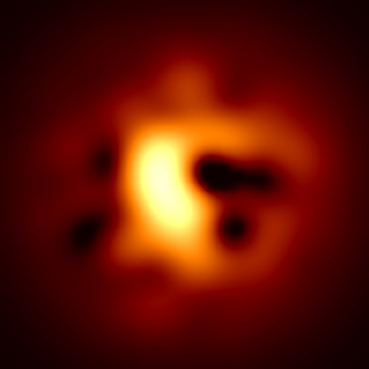}%
\includegraphics[width=17mm]{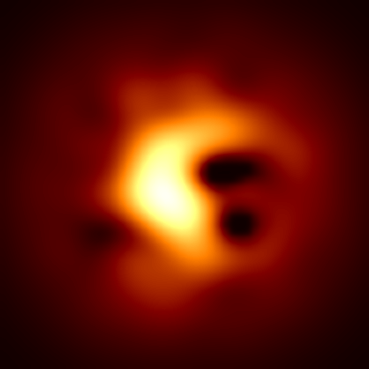}%
\includegraphics[width=17mm]{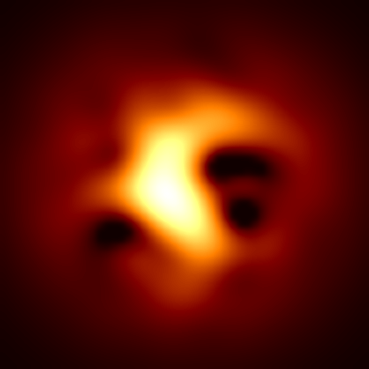}%
\includegraphics[width=17mm]{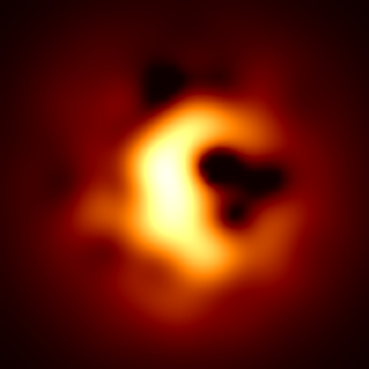}% 
\includegraphics[width=17mm]{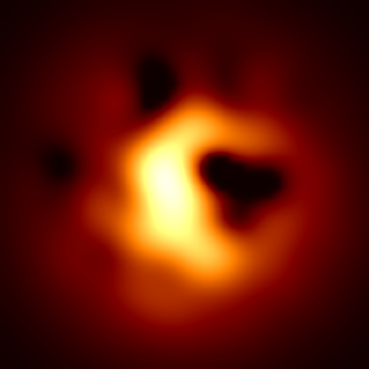}% 
\includegraphics[width=17mm]{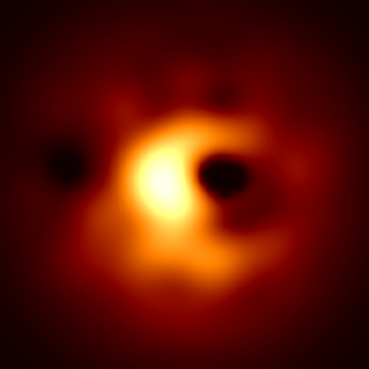}% 
\includegraphics[width=17mm]{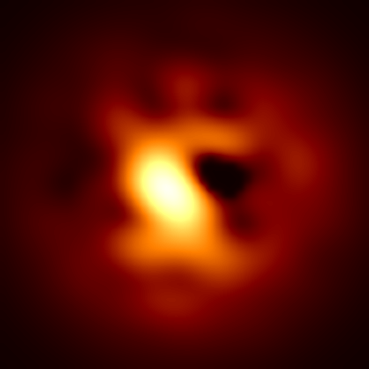}% \\

\end{adjustwidth}
  \caption{Challenge 2 Sgr A* GRMHD 230 GHz submissions. Images are shown on a square root scale, which is normalized to the brightest pixel value across each submitted set of movie frames, on a field of view of 126 $\mu$as.}
     \label{fig:ch2_sgra_grmhd}
\end{figure*}

Finally, Figure \ref{fig:ch2_sgra_grmhd} shows eleven frames of all Sgr A* GRMHD submissions at 230 GHz, spanning the best times window (11.3 - 13.5 h UT). Like for the Sgr A* RIAF+hotspot model, the EHT2022 reconstructions are static. In the StarWarps reconstruction, the ring morphology is recovered, but the detailed emission along the ring is not. The ngEHT1 reconstructions are generally much sharper, and the azimuthal brightness variations are reconstructed accurately, with the {\tt ehtim-di} and StarWarps submissions showing the best quality metric values (Table \ref{tab:metrics_challenge2_M87}). Due to the relatively stable and turbulent nature of the variability in this model, the reconstruction of temporal variations is more difficult to assess than for the other source models.

\begin{table*}[h]
\caption{Reconstruction quality metrics for Challenge 2. Tabulated values are medians across the reconstructed frames except for the {\tt resolve-mf} reconstructions, which were only done for the first movie frame. For the Sgr A* models, the metrics were evaluated on a common UT range for all reconstructions (see text for details).}
\begin{adjustwidth}{-\extralength}{0cm}
\resizebox{\fulllength}{!}{\begin{tabular}{lllll|llllll}
Model&Array&$\nu$ (GHz)&Submitter&Method&$\chi^2_{\mathrm{cp}}$&$\chi^2_{\mathrm{lcamp}}$&$\rho_{\mathrm{NX}}$&$\rho_{\mathrm{NX,log}}$&$\theta_{\mathrm{eff}}$&$\mathcal{D}_{0.1}$ \\
\hline
M87 GRMHD&EHT2022&86&P. Arras, J. Knollmüller&{\tt resolve}&1.94&2.01&0.83&0.92&24.5&1156 \\
M87 GRMHD&EHT2022&86&P. Arras, J. Knollmüller&{\tt resolve-mf}&1.71&4.82&0.96&0.97&7.0&3970 \\
M87 GRMHD&EHT2022&86&N. Kosogorov&{\tt ehtim}&7.16&2.69&0.8&0.82&32.0&585 \\
M87 GRMHD&ngEHT1&86&P. Arras, J. Knollmüller&{\tt resolve}&1.45&1.4&0.85&0.96&21.2&3054 \\
M87 GRMHD&ngEHT1&86&P. Arras, J. Knollmüller&{\tt resolve-mf}&1.43&1.73&0.95&0.99&8.2&7248 \\
M87 GRMHD&ngEHT1&86&R. Emami&{\tt ehtim}&1.84&1.69&0.8&0.89&30.4&1315 \\
M87 GRMHD&ngEHT1&86&N. Kosogorov&{\tt ehtim}&2.06&1.58&0.8&0.93&30.4&919 \\
M87 GRMHD&ngEHT1&86&N. Kosogorov&{\tt CLEAN}&193.55&10266.39&0.75&0.74&46.8&749 \\
M87 GRMHD&EHT2022&230&P. Arras, J. Knollmüller&{\tt resolve}&2.03&3.25&0.92&0.96&7.3&3881 \\
M87 GRMHD&EHT2022&230&P. Arras, J. Knollmüller&{\tt resolve-mf}&2.03&6.67&0.93&0.97&7.1&6424 \\
M87 GRMHD&EHT2022&230&N. Kosogorov&{\tt ehtim}&4.16&3.15&0.88&0.54&12.6&429 \\
M87 GRMHD&ngEHT1&230&P. Arras, J. Knollmüller&{\tt resolve}&2.53&2.35&0.92&0.98&6.6&8742 \\
M87 GRMHD&ngEHT1&230&P. Arras, J. Knollmüller&{\tt resolve-mf}&2.57&3.12&0.93&0.99&7.1&12154 \\
M87 GRMHD&ngEHT1&230&J. Vega&{\tt ehtim}&2.55&2.3&0.91&0.97&8.1&4807 \\
M87 GRMHD&ngEHT1&230&R. Emami&{\tt ehtim}&2.55&2.47&0.89&0.83&10.9&2061 \\
M87 GRMHD&ngEHT1&230&N. Kosogorov&{\tt ehtim}&2.85&2.81&0.89&0.71&11.4&1060 \\
M87 GRMHD&ngEHT1&230&N. Kosogorov&{\tt CLEAN}&325.47&385.28&0.79&0.6&22.6&226 \\
M87 GRMHD&EHT2022&345&P. Arras, J. Knollmüller&{\tt resolve-mf}&5.26&5.79&0.93&0.97&7.2&6994 \\
M87 GRMHD&ngEHT1&345&P. Arras, J. Knollmüller&{\tt resolve-mf}&6.39&6.89&0.92&0.98&7.3&9732 \\
M87 GRMHD&ngEHT1&345&R. Emami&{\tt ehtim}&6.14&5.38&0.59&0.42&61.8&61 \\
M87 GRMHD&ngEHT1&345&N. Kosogorov&{\tt ehtim}&5.99&4.94&0.81&0.47&16.6&563 \\
M87 GRMHD&ngEHT1&345&N. Kosogorov&{\tt CLEAN}&12.41&16.28&0.83&0.67&14.4&1157 \\
Sgr A* RIAFSPOT&EHT2022&230&A. Fuentes&{\tt StarWarps}&1.85&1.78&0.83&-&37.3&- \\
Sgr A* RIAFSPOT&EHT2022&230&H. Müller&{\tt DoG-HiT}&5.61&5.12&0.77&-&56.8&- \\
Sgr A* RIAFSPOT&ngEHT1&230&M. Johnson&{\tt ehtim-di}&7.39&11.78&0.87&-&24.4&- \\
Sgr A* RIAFSPOT&ngEHT1&230&A. Fuentes&{\tt StarWarps}&4.24&3.05&0.89&-&23.0&- \\
Sgr A* RIAFSPOT&ngEHT1&230&R. Emami&{\tt StarWarps}&6.87&11.98&0.83&-&43.3&- \\
Sgr A* RIAFSPOT&ngEHT1&230&H. Müller&{\tt DoG-HiT}&33.31&38.91&0.84&-&33.0&- \\
Sgr A* RIAFSPOT&ngEHT1&345&A. Fuentes&{\tt StarWarps}&5.37&3.63&0.85&-&28.6&- \\
Sgr A* RIAFSPOT&ngEHT1&345&R. Emami&{\tt StarWarps}&5.7&3.86&0.74&-&56.8&- \\
Sgr A* GRMHD&EHT2022&230&A. Fuentes&{\tt StarWarps}&9.49&3.61&0.68&-&56.0&- \\
Sgr A* GRMHD&EHT2022&230&H. Müller&{\tt DoG-HiT}&153.81&32.15&0.68&-&57.4&- \\
Sgr A* GRMHD&ngEHT1&230&M. Johnson&{\tt ehtim-di}&3.99&7.14&0.87&-&18.4&- \\
Sgr A* GRMHD&ngEHT1&230&A. Fuentes&{\tt StarWarps}&3.97&7.47&0.85&-&21.1&- \\
Sgr A* GRMHD&ngEHT1&230&R. Emami&{\tt StarWarps}&4.0&6.91&0.87&-&17.5&- \\
Sgr A* GRMHD&ngEHT1&230&H. Müller&{\tt DoG-HiT}&13.88&8.18&0.8&-&29.0&- \\
Sgr A* GRMHD&ngEHT1&230&P. Arras, J. Knollmüller&{\tt resolve}&5.57&4.52&0.84&-&21.9&- \\
Sgr A* GRMHD&ngEHT1&345&R. Emami&{\tt StarWarps}&4.94&4.19&0.61&-&56.9&- \\
\hline
\end{tabular}}
\end{adjustwidth}
\label{tab:metrics_challenge2_M87}
\end{table*}

\section{Conclusions and outlook}
\label{sec:conclusions}
The first two ngEHT Analysis Challenges have provided a number of useful insights. First, current imaging algorithms are capable of reconstructing high-resolution and high-fidelity movies of M87* and Sgr A*, revealing the jet dynamics of M87* on mas scales down to the event horizon, and intra-hour dynamics of the Sgr A* accretion flow on event-horizon scales. This work provides high-quality reconstructions from synthetic data that includes fully realistic observation and calibration effects under mediocre weather conditions (median conditions for April), showing excellent prospects for ngEHT performance. 

The sources have been reconstructed with a breadth of traditional (EHT) imaging algorithms and newer algorithms that are under active development, such as {\tt resolve} and {\tt DoG-HiT}. Different algorithms showed different performance for different datasets. For example, the {\tt StarWarps} and {\tt eht-imaging} algorithms showed strong performance on dynamical reconstructions of Sgr A* \citep[see][for other examples]{Emami2022, LaBella2022}, while the {\tt resolve} algorithm did remarkably well on recovering the extended jet structure of M87. Multi-frequency reconstructions gave the best M87 jet reconstruction results for both challenges, providing an opportunity to produce high-resolution 86 GHz images showing the central brightness depression related to the black hole (inner) shadow \citep{Chael2021, Bronzwaer2021}, and faithful reconstructions of the extended jet at 345 GHz. Both these features are difficult to reconstruct using the individual frequencies alone. {\tt DoG-HiT} reconstructions of the Sgr A* RIAF+hotspot model resolved the hotspot orbit and shearing, albeit with a lower quality than with other methods. Since {\tt DoG-HiT} is the most recently developed algorithm used in the challenges and completely automatic without special manual adaption to the data sets, these results are promising and can inform further development. {\tt SMILI} and {\tt CLEAN} have been applied to Challenge 1 data only, where they did not perform as well as {\tt eht-imaging} in reconstructing the extended M87 jet.

The reconstructions from any algorithm do not necessarily show its maximum potential performance. Between algorithms, there are differences in the freedom that the user has to steer the reconstruction process. {\tt CLEAN} traditionally requires significant user input and steering (e.g., defining {\tt CLEAN} windows), but has been adapted to a more automated approach for EHT analysis \citep{PaperIV}. RML methods such as {\tt eht-imaging} and {\tt SMILI} require setting regularizers and weights, but also allow some input on the reconstruction procedure by setting, e.g., convergence criteria and blurring steps between image rounds. On the other hand {\tt DoG-HiT} depends on just one free parameter. The outcome of Bayesian methods generally depends on the set priors. From the results of these challenges, each method's dependence on user input is difficult to assess and would require dedicated parameter surveys. Based on, e.g., the {\tt eht-imaging} submissions, the results can certainly depend strongly on the user. However, for submissions reconstructed with the same method but with strongly different resulting images, the $\chi^2$ are a good indicator of the reconstruction quality. For the lower-quality image reconstructions, either the used parameters or the script setup often did not allow a good fit to the data.

Regarding data generation, one lesson learned is that the used schedule of 10-minute scans and 10-minute gaps makes reconstructing the rapid variability of Sgr A* challenging, and in fact it is remarkable that dynamical imaging algorithms were able to reconstruct the 1-hour period and rapidly shearing Sgr A* hotspot orbit with just three scans and a duty cycle of 50\%. A denser schedule with shorter gaps could potentially improve these reconstructions significantly, and also help reconstructing the rapid variability from GRMHD simulations. Furthermore, the Challenge 2 345 GHz data has proven difficult to image, which is likely attributable to the severe atmospheric effects considering the weather parameters were medians for April at all sites. Since in reality 345 GHz observations would likely only be scheduled on days with excellent weather at suitable sites, a next challenge should be done with more optimistic 345 GHz weather conditions. The experience from these challenges has shown that both {\tt eht-imaging} and {\tt SYMBA} are viable and well-performing pathways for generating synthetic ngEHT data. User-friendly tools to generate synthetic ngEHT data from a centralized repository of instrument and weather parameters using both pathways are under development \citep[][]{Doeleman2022}.

ngEHT Analysis Challenge 3\footnote{\url{https://challenge.ngeht.org/challenge3/}} is an extension of Challenge 2 to full Stokes, and will show how well various algorithms can reconstruct dynamics in polarization. Challenge 4 will focus on more specific ngEHT science goals such as measuring the photon ring size and black hole spin, involving modeling methods as well \citep[e.g.][]{Tiede2022}. The merit of simultaneous multi-frequency observations allowing for frequency phase transfer \citep[e.g.][]{Dodson2017, Rioja2020} will be explored in this challenge as well \citep[see also][]{Issaoun2022}. Future challenges could also involve varying the number and locations of stations. While the impact of a single station's presence or location diminishes as the array grows and becomes more robust against station losses, the effect of using partial instead of the full array or using different sets of new sites could be tested in end-to-end simulations \citep[see also][]{Doeleman2022}.

The ngEHT Analysis Challenges have brought together expertise in theoretical modeling, synthetic data generation, and image reconstruction, spurring development in all these areas. The continued challenges will involve polarization, model fitting, and science interpretation, to form a complete and end-to-end process of ngEHT simulations which helps maximizing the science potential of the array.

%%%%%%%%%%%%%%%%%%%%%%%%%%%%%%%%%%%%%%%%%%
\vspace{6pt} 

%%%%%%%%%%%%%%%%%%%%%%%%%%%%%%%%%%%%%%%%%%
%% optional
%\supplementary{The following are available online at \linksupplementary{s1}, Figure S1: title, Table S1: title, Video S1: title.}

% Only for the journal Methods and Protocols:
% If you wish to submit a video article, please do so with any other supplementary material.
% \supplementary{The following are available at \linksupplementary{s1}, Figure S1: title, Table S1: title, Video S1: title. A supporting video article is available at doi: link.} 

%%%%%%%%%%%%%%%%%%%%%%%%%%%%%%%%%%%%%%%%%%
\authorcontributions{Conceptualization, L.B., S.D., M.J., and F.R.; methodology, P.A., L.B., K.C., R.E., C.F., A.F., M.J., J.K., N.K., H.M., N.P., A.R., F.R., P.T., T.T., and J.V.; software, G.L., P.A., K.C., C.F., J.K., H.M., M.J., and F.R.; validation, L.B. and F.R.; formal analysis, P.A., L.B., K.C., R.E., C.F., A.F., M.J., J.K., N.K., H.M., N.P., A.R., F.R., P.T., T.T., and J.V.; data curation, L.B., G.L., and F.R.; writing---original draft preparation, P.A., R.E., M.J., H.M., F.R., and T.T.; writing---review and editing, P.A., L.B., K.C., S.D., R.E., C.F., A.F., M.J., J.K., N.K., G.L., H.M., N.P., A.R., F.R., P.T., T.T., and J.V.; visualization, F.R.; supervision, S.D.; funding acquisition, S.D. All authors have read and agreed to the published version of the manuscript.}

%For research articles with several authors, a short paragraph specifying their individual contributions must be provided. The following statements should be used ``Conceptualization, X.X. and Y.Y.; methodology, X.X.; software, X.X.; validation, X.X., Y.Y. and Z.Z.; formal analysis, X.X.; investigation, X.X.; resources, X.X.; data curation, X.X.; writing---original draft preparation, X.X.; writing---review and editing, X.X.; visualization, X.X.; supervision, X.X.; project administration, X.X.; funding acquisition, Y.Y. All authors have read and agreed to the published version of the manuscript.'', please turn to the  \href{http://img.mdpi.org/data/contributor-role-instruction.pdf}{CRediT taxonomy} for the term explanation. Authorship must be limited to those who have contributed substantially to the work~reported.}

%\funding{Please add: ``This research received no external funding'' or ``This research was funded by NAME OF FUNDER grant number XXX.'' and  and ``The APC was funded by XXX''. Check carefully that the details given are accurate and use the standard spelling of funding agency names at \url{https://search.crossref.org/funding}, any errors may affect your future funding.}

\funding{This research was supported by NSF grants AST-1935980 and AST-2034306. This work was supported by the Black Hole Initiative at Harvard University, made possible through the support of grants from the Gordon and Betty Moore Foundation and the John Templeton Foundation. The opinions expressed in this publication are those of the author(s) and do not necessarily reflect the views of the Moore or Templeton Foundations. Hendrik M\"{u}ller received financial support for this research from the International Max Planck Research School (IMPRS) for Astronomy and Astrophysics at the Universities of Bonn and Cologne. This research is supported by the DFG research grant ``Jet physics on horizon scales and beyond" (Grant No.  FR 4069/2-1), the ERC synergy grant ``BlackHoleCam: Imaging the Event Horizon of Black Holes" (Grant No. 610058) and ERC advanced grant ``JETSET: Launching, propagation and emission of relativistic jets from binary mergers and across mass scales" (Grant No. 884631). Jakob Knollm\"{u}ller acknowledges funding by the Deutsche Forschungsgemeinschaft (DFG, German Research Foundation) under Germany´s Excellence Strategy – EXC 2094 – 390783311. Razieh Emami acknowledges the support by the Institute for Theory and Computation at the Center for Astrophysics as well as grant numbers 21-atp21-0077, NSF AST-1816420 and HST-GO-16173.001-A for very generous supports. }

%\dataavailability{In this section, please provide details regarding where data supporting reported results can be found, including links to publicly archived datasets analyzed or generated during the study. Please refer to suggested Data Availability Statements in section ``MDPI Research Data Policies'' at \url{https://www.mdpi.com/ethics}. You might choose to exclude this statement if the study did not report any data.} 

%\acknowledgments{In this section you can acknowledge any support given which is not covered by the author contribution or funding sections. This may include administrative and technical support, or donations in kind (e.g., materials used for experiments).}

\conflictsofinterest{The authors declare no conflict of interest. The funders had no role in the design of the study; in the collection, analyses, or interpretation of data; in the writing of the manuscript, or in the decision to publish the~results.} 

%%%%%%%%%%%%%%%%%%%%%%%%%%%%%%%%%%%%%%%%%%
%% Only for journal Encyclopedia
%\entrylink{The Link to this entry published on the encyclopedia platform.}

%%%%%%%%%%%%%%%%%%%%%%%%%%%%%%%%%%%%%%%%%%
%% Optional
%\abbreviations{Abbreviations}{
%The following abbreviations are used in this manuscript:\\

%\noindent 
%\begin{tabular}{@{}ll}
%MDPI & Multidisciplinary Digital Publishing Institute\\
%DOAJ & Directory of open access journals\\
%TLA & Three letter acronym\\
%LD & Linear dichroism
%\end{tabular}}

%%%%%%%%%%%%%%%%%%%%%%%%%%%%%%%%%%%%%%%%%%
\begin{adjustwidth}{-\extralength}{0cm}
%\printendnotes[custom] % Un-comment to print a list of endnotes

\reftitle{References}

% Please provide either the correct journal abbreviation (e.g. according to the “List of Title Word Abbreviations” http://www.issn.org/services/online-services/access-to-the-ltwa/) or the full name of the journal.
% Citations and References in Supplementary files are permitted provided that they also appear in the reference list here. 

%=====================================
% References, variant A: external bibliography
%=====================================
\bibliography{main.bib}

\end{adjustwidth}
\end{document}